# THEORETICAL PLASMA PHYSICS

*Allan N. Kaufman*

*Physics Department, University of California*
*Berkeley, California*

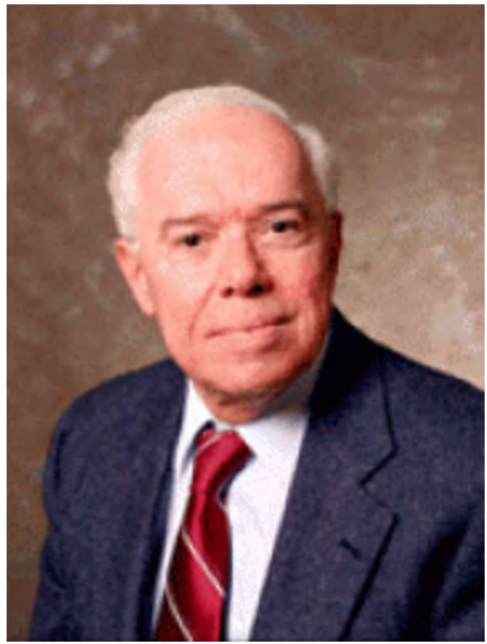

*Lecture Notes for Physics 242A, B, C and Physics 250*
*1971-1972*

*Transcribed, edited, and graphics added by Bruce I. Cohen*

# Table of Contents

# Foreword

Allan Kaufman (b. 1927) grew up in the Hyde Park neighborhood of Chicago not far from the University of Chicago.  Allan attended the University of Chicago for both his undergraduate and doctoral degrees in physics.  Chicago was replete with physics luminaries on its faculty and future luminaries among the doctoral students.  Allan's doctoral thesis advisor was Murph Goldberger who was relatively new to the faculty at Chicago and just five years older than Allan.  Allan did a theoretical thesis on a strong-coupling theory of meson-nucleon scattering. Allan published an autobiographical article entitled "A half-century in plasma physics" in A. N. Kaufman, Journal of Physics: Conference Series **169** (2009) 012002.

Allan worked at Lawrence Livermore Laboratory from June 1953 through 1963.  While at Livermore Laboratory he taught the one-year graduate course in electricity and magnetism in 1959-1963 at UC Berkeley.  In 1963 he first taught the first semester of the graduate course in Theoretical Plasma Physics 242A at Berkeley.  He taught the plasma theory course at UCLA in the 1964-1965 school year while on leave from Livermore before joining the faculty at UC Berkeley in the 1965 school year.  Allan frequently taught the graduate plasma theory course and the graduate statistical mechanics course until his retirement from teaching in 1998.

The lecture notes from Kaufman's graduate plasma theory course and a follow-on special topics course presented here were from the 1971-1972 academic year and the first quarter of the 1972-1973 academic year.  The notes follow the chronological order of the lectures as they were presented.  The equations and derivations are as Kaufman presented, but the text is a reconstruction of Kaufman's discussion and commentary.  The content of Kaufman's graduate plasma theory courses evolved over time motivated by new developments in plasma theory.  Thus, the material reported here does not represent the totality of Kaufman's lecture notes on plasma theory.  Some of the graphics have been downloaded from material posted as open access on the Internet, typically with attributions to the source.

I joined Kaufman's research group during the 1971-1972 academic year.  At that time Allan's group included doctoral students Dwight Nicholson, Michael Mostrom, Gary Smith, and myself.   Claire Max was a post-doctoral research physicist associated with the group for part of this period.  I graduated in August 1975.  Harry Mynick, John Cary, and Robert Littlejohn did their doctoral theses with Allan shortly thereafter.  One can see the influence of Allan Kaufman's formulation of plasma theory in the late Dwight Nicholson's fine textbook *Introduction to Plasma Theory* (John Wiley & Sons, 1983).

I am very grateful to Allan Kaufman for his encouragement, interest, and feedback as I prepared these lecture notes and to Alain Brizard for reviewing the manuscript and making suggestions, and corrections.  I also thank Gene Tracy, Robert Littlejohn, and Jonathan Wurtele for their interest and encouragement.

Bruce I. Cohen
December 2018
Orinda, California

# PART 1

## 1. Introduction to plasma dynamics

[*Editor's note: In the first lecture of Physics 242A Kaufman discussed the syllabus for Physics 242A,B, and C. Kaufman used CGS units throughout his notes. The textbook used as a general resource for the class at that time was P. C. Clemow and J. P. Dougherty, <u>The Electrodynamics of Particles and Plasmas</u>, Addison-Wesley (1969).*]

### 1.A Basic assumptions, definitions, and restrictions on scope

<u>Definition</u>: An ideal plasma is a charged gas wherein no bound states exist (a "mythical beast").

<u>Postulate</u>: We exclude the sufficiently dense plasma that requires quantum effects: $\hbar \to 0$ here.

<u>Postulate</u>: We further ignore special relativity: $\beta \equiv v/c \ll 1$

For purposes of an introductory study of plasma dynamics we initially assume no applied magnetic field **B**=0 and dispense with the generality of Maxwell's equations in favor of retaining only Coulomb interactions. We assume a gas of *N* charged particles. Then the force on particle *i* due to all the other particles is given by

$$m_i \dot{\mathbf{v}}_i = e_i \sum_{j(\neq i)}^{N} \hat{\mathbf{r}}_{ij} \frac{e_j}{r_{ij}^2} \tag{1.A.1}$$

and there are *N* such equations. We require an approximation method to solve this system of nonlinear equations. The charges and masses are parameters that have explicit dimensions. We also require initial conditions on particle positions and velocities, and need a statistical approach because *N* is large.

<u>Definition</u>: $\ell_0$ is the average distance between nearest neighbors; $n \approx 1/\ell_0^3$ is the number density of particles; and $\bar{v}$ is an average velocity. These define the state of the plasma, statistically.

### 1.B Definition of a plasma

Form the dimensionless quantity $e^2/m\ell_0 \bar{v}^2$. The classical electron radius is $r_e = e^2/mc^2$; so divide by another length $\ell_0$ to form a dimensionless quantity:

Definition: $\dfrac{e^2}{m\ell_0 \bar{v}^2} = \dfrac{e^2/\ell_0}{m\bar{v}^2} = \dfrac{\text{avg. nearest neighbor interaction energy}}{\text{avg. kinetic energy}} = \dfrac{1}{\Lambda^*}$

Thus, we are comparing the interaction energy to the kinetic energy in the plasma; and we treat the interaction energy as a perturbation. The plasma is said to be weakly coupled.

Postulate: In our plasmas $N$ and $\Lambda^* \gg 1$, equivalently $m\bar{v}^2 \sim k_B T \gg e^2/\ell_0$

We are **not** assuming that the total kinetic energy $Nk_B T \gg N^2 e^2/\ell_0$. [*Editor's note: In what follows, units are employed for the temperature T such that $k_B \equiv 1$.*] There are some plasmas in which $\Lambda^* \leq O(1)$, for instance in a metal where the interaction and the Fermi energies are comparable; and a quantum mechanical treatment is then necessary. In ionic crystals $\Lambda^* \ll 1$ is possible.

Exercise: i) Find the region in temperature $T$ and density $n$ parameter space such that $\Lambda^* \gg 1$. ii) Impose the additional constraints v/c<<1 and $n\lambda_{\text{deBroglie}}^3 = n(h/mv)^3 \ll 1$.

Definition: The collision frequency is $v \sim n\sigma v$ where $\sigma \sim (e^2/mv^2)^2$, and the plasma frequency is $\omega_{\text{pe}} \sim (4\pi n e^2/m)^{1/2}$. Then $v/\omega_{\text{pe}} \sim (\Lambda^*)^{-3/2} \ll 1$, i.e., the relative collisionality of the plasma is weak.

Definition: The Debye length $\lambda_D \equiv \bar{v}/\omega_{pe} = (T/4\pi n e^2)^{1/2}$ is the characteristic shielding length, i.e., the effective interaction distance. The shielded potential from a test particle is $V \sim (e/r)\exp(-r/\lambda_D)$, and the number of particles in a region around a test particle of order the Debye length in dimension is then $\Lambda \sim n\lambda_D^3$. We must require that $\Lambda \gg 1$ for the validity of a statistical approach.

Theorem: $\Lambda \sim (\Lambda^*)^{3/2}$ so that the conditions of weak collisionality and weak interaction energy are closely related. We will use $\Lambda \gg 1$ exclusively and call it the plasma parameter. [Note: Sometimes the plasma parameter is defined as $\Lambda \equiv 4\pi n \lambda_D^3$.]

We note it is a very good assumption for most plasmas to assume that the Debye length $\lambda_D$ is small compared to the plasma macroscopic dimension $L$, so that $N \sim nL^3 \gg \Lambda \sim n\lambda_D^3$.

## 2. Vlasov-Poisson equation formulation for a collisionless plasma

## 2.A Equations of motion in phase space, Poisson equation, and definition of distribution function

Consider the group collective electric field **E** and the equations of motion

$$m_i \dot{\mathbf{v}}_i = e_i \mathbf{E}^i = e_i \sum_{j(i \neq i)} \hat{\mathbf{r}}_{ij} \frac{e_j}{r_{ij}^2} \tag{2.A.1}$$

We coarse-grain average the point charges to smear and smooth the collective electric field:

$$m_i \dot{\mathbf{v}}_i = e_i \mathbf{E}^i(\mathbf{r}_i) \rightarrow e_i \bar{\mathbf{E}}(\mathbf{r}_i) \tag{2.A.2}$$

The six-dimensional phase-space equations of motion are then

$$\left. \begin{array}{l} m_s \dot{\mathbf{v}}_s = e_s \bar{\mathbf{E}}(\mathbf{r}) \\ \dot{\mathbf{r}} = \mathbf{v} \end{array} \right\} \quad \frac{d}{dt}(\mathbf{r}, \mathbf{v}) = \left( \mathbf{v}, \frac{e_s}{m_s} \bar{\mathbf{E}}(\mathbf{r}) \right) \tag{2.A.3}$$

This phase space is not the same as the Gibbs phase space in statistical mechanics.

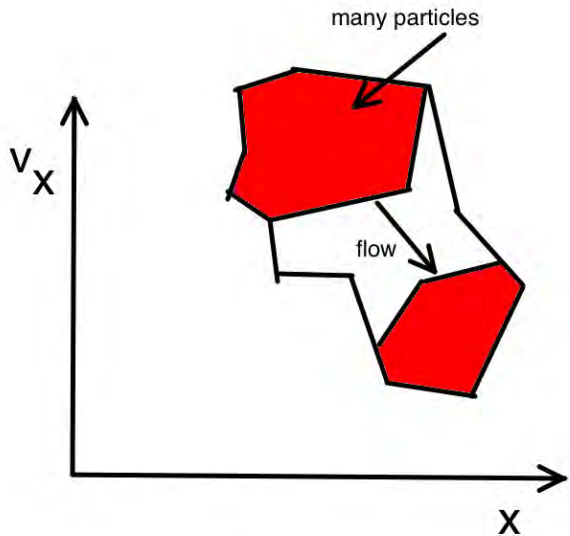

Fig. 2.A.1  Flow in phase space (cartoon)

<u>Theorem</u>: Poisson's equation is

$$\nabla \cdot \mathbf{E} = 4\pi \rho(\mathbf{r}) \quad \nabla \times \mathbf{E}(\mathbf{r}) = 0 \quad \text{where } \rho(\mathbf{r}) = \sum_i e_i \delta(\mathbf{r} - \mathbf{r}_i) \tag{2.A.4}$$

with the electrostatic constraint on **E** and the charge density $\rho(\mathbf{r})$ needs to be smoothed.

<u>Definition</u>: $f_s(\mathbf{r},\mathbf{v})$ is the mean density of particles of a species $s$ in six-dimensional phase space; then

$$\bar{\rho}(\mathbf{r}) \equiv \sum_s e_s \int d^3v f_s(\mathbf{r},\mathbf{v}) \tag{2.A.5}$$

The smoothed version of Eq.(2.A.4) for $\bar{\mathbf{E}}$ becomes

$$\nabla \cdot \bar{\mathbf{E}} = 4\pi \sum_s e_s \int d^3v f_s(\mathbf{r},\mathbf{v}) \qquad \nabla \times \bar{\mathbf{E}} = 0 \tag{2.A.6}$$

$f_s(\mathbf{r},\mathbf{v})$ evolves in time: what is the equation of evolution for $f_s(\mathbf{r},\mathbf{v};t)$ in time? Introduce $\mathbf{x}=(\mathbf{r},\mathbf{v})$ and $\dfrac{d}{dt}\mathbf{x} \equiv \mathbf{X}(\mathbf{x},t)$; then $f_s(\mathbf{r},\mathbf{v};t) \equiv f_s(\mathbf{x};t)$.

<u>Theorem</u>: The number of particles $N_V$ for any species in a volume V is

$$N_V(t) = \int_V d^6\mathbf{x} f(\mathbf{x};t) \tag{2.A.7}$$

Because the number of particles in the volume is conserved, except for net fluxes into or out of the surfaces bounding the volume, it follows that

$$\frac{dN_V}{dt} = \int_V d^6\mathbf{x} \frac{\partial f(\mathbf{x};t)}{\partial t} = -\oint_{surfaces} d\hat{\sigma} \cdot \mathbf{X} f = -\int_V d^6\mathbf{x} \nabla \cdot (\mathbf{X} f) \tag{2.A.8}$$

where $\hat{\sigma}$ points out of the volume and the divergence theorem has been used. Given that the volume integrals in Eq.(2.A.8) are equal for whatever subdomain of phase-space is enclosed in V, the integrands must be equal; and we arrive at the Vlasov equation.

<u>Theorem</u>: Vlasov Equation

$$\begin{aligned}\frac{\partial f_s(\mathbf{x};t)}{\partial t} &= -\nabla \cdot (\mathbf{X} f_s) = -\frac{\partial}{\partial \mathbf{r}}(\mathbf{v} f_s) - \frac{\partial}{\partial \mathbf{v}}\left(\frac{e_s}{m_s}\bar{\mathbf{E}}(\mathbf{r},t) f_s\right) \\ &= -\mathbf{v} \cdot \frac{\partial}{\partial \mathbf{r}} f_s - \frac{e_s}{m_s}\bar{\mathbf{E}}(\mathbf{r},t) \cdot \frac{\partial}{\partial \mathbf{v}} f_s\end{aligned} \tag{2.A.9}$$

which can be rewritten as

$$\frac{\partial f_s(\mathbf{x};t)}{\partial t} + \mathbf{v} \cdot \frac{\partial}{\partial \mathbf{r}} f_s + \frac{e_s}{m_s}\bar{\mathbf{E}}(\mathbf{r},t) \cdot \frac{\partial}{\partial \mathbf{v}} f_s = 0 \tag{2.A.10}$$

In the presence of volumetric sources and sinks, e.g., ionization and recombination, and/or collisions, the right side of Eq.(2.A.10) is no longer zero.

## 2.B Continuity equation in phase space – Liouville theorem

A number of observations can be made immediately on inspecting the derivation of the Vlasov equation. From Eqs.(2.A.7), (2.A.8), and $d\mathbf{x}/dt \equiv \mathbf{X}(\mathbf{x},t)$ we have

$$\frac{\partial f(\mathbf{x};t)}{\partial t} = -\frac{\partial}{\partial \mathbf{x}} \cdot (\mathbf{X}f) = -\mathbf{X} \cdot \frac{\partial f}{\partial \mathbf{x}} - f \frac{\partial}{\partial \mathbf{x}} \cdot \mathbf{X}$$
(2.B.1)

and hence,

$$\left(\frac{\partial}{\partial t} + \dot{\mathbf{x}} \cdot \frac{\partial}{\partial \mathbf{x}}\right) f(\mathbf{x};t) = -f(\mathbf{x};t) \frac{\partial}{\partial \mathbf{x}} \cdot \mathbf{X} = -f(\mathbf{x};t) \nabla \cdot \mathbf{X}$$
(2.B.2)

Eq.(2.B.2) is a phase-space continuity equation. The left side of this equation is just a convective derivative, and the right side allows for compressibility. If $\nabla \cdot \mathbf{X} < 0$ then $Df/Dt > 0$, and $Df/Dt < 0$ if $\nabla \cdot \mathbf{X} > 0$. We note that as an almost trivial consequence of the independent phase space variables:

$$\frac{\partial}{\partial \mathbf{r}} \cdot \mathbf{v} = 0 \quad \frac{\partial}{\partial \mathbf{v}} \cdot \left(\frac{e}{m}\left(\mathbf{E} + \frac{\mathbf{v}}{c} \times \mathbf{B}\right)\right) = 0 \quad \Rightarrow \nabla \cdot \mathbf{X} = \frac{\partial}{\partial \mathbf{r}} \cdot \mathbf{v} + \frac{\partial}{\partial \mathbf{v}} \cdot \dot{\mathbf{v}} = 0$$
(2.B.3)

<u>Theorem</u>: (Liouville theorem) If $\nabla \cdot \mathbf{X} = 0$, then the right side of Eq.(2.B.2) is zero and (2.8.2) corresponds exactly to the Liouville theorem for Hamiltionian systems:

$$\left(\frac{\partial}{\partial t} + \dot{\mathbf{x}} \cdot \frac{\partial}{\partial \mathbf{x}}\right) f(\mathbf{x};t) = 0$$
(2.B.4)

In this limit the phase-space flow is "incompressible" and $Df/Dt = 0$, i.e., $f$ is conserved along the phase-space trajectories. If the number of particles per unit volume $f$ is conserved then so is $1/f$, which is the differential volume element per unit particle, i.e., the phase-space volume element is also conserved (although its shape may deform).

## 2.C Nonlinear Vlasov equation with self-consistent fields

<u>Theorem</u>: Vlasov-Maxwell equations in a plasma with self-consistent fields

$$\left(\frac{\partial}{\partial t} + \mathbf{v} \cdot \frac{\partial}{\partial \mathbf{r}} + \frac{e_s}{m_s}\left(\mathbf{E} + \frac{\mathbf{v}}{c} \times \mathbf{B}\right) \cdot \frac{\partial}{\partial \mathbf{v}}\right) f_s(\mathbf{r},\mathbf{v};t) = 0$$
(2.C.1)

$$\nabla \cdot \mathbf{E} = 4\pi \sum_s e_s \int d^3\mathbf{v} f_s(\mathbf{r},\mathbf{v};t)$$
(2.C.2)

$$\nabla \times \mathbf{B} - \frac{1}{c}\frac{\partial \mathbf{E}}{\partial t} = \frac{4\pi}{c}\sum_s e_s \int d^3\mathbf{v}\, \mathbf{v} f_s(\mathbf{r},\mathbf{v};t) + \frac{4\pi}{c}\mathbf{j}_{ext}$$
(2.C.3)

$$\nabla \times \mathbf{E} + \frac{1}{c}\frac{\partial \mathbf{B}}{\partial t} = 0$$
(2.C.4)

$$\nabla \cdot \mathbf{B} = 0$$
(2.C.5)

and we can include the gravitational Poisson equation:

$$\nabla^2 \phi_g = 4\pi G \rho_m = 4\pi G \sum_s m_s \int d^3\mathbf{v}\, f_s$$
(2.C.6)

where the gravitational field $\mathbf{g}=-\nabla\phi_g$, $\phi_g$ is the gravitational potential, $\rho_m$ is the mass density, and $G$ is the universal gravitational constant.

In a Hamiltonian system one can introduce the notation

$$\mathbf{x} = (q_i, p_i) \qquad \mathbf{X} = \left(\frac{\partial H}{\partial p_i}, -\frac{\partial H}{\partial q_i}\right)$$
(2.C.7)

where H is the particle Hamiltonian and the i index represents a phase-space degree of freedom. The Vlasov equation then can be written as

$$\frac{\partial f}{\partial t} + \sum_i \left(\frac{\partial f}{\partial q_i}\frac{\partial H}{\partial p_i} - \frac{\partial f}{\partial p_i}\frac{\partial H}{\partial q_i}\right) = \frac{\partial f}{\partial t} + \{f, H\} = 0$$
(2.C.8)

where {f,H} denotes the Poisson bracket.

The Vlasov-Poisson equations can be written as

$$\mathbf{E} = -\nabla\phi \qquad \nabla^2\phi = -4\pi\rho_c \qquad \phi(\mathbf{r},t) = \int d^3\mathbf{r}'\frac{\rho_c(\mathbf{r}',t)}{|\mathbf{r}-\mathbf{r}'|}$$

$$\frac{\partial f_s}{\partial t} + \mathbf{v}\cdot\frac{\partial f_s}{\partial \mathbf{r}} + \frac{e_s}{m_s}\frac{\partial f_s}{\partial \mathbf{v}}\cdot\left(-\frac{\partial}{\partial \mathbf{r}}\right)\int d^3\mathbf{r}'\frac{4\pi\sum_{s'}e_{s'}\int d^3\mathbf{v}'f_{s'}(\mathbf{r}',\mathbf{v}',t)}{|\mathbf{r}-\mathbf{r}'|} = 0$$
(2.C.9)

We next consider the qualitative properties of the nonlinear self-consistent Vlasov equation in Eq.(2.C.9). The relative orders of the three terms are $f/\tau$: $vf/\lambda$ : $f^2e^2\lambda v^2/m$. i) Balancing the first two terms in the Vlasov equations yields $\tau \sim \lambda/v$ or $\omega/k \sim v$, i.e., $v_p \sim v$, where $v_p = \omega/k$, $\omega$ is a characteristic frequency and $k$ is a wavenumber. ii) Balancing the second and third terms yields $\omega \sim$ $4\pi(e^2/m)fv^3\lambda/v \sim (4\pi ne^2/m)\lambda/v$ which using $\omega\lambda \sim v$ leads to $\omega^2 \sim 4\pi ne^2/m$, which is the plasma frequency squared. iii) With $\lambda \sim v/\omega$ and setting $v \sim v_{th}=(T/m)^{1/2}$ the electron thermal velocity, then $\lambda \sim \lambda_D \sim (T/4\pi ne^2)^{1/2}$ where $\lambda_D$ is the Debye length. We will see that many plasma phenomena can be characterized in terms of important dimensionless variables, for example, $\omega/\omega_{pe}$, $\lambda/\lambda_D$, $m_e/m_i$, $T_e/T_i$, $\omega/kv$, $L/\lambda_D$, $\omega_{ps}/\omega_{cs}$, $\Delta\omega/\omega$, $v_{th}/c$, and the ratios of $E^2$ to $B^2$ and to $nmv^2$. There are also plasma attributes and phenomena associated with nonuniformity and anisotropy.

## 2.D Moment equations

### 2.D.a Conservation of mass density, momentum density, energy density

Rewrite the collisionless Vlasov equation in the alternative form

$$\frac{\partial}{\partial t} f_s + \frac{\partial}{\partial \mathbf{r}} \cdot (\mathbf{v} f_s) + \frac{\partial}{\partial \mathbf{v}} \cdot (\mathbf{a} f_s) = 0 \tag{2.D.1}$$

where the acceleration **a** is

$$\mathbf{a} = \frac{e_s}{m_s}\left(\mathbf{E} + \frac{1}{c}\mathbf{v} \times \mathbf{B}\right) \tag{2.D.2}$$

and define the number density $n_s$

$$n_s(\mathbf{r},t) = \int d^3\mathbf{v}\, f_s(\mathbf{r},\mathbf{v};t) \tag{2.D.3}$$

Defintion: Moments of the velocitiy distribution are constructed from

$$\langle A \rangle_s (\mathbf{r};t) \equiv \frac{\int d^3\mathbf{v}\, A f_s(\mathbf{r},\mathbf{v};t)}{\int d^3\mathbf{v}\, f_s(\mathbf{r},\mathbf{v};t)} \tag{2.D.4}$$

Examples:
1. $A=1$ → Identity operation
2. $A=\mathbf{v}$ → $\mathbf{u} \equiv \langle\mathbf{v}\rangle$ the average velocity
3. $A=(\mathbf{v}-\mathbf{u})(\mathbf{v}-\mathbf{u})$ → $nm\langle(\mathbf{v}-\mathbf{u})(\mathbf{v}-\mathbf{u})\rangle \equiv P(\mathbf{r},t)$ the pressure tensor
4. $A=e$ then $\sum_s e_s n_s \equiv \rho(\mathbf{r};t)$ the charge density using (2.D.3)
5. $A=\tfrac{1}{2}mv^2$ → $K(\mathbf{r};t) \equiv \int d^3\mathbf{v}\, \tfrac{1}{2}mv^2 f = n\langle\tfrac{1}{2}mv^2\rangle$ the kinetic energy density

Theorem: A generalized moment equation can be derived directly from the Vlasov equation (the species index $s$ is understood):

$$\frac{\partial}{\partial t}\left[n\langle A\rangle\right] = \int d^3\mathbf{v}\left(\frac{\partial f}{\partial t}A + \frac{\partial A}{\partial t}f\right) = n\left\langle\frac{\partial A}{\partial t}\right\rangle + \int d^3\mathbf{v}\, A\left[-\frac{\partial}{\partial \mathbf{r}}\cdot(\mathbf{v}f) - \frac{\partial}{\partial \mathbf{v}}\cdot(\mathbf{a}f)\right]$$

$$= n\left\langle\frac{\partial A}{\partial t}\right\rangle - \nabla\cdot\left[n\langle A\mathbf{v}\rangle\right] + n\left\langle\mathbf{a}\cdot\frac{\partial A}{\partial \mathbf{v}}\right\rangle \tag{2.D.5}$$

Examples:
1. $A=1 \to$ continuity equation

$$\frac{\partial n(\mathbf{r};t)}{\partial t} = -\nabla \cdot (n\mathbf{u}) \tag{2.D.6}$$

or

$$\frac{d}{dt} n(\mathbf{r};t) = \left(\frac{\partial}{\partial t} + \mathbf{u}\cdot\nabla\right) n = -n\nabla\cdot\mathbf{u} \tag{2.D.7}$$

2. $A=m\mathbf{v} \to$ momentum conservation

$$\frac{\partial}{\partial t}\langle nm\mathbf{v}\rangle = \frac{\partial}{\partial t}nm\mathbf{u} = -\nabla\cdot(nm\langle\mathbf{vv}\rangle) + ne\left(\mathbf{E}+\frac{1}{c}\mathbf{u}\times\mathbf{B}\right) \tag{2.D.8}$$

We can use the identity <**vv**>=<(**v**-**u**)(**v**-**u**)>+**uu**, the continuity equation Eq.(2.D.7), and the definition of the pressure tensor *P* in conjunction with Eq.(2.D.8) to derive the fluid momentum balance equation:

$$nm\frac{d}{dt}\mathbf{u} = -\nabla\cdot\mathbf{P} + ne\left(\mathbf{E}+\frac{1}{c}\mathbf{u}\times\mathbf{B}\right) \tag{2.D.9}$$

We can consider the Coulomb case, assume there is no magnetic field, sum over species, and integrate Eq.(2.D.8) over all space to demonstrate the total particle momentum is a constant:

$$\int d^3\mathbf{r}\left[\frac{\partial}{\partial t}\sum_s\langle n_s m_s \mathbf{v}\rangle\right] = \int d^3\mathbf{r}\left[\sum_s(-\nabla\cdot\mathbf{P}_s + n_s e_s \mathbf{E})\right] = \int d^3\mathbf{r}\left[\sum_s(-\nabla\cdot\mathbf{P}_s) + \rho\mathbf{E}\right]$$

$$= \int d^3\mathbf{r}\left[\sum_s(-\nabla\cdot\mathbf{P}_s) + \frac{1}{4\pi}(\nabla\cdot\mathbf{E})\mathbf{E}\right] = \int d^3\mathbf{r}\left[\sum_s(-\nabla\cdot\mathbf{P}_s) + \nabla\cdot\left(\frac{\mathbf{EE}}{4\pi}-\frac{E^2\mathbf{I}}{8\pi}\right)\right] \tag{2.D.10}$$

$$= \int d^3\mathbf{r}\nabla\cdot\left[\sum_s\left(-\mathbf{P}_s - \frac{\mathbf{EE}}{4\pi}+\frac{E^2\mathbf{I}}{8\pi}\right)\right] = \oint d\mathbf{S}\cdot\left(-\mathbf{P}_s - \frac{\mathbf{EE}}{4\pi}+\frac{E^2\mathbf{I}}{8\pi}\right)$$

where *d***S** is directed out of volume on its surface and we have used the divergence theorem and Gauss' law, and assumed the fields and the velocity distributions vanish at infinity. We can include field stresses and field momentum using Maxwell's equations as in Sec. 6.7 of Jackson's *Classical Electrodynamics* textbook to demonstrate that the total particle and field momentum is conserved.

3. $A=\frac{1}{2}mv^2 \to$ energy conservation

$$\frac{\partial}{\partial t}K_s = -\nabla\cdot n_s\left\langle\frac{1}{2}m_s v^2\mathbf{v}\right\rangle + n_s\langle e_s\mathbf{v}\rangle\cdot\mathbf{E} = -\nabla\cdot\mathbf{S}_s^K + \mathbf{j}_s\cdot\mathbf{E} \tag{2.D.11}$$

We note that the magnetic field does no work and Eq.(2.D.11) can be summed over species to obtain the equation for energy conservation of all particle species. $K$ is an energy density. Equation(2.D.11) summed over species is extended to include the electromagnetic field energy density as follows:

$$\frac{\partial}{\partial t}\left(K+\frac{E^2+B^2}{8\pi}\right)=\frac{\partial}{\partial t}(K+\mathrm{E})=-\nabla\cdot\left(\mathbf{S}^K+\frac{c}{4\pi}\mathbf{E}\times\mathbf{B}\right)=-\nabla\cdot\left(\mathbf{S}^K+\mathbf{S}^{EM}\right) \quad (2.D.12)$$

where we recognize $\mathbf{S}^{EM}$ as the electromagnetic Poynting flux and note that $\mathbf{J}\cdot\mathbf{E}$ terms identically cancel. By integrating Eq.(2.D.12) over volume, using the divergence theorem, and assuming all quantities vanish at infinity, we can demonstrate that total energy is conserved.

4. $A=mr^2 \rightarrow$ moment of inertia

$$I(t)=\frac{1}{2}\int d^3\mathbf{r}\sum_s m_s r^2 n_s(\mathbf{r};t)$$

$$\frac{\partial}{\partial t}I=-\frac{1}{2}\int d^3\mathbf{r}\sum_s m_s r^2 \nabla\cdot(n_s\mathbf{u}_s)=\int d^3\mathbf{r}\sum_s m_s n_s \mathbf{u}_s\cdot\mathbf{r} \quad (2.D.13)$$

$I$ is a global scalar quantity that only has a time variation. Equation(2.D.13) is derived using the continuity equation (2.D.6), integrating by parts, and using $\mathbf{u}\cdot\nabla r^2 = 2\mathbf{u}\cdot\mathbf{r}$.

### 2.D.b Virial theorem

<u>Virial theorem</u>: We begin by deriving the electromagnetic momentum conservation law from Maxwell's equations:

$$-\frac{\partial}{\partial t}\left(\frac{\mathbf{E}\times\mathbf{B}}{4\pi c}\right)=\rho\mathbf{E}+\frac{1}{c}\mathbf{j}\times\mathbf{B}-\frac{1}{4\pi}\nabla\cdot\left(\mathbf{E}\mathbf{E}-\tfrac{1}{2}E^2\mathbf{I}+\mathbf{B}\mathbf{B}-\tfrac{1}{2}B^2\mathbf{I}\right)$$

$$=\rho\mathbf{E}+\frac{1}{c}\mathbf{j}\times\mathbf{B}-\frac{1}{4\pi}\nabla\cdot\mathbf{P}^{EM} \quad (2.D.14)$$

We then take another time derivative in Eq.(2.D.13), use the momentum conservation equation Eq. (2.D.8), and include the $\mathbf{j}\times\mathbf{B}/c$ force in (2.D.10). We next eliminate $\rho\mathbf{E}+\mathbf{j}\times\mathbf{B}/c$ using Eq.(2.D.14) to obtain

$$\ddot{I}=\int d^3\mathbf{r}\,\mathbf{r}\cdot\frac{\partial}{\partial t}\sum_s n_s m_s \mathbf{u}_s = \int d^3\mathbf{r}\,\mathbf{r}\cdot\left[-\nabla\cdot\left(\mathbf{P}^K+\mathbf{P}^{EM}\right)-\frac{\partial}{\partial t}\left(\frac{\mathbf{E}\times\mathbf{B}}{4\pi c}\right)\right]$$

We add the term involving the electromagnetic momentum to both sides of the equation to obtain

$$\ddot{I} + \frac{d}{dt}\int d^3\mathbf{r}\,\mathbf{r}\cdot\left(\frac{\mathbf{E}\times\mathbf{B}}{4\pi c}\right) = \int d^3\mathbf{r}\,\mathbf{I}:\left(\mathbf{P}^K + \mathbf{P}^{EM}\right) = \int d^3\mathbf{r}\,\mathbf{I}:\left(2K + E^{EM}\right) > 0 \qquad (2.D.15)$$

where an integration by parts has been performed and **I:** denotes the resulting double dot product of the identity tensor with the tensor(s) following it. In the absence of the radiation flux $\mathbf{S}^{EM}$, one concludes that $\ddot{I} > 0$ because the right side of (2.D.14) is positive. In this limit the moment of inertia can only increase: the system cannot contain itself. With a finite radiation flux, the system can collapse and radiate energy away. An important caveat that limits these conclusions is that we did *not* include gravitation, which would introduce a negative term on the right side.

Exercise: Verify Eq.(2.D.14) and the double dot product in Eq.(2.D.15).

## 2.E Linear analysis of the Vlasov equation for small-amplitude disturbances in a uniform plasma

We can obtain exact solutions of the linearized Vlasov equation for infinitesimal amplitude perturbations.

Definition: Static solutions correspond to $\partial f/\partial t = 0$. This situation applies to a very small, but important, class of solutions. The time-independent Vlasov equation is

$$\mathbf{v}\cdot\frac{\partial}{\partial \mathbf{r}}f + \frac{e}{m}\left(\mathbf{E} + \frac{1}{c}\mathbf{v}\times\mathbf{B}\right)\cdot\frac{\partial}{\partial \mathbf{v}}f = 0 \qquad (2.E.1)$$

Uniform solutions correspond to $\partial f/\partial \mathbf{r} = 0$.

In the absence of electric and magnetic fields there exists a solution for a spatially uniform $f(\mathbf{v})$ that can be an arbitrary function of velocity. For this simple case the solution of the time-dependent Vlasov equation is that $f$ is a constant along the phase-space trajectories and remains fixed at its initial, arbitrary function of velocity:

$$\frac{d}{dt}f(\mathbf{v}) = \frac{d\mathbf{v}}{dt}\cdot\frac{\partial}{\partial \mathbf{v}}f(\mathbf{v}) = 0 \qquad (2.E.2)$$

This almost trivial result has utility in that we now add a small-amplitude perturbation that can depend on time and space. For simplicity we restrict consideration to the case of a Coulomb model. The equation set for the linearized system is as follows:

$$f_s(\mathbf{r},\mathbf{v};t) = f_{0s}(\mathbf{v}) + \delta f_s(\mathbf{r},\mathbf{v}:t) \tag{2.E.3a}$$

$$\mathbf{E}(\mathbf{r};t) = 0 + \delta\mathbf{E}(\mathbf{r};t) \quad \nabla\times\mathbf{E}=0 \tag{2.E.3b}$$

$$\nabla\cdot\mathbf{E} = 4\pi\rho(\mathbf{r};t) = 4\pi\sum_s e_s \int d^3\mathbf{v} f_s \tag{2.E.3c}$$

$$\frac{\partial \delta f_s}{\partial t} + \mathbf{v}\cdot\frac{\partial \delta f_s}{\partial \mathbf{r}} + \frac{e_s}{m_s}\delta\mathbf{E}\cdot\frac{\partial}{\partial \mathbf{v}}\left(f_{0s} + \delta f_s\right) = 0 \rightarrow$$

$$\frac{\partial \delta f_s}{\partial t} + \mathbf{v}\cdot\frac{\partial \delta f_s}{\partial \mathbf{r}} = -\frac{e_s}{m_s}\delta\mathbf{E}\cdot\frac{\partial f_{0s}}{\partial \mathbf{v}} \tag{2.E.3d}$$

The term on the right side of (2.E.3d) that is nonlinear in the product of $\delta\mathbf{E}$ and $\delta f_s$, is dropped because of the linearization. One must be careful with the vector calculus in Eq.(2.E.3d) when using non-Cartesian coordinates, and canonical coordinates can prove useful. We introduce the perturbed electric potential such that $\delta\mathbf{E} = -\nabla\delta\phi$ and follow the prescription: 1) Solve for $\delta f_s$ in terms of $\delta\phi$ using Eq.(2.E.3d). 2) Construct the linearly perturbed charge density $\delta\rho$ from $\delta f_s$ using Eq.(2.E.3c). 3) Solve for $\delta\phi$ using Poisson's equation derived from Eq.(2.E.3b) using suitable boundary conditions.

### 2.E.a Causality, stationarity, and uniformity in the dielectric kernel

$\delta\rho$ depends linearly on $\delta\phi$. The most general linear relation can be represented as

$$\delta\rho(\mathbf{r};t) = \int d^3\mathbf{r}' \int dt'\, \chi(\mathbf{r},\mathbf{r}';t,t')\delta\phi(\mathbf{r}';t') \tag{2.E.4}$$

The representation allows $\chi$ to be a generalized function. In fact, it can have some unusual properties, viz., including being a derivative of a delta function. However, $\chi$ is subject to at least three important constraints:

1) Causality: A perturbation in $\delta\phi$ will cause a later perturbation in $\delta\rho$.
2) Stationarity: The effect of $\delta\phi$ on $\delta\rho$ can depend only on the time interval between cause and effect $(t-t')>0$, and cannot depend on absolute time. This is a consequence of the underlying unperturbed system being stationary, i.e., time independent.
3) Uniformity: $\chi$ can only depend spatially on $\mathbf{r}-\mathbf{r}'$ (isotropy would imply $|\mathbf{r}-\mathbf{r}'|$) because the underlying unperturbed system has no dependence on spatial coordinate.

<u>Theorem</u>: We introduce $\tau \equiv t-t'$ and $\mathbf{s} \equiv \mathbf{r}-\mathbf{r}'$, and express $\delta\rho$ in terms convolution integrals.

$$\delta\rho(\mathbf{r};t) = \int d^3\mathbf{s} \int_0^\infty d\tau\, \chi(\mathbf{s};\tau)\delta\phi(\mathbf{r}-\mathbf{s};t-\tau) \tag{2.E.5}$$

## 2.E.b Solution of the dielectric function via Fourier transform in time

We introduce the Fourier transform in time.

<u>Definition</u>: The Fourier transform of $g(t)$ is

$$g(\omega) = \int_{-\infty}^{\infty} dt\, g(t) \exp(i\omega t) \tag{2.E.6}$$

We need to impose initial conditions on the linear perturbations $\{\delta\phi, \delta f, \delta\rho^{ext}\}$ to calculate the Fourier transform: $g(t)=0$ for $t<0$ and $g(\omega) = \int_{0}^{\infty} dt\, g(t)\exp(i\omega t)$. $g(t)$ must die out with time in order that $g(\omega)$ converges. However, in general $g(t)$ does not die out and may even grow. If $g(t)$ does not die, then $g(\omega)$ can be made to converge if $\omega$ is complex, i.e., $\omega=\omega'+i\omega''$ with $\omega''>0$. $g(\omega)$ will converge even if $g(t)$ is growing exponentially. For exponential growth $\omega'' \geq 1/(\text{growth time})$. If $g(t)$ grows faster than exponentially, no convergence is possible. The integration contour for the Fourier transform is shown in Fig. 2.E.1

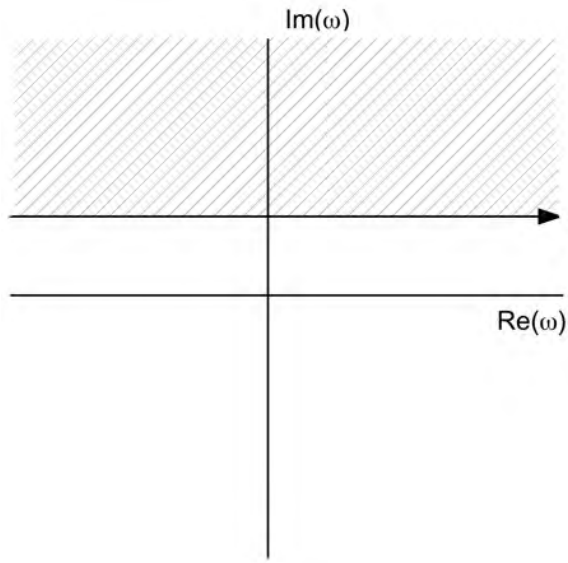

Fig. 2.E.1 Fourier transform integration contour in the complex $\omega$ plane.

We next calculate the Fourier transform of Eq.(2.E.5) and use the convolution theorem for Fourier transforms to obtain:

$$\delta\rho(\mathbf{r},\omega) = \int d^3s\, \chi(\mathbf{s},\omega)\delta\phi(\mathbf{r}-\mathbf{s},\omega) \Rightarrow \delta\rho(\mathbf{k},\omega) = \chi(\mathbf{k},\omega)\delta\phi(\mathbf{k},\omega) \tag{2.E.7}$$

where we have also Fourier transformed in space: $g(k) = \int d^3\mathbf{r}\, g(\mathbf{r})\exp(-i\mathbf{k}\cdot\mathbf{r})$

Theorem: Inverse Fourier transforms

$$g(\mathbf{r}) = \frac{1}{(2\pi)^3} \int d^3\mathbf{k}\, \exp(i\mathbf{k}\cdot\mathbf{r}) g(\mathbf{k}) \quad g(t) = \frac{1}{(2\pi)} \int d\omega\, \exp(-i\omega t) g(\omega) \quad (2.\text{E}.8)$$

The Fourier transformed Poisson equation is

$$-\nabla^2 \delta\phi(\mathbf{r},t) = 4\pi\left[\delta\rho(\mathbf{r},t) + \delta\rho^{ext}(\mathbf{r},t)\right] \Rightarrow k^2 \delta\phi(\mathbf{k},\omega) = 4\pi\left[\delta\rho(\mathbf{k},\omega) + \delta\rho^{ext}(\mathbf{k},\omega)\right] \quad (2.\text{E}.9)$$

and we then use (2.E.7) to obtain an equation for $\delta\phi(\omega,\mathbf{k})$:

$$k^2\left[1 - \frac{4\pi}{k^2}\chi(\mathbf{k},\omega)\right]\delta\phi(\mathbf{k},\omega) = 4\pi\delta\rho^{ext}(\mathbf{k},\omega) = k^2\delta\phi^{ext}(\mathbf{k},\omega) \quad (2.\text{E}.10)$$

Recall the conventional formulation of Gauss' law in a dielectric medium with free (external) charges present:

$$\nabla\cdot\mathbf{D} = 4\pi\rho^{ext} \quad \mathbf{D} = \mathbf{E} + 4\pi\mathbf{P} = \varepsilon\mathbf{E} \quad \Rightarrow \quad -\nabla\cdot(\varepsilon\nabla\phi) = 4\pi\rho^{ext} \quad (2.\text{E}.11)$$

In a **uniform** plasma $\varepsilon$ is spatially uniform and Eq.(2.E.11) yields $-\varepsilon\nabla^2\phi = 4\pi\rho^{ext}$ which combined with Eq.(2.E.10) leads to the following result.

Theorem: Poisson's equation and plasma dielectric response

$$\varepsilon k^2 \delta\phi(\mathbf{k},\omega) = 4\pi\rho^{ext}(\mathbf{k},\omega) \rightarrow \left[1 - \frac{4\pi}{k^2}\chi(\mathbf{k},\omega)\right] \equiv \varepsilon(\mathbf{k},\omega) \rightarrow \delta\phi(\mathbf{k},\omega) = \frac{\delta\phi^{ext}(\mathbf{k},\omega)}{\varepsilon(\mathbf{k},\omega)}$$

(2.E.12)

Corollary: $\quad \varepsilon(\mathbf{k},\omega) = \frac{k^2 \delta\phi^{ext}}{k^2 \delta\phi} = \frac{\delta\rho^{ext}}{\delta\rho^{tot}} = 1 - \frac{\delta\rho(\mathbf{k},\omega)}{\delta\rho^{tot}(\mathbf{k},\omega)}, \quad \delta\rho^{tot} = \delta\rho + \delta\rho^{ext} \quad (2.\text{E}.13)$

Fourier transform the linearized Vlasov equation:

$$\left(\frac{\partial}{\partial t} + \mathbf{v}\cdot\frac{\partial}{\partial \mathbf{r}}\right)\delta f(\mathbf{r},\mathbf{v};t) = \frac{e}{m}(\nabla\delta\phi)\cdot\frac{\partial f_0}{\partial \mathbf{v}} \rightarrow \left(-i\omega + i\mathbf{k}\cdot\mathbf{v}\right)\delta f(\mathbf{k},\omega;\mathbf{v}) = \frac{e}{m}\delta\phi(\mathbf{k},\omega) i\mathbf{k}\cdot\frac{\partial f_0}{\partial \mathbf{v}}$$

$$\delta f(\mathbf{k},\omega;\mathbf{v}) = \frac{\mathbf{k}\cdot\dfrac{\partial f_0}{\partial \mathbf{v}}}{(\mathbf{k}\cdot\mathbf{v} - \omega)} \frac{e}{m} \delta\phi(\mathbf{k},\omega)$$

(2.E.14)

Theorem: Linear susceptibility

$$\chi(\mathbf{k},\omega) = \frac{\delta\rho(\mathbf{k},\omega)}{\delta\phi(\mathbf{k},\omega)} = \sum_s \frac{e_s^2}{m_s} \int d^3\mathbf{v} \frac{\mathbf{k} \cdot \frac{\partial f_0^s(\mathbf{v})}{\partial \mathbf{v}}}{\mathbf{k} \cdot \mathbf{v} - \omega} \quad (2.E.15)$$

<u>Definitions</u>: $n_0^s = \int d^3\mathbf{v} f_0^s(\mathbf{v})$, $g^s(\mathbf{v}) \equiv f_0^s(\mathbf{v})/n_0^s$, $\int d^3\mathbf{v} g^s(\mathbf{v}) = 1$

<u>Theorem</u>: From Eqs.(2.E.12-15) we obtain the linear dielectric function

$$\varepsilon(\mathbf{k},\omega) = 1 - \frac{4\pi}{k^2}\chi(\mathbf{k},\omega) = 1 - \sum_s \frac{\omega_s^2}{k^2} \int d^3\mathbf{v} \frac{\mathbf{k} \cdot \frac{\partial g^s}{\partial \mathbf{v}}}{\mathbf{k} \cdot \mathbf{v} - \omega}, \quad \text{Im}\,\omega > 0 \quad (2.E.16)$$

where $\omega_s^2 \equiv \frac{4\pi n_0^s e_s^2}{m_s}$

From Eq.(2.E.12) one solves for $\delta\phi(\mathbf{k},\omega) = \varepsilon^{-1}(\mathbf{k},\omega)\delta\phi^{ext}(\mathbf{k},\omega)$ from which we obtain the following using the convolution theorem

$$\delta\phi(\mathbf{k},t) = \int_{-\infty}^{\infty} d\tau \varepsilon^{-1}(\mathbf{k},\tau)\delta\phi^{ext}(\mathbf{k},t-\tau) \text{ where } \varepsilon^{-1}(\mathbf{k},\tau) = \int d\omega \exp(-i\omega\tau)\varepsilon^{-1}(\mathbf{k},\omega) \quad (2.E.17)$$

**2.E.c Stable and unstable waves/disturbances**

Suppose $\delta\phi^{ext}(\mathbf{k},t) = a(\mathbf{k})\delta(t)$ for a pulse-type forcing function without specifying $a(\mathbf{k})$ yet. Hence, $\delta\phi(\mathbf{k},t) = a(\mathbf{k})\varepsilon^{-1}(\mathbf{k},t)$. Consider a specific representation for $a(\mathbf{k})$ so that the external forcing function is monochromatic: $a(\mathbf{k}) = a\delta(\mathbf{k}-\mathbf{k}_0)(2\pi)^3$ and $\delta\phi^{ext}(\mathbf{r},t) = ae^{i\mathbf{k}_0 \cdot \mathbf{r}}\delta(t)$.

<u>Theorem</u>: The pulse-response solution for the perturbed electric potential using Eq.(2.E.17) is then

$$\delta\phi(\mathbf{r},t) = \frac{1}{(2\pi)^3}\int d^3\mathbf{k}\, e^{i\mathbf{k}\cdot\mathbf{r}} a(\mathbf{k})\varepsilon^{-1}(\mathbf{k},t) = ae^{i\mathbf{k}_0\cdot\mathbf{r}}\varepsilon^{-1}(\mathbf{k}_0,t) \quad (2.E.18)$$

The stability or instability of the pulse response is determined by $\varepsilon^{-1}(\mathbf{k},t)$. The contour integration in (2.E.17) is illustrated in Fig. 2.E.2. We make use of analytic continuation and depress the integration contour: $\omega = \omega' + ib$ to remove the line integration and only leave the poles. In so doing, we hope that there are no vertical branch cuts. The integration of the deformed contour integration becomes

$$\varepsilon^{-1}(\mathbf{k},t) = \int_{-\infty}^{\infty} \frac{d\omega'}{2\pi} \frac{\exp(-i\omega't - bt)}{\varepsilon(\mathbf{k},\omega'-ib)} - \frac{2\pi i}{2\pi} \sum_\ell \frac{\exp(-i\omega_\ell(\mathbf{k})t)}{\left.\frac{\partial\varepsilon(\mathbf{k},\omega)}{\partial\omega}\right|_{\omega_\ell}} \quad (2.E.19)$$

Consider a perturbative solution of the zeros of the linear dielectric function:

$$\varepsilon(\mathbf{k},\omega) \approx \varepsilon(\mathbf{k},\omega_\ell) + (\omega - \omega_\ell)\frac{\partial \varepsilon}{\partial \omega}\bigg|_{\omega_\ell} \quad \text{where} \quad \varepsilon(\mathbf{k},\omega_\ell) = 0 \text{ defines the pole at } \omega_\ell.$$

Define the complex frequency at the pole as $\omega_\ell = \Omega_\ell + i\gamma_\ell$ so that $\exp(-i\omega_\ell(\mathbf{k})t) = \exp(-i\Omega_\ell(\mathbf{k})t + \gamma_\ell t)$.

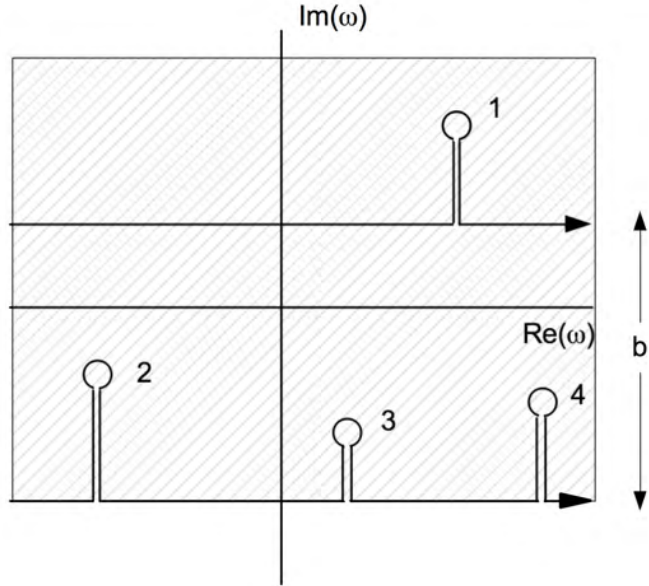

Fig. 2.E.2 Contour integration for pulse response showing the depressed contour and poles of $\varepsilon^{-1}$ in the region of analytic continuation.

Theorem: A pole in the upper half-plane corresponds to instability $\gamma_\ell > 0$, and a pole in the lower half-plane is a stable root $\gamma_\ell < 0$.

For $b > \gamma$ the first term on the right side of Eq.(2.E.19) damps out after a suitable length of time, which leads to

$$\varepsilon^{-1}(\mathbf{k},t) \Rightarrow -i \sum_\ell \frac{\exp(-i\omega_\ell(\mathbf{k})t)}{\frac{\partial \varepsilon(\mathbf{k},\omega)}{\partial \omega}\bigg|_{\omega_\ell}} \quad (2.E.20)$$

Definition: $g(\mathbf{v}) \equiv \sum_s \omega_s^2 g^s(\mathbf{v}) / \sum_s \omega_s^2$ and $\omega_p^2 \equiv \sum_s \omega_s^2$

The linear dielectric function becomes

$$\varepsilon(\mathbf{k},\omega)=1-\frac{\omega_p^2}{k^2}\int d^3\mathbf{v}\frac{\mathbf{k}\cdot\frac{\partial g}{\partial \mathbf{v}}}{\mathbf{k}\cdot\mathbf{v}-\omega}=1-\frac{\omega_p^2}{k^2}\int d^3\mathbf{v}\frac{\hat{\mathbf{k}}\cdot\frac{\partial g}{\partial \mathbf{v}}}{\hat{\mathbf{k}}\cdot\mathbf{v}-\omega/k} \qquad (2.\text{E}.21)$$

We note the integral on the right side of Eq.(2.E.21) can be simplified using the definition $u\equiv\hat{\mathbf{k}}\cdot\mathbf{v}$ so that

$$\hat{\mathbf{k}}\cdot\frac{\partial g}{\partial \mathbf{v}}=\frac{\partial g(u,\mathbf{v}_2,\mathbf{v}_3)}{\partial u}$$

This allows the velocity space integral over two of the three dimensions in (2.E.21) to be done immediately:

$$g(u)=\int d\mathrm{v}_2 d\mathrm{v}_3 g(u,\mathrm{v}_2,\mathrm{v}_3)\ \text{where}\ g(u)\equiv\int d^3\mathbf{v}\,g(\mathbf{v})\,\delta(\mathbf{k}\cdot\mathbf{v}-\omega)\ \text{and}\ \int_{-\infty}^{\infty}du\,g(u)=1$$

Eq.(2.E.16) then becomes

$$\varepsilon(\mathbf{k},\omega)=1-\frac{\omega_p^2}{k^2}\int_{-\infty}^{\infty}du\frac{g'(u)}{u-\mathrm{v}_p},\quad \mathrm{v}_p=\frac{\omega}{k} \qquad (2.\text{E}.22)$$

where $\mathrm{v}_p=\omega/k$ is the phase velocity; and in the following few expressions we drop the "$p$" subscript.

Theorem: The eigenvalues of the linearized Vlasov-Poisson system are determined by the roots of the linear dielectric function $\varepsilon(\mathbf{k},\omega_\ell)=0\to\omega_\ell(\mathbf{k})$, which are the eigenfrequencies. This follows from $\varepsilon(\mathbf{k},\omega)\phi(\mathbf{k},\omega)=\phi^{ext}(\mathbf{k},\omega)=0$ and corresponds to free, undriven oscillations.

Definition: The Hilbert transform of $g$ is

$$\int_{-\infty}^{\infty}du\frac{g_{\hat{\mathbf{k}}}(u)}{u-\mathrm{v}}\equiv\text{Hilbert transform of } g \equiv Z_{\hat{\mathbf{k}}}(\mathrm{v})\quad(\text{Im}\,\mathrm{v}>0,\,k\equiv|\mathbf{k}|>0) \qquad (2.\text{E}.23)$$

One can differentiate the Hilbert transform with respect to v to obtain:

$$\frac{d}{d\mathrm{v}}Z_{\hat{\mathbf{k}}}=\int_{-\infty}^{\infty}du\,g_{\hat{\mathbf{k}}}(u)\frac{\partial}{\partial \mathrm{v}}\frac{1}{u-\mathrm{v}}\equiv\int_{-\infty}^{\infty}du\frac{g_{\hat{\mathbf{k}}}'(u)}{u-\mathrm{v}},\quad g_{\hat{\mathbf{k}}}'\equiv\frac{d}{du}g_{\hat{\mathbf{k}}}(u) \qquad (2.\text{E}.24)$$

using $(\partial/\partial\mathrm{v})(u-\mathrm{v})^{-1}=-(\partial/\partial u)(u-\mathrm{v})^{-1}$ and integrating by parts. Equation (2.E.22) then yields the following.
Theorem: The dielectric function is

$$\varepsilon(\mathbf{k},\omega)=1-\frac{\omega_p^2}{k^2}Z_{\hat{\mathbf{k}}}'(\mathrm{v}) \qquad (2.\text{E}.25)$$

## 2.E.d Examples of linear dielectrics for a few simple velocity distributions

Table 2.E.1 presents examples of five model velocity distributions and the corresponding dielectric responses. We note that the Cauchy velocity distribution does not exist in nature because its energy moment diverges.

| Distribution → <br> Dielectric Attributes ↓ | COLD | COLD BEAM | SQUARE | CAUCHY | 2 SPECIES hot e & cold i |
|---|---|---|---|---|---|
| $g(u)$ | $\delta(u)$ 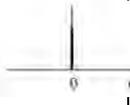 | $\delta(u-u_0)$ 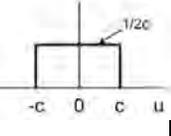 | $\dfrac{1}{2c},\ -c \le u \le c$ 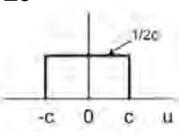 | $\dfrac{c}{\pi}\dfrac{1}{u^2+c^2}$ 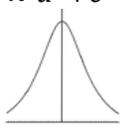 | hot electrons + cold ions 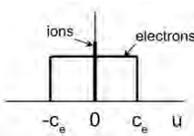 |
| $Z(v)$ | $-\dfrac{1}{v}$ | $-\dfrac{1}{v-u_0}$ | $\dfrac{1}{2c}\ln\left(\dfrac{c-v}{-c-v}\right)$ | $-\dfrac{1}{v+ic}$ | hot electrons + cold ions |
| $Z'(V)$ | $\dfrac{1}{v^2}$ | $\dfrac{1}{(v-u_0)^2}$ | $\dfrac{1}{v^2-c^2}$ | $\dfrac{1}{(v+ic)^2}$ | hot electrons + cold ions |
| $\varepsilon(\mathbf{k},\omega)$ | $1-\dfrac{\omega_p^2}{\omega^2}$ | $1-\dfrac{\omega_p^2}{(\omega-ku_0)^2}$ | $1-\dfrac{\omega_p^2}{\omega^2-k^2c^2}$ | $1-\dfrac{\omega_p^2}{(\omega+ikc)^2}$ | $1-\dfrac{\omega_e^2}{\omega^2-k^2c^2}-\dfrac{\omega_i^2}{\omega^2}$ |
| $\omega_\ell$ | $\omega_\ell = \pm\omega_p$ | $\omega_\ell = \pm\omega_p + ku_0$ | $\omega_\ell = \pm\sqrt{\omega_p^2+k^2c^2}$ | $\omega_\ell = \pm\omega_p - ikc$ | $\omega_\ell = \pm\sqrt{\omega_\ell^2},\ \omega_p^2=\sum_s \omega_s^2$ <br> $\omega_\ell^2 = \dfrac{1}{2}\left[k^2c^2+\omega_p^2 \pm \sqrt{(k^2c^2+\omega_p^2)^2 - 4\omega_i^2 k^2 c^2}\right]$ |
| $v_g = \dfrac{d}{dk}\mathrm{Re}\,\omega_\ell$ | 0 | $u_0$ | $c^2k/\omega$ | 0 | $c^2k/\omega$ |

Table 2.E.1 Examples of velocity distributions and resulting dielectric responses

## 2.E.f Inverse Fourier transform to obtain spatial and temporal response – Green's function for pulse response and cold plasma example

From Eq.(2.E.17) we have an integral relation for $\delta\phi(\mathbf{k},t)$, i.e.,

$$\delta\phi(\mathbf{k},t) = \int_{-\infty}^{\infty} d\tau \, \varepsilon^{-1}(\mathbf{k},\tau) \delta\phi^{ext}(\mathbf{k},t-\tau)$$

In configuration space, Eq.(2.E.17) becomes

$$\delta\phi(\mathbf{r},t) = \int d^3s \int_{-\infty}^{\infty} d\tau \, \varepsilon^{-1}(\mathbf{s},\tau) \delta\phi^{ext}(\mathbf{r}-\mathbf{s},t-\tau)$$

$$\text{where } \varepsilon^{-1}(\mathbf{s},\tau) = \int \frac{d^3k}{(2\pi)^3} \int \frac{d\omega}{2\pi} \varepsilon^{-1}(\mathbf{k},\omega) \exp(i\mathbf{k}\cdot\mathbf{s} - i\omega\tau) \tag{2.E.26}$$

Exercise: Prove $\varepsilon^{-1}(\mathbf{k},\tau)=0$ for $\tau<0$. See Sec. 2.E.c

Exercise: Simplify $\varepsilon(k,\omega)$ for hot electrons and cold ions in the limit that $\omega/k = V_p \ll c_e$. Introduce the definition of the Debye length $\lambda_e \equiv c_e/\omega_e$. Derive the dispersion relation for the ion acoustic wave:

$$\omega^2 = \frac{\omega_i^2}{1 + \frac{1}{k^2\lambda_e^2}} = \frac{k^2 c_s^2}{1 + k^2\lambda_e^2}, \quad c_s^2 \equiv \frac{m_e}{m_i} c_e^2 = \frac{k_B T_e}{m_i} \tag{2.E.27}$$

where $c_s$ is the ion sound speed for cold ions. The ions provide the inertia while the electrons provide pressure, and the electric field binds the motion of the two species.

We return to Eq.(2.E.26) to calculate the Green's function for the pulse response of the electric potential in a cold plasma with dielectric function $\varepsilon(\mathbf{k},\omega) = 1 - \omega_p^2/\omega^2$. If there is no $\mathbf{k}$ dependence in $\varepsilon(\mathbf{k},\omega)$ then

$$\varepsilon^{-1}(\mathbf{s},\tau) = \delta(\mathbf{s}) \int \frac{d\omega}{2\pi} \exp(-i\omega\tau) \left(1 + \frac{\omega_p^2}{(\omega-\omega_p)(\omega+\omega_p)}\right)$$

$$= \delta(\mathbf{s}) \left[ \delta(\tau) + \omega_p^2 \times \{0 \quad \tau<0, \quad -\frac{1}{\omega_p}\sin\omega_p\tau \quad \tau>0\} \right] \tag{2.E.28}$$

The integral over $\omega$ has poles at $\pm\omega_p$. The term involving $\delta(\tau)$ is the vacuum response, and the rest is associated with the plasma shielded response.

Theorem: Using Eq.(2.E.28), Eq.(2.E.26) becomes

$$\delta\phi(\mathbf{r},t) = \delta\phi^{ext}(\mathbf{r},t) - \int_0^{\infty} d\tau \, \omega_p \sin(\omega_p \tau) \delta\phi^{ext}(\mathbf{r},t-\tau) \tag{2.E.29}$$

The excitation in $\delta\phi(\mathbf{r},t)$ depends on all the other excitations and the external potential at $\mathbf{r}$ from earlier times. This is partly due to the absence of damping in the cold plasma.

Example: The impulse response for $\delta\phi^{ext}=\delta(\mathbf{r})\delta(t)$ is $\delta\phi(\mathbf{r},t)= -\omega_p\delta(\mathbf{r})\sin(\omega_p t)$ for $t >0$. For this response $\omega(\mathbf{k})=\omega_p$ and $d\omega/d\mathbf{k}=0$. The plasma response to the $\delta\phi^{ext}$ impulse is negative over the first half cycle of the plasma oscillation as the plasma tries to neutralize $\delta\phi^{ext}$ (Fig. 2.E.3)

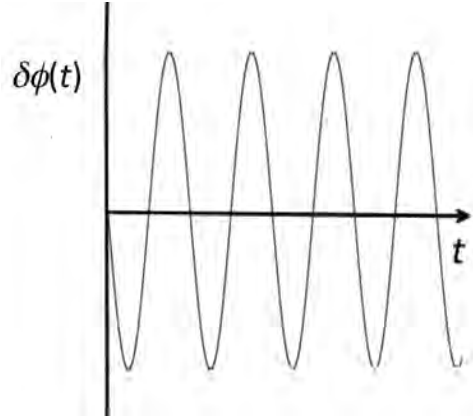

Fig. 2.E.3 Impulse response $\delta\phi(t)$ for $\delta\phi^{ext}=\delta(\mathbf{r})\delta(t)$

Example: $\delta\phi(\mathbf{r},t)$ due to a moving test particle. Consider the external potential for a moving charge: $\delta\phi^{ext}(\mathbf{r},t)=e_0/(\mathbf{r}-\mathbf{r}_0(t))$ with $\mathbf{r}_0(t)=\mathbf{v}_0 t$ and charge $e_0>0$. From Poisson's equation one obtains the total charge density:

$$\delta\rho(\mathbf{r},t)=-\left(\frac{1}{4\pi}\right)\nabla^2\delta\phi(\mathbf{r},t)=-\left(\frac{1}{4\pi}\right)\nabla^2\left[-\omega_p\int_0^\infty d\tau\sin(\omega_p\tau)\frac{e_0}{\mathbf{r}-\mathbf{r}_0(t-\tau)}+\delta\phi^{ext}(\mathbf{r},t)\right]$$

$$=\delta(x)\delta(y)\left(-\frac{e_0\omega_p}{v_0}\right)\sin\left(-\frac{\omega_p z}{v_0}\right)+e_0\delta(x)\delta(y)\delta(z-v_0 t)$$

(2.E.30)

A schematic of the one-dimensional charge density and the local charge under the curve in the wake of the moving test particle is presented in Fig. 2.E.4. The total charge exclusive of the test particle is $-e_0$ as can be shown from the integral of $\delta\rho$

$$\lim_{\lambda \to 0} \int_{-\infty}^{0} dz' \left(-\frac{e_0 \omega_p}{v_0}\right) \sin\left(\frac{\omega_p z'}{v_0}\right) \exp(-\lambda z') = -e_0 \tag{2.E.31}$$

Thus, the total charge is zero. We can also calculate the dipole moment **P** of the plasma response similarly by weighting the integrand in (2.E.31) with $z$ to show that **P**=0.

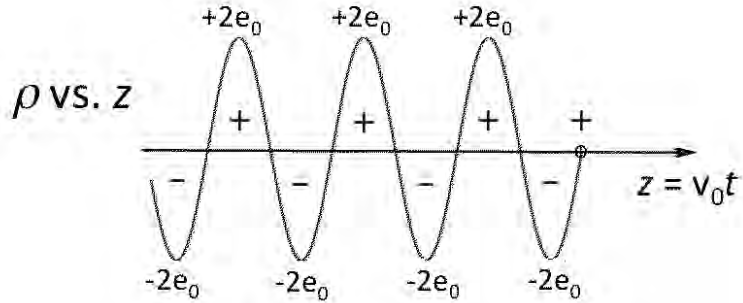

Fig. 2.E.4 One-dimensional charge density
in the wake of a moving test-particle with charge $e_0$

Example: For the more general case where $\varepsilon(\mathbf{k},\omega)$ has **k** dependence, we use Cauchy's theorem and Eqs(2.E.19), (2.E.20), and (2.E.26) to obtain

$$\varepsilon^{-1}(\mathbf{s},\tau) = -i \int \frac{d^3\mathbf{k}}{(2\pi)^3} \sum_{\ell} \frac{\exp\left(i\mathbf{k}\cdot\mathbf{s} - i\omega_\ell(\mathbf{k})\tau\right)}{\left.\frac{\partial \varepsilon(\mathbf{k},\omega)}{\partial \omega}\right|_{\omega_\ell(\mathbf{k})}} \tag{2.E.32}$$

However, for warm plasmas it is difficult to obtain explicit formulae because we cannot do the integrals in most cases of physical interest.

## 2.F Streaming instabilities

Consider an infinite uniform plasma with two or more beams:

$$f_s^0(\mathbf{v}) = n_s \delta(\mathbf{v} - \mathbf{u}_s) \text{ and } \varepsilon(\mathbf{k},\omega) = 1 - \sum_s \frac{\omega_s^2}{(\omega - \mathbf{k}\cdot\mathbf{u}_s)^2}, \text{ with } \omega_s^2 \equiv \frac{4\pi n_s e_s^2}{m_s}$$

Restrict this linear analysis to one spatial dimension: $k\hat{\mathbf{k}} \cdot \mathbf{u}_s = k u_s$. Then the linear dielectric can be rewritten as

$$\varepsilon(k,\omega) = 1 - \sum_s \frac{\omega_s^2}{(\omega - ku_s)^2} = 1 - \frac{1}{k^2} \sum_s \frac{\omega_s^2}{(V - u_s)^2}, \quad V \equiv \omega/k \tag{2.F.1}$$

The solution of $\varepsilon(k,\omega)=0$ for the normal modes $\omega_k$ is then determined by solving the following order $2N$ polynomial equation for $V=\omega/k$:

$$k^2 = F(V) = \sum_{s=1}^{N} \frac{\omega_s^2}{(V - u_s)^2} \tag{2.F.2}$$

Example: Consider the graphical solution of Eq.(2.F.2) for N=3.

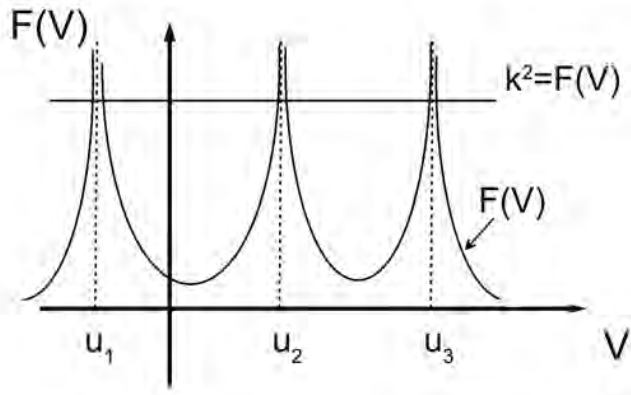

Fig. 2.F.1 Schematic of solution of Eq.(2.F.2) $k^2=F(V)$ for N=3

There are resonances at $V=u_i$, i=1, 2 and 3 and three branches of the dispersion relation with Re$\omega \sim ku_i \pm \omega_s$. For very large values of $k^2$ there are $2N=6$ intersections of $k^2$ with $F(V)$, i.e., 6 stable roots. If for smaller values of $k^2$ there are fewer than $2N$ intersections, then there are one or more pairs of complex-conjugate roots. The complex root with Im$\omega>0$ is the unstable root, and the conjugate root with Im$\omega<0$ is damped normal mode. In a finite plasma, boundary conditions must be defined; and there is a lower limit on $k$ corresponding to $2\pi/L$, where $L$ is the length of the plasma.

### 2.F.a Examples – two-stream and weak-beam instabilities

Example: N=2, two-stream instability. Suppose $\omega_1=\omega_2$ and select a reference frame with $u_1=-u_2$. The infinite-medium normal modes are determined by the solutions of the dispersion relation:

$$\varepsilon(k,\omega_\ell) = 1 - \frac{\omega_1^2}{(\omega - ku_1)^2} - \frac{\omega_1^2}{(\omega + ku_1)^2} = 0$$

$$\rightarrow \frac{\omega_\ell^2}{\omega_p^2} = \frac{1}{2}\left[1 + 2\frac{k^2 u_1^2}{\omega_p^2} \pm \sqrt{1 + \frac{8k^2 u_1^2}{\omega_p^2}}\right] \quad \text{with } \omega_p^2 = \omega_1^2 + \omega_2^2 = 2\omega_1^2 \tag{2.F.3}$$

Exercise: Sketch $\omega^2$ vs. $k^2$, Re$\omega$ vs. $k$, and Im$\omega$ vs. $k$. Show that for $k < k_c$ there is instability; what is $k_c$? Show that max(Im$\omega$)=$\omega_p/\sqrt{8}$ for $k_{max}=(3/8)^{1/2}\omega_p/u_1$. Evaluate $\varepsilon^{-1}(x,t)$ approximately and sketch. Show that $\varepsilon^{-1}(x,t)=0$ for $|x/t|>u_1$ using analyticity.

Example: $N=2$, weak-beam instability. Assume $\omega_2=\omega_b<<\omega_2=\omega_b<<\omega_1=\omega_p$ where the "first" component is the "plasma." Select the frame in which the plasma component is at rest. The weak-beam instability is diagrammed in Figure 2.F.2 The dispersion

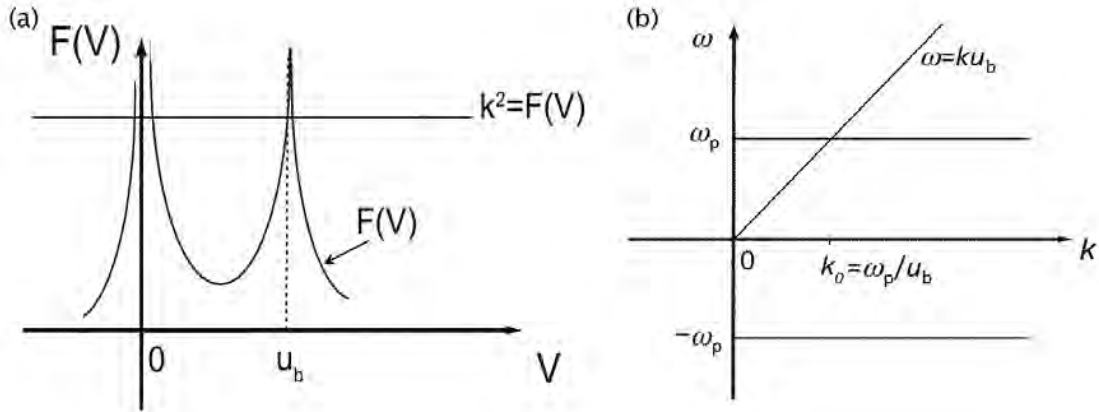

Fig. 2.F.2 (a) Schematic of the solution of $k^2=F(V)$ for $N=2$ and $\omega_2=\omega_b<<\omega_2=\omega_b<<\omega_1=\omega_p$, weak-beam instability. (b) Schematic showing the crossing of the plasma frequency and beam branches.

relation for the normal modes is given by the solution of

$$\varepsilon(k,\omega) = 1 - \frac{\omega_p^2}{\omega^2} - \frac{\omega_b^2}{(\omega-ku_b)^2} = \varepsilon_p(k,\omega) - \frac{\omega_b^2}{(\omega-ku_b)^2} = 0 \qquad (2.F.4)$$

Where the beam and plasma wave branches cross, one of the real roots can disappear giving rise to a pair of complex-conjugate roots and instability. If the plasma is warm, the plasma branches acquire curvature. In general, one uses the warm plasma dielectric for $\varepsilon_p$. For $\omega_b^2/\omega_p^2<<1$ we solve Eq.(2.F.4) perturbatively:

$$0 = \varepsilon_p(k,\omega) - \frac{\omega_b^2}{(\omega-ku_b)^2}$$

$$= \varepsilon_p(k,\omega_0) + \delta\omega \left.\frac{\partial \varepsilon_p}{\partial \omega}\right|_{\omega_0,k_0} + \delta k \left.\frac{\partial \varepsilon_p}{\partial k}\right|_{\omega_0,k_0} + \frac{1}{2}\delta\omega^2 \left.\frac{\partial^2 \varepsilon_p}{\partial \omega^2}\right|_{\omega_0,k_0} + ... - \frac{\omega_b^2}{(\omega_0+\delta\omega-k_0u_b-\delta k u_b)^2},$$

$$\varepsilon_p(k,\omega_0) = 0, \quad \omega = \omega_0 + \delta\omega, \quad k = k_0 + \delta k$$

(2.F.5)

At resonance $\omega_0-k_0 u_b=0$, and Eq.(2.F.5) becomes

$$(\delta\omega-u\delta k)^2(\delta\omega\varepsilon_\omega+\delta k\varepsilon_k+O(\delta^2))=\omega_b^2, \quad \varepsilon_\omega\equiv\partial\varepsilon/\partial\omega\big|_{\omega_0,k_0} \quad (2.F.6)$$

i) For $\delta k=0$: $\delta\omega^3+O(\delta\omega^4)=\varepsilon_\omega^{-1}\omega_b^2$  For small $\delta\omega$ small,

$$\left(\frac{\delta\omega}{\omega_0}\right)^3=\frac{\omega_b^2}{\omega_0^2}\frac{1}{\omega_0\varepsilon_\omega}\equiv\eta\approx\frac{1}{2}\frac{\omega_b^2}{\omega_0^2}<<1$$

$$\rightarrow\frac{\delta\omega}{\omega_0}=\eta^{1/3}=|\eta^{1/3}|\left\{1,-\frac{1}{2}\pm i\frac{\sqrt{3}}{2}\right\} \quad (2.F.7)$$

We note that for a cold plasma, $\varepsilon_p=1-\omega_p^2/\omega^2$ and $\varepsilon_\omega=2/\omega_p$.  There are three solutions for the frequency shift $\delta\omega$ in Eq.(2.F.7): the coupling of the plasma wave to the weak beam produces a small frequency shift, a damped mode, and a weak instability.

ii) For $\delta k\neq 0$, Eq.(2.F.6) is solved.  In the beam frame $\omega'\equiv\omega-ku_b=\delta\omega-\delta k u_b$

In a cold plasma the solution for $\omega'$ is given by

$$\omega'=\omega_p\left[|\eta^{1/3}|\exp\left(i\frac{2\pi}{3}\right)-\frac{1}{3}\frac{\delta k}{k_0}+\frac{1}{9}\left(\frac{\delta k}{k_0}\right)^2|\eta^{1/3}|\exp\left(-i\frac{2\pi}{3}\right)+O(\delta k^3)\right]$$

$$\text{Re}\,\omega'=\omega_p\left[-\frac{1}{2}|\eta^{1/3}|-\frac{1}{3}\frac{\delta k}{k_0}-\frac{1}{18}\left(\frac{\delta k}{k_0}\right)^2|\eta^{1/3}|+...\right] \quad (2.F.8)$$

$$\text{Im}\,\omega'=\omega_p\left[\frac{\sqrt{3}}{2}|\eta^{1/3}|+\frac{1}{9}\left(-\frac{\sqrt{3}}{2}\right)\left(\frac{\delta k}{k_0}\right)^2|\eta^{1/3}|+...\right]$$

In the beam frame there is a small negative shift of the phase velocity given by $\text{Re}\,\omega'/k_0 = -(1/2)|\eta|^{1/3}u_b$  There is also a shift in the group velocity of the plasma wave due to the coupling with the beam, which is given by $d\text{Re}\,\omega'/dk= d\text{Re}\,\omega'/d\delta k=-(1/3)u_b$, which is also negative but is not small.  We note that the dispersion in the real frequency of the instability is given by $d^2\text{Re}\,\omega'/d\delta k^2=-(1/9)|\eta|^{1/3}(\omega_p/k_0^2)$, which is small in $|\eta|^{1/3}$.  From the solution for $\text{Im}\,\omega'$ we see that the growth rate is peaked at a value of $(\sqrt{3}/2)\eta^{1/3}\omega_p$ for $\delta k=0$; and

$$\gamma_{kk}\equiv\left|\frac{d^2\gamma}{dk^2}\right|=\frac{\gamma_0}{k_0^2}\frac{2}{9}|\eta^{-2/3}|, \quad \gamma_0=\frac{\sqrt{3}}{2}\omega_p|\eta^{1/3}|$$

From the square root of the ratio of the peak growth rate to $-d^2\text{Im}\,\omega'/d\delta k^2$ at $k_0$ we can estimate the half-width in $\delta k_{1/2}$ of the peak in the growth rate, which has a scaling $\delta k_{1/2}/k_0\sim|\eta^{1/3}|<<1$.  Thus, a very small range in $k$-space is involved in the weak-beam instability; and a long wavepacket can be formed that remains coherent over a long distance.

All of the results so far in Sec. 2.F were derived for a cold plasma. For a "hot" plasma only numerical coefficients will change for the examples considered. The generalization of the results will involve formulae like the following:

$$\text{Re}\,\omega' = k_0 V_p' + \delta k V_g' + \frac{1}{2}\delta k^2 \frac{d^2}{d\delta k^2}\omega' \quad \gamma \equiv \text{Im}\,\omega' = \gamma_0 + \frac{1}{2}\delta k^2 \frac{d^2}{d\delta k^2}\gamma \quad (2.\text{F}.9)$$

where $V_p'=\omega'(k_0)/k_0$, $V_g'=d\omega'/dk$ at $k_0$, and $\gamma_0$ is the peak growth rate occurring at $k_0$. Finite-temperature effects naturally lead to dispersion, i.e., the phase velocity $V_p$ is a function of wavenumber $k$. Thus, in a hot plasma we expect that the *x-t* response of a growing disturbance will exhibit a growing and spreading wave packet traveling at the group velocity $V_g$. The *x-t* response depends on the content of Eq.(2.F.9) and makes explicit how fast the growing wave packet spreads compared to its advection. The next sub-section elaborates the *x-t* response of an unstable disturbance.

### 2.F.b Definition of convective and absolute instability

How fast a growing wave packet spreads compared to how fast it advects past a fixed observation point is an important distinction.

<u>Definition</u>: Absolute vs. convective instability. If an unstable wave packet advects faster than it spreads, an observer at a fixed observation point will see a growing signal as the front of the wave packet passes followed by peaking and then decay of the signal. As the packet advects the peak signal continues to grow exponentially in time. The foregoing corresponds to a *convective instability*. An *absolute instability* corresponds to when the spreading of the growing response exceeds the advection at the group velocity $V_g$ so that the signal at a fixed observation point continues to grow without cessation (until the linear assumption fails and nonlinear effects may come into the problem).

<u>Exercise</u>: Sketch a growing and advecting pulse at two distinct times in one spatial dimension for a convective instability, and make the corresponding sketch for an absolute instability.

The dielectric pulse response in one spatial dimension for an unstable root of the dispersion relation follows from Eq.(2.E.32):

$$\varepsilon^{-1}(x,t) = \int_{-\infty}^{\infty}\frac{dk}{2\pi}\int_{-\infty}^{\infty}\frac{d\omega}{2\pi}\frac{\exp(ikx-i\omega t)}{\varepsilon} + c.c. = -i\int_{-\infty}^{\infty}\frac{dk}{2\pi}\frac{\exp(ikx-i\omega t)}{\partial\varepsilon/\partial\omega\big|_{\omega_k(k)}} + c.c.$$

$$\approx -i\frac{1}{\frac{\partial\varepsilon}{\partial\omega}\big|_{\omega_k(k_0)}}\int_{-\infty}^{\infty}\frac{d(\delta k)}{2\pi}\exp\left[ik_0 x + i\delta k x - i\left(k_0 V_p + \delta k V_g + \tfrac{1}{2}\omega_{kk}\delta k^2\right)t + \left(\gamma_0 - \tfrac{1}{2}|\gamma_{kk}|\delta k^2\right)t\right] + c.c.$$

$$\approx -i\frac{\exp(ik_0(x-V_p t))\exp\gamma_0 t}{\varepsilon_\omega\big|_{\omega_k(k_0)}}\int_{-\infty}^{\infty}\frac{d(\delta k)}{2\pi}\exp\left[i\delta k(x-V_g t)\right]\exp\left[-\tfrac{1}{2}\delta k^2(|\gamma_{kk}|+i\omega_{kk})t\right] + c.c.$$

$$\approx -i\frac{\exp(ik_0(x-V_p t))\exp\gamma_0 t}{\varepsilon_\omega\big|_{\omega_k(k_0)}}\frac{1}{\sqrt{2\pi}}\frac{1}{\sqrt{(|\gamma_{kk}|+i\omega_{kk})t}}\exp\left[\frac{-\tfrac{1}{2}(x-V_g t)^2}{(|\gamma_{kk}|+i\omega_{kk})t}\right] + c.c.$$

$$\approx -i\frac{\exp(ik_0(x-V_p t))\exp\gamma_0 t}{\varepsilon_\omega\big|_{\omega_k(k_0)}}\frac{1}{\sqrt{2\pi}}\frac{1}{\sqrt{(|\gamma_{kk}|+i\omega_{kk})t}}\exp\left[\frac{-\tfrac{1}{2}(x-V_g t)^2|\gamma_{kk}|}{(|\gamma_{kk}|^2+\omega_{kk}^2)t} + i\frac{\tfrac{1}{2}(x-V_g t)^2|\omega_{kk}|}{(|\gamma_{kk}|^2+\omega_{kk}^2)t}\right] + c.c.$$

(2.F.10)

We note that the spatial width of the pulse scales as $\Delta x \sim t^{1/2}(|\gamma_{kk}|^2 + \omega_{kk}^2)^{1/2}/|\gamma_{kk}|$ Hence, the larger (and steeper) $|\gamma_{kk}|$, the faster the spreading in configuration space. To obtain the results in Eq.(2.F.9) we made use of the Taylor-series expansion of the dispersion relation as given in Eq.(2.F.8) for small $\delta k$. Because we did the integral with respect to $\delta k$ by the method of steepest descents and took advantage of the rapid convergence of that integral with respect to large $\delta k$ due to the term $\exp(-\tfrac{1}{2}\delta k^2|\gamma_{kk}|t)$, there is no conflict between the limits of the $\delta k$ integration being $(-\infty,\infty)$ and Taylor-series expansion in small $\delta k$. Because the pulse width in the space-time domain is increasing as $t^{1/2}$, the pulse in the frequency-wavenumber dual domain is decreasing as $t^{-1/2}$; hence, the wave packet is lengthening in space-time and becoming purer in its spectral content as it advects and grows.

To determine whether an instability is absolute or convective, we must compare the spreading $\sqrt{Dt} = \sqrt{(|\gamma_{kk}|^2+\omega_{kk}^2)t/\gamma_{kk}}$ and the exponential growth $\exp(\gamma_0 t)$ with the advection of the pulse at the group velocity $V_g$. In the frame advecting with the wave packet, $|\varepsilon^{-1}(x,t)| \propto \exp(\gamma_0 t - x^2/2Dt)$; and the pulse grows exponentially and spreads. In the plasma frame one obtains

$$|\varepsilon^{-1}(x,t)| \propto \exp\left(\gamma_0 t - \frac{(x-V_g t)^2}{2Dt}\right) \equiv \exp\left(\tau - \frac{(\xi-\sigma\tau)^2}{2\tau}\right), \xi \equiv \sqrt{\frac{\gamma_0}{D}}x, \tau \equiv \gamma_0 t, \sigma \equiv V_g\sqrt{\frac{1}{\gamma_0 D}}$$

$$\rightarrow \ln|\varepsilon^{-1}(x,t)| \sim \gamma_0 t - \frac{(x-V_g t)^2}{2Dt} = \tau - \frac{(\xi-\sigma\tau)^2}{2\tau}$$

(2.F.11)

Theorem: For fixed $\xi$, the asymptotics of $|\varepsilon^{-1}(x,t)|$ in Eq.(2.F.11) as $\tau \to \infty$ determines whether the pulse is growing absolutely or convectively, i.e.,

$$\lim_{\tau \to \infty}\left[\ln|\varepsilon^{-1}(x,t)| \sim \tau - \frac{(\xi-\sigma\tau)^2}{2\tau}\right] \to \tau(1-\frac{\sigma^2}{2}) \qquad (2.F.12)$$

For $\sigma > \sqrt{2}$, $\ln|\varepsilon^{-1}| \to -\infty$ as $\tau \to \infty$, i.e., the instability is convective; and the disturbance grows at first and then dies away. For $\sigma < \sqrt{2}$, $\ln|\varepsilon^{-1}| \to \infty$ as $\tau \to \infty$, i.e., the instability is absolute and continues to grow exponentially at any fixed position. The condition for absolute (or convective) instability is then

$$V_g < (>)\sqrt{2\gamma_0 D} = \sqrt{2\gamma_0 \left(\frac{|\gamma_{kk}|^2 + \omega_{kk}^2}{|\gamma_{kk}|}\right)} \qquad (2.F.13)$$

Exercise: Show that in the plasma frame the group velocity for the weak beam in a cold plasma instability is $(2/3)u_b$ and the instability is convective using Eq.(2.F.13) and the analysis in Sec. 2.F.a.

### 2.F.c Peter Sturrock's method for analyzing absolute instability (reference)

Stanford Professor P. A. Sturrock introduced a method for analyzing whether a growing instability is convective or absolute.[1] Sturrock's method is reviewed in P. C. Clemow and J. P. Dougherty, *The Electrodynamics of Particles and Plasmas*, Addison-Wesley (1969) and was assigned as reading but not covered in class lectures.

### 2.F.d Bers and Briggs' method for analyzing absolute instability

A. Bers[2] and R. J. Briggs[3] derived a method for calculating the impulse response using contour integration and analytic continuation, from which absolute and convective instability can be distinguished. The calculation begins with consideration of Eq.(2.F.9) in one spatial dimension transformed to the moving reference frame $x=wt$:

$$\varepsilon^{-1}(x,t) = \int \frac{dk}{2\pi} \int \frac{d\omega}{2\pi} \frac{\exp(ikx-i\omega t)}{\varepsilon(k,\omega)} = \int \frac{dk}{2\pi} \int \frac{d\omega}{2\pi} \frac{\exp(i\omega_w t)}{\varepsilon(k,\omega_w+kw)} \qquad (2.F.14)$$

---

[1] P. A. Sturrock, Phys. Rev. **112**, 1488 (1958).
[2] A. Bers, "Theory of Absolute and Convective Instabilities" in G. Auer and F. Cap, International Congress on Waves and Instabilities in Plasma (Innsbruck, Austria, April 1973), pp. B1-B52.
[3] R. J. Briggs, *Electron-Stream Interaction with Plasma* (MIT Press, Cambridge, MA, 1964).

where $\omega_w = \omega - kw$ which is the Doppler-shifted frequency. There are two contour integrals to perform in Eq.(2.F.14). These are diagrammed in Fig. 2.F.1 for a convective instability and in Fig. 2.F.2 for an absolute instability. In the complex $k$ plane the locus of poles in $k$ of the integrand in Eq.(2.F.14) for fixed $\omega$ is shown by $C_\omega$, and the integration contour is $C_k$. In the complex $\omega$ plane the locus of poles in $\omega$ of the integrand in Eq.(2.F.14) for fixed $k$ is shown by $C_k$, and the integration contour is $C_\omega$. Making use of analytic continuation, the integration contour in the complex $\omega$ plane is depressed to lower values on the imaginary axis as shown in Fig. 2.F.1 until poles on $C_k$ from below the $C_\omega$ contour are encountered. The poles $C_k$ in the complex $\omega$ plane are parametrized by the complex value of $k$ swept out by the $C_k$ contour in the complex $k$ plane. In order to depress the $C_\omega$ contour to lower values of Im $\omega$, we must move the $C_k$ contour down in the complex $k$ plane. In the complex $k$ plane there are a loci of poles $C_\omega$ above and below the $C_k$ integration contour. For the convectively unstable case, the $C_\omega$ contour in the complex $\omega$ plane can be depressed to values of Im $\omega < 0$; and $\varepsilon^{-1}(x,t)$ decays (Fig. 2.F.3). In the absolutely unstable case, the $C_\omega$ contour in the complex $\omega$ plane can be depressed and deformed to lower values of Im $\omega$ everywhere except as one approaches the pinch point $P$ in Fig. 2.F.4c because the $C_k$ integration contour in the $k$ plane becomes trapped ("pinched") between the $C_\omega$ contours above and below it; and part of the $C_k$ loci of poles has some values Im $\omega_k > 0$ in Fig. 2.F.4d. The asymptotic response for $\varepsilon^{-1}(x,t)$ exhibits exponential growth at the value of Im $\omega_k > 0$ at the pinch point.

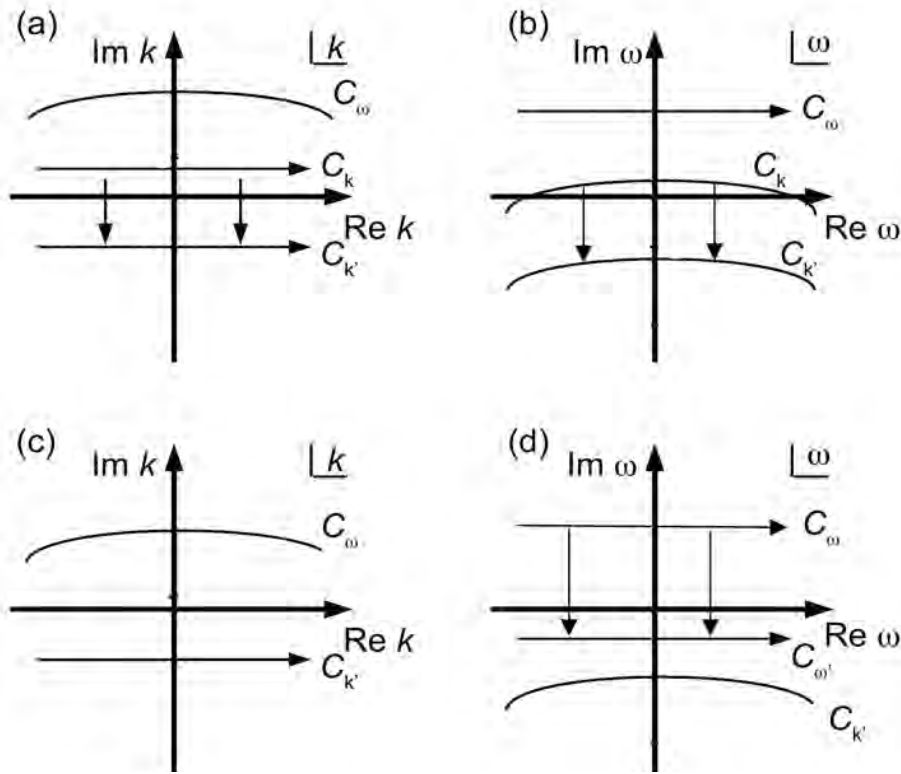

Fig. 2.F.3 Convective instability: diagrams of contour integral paths $C_k$ in complex $k$ plane (a) and (c), and contour integral paths $C_\omega$ in complex $\omega$ plane (b) and depressed in (d) showing the loci of poles of the integrand in Eq.(2.F.13).

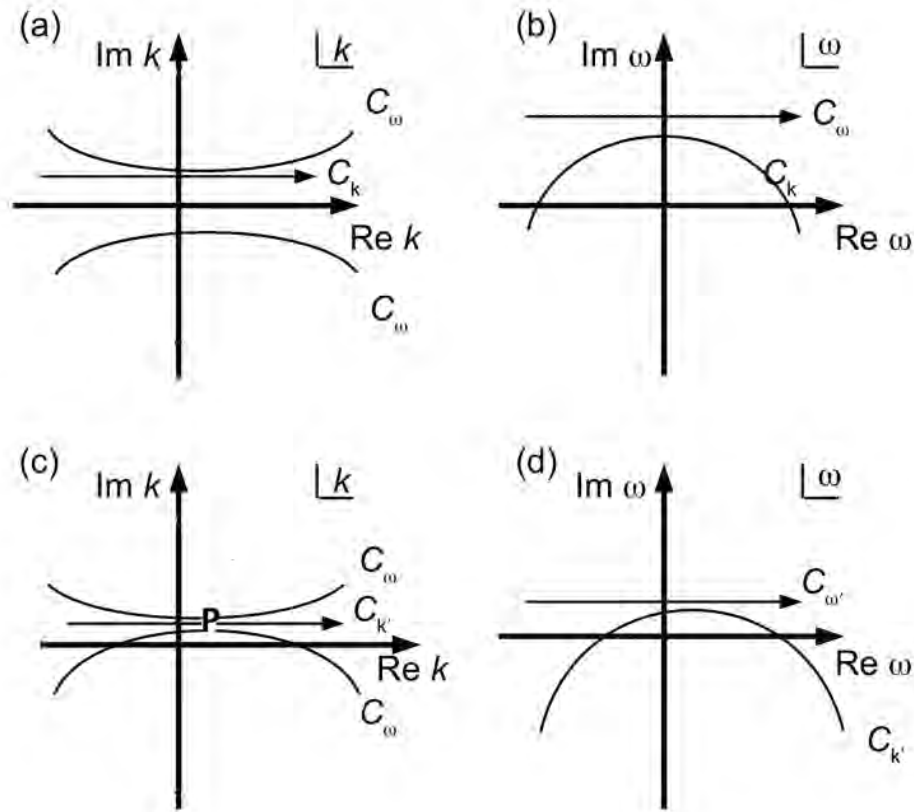

Fig. 2.F.4 Absolute instability: diagrams of contour integral paths $C_k$ in complex $k$ plane (a) and (c), and contour integral paths $C_\omega$ in complex $\omega$ plane (b) and depressed in (d) showing the loci of poles of the integrand in Eq.(2.F.13). A pinching of roots occurs at point $P$ in (c).

Exercise: Examine the conditions for absolute vs. convective instability for the cold beam – cold plasma instability investigated in Sec. 2.F.a in the (a) plasma frame in which $\varepsilon(k,\omega) = 1 - \dfrac{\omega_p^2}{\omega^2} - \dfrac{\omega_b^2}{(\omega - ku_b)^2}$ and (b) the beam frame in which $\varepsilon(k,\omega) = 1 - \dfrac{\omega_p^2}{(\omega + ku_b)^2} - \dfrac{\omega_b^2}{\omega^2}$, and there is absolute instability.

Theorem: The condition for absolute instability is equivalent to obtaining two pinching roots for ($\omega$,k) from the simultaneous solution of $\varepsilon(\omega,k)=0$ and $\partial\varepsilon(\omega,k)/\partial k=0$ with Im $\omega > 0$, provided that $\partial\varepsilon(\omega,k)/\partial\omega \neq 0$.
Exercise: Prove this theorem.

## 2.G Linear steady-state response to a fixed frequency disturbance

We consider here the response of a plasma to a steady-state force at a fixed frequency. We assume that the plasma is quiescent and non-turbulent. There may be transient convective instabilities but no absolute instability. We implant a localized fixed-frequency disturbance and derive the plasma response. Let the implanted disturbance by a planar disturbance, for example a biased grid connected to an oscillator:

$$\delta\phi^{ext}(\mathbf{r},t) = \{0 \quad t<0, \quad \delta(z)\sin(\omega_0 t) \quad t>0\}$$
$$\rightarrow \quad \delta\phi^{ext}(\mathbf{k},\omega) = -(2\pi)^2 \delta(k_x)\delta(k_y)\frac{\omega_0}{\omega^2 - \omega_0^2} \quad (2.G.1)$$

The plasma potential is then given by Eqs.(2.E.12) and (2.G.1)

$$\delta\phi(\mathbf{k},\omega) = \frac{\delta\phi^{ext}(\mathbf{k},\omega)}{\varepsilon(\mathbf{k},\omega)}$$

$$\rightarrow \quad \delta\phi(\mathbf{r},t) = -\int\frac{d^3k}{(2\pi)^3}\int\frac{d\omega}{2\pi}\exp i(k_x x + k_y y + k_z z - \omega t)\frac{(2\pi)^2 \delta(k_x)\delta(k_y)\frac{\omega_0}{\omega^2 - \omega_0^2}}{\varepsilon(\mathbf{k},\omega)}$$

$$\rightarrow -\int\frac{dk_z}{2\pi}\int\frac{d\omega}{2\pi}\exp(ik_z z - i\omega t)\frac{1}{\varepsilon(\mathbf{k},\omega)}\frac{\omega_0}{\omega^2 - \omega_0^2}$$
(2.G.2)

We let any transients associated with initial conditions and any instabilities convect or decay away.

Theorem: Given that the dielectric is an even function of z, then
$\varepsilon(-k_z, -\omega_r + i\omega_i) = \varepsilon(k_z, \omega_r + i\omega_i)$.

## 2.G.a Response for a sinusoidally driven stable system

The contour integrations in Eq.(2.G.2) are diagrammed in Fig. 2.G.1. For every value of k the $C_k$ contour in Fig. 2.G.1b is below the Re $\omega$ axis, which dictates that $\delta\phi(r,t)$ is stable. The $\omega$ integration in Eq.(2.G.2) is performed using Cauchy's theorem to obtain:

$$\delta\phi(\mathbf{r},t) = \frac{i}{2}\int \frac{dk_z}{2\pi} \frac{\exp(ik_z z - i\omega_0 t)}{\varepsilon(k_z,\omega_0)} + c.c. \qquad (2.G.3)$$

and transients that are evanescent. The $k_z$ integration is now performed.

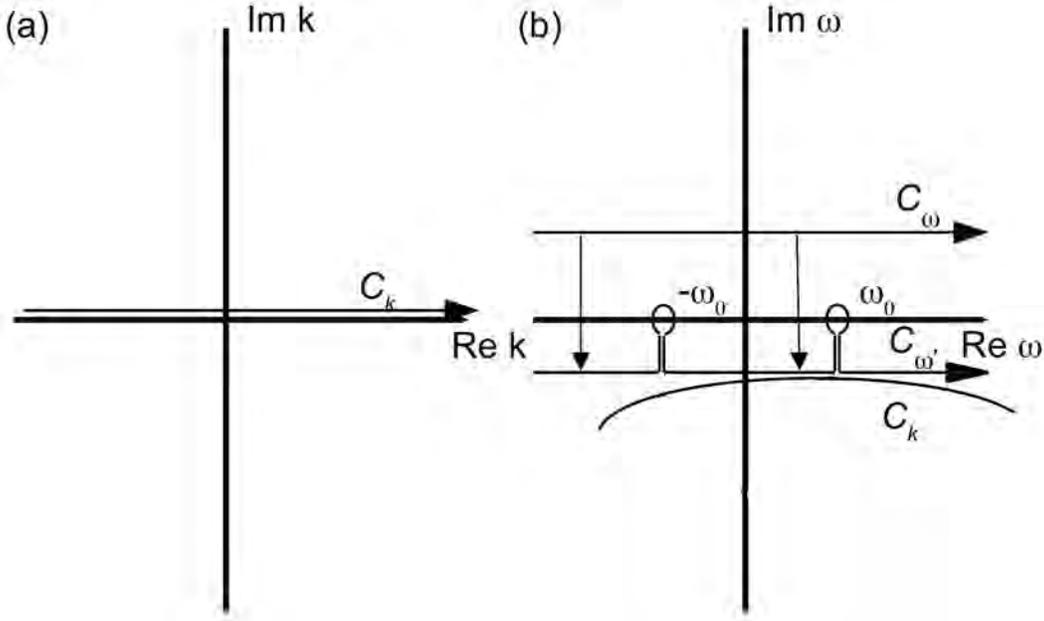

Fig. 2.G.1 Steady-state response to a fixed-frequency disturbance: diagrams of the contour integral path $C_k$ in complex $k$ plane (a) and the contour integral paths $C_\omega$ in complex $\omega$ plane (b) showing the loci of poles of the integrand in Eq.(2.G.2).

Consider the poles of the integrand in Eq.(2.G.3), i.e., the solutions of $\varepsilon(k_z,\omega_0) = 0$. For the example of a velocity distribution that is a square in velocity space (see Table 2.E.1) then $\omega_0^2 = \omega_p^2 + k_z^2 c^2 \to k_z = \pm\sqrt{\omega_0^2 - \omega_p^2}/c$, which roots lie on the real $k_z$ axis. A real physical system will have some finite dissipation. Hence, $\omega(k) = \pm\sqrt{\omega_p^2 + k^2 c^2} - i\nu$ and $(\omega + i\nu)^2 = \omega_p^2 + k^2 c^2$, where $\nu$ is a damping rate, e.g., due to weak collisions, so that free oscillations are damped and transients will indeed die away. In this circumstance the solutions for $k_z$ become $k_z = \pm\sqrt{(\omega_0 + i\nu)^2 - \omega_p^2}/c$. The sign of $k_z$ is selected based on whether values of $z$ are negative or positive so that the response of the system dies away for $|z|\to\infty$. For $\omega_0 < \omega_p$ $k_z$ is purely imaginary, and the plasma response is evanescent; and modes do not propagate.

For $z > 0$ the contour integral in Eq.(2.G.3) is closed counter-clockwise in the upper half-plane in Fig. 2.G.1a, and we sum over pole contributions:

$$\delta\phi(\mathbf{r},t) = -\frac{1}{2}\sum_{\ell}\exp(ik'_{zl}z - k''_{zl}z - i\omega_0 t)\frac{1}{\left.\frac{\partial\varepsilon(k_z,\omega_0)}{\partial k_z}\right|_{\ell}} + c.c. \qquad (2.G.4)$$

where $k_{zl} = k_{zl}' + i\,k_{zl}''$ is a solution of $\varepsilon(k_{zl},\omega_0)=0$ with $k_{zl}'' > 0$. For $z < 0$ the contour integral is closed clockwise in the lower half-plane with $k_{zl}'' < 0$; and the overall sign of the right side of Eq.(2.G.4) changes. For $z$ positive or negative the plasma response is sinusoidal in time with frequency $\omega_0$ and damps away from the origin in $z$.

### 2.G.b Response for a sinusoidally driven convectively unstable system

One can compute the sinusoidally driven response for a convectively unstable system in the same manner as in Sec. 2.G.a. Because the system is convectively unstable, the integration contour $C_\omega$ can be depressed as in Fig. 2.G.1; and the results in Eqs.(2.G.3) and (2.G.4) pertain. However, the solutions of $\varepsilon(k_z,\omega_0)=0$ for $k_z$ can have Im $k_z < 0$ for $z > 0$ (and Im $k_z > 0$ for $z < 0$) leading to spatially growing solutions away from the origin in in $z$.

Exercise: Consider the beam-plasma system in the plasma reference frame (Sec. 2.F.a) which is convectively unstable. Solve $\varepsilon(k_z,\omega_0)=0$ based on Eq.(2.F.4) for $k_z$ and evaluate Eq.(2.G.4). Contrast this to the stable case of a beaming plasma:

$$\varepsilon(k,\omega) = 1 - \frac{\omega_p^2}{(\omega+i\nu-ku)^2} = 0$$

Exercise: Consider an external driving potential that is a spherical waveform,
$$\delta\phi^{ext}(\mathbf{r},t) = \{0 \quad t<0,\ \delta(r)\sin(\omega_0 t) \quad t>0\}$$
in the rest frame of a cold plasma. Find the response for $\omega_0 > \omega_p$ and $\omega_0 < \omega_p$. Contrast this to the case of a mono-energetic velocity distribution: $f_0 = \delta(|\mathbf{v}|-v_0)/\text{norm}$.

### 2.H Linear stability or instability for a few simple velocity distributions

We return to the consideration Vlasov stability for more general velocity distributions based on Eq.(2.E.25):

$$\varepsilon(\mathbf{k},\omega) = 1 - \frac{\omega_p^2}{k^2}Z'(\text{v}) = 1 - \frac{\omega_p^2}{k^2}\int du\,\frac{g_{\hat{\mathbf{k}}}'(u)}{u-\text{v}} \qquad (2.H.1)$$

where $\text{v}=\omega/k$ and $\int du\,g(u)=1$. In Fig. 2.H.1 are shown four examples of velocity distributions. Figure 2.H.1a depicts two warm beams, which is always linearly unstable. A single-humped velocity distribution is shown in Fig. 2.H.1b, which is stable. The velocity distribution function shown in Fig. 2.H.1c can be stable or unstable depending on the relative heights $b$ and $a$. In the limit $b \to 0$, the velocity distribution is single-humped and stable. In the limit $b \to a$, the velocity distribution corresponds to two counter propagating warm beams and is unstable. Figure 2.H.d depicts a double Cauchy beam. When the thermal spread $c$ is larger than the beam centroid velocity $u_0$, the plasma is stable; and when $c < u_0$, the plasma is unstable. The results of these examples give one a hint toward a more general stability condition. A general condition is that it is *necessary* for $g(u)$ to have a minimum for instability.

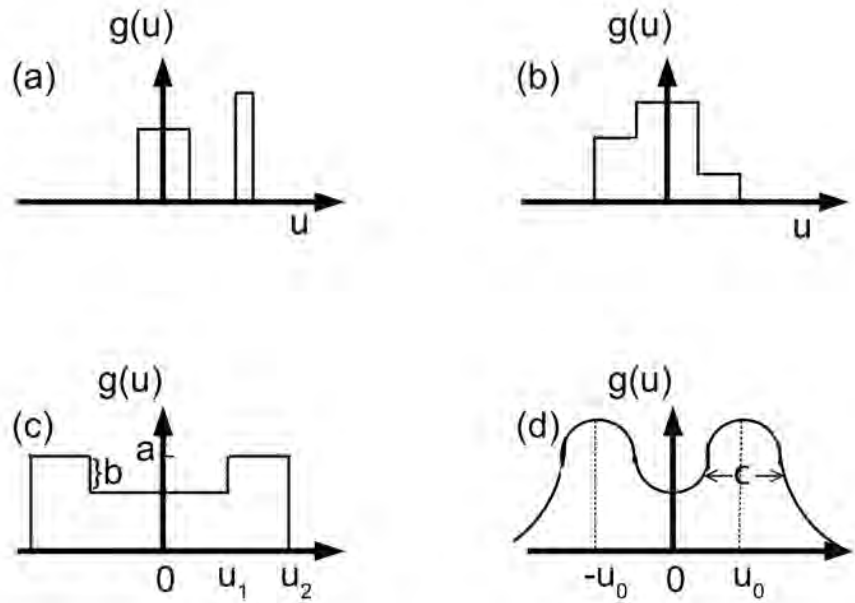

Fig. 2.H.1 Four examples of finite-temperature velocity distribution functions.

Exercise: Show that $\omega(k,c) = \omega(k,0) - ikc$ for the double Cauchy beam, which illustrates the stabilizing influence of the thermal spread.

## 2.I General analysis of the dielectric response

### 2.I.a Perturbative expansion for a fast wave

Equation (2.H.1) can be evaluated easily when thermal effects are weak. For a "fast" wave, $V \gg v_{thermal}$, one can expand

$$\frac{1}{u-v} = -\frac{1}{v}\left(\frac{1}{1-\frac{u}{v}}\right) = -\frac{1}{v}\left(1 + \frac{u}{v} + \frac{u^2}{v^2} + \frac{u^3}{v^3} + \ldots\right) \qquad (2.H.2)$$

Theorem: From Eqs.(2.E.23) and (2.H.2) the first thermal correction to Z(v) is then

$$Z(v) = -\frac{1}{v}\left(1 + \frac{v_{th}^2}{v^2} + O\left(\frac{v_{th}^3}{v^3}\right)\right) \quad (2.H.3)$$

where $v=\omega/k$ and $[1, 0, v_{th}^2] = \int du\, [1, u, u^2]\, g(u)$, and

$$Z'(v) = \frac{1}{v^2} + \frac{3v_{th}^2}{v^4} + \ldots \quad (2.H.4)$$

Hence,

$$\varepsilon(k,\omega) = 1 - \frac{\omega_p^2}{\omega^2} - \frac{\omega_p^2 3k^2 v_{th}^2}{\omega^4} + O(\omega^{-5}) \quad (2.H.5)$$

and the Bohm-Gross dispersion relation for the electron plasma wave results from $\varepsilon(k,\omega)=0$:

$$\omega^2 = \omega_p^2 + \frac{3k^2 v_{th}^2}{\omega^2}\omega_p^2 + \ldots \approx \omega_p^2 + 3k^2 v_{th}^2 \quad (2.H.6)$$

With the inclusion of thermal effects the electron plasma wave has acquired dispersion.

Theorem: The product of the phase and group velocities for the electron plasma wave is given by $v_p v_g = 3v_{th}^2$ for $kv_{th} \ll \omega_p$, i.e., $k\lambda_D \ll 1$, where $\lambda_D = v_{th}/\omega_p$ is the Debye length. With $v_p \gg v_{th}$ the group velocity satisfies $v_g \ll v_{th}$.

### 2.I.b Use of the Hilbert transform in deriving the dielectric function

We return to the consideration of Z(v), Eq.(2.E.23),

$$Z(v) = \int_{-\infty}^{\infty} du \frac{g_{\hat{k}}(u)}{u - v_R - iv_I} = \int_{-\infty}^{\infty} du \frac{g_{\hat{k}}(u)}{(u - v_R)^2 + v_I^2}\left[u - v_R + iv_I\right], \quad v = v_R + iv_I, \quad v_I > 0$$

$$\rightarrow \operatorname{Re} Z(v) = \int_{-\infty}^{\infty} du \frac{g_{\hat{k}}(u)}{(u - v_R)^2 + v_I^2}(u - v_R), \quad \operatorname{Im} Z(v) = v_I \int_{-\infty}^{\infty} du \frac{g_{\hat{k}}(u)}{(u - v_R)^2 + v_I^2} \quad (2.I.1)$$

One notes that Re Z(v) has odd symmetry with respect to $u-v_R$ and goes to zero for $v_I=0$ and for $v_R \to \pm\infty$.

Theorem: As $v_I \to 0$ there is a near singularity in Re Z(v) at $v_R=0$ while Im Z(v) $\propto$ $v_I \to 0$. In the limit $v_I \to 0$, Z(v) takes on the following special forms:

$$Z(v_R) = P\int du \frac{g(u)}{u - v_R} + i\int du\, g(u)\pi\delta(u - v_R) \quad (2.I.2)$$

where $P$ denotes the principal value of the integral that immediately follows it. The development of Eq.(2.I.2) has a basis in distribution theory:[4]

$$\lim_{y \to 0\pm} \frac{1}{x+iy} = P\left(\frac{1}{x}\right) \mp i\pi\delta(x) \tag{2.I.3}$$

Corollary: Hilbert transform

$$Z(v) = i\int_0^\infty ds \exp(ivs) \int_{-\infty}^\infty du\, g(u)\exp(-ius) = \int_{-\infty}^\infty du\, \frac{g(u)}{u-v} \tag{2.I.4}$$

where the first integral is a Laplace transform and the second integral is a Fourier transform.

Exercise: Invert the Hilbert transform to obtain $g(u)$ given $Z(v)$.

Theorem: Relation of the dielectric function to $Z'(v)$ from Eq.(2.H.1)

$$\varepsilon(\mathbf{k},\omega) = 1 - \frac{\omega_p^2}{k^2} Z'(v) = 1 - \frac{\omega_p^2}{k^2} \int du\, \frac{g_{\hat{\mathbf{k}}}'(u)}{u-v}$$

$$= 1 - \frac{\omega_p^2}{k^2} \int du\, g_{\hat{\mathbf{k}}}'(u) \left[\frac{P}{u-v} + i\pi\delta(u-v)\right], \quad v \equiv \frac{\omega}{k} \tag{2.I.5}$$

Exercise: Show that $\varepsilon' = \mathrm{Re}\,\varepsilon$ is an even function of $\omega$ and $\varepsilon'' = \mathrm{Im}\,\varepsilon$ is an odd function of $\omega$.

### 2.I.c Dispersion relation for weak damping or growth rate compared to the real part of the frequency

Consider $\omega = \omega_R + i\gamma$ and conditions such that $|\gamma| = |\mathrm{Im}\,\omega| \ll |\mathrm{Re}\,\omega|$ and Taylor-series expand $\varepsilon(\mathbf{k},\omega) = 0$ for small $\gamma$:

$$\varepsilon(\mathbf{k},\omega_k) = \varepsilon'(\mathbf{k},\omega_k) + i\varepsilon''(\mathbf{k},\omega_k) + i\gamma \left.\frac{\partial}{\partial\omega}\varepsilon(\mathbf{k},\omega)\right|_{\omega_R} + O(\gamma^2) = 0$$

$$\to \quad \varepsilon'(\mathbf{k},\omega_k) - \gamma \left.\frac{\partial}{\partial\omega}\varepsilon''(\mathbf{k},\omega)\right|_{\omega_R} + O(\gamma^2) = 0 \quad \varepsilon''(\mathbf{k},\omega_k) + \gamma \left.\frac{\partial}{\partial\omega}\varepsilon'(\mathbf{k},\omega)\right|_{\omega_R} + O(\gamma^2) = 0$$

$$\tag{2.I.6}$$

Theorem: For weak damping or weak growth rates the solution of $\varepsilon(\mathbf{k},\omega) = 0$ yields solutions for $\omega_R(\mathbf{k})$ and $\gamma(\mathbf{k})$:

---

[4] M. J., Lighthill, *Introduction to Fourier analysis and generalised functions*, New York: Cambridge University Press, (1958). ISBN 0-521-05556-3.

$$\varepsilon'(\mathbf{k},\omega_R)=0 \to \omega_R = \ldots \quad \gamma(\mathbf{k}) = -\frac{\varepsilon''(\mathbf{k},\omega_R)}{\left.\frac{\partial}{\partial \omega}\varepsilon'(\mathbf{k},\omega)\right|_{\omega_R}} \quad (2.\mathrm{I}.7)$$

The ratio of the damping rate to the frequency $\gamma(\mathbf{k})/\omega_R(\mathbf{k})$ is a measure of the ratio of the dissipation (negative or positive) to the wave energy:

$$\frac{\gamma(\mathbf{k})}{\omega_R} = -\frac{\pi g'(v)}{\frac{\omega_R}{k}P\int du \frac{g''(u)}{u-v}} = -\frac{\pi g'(v)}{vZ''_R(v)} \quad (2.\mathrm{I}.8)$$

Exercise: Derive the derivative of the principal part of the Hilbert transform.

### 2.I.d Maxwellian velocity distribution function – electron Landau damping in fast and slow waves

We next construct the dispersion relation for a Maxwellian electron velocity distribution function:

$$g(u) = \frac{1}{\sqrt{2\pi v_{th}^2}}\exp\left(-\frac{u^2}{2v_{th}^2}\right), \quad g'(u) = -\frac{u}{v_{th}^3\sqrt{2\pi}}\exp\left(-\frac{u^2}{2v_{th}^2}\right), \quad v_{th}^2 = \frac{T_e}{m_e} \quad (2.\mathrm{I}.9)$$

Waves in a warm plasma can be classified in three categories: fast, intermediate, and slow depending on the ratio of the phase velocity to the thermal velocity. For a fast wave $v=\omega/k \gg v_{th}$, $\gamma \propto \exp(-v^2/2v_{th}^2)$ is exponentially small. For intermediate waves $v \sim v_{th}$, $\gamma \sim \omega_k$; and $\gamma$ is relatively large, which invalidates the Taylor series expansion in Eq.(2.I.6) leading to Eqs.(2.I.7) and (2.I.8). For slow waves $v \ll v_{th}$, $\gamma \propto v/v_{th}$ and is linearly small.

Example: Electron plasma wave. For a fast wave in a Maxwellian plasma, the phase velocity falls far out on the high energy tail of the velocity distribution function. From Eqs.(2.I.7-2.I.9) one obtains

$$\varepsilon' \to 1-\frac{\omega_p^2}{\omega^2}, \quad \omega\frac{\partial \varepsilon'}{\partial \omega} \approx 2\frac{\omega_p^2}{\omega^2} \approx 2, \quad \frac{\gamma(\mathbf{k})}{\omega_R} = -\sqrt{\frac{\pi}{8}}\left(\frac{v}{v_{th}}\right)^3 \exp\left(-\frac{v^2}{2v_{th}^2}\right) \quad (2.\mathrm{I}.10)$$

The damping rate given in Eq.(2.I.10) is the Landau damping rate for the electron plasma wave in a Vlasov plasma (collisionless). Equation (2.H.6) gives the thermal correction to the real part of the frequency for the electron plasma wave. For small $k\lambda_D \ll 1$, the phase velocity $v=\omega/k \to \infty$ as $k \to 0$, while $v_g \to 0$; and $\gamma/\omega \to 0$ because the number of resonant particles in the tail is exponentially small. For increasing $k$, the wave frequency increases while the phase velocity decreases, which increases the damping rate and ultimately invalidates the fast-wave assumption.

Example: Ion acoustic wave. For a slow wave in a Maxwellian plasma, $v \ll v_{the}$, the phase velocity falls near the peak of the electron velocity distribution at low velocities. We assume that the ions are singly charged and relatively cold, $T_i \ll T_e$, and $v_i \ll v \ll v_{the}$. In these limits $v = c_s = (T_e/m_i)^{1/2}$, ion Landau damping is exponentially small, and the electron Landau damping is linearly small. In this two-species plasma $g(u) = \sum_s \omega_s^2 g_s(u)/\omega_p^2$. For cold ions the ion acoustic wave dispersion relation is derived from

$$\varepsilon'(\mathbf{k},\omega) = 1 - \frac{\omega_{pi}^2}{\omega^2} + \frac{1}{k^2 \lambda_e^2} = 0 \rightarrow \omega_k^2 = \frac{k^2 \lambda_e^2 \omega_{pi}^2}{1+k^2 \lambda_e^2} = \frac{k^2 c_s^2}{1+k^2 \lambda_e^2}$$

$$\omega \frac{\partial \varepsilon'}{\partial \omega}\bigg|_{\omega_k} = 2\left(1+\frac{1}{k^2 \lambda_e^2}\right) \qquad \frac{\gamma(k)}{\omega(k)} = -\sqrt{\frac{\pi}{8}}\left(\frac{m_e}{m_i}\right)^{1/2}\left(\frac{1}{1+k^2 \lambda_e^2}\right)^{3/2} \qquad (2.\text{I}.11)$$

The electron Landau damping for the ion acoustic wave is small for all values of $k\lambda_e$.

### 2.I.e Bump-on-tail velocity distribution and resonant instability

The bump-on-tail velocity tail velocity distribution is diagrammed in Fig. 2.I.1 This velocity distribution is obviously related to the weak beam case studied in Sec. 2.F. However, here both the plasma and the beam have acquired finite thermal spreads. If the phase velocity of an electron plasma wave falls on an interval of the velocity distribution of the beam with positive slope there can be instability. Equations (2.I.7) and (2.I.8) can be used to compute the growth or damping rate for

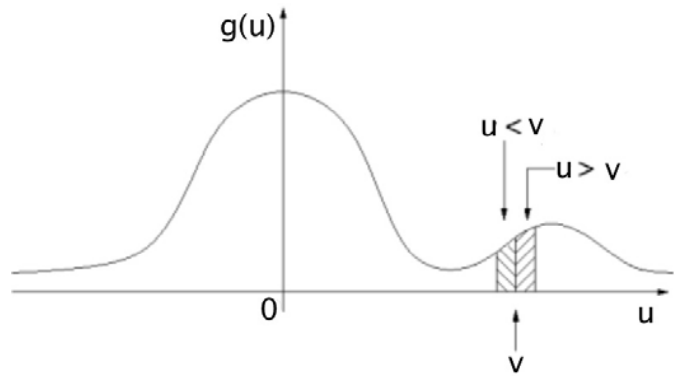

Fig. 2.I.1 Velocity distribution function for the bump-on-tail instability

the bump-on-tail distribution:

$$\frac{\gamma}{\omega_k} \approx \frac{\omega_p^2}{k^2}\frac{\pi g'(v)}{2} \tag{2.I.12}$$

where we have used $\omega_k \partial \varepsilon'/\partial \omega = 2\omega_p^2/\omega_k^2 \sim 2$, which is good for a relatively small bump on the tail. If $g'(v)>0$ then $\gamma>0$, i.e., there is instability.

<u>Definition</u>: Resonant instability or damping.  When there are particles whose unperturbed velocity is coincident or nearly coincident with the phase velocity of a wave (velocity resonance), the particles can resonantly interact with the wave and exchange energy.  In the wave frame, $\omega-\mathbf{k}\cdot\mathbf{v}=0$, the resonant particle sees a constant phase and can be steadily accelerated or decelerated by the wave's electric field.  When there are more resonant particles giving energy to the wave due to deceleration than particles extracting energy from the wave due to acceleration, the wave can grow; and there is resonant instability.  The slope of the velocity distribution dictates whether there are more or fewer particles faster or slower than the wave phase velocity.   If the slope of the velocity distribution is negative, then there are more particles with velocities slower than the wave phase velocity that are accelerated by the wave; and the wave experiences resonant (Landau) damping.

We can consider the resonant particle interaction with a wave from the quantum mechanical perspective.  Suppose there are exchanges of momentum and energy with the wave given by $\Delta \mathbf{p} = \pm \hbar \mathbf{k}$  and $\Delta \mathcal{E} = \pm \hbar \omega$ that are small compared to the particle's momentum and energy ($h$ is Planck's constant, and the over bar indicates $h/2\pi$).  Then $\Delta \mathcal{E} = m\mathbf{v}\cdot \Delta \mathbf{v} = \mathbf{v}\cdot \Delta \mathbf{p}$, and substituting the expressions for $\Delta \mathbf{p}$ and $\Delta \mathcal{E}$ one immediately obtains $\omega=\mathbf{k}\cdot\mathbf{v}$, which is the resonance condition.

## 3. Vlasov-Maxwell plasma formulation

In this section the totality of Maxwell's equations are introduced in the context of a collisionless Vlasov plasma theory.

### 3.A Wave energy and Poynting theorem

Maxwell's equations in a plasma take the form

$$\nabla\times\mathbf{B} - \frac{1}{c}\frac{\partial \mathbf{E}}{\partial t} = \frac{4\pi}{c}\mathbf{j} \qquad \nabla\times\mathbf{E} + \frac{1}{c}\frac{\partial \mathbf{B}}{\partial t} = 0 \tag{3.A.1}$$

where the current **j** is the sum of externally applied currents and the currents due to free charges in the plasma.  We will ignore gravity here but otherwise will be as general as possible. We introduce a notation to emphasize that the wave phenomena are spatially and temporally varying perturbations $\delta\mathbf{B}$, $\delta\mathbf{E}$, and $\delta\mathbf{j}$. If we compute the dot product of $c\delta\mathbf{E}$ with the perturbed Ampere's law and combine with the dot product of $c\delta\mathbf{B}$ with Faraday's law, we obtain

$$-\nabla \cdot \left( \frac{c}{4\pi} \delta \mathbf{E} \times \delta \mathbf{B} \right) = \frac{\partial}{\partial t} \left( \frac{|\delta \mathbf{E}|^2}{8\pi} + \frac{|\delta \mathbf{B}|^2}{8\pi} \right) + \delta \mathbf{j} \cdot \delta \mathbf{E} \quad (3.\text{A}.2)$$

The left side of Eq.(2.A.2) is the divergence of the Poynting flux. The last term on the right is the rate of work done by the electromagnetic fields on the particles. In obtaining Eq.(2.A.2) only terms that are bilinear in the perturbed fields and currents are retained, and linear terms like $\delta \mathbf{B} \cdot \mathbf{B}_0$ are dropped because they have no finite average value when time averaged over the cycle of the perturbed fields.

Compute the volume integral of Eq.(2.A.2) with vanishing surface terms at ∞ to obtain

$$\frac{d}{dt} \int d^3\mathbf{r} \frac{|\delta \mathbf{E}|^2 + |\delta \mathbf{B}|^2}{8\pi} = -\int d^3\mathbf{r}\, \delta \mathbf{j} \cdot \delta \mathbf{E} \quad (3.\text{A}.3)$$

Equation (3.A.3) describes a balance between the change of stored electromagnetic field energy and work done by the electric field on the plasma.

## 3.B Conductivity tensor

The conductivity tensor relates the perturbed current $\delta \mathbf{j}$ to the perturbed electric field $\delta \mathbf{E}$. This is a linear relation. The analytic construction of the conductivity tensor typically begins with the assumption that the unperturbed system is stationary, but there are notable exceptions, e.g., when there are slowly varying background fields as in the solar wind and the ionosphere, or when there are rapidly oscillatory finite-amplitude fields as in laser-plasma interactions. As in the development of the dielectric response in Sec. 2.E, we begin by employing causality and writing down the general expression

$$\delta \mathbf{j}(r,t) = \int_0^\infty d\tau \int d^3\mathbf{r}'\, \vec{\sigma}(\mathbf{r};\mathbf{r}';\tau) \cdot \delta \mathbf{E}(\mathbf{r}',t-\tau) \quad (3.\text{B}.1)$$

Next we assume that perturbed electric field can be represented as a superposition of normal modes:

$$\delta \mathbf{E}(\mathbf{r},t) = \sum_\ell \exp(-i\omega_\ell t) \mathbf{E}_\ell(\mathbf{r},t) \quad (3.\text{B}.2)$$

where $\omega_\ell$ is real, $\mathbf{E}_\ell$ is slowly varying in time, $\omega_{-\ell}=-\omega_\ell$, and $\mathbf{E}_{-\ell}=-\mathbf{E}_\ell$. Use of Eq.(3.B.2) in (3.B.1) yields

$$\delta \mathbf{j}(r,t) = \sum_\ell \exp(-i\omega_\ell t) \int_0^\infty d\tau \int d^3\mathbf{r}'\, \vec{\sigma}(\mathbf{r};\mathbf{r}';\tau) \cdot \mathbf{E}_\ell(\mathbf{r}',t-\tau) \exp(i\omega_\ell \tau)$$

$$= \sum_\ell \exp(-i\omega_\ell t) \int_0^\infty d\tau \int d^3\mathbf{r}'\, \vec{\sigma}(\mathbf{r};\mathbf{r}';\tau) \cdot \left[ \mathbf{E}_\ell(\mathbf{r}',t) - \tau \frac{\partial}{\partial t} \mathbf{E}_\ell(\mathbf{r}',t) + O(\tau^2 \ddot{\mathbf{E}}_\ell) \right] \exp(i\omega_\ell \tau)$$

$$= \sum_\ell \exp(-i\omega_\ell t) \int d^3\mathbf{r}' \left[ \vec{\sigma}(\mathbf{r};\mathbf{r}';\omega_\ell) \cdot \mathbf{E}_\ell(\mathbf{r}',t) + i \left. \frac{\partial \vec{\sigma}(\mathbf{r};\mathbf{r}';\omega)}{\partial \omega} \right|_{\omega_\ell} \cdot \frac{\partial}{\partial t} \mathbf{E}_\ell(\mathbf{r}',t) + ... \right]$$

$$(3.\text{B}.3)$$

where we have made use of the following definition and identity.

Definition: The Fourier transformed conductivity tensor is given by

$$\vec{\sigma}(\mathbf{r},\mathbf{r}';\omega) = \int_0^\infty d\tau\, \vec{\sigma}(\mathbf{r},\mathbf{r}';\tau) \exp(i\omega\tau) \quad \text{with } \vec{\sigma}(\mathbf{r},\mathbf{r}';\tau<0) = 0$$

Theorem:

$$\frac{\partial}{\partial \omega} F_{\tau \to \omega}\big(g(\tau)\big) = F_{\tau \to \omega}\big(i\tau g(\tau)\big) \quad \text{where } F_{\tau \to \omega}\big(g(\tau)\big) \text{ is the Fourier transform from } \tau \text{ to } \omega.$$

The conductivity is a three-dimensional tensor and a kernel, but it is not Hermitian in general. One can decompose any tensor into Hermitian and anti-Hermitian parts:

Theorem: $\mathbf{A} = \mathbf{A}^t + i\mathbf{A}^u$ $\quad A^t_{\mu\nu} = \frac{1}{2}(A_{\mu\nu} + A^*_{\mu\nu})$ $\quad A^t_{\mu\nu} = \big(A^t_{\nu\mu}\big)^*$ $\quad A^u_{\mu\nu} = \big(A^u_{\nu\mu}\big)^*$ For real $\omega$,

$$\sigma^t_{\mu\nu}(\mathbf{r},\mathbf{r}';\omega) \equiv \frac{1}{2}\big(\sigma_{\mu\nu}(\mathbf{r},\mathbf{r}';\omega) + \sigma_{\nu\mu}(\mathbf{r}',\mathbf{r};\omega)\big)$$

Use of Eq.(3.B.3) in Eq.(3.A.3) yields

$$\frac{d}{dt}\int d^3\mathbf{r}\, \frac{|\delta\mathbf{E}|^2 + |\delta\mathbf{B}|^2}{8\pi} = -\int d^3\mathbf{r}\, \delta\mathbf{j}\cdot\delta\mathbf{E}$$

$$= -\int d^3\mathbf{r}\int d^3\mathbf{r}' \sum_\ell \exp(i\omega_\ell t)\mathbf{E}^*_\ell(\mathbf{r},t) \cdot \sum_{\ell'} \exp(-i\omega_{\ell'}t)\left[\vec{\sigma}(\mathbf{r};\mathbf{r}';\omega_{\ell'})\cdot\mathbf{E}_{\ell'}(\mathbf{r}',t) + i\left.\frac{\partial \vec{\sigma}(\mathbf{r};\mathbf{r}';\omega)}{\partial \omega}\right|_{\omega_{\ell'}} \cdot \frac{\partial}{\partial t}\mathbf{E}_{\ell'}(\mathbf{r}',t)\right]$$

(3.B.4)

If we perform a coarse-grain time average of Eq.(3.B.4), i.e., we average over the characteristic time scales $\sim 1/\omega_\ell$, but retain the slow time scale variations, and make use of

$$\big\langle \exp(-i(\omega_\ell - \omega_{\ell'})t)\big\rangle = \delta_{\ell\ell'}$$

we can delete the sum over $\ell'$ and set $\ell = \ell'$. Thus, Eq.(3.B.4) simplifies to

$$\frac{d}{dt}\int d^3\mathbf{r}\frac{|\delta\mathbf{E}|^2+|\delta\mathbf{B}|^2}{8\pi}=-\int d^3\mathbf{r}\,\delta\mathbf{j}\cdot\delta\mathbf{E}$$

$$=-\int d^3\mathbf{r}\int d^3\mathbf{r}'\sum_\ell \mathbf{E}_\ell^*(\mathbf{r},t)\cdot\left[\vec{\sigma}(r,r';\omega_\ell)\cdot\mathbf{E}_\ell(\mathbf{r}';t)+i\left.\frac{\partial\vec{\sigma}(\mathbf{r};\mathbf{r}';\omega)}{\partial\omega}\right|_{\omega_\ell}\cdot\frac{\partial}{\partial t}\mathbf{E}_\ell(\mathbf{r}',t)\right]$$

$$=-\int d^3\mathbf{r}\int d^3\mathbf{r}'\sum_\ell \mathbf{E}_\ell^*(\mathbf{r},t)\cdot\left[\vec{\sigma}'(r,r';\omega_\ell)\cdot\mathbf{E}_\ell(\mathbf{r}';t)-\left.\frac{\partial\vec{\sigma}''(\mathbf{r};\mathbf{r}';\omega)}{\partial\omega}\right|_{\omega_\ell}\cdot\frac{\partial}{\partial t}\mathbf{E}_\ell(\mathbf{r}',t)\right]$$

$$=-\int d^3\mathbf{r}\int d^3\mathbf{r}'\sum_\ell \mathbf{E}_\ell^*(\mathbf{r},t)\cdot\vec{\sigma}'(r,r';\omega_\ell)\cdot\mathbf{E}_\ell(\mathbf{r}';t)$$

$$+\frac{1}{2}\frac{d}{dt}\int d^3\mathbf{r}\int d^3\mathbf{r}'\sum_\ell \mathbf{E}_\ell^*(\mathbf{r},t)\cdot\left.\frac{\partial\vec{\sigma}''(\mathbf{r};\mathbf{r}';\omega)}{\partial\omega}\right|_{\omega_\ell}\cdot\mathbf{E}_\ell(\mathbf{r}',t)$$

(3.B.5)

where $\sigma = \sigma' + i\sigma''$

Exercise: Prove Eq.(3.B.5) using the hermiticity of $\sigma'$ and $\sigma''$, and the right side of (3.B.5) is necessarily real.

## 3.C Energy conservation

Theorem: Eq.(3.B.5) can be rewritten in the form that expresses energy conservation between the electromagnetic field energy and the plasma

$$\frac{d}{dt}\sum_\ell U_\ell(t)=-\sum_\ell Q_\ell \rightarrow \frac{d}{dt}U_\ell(t)=-Q_\ell$$

(3.C.1)

using mode by mode equivalence. Equation (3.B.5) becomes

$$\frac{d}{dt}\left[\int d^3\mathbf{r}\sum_\ell\frac{|\mathbf{E}_\ell|^2+|\mathbf{B}_\ell|^2}{8\pi}-\frac{1}{2}\int d^3\mathbf{r}\int d^3\mathbf{r}'\sum_\ell \mathbf{E}_\ell^*(\mathbf{r},t)\cdot\left.\frac{\partial\vec{\sigma}''(\mathbf{r};\mathbf{r}';\omega)}{\partial\omega}\right|_{\omega_\ell}\cdot\mathbf{E}_\ell(\mathbf{r}',t)\right]$$

$$=-\int d^3\mathbf{r}\int d^3\mathbf{r}'\sum_\ell \mathbf{E}_\ell^*(\mathbf{r},t)\cdot\vec{\sigma}'(r,r';\omega_\ell)\cdot\mathbf{E}_\ell(\mathbf{r}';t)$$

(3.C.2)

and the equality also holds in a mode by mode sense. The left side of Eq.(3.C.2) is the time derivative of the stored energy summed over the electromagnetic field energy and the plasma kinetic sloshing energy in the waves. The right side of Eq.(3.C.2) is the rate of energy lost or gained due to resistive effects (loss in stable plasma). The imaginary part of the conductivity $\sigma''$ is the reactive component, and the real part $\sigma'$ is the resistive component.

Theorem: In a uniform medium Eqs.(3.C.1) and (3.C.2) can be Fourier transformed to **k**-space:

$$\frac{d}{dt}\int \frac{d^3\mathbf{k}}{(2\pi)^3}\sum_\ell U_\ell(\mathbf{k},t) = -\int \frac{d^3\mathbf{k}}{(2\pi)^3}\sum_\ell Q_\ell(\mathbf{k},t) \tag{3.C.3a}$$

where

$$U_\ell(\mathbf{k},t) = \frac{|\mathbf{E}_\ell|^2 + |\mathbf{B}_\ell|^2}{8\pi} - \mathbf{E}^*_\ell(\mathbf{k},t)\cdot\left.\frac{\partial\vec{\sigma}''(\mathbf{k};\omega)}{2\partial\omega}\right|_{\omega_\ell(\mathbf{k})}\cdot \mathbf{E}_\ell(\mathbf{k},t) \tag{3.C.3b}$$

$$Q_\ell(\mathbf{k},t) = \mathbf{E}^*_\ell(\mathbf{k},t)\cdot\vec{\sigma}'(\mathbf{k};\omega_\ell(\mathbf{k}))\cdot \mathbf{E}_\ell(\mathbf{k},t)$$

$U_\ell(\mathbf{k},t)$ is the wave energy, which can be positive or negative depending on the particle energy contribution. $Q_\ell(\mathbf{k},t)$ can be positive or negative.

The continuity equation (2.D.6) relates the charge density and the current, while the charge density and the electric field are related in Eq.(2.E.5). The susceptibility $\chi(\mathbf{k},\omega)$ is related to the dielectric function $\varepsilon(\mathbf{k},\omega)$ in Eq.(2.E.16). Using these relations the dielectric function and the conductivity are related as follows:

$$\vec{\varepsilon}(\mathbf{k},\omega) = \vec{\mathbf{I}} + i\frac{4\pi}{\omega}\vec{\sigma}(\mathbf{k},\omega) = \vec{\varepsilon}' + i\vec{\varepsilon}'' = \vec{\mathbf{I}} + i\frac{4\pi}{\omega}(\vec{\sigma}' + i\vec{\sigma}'')$$

$$\vec{\varepsilon}' = \vec{\mathbf{I}} - \frac{4\pi}{\omega}\vec{\sigma}'' \qquad \vec{\varepsilon}'' = \frac{4\pi}{\omega}\vec{\sigma}'' \qquad \frac{\partial}{\partial\omega}(\omega\vec{\varepsilon}') = \vec{\mathbf{I}} - 4\pi\frac{\partial}{\partial\omega}\vec{\sigma}'' \tag{3.C.4}$$

Theorem: Eqs.(3.C.3) and (3.C.4) can be combined to yield the following alternative form for $U_\ell(\mathbf{k},t)$ in terms of the real part of $\varepsilon(\mathbf{k},\omega)$:

$$U_\ell(\mathbf{k},t) = \frac{|\mathbf{B}_\ell|^2}{8\pi} + \frac{\mathbf{E}^*_\ell(\mathbf{k},t)\cdot\left.\frac{\partial}{\partial\omega}(\omega\vec{\varepsilon}'(\mathbf{k};\omega))\right|_{\omega_\ell(\mathbf{k})}\cdot \mathbf{E}_\ell(\mathbf{k},t)}{8\pi} \tag{3.C.5}$$

## 3.D Coulomb model examples – cold plasma, Vlasov plasma, beam in hot plasma

In this section we simplify the analysis to the Coulomb model and analyze a few examples. In a Coulomb model the waves are longitudinal, i.e., the electric field is given by the gradient of a scalar potential:

$$\mathbf{E} = -\nabla\phi \qquad \mathbf{E}(\mathbf{k},\omega) = -i\mathbf{k}\phi(\mathbf{k},\omega) \qquad \nabla\times\mathbf{E} = 0$$

$$U_\ell(\mathbf{k},t) = \frac{k^2}{8\pi}\left.\frac{\partial}{\partial\omega}(\omega\hat{\mathbf{k}}\cdot\vec{\varepsilon}'(\mathbf{k};\omega)\cdot\hat{\mathbf{k}})\right|_{\omega_\ell(\mathbf{k})}|\phi_\ell(\mathbf{k},t)|^2 \tag{3.D.1}$$

Definition: $\varepsilon^L(\mathbf{k},\omega) \equiv \mathbf{k}\cdot\hat{\varepsilon}\cdot\mathbf{k}$

Exercise: Prove $\varepsilon^L$ is equivalent to $\varepsilon(\mathbf{k},\omega)$ in Sec. 2.E

Corollary:
$$U_\ell(\mathbf{k},t) = \frac{k^2|\phi_\ell(\mathbf{k},t)|^2}{8\pi} \frac{\partial}{\partial \omega}(\omega \varepsilon')\bigg|_{\omega_\ell(\mathbf{k})} \qquad Q_\ell(\mathbf{k},t) = \frac{\omega k^2 |\phi_\ell(\mathbf{k},t)|^2}{8\pi} \varepsilon''(\mathbf{k};\omega_\ell(\mathbf{k})) \qquad (3.D.2)$$

Theorem: For $U_\ell \propto \exp(2\gamma_\ell(\mathbf{k})t)$ then $\gamma_\ell(\mathbf{k})$ is determined by the ratio of the dissipation to the wave energy:

$$\gamma_\ell(\mathbf{k}) = -\frac{1}{2}\frac{Q_\ell(\mathbf{k})}{U_\ell(\mathbf{k})} = -\frac{\omega \varepsilon''}{\frac{\partial}{\partial\omega}(\omega\varepsilon')} = -\frac{\varepsilon''}{\partial\varepsilon'/\partial\omega}, \quad \varepsilon'(\omega_\ell(\mathbf{k}))=0 \qquad (3.D.3)$$

Example: In a Vlasov plasma $\varepsilon'(\mathbf{k},\omega) = 1 - \frac{\omega_p^2}{k^2}Z'(v) \equiv 1 - \frac{F(v)}{k^2} \quad v \equiv \frac{\omega}{k}$. The eigenfrequencies are determined by $\varepsilon'=0$, i.e., $F(v)=k^2$ introduced in Eq.(2.F.2). We also have the following useful relations

$$\frac{\partial \varepsilon'}{\partial \omega} = -\frac{1}{k^3}\frac{dF}{dv} \qquad \frac{dF}{dv} = 2k\frac{dk}{dv} \qquad \frac{\partial \varepsilon'}{\partial \omega} = -\frac{2}{k^2}\left(\frac{dv}{dk}\right)^{-1}$$

$$\gamma = \frac{-\varepsilon''}{\partial\varepsilon'/\partial\omega} = -\frac{\pi}{2}\omega_p^2 g'(v)\frac{dv}{dk} \qquad \frac{dv}{dk} = \frac{1}{k}(v_g - v) \qquad v_g \equiv \frac{d\omega}{dk} \qquad (3.D.4)$$

Theorem: Examination of Eq.(3.D.1) – (3.D.4) shows that the wave energy $U_\ell$ and the growth/damping rate $\gamma$ are independent of reference frame under Galilean transformation ($v_g$-v is invariant under Galilean transformation).

Usually $v_g<v$, which is referred to as normal dispersion, and then $\frac{\partial \varepsilon'}{\partial \omega} > 0$ so that the wave energy is positive. We next summarize some results for a few examples.

Examples
1. Electron plasma wave (Langmuir wave): $\omega\frac{\partial \varepsilon'}{\partial \omega} = 2$ and, hence, $U=2|\mathbf{E}|^2/8\pi$. Energy is partitioned equally between the field energy and the plasma kinetic sloshing energy.
2. Ion-acoustic wave: $\omega\frac{\partial \varepsilon'}{\partial \omega} = 2\left(1 + \frac{1}{k^2\lambda_e^2}\right) \to \infty$ for $k^2\lambda_e^2 \to 0$. The electrons and ions are in phase with one another, and the plasma kinetic sloshing energy is much larger than the field energy.

3. Cold beam: $\epsilon = 1 - \omega_b^2/(\omega - ku_b)^2$ and $\omega \frac{\partial \epsilon'}{\partial \omega} = \pm 2ku_b/\omega_b$. The two branches of the dispersion relation are the fast wave (positive energy wave) and the slow wave (negative energy wave).
4. Beam through a hot plasma: $\epsilon = 1 - (\omega_p^2/k^2) Z'(v) - \omega_b^2/(\omega - ku_b)^2$. For $ku_b \gg \omega_p$, there are fast and slow beam modes with approximate eigenfrequencies $\omega = ku_b \pm \omega_b$ and $v = u_b \pm \omega_b/k$. The beam component is a $\delta$-function in velocity, while the plasma component is a Maxwellian centered at zero velocity. The slope of the velocity distribution for the plasma component is negative at $u_b \pm \omega_b/k$ and the resonant particles damp the fast wave. However, the wave energy for the slow mode is negative; and the resonant particles destabilize the slow mode in consequence of Eq.(3.B.13). The dissipation $\epsilon''$ is positive for both the fast and slow mode, but dissipation further decreases the wave energy for negative energy waves which increases the magnitude of the slow-wave amplitude.

### 3.E Penrose criterion for instability and examples

We note that $F(V)=k^2$ in Eq.(2.F.2) generalized to (2.H.2) depends on the principal value integral of $g'$; and the expression for $\gamma$ in Eq.(3.B.14) depends directly on $g'(v)$:

$$\varepsilon(\mathbf{k},\omega) = 0 \rightarrow k^2 = \omega_p^2 Z'(V) \equiv F(V) = \omega_p^2 \int du \frac{g'(u)}{u-V}, \quad \text{Im } V > 0$$

$$\gamma(\mathbf{k}) = -\frac{\pi}{2} \omega_p^2 \frac{dV}{dk} g'(V) \tag{3.E.1}$$

We have analyzed the stability of several examples of Vlasov plasmas in the Coulomb model, e.g., two stream, plasma beam, beam and plasma, bump on tail, and Maxwellian plasma. Can we look at $g(u)$ and decide whether there is instability? Penrose answered this question in the affirmative.[5]

Theorem: (Penrose criterion) The condition for instability is that as one varies $k$, for some real $k>0$, Im$(\omega/k)>0$

Examine Eq.(3.E.1) in the complex V plane. We are looking for unstable solutions for the complex phase velocity V as a function of $k^2$ with Im V>0. For Im V>0, $F$ in Eq.(3.E.1) is well behaved and cannot go to infinity for any value of $k^2$. As $k^2 \rightarrow \infty$, there are no roots for Im V > 0, i.e., there are no unstable roots for $k=\pm\infty$; and $F$ goes to zero for $k=\pm\infty$. Now consider solutions of Eq.(3.E.1) for V with finite $k$ where V crosses the real V axis from below as we vary $k$. Label the point as $k_0$ where $\gamma_0(k_0)=0$, which implies $g'=0$ from Eq.(3.E.1). Consider a positive energy wave $dV/dk<0$. As $k$ decreases differentially from $k_0$, V increases differentially from its finite value at $k_0$; and $\gamma$ transitions from $\gamma<0$ to $\gamma=0$, i.e., $d\gamma/dk<0$ at $k_0$ is required for instability. In consequence of $d\gamma/dk<0$ and Eq.(3.E.1), $g'$ must increase and become positive as $k$ is reduced below $k_0$ to support instability, which requires that $g''>0$ where $g'=0$ at $k_0$, i.e., $g$ must have a minimum with respect to V. For a negative energy wave $dV/dk>0$, and V decreases differentially from its finite value at $k_0$ as $k$ is reduced below $k_0$. However, $d\gamma/dk \propto -(dV/dk)^2 g''$ remains negative if $g''>0$, as required for instability. We note that with $d\gamma/dk<0$ for either sign of $dV/dk$, a differential increase in $k$ from $k_0$ leads to $\gamma<0$.

---

[5] O. Penrose, Phys. Fluids, **3**(2), 258–265 (1960).

Theorem: A necessary condition for instability is a minimum in $g(u)$. Penrose shows sufficiency as well if i) $g'=0$, ii) $g''>0$, and iii) $\int du\, g'(u)/(u-V) >0$. The implication of this is that a weak minimum may not be unstable, but a deep minimum probably will be unstable.

[*Editor's note: There are good discussions of the Penrose criterion in textbooks such as in Sec. 9.6 of Nicholas A. Krall and Alvin W. Trivelpiece, Principles of Plasma Physics, McGraw-Hill, 1973, and in Sec. 5.3 of P. M. Bellan, Fundamentals of Plasma Physics, Cambridge University Press, 2008 where a nice mathematical argument is given making use of Nyquist's theorem.*]

Examples:
i. A single-hump velocity distribution is always stable.
ii. A double Cauchy beam (Sec. 2.H) can be unstable if the beam velocity $u_0$ exceeds the thermal spread $v_{th}$.
iii. Several cold beams are always unstable.
iv. A cold beam in a hot plasma is unstable.
v. An isotropic single-hump distribution of speed is stable. Some spherical shell distributions of speed are stable.
vi. Electrons and ions with a finite-temperature Maxwellian velocity distribution and a relative drift between the species can be unstable if the relative drift is sufficiently large (ion acoustic instability).

Example: The velocity distributions for the ion acoustic instability are diagrammed in Fig. 3.E.1 The electrons have a finite temperature $T_e$, and the ion temperature is $T_i$. The singly charged ions have a drift $u_d$ relative to the electrons. In order to have a minimum in the composite distribution $g(u)$, $g(u) = \sum_s \omega_s^2 g_s(u) / \sum_s \omega_s^2$, there is some minimum value of $u_d$. The condition $\int du\, g'(u)/(u-V) >0$ requires an even larger value of $u_d$, which we define as $u_{d,crit}$ which is a function of $T_e$, $T_i$, and $m_e/m_i$.

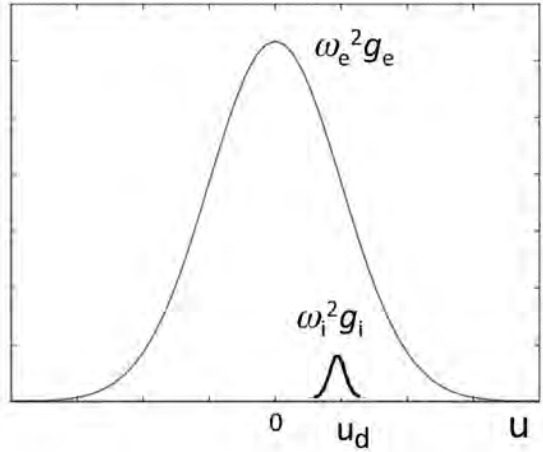

Fig.3.E.1 Hot electron and ion velocity distributions with a relative drift $u_d$.

Calculations published by Bernstein and Kulsrud[6] and Fried and Gould[7] obtained the following results:

$$\frac{u_{d,crit}}{v_i} \approx \left\{ 4 \quad \text{for } \frac{T_e}{T_i} \geq 20, \quad \left(\frac{T_e}{2T_i}\right)^{1/2}\left[1+\sqrt{\frac{m_i}{m_e}}\left(\frac{T_e}{T_i}\right)^{3/2}\exp(-\frac{T_e}{2T_i})\right] \quad \text{for } 1 \leq \frac{T_e}{T_i} \leq 20 \right\} \quad (3.E.2)$$

$u_{d,crit}/v_i$ as a function of $T_i/T_e$ steadily increases from a value equal to 4 for $T_i/T_e=0$ with $u_{d,crit}$ saturating at a value $O(v_e)$ for $T_i/T_e=O(1)$, and the instability evolves into a warm two-stream instability.

## 3.F Wave momentum

In Sec. 3.A a Poynting theorem was derived that describes the dynamics of field and plasma energy exchange. Here we examine the exchange of momentum between fields and plasma. The starting point is Eq.(2.D.9) summed over species, which we then coarse-grain time average over high frequencies having decomposed all fluid quantities and fields in a linear superposition of modes as in Eq.(3.B.2). Only bilinear terms survive the averaging, and we determine that the irreversible momentum transfer from waves to plasma per unit time and per volume is given by

$$\langle\delta\rho\delta\mathbf{E}\rangle+\frac{1}{c}\langle\delta\mathbf{j}\times\delta\mathbf{B}\rangle=\langle\delta\rho\delta\mathbf{E}\rangle+\left\langle\frac{1}{\omega}\delta\mathbf{j}\times(\mathbf{k}\times\delta\mathbf{E})\right\rangle$$

$$=\langle\delta\rho\delta\mathbf{E}\rangle+\left\langle\frac{1}{\omega}\mathbf{k}(\delta\mathbf{j}\cdot\delta\mathbf{E})-\frac{1}{\omega}\delta\mathbf{E}(\mathbf{k}\cdot\delta\mathbf{j})\right\rangle=\langle\delta\rho\delta\mathbf{E}\rangle+\left\langle\frac{1}{\omega}\mathbf{k}(\delta\mathbf{j}\cdot\delta\mathbf{E})-\frac{1}{\omega}\delta\mathbf{E}(\omega\delta\rho)\right\rangle \quad (3.F.1)$$

$$=\frac{\mathbf{k}}{\omega}\langle\delta\mathbf{j}\cdot\delta\mathbf{E}\rangle$$

in a mode by mode sense, and we have made use of Faraday's law $-i\omega\delta\mathbf{B}=-c i\mathbf{k}\times\delta\mathbf{E}$ and the continuity equation $-i\omega\delta\rho=-i\mathbf{k}\cdot\delta\mathbf{j}$.

---

[6] I. B. Bernstein and R. M. Kulsrud, Phys. Fluids **3**, 937 (1960).
[7] B. D. Fried and R. W. Gould, Phys. Fluids **4,** 139 (1961).

Theorem: The irreversible momentum transfer from waves to the plasma is given by $\frac{\mathbf{k}}{\omega}$ times the irreversible energy transfer, i.e., [8]

$$\frac{\mathbf{g}_{wave}}{U_{wave}}=\frac{\mathbf{k}}{\omega} \quad \rightarrow \quad \frac{d}{dt}\mathbf{g}_{wave}=\frac{\mathbf{k}}{\omega}\frac{d}{dt}U_{wave} \quad (3.F.2)$$

There is an analogy here to quantum mechanics wherein $U=N\hbar\omega$ and $\mathbf{g}=N\hbar\mathbf{k}$.

Theorem: ($U_{wave}$, $\mathbf{g}_{wave}$) are related as ($E$, $\mathbf{p}$), i.e., these are four-vectors. Thus, if we consider a transformation to a reference frame translating with relative velocity $\mathbf{w}$, then $\omega^{(1)}=\omega^{(0)}-\mathbf{k}\cdot\mathbf{w}$, $U_{wave}^{(1)}=U_{wave}^{(0)}-\mathbf{w}\cdot\mathbf{g}$, and $\left(\omega\,\partial\epsilon/\partial\omega\right)^{(1)}=\left(\omega\,\partial\epsilon/\partial\omega\right)^{(0)}\left(1-\mathbf{k}\cdot\mathbf{w}/\omega^{(0)}\right)$, where the dielectric function $\varepsilon$ is an invariant scalar.

# 4. Magnetic fields

In this section we examine the solution of the Vlasov-Maxwell system including both externally applied magnetic fields and perturbed magnetic fields associated with electromagnetic waves.

## 4.A Response tensor for Vlasov-Maxwell system and general dispersion relation

We Fourier transform Faraday's law and Ampere's equations in space and time to obtain the following relations:

---

[8] R. Dewar, Phys. Fluids **13**, 2710 (1970).

Faraday's law: $\dfrac{\partial \delta \mathbf{B}}{\partial t} = -c\nabla \times \delta \mathbf{E} \;\rightarrow\; -i\omega \delta \mathbf{B} = -ic\mathbf{k}\times\delta\mathbf{E} \;\rightarrow\; \delta\mathbf{B}(\mathbf{k},\omega) = \mathbf{n}\times\delta\mathbf{E}(\mathbf{k},\omega)\quad \mathbf{n}\equiv \dfrac{\mathbf{k}c}{\omega}$

Ampere's law: $\nabla\times\delta\mathbf{B} - \dfrac{1}{c}\dfrac{\partial \delta\mathbf{E}}{\partial t} = \dfrac{4\pi}{c}\left(\delta\mathbf{j}+\delta\mathbf{j}^{ext}\right) \rightarrow i\mathbf{k}\times\delta\mathbf{B}+\dfrac{i\omega}{c}\delta\mathbf{E} = \dfrac{4\pi}{c}\left(\delta\mathbf{j}+\delta\mathbf{j}^{ext}\right) = \dfrac{4\pi}{c}\left(\vec{\sigma}(\mathbf{k},\omega)\cdot\delta\mathbf{E}+\delta\mathbf{j}^{ext}\right)$

(4.A.1)

<u>Theorem</u>: From Faraday's law the transverse (divergence-free) component of the linearized electric field $\delta\mathbf{E}^T$ satisfies $\delta\mathbf{E}^T = -\dfrac{1}{c}\mathbf{V}\times\delta\mathbf{B}$ where $\mathbf{V}\equiv \hat{\mathbf{k}}\omega/k$ is the wave phase velocity. For slow waves V/c<<1, the transverse waves are mostly magnetic, $|\delta\mathbf{B}|/|\delta\mathbf{E}|>>1$.

<u>Theorem</u>:

$$i\mathbf{k}\times(\mathbf{n}\times\delta\mathbf{E}) + \dfrac{i\omega}{c}\left[\vec{\mathbf{I}}+\dfrac{4\pi}{-i\omega}\vec{\sigma}(k,\omega)\right]\cdot\delta\mathbf{E} = i\mathbf{k}\times(\mathbf{n}\times\delta\mathbf{E}) + \dfrac{i\omega}{c}\vec{\varepsilon}(k,\omega)\cdot\delta\mathbf{E} = \dfrac{4\pi}{c}\delta\mathbf{j}^{ext}$$

(4.A.2)

$$\rightarrow \vec{\varepsilon}(k,\omega)\cdot\delta\mathbf{E}+\mathbf{n}\times(\mathbf{n}\times\delta\mathbf{E}) = \dfrac{4\pi}{i\omega}\delta\mathbf{j}^{ext} \;\rightarrow\; \left[\vec{\varepsilon}(k,\omega) - n^2\left(\vec{\mathbf{I}}-\hat{\mathbf{k}}\hat{\mathbf{k}}\right)\right]\cdot\delta\mathbf{E} = \dfrac{4\pi}{i\omega}\delta\mathbf{j}^{ext}$$

<u>Definition</u>: $\vec{\mathbf{I}}^T \equiv \vec{\mathbf{I}}-\hat{\mathbf{k}}\hat{\mathbf{k}}$ projects out the transverse part of the vector field.

<u>Definition</u>: The response tensor is defined by Eq.(4.A.2): $\vec{\kappa}\equiv \vec{\varepsilon}-n^2\vec{\mathbf{I}}^T(\hat{\mathbf{k}})$ and hence, $\vec{\kappa}\cdot\delta\mathbf{E}(\mathbf{k},\omega) \equiv (4\pi/i\omega)\delta\mathbf{j}^{ext}(\mathbf{k},\omega)$

<u>Theorem</u>: (i) Normal modes are determined by the solution of Eq.(4.A.2) with $\mathbf{j}^{ext}=0$. From $\vec{\kappa}\cdot\delta\mathbf{E}(\mathbf{k},\omega)=0$ one can make some statements regarding polarization. The dispersion relation for normal modes is obtained from $D(\mathbf{k},\omega) = \det|\vec{\kappa}|=0 \;\rightarrow\; \omega_\ell(\mathbf{k})$ or $k_\ell(\hat{\mathbf{k}},\omega)$.

(ii) With an external source $\delta\mathbf{j}^{ext}$, $\delta\mathbf{E}(\mathbf{k},\omega)$ is determined by matrix inversion:

$$\delta\mathbf{E}(\mathbf{k},\omega) \equiv \dfrac{4\pi}{i\omega}\vec{\kappa}^{-1}\cdot\delta\mathbf{j}^{ext}(\mathbf{k},\omega) \quad \vec{\kappa}^{-1} = \dfrac{\vec{\kappa}^A}{D(\mathbf{k},\omega)}$$

(4.A.3)

where $\vec{\kappa}^A$ is the adjoint of the response tensor. To compute the inverse Fourier-Laplace inverse transform of Eq.(4.A.3) we will need to know the poles of the integrand, i.e., the roots of $D(\mathbf{k},\omega)=0$.

So far the discussion was completely general until we introduced the conductivity tensor and certain simplifying assumptions (see Sec. 3.B). In the rest of Sec. 4 we will include an externally applied magnetic field and analyze particle motion and waves in a magnetized plasma, and introduce the study of waves in a nonuniform medium. Topics such as magnetic confinement of a plasma, macro-instabilities, and drift waves are addressed in Sec. 8.

## 4.B Waves propagating parallel to a magnetic field in a cold, uniformly magnetized plasma with collisions

Consider the linearized fluid equations for cold electrons in a uniform magnetic field:

$$m\dot{\mathbf{v}} = e\left(\delta\mathbf{E} + \frac{1}{c}\delta\mathbf{v}\times\mathbf{B}_0\right) - m\nu_c \delta\mathbf{v} \Rightarrow -i(\omega + i\nu_c)\delta\mathbf{v}(\omega) = \frac{e}{m}\delta\mathbf{E}(\omega) \mp (\delta\mathbf{v}\times\mathbf{\Omega}) \quad (4.B.1)$$

where $\nu_c$ is the collision frequency and $\mathbf{\Omega} = -eB_0\hat{\mathbf{z}}/mc$. Negatively charged electrons gyrate counterclockwise around the magnetic field, and ions gyrate clockwise.

<u>Definition</u>: (Circular polarization eigenvectors) For a general vector field define

$$A_0 = A_z \quad A_\pm \equiv \frac{1}{\sqrt{2}}(A_x \mp iA_y) \quad \hat{\mathbf{e}}_\pm \equiv \frac{1}{\sqrt{2}}(\hat{\mathbf{e}}_x \mp i\hat{\mathbf{e}}_y) \quad \hat{\mathbf{e}}_0 \equiv \hat{\mathbf{e}}_z \quad A_l \equiv \hat{\mathbf{e}}_l^* \cdot \mathbf{A} \quad \hat{\mathbf{e}}_l^* \cdot \hat{\mathbf{e}}_j = \delta_{lj} \quad (4.B.2)$$

The linear eigenmodes of any field can be decomposed by using the definitions in Eq.(4.B.2). In this representation $\mathbf{A}\cdot\mathbf{B} = A_+B_- + A_-B_+ + A_0B_0$.

<u>Theorem</u>: $-i(\omega + i\nu_c - \ell\Omega)\delta v_\ell = \frac{e}{m}\delta E_\ell$, $\ell = 0, \pm 1$ in $\omega$ space and this is diagonal in this representation. The conductivity is

$$\sigma_{\ell\ell} = \frac{\delta j_\ell}{\delta E_\ell} = \frac{\sum_s n_s e_s \delta v_\ell^s}{\delta E_\ell} = i\sum_s \frac{n_s e_s^2 \delta v_\ell^s}{m_s} \frac{1}{\omega + i\nu_{cs} - \ell\Omega_s} \quad (4.B.3)$$

We note that for electrons and $\ell = +1$ with $\Omega_e > 0$, then $\sigma$ can diverge when $\omega + i\nu_{ce} - \Omega_e = 0$. With $E_\pm \propto \exp(-i\Omega_e t) = \cos(\Omega_e t) - i\sin(\Omega_e t) \propto E_x - iE_y$ and $\ell = +1$ corresponds to *right-circular polarization*. For ions and $\ell = -1$ with $\Omega_i < 0$, $\sigma$ can diverge for $\omega + i\nu_{ci} + \Omega_i = 0$ and $\ell = -1$ corresponds to *left-circular polarization*.

<u>Theorem</u>: The dielectric function resulting from the conductivity in Eq.(4.B.3) is

$$\varepsilon_\ell(\omega) \equiv 1 + \frac{4\pi i}{\omega}\sigma_\ell(\omega) = 1 - \sum_s \frac{\omega_s^2}{\omega(\omega + i\nu_{cs} - \ell\Omega_s)} \quad (4.B.4)$$

<u>Example</u>: For a two-species plasma with $n_e = n_i$ and $\nu = 0$, then $\omega_p^2 = \omega_e^2 + \omega_i^2$ and

$$\varepsilon_\ell(\omega) = 1 - \frac{\omega_p^2}{(\omega - \ell\Omega_e)(\omega - \ell\Omega_i)} \quad (4.B.5)$$

We examine Eq.(4.B.5) in two limits, viz., low and high frequency:

$$\varepsilon_{\pm}(\omega=0)=1-\frac{\omega_p^2}{\Omega_e\Omega_i}=1+\frac{4\pi\rho c^2}{B_0^2}=\varepsilon_{\perp} \quad \text{and} \quad \varepsilon(\omega=\infty)=1 \qquad (4.B.6)$$

At this point we recall certain definitions and expressions derived earlier.

<u>Definitions</u>: $\vec{\varepsilon}(\mathbf{k},\omega)\equiv\vec{\mathbf{I}}+\frac{4\pi i}{\omega}\vec{\sigma}(\mathbf{k},\omega)$ and $\vec{\mathbf{I}}^T(\mathbf{k})=\vec{\mathbf{I}}-\hat{\mathbf{k}}\hat{\mathbf{k}}$ (4.B.7)

Plasma electrodynamics:

$$\mathbf{K}(\mathbf{k},\omega)=\vec{\varepsilon}(\mathbf{k},\omega)-n^2(\mathbf{k},\omega)\vec{\mathbf{I}}^T(\mathbf{k})$$
$$\mathbf{K}(\mathbf{k},\omega)\cdot\delta\mathbf{E}(\mathbf{k},\omega)=\frac{4\pi}{i\omega}j^{ext}(\mathbf{k},\omega) \qquad (4.B.8)$$
$$D(\mathbf{k},\omega)=\det|\mathbf{K}(\mathbf{k},\omega)|=0$$

An analysis of Eqs.(4.B.7) and (4.B.8) in a cold plasma will elucidate the directional and $\omega$ dependence of $n(\mathbf{k},\omega)=kc/\omega\Leftrightarrow v_\phi$ and $n(\theta,\omega)$ where $\theta$ is the angle between $\mathbf{k}$ and $\mathbf{B}_0$. In a cold plasma the frequencies $\omega_{ps}^2=4\pi n_s e^2/m_s$ and $\Omega_s=e_s B_0/m_s c$ are independent of energy, which simplifies the frequency dependence in the linear dielectric functions and dispersion relations. In addition, the Debye length is zero and the wavenumber dependences of the dielectric tensor and the dispersion relation are further simplified. When the phase velocity of a wave in the presence of a beam is comparable to the beam velocity, the physics becomes sensitive to thermal effects. The cold-plasma limit removes the electron pressure and ion waves; and resonant particle effects vanish (Landau damping and cyclotron damping). Cyclotron harmonics are also gone.

<u>Example</u>: Consider wave propagation parallel to an applied magnetic field, $\hat{\mathbf{k}}=\hat{\mathbf{B}}_0=\hat{\mathbf{z}}=\hat{\mathbf{e}}_0$. Using $\vec{\mathbf{I}}^T=\hat{\mathbf{e}}_+\hat{\mathbf{e}}_+^*+\hat{\mathbf{e}}_-\hat{\mathbf{e}}_-^*$ to project field vectors onto a plane transverse to the magnetic field, one can show that

$$\mathbf{K}(\mathbf{k},\omega)=\vec{\varepsilon}(\mathbf{k},\omega)-n^2(\mathbf{k},\omega)\vec{\mathbf{I}}^T(\mathbf{k}) \rightarrow \mathbf{K}=\left[\varepsilon_+-n^2\right]\hat{\mathbf{e}}_+\hat{\mathbf{e}}_+^*+\left[\varepsilon_--n^2\right]\hat{\mathbf{e}}_-\hat{\mathbf{e}}_-^*+\varepsilon_0\hat{\mathbf{e}}_0\hat{\mathbf{e}}_0$$
$$D(\mathbf{k},\omega)=\det|\mathbf{K}(\mathbf{k},\omega)|=0 \rightarrow D=\left(\varepsilon_+-n^2\right)\left(\varepsilon_--n^2\right)\varepsilon_0=0 \qquad (4.B.9)$$

**4.B.a Right-hand circularly polarized waves: whistler, magnetosonic and extraordinary waves**

Consider a right-circularly polarized wave, $\ell=+$, with $\omega>0$ and no collisions $\nu=0$: the dispersion relation is

$$n_R^2(\omega)=\varepsilon_+(\omega)=1-\frac{\omega_p^2}{(\omega-\Omega_e)(\omega+|\Omega_i|)} \qquad (4.B.10)$$

<u>Definitions</u>: There is a *resonance* at $\omega = \Omega_e$ where $n=\infty$ and $v_\phi=0$. As a function of frequency there is an abrupt change at $\omega = \Omega_e$ from propagation ($\omega < \Omega_e$) to evanescence ($\omega > \Omega_e$). There is a *cutoff* where $n=0$, $v_\phi=\infty$, and

$$\omega = \omega_{RC} = \frac{\Omega_e}{2} + \sqrt{\left(\frac{\Omega_e}{2}\right)^2 + \omega_p^2}$$ (ignoring ion contributions to Eq.(4.B.10)).

Figures 4.B.1 and 4.B.2 sketch solutions for the dispersion relation for right - circularly polarized waves propagating parallel to a uniform applied magnetic field in a cold plasma.

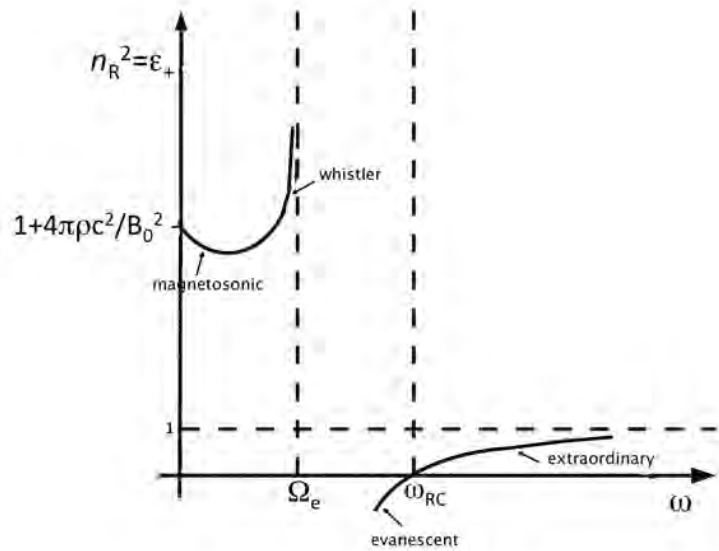

Fig. 4.B.1 Schematic of the solutions of Eq. (4.B.10) for waves right-circularly polarized wave propagating parallel to **B**$_0$.

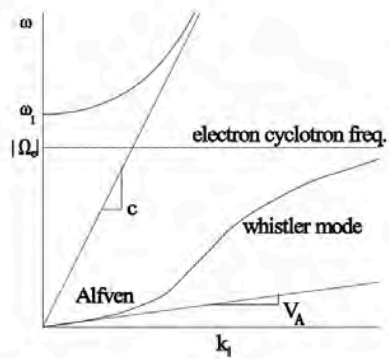

Fig. 4.B.2 Parallel modes with positive helicity. Electron cyclotron resonance ($k_\parallel \to \infty$) occurs at $\omega \lesssim |\Omega_e|$. The non-dispersive low frequency mode $\omega \ll \Omega_i$ ($\ll |\Omega_e|$) is the Alfven mode described by $\omega = k_\parallel V_A$. (due to A. Hirose)

A.Hirose, Physics 862 lecture notes, Chapt. 6, University of Saskatchewan; http://physics.usask.ca/~hirose/P862/notes.htm

Exercise: Include a finite, but small collision frequency in Eq. (4.B.10). Consider frequencies of the order of $\Omega_e$. Sketch the real and imaginary parts of $\varepsilon_+$. Calculate and sketch the real and imaginary parts of $k$ for real $\omega$. Calculate Im($\omega$) as a function of real $k$.

### 4.B.b Left-hand circularly polarized waves: Alfvén, ion cyclotron, and ordinary

Consider left-circularly polarized waves propagating parallel to the magnetic field. Equation (4.B.5) yields the following dispersion relation.

Theorem: Left-circularly polarized waves, $\ell = -$, with $\omega > 0$ and no collisions $\nu = 0$, resonate with ions and satisfy the dispersion relation

$$n_R^2(\omega) = \varepsilon_-(\omega) = 1 - \frac{\omega_p^2}{(\omega + \Omega_e)(\omega - |\Omega_i|)} \qquad (4.B.11)$$

Figures 4.B.3 and 4.B.4 sketch the solutions of the dispersion relation in Eq. (4.B.11).

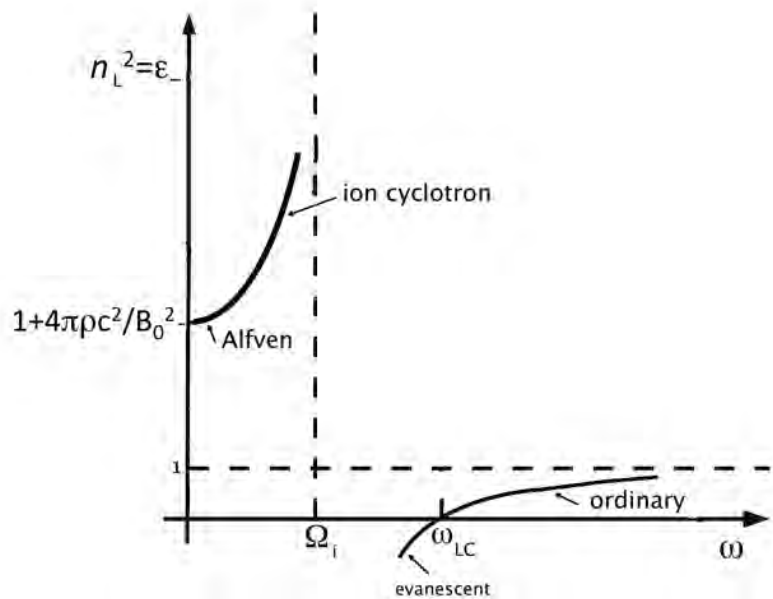

Fig. 4.B.3 Schematic of the solutions of Eq. (4.B.11) for waves left-circularly polarized wave propagating parallel to $\mathbf{B}_0$.

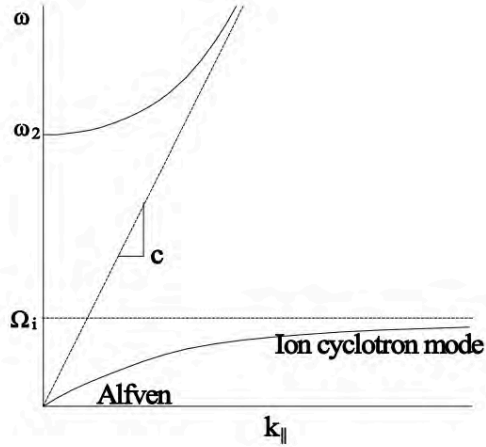

Fig. 4.B.4 Parallel modes with negative helicity. The Alfven mode $\omega = V_A k_\parallel$ with negative helicity exists in the low frequency region $\omega \ll \Omega_i$. The ion cyclotron resonance occurs at $\omega \lesssim \Omega_i$. The cutoff frequency $\omega_2$ is given by $\omega_2 = \left(\sqrt{4\omega_{pe}^2 + \Omega_e^2} - |\Omega_e|\right)/2$. (due to A. Hirose)

A.Hirose, Physics 862 lecture notes, Chapt. 6, University of Saskatchewan;
http://physics.usask.ca/~hirose/P862/notes.htm

### 4.B.c Electron plasma waves

With $\ell = 0$ the solution of Eq. (4.B.5) yields longitudinal waves with $\delta \mathbf{E} || \mathbf{B}_0, \mathbf{k}$ and $\varepsilon_0(\omega) = 0 = 1 - \omega_p^2/\omega^2 \;\rightarrow\; \omega = \pm \omega_p$, i.e., electron plasma waves.

### 4.C Waves propagating at an angle with respect to the magnetic field

Here we generalize the consideration of waves propagating at an arbitrary angle with respect to an applied magnetic field. Define a polar angle $\theta$ between the vector $\mathbf{k}$ and $\mathbf{B}_0$ with an azimuthal angle $\phi$ with respect to the magnetic field aligned with $\mathbf{z}$. Then $\hat{\mathbf{k}} = \cos\theta \hat{\mathbf{z}} + \sin\theta(\cos\phi \hat{\mathbf{x}} + \sin\phi \hat{\mathbf{y}})$, and we express this in terms of $\hat{\mathbf{e}}_\pm = (\hat{\mathbf{x}} \pm \hat{\mathbf{y}})/\sqrt{2}$ and $\hat{\mathbf{e}}_0 = \hat{\mathbf{z}}$ to evaluate $\hat{\mathbf{k}}\hat{\mathbf{k}}$ in Eqs.(4.B.7) and (4.B.8). The dispersion relation determined by Eq.(4.B.8) becomes

$$\det(\ddot{\mathbf{K}}) = D(\omega, n^2, \theta) = n^4 A(\omega, \theta) - n^2 B(\omega, \theta) + C(\omega) \qquad (4.C.1)$$

which does not depend on the angle $\phi$. We recall the definitions of $\varepsilon_0$, $\varepsilon_+$ and $\varepsilon_-$ in Sec. 4.B and introduce $\varepsilon_1 \equiv \frac{1}{2}(\varepsilon_+ + \varepsilon_-) = 1 - \sum_s \frac{\omega'}{\omega} \frac{\omega_s^2}{\omega'^2 - \Omega_s^2}, \omega' = \omega + i\nu$ to obtain

$$A(\omega,\theta) = \varepsilon_1(\omega)\sin^2\theta + \varepsilon_0(\omega)\cos^2\theta$$

$$B(\omega,\theta) = \varepsilon_+\varepsilon_-\sin^2\theta + \varepsilon_1\varepsilon_0(1+\cos^2\theta)$$

$$C(\omega,\theta) = \det(\ddot{\mathbf{e}}) = \varepsilon_+\varepsilon_-\varepsilon_0 \qquad (4.C.2)$$

$$n^2(\omega,\theta) = \frac{B \pm \sqrt{B^2 - 4AC}}{2A}$$

Exercise: With $\nu=0$ show that $B^2-4AC>0$ which implies that Eqs.(4.C.1) and (4.C.2) have real roots for $\omega$.

We can examine the solutions of Eq.(4.C.1) for $n^2$ as a function of $\omega$ for finite $\theta$, which is just a quadratic. First one looks for resonances ($n^2\to\infty$). One class of resonances corresponds to $A\to 0$, for which $n_+^2\to\infty$ and $n_-^2=C/B$. We introduce the following useful definitions.

Definitions: The upper and lower hybrid frequencies are

$$\omega_{uh}^2 = \omega_{pe}^2 + \Omega_e^2 \quad \text{and} \quad \omega_{lh}^2 = \frac{\omega_{pi}^2 + \Omega_i^2}{1 + \omega_{pe}^2/\Omega_e^2} \qquad (4.C.3)$$

In many laboratory experiments $\Omega_i^2 \ll \omega_{pi}^2$ in the lower hybrid frequency. A schematic of the resonances determined by the solutions of $A=0$ as a function of angle $\theta$ for $\omega_{pe}>\Omega_e$ is shown in Fig. 4.C.1. Figure 4.C.2 is a schematic of the dispersion relation $n^2$ vs. $\omega$ for $\theta \neq 0$ determined by solutions of Eq.(4.C.1). Figure 4.C.2 is a schematic for the solution of $n^2$ vs. $k$ for $\theta \neq 0$, and Fig. 4.C.3 is a schematic for the solution of $\omega$ vs. $k$.

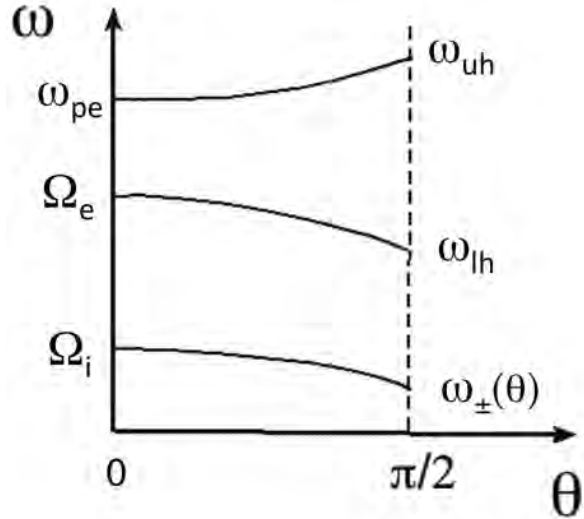

Fig. 4.C.1 Schematic of the solutions of $A=0$ in Eq.(4.C.2) for frequencies $\omega$ vs. angle $\theta$ where resonances $n_-^2\to\infty$ occur.

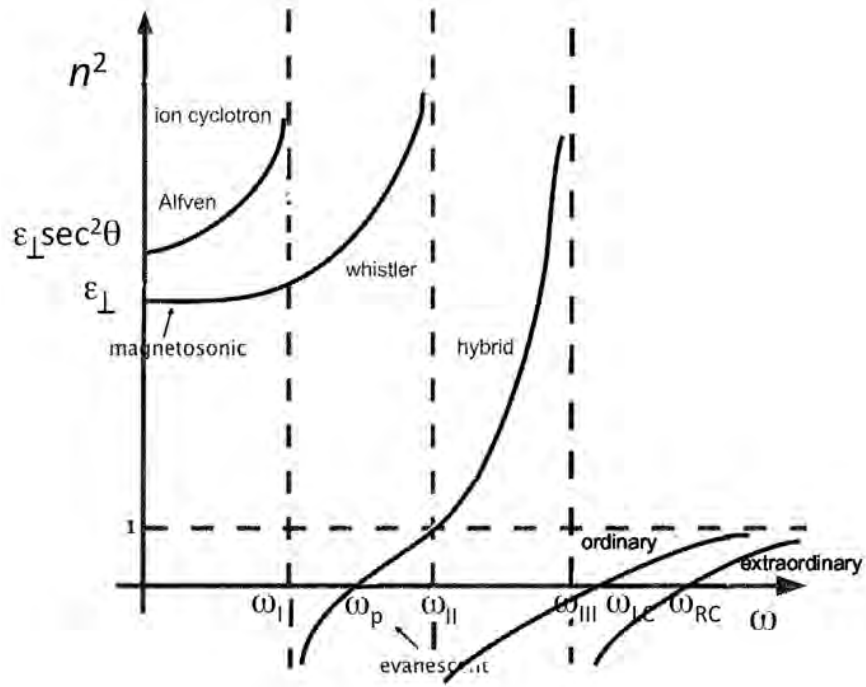

Fig. 4.C.2 Schematic of the solutions of Eq.(4.C.1) for $n^2$ vs. $\omega$ with $\theta \neq 0$.

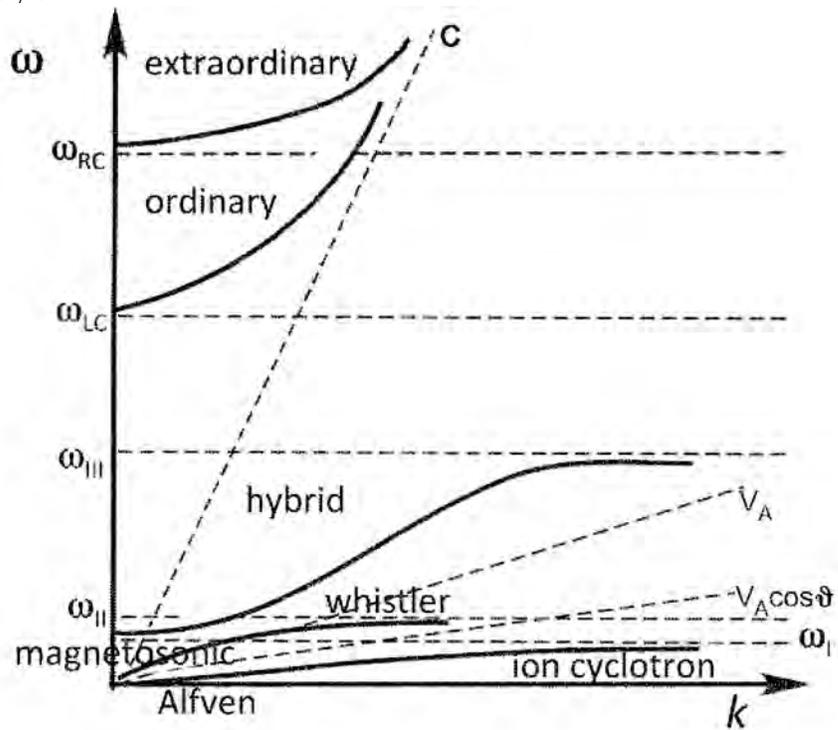

Fig. 4.C.3 Schematic of solutions of Eq.(4.C.1) for $\omega$ vs. $k$ with $\theta \neq 0$.

Theorem: In a cold plasma at low frequencies $\varepsilon_\pm \Rightarrow 1 + 4\pi\rho c^2/B_0^2$, $\varepsilon_0 = 1 - \omega_p^2/\omega^2$, $n_+^2 \to \varepsilon_\perp \sec^2\theta$ and $n_-^2 \to \varepsilon_\perp$. For any $k$ there are five branches for $\omega$. There are cutoffs, $n^2=k=0$, when $C=0$, i.e., $\omega = \omega_{RC}$ where $\varepsilon_+ = 0$, $\omega = \omega_{LC}$ where $\varepsilon_- = 0$, and $\omega = \omega_{pe}$ where $\varepsilon_0 = 0$. Resonances are determined by $n=k=\infty$ which derive from the poles of $\varepsilon_\pm$. The polarization of the waves is determined by the solutions of **K**•δ**E**=0 and det**K**=0 for the eigenvectors

Examples: **i.** What distinguishes the Alfvén from the magnetosonic wave? An Alfvén wave can be viewed as a plucked magnetic field line. The tension in the field line is given by $B_0^2/4\pi$. The dispersion relation is $\omega = kV_A \cos\theta = k_z V_A$, and the group velocity $\mathbf{V}_g = \partial\omega/\partial\mathbf{k} = \hat{\mathbf{z}} V_A$. The oscillation propagates along the field line. The magnetosonic wave is a solution of $n_-^2 = \varepsilon_\perp$: $\omega = kV_A$. In a cold plasma the magnetosonic wave is dispersionless: the phase and group velocities are equal to $V_A$, and the wave is isotropic. **ii.** The ordinary (O) and extraordinary (X) modes exhibit characteristic birefringence, i.e., their index of refraction $n$ depends on their polarization. At low plasma density, the O and X modes become light waves.

Exercise: How does $n^2$ depend on $\theta$ for fixed $\omega$? Look at textbooks such T. J. M. Boyd and J. J. Sanderson, *Plasma Dynamics* (1969) or T. H. Stix, *The Theory of Plasma Waves* (1962). Of particular interest is the concept of the ray velocity surface (wave normal surface) and the Clemmow-Mullaly-Allis (CMA) diagram. At resonances $n=k=\infty$ show that δ**E** is parallel to **k**. Consider the dispersion relation for the whistler wave in the frequency range $\Omega_i \ll \omega << \omega_{pe}, \Omega_e$ in a cold plasma, which is $n^2(\omega,\theta) = k^2 c^2/\omega^2 = \omega_{pe}^2/(\omega\Omega_e \cos\theta)$. Calculate the group velocity **V**g=$d\omega/d$**k** and plot it as a function of $\theta$. The group velocity **V**g makes and angle $\gamma$ with respect to the phase velocity **V**ϕ=ω/k. Show that $\tan\gamma = (\partial/\partial\theta)\ln k(\omega,\theta)$. Show that the energy in a whistler wave is within a 20° angle with respect to **B**0. Make a polar plot of the phase velocity **V**ϕ, as a function of $\theta$, which when rotated around the axis of symmetry is the so-called wave normal surface in Fig. 4.C.4a The refraction index surface is shown in Fig. 4.C.4b [figure and caption due to R. D. Blandford and K. S. Thorne, http://www.pmaweb.caltech.edu/Courses/ph136/yr2012/, Sec. 21.7; K. S. Thorne and R. D. Blandford, *Modern Classical Physics* (Princeton University Press, 2017), Fig. 21.6].

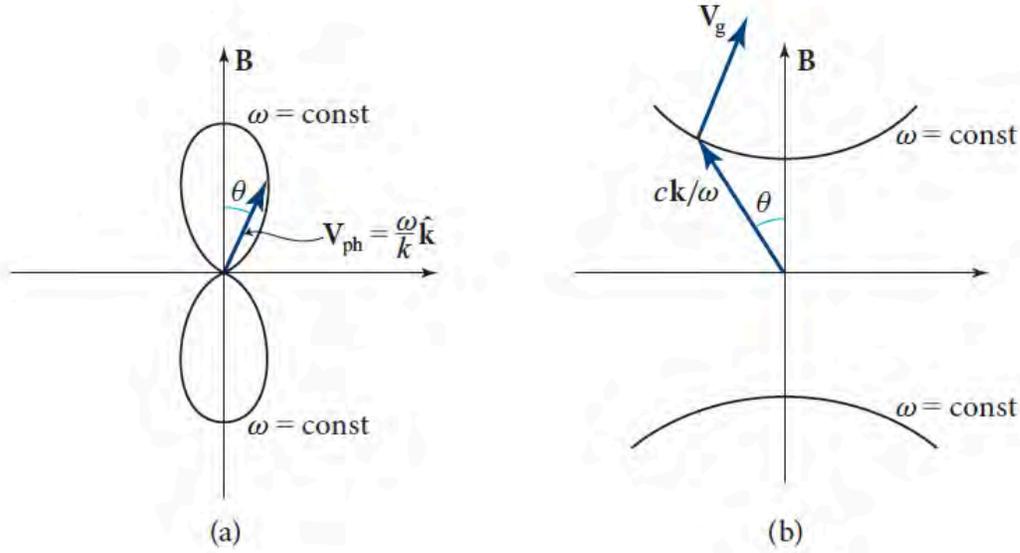

Fig. 4.C.4 (a) Wave normal surface for a whistler mode propagating at an angle $\theta$ with respect to the magnetic field direction. The phase velocity $\mathbf{V}_{ph} = (\omega/k)\hat{\mathbf{k}}$ as a vector from the origin, with the direction of the magnetic field chosen upward. When we fix the frequency ω of the wave, the tip of the phase velocity vector sweeps out the figure-8 curve as its angle $\theta$ to the magnetic field changes. This curve should be thought of as rotated around the vertical (magnetic field) direction to form the figure-8 "wave-normal" surface. Note that there are some directions where no mode can propagate. (b) *Refractive index surface* for the same whistler mode. Here we plot $c\mathbf{k}/\omega$ as a vector from the origin, and as its direction changes with fixed ω, this vector sweeps out the two hyperboloid-like surfaces. Since the length of the vector is $\frac{ck}{\omega} = \mathbf{n}$, this figure can be thought of as a polar plot of the refractive index **n** as a function of the wave propagation direction $\theta$ for fixed ω; hence the name "refractive index surface". The group velocity $\mathbf{V}_g$ is orthogonal to refractive-index surface. Note that for this whistler mode, the energy flow (along $\mathbf{V}_g$) is focused toward the direction of the magnetic field [from R.D. Blandford and K.S. Thorne, ph136/yr2012; K.S. Thorne and R.D. Blandford, *Modern Classical Physics* (Princeton University Press, 2017), Fig. 21.6].

### 4.D Energy transport

Consider the energy transport associated with waves in a magnetized plasma.

<u>Definition</u>: Let U be defined as the energy density of a wave packet and $\mathbf{S} = U\mathbf{V_g}$ is the energy flux density.

<u>Theorem</u>: It can be shown that

$$\mathbf{S} = -\frac{\omega}{8\pi}\delta\mathbf{E}^* \cdot \frac{\partial \mathbf{K}}{\partial \mathbf{k}} \cdot \delta\mathbf{E} \tag{4.D.1}$$

given $U = \frac{|\delta\mathbf{B}|^2}{8\pi} + \frac{\omega\delta\mathbf{E}^* \cdot \dfrac{\partial \mathbf{K}}{\partial \omega} \cdot \delta\mathbf{E}}{8\pi}$, $\mathbf{K} = \vec{\vec{\varepsilon}} - n^2(\vec{\mathbf{I}} - \hat{\mathbf{k}}\hat{\mathbf{k}})$, $\mathbf{V}_g = \frac{\partial \omega}{\partial \mathbf{k}}$, $\mathbf{K} \cdot \delta\mathbf{E} = 0$ and Maxwell's equations.

Theorem: It then follows from the definition of **K** and Maxwell's equations that

$$\mathbf{S} = \frac{c}{4\pi}\mathrm{Re}\,\delta\mathbf{E}^* \times \delta\mathbf{B} - \frac{\omega}{8\pi}\delta\mathbf{E}^* \cdot \frac{\partial \vec{\vec{\varepsilon}}}{\partial \mathbf{k}} \cdot \delta\mathbf{E} \qquad (4.\mathrm{D}.2)$$

We note that the first term on the right side of (4.D.2) is the electromagnetic Poynting flux and the second term is the energy flux density associated with particle motion.

Exercise: In a Coulomb plasma evaluate the dielectric function $\varepsilon(\mathbf{k},\omega)$ and **S**, and apply in a few limiting cases.

## 4.E Wave propagation in an inhomogeneous cold plasma

Consider small-amplitude waves in an inhomogeneous plasma. We begin with the Maxwell equations and linearize the fields and currents with respect to a field-free and current-free unperturbed system:

$$\nabla \times \mathbf{B} - \frac{1}{c}\frac{\partial \mathbf{E}}{\partial t} = \frac{4\pi}{c}\mathbf{j} \qquad \nabla \times \mathbf{E} + \frac{1}{c}\frac{\partial \mathbf{B}}{\partial t} = 0 \quad \Rightarrow$$

$$\nabla \times \delta\mathbf{B} - \frac{1}{c}\frac{\partial \delta\mathbf{E}}{\partial t} = \frac{4\pi}{c}\delta\mathbf{j} \qquad \nabla \times \delta\mathbf{E} + \frac{1}{c}\frac{\partial \delta\mathbf{B}}{\partial t} = 0 \quad \frac{\partial}{\partial t} \to -i\omega \qquad (4.\mathrm{E}.1)$$

Theorem: Upon Fourier transforming the time dependence we obtain

$$\nabla \times \left(\nabla \times \delta\mathbf{E}(\mathbf{x},\omega)\right) - \frac{\omega^2}{c^2}\delta\mathbf{E}(\mathbf{x},\omega) = 4\pi\frac{i\omega}{c^2}\delta\mathbf{j}(\mathbf{x},\omega) \qquad \delta\mathbf{j}(\mathbf{x},\omega) = \vec{\vec{\sigma}}(\mathbf{x},\omega) \cdot \delta\mathbf{E}(\mathbf{x},\omega)$$

$$\nabla \times \left(\nabla \times \delta\mathbf{E}(\mathbf{x},\omega)\right) = \frac{\omega^2}{c^2}\vec{\vec{\varepsilon}} \cdot \delta\mathbf{E}(\mathbf{x},\omega) \qquad \vec{\vec{\varepsilon}} = \vec{\vec{\mathbf{I}}} + \frac{4\pi i}{\omega}\vec{\vec{\sigma}} \qquad (4.\mathrm{E}.2)$$

in which we have introduced the conductivity and dielectric tensors. In some special cases the physical system may possess sufficient symmetry to allow separation of the dependence of the fields on the spatial variables. In the most general circumstances one cannot separate the spatial variables and approximate methods are employed.

### 4.E.a WKB eikonal method

In the WKB (Wentzel–Kramers–Brillouin or Wentzel–Kramers–Brillouin-Jeffreys) method, all of the field quantities are decomposed in terms of a slowly varying amplitude and a rapidly varying phase factor:

Definitions:

$$\delta\mathbf{E}(\mathbf{x}) = A(\mathbf{x})\hat{\mathbf{e}}(\mathbf{x})\exp(i\Phi(\mathbf{x}))$$

i. $\hat{\mathbf{e}}(\mathbf{x})$ is the complex-valued polarization vector and is slowly varying
ii. $A(\mathbf{x})$ is a slowly varying amplitude  (4.E.3)
iii. $\exp(i\Phi(\mathbf{x}))$ is a phase factor that is a rapidly varying function of $\mathbf{x}$
iv. The slowly varying spatial scale is $L \gg \lambda$ the rapidly varying spatial scale.

Using the definitions above we obtain

$$\nabla \times \delta\mathbf{E} \approx i\nabla\Phi \times \delta\mathbf{E} = i\mathbf{k}(\mathbf{x}) \times \delta\mathbf{E}(\mathbf{x})$$
$$\nabla \times \nabla \times \delta\mathbf{E} = \nabla(\nabla \cdot \delta\mathbf{E}) - \nabla^2 \delta\mathbf{E} \approx i\nabla\Phi \times (i\mathbf{k} \times \delta\mathbf{E}) \quad (4.E.4)$$
$$= -\mathbf{k} \times \mathbf{k} \times \delta\mathbf{E} = -\mathbf{k}\,\mathbf{k} \cdot \delta\mathbf{E} + k^2 \delta\mathbf{E}$$

Theorem: Using Eqs.(4.5.2), (4.5.3), and (4.5.4), it follows that

$$\left[\ddot{\varepsilon}(\omega,\mathbf{x}) - n^2(\mathbf{x})(\ddot{\mathbf{I}} - \hat{\mathbf{k}}(\mathbf{x})\hat{\mathbf{k}}(\mathbf{x}))\right] \cdot \delta\mathbf{E}(\mathbf{x}) \equiv \ddot{\mathbf{K}}(\omega,\mathbf{k}(\mathbf{x}),\mathbf{x}) \cdot \delta\mathbf{E}(\mathbf{x}) = 0 \quad (4.E.5)$$

where $n^2 \equiv k^2 c^2/\omega^2$ and we require that $D(\omega, \mathbf{k}(\mathbf{x}), \mathbf{x}) \equiv \det|\vec{\mathbf{K}}| = 0$.

Theorem: $D(\omega, i\nabla\Phi(\mathbf{x}), \mathbf{x}) = 0$ is a differential equation.

Example: In a plane-stratified medium with $\omega_p^2(z)$ and $B_0 = 0$, $D = 0 \rightarrow \omega^2 = \omega_p^2(z) + |\nabla\Phi|^2 c^2$ and $\mathbf{k} = \nabla\Phi$. $\hat{\mathbf{k}}$ is parallel to $\hat{\mathbf{n}}$, and both are normal to constant phase surface. Given the solution $\omega(\mathbf{k}, \mathbf{x})$ to $D=0$, wave packets convect at the group velocity $V^g = \frac{\partial\omega(\mathbf{k},\mathbf{x})}{\partial \mathbf{k}}|_\mathbf{x} = \frac{d\mathbf{x}}{dt} = \frac{\mathbf{k}}{\omega}c^2 = c\sqrt{1 - \frac{\omega_p^2}{\omega^2}}$. In the atmosphere the electron charge density increases with vertical altitude going up into the ionosphere above 85 km and then decreases above 300 km. If a wave is launched into the ionosphere at a fixed frequency $\omega_0$ then the equation $D=0$ can be solved for $\mathbf{k}(\mathbf{x})$. The wave will exhibit reflection when $\omega_p(\mathbf{x}) = \omega_0$ and will refract where the wave is not evanescent.

Introduce a variation in the transverse direction $x$ in the example of the plane-stratified medium considered above. Fourier analyze in $x$ and introduce $k_x$. From the eikonal dispersion relation one obtains

$$\frac{dz}{dt} = \frac{\partial\omega}{\partial k_z} = \frac{k_z(z)c^2}{\omega} = c\sqrt{1 - \frac{\omega_p^2(z)}{\omega^2} - \frac{k_x^2 c^2}{\omega^2}}$$

$$\frac{dx}{dt} = \frac{\partial\omega}{\partial k_x} = \frac{k_x c^2}{\omega} = c\sqrt{1 - \frac{\omega_p^2(z)}{\omega^2} - \frac{k_z^2 c^2}{\omega^2}} \quad (4.E.6)$$

$$\frac{dz}{dx} = \frac{k_z(z)}{k_x}$$

Theorem: The eikonal dispersion relation $D(\omega, \mathbf{k}(\mathbf{x}), \mathbf{x})=0$ is equivalent to $\omega(\mathbf{k}, \mathbf{x}) = \omega_0$ and one can insist the total time derivative of $\omega$ vanishes:

$$0 = \frac{d\omega(\mathbf{k},\mathbf{x})}{dt} = \frac{\partial\omega}{\partial \mathbf{k}} \cdot \frac{d\mathbf{k}}{dt} + \frac{\partial\omega}{\partial \mathbf{x}} \cdot \frac{d\mathbf{x}}{dt} \rightarrow \frac{\partial\omega}{\partial \mathbf{k}} \cdot \left(-\frac{d\omega}{d\mathbf{x}}\right) + \frac{\partial\omega}{\partial \mathbf{x}} \cdot \frac{\partial\omega}{\partial \mathbf{k}} \equiv 0 \quad (4.\text{E}.7)$$

if $\dfrac{d\mathbf{x}}{dt} = \dfrac{\partial\omega}{\partial \mathbf{k}}$ and $\dfrac{d\mathbf{k}}{dt} = -\dfrac{d\omega}{d\mathbf{x}}$, the WKBJ wave-packet equations.

For only $z$ dependence, $\dfrac{d\mathbf{k}}{dt} = \dfrac{dk_z}{dt}\hat{\mathbf{k}} = -\dfrac{d\omega}{dz}\hat{\mathbf{k}}$. Because $\dfrac{d\omega}{dx} = 0$ then $\dfrac{dk_x}{dt} = 0$, i.e., $k_x$ is a constant. The equations in (4.E.7) are Hamilton's equations. We note that by multiplying the wave packet equations of motion through by $\hbar$ we have the equations of motion for quantum mechanical wave packets. If we identify the Hamiltonian $H = \hbar\omega$, we note that $\dfrac{dH}{dt} = \dfrac{\partial H}{\partial t} = \hbar\dfrac{\partial\omega(\mathbf{k},\mathbf{x},t)}{\partial t}$. There is a nice discussion of the WKBJ wave-packet equations in T. H. Stix, *The Theory of Plasma Waves,* 1962.

Consider an infinite-length wave train in the absence of sources and sinks. Then energy flux conservation dictates:

$$\frac{\partial}{\partial t}U(z) = -\frac{\partial}{\partial z}S_z = 0, \quad S_z = const. = U(z)\frac{\partial\omega}{\partial k}(z) \quad (4.\text{E}.8)$$

where $U(z)$ is is the wave energy density, and $S_z$ is the $z$-component of the wave energy flux. In consequence of Eq.(4.E.8), the square of the wave packet amplitude scales as $A^2 \sim U(z) \sim S/\left|\partial\omega/\partial k_z\right| \sim \text{const}/\left|\partial\omega/\partial k_z\right|$; thus, $A(z) \sim 1/\sqrt{\left|\partial\omega/\partial k_z\right|}$, which is useful up to a reflection point where $S_z$ and $\partial\omega/\partial k_z$ vanish.

### 4.E.b Reflection, refraction, turning points, Bohr-Sommerfeld quantization

In WKB either reflection or refraction only can occur. Partial reflection, refraction, and transmission at a sharp boundary like a glass plate cannot be described by WKB. How good is WKB? WKB can be applied to a medium with spatial variation that is slow on the scale of the characteristic wavelength of the wave. Consider a one-dimensional example (ref. Landau and Lifschitz, Quantum Mechanics, Sec. 23, Problem 3) wherein there is a transition in the square of the index of refraction from a higher value $n_0$ to a lower value $n_1$ over a length scale $L$. Assume the dispersion relation for an electromagnetic wave in an unmagnetized spatially varying plasma dielectric varies as

$$\varepsilon(x) = \frac{n_0^2 + n_1^2 e^{x/L}}{1 + e^{x/L}} \quad (4.\text{E}.9)$$

<u>Theorem</u>: For a wave packet incident from $+\infty$ in the profile in (4.E.9), the reflection coefficient, defined as the reflected power divided by the incident power, is given by[9]

$$R = \frac{\sinh^2\left[\pi \frac{L}{\bar{\lambda}}(n_0 - n_1)\right]}{\sinh^2\left[\pi \frac{L}{\bar{\lambda}}(n_0 + n_1)\right]}, \quad \bar{\lambda} \equiv \frac{c}{\omega} \qquad (4.E.10)$$

The WKB limit corresponds to $L \gg \bar{\lambda}$, and $R \to e^{-4\pi L/\bar{\lambda}}$; so the reflectivity is exponentially small in this limit. In the sharp-boundary limit, $L \ll \bar{\lambda}$, $R \to (n_0 - n_1)^2/(n_0 + n_1)^2$; and the WKB estimate is a very poor approximation.

There are many treatments of reflection of waves in spatially varying media. There are elegant discussions in J. Heading, *Introduction to Phase-Integral Methods* (Methuen, 1962); K. G. Budden, *Radio Waves in the Ionosphere* (Dover, 1961); and E. R. Tracy, A. J. Brizard, A. S. Richardson, and A. N. Kaufman, *Ray Tracing and Beyond: Phase-space Methods in Plasma Wave Theory* (Cambridge University Press, 2014). A simpler approach is presented in J. Mathews and R. Walker, *Mathematical Methods of Physics*, 2nd ed. (Benjamin, 1970). Here we follow the treatment in Mathews and Walker. Consider the one-dimensional limit of Eqs.(4.E.4) and (4.E.5):

$$\nabla \times \left(\nabla \times \delta \mathbf{E}(\mathbf{x})\right) = \frac{\omega^2}{c^2} \bar{\bar{\varepsilon}}(\mathbf{x}) \cdot \delta \mathbf{E}(\mathbf{x}) = \nabla \nabla \cdot \delta \mathbf{E}(\mathbf{x}) - \nabla^2 \delta \mathbf{E} \qquad (4.E.11)$$

Consider a purely transverse wave $\nabla \cdot \delta \mathbf{E} = 0$ in one dimension and introduce a specific polarization to obtain a scalar wave equation:

$$-\frac{d^2}{dz^2}\delta E(z) = \frac{\omega^2}{c^2}\varepsilon(z)\delta E(z) \to k(z) = \frac{2\pi}{\lambda(z)} = \frac{\omega}{c}\sqrt{\varepsilon(z)} \qquad (4.E.12)$$

Figure 4.E.1 illustrates an electromagnetic wave incident from the right in a spatially varying medium with plasma cut-off at $z=0$. Regions away from $z=0$ are accurately described by WKB. Near $z=0$ if the dielectric $\varepsilon$ goes through zero linearly in $z$, solving the differential equation in Eq.(4.E.12) yields an Airy function solution, and the Airy function is related to Bessel functions and modified Bessel functions of fractional order 1/3.

<u>Example</u>: Consider $\epsilon(z) = 1 - \omega_p^2(z)/\omega^2$ and assume for a model that the WKB dispersion relation is $\omega_p^2 = -n_1^2 \tanh\left(\frac{z}{L}\right)\omega^2 + \omega^2$ For the moment select units such that $\frac{\omega}{c} = 1$ and recall that the WKB solution for the phase function satisfies $d\Phi/dz \equiv k(z)$. The WKB solutions of Eq.(4.E.12) are then given by

$$W_\pm(z) = \frac{1}{\varepsilon^{1/4}(z)} \exp\left(\pm i \int_0^z dz' \sqrt{\varepsilon(z')}\right) \qquad (4.E.13)$$

---

[9] C. Eckart, Phys. Rev. 36, 878 (1930).

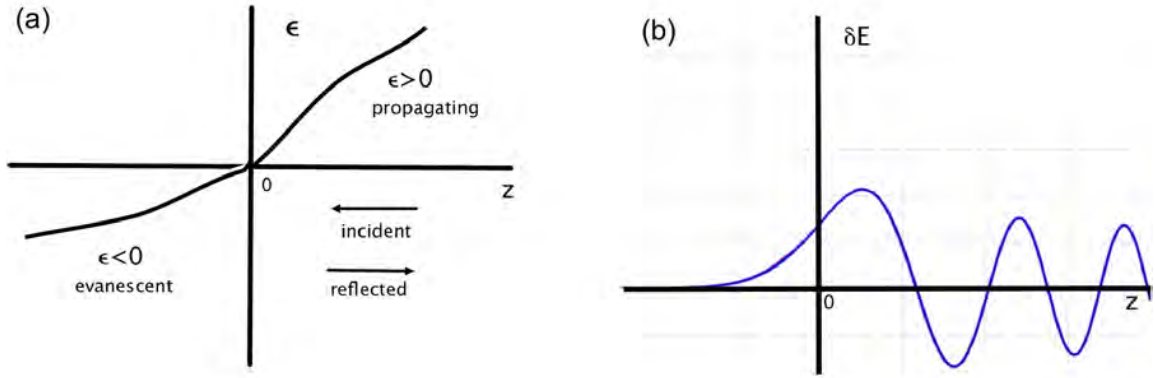

Fig. 4.E.1 (a) Schematic of the dielectric function as function of position z for a wave incident from the right with cutoff at z=0. (b) Schematic of electromagnetic wave amplitude for wave incident from the right with cutoff at z=0.

The asymptotic WKB solutions for the wave amplitude and phase far from z=0 are given by

$$z \geq \lambda: \quad \delta E(z) = \frac{\beta}{\varepsilon^{1/4}(z)} \cos\left(\frac{\omega}{c}\int_0^z dz' \sqrt{\varepsilon(z')} - \alpha\right)$$

$$z \leq -\lambda: \quad \delta E(z) = \frac{1}{\varepsilon^{1/4}(z)} \exp\left(-\frac{\omega}{c}\int_z^0 dz' \sqrt{|\varepsilon(z')|}\right)$$
(4.E.14)

The constants $\alpha$ and $\beta$ are to be determined. The normalization of the wave amplitude is arbitrary. We rewrite the differential equation (4.E.12) as

$$\left[-\frac{d^2}{dz^2} + \frac{\omega^2}{c^2}(-\varepsilon(z))\right]\delta E(z) = 0 \tag{4.E.15}$$

which is analogous to the quantum mechanical problem:

$$[H]\psi = [K+V]\psi = 0 \tag{4.E.16}$$

With $V \propto -\varepsilon$, Fig. 4.E.1a can be used as a diagram for V(z) (with the sign flipped from that of $\varepsilon(z)$). With total energy E=0, the kinetic energy K is positive for z>0; and there is a turning point at z=0 for V(0)=0. We return to the WKB solution of Eq.(4.E.12) with units such that $\omega/c = 1$. Assume that

$$\delta E(z) = A_+(z)W_+(z) + A_-(z)W_-(z)$$

$$W_\pm \propto \exp\mp \int_0^z dz' |\varepsilon|^{1/2}$$
(4.E.17)

$W_+$ is exponentially large in the evanescent region z<0, and $W_+$ is exponentially small in the evanescent region. $A_\pm$ are rapidly varying in the neighborhood of the turning point. We constrain $A_+'W_+ + A_-'W_- = 0$ so that $(d/dz)\delta E(z) = A_+W_+' + A_-W_-'$. Now substitute Eq.(4.E.17) into Eq.(4.E.12) to obtain

$$\frac{dA_+}{dz} = -\frac{i}{2}\frac{\Delta(z)}{\varepsilon^{1/2}(z)}\left[A_+ + A_-\frac{W_-}{W_+}\right] \text{ and } \frac{dA_-}{dz} = \frac{i}{2}\frac{\Delta(z)}{\varepsilon^{1/2}(z)}\left[A_- + A_+\frac{W_+}{W_-}\right] \quad (4.E.18)$$

$$\text{where } \Delta(z)^2 \equiv \frac{1}{4}\frac{d^2\varepsilon/dz^2}{\varepsilon} - \frac{5}{16}\left(\frac{d\varepsilon/dz}{\varepsilon}\right)^2$$

Consider integration of the first-order differential equations in Eq.(4.E.18) in the complex $z$ plane making use of analytic continuation. Far from the turning point at $z=0$ in the evanescent region select a point #1 on the real $z$ axis where $A_+=0$ and $A_-=1$, while in the propagation region $z>0$ far from $z=0$, select a point #2 where $A_+=1$ and $A_-=0$. We can integrate $dA_+/dz$ from #1 on an arc approaching the imaginary axis, and it remains exponentially small. However, $dA_-/dz \approx i\Delta/(2\varepsilon^{1/2}(z))$ is small, but not exponentially so. Near the imaginary $z$ axis, $z=i|z|$, $\varepsilon \sim z$ and $W_\pm \sim \exp(\pm z^{3/2}) \sim \exp(\pm i^{3/2}|z|^{3/2})$. Thus, $W_+$ is exponentially small near the imaginary axis; and $W_-$ is exponentially large near the imaginary axis. The ratio $W_-/W_+$ remains exponentially large until the real axis. We can examine the integration of (4.E.18):

$$\frac{dA_+}{dz} = -\frac{i}{2}\frac{\Delta(z)}{\varepsilon^{1/2}(z)}\left[A_+(\text{starts small}) + A_-(\sim 1)\frac{W_-}{W_+}(\text{exponentially large})\right] \quad (4.E.19a)$$

$$\frac{dA_-}{dz} = \frac{i}{2}\frac{\Delta(z)}{\varepsilon^{1/2}(z)}\left[A_-(\sim 1) + A_+(\text{exponentially small})\frac{W_+}{W_-}(\sim \text{exponentially small})\right] \quad (4.E.19b)$$

The coefficient in front of the bracket in Eq.(4.E.19a) is small and $A_+$ starts small, but has a growth rate that is the product of a small quantity and an exponentially large quantity; so $A_+$ becomes finite, and a more careful analysis is required. $A_-$ has a small growth rate and remains near unity in WKB. To resolve this, we pick an integration contour from #1 to #2 following a half-circle in the upper $z$ half-plane back to the real axis where there is a branch cut and a jump condition (rf. Heading). Mathews and Walker's treatment gives the following. In the evanescent region, the solution for wave amplitude is $\delta E \sim W_-(z)$, while in the propagation region the wave amplitude is the sum of the incident and reflected wave: $\delta E \sim W_-(z) + A_+ W_+(z)$. In the propagation region, the incident wave amplitude physically is $|A_+|=1$; and the reflected wave amplitude should be the same, but the phase needs to be determined. The expressions for $\delta E$ coming from either direction must match where the arc crosses the imaginary $z$ axis:

$$\frac{e^{-\int_0^{|z|} dz'|\varepsilon|^{1/2}(z')}}{e^{i\pi/4}|\varepsilon|^{1/4}(z)} \rightarrow \frac{e^{-i\int_0^{z} dz'\varepsilon^{1/2}(z')} + A_+ e^{+i\int_0^{z} dz'\varepsilon^{1/2}(z')}}{\varepsilon^{1/4}(z)} \quad (4.E.20)$$

We multiply both sides of Eq.(4.E.20) by $e^{i\pi/4}$

$$\frac{e^{-\frac{\omega}{c}\int_0^{|z|}dz'|\varepsilon|^{1/2}(z')}}{|\varepsilon|^{1/4}(z)} \rightarrow \frac{e^{i\pi/4}e^{-i\frac{\omega}{c}\int_0^z dz'\varepsilon^{1/2}(z')} + A_+ e^{i\pi/4}e^{+i\frac{\omega}{c}\int_0^z dz'\varepsilon^{1/2}(z')}}{\varepsilon^{1/4}(z)} \rightarrow \frac{2\cos\left[\frac{\omega}{c}\int_0^z dz'\varepsilon^{1/2}(z') - \pi/4\right]}{\varepsilon^{1/4}(z)}$$

(4.E.21)

Because the left side of (4.E.21) is real, the two terms in the intermediate expression must be complex conjugates, which leads to the final expression on the right side. We see that the phase factor $\pi/4$ has emerged. Figure 4.E.1b captures the qualitative behavior of the solution in Eq.(4.E.21) characteristic of single turning-point behavior. The peak of the wave at relative value of 2 occurs at $z \approx (\pi/2)c/\omega$.

Exercise: For a dielectric function $\varepsilon$ with linear variation in $z$, show that the exact solution of Eq.(4.E.15) is given by the Airy function and compare the Airy function to the WKB solution in Eq.(4.E.21).

If the variation of the dielectric function is such that propagating waves can be trapped between two turning points, the analysis corresponding to Eq.(4.E.21) leads to an eigenvalue problem for the frequency $\omega$. Given a length $L$ between the turning points, only particular values of $\omega$ will allow the wave forms from either side to link up. Imagine picking a value of $z$ and reflecting the solution for $\delta E(z)$ symmetrically in Figure 4.E.1b so that $\delta E(z)$ and its spatial derivative are continuous. This can only be satisfied if the spatial derivative of $\delta E(z)$ vanishes at the mirror point, which quantizes the relationship between the values of $\omega$ and $L$. For the two-turning-point problem we take the solution for the wave amplitude on the right-side of Eq.(4.E.21) associated with the turning point at $z_0$ and match it to the corresponding mirror image waves associated with the turning point at $z_1$ and introduce constant $C$:

$$C\frac{2\cos\left[\frac{\omega}{c}\int_z^{z_1}dz'\varepsilon^{1/2}(z') - \pi/4\right]}{\varepsilon^{1/4}(z)} = \frac{2\cos\left[\frac{\omega}{c}\int_0^z dz'\varepsilon^{1/2}(z') - \pi/4\right]}{\varepsilon^{1/4}(z)} \qquad (4.E.22)$$

For the left and right sides of Eq.(4.E.22) to be equal, the arguments of the cosines must be equal or the negative of one another so that $C=1$ or $C=-1$. Hence, we arrive at the relations:

$$\frac{\omega}{c}\int_z^{z_1} dz'\varepsilon^{1/2}(z') - \pi/4 = -\left[\frac{\omega}{c}\int_0^z dz'\varepsilon^{1/2}(z') - \pi/4\right] + 2\pi\ell, \quad C=1, \ell=0,1,2,...$$

$$\rightarrow \frac{\omega}{c}\int_0^{z_1} dz'\varepsilon^{1/2}(z') = \pi/2 + 2\pi\ell \tag{4.E.23}$$

$$\text{or } \frac{\omega}{c}\int_0^{z_1} dz'\varepsilon^{1/2}(z') = \pi/2 + (2\ell+1)\pi, \quad C=-1$$

$$\Rightarrow \frac{\omega}{c}\int_0^{z_1} dz'\varepsilon^{1/2}(\omega,z') = \left(\ell+\tfrac{1}{2}\right)\pi \text{ or } \frac{\omega}{c}\oint dz'\varepsilon^{1/2}(\omega,z') = (2\ell+1)\pi$$

The final integral in Eq.(4.E.23) is from the turning points $\varepsilon=0$ $z=0$ to $z=z_1$ and return. We recognize this expression as the Bohr-Sommmerfeld quantization rule. Eq.(4.E.23) is an eigenequation that determines the eigenvalue $\omega$ for each value of $\ell$.

Example: Consider an unmagnetized plasma with dielectric function:
$\varepsilon(\omega,z) = 1 - \omega_p^2(z)/\omega(\omega+i\nu)$ and $\omega_p^2 = \alpha^2 z^2$ where $z=0$ is a reference altitude. Then $\text{Re}\varepsilon=0$ has two symmetric turning points in $z$, and the eigenequation obtained from Eq.(4.E.23) is

$$\pi\frac{\omega}{c}\frac{\alpha}{\sqrt{\omega(\omega+i\nu)}}\frac{\omega(\omega+i\nu)}{\alpha^2} = \pi\frac{\omega}{c}\frac{\sqrt{\omega(\omega+i\nu)}}{\alpha} = (2\ell+1)\pi \tag{4.E.24}$$

For $\nu \rightarrow 0$, $\omega_\ell = \sqrt{2\ell+1}\sqrt{\alpha c}, \ell = 0, 1, 2, ...$

WKB breaks down near resonances as well as near turning points. At resonance $n = kc/\omega \rightarrow \infty$, i.e., $k\rightarrow\infty$ and $v_\phi \equiv \omega/k \rightarrow 0$ To resolve the difficulties posed by resonances we will abandon the cold plasma restriction and include warm or hot plasma effects.

### 4.F Vlasov-Maxwell equations – linear electrodynamics

In this section we return to the examination of the Vlasov-Maxwell equations including consideration of waves and instabilities in a warm, magnetized plasma, e.g., Bernstein waves, a general dispersion relation and survey of waves, the Harris instability, and the effects of gyro-resonance.

### 4.F.a Bernstein waves (electrostatic model)

Bernstein waves named after Ira Bernstein are an example of a wave in a warm magnetized plasma. The simplest dispersion relation for Bernstein waves satisfies the electrostatic approximation, i.e., the perturbed electric field is oriented parallel to the wave vector and there is no magnetic perturbation associated with the wave. Most generally we can always represent the perturbed electric field as follows.

<u>Definition</u>: $\delta\mathbf{E} = \delta E^L \hat{\mathbf{k}} + \delta\mathbf{E}^T$ where "L" indicates the "longitudinal" component which is parallel to the wave vector $\hat{\mathbf{k}}$ and "T" indicates the "transverse" component relative to the wave vector. [Note that "perpendicular" will mean at right angles to the applied magnetic field $\mathbf{B}_0$.]

The general linear dispersion relation derived in Sec. 4.B Eq.(4.B.8) is

$$\left(\vec{\vec{\varepsilon}} - n^2 \vec{\vec{I}}^T\right) \cdot \delta\mathbf{E} = 0, \quad n^2 = \frac{k^2 c^2}{\omega^2} \tag{4.F.1}$$

What are the conditions such that $\delta E^L \gg \delta E^T$? Consider a nonrelativistic plasma with $\beta \equiv \frac{v_{th}}{c} \ll 1$. If $\omega/k \ll c$ so that $n \gg 1$ and $\varepsilon \sim O(1)$, then

$$\left|\frac{\delta E^T}{\delta E^L}\right| \approx \frac{1}{n^2} = \frac{\omega^2}{k^2 c^2} \ll 1 \tag{4.F.2}$$

The electrostatic dispersion relation for electron Bernstein waves propagating perpendicular to a uniform, applied magnetic field in a warm plasma ignoring ion motion is shown in Fig. 4.F.1 below.[10] Here $\Omega_e$ is the electron cyclotron frequency and $\omega_{uh} = (\Omega_e^2 + \omega_{pe}^2)^{1/2}$ is the upper hybrid frequency. In the example shown in Fig. 4.F.1, $\omega_{pe} \approx 2.4 \Omega_e$. For the higher harmonics of the electron cyclotron frequency inevitably $\omega/k > c$, and the electrostatic approximation fails, invalidating the solution of the dispersion relation. We recall the solution of the cold plasma electromagnetic dispersion relation plotted in Fig. 4.B.1

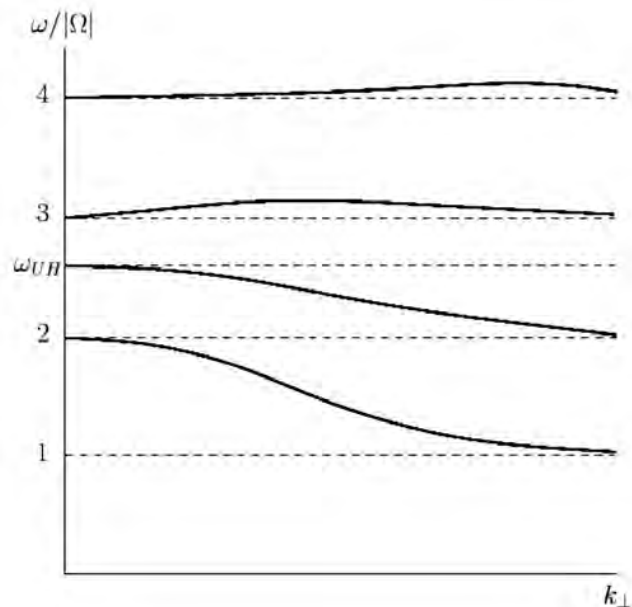

Fig. 4.F.1 Electron Bernstein waves $\omega/\Omega_e$ vs. $k_\perp$
[ref. R. Fitzpatrick, U. Texas Austin[11]]

---

[10] I. Shkarofsky, Phys. Fluids **9**, 570 (1966).
[11] https://farside.ph.utexas.edu/teaching/plasma/lectures1/node91.html; R. Fitzpatrick, *Introduction to Plasma Physics* (CRC Press, 2014), Chapt. 5.

The longitudinal dielectric function for perpendicular propagation in a warm plasma is derived in many textbooks, e.g., Eq.(9.36) of Clemow and Dougherty:

$$0 = \varepsilon(\omega, k_\perp) = 1 - \frac{\omega_p^2}{\Omega_e^2} \sum_{n=-\infty}^{\infty} \frac{e^{-\lambda} I_n(\lambda)/\lambda}{\frac{\omega}{n\Omega_e} - 1}, \quad \lambda \equiv (k_\perp r_e)^2, \quad r_e \equiv \frac{v_{th}}{\Omega_e} = \left(\frac{T_e}{m_e \Omega_e}\right)^{1/2} \quad (4.F.3)$$

where $I_n$ are Bessel functions of complex argument and have limiting forms

$$I_n(\lambda) \xrightarrow[\lambda \ll 1]{} \frac{1}{n!}\left(\frac{1}{2}\lambda\right)^n \quad \text{and} \quad I_n(\lambda) \xrightarrow[\lambda \gg 1]{} \frac{e^\lambda}{\sqrt{2\pi\lambda}} \quad (4.F.4)$$

For frequencies near a particular cyclotron harmonic $\omega = \omega_n$ the sum over harmonics in Eq.(4.F.2) can be reduced to the one "resonant" term in the series; then

$$\frac{\omega_n - n\Omega_e}{n\Omega_e} = \frac{\omega_p^2}{\Omega_e^2} \frac{e^{-\lambda} I_n(\lambda)}{\lambda} \quad (4.F.5)$$

<u>Example</u>: Consider launching a Bernstein wave by inserting a grid attached to power supply driven at a given frequency near the 3rd electron cyclotron harmonic. Furthermore, assume that $\nabla B_0 < 0$ for increasing $x$ propagating into the plasma. Because the cyclotron frequency is decreasing, $\omega/\Omega_e$ and $k$ increase; and the group velocity $v_g \propto \frac{d\omega/\Omega_e}{dk_\perp} > 0$. As $x$ continues to increase, $k$ continues to increase on the branch and $\omega/\Omega_e$ reaches a maximum where the group velocity vanishes, at which point the wave packet must turn around. This is akin to ionospheric reflection of an electromagnetic wave (see Secs. 4.E.a and 4.E.b). These considerations are significantly relevant to applications such as wave heating of a magnetically confined plasma.

### 4.F.b Instabilities, e.g., beam-cyclotron instability, ion acoustic, Dory-Guest-Harris

In the preceding subsection the electron Bernstein wave had a real frequency for a given $k_\perp$. How might the Bernstein wave be destabilized? Consider injection of a beam resonating with an electron Bernstein wave. For this example consider an injected ion beam. Because the ion cyclotron frequency is so low $\Omega_i \ll \Omega_e, \omega, \gamma$, the unperturbed ion motion can be treated as unmagnetized. As a matter of consistency with the assumption that the ions are unmagnetized the ions should satisfy $r_i \equiv \frac{u_i}{\Omega_i} \gg 1/k$. Where the beam velocity $u_i$ equals the Bernstein phase velocity $\omega/k$, we expect the possibility of complex conjugate roots of the dispersion relation, i.e., one of the roots is unstable. Solution of the dispersion relation yields

$$\frac{\delta\omega + i\gamma}{\omega^{(0)}} = |\eta|^{1/3}\left(-\frac{1}{2} + i\frac{\sqrt{3}}{2}\right), \quad \eta = \frac{\omega_b^2}{\omega_p^2} \frac{1}{\omega^{(0)} \frac{\partial \varepsilon}{\partial \omega}} \quad (4.F.6)$$

in analogy to the weak beam instability analysis in Sec. 2.F.a. Here $\omega_b$ is the plasma frequency for the beam ions.

Exercise: Provide the details in deriving Eq.(4.F.6) and show that

$$\eta = \frac{\omega_b^2 n I_n(\lambda)}{\lambda e^{\lambda} \Omega_e \left( n\Omega_e + \frac{n\omega_p^2}{\Omega_e} \frac{e^{-\lambda} I_n(\lambda)}{\lambda} \right)}, \quad \lambda \equiv (k_\perp r_e)^2, \quad r_e \equiv \frac{v_{th}}{\Omega_e} = \left( \frac{T_e}{m_e \Omega_e} \right)^{1/2} \quad (4.F.7)$$

Exercise: (Magnetized ion acoustic instability) Consider ions with velocity relative to the electrons $u_i$ satisfying the inequalities $c_s < u_i < v_{the}$ The beat of the ion acoustic wave with the beam modes produce fast and slow beam modes $u_i \pm c_s$ From the general analysis leading to Eq.(2.I.7) evaluate the expression

$$\gamma(\mathbf{k}) = -\frac{\varepsilon''(\mathbf{k}, \omega_R)}{\left. \frac{\partial}{\partial \omega} \varepsilon'(\mathbf{k}, \omega) \right|_{\omega_R}} \sim \frac{\pi \omega_p^2}{\sqrt{2\pi} k^2 v_{the}^2} \frac{\omega_p^2}{\Omega_e} \frac{n I_n(\lambda)}{\lambda e^{\lambda}}$$

where in the numerator we use the contribution from the unmagnetized beam ions and the electron Bernstein formula is used in the denominator.

Example: (Dory-Guest-Harris modes[12]) Suppose the velocity distribution $f(v_\perp)$ is not a Maxwellian. What happens to the Bernstein modes? Consider a family of velocity distributions: $f \sim (v_\perp^2)^m e^{-v_\perp^2/v_{th}^2}$ Analysis of the electrostatic dispersion relation yields the following results:
1. $m = 1,2$ remains stable
2. $m = 3$ unstable if $\frac{\omega_p}{\Omega_e} > 28$ and purely growing (Re$\omega$=0)
3. $m = 4$ unstable if $\frac{\omega_p}{\Omega_e} > 11$ and purely growing (Re$\omega$=0)
4. $m = 6$ unstable if $\frac{\omega_p}{\Omega_e} > 10$ and purely growing (Re$\omega$=0); also unstable at Re$\omega$=1.2$\Omega_e$.
5. $m = \infty$ $f \to \delta(v_\perp - v_0)$

Example: $f \sim \delta(v_\perp - v_0)$ For a $\delta$-function the longitudinal dielectric is[13]

$$\varepsilon(\omega, k_\perp) = 1 - \frac{\omega_p^2}{\Omega_e^2} \sum_{n=-\infty}^{\infty} \frac{\frac{1}{b}\frac{d}{db} J_n^2(b)}{\frac{\omega}{n\Omega_e} - 1}, \quad b \equiv \frac{k_\perp v_0}{\Omega_e} \quad (4.F.8)$$

with solution to $\varepsilon=0$ for $\omega$ near a cyclotron harmonic:

$$\frac{\omega - n\Omega_e}{n\Omega_e} = \frac{\omega_p^2}{\Omega_e^2} \frac{1}{b} \frac{d}{db} J_n^2(b) \quad (4.F.9)$$

We note that the wave energy is $\propto \omega \, \partial \epsilon / \partial \omega$ can have either sign depending on the nearest harmonic $n$ and the wavenumber. As the left side of (4.F.9) departs from zero so that the *n* and *n*-1 neighboring branches can merge into one another, and complex conjugate roots appear. A very strong interaction can take place, particularly if the wave energies have opposite signs, which leads to instability. Increasing the plasma density increases $\omega_p^2/\Omega_e^2$ and the magnitude of the right side of Eq.(4.F.8), which exacerbates the tendency to instability.

---

[12] R.A. Dory, G.E. Guest, E.G. Harris, Phys. Rev. Lett. **14**, 131 (1965).
[13] F.W. Crawford and J.A. Tataronis, J. Appl. Phys. 36, 293 (1965).

Example: Ordinary mode destabilized by anisotropy – Consider an electromagnetic mode propagating across the magnetic field with electric field perturbation parallel to a uniform, applied magnetic field. With Maxwellian electrons the solution of Eq.(4.F.1) is

$$\omega^2 = k^2 c^2 + \omega_p^2 \left(1 - \tau \left[1 - \omega^2 \sum_{\ell=-\infty}^{\infty} \frac{e^{-\lambda} I_\ell(\lambda)}{\omega^2 - \ell^2 \Omega_e^2}\right]\right), \quad \tau \equiv \frac{T_\parallel}{T_\perp}, \quad \lambda = \left(\frac{k_\perp v_\perp^{th}}{\Omega_e}\right)^2 = (k_\perp r_e)^2 \quad (4.F.10)$$

The dispersion relation in Eq.(4.F.10) is solved numerically. One can assume that $\omega$ is complex and solve for $\tau$:

$$\tau = \frac{1 + \dfrac{k^2 c^2 - \omega^2}{\omega_p^2}}{1 + \omega^2 e^{-\lambda} \sum_{\ell=-\infty}^{\infty} \dfrac{I_\ell(\lambda)}{\ell^2 \Omega_e^2 - \omega^2}} \rightarrow \frac{1 + \dfrac{k^2 c^2 + \gamma^2}{\omega_p^2}}{1 - \gamma^2 e^{-\lambda} \sum_{\ell=-\infty}^{\infty} \dfrac{I_\ell(\lambda)}{\ell^2 \Omega_e^2 + \gamma^2}} \quad (4.F.11)$$

for $\omega^2 \to -\gamma^2$, a non-resonant, absolute instability. Furth analyzed this instability from the perspective of the self-pinching of perturbed currents, calling the instability the "prevalent instability."[14]

### 4.F.c Additional examples of waves in a plasma

Here we return to the subject matter introduced earlier in Sec. 4 and present a few additional examples of waves in a warm plasma. Small amplitude waves in a plasma are either longitudinal ($\nabla \times \delta \mathbf{E} = 0$) or not longitudinal ($\nabla \times \delta \mathbf{E} \neq 0$).

The dispersion relation for longitudinal waves in a uniform magnetic field was previously derived in Secs. 2, 3, and 4. Using $\frac{v}{c} \ll 1$, $\frac{\omega}{k} \ll c$, and $\mathbf{k} || \delta \mathbf{E}$, the roots of the dispersion relation $\omega(k_\parallel, k_\perp)$ are determined by the solution of

$$\varepsilon \equiv \hat{\mathbf{k}} \cdot \bar{\varepsilon} \cdot \hat{\mathbf{k}} = 1 + \sum_s \chi_s(\mathbf{k}, \omega) = 0 \quad (4.F.12a)$$

$$\chi_s(\mathbf{k}, \omega) = -\frac{\omega_s^2}{k^2} \int d^3 v \sum_{\ell=-\infty}^{\infty} J_\ell^2\left(\frac{k_\perp v_\perp}{\Omega_s}\right) \frac{k_z \dfrac{\partial g}{\partial v_z} + \dfrac{\ell \Omega_s}{v_\perp} \dfrac{\partial g}{\partial v_\perp}}{k_z v_z - (\omega - \ell \Omega_s)} \quad (4.F.12b)$$

with $\int d^3 v g(v_\parallel, v_\perp) = 1$. It is often assumed that $g$ is separable: $g(v_\parallel, v_\perp) = g(v_\parallel)g(v_\perp)$, e.g., $g(v_\perp) = (2\pi v_{th})^{-1} exp(-v_\perp^2/2v_{th}^2)$ a Gaussian, or $(v_\perp)$ a non-Gaussian as in the Dory-Guest-Harris instability.

---

[14] H.P. Furth, Phys. Fluids **6**, 48 (1963).

For non-longitudinal waves, **k** and **δE** are not parallel. Much of the existing analysis in the literature treats the plasma as cold with thermal corrections. There are many interesting examples. Calculations have established a relation between either whistler or ordinary waves and pinching instabilities, and a relation between Alfvèn waves and the firehose instability.

For $g$ a separable velocity distribution function and $g(v_\perp)$ a Gaussian, the susceptibility $\chi_s$ can be partially evaluated and reduced to

$$\chi_s(\mathbf{k},\omega) = -\frac{\omega_s^2}{k^2}\int d^3v \sum_{\ell=-\infty}^{\infty} J_\ell^2\left(\frac{k_\perp v_\perp}{\Omega_s}\right) \frac{k_z \frac{\partial g}{\partial v_z} + \frac{\ell\Omega_s}{v_\perp}\frac{\partial g}{\partial v_\perp}}{k_z v_z - (\omega - \ell\Omega_s)}$$

$$= -\frac{\omega_s^2}{k^2}\int dv_z \sum_{\ell=-\infty}^{\infty} e^{-b} I_\ell(b) \frac{k_z \frac{\partial g}{\partial v_z} - \frac{\ell\Omega_s}{v_{th}^2} g_\parallel(v_z)}{k_z v_z - (\omega - \ell\Omega_s)} \qquad (4.F.13)$$

where $b = (k_\perp v_{\perp th})^2 = (k_\perp r_\perp)^2$. We see the cyclotron resonance in the denominator on the right side of Eq.(4.F.13). Recalling that $Z(V) = \int dv_z g_\parallel /(v_z - V)$ then Eq.(4.F.13) becomes

$$\chi_s(\mathbf{k},\omega) = -\frac{\omega_s^2}{k^2}\int dv_z \sum_{\ell=-\infty}^{\infty} e^{-b} I_\ell(b) \frac{k_z \frac{\partial g}{\partial v_z} - \frac{\ell\Omega_s}{v_{th}^2} g_\parallel(v_z)}{k_z v_z - (\omega - \ell\Omega_s)}$$

$$= -\frac{\omega_s^2}{k^2}\sum_{\ell=-\infty}^{\infty} e^{-b} I_\ell(b)\left[Z'(V_\ell) - \frac{\ell\Omega_s}{k_z v_{th}^2} Z(V_\ell)\right], \qquad V_\ell = \frac{(\omega - \ell\Omega_s)}{k_z} \qquad (4.F.14)$$

Some useful asymptotic forms for further evaluation of Eq.(4.F.13) are given in Eq.(4.F.3) for $I_n(b)$ and in Eqs.(2.H.3) and (2.H.4) for ReZ'(V) and ReZ(V) for V/$v_{th}$>>1.

Examples

i. For Maxwellian $g_\parallel$ and no drift, V/$v_{th}$→0, ReZ(V) →0 and ReZ'(V) →-1/$v_{th}^2$.
ii. As Ω→0 for weak magnetic fields, we recover the results in Sec. 2.E
iii. As the temperature goes to zero, the cold plasma limit of the longitudinal susceptibility is recovered:

$$\chi_s \equiv \hat{\mathbf{k}} \cdot \ddot{\chi}_s \cdot \hat{\mathbf{k}} = -\frac{\omega_s^2}{\omega^2}\frac{\omega - \Omega_s^2 \cos^2\theta}{\omega^2 - \Omega_s^2} \qquad (4.F.15)$$

iv. For ion waves or other very low frequency waves $\omega$→0, the limit in which the electron susceptiblility is $\chi_e = 1/k^2\lambda_e^2$ can be recovered with some labor.
v. In the strong field limit, $\Omega \gg \omega$. In this limit with V large, the $\ell=0$ term is dominant in the infinite sum in Eq.(4.F.14); and one obtains

$$\chi_s(\mathbf{k},\omega) \approx -\frac{\omega_s^2}{k^2}\sum_{\ell=-\infty}^{\infty} e^{-b}I_0(b)\frac{k_z^2}{\omega^2}\left(1+3\frac{v_{\|th}^2}{V^2}+\ldots\right) \qquad (4.F.16)$$

Exercises:
i. Derive the dispersion relation for the electron plasma wave in the strong magnetic field limit: $\omega^2 = \omega_p^2\cos^2\theta + 3k_z^2 v_{\|th}^2$
ii. Derive the electron plasma wave dispersion relation for a finite magnetic field with $\omega \sim \Omega$ for $T_e \to 0$ and compare to electron Bernstein wave dispersion relation.
iii. Derive the dispersion relation for the ion acoustic wave in the strong magnetic field limit using the low-frequency limit for $\chi_e = 1/k^2\lambda_e^2$:

$$\omega^2 = \omega_i^2\left(1+\frac{1}{k^2\lambda_e^2}\right)^{-1}\cos^2\theta$$

iv. Derive the dispersion relation for longitudinal ion cyclotron waves $\omega \sim \ell\Omega_i$ using the low-frequency limit for $\chi_e = 1/k^2\lambda_e^2$. Assume that the wave phase velocity $V_\ell \gg v_{th,i}$ to obtain

$$\frac{\omega - \ell\Omega_i}{\ell\Omega_i} = \frac{T_e}{T_i}e^{-b_i}I_\ell(b_i)$$

# LECTURES ON THEORETICAL PLASMA PHYSICS – PART 2

*Allan N. Kaufman*

## 5. Miscellaneous topics

In this section six miscellaneous topics are covered, some of which were introduced earlier: (i) the Nyquist method for solving dispersion relations, (ii) analytic continuation in applications to solving for the plasma response to an electric field, (iii) the plasma dispersion function, (iv) nonanalytic velocity distribution functions, (v) initial-value problems – linear response function, and (vi) Van Kampen modes.

### 5.A Nyquist method for solving dispersion relations

One is frequently faced with the task of solving for the roots of a dispersion relation: $D(\omega) = 0$, which may be a complicated function of $\omega$. Sometimes it is sufficient just to determine that there are no roots $\omega_\ell$ with Im $\omega_\ell > 0$. The Nyquist stability criterion is a graphical methodology that only requires evaluating the dispersion function $D$ as a function of complex $\omega$. Figure 5.A.1 illustrates the Nyquist stability criteria. In Fig. 5.A.1(a) there is a contour in the complex $\omega$ plane that is closed in the upper half-plane enclosing the possible roots of $D(\omega) = 0$. The corresponding contour $D(\omega)$ is drawn in Fig. 5.A.1(b). In this example any value of $D$ in the region of the complex $D$ encircled by just one loop that encircles the origin maps to just one value of complex $\omega$ within the $\omega$ contour. Any value of $D$ in the region encircled by two loops maps to two values of complex $\omega$ within the $\omega$ contour.

Theorem: (Nyquist criteria) The number of complex $\omega$ roots in the upper half-plane equals the number of times the origin is encircled by loops in the $D$ plane. This is the number of unstable roots.

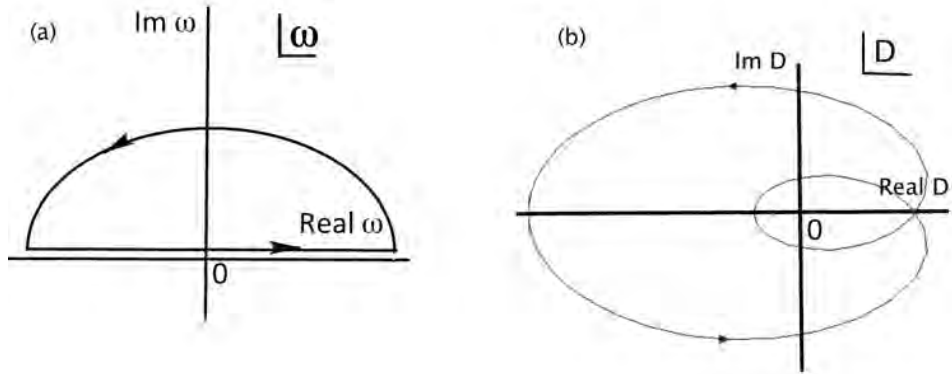

Figure 5.A.1  (a) Contour in the complex $\omega$ plane encircling all roots $\omega_\ell$ with Im $\omega_\ell > 0$.  (b) Corresponding contour $D(\omega)$ in the complex $D$ plane encircling the origin twice.

## 5.B Analytic continuation

Analytic continuation was introduced in Sec. 2E in the context of analyzing the electrostatic plasma response:

$$\varepsilon(k,\omega) = 1 - \frac{\omega_p^2}{k^2} Z'(v) \quad (v \equiv \omega/k)$$

$$Z(v) = \int_{-\infty}^{\infty} du \frac{g(u)}{u-v} \quad (\text{Im}\, v = \text{Im}\, \omega/k > 0) \tag{5.B.1}$$

$$\varepsilon^{-1}(k,t) = \int \frac{d\omega}{2\pi} \frac{e^{-i\omega t}}{\varepsilon(k,\omega)}$$

Figure 2.E.2 displays the contour integration for the pulse response showing the depressed contour and poles of $\varepsilon^{-1}$ in the region of analytic continuation.  When v descends in the lower $\omega$ half-plane we depress the integration contour in $u$ down to loop around the pole in the region of analyticity to calculate $Z(v)$.  $g(u)$ must be given as an analytic function.  There are three cases to consider: (i) v in the upper half-plane for which the integration contour remains on the real $u$ axis; (ii) v on the real $u$ axis for which the integration contour takes a semi-circular loop below the pole and picks up a $i\pi g(v)$ contribution in addition to the principal value of the contour integral; (iii) v below the real $u$ axis for which the integration contour takes a full circular loop around the pole (as in Fig. 2.E.2) and picks up a $i2\pi g(v)$ contribution in addition to the contour integral.

Exercise: Show that for $g(u)$ a rational function, one can calculate $Z(v)$ in the upper $\omega$ half-plane and then analytically continue to the lower half-plane.

## 5.C Plasma dispersion function – Fried-Conte function

Consider the longitudinal dielectric response when the velocity distribution is a Gaussian. Fried and Conte treated this subject in their 1961 book.[15] We introduce the following notation and equations:

$$g(u) = \frac{1}{\sqrt{2\pi}v_{th}} e^{-\frac{u^2}{2v_{th}^2}} = \frac{1}{\sqrt{\pi}\bar{c}} e^{-\eta^2}, \quad \eta = \frac{u}{\sqrt{2}v_{th}} = \frac{u}{\bar{c}}, \quad \bar{c} = \sqrt{2}v_{th}, \quad \zeta = \frac{v}{\bar{c}} \quad (5.C.1)$$

The Fried-Conte function or plasma dispersion function $z(\zeta)$ is then given by

$$z(v) = \frac{1}{\bar{c}} \left[ \frac{1}{\sqrt{\pi}} \int_{-\infty}^{\infty} d\eta \frac{e^{-\eta^2}}{\eta - \zeta} \right] = \frac{1}{\bar{c}} z(\zeta) \quad (5.C.2)$$

Limiting expressions are as follows:

$$\text{For real } \zeta: \quad \text{Im } z = \sqrt{\pi} e^{-\zeta^2} \quad \text{Im} z' = -2\sqrt{\pi}\zeta e^{-\zeta^2}$$

$$\text{Re} z = \frac{1}{\sqrt{\pi}} P \int_{-\infty}^{\infty} d\eta \frac{e^{-\eta^2}}{\eta - \zeta}$$

$$\rightarrow -2\zeta\left(1 - \tfrac{2}{3}\zeta^2 + \ldots\right), \quad \zeta \rightarrow 0$$

$$\rightarrow -\frac{1}{\zeta}\left(1 + \frac{1}{2\zeta^2} + \ldots\right), \quad \zeta \rightarrow \infty \quad (5.C.3)$$

$$\text{Re} z' = \frac{1}{\sqrt{\pi}} P \int_{-\infty}^{\infty} d\eta \frac{-2\eta e^{-\eta^2}}{\eta - \zeta}$$

$$\rightarrow -2 + 4\zeta^2, \quad \zeta \rightarrow 0$$

$$\rightarrow \frac{1}{\zeta^2}\left(1 + \frac{3}{2\zeta^2} + \ldots\right), \quad \zeta \rightarrow \infty$$

## 5.D Non-analytic velocity distribution functions

Here we consider two examples of velocity distribution functions that are not analytic.

Example: Consider a warm plasma with sharp cut-offs in velocity (Fig. 5.D.1).

---

[15] B. D. Fried and S. D. Conte, *The Plasma Dispersion Function - The Hilbert Transform of a Gaussian* (Academic Press), 1961.

$$g(u) = \frac{3}{4c^3}(c^2 - u^2), \quad |u| < c$$
$$= 0, \quad\quad |u| > c$$
$$g'(u) = -\frac{3u}{2c^3}, \quad |u| < c$$
$$= 0, \quad\quad |u| > c$$
(5.D.1)

The plasma dielectric response then derives from

$$\varepsilon(k,\omega) = 1 - \frac{\omega_p^2}{k^2} Z'(v)$$

$$Z'(v) = \int_{-\infty}^{\infty} du \frac{g'(u)}{u-v} = -\frac{3}{2c^3} \int_{-c}^{c} du \frac{u}{u-v} = -\frac{3}{2c^3}\left[2c + v \ln \frac{c-v}{-c-v}\right]$$
(5.D.2)

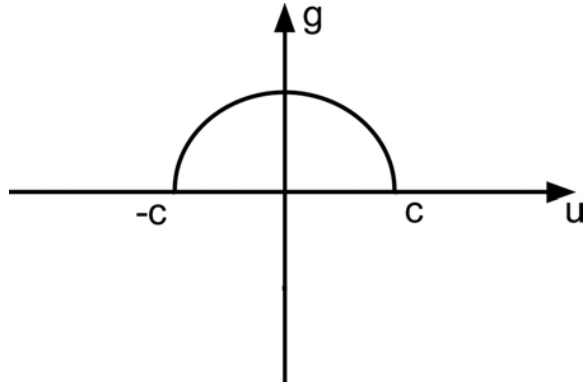

Figure 5.D.1 Warm velocity distribution function with sharp cutoffs.

The dielectric function in Eq.(5.D.2) has singularities at v=±c and is symmetric with respect to changing the sign of v=$\omega$/k. The pulse response derives from

$$\varepsilon^{-1}(k,t) = \int \frac{d\omega}{2\pi} \frac{e^{-i\omega t}}{\varepsilon(k,\omega)}$$
(5.D.3)

The contour integral in Eq.(5.D.3) is depicted in Fig. 5.D.2 where we show how the contour is depressed in the complex $\omega$ plane. The contributions on the depressed horizontal contour segments contribute negligibly because they rapidly damp, which leaves the contributions from the branch cuts and the two poles corresponding to roots of $\epsilon(\omega_\ell, k) = 0$.

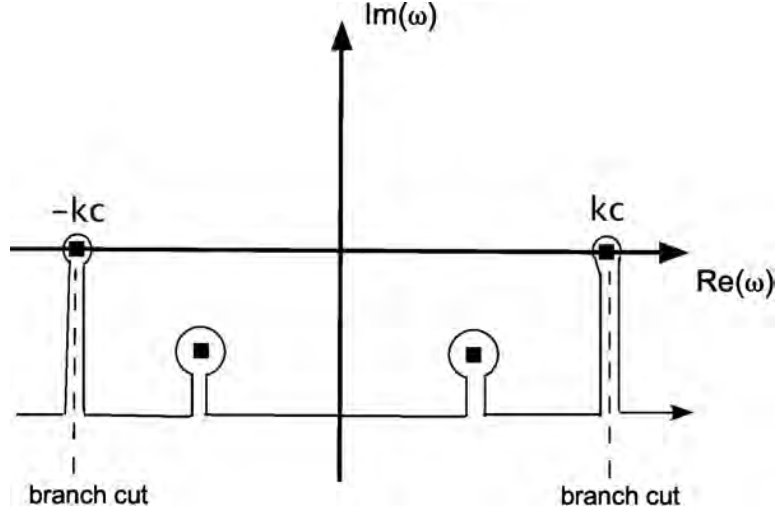

Figure 5.D.2 Diagram of contour integration in Eq.(5.D.3) after depressing the contour.

Example: Consider a relativistic distribution function that is a Gaussian in energy:
$$g(u) \propto \exp\left(\frac{-\beta m_e c^2}{\sqrt{1-u^2/c^2}}\right).$$
This distribution function has an essential singularity at $u=c$.

**5.E Initial-value problem – response function**

Consider the linear response to an external charge density perturbation $\rho^{ext}(\mathbf{k}, t > 0)$ given the initial condition on the velocity distribution function $\delta f(\mathbf{k}, \mathbf{v}; t = 0)$. We calculate $\delta f(\mathbf{k}, \mathbf{v}; t > 0)$ from the solution of the linearized Vlasov equation:

$$\frac{\partial}{\partial t}\delta f_s(\mathbf{k},\mathbf{v};t) + i\mathbf{k}\cdot\mathbf{v}\delta f_s = \frac{e_s}{m_s} i\phi(\mathbf{k};t)\mathbf{k}\cdot\frac{\partial f_{0,s}}{\partial \mathbf{v}} \tag{5.E.1}$$

Definition: The Laplace transform is defined as $g(\omega) = \int_0^\infty dt\, e^{i\omega t} g(t)$, $\text{Im}\,\omega > \gamma_{max}$. The inverse Laplace transform is $g(t) = \frac{1}{2\pi}\int d\omega\, e^{-i\omega t} g(\omega)$, $\text{Im}\,\omega > \gamma_{max}$ integrated along a contour $C$ in the upper $\omega$ half-plane.

Theorem: Integrating by parts and Laplace transforming the Vlasov equation,

$$\int_0^\infty e^{i\omega t} d(\delta f) = \left(e^{i\omega t}\delta f\right)\Big|_0^\infty - \int_0^\infty \delta f\, d(e^{i\omega t}) \Rightarrow$$

$$-\delta f(\mathbf{k},\mathbf{v};t=0) - i\omega\delta f(\mathbf{k},\mathbf{v};\omega) + i\mathbf{k}\cdot\mathbf{v}\delta f(\mathbf{k},\mathbf{v};\omega) = -\frac{e}{m}\delta\mathbf{E}(\mathbf{k},\omega)\cdot\frac{\partial f_0}{\partial \mathbf{v}} \tag{5.E.2a}$$

$$\Rightarrow \delta f(\mathbf{k},\mathbf{v};\omega) = \frac{i\delta f(\mathbf{k},\mathbf{v};t=0)}{\omega - \mathbf{k}\cdot\mathbf{v}} - \frac{e}{m}\phi(\mathbf{k},\omega)\frac{\mathbf{k}\cdot\frac{\partial f_0}{\partial \mathbf{v}}}{\omega - \mathbf{k}\cdot\mathbf{v}} \quad (5.\text{E}.2\text{b})$$

If $\rho^{ext} = 0$ then Poisson's equation is

$$k^2 \phi(\mathbf{k},\omega) = 4\pi e \int d^3\mathbf{v}\, \delta f(\mathbf{k},\mathbf{v},\omega) \quad (5.\text{E}.3)$$

We combine Eqs.(5.E.2b) and (5.E.3) to obtain

$$\varepsilon(\mathbf{k},\omega)\phi(\mathbf{k},\omega) = i\frac{4\pi e}{k^2}\int d^3\mathbf{v}\frac{\delta f(\mathbf{k},\mathbf{v};t=0)}{\omega - \mathbf{k}\cdot\mathbf{v}}, \quad \text{Im}\,\omega > 0 \quad (5.\text{E}.4)$$

The solution of Eq.(5.E.4) for $\phi(\mathbf{k},\omega)$ is obtained by dividing through by $\varepsilon$; and the solution for $\phi(\mathbf{k},t)$ is obtained by taking the inverse Laplace transform of $\phi(\mathbf{k},\omega)$. The contour integration for the inverse Laplace transform involves depressing the integration contour and looking for poles inside the integral, e.g., from the roots of $\varepsilon(\mathbf{k},\omega)$ in Eq.(5.D.3).

Examples:
1. $\delta f(k,u;t=0)$ an entire function of $u$, for example, $ue^{-u^2/2a^2}$  The velocity integral in Eq.(5.E.4) yields a principal value contribution and a residue from the simple pole at $\omega - k\mathrm{v} = 0$. In the inverse Laplace transform the contour is depressed so that the dominant contributions arise from the residues due to the poles corresponding to roots of $\varepsilon(\mathbf{k},\omega)$.
2. $\delta f(k,u;t=0) \sim \delta(u-u_0)$  The velocity integral in Eq.(5.E.4) yields $1/(\omega - ku_0)$ which is a simple pole on the real $\omega$ axis and never decays. We also note that $\delta f(x,u;t=0) \sim \delta(u-u_0)e^{i\mathbf{k}\cdot\mathbf{x}}$ Thus, how a disturbance is initially excited can be essentially important, because in this example the initial condition does not decay; it persists forever. The inclusion of collisions is very effective in causing otherwise undamped modes to die off.
3. $\delta f(k,u;t=0)$ is a square function of $u$ extending from $u=u_1$ to $u=u_2$. In the complex $\omega$ plane there are two branch cuts extending vertically down from $\omega=ku_1$ and $\omega=ku_1$ that affect $\phi(\mathbf{k},t)$, and the contour integration must go along the branch cuts.

We note that the inverse Laplace transform of Eq.(5.E.2b) has the following form:

$$\delta f(\mathbf{k},\mathbf{v};t) = e^{-i\mathbf{k}\cdot\mathbf{v}t}\delta f(\mathbf{k},\mathbf{v};t=0) + \frac{e^{-i\omega_\ell t}}{\omega_\ell - \mathbf{k}\cdot\mathbf{v}}\frac{\omega_p^2}{k^2}\frac{\hat{\mathbf{k}}\cdot\frac{\partial f_0}{\partial \mathbf{v}}\int_{-\infty}^{\infty}du\frac{\delta g(\mathbf{k},u;t=0)}{u-\omega_\ell/k}}{\left.\frac{\partial \varepsilon(\mathbf{k},\omega)}{\partial \omega}\right|_{\omega_\ell}} + \ldots \quad (5.\text{E}.5)$$

plus terms arising from initial conditions and oscillatory terms. If there are unstable roots $\omega_\ell$, the second term in (5.E.5) will overtake the first term and dominate after a short time. In contrast, damped roots will cause the second term to decay exponentially in time; and undamped oscillatory terms associated with initial conditions will survive.

Consider the second moment of $\delta f(\mathbf{k}, v; t)$ for the example of $\delta f(k, v; t = 0) = v e^{-v^2/2a^2}$:

$$\int_{-\infty}^{\infty} dv\, v^2 e^{-ikvt} v e^{-v^2/2a^2} \propto t^n e^{-k^2 a^2 t^2 / 2} \tag{5.E.6}$$

As time progresses the exponential in Eq.(5.E.6) dies off faster than does any power of $t$. This result demonstrates that oscillatory terms when integrated over velocity or $k$ fall off faster than do terms coming from Landau poles which lead to terms $e^{-(\,)t}$ that evidence Landau damping. This result is referred to as phase mixing and destructive interference. These oscillatory terms can be observed through nonlinear effects as in the echo phenomenon.[16]

## 5.F Van Kampen modes

Van Kampen modes are a special class of normal modes in a plasma. Consider distribution functions and electric fields satisfying the linearized Vlasov and Poisson equations:

$$\delta f(\mathbf{k}, \mathbf{v}:t) = \delta f_\omega(\mathbf{k}, \mathbf{v}) e^{-i\omega t} \quad \mathbf{E}(\mathbf{k}:t) = \mathbf{E}(\mathbf{k})_\omega e^{-i\omega t} = -i\mathbf{k}\phi_\omega e^{-i\omega t} \tag{5.F.1}$$

From the Vlasov-Poisson equations for a single species, e.g., electrons with infinitely massive ions,

$$(\omega - \mathbf{k}\cdot\mathbf{v})\delta f(\mathbf{k}, \omega; \mathbf{v}) = -\frac{e}{m}\phi(\mathbf{k}, \omega)\mathbf{k}\cdot\frac{\partial f_0}{\partial \mathbf{v}} \tag{5.F.2a}$$

$$k^2 \phi(\mathbf{k}, \omega) = 4\pi e \int d^3 v\, \delta f(\mathbf{k}, \omega; \mathbf{v}) \tag{5.F.2b}$$

We assume that $\omega$ is not real and divide (5.F.2a) by $\omega - \mathbf{k}\cdot\mathbf{v} \neq 0$ to obtain

$$\delta f(\mathbf{k}, \omega; \mathbf{v}) = (P)\frac{1}{\omega - \mathbf{k}\cdot\mathbf{v}}\left(-\frac{e}{m}\right)\phi(\mathbf{k}, \omega)\mathbf{k}\cdot\frac{\partial f_0}{\partial \mathbf{v}} + \lambda(\mathbf{k}, \omega)\delta(\omega - \mathbf{k}\cdot\mathbf{v}) \tag{5.F.3}$$

where $(P)$ denotes the principal value. Note that the inclusion of the $\delta$-function term still satisfies the Vlasov equation because $c(\omega - \mathbf{k}\cdot\mathbf{v})\delta(\omega - \mathbf{k}\cdot\mathbf{v}) = 0$. This is a singular solution for the distribution function that violates linearization. However, superposition of these $\omega$ Fourier components can eliminate the singularity.

We next substitute (5.F.3) in (5.F.2b, integrate over the two velocity directions perpendicular to $\mathbf{k}$, and go to one spatial dimension:

$$k^2 \phi(k, \omega) = -\frac{4\pi e^2}{m}\phi(k, \omega)(P)\int dv \frac{k\frac{\partial f_0}{\partial v}}{\omega - kv} + 4\pi e\lambda(k, \omega)\int dv\, \delta(\omega - kv) \tag{5.F.4}$$

The integral of the $\delta$-function gives $\int dv\, \delta(\omega - kv) = 1/k$. We then solve for $\lambda(k, \omega)$:

---

$$\lambda(k,\omega) = \frac{k^3}{4\pi e}\phi(k,\omega)\left[1 - \frac{\omega_p^2}{k^2}(P)\int dv \frac{g'(v)}{v-\omega/k}\right]$$

$$= \frac{k^3}{4\pi e}\phi(k,\omega)\left[1 - \frac{\omega_p^2}{k^2}\text{Re}\,Z'(v)\right] = \frac{k^3}{4\pi e}\phi(k,\omega)\text{Re}\,\varepsilon(k,\omega) \quad (5.F.5)$$

Using Eqs.(5.F.3) and (5.F.5) we obtain the distribution function for the Van Kampen modes:

$$\delta f(k,\omega;v) = \frac{k^3}{4\pi e}\phi(k,\omega)\left[\text{Re}\,\varepsilon(k,\omega)\delta(\omega - kv) - \frac{\omega_p^2}{k^2}(P)\frac{g'(v)}{\omega - kv}\right] \quad (5.F.6)$$

for real $\omega$. We note that there is no eigenvalue for $\omega$; all real values of $\omega$ are allowed. Moreover, $k$ is just a parameter. We can arbitrarily assign the amplitude of a specific Fourier component of $\phi(k,\omega) = 1$. The Van Kampen modes are a continuous spectrum in $\omega$ of singular normal modes described by Eq.(5.F.6). A superposition of Van Kampen modes can be non-singular and damp due to phase mixing and destructive interference.

Consider the special case that $\omega$ is complex. In this case $\delta(\omega - \mathbf{k}\cdot\mathbf{v}) = 0$; we can set $\lambda(k,\omega)=0$ and from the right side of Eq.(5.F.5) we recover the familiar dielectric function and dispersion relation for Im$\omega$ >0:

$$\left[1 - \frac{\omega_p^2}{k^2}(P)\int dv \frac{g'(v)}{v-\omega/k}\right] = \left[1 - \frac{\omega_p^2}{k^2}\text{Re}\,Z'(v)\right] = 0 \quad (5.F.7)$$

For Im$\omega$ <0 there are difficulties with analytic continuation in the lower half $\omega$ plane, and there is no proper dielectric function. For Im$\omega$ >0 we have a discrete set of normal modes $\omega_\ell$ as a function of $k$.

# 6. Nonlinear Vlasov plasma

### 6.A Vlasov-Poisson system in one dimension

#### 6.A.a Stationary nonlinear solutions – BGK modes

Our consideration of nonlinear phenomena begins with the Coulomb model, i.e., the Vlasov-Poisson equations model. The Vlasov-Maxwell equations system is significantly more complex.

Definition: The velocity distribution function is $f(x, y, z, v_x, v_y, v_z)$

Theorem: The Vlasov-Poisson equations are

$$\left(\frac{\partial}{\partial t} + \mathbf{v}\cdot\nabla_{\mathbf{x}} + \frac{e}{m}\mathbf{E}\cdot\nabla_{\mathbf{v}}\right)f(\mathbf{x},\mathbf{v};t) = 0 \tag{6.A.1a}$$

$$\mathbf{E} = -\nabla\phi \qquad \nabla^2\phi = -4\pi\int d^3\mathbf{v}\sum_s e_s f_s(\mathbf{x},\mathbf{v};t) \tag{6.A.1b}$$

The second step of simplification is to reduce the number of variables by going to one dimension:

$$\left(\frac{\partial}{\partial t} + v_x\frac{\partial}{\partial x} + \frac{e}{m}E_x\frac{\partial}{\partial v_x}\right)f(x,v_x;t) = 0 \tag{6.A.2a}$$

$$E_x = -\frac{\partial}{\partial x}\phi \qquad \frac{\partial^2}{\partial x^2}\phi = -4\pi\int dv_x \sum_s e_s f_s(x,v_x;t) \tag{6.A.2b}$$

where we have integrated over the ignorable velocity dimensions $v_y$ and $v_z$. This is the nonlinear system we hope to solve.

We cannot solve Eqs.(6.A.2) analytically for arbitrary initial conditions. However, there are some specific examples where analytic solutions do exist. One such example is due to Bernstein, Greene, and Kruskal (BGK).[17] BGK constructed solutions of the one-dimensional nonlinear Vlasov-Poisson equations that are stationary in time: $\phi(x,t) = \Phi(x - Vt)$. In the wave frame $x - Vt = x'$ the solutions for $f$ and $\Phi$ are stationary. Solution of the nonlinear Vlasov equation is obtained by the method of characteristics. In the wave frame the Vlasov equation becomes

$$\left(v_x\frac{\partial}{\partial x} + \frac{e_s}{m_s}E_x\frac{\partial}{\partial v_x}\right)f_s(x,v_x;t) = 0 \tag{6.A.3}$$

A constant of the particle motion is $H = \frac{1}{2}mv^2 + e\phi$. The most general solution of the Vlasov equation is any function of the constants of the motion: $f_s(x,v) = F_\pm^s(H)$. The ± subscript in $F_\pm^s$ denotes velocity directionality. For trapped particles ($H<0$) there is only one solution for $F$, while there are two solutions for untrapped (passing) particles ($H>0$). The Poisson equation becomes

$$\frac{\partial^2}{\partial x^2}\phi = -4\pi\int dv_x\sum_s e_s f_s(x,v_x;t) = -4\pi\sum_s e_s \sum_\pm \int_{e\phi(x)}^\infty dH \frac{F_\pm^s(H)}{\sqrt{2m_s(H - e_s\phi(x))}} \tag{6.A.4a}$$

where $dH\big|_{x\,\text{const}} = mvdv$ and $dv = \frac{dH}{mv} = \frac{dH}{\sqrt{2m_s(H-e_s\phi(x))}}$ (6.A.4b)

We now limit ourselves to a periodic potential over a length $\lambda$. Our Fourier representation will retain all harmonics of $k = 2\pi/\lambda$. The frequency in the laboratory (L) frame is related to the frequency in the wave (W) frame by $\omega^{(L)} =$ $\omega^{(W)} + kV$. In the notation used here $e \equiv$ the charge of the electron=-4.8×10⁻¹⁰ esu, and singly charged ions have the opposite sign charge. The number density is computed by integrating and summing over ± to obtain:

$$n_s = \int_{e\phi(x)}^{\infty} dH \frac{F^s(H)}{\sqrt{2m_s(H - e_s\phi(x))}} \tag{6.A.5}$$

---

[17] Ira B. Bernstein, John M. Greene, and Martin D. Kruskal. Phys. Rev. **108**, 546 (1957).

Example: Consider a velocity distribution function that is constant in a defined band of energy, $F_{\pm}^s(H) = C, H_1 \leq H \leq H_2$ for electrons (flip order $H_1$ and $H_2$ for ions), and select the electron energy band so that the electrons are untrapped. For the density computation, $F^s = \sum_{\pm} F_{\pm}^s$, and we obtain

$$n_s(\phi) = C \int_{H_1}^{H_2} \frac{dH}{\sqrt{2m(H - e\phi(x))}} \tag{6.A.6}$$

with x fixed so that the number density is a function of electric potential $\phi(x)$:

$$\frac{d^2\phi}{dx^2} = -4\pi\rho(\phi) \quad \rho(\phi) \equiv \sum_s e_s n_s(\phi) \tag{6.A.7}$$

At this point, we identify the potential $\phi$ as the pseudo-position $X$ and $x$ as the pseudo-time $T$ for a pseudo-particle in a pseudo-potential $U$:

$$\frac{d^2X}{dT^2} = -4\pi\rho(X) = -\frac{dU}{dX} \tag{6.A.8}$$

Definition: The pseudo-potential is $U(X) = 4\pi \int dX' \rho(X') = 4\pi \int d\phi' \rho(\phi')$
We solve Eq.(6.F.8) by introducing the energy

$$E = \frac{1}{2}\left(\frac{dX}{dT}\right)^2 + U(X), \tag{6.A.9}$$

solving algebraically for $dX/dT = d\phi/dx$, separating variables, and integrating to obtain

$$T = \int^X \frac{dX'}{\sqrt{2(E - U(X'))}} \quad x = \int^\phi \frac{d\phi'}{\sqrt{2(E - U(\phi'))}} \tag{6.A.10}$$

What is the physical significance of E? The term $\frac{1}{2}\left(\frac{dX}{dT}\right)^2 = \frac{1}{2}\left(\frac{d\phi}{dx}\right)^2$ is the square of the electric field. The second term
$U(\phi) = \sum_s 4\pi \int \frac{dHF(H)}{\sqrt{2m}} \int d\phi \frac{1}{\sqrt{H-e\phi}} \propto -\sum_s 4\pi \int \frac{dHF(H)}{\sqrt{2m}} \sqrt{H - e\phi} \propto$
$-\sum_s \int mv^2 f(x,v) dv \sim -p$ which is the pressure or kinetic energy density, if we recall our earlier consideration of the moment equations in Sec. 2.D and Eqs.(2.D.10-12).

Theorem: $\quad \frac{E}{4\pi} = \frac{E^2}{8\pi} - \int \rho d\phi = \frac{E^2}{8\pi} - p(x) \tag{6.A.11}$

Example: Consider a situation in which the particles are relatively cold and the potential is weak so that in the wave frame $H_s = \frac{1}{2}m_s V^2 + const$ and $F^s(H) = n_0\sqrt{2m_s H_s}\delta(H - H_s)$ . The electrons are untrapped. Then $p_s(\phi) = \int \frac{dH}{\sqrt{2m(H-e_s\phi)}} 2(H - e_s\phi)F(H) \propto \sum_s 2n_0 \sqrt{H_s(H_s - e_s\phi)}$, while the species number density is given by $n_s(\phi) = \int \frac{dH}{\sqrt{2m(H-e_s\phi)}} F(H) \propto \sum_s n_0 \sqrt{H_s/(H_s - e_s\phi)}$
There are integration constants in Eqs.(6.A.9) and (6.A.10), which are resolved by boundary and initial conditions, and physical arguments.[12] For small $\phi$ we expand $p_s(\phi)$ and $n_s(\phi)$ in power series in $e_s\phi$ through $\phi^2$. Because $e_s$ changes sign between ions and electrons, the coefficient of the linear term in $e_s\phi$ in $n_s$ must vanish (and, hence, in $p_s$ also) to guarantee overall charge density neutrality, which constrains an integration constant.

<u>Theorem</u>: Using $U(\phi) = -4\pi p(\phi) + const.$ it can be shown that $U(\phi) = \frac{1}{2}\left(\frac{\omega_p}{V}\right)^2 \phi^2 + O(\phi^3) + const.$ With $U$ a quadratic in $\phi$, Eq.(6.A.8) is the equation for a simple harmonic oscillator with pseudo-frequency $\Omega$ given by $\Omega=\omega_p/V$ in the wave frame, which corresponds to a period (pseudo wavelength) $\lambda=2\pi/\Omega=2\pi V/\omega_p$ or $k=2\pi/\lambda=\omega_p/V$. The Doppler-shifted frequency in the laboratory frame is then $\omega = kV = \omega_p$ regardless of wavelength because the solution of the Poisson-Vlasov equation was stationary in the wave frame. This recovers the standard result for a small amplitude wave in a cold plasma and lays the foundation for using the same machinery for the fully nonlinear problem.

### *6.A.b Nonlinear electron wave*

Consider the case of a nonlinear electron wave. To simplify, consider infinitely massive, stationary ions: $m_i \to \infty$ and finite mass electrons. Although the ions are stationary, they are charged and contribute to $U(\phi)$: $U(\phi) = 4\pi n_0 e\phi - 8\pi n_0 \sqrt{H_e(H_e - e\phi)}$, $e = -4.8\times 10^{-10}$ esu. We next introduce a change of variables:

$$\Phi \equiv \frac{e\phi}{H_e} \quad \psi \equiv 1-\Phi \quad \sigma \equiv \sqrt{\psi} \quad \tau \equiv 1+\sigma$$

$$\alpha \equiv \frac{8\pi n_0}{m_e V^2} \quad E'=E/\alpha \quad E''=E'-1$$
(6.A.12)

We substitute (6.A.12) into (6.A.10) and integrate to obtain an elementary integral whose result is:

$$x(\phi) = \sqrt{\frac{2}{\alpha}\frac{H_e}{e}} \left[\sqrt{E''-\tau^2} + \sin^{-1}\left(\frac{\tau}{\sqrt{E''}}\right)\right]$$
(6.A.13)

where E is the energy integral of the pseudo particle above the bottom of the minimum in $U$ as a function of $e\phi$ and is chosen so that at $e\phi=H_e$, $E=U(e\phi=H_e)$. The amplitude of $e\phi$ is limited by $H_e$, i.e., $\Phi \le 1$. Near $\Phi=1$ for this choice of E, the variation of $\Phi$ as a function of $x$ is cusp-like, $\delta\Phi \sim -(\delta x)^{2/3}$.

Exercise: From Eq.(6.A.13) calculate the periodicity length $\lambda(\Delta\phi)$ for V as $\Delta\phi \to 0$ and recover the earlier result in the laboratory frame $\omega=\omega_p$ independent of wave amplitude. In a one-dimensional cold plasma with charge sheets, there is no change in the characteristic oscillation frequency from $\omega=\omega_p$ as long as there is no crossing of the sheets. Sort out the units in Eqs.(6.A.8-6.A.13).

Again consider a velocity distribution function that is constant in a defined band of energy, $F_\pm^S(H) = C, H_1 \leq H \leq H_2$ for electrons and zero otherwise (with cold, infinitely massive ions). The electric potential is small, but finite; and the electrons are untrapped. For the density computation, $F^S = \sum_\pm F_\pm^S$; and we determine $U$ from Eqs.(6.A.5-6.A.8). On expanding in a Taylor series in powers of $\phi$ we obtain

$$U(\phi) = \frac{1}{2}\Omega_0^2\phi^2 + \frac{1}{3}\alpha\phi^3 + \frac{1}{4}\beta\alpha\phi^4 + ... \tag{6.A.14}$$

where the coefficients in the power series are

$$\Omega_0^2 = \frac{2\pi n_0 e^2}{\sqrt{H_1 H_2}} \tag{6.A.15a}$$

$$\alpha = \frac{\pi n_0 e^3}{2(H_1 H_2)^{3/2}}\left(H_1 + H_2 + \sqrt{H_1 H_2}\right) \tag{6.A.15b}$$

$$\beta = \frac{\pi n_0 e^4}{4(H_1 H_2)^{5/2}}\left(H_1^2 + H_1 H_2 + H_2^2 + (H_1 + H_2)\sqrt{H_1 H_2}\right) \tag{6.A.15c}$$

The presence of higher order terms in $\phi$ means that the pseudo potential $U$ is anharmonic. The solutions for the anharmonic oscillator are still given in terms of elliptic functions. From Landau and Lifschitz, *Mechanics*, 3rd Edition, Eq.(28.13), the characteristic frequency is given by

$$\Omega = \Omega_0 + \left(\frac{3}{8}\frac{\beta}{\Omega_0} - \frac{5}{12}\frac{\alpha^2}{\Omega_0^3}\right)\phi_0^2 \tag{6.A.16}$$

where the turning points are approximately -$\phi_0$ and $\phi_0$. In the laboratory frame $\omega = kV = \frac{2\pi}{\lambda} = \Omega V$ where we let $\frac{H_1+H_2}{2} = \frac{1}{2}m_e V^2$ assuming that $\Delta H \equiv H_2 - H_1 = mu\Delta u \ll H$. Finally, we obtain the frequency of the BGK mode in the laboratory frame

$$\omega^2 = \omega_p^2\left\{1 + \frac{k^2\Delta u^2}{4\omega_p^2}\left[1 + \frac{5}{2}\left(\frac{e\phi_0}{m_e V^2}\right)^2\right]\right\} \tag{6.A.17}$$

Recall that the linear dispersion relation for an infinitesimal amplitude electron wave with a square velocity distribution of electrons from $u=-c$ to $u=c$ is $\omega^2 = \omega_p^2 + k^2 c^2$ Different frequency shifts in the dispersion relation in Eq.(6.A.17) will be produced depending on the distribution of particles relative to the bottom of the periodic potential wells. For example, trapped electrons in the wave potential produce a negative frequency shift.

### 6.A.c Nonlinear ion wave and solitary pulse solutions

Consider the case of a nonlinear ion wave. We model the electrons to be in thermal equilibrium and responding adiabatically to the low frequency ion wave. We further assume $\omega \lesssim \omega_i$ We recall that from linear theory $\omega^2 = k^2 c_s^2/(1 + k^2 \lambda_D^2)$ where $c_s^2 = T_e/m_i$ for $T_i \ll T_e$ and $\lambda_D^2 = T_e/4\pi n e^2$. Here we adopt the convention that the electron charge is $-e$ and the ion charge is $+e$. One can show that

$$p(\phi) = n_0 \left[ T_e e^{\beta e \phi} + 2\sqrt{H_i(H_i - e\phi)} \right]$$

$$U(\phi) = const + \alpha \phi^2 + \beta \phi^3 + ...$$

(6.A.18)

In the acoustic range $k^2 \lambda_D^2 \ll 1$ with $|e\phi/T_e| k^2 \lambda_D^2 \ll 1$, then from Eq.(6.A.18) following the same methodology as in the nonlinear electron wave one obtains the nonlinear dispersion relation for the ion wave:

$$\omega^2(k,\phi_0) = k^2 c_s^2 - \frac{5}{6} \omega_i^2 \left| \frac{e\phi_0}{T_e} \right|^2$$

(6.A.19)

Example: Consider a potential pulse moving at a velocity near the ion sound speed, with $\phi \to 0$ for $x = \pm\infty$ in the frame moving with the pulse and a maximum $\phi = \phi_0$ at $x = 0$. Compute $U(\phi)$ as in Eqs.(6.A.18). From Eq.(6.A.10) one can compute an elementary integral leading to the results

$$\phi(x) = \phi_0 \operatorname{sech}^2(x/x_0) \quad \text{with } x_0 = \lambda_D \sqrt{\frac{3T_e}{|e\phi_0|}}$$

$$\frac{V^2}{c_s^2} = 1 + \frac{2}{3} \frac{|e\phi_0|}{T_e} > 1$$

(6.A.20)

Definition: This corresponds to a small-amplitude solitary pulse, called a small-amplitude soliton. This is a particular limiting case of a wave train with $\lambda \to \infty$.

*[Editor's note: At the time of these lectures, there had not been much attention given to electromagnetic BGK modes.]*

### 6.B Nonlinear Landau damping

### 6.B.a Phase-space dynamics

We assume there is a very weakly damped small-amplitude electron wave, $|\gamma| \ll \omega$ and $|e\phi/T_e| \ll 1$. In the wave frame we also assume the wave is a stationary BGK mode. At very small amplitude the wave is close to sinusoidal.

Particles with velocities close to the phase velocity of the wave are Landau damped. We focus attention on these resonant particles. A schematic of the (x',v') phase-space particle orbits for both trapped and untrapped particles in the wave frame is shown in Fig. 6.B.1 For an arbitrarily small-amplitude wave the trapped particle orbits are elliptic. Finite-amplitude effects introduce anharmonic effects that distort the ellipses.

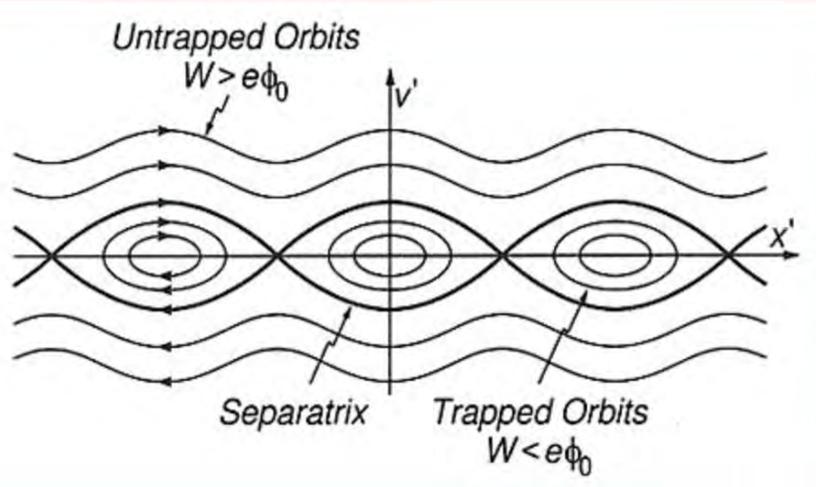

Figure 6.B.1 Phase-space orbits in the wave frame for untrapped ($W > e\phi_0$) and trapped ($W < e\phi_0$) orbits, where $W$ is the particle energy. The "separatrix" denotes where the particle energy $W = e\phi_0$.

### 6.B.b Quasilinear analysis – linear waves and nonlinear particle orbits

Consider a particle with energy near the top of the wave trough, i.e., $H \sim e\phi_0$ for $x \sim x_0$ where $\phi(x_0) = -\phi_0$. The range of velocities around v=V in the laboratory frame in which particles interact strongly with the wave scales as $\Delta v \sim \sqrt{e\phi_0/m}$. We note that along any orbit segment in which the magnitude of v grows the particle is extracting energy from the wave, but there is a compensating loss of energy as the magnitude of v decreases. In consequence, there is no net energy exchange in a stationary wave over time to leading order. We proceed with a direct calculation of the energy exchange of a particle with a wave:

$$\Delta v(t) = \int_0^t dt' \dot{v}(t') dt' = \frac{e}{m} \int_0^t dt' E(x') = \frac{e}{m} k\phi_0 \int_0^t dt' \sin kx', \quad x' \equiv x(t')$$

$$x' = x_0 + \int_0^{t'} dt'' v(t'') dt'' = x_0 + \int_0^{t'} dt'' \left[ v_0 + \frac{e}{m} k\phi_0 \int_0^{t''} dt''' \sin kx''' \right] \quad (6.B.1)$$

$$x''' \approx x_0 + v_0 t'''$$

After making substitutions, performing the nested time integrations, and expanding for small values of the electric field as it influences x' in sin(kx'), we average with respect to $x_0$ over a wavelength to obtain

$$\langle \Delta v(t) \rangle_{x_0} = \frac{1}{2}\left(\frac{e}{m}E_0\right)^2 \frac{1}{k^2 v_0^3}\left[2(\cos kv_0 t - 1) + kv_0 t \sin kv_0 t\right] \quad (6.B.2)$$

In the limit $kv_0 t \ll 1$, the expression in the square bracket on the right side of Eq.(6.B.2) [...] $\to -\frac{1}{12}(kv_0 t)^4$; and $\langle \Delta v \rangle_{x_0} \propto -v_0 t^4$ Thus, in the wave frame a resonant particle moving faster than the wave ($v_0 > 0$) is decelerated; and a resonant particle moving slower than the wave ($v_0 < 0$) is accelerated at early times. Recall that for a stable velocity distribution that damps the wave, there are more particles slower than the wave phase velocity than particles faster than the wave. The wave tends to drag resonant particles to its velocity irrespective of whether the particle is trapped or untrapped.

Exercise: Verify the calculation leading to Eq.(6.B.2).

Next consider the time dependence of the wave momentum density. The momentum density $P_W$ of the wave is related to the energy density $E_W$ of a wave:

$$P_W = \frac{k}{\omega}E_W = k\frac{\partial \varepsilon}{\partial \omega}\left\langle \frac{|\mathbf{E}|^2}{8\pi}\right\rangle_x = \frac{k}{\omega_p}\frac{E_0^2}{8\pi} \qquad \varepsilon = 1 - \frac{\omega^2}{\omega_p^2} \quad (6.B.3)$$

in the laboratory frame. The derivation of the expression in Eq.(6.B.3) is Galilean invariant and valid in any frame. If the wave is exponentially damped, $E_0(t) = E_0(t)e^{\gamma t}, \gamma < 0$; and the wave momentum and energy densities inherit this time dependence. At early time $e^{\gamma t} \approx 1 + \gamma t$, and the wave momentum and energy densities vary linearly in time:

$$\Delta P_W(t) \approx P_W(0)2\gamma t \quad (6.B.4)$$

We now calculate whether the sum of the wave and particle momentum densities is conserved. We integrate over all possible initial velocities of the resonant particles in the wave frame:

$$\Delta P|_{resonant} = \int_{resonant} dv_0 f_0(v_0) m \langle \Delta v(t) \rangle_{x_0}$$

$$= \int_{resonant} dv_0 \left[ f_0(0) + v_0 f_0'(0) + \ldots \right] m \langle \Delta v(t) \rangle_{x_0}$$

$$= \frac{m}{k^2} \int_{resonant} dv_0 \frac{1}{v_0^3} \left[ f_0(0) + v_0 f_0'(0) + \ldots \right] \frac{1}{2} \left( \frac{e}{m} E_0 \right)^2 \left[ 2(\cos\eta - 1) + \eta\sin\eta \right]$$

$$\approx \frac{m}{k^2} \frac{1}{2} \left( \frac{e}{m} E_0 \right)^2 \int_{-\infty}^{\infty} dv_0 \frac{1}{v_0^3} v_0 f_0'(0) \left[ 2(\cos\eta - 1) + \eta\sin\eta \right]$$

$$= \frac{m}{k^2} \frac{1}{2} \left( \frac{e}{m} E_0 \right)^2 f_0'(0) kt \int_{-\infty}^{\infty} d\eta \frac{1}{\eta^2} \left[ 2(\cos\eta - 1) + \eta\sin\eta \right] = -\frac{\pi}{2} \frac{m}{k^2} \left( \frac{e}{m} E_0 \right)^2 f_0'(0) kt$$

(6.B.5)

where $\eta = kv_0 t$

We can evaluate the $\gamma$ in Eq.(6.B.4) using $\frac{\gamma}{\omega_p} = \frac{\pi}{2} \left( \frac{\omega_p}{k} \right)^2 g'(v) = \frac{\pi}{2} \left( \frac{\omega_p}{k} \right)^2 \frac{f'(0)}{n_0}$.
Substituting this into Eqs.(6.B.3) and (6.B.4) we deduce that

$$\Delta P_W + \Delta P|_{resonant} = 0 \quad (6.B.6)$$

### 6.B.c Trapped and untrapped particles – evolution of the distribution function

We next consider the trajectories of the trapped and untrapped (passing) particles. In the wave frame, if particle trajectories are localized to a small excursion distance relative to the bottom of the potential well, $\phi(x_0) = -\phi_0$, then the potential can be expanded around $x_0$:

$$e\phi(x) = e\phi_0 \cos kx \to e\phi_0 \left( -1 + \frac{1}{2} k^2 \delta x^2 + \ldots \right)$$

$$\to H = \frac{1}{2} mv^2 + e\phi_0 \frac{1}{2} k^2 \delta x^2 + const$$

(6.B.7)

Equation (6.B.7) is the Hamiltonian for a simple harmonic oscillator whose characteristic oscillation frequency (bounce frequency $\omega_B$) is

$$\omega_B = \sqrt{\frac{k^2 e\phi_0}{m}} \quad (6.B.8)$$

with the restriction that the electric field amplitude temporal variation is negligible on the time scale $1/\omega_B$, i.e., $\gamma/\omega_B \ll 1$. We recall that our analysis also requires that $e\phi_0 \ll T$ and, hence, $\Delta v \ll v_{th}$ and $\omega_B \ll kv_{th}$. Initially the wave tranfers energy and momentum to the resonant particles, but a little later (on the time scale of the trapped particle bounce motion), energy and momentum are transferred back from the particles to the wave.[18] The general solutions of the equations of motion resulting from Eq.(6.B.7) for $e\phi(x) = e\phi_0 \cos kx$ are given in terms of Jacobi elliptic functions. A useful reference for elliptic functions is Byrd and Friedman.[19]

---

[18] T. O'Neil, Physics of Fluids **8**, 2255 (1965); G. J. Morales and T. M. O'Neil, Phys. Rev. Lett. **28**, 417 (1972).

For untrapped particles, $H \geq e\phi_0$, the solution of the equation of motion is given by

$$v(t) = \pm \sqrt{\frac{2}{m}(H + e\phi_0)} \, \text{dn}\left(\frac{\omega_B t}{\kappa}, \kappa\right), \quad \text{where } \kappa \equiv \sqrt{\frac{2e\phi_0}{H + e\phi_0}} < 1 \qquad (6.B.8)$$

without being precise about the initial conditions for the velocity of the particle. The passing particle velocity has a periodic variation with period given by

$$\tau_t = \frac{2\kappa K(\kappa)}{\omega_B} \qquad (6.B.9)$$

where $K$ is one of the Jacobi elliptic functions:

$$K(\kappa) \;\rightarrow\; \ln \frac{4}{\sqrt{1-\kappa^2}}, \text{ for } \kappa \rightarrow 1$$
$$\rightarrow \;\; \frac{\pi}{2}, \text{ for } \kappa \rightarrow 0 \qquad (6.B.10)$$

The Jacobi elliptic function $\text{dn}\left(\frac{\omega_B t}{\kappa}, \kappa\right) = 1$ for $t=0$ and is bounded by

$$\sqrt{1-\kappa^2} \leq \text{dn}\left(\frac{\omega_B t}{\kappa}, \kappa\right) \leq 1 \qquad (6.B.11)$$

For trapped particles $H < e\phi_0$, the solution of the equation of motion is given by

$$v(t) = \pm \sqrt{\frac{2}{m}(H + e\phi_0)} \, \text{cn}(\omega_B t, \kappa), \quad \text{where } \kappa \equiv \sqrt{\frac{H + e\phi_0}{2e\phi_0}} < 1 \qquad (6.B.12)$$

The trapped particle velocity has a periodic variation with period given by

$$\tau_t = \frac{4K(\kappa)}{\omega_B} \;\rightarrow\; \frac{2\pi}{\omega_B} \text{ for } \kappa \rightarrow 0 \qquad (6.B.13)$$

The Jacobi elliptic function $\text{cn}(\omega_B t, \kappa)$ is bounded by $\pm 1$. A distribution of particles sharing the same energy, but distributed in space, will have different relative phases in the wave frame: $t \rightarrow t - t_1$. A distribution of particle energies can be parameterized in velocity by their distribution of initial speeds (initial conditions are trivial constants of the motion).

Figures 6.B.2, 6.B.3, and 6.B.4 illustrate the particle orbits and time evolution of quantities associated with the particles and the wave. We know from Eqs.(6.B.5) and (6.B.6) that the momentum in the resonant particles initially grows at the expense of the wave amplitude. Thus, the evolution of the wave amplitude shown in Fig. 6.B.3 and the momentum in the resonant particles mirror one another. The wave amplitude decreases and then increases, but not back to its initial amplitude; and the momentum in resonant particles first increases and then decreases, but not back to its initial amplitude. The period of the oscillations is approximately given by

---

[19] P. F. Byrd and M. D. Friedman, *Handbook of Elliptic Integrals for Engineers and Physicists*, (Springer-Verlag, 1971 & Springer, 2013).

the trapped particle bounce period, Eq.(6.B.13). The oscillations in the damping rate are at the bounce frequency and phase mix to zero, while the oscillations in the frequency shift are at twice the bounce frequency and asymptote to a finite value (Fig. 6.B.4).

[*Editor' note: There were several other contemporaneous papers in the literature addressing the nonlinear frequency shift of an electron plasma wave in addition to the work of Morales and O'Neil, e.g., R. L Dewar (1972), Manheimer and Flynn (1971), and Lee and Pocobelli (1972).*]

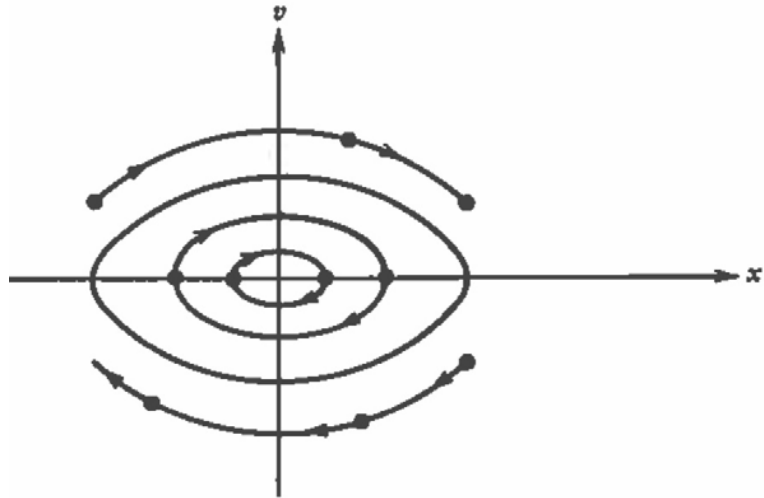

Fig. 6.B.2 Initial particle orbits in phase space in wave frame (ref. Fig. 6.14 in D. R. Nicholson, *Introduction to Plasma Theory* (Wiley, 1983)).

<u>Theorem</u>: Consider a velocity distribution function $f$ that is initially uniform in position $x$ and decreasing in $|v|$. $f \to f_\pm(H)$ where $H = \frac{1}{2}mv^2 + e\phi(x)$ is a contant of the motion. Hence, $f$ will be a constant of the motion on the phase-space orbits.

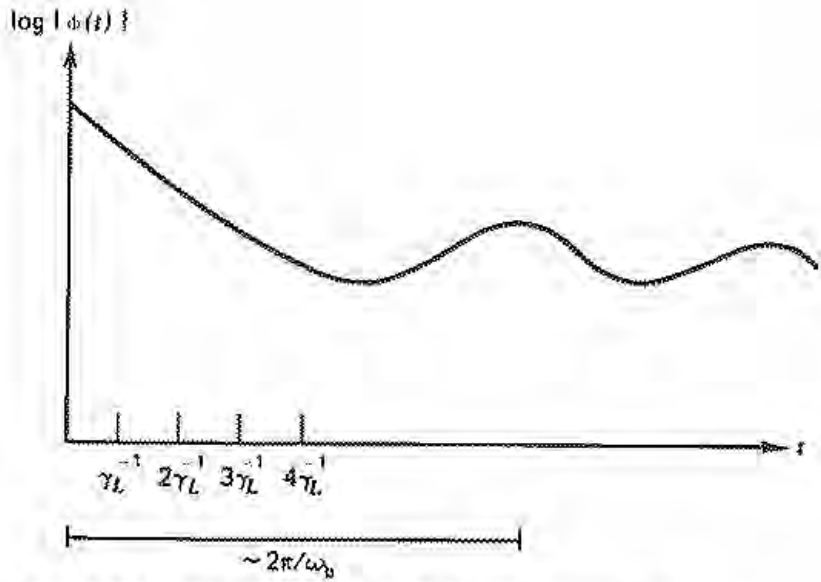

Fig. 6.B.3 Langmuir wave amplitude versus time
(ref. Fig. 6.17 in D. R. Nicholson, *Introduction to Plasma Theory* (Wiley, 1983)).

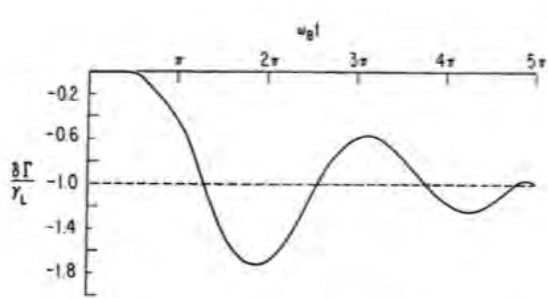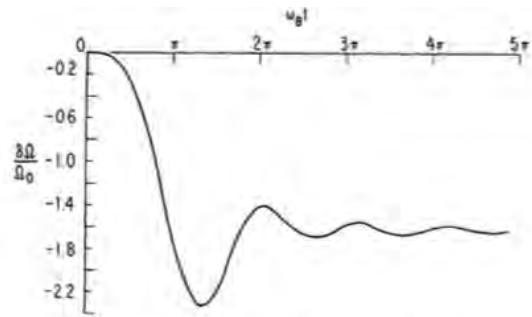

Fig. 6.B.4 Figures 1 and 2 from G. J. Morales and T. M. O'Neil,
Phys. Rev. Lett. **28**, 417 (1972) illustrating the self-consistent
effects of nonlinear Landau damping on the total damping rate
and frequency shift of the electrostatic wave.

Exercise: Read O'Neil's and Morales and O'Neil's seminal papers, Ref. 13. The Morales and O'Neil paper has an illuminating discussion of the relationship between the nonlinear increments in the damping rate and frequency, and the excursions in the momentum and energy in the wave and the particles.

## 6.B.d Effects of trapped particles on longitudinal plasma waves and saturation of intstabilities

The interaction of a finite-amplitude wave with a velocity distribution that is decreasing with respect to velocity in the neighborhood of the velocity equal to the phase velocity of the wave is depicted in Fig. 6.B.5 Particles traveling slower than the wave are accelerated, while particles traveling faster than the wave are decelerated. Because there are more particles slower than the phase velocity, there is net momentum and energy transferred from the wave to the particles. The region of flattening in velocity is centered at the phase velocity of the wave and the width of the region is associated with the trapping width $\sim \sqrt{e\phi_0/m}$. An estimate of the asymptotic field amplitude of the electron plasma wave can be computed from

$$E_0(t) = E_0(0)\exp\left(\int_0^t dt' \gamma(t')\right) \sim E_0(0)\left(1-|\gamma_L|\tau_T\right) \quad (6.B.14)$$

where $\gamma$ is the total growth rate ($\gamma<0$ due to Landau damping, $\gamma_L < 0$) and is given by $\gamma = \gamma_L + \delta\Gamma$ with $\delta\Gamma$ given in Fig. 6.B.4, $\gamma_L \sim \left(\frac{\omega_p^3}{k^2}\right)\left(\frac{f_0'}{n_0}\right)$, and $\tau_T \sim \sqrt{m/e\phi_0}(1/k)$. The sign of $\gamma$ is opposite the sign of the time derivative of the resonant particles, Eqs.(6.B.5) and (6.B.6). Thus, the wave settles down to a BGK mode of constant amplitude over a few bounce times.

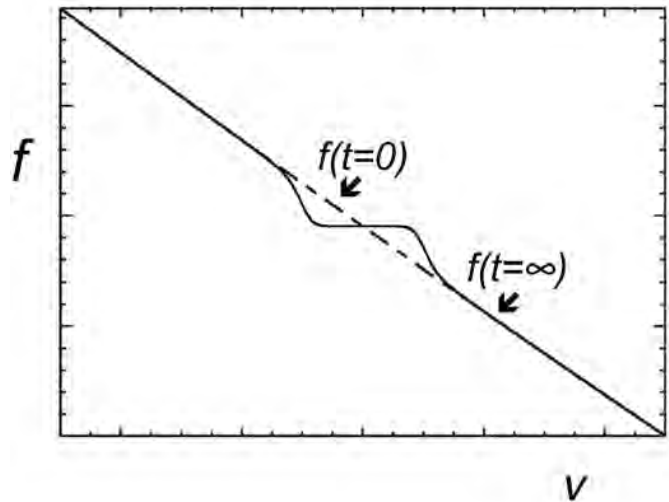

Fig. 6.B.5 Schematic showing flattening of the velocity distribution function due to particle trapping.

The interaction of an unstable wave(s) with resonant particles is expected to evolve as follows. Consider a bump-on-tail velocity distribution function (Fig. 2.I.1). The velocity inversion in the distribution function is a source of free energy to destabilize a wave with phase velocity v. There are more resonant particles with u> $v_{ph}$ than for u< $v_{ph}$, which allows the wave to grow at the expense of the energy in the resonant particles which decreases (more particles are decelerated than are accelerated). Thus, wave-particle interactions lead to a flattening of the distribution function near the phase velocity of the wave. In contrast to Fig. 6.B.3, the plot of the logarithm of the wave energy vs. time for the unstable case shows linear growth in time until there is sufficient flattening of the velocity distribution function (due to mixing) stabilizes the plasma. The wave energy no longer grows, but we expect decaying oscillations similar to those in Fig. 6.B.3. If the wave-particle interaction is relatively coherent because the wave spectrum is dominated by a single wave, then the relative peaks in the oscillations of the wave energy will be separated by the characteristic trapped particle bounce time $\tau_T$.

We expect that the growing wave will approach saturation when the trapped particles accumulate $\pi$ radians of phase in the growing wave:

$$\pi = \int_0^{t_{sat}} \omega(t)dt = k\sqrt{\frac{e}{m}} \int_0^{\phi_{sat}} \frac{d\phi_0}{d\phi_0/dt = \gamma(\phi_0)\phi_0} \phi_0^{1/2}(t) = k\sqrt{\frac{e}{m}} \int_0^{\phi_{sat}} \frac{d\phi_0}{\phi_0^{1/2}\gamma(\phi_0)} \approx 2k\sqrt{\frac{e}{m}} \frac{\phi_{sat}^{1/2}}{\gamma_L} \quad (6.B.15)$$

Hence,

$$\frac{\omega_T(\phi_0^{sat})}{\gamma_L} \approx \frac{\pi}{2} \quad \text{and} \quad \left|e\phi_0^{sat}\right| = m\left(\frac{\pi}{2}\frac{\gamma_L}{k}\right)^2 \quad (6.B.16)$$

Eq.(6.B.15) is derived assuming the growth rate $\gamma$ is well approximated by the linear growth rate $\gamma_L$ up to the point of saturation. This assumption is reasonably good until the last e-folding or two of the growing wave, when the growth rate starts to decrease to zero. Research by Fried, Liu, Means, and Sagdeev[20] showed numerically that for a broad range of parameters $\omega_T(\phi_0^{sat})/\gamma_L \approx 3.2$ We note that this analysis has assumed that a single wave grows up to dominate the spectrum so that the wave-particle interaction is coherent. This assumption has eliminated the noise of other waves that might otherwise be expected to grow.

Exercise: Using energy conservation, derive a relation between the width in velocity over which the velocity distribution is flattened and the wave amplitude at saturation; and using Eq.(6.B.16) and the results in Sec. 2.F.a relate the velocity width to the linear growth rate and the plasma parameters for the weak beam-plasma instability. Show for the bump-on-tail velocity distribution subject to flattening, that the distance $\Delta v$ between the two locations where $g' = 0$ scales as

$$\frac{v_T}{\Delta v} \sim \frac{n_{bump}}{n_0} \frac{V}{v_{th}} \left(\frac{V}{\Delta v}\right)^2 \quad (6.B.17)$$

where V is the phase velocity and $v_T=(e\phi/m_e)^{1/2}$ is the trapping velocity. We note that $n_{bump}/n_0 \ll 1$, $V/v_{th} > 1$, and $(V/\Delta v)^2 \gg 1$, which leaves a good deal of freedom in determining $v_T/\Delta v$ in Eq.(6.B.17).

---

[20] B. D. Fried, C. S. Liu, R. W. Means, R. Z. Sagdeev, "Nonlinear Evolution and Saturation of an Unstable Electrostatic Wave," UCLA Plasma Physics Technical Report, 1971; http://www.dtic.mil/get-tr-doc/pdf?AD=AD0730123.

The nonlinear Landau damping of an ion wave ought to be similar to that of the damping of an electron wave considered already. However, in the case of the ion wave one should consider the resonant interactions of both electrons and ions.

There are many calculations and simulations of the saturation of instabilities in a collisionless plasma.  Here is brief list (circa 1971):

1. Bump-on-tail, which leads to a BGK mode
2. Weak-beam-instability (linear evolution discussed in Sec. 2.F.a), which leads to a BGK mode[21]
3. Electron-ion-two-stream instability, which leads to a chaotic state[22]
4. Two-stream instability (equal beams), which leads to a chaotic state[23]
5. Ion-acoustic instability[24]
6. Cyclotron instability[25]

The references given here are only representative and by no means complete.

Example: Weak-beam instability -- We can derive additional results for the saturation of the weak-beam instability. The linear attributes of the weak-beam instability were presented in Sec. 2.F.a . In the weak-beam instability there is an approximate resonance between the beam velocity $u_b$ and the unstable wave phase velocity $v_\phi$. However, a careful analysis shows that $v_\phi < u_b$ for instability. The results for the dispersion relation of the weak-beam instability in Eq.(2.F.7) are

$$\frac{\delta\omega}{\omega_0} = \eta^{1/3} = |\eta^{1/3}|\left\{1, -\frac{1}{2} \pm i\sqrt{\frac{3}{2}}\right\}, \qquad \eta \approx \frac{1}{2}\frac{\omega_b^2}{\omega_p^2} \qquad (6.B.18)$$

where $\omega_0 = k_0 u_b$ and $\omega = \omega_0 + \delta\omega$, i.e., the unstable wave frequency is slightly downshifted. The wave energy and the wave momentum densities are given by

$$E_w = \omega \frac{\partial \varepsilon}{\partial \omega} \frac{|E|^2}{8\pi} = \omega \frac{2}{\omega_p} \frac{|E|^2}{8\pi} \qquad P_w = k \frac{2}{\omega_p} \frac{|E|^2}{8\pi} \qquad (6.B.19)$$

where $v_\phi = \omega/k$. The wave momentum density grows at the expense of the momentum density in the weak beam, and the wave energy density grows as the momentum density grows. Consider a single wave with $k=k_0$ the most unstable wave such that

$$\phi(x,t) = \phi_0(t)e^{ik_0 x} + c.c. \qquad \phi_0(t) = \phi_0(0)\exp\left(-i\int_0^t dt'\,\omega(t')\right) \qquad (6.B.20)$$

---

[21] T. M. O'Neil, J. H. Winfrey, J. M. Malmberg, Phys. Fluids **14**, 1204 (1971).
[22] A. Hirose, Plasma Phys. 20, 481 (1978).
[23] J. P. Freidberg and B. M. Marder, Phys. Rev. A **4**, 1549 (1971).
[24] K. Nishikawa and C. S. Wu, Phys. Rev. Lett. 23, 1020 (1969).
[25] R. E. Aamodt and S. E. Bodner, Phys. Fluids **12**, 1471 (1969).

The beam can be represented as a sum of particles:

$$x_i(t) \quad i=1,\ldots,N_B \qquad n_B(x,t) = \sum_{i=1}^{N_B} \delta(x - x_i(t))$$

$$n_B(k_0,t) \equiv \int dx\, e^{-ik_0 x} n_B(x,t) = \sum_{i=1}^{N_B} e^{-ik_0 x_i(t)} \qquad (6.B.21)$$

The particle simulation then solves the nonlinear equations of motion given by

$$m\ddot{x}_i(t) = e(-ik_0)\phi(x_i,t) \qquad (6.B.22)$$

and the Poisson equation to determine the electric potential $\phi_0$ self-consistently:

$$k_0^2 \phi(k_0,t) = \frac{4\pi e n_b(k_0,t)}{\varepsilon(k_0,\omega)} \qquad \varepsilon(k_0,\omega) = 1 - \frac{\omega_p^2}{\omega^2} - \frac{\omega_B^2}{(\omega - k_0 u_b)^2} \qquad (6.B.23)$$

In the beam frame the dielectric function becomes

$$\varepsilon(k_0,\omega) = 1 - \frac{\omega_p^2}{(\omega + k_0 u_b)^2} - \frac{\omega_b^2}{\omega^2} \approx 2\frac{\omega}{\omega_p} \qquad (6.B.24)$$

at resonance using $\omega_b^2 \ll \omega_p^2$. If we use Eq.(6.B.24) in the Poisson equation, Eq.(6.B.23), and Fourier transform from $\omega$ to $t$, one obtains that the wave amplitude grows as the beam density perturbation grows:

$$ik_0^2 \frac{d\phi_0(x_0,t)}{dt} = -\frac{1}{2}\omega_p 4\pi e n_b(k_0,t) \qquad (6.B.25)$$

The simulation then advances the nonlinear equations of motion Eqs.(6.B.21) and (6.B.22), and the Poisson equation (6.B.25). A plot of $\ln\phi_0$ vs. time from simulations shows initial linear growth consistent with (6.B.18) and then saturation with amplitude oscillations having a period consistent with the trapping frequency $\omega_T = k_0|e\phi_0/m|^{1/2}$. The saturation amplitude and saturation time observed in the simulations scale as

$$\frac{|E_0^{sat}|^2}{8\pi} \approx n_b m u_b^2 |\eta^{1/3}| \qquad \omega_0 t^{sat} \approx \eta^{-1/3} \gg 1 \qquad (6.B.26)$$

This scaling of the saturation amplitude for the weak-beam instability in Eq.(6.B.26) differs significantly from the scaling in Eq.(6.B.16) for the bump-on-tail instability.

[*Editor's note: There is a discussion of the saturation of the weak-beam instability based on particle simulations in Section 5-11 of C. K. Birdsall and A. B. Langdon, <u>Plasma Physics via Computer Simulation</u> (McGraw-Hill, New York, 1985).*]

## 6.C Stability of electrostatic BGK modes – sideband instability of Kruer, Dawson, and Sudan

Kruer, Dawson, and Sudan[26] proposed a theory to explain the observation of satellite frequencies in an experiment[27] in which a large amplitude electron plasma wave trapped electrons. The instability grows up from noise and results from the interaction of the trapped particles and the large-amplitude primary wave. The derivation considers particles trapped near the bottom of the wave troughs. The electrons oscillate back and forth at the trapping frequency $\omega_T = \omega_B = k_0 |e\phi_0/m|^{1/2}$. Consider the electron equation of motion in one spatial dimension for the trapped electrons as perturbed by a small-amplitude disturbance:

$$\ddot{x}_j = \frac{e}{m}\delta E(x_j, t) - \omega_B^2 x_j \tag{6.C.1}$$

whose Fourier transform from time to frequency is

$$-\omega^2 x_j(\omega) = \frac{e}{m}\delta E(x_j, \omega) - \omega_B^2 x_j(\omega) \tag{6.C.2}$$

and

$$x_j(\omega) = \frac{1}{\omega_B^2 - \omega^2}\frac{e}{m}\delta E(x_j, t) \tag{6.C.3}$$

The charge density for the trapped electrons and the resulting Poisson equation are

$$n_T(x,t) = \sum_j \delta\bigl(x - x_j(t)\bigr) N_j \qquad ik\delta E(k,\omega) = \frac{4\pi e \delta n_T(k,\omega)}{\varepsilon(k,\omega)} \tag{6.C.4}$$

where $N_j$ is the number of trapped electrons in the $j$th equally spaced wave trough. The spatially averaged unperturbed trapped particle number density is then $n_T^o = \frac{N_1}{\lambda_o}$. The Fourier transform of $n_T(x,t)$ from $x$ to $k$ space is

$$n_T(k,t) = \lambda_o n_T^o \sum_j \int dx\, e^{-ikx}\delta\bigl(x - x_j(t)\bigr) = \lambda_o n_T^o \sum_j e^{-ikx_j(t)} \approx \lambda_o n_T^o \sum_j e^{-ikx_j^o(t)}\bigl[1 - ik\delta x_j(t)\bigr] \tag{6.C.5}$$

where $x_j(t) = x_j^o + \delta x_j(t)$ has been linearized for small oscillations. From Eqs.(6.C.4) and (6.C.5)

$$\delta E(k,\omega) = -\frac{4\pi e}{\varepsilon(k,\omega)}\left[\frac{2\pi}{k_0}n_T^o \sum_j e^{-ikx_j^o}\delta x_j(\omega)\right] \tag{6.C.6}$$

from which one obtains

---

[26] W. L. Kruer, J. M. Dawson, and R. N. Sudan, Phys. Rev. Lett. **23**, 838 (1969).
[27] C. B. Wharton, J. H. Malmberg, and T. M. O'Neil, Phys. Fluids **11**, 1761 (1968).

$$\delta E(k,\omega) = \frac{4\pi n_T^o e^2}{m} \frac{1}{\left(\omega^2 - \omega_B^2\right)\varepsilon(k,\omega)} \frac{2\pi}{k_0} \sum_j e^{-ikx_j^o} \int \frac{dk'}{2\pi} \delta E(k',\omega) e^{ik'x_j^o} \quad (6.C.7)$$

where the "plasma frequency" associated with the trapped electrons is $\omega_{pT}^2 \equiv 4\pi n_T^o e^2/m \ll \omega_B^2$. Equation (6.C.7) can be rewritten using the identity $k_0^{-1} \sum_\ell e^{i(k'-k)2\pi\ell/k_0} = \sum_{\ell=-\infty}^{\infty} \delta(k' - k - \ell k_0)$:

$$\left(\omega^2 - \omega_B^2\right)\delta E(k,\omega) = \frac{4\pi n_T^o e^2}{m} \frac{1}{\varepsilon(k,\omega)} \sum_{\ell=-\infty}^{\infty} \int dk' \delta E(k',\omega) \delta(k'-k-\ell k_0) \quad (6.C.8)$$

One $k$ mode is coupled to all $k + \ell k_o$ where $\ell$ is an integer and $k_0$ is the effective lattice constant. The integral on the right side of Eq.(6.C.8) is trivial, and one obtains

$$\frac{\left(\omega^2 - \omega_B^2\right)}{\omega_{pT}^2} \delta E(k,\omega) = \frac{1}{\varepsilon(k,\omega)} \sum_{\ell=-\infty}^{\infty} \delta E(k+\ell k_0,\omega)$$

$$\Rightarrow \sum_{\ell'} \delta E(k+\ell' k_0,\omega) = \sum_{\ell'} \left[ \frac{\omega_{pT}^2}{\left(\omega^2 - \omega_B^2\right)} \frac{1}{\varepsilon(k+\ell' k_0,\omega)} \sum_{\ell} \delta E(k+\ell k_0,\omega) \right] \quad (6.C.9)$$

$$\Rightarrow 1 = \sum_{\ell'} \left[ \frac{\omega_{pT}^2}{\left(\omega^2 - \omega_B^2\right)} \frac{1}{\varepsilon(k+\ell' k_0,\omega)} \right]$$

An example of a warm fluid dielectric function in Eq.(6.C.9) in the laboratory frame is

$$\varepsilon^{(L)}(k,\omega) = 1 - \frac{\omega_p^2}{\omega^2 - 3k^2 v_{th}^2} \quad (6.C.10)$$

In the wave frame this fluid dielectric becomes

$$\varepsilon^{(W)}(k,\omega) = 1 - \frac{\omega_p^2}{\left(\omega+kV_0\right)^2 - 3k^2 v_{th}^2} \quad (6.C.11)$$

The dispersion relation for the sideband instability is given by the last expression in Eq.(6.C.9), which can be rewritten as

$$\frac{\left(\omega^2 - \omega_B^2\right)}{\omega_{pT}^2} = \sum_{\ell} \left[ \frac{1}{\varepsilon(k+\ell k_0,\omega)} \right] \quad (6.C.12)$$

The left side of Eq.(6.C.12) has a small divisor; thus, we look for terms in the sum on the right side that are near resonant, i.e., values of $k$ and $\ell$ such that $\epsilon(k + \ell k_0, \omega) = 0$. For $k = k' + \ell k_0 = \{k_0 + \delta k, -k_0 + \delta k\}$ and $\delta k$ small, the two terms $\ell = 0, -2$ dominate the sum on the right side of (6.C.12):

$$\frac{(\omega^2 - \omega_B^2)}{\omega_{pT}^2} = \frac{1}{\varepsilon(k,\omega)} + \frac{1}{\varepsilon(k-2k_0,\omega)} \tag{6.C.13}$$

The dispersion relation in Eq.(6.C.13) using Eq.(6.C.11) in the wave frame is a cubic, which we expand for small $\omega$ and $\delta k$ to quadratic order in small quantities $\omega/\omega_B$ and $\delta k/k_0$:

$$\frac{\omega}{\omega_p} = \pm \frac{\delta k}{k_0}\left[-1 + \left\{\frac{\lambda[(1+\lambda)(2\beta+4)+1] \pm i\sqrt{\lambda(\lambda+2)(2\beta+3)}}{(1+\lambda)^2(2\beta+4)-1}\right\}\right] \tag{6.C.14}$$

where $\beta \equiv \omega_B^2/\omega_{pT}^2 \gg 1$, $\lambda \equiv (\omega_o^L - \omega_p)/\omega_p = 3v_{th}^2/2V_\phi^2 \ll 1$, and $\omega_o^L$ is the frequency of the large-amplitude wave in the laboratory frame (the Bohm-Gross frequency in this case). $\lambda$ must be small, else the primary wave and its sidebands are strongly Landau damped. Continuing the expansion of (6.C.14) for large $\beta$ and small $\lambda$, we obtain the simpler expression:

$$\frac{\omega}{\omega_p} = \pm \frac{\delta k}{k_0}\left[-1 + \lambda \pm i\sqrt{\frac{\lambda}{\beta}}\right] \tag{6.C.15}$$

In the laboratory frame:

$$\omega^{(L)} = \omega^{(W)} + (k_0 + \delta k)V_0 \approx \omega_0(k_0) + \delta k V_g^o\left(1 + \frac{1}{2\beta}\right) + iv_{th}|\delta k|\sqrt{\frac{3}{4\beta}} \tag{6.C.16}$$

The symmetric satellite waves in $k$ grow in time and modulate the envelope of the large amplitude wave. The growth rate scales as $\sqrt{n_T^o}$. The publication of Kruer, Dawson, and Sudan presents numerical solutions of the dispersion relation Eq.(6.C.13) for parameters motivated by the Wharton *et al.* experiment.

Exercise: Why is the sideband instability equivalent to unstable amplitude modulation? Motivated by the structure of Eq.(6.C.16) consider a traveling wave with growing noise superposed on the wave:

$$\phi(x,t) = \phi_0\left[e^{i(k_0 x - \omega_0 t)} + \int dk A(k)e^{\gamma(k)t}e^{i(k_0+\delta k)x - (\omega+\delta\omega)t}\right]$$

$$= \phi_0 e^{i(k_0 x - \omega_0 t)}\left[1 + \int dk A(k)e^{\gamma(k)t}e^{i(\delta k x - \delta\omega t)}\right] = \phi_0 e^{i(k_0 x - \omega_0 t)}\left[1 + \int dk A(k)e^{\gamma(k)t}e^{i\delta k(x - V_g^o t)}\right]$$

$$= \phi_0 e^{i(k_0 x - \omega_0 t)} a(x,t)$$

## 6.D Example of the saturation of the two-stream instability due to trapping

A two-stream instability for the simple case of two streams of the same charge and mass is a strong instability with growth rate comparable to the plasma frequency. The wave and center-of-mass frames coincide. For a one-dimensional periodic simulation model with a system length such that only the fundamental is unstable, we expect that the single unstable mode will grow to a large amplitude such that the wave can trap the beams: $|e\phi_0| \sim \frac{1}{2} m v_0^2$. This estimate is borne out in direct kinetic simulations that illustrate the growth and saturation of the linearly unstable fundamental. The second harmonic is linearly stable but is excited nonlinearly. The evolution of the phase space in a particle simulation of the relativistic two-stream instability is illustrated in Fig. 6.D.1 Berk and Roberts[28] employed a simplified "water-bag" model of the two-stream instability, followed the motion of phase-space boundaries, and observed that a large-scale nonlinear wave evolves accompanying the condensation of holes in phase space.

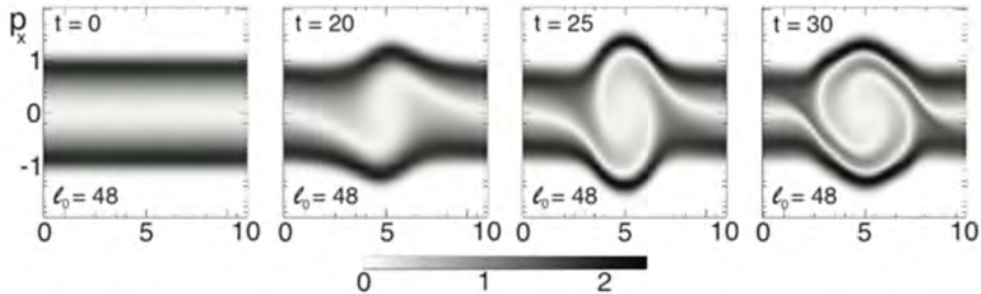

Fig. 6.D.1 The evolution of the relativistic two-stream instability in the frame of the unstable wave is shown from a simulation with OSHUN produced by Michail Tzoufras, UCLA Particle-in-Cell (PIC) and Kinetic Simulation Software Center (PICKSC), https://picksc.idre.ucla.edu

### 6.E Quasi-linear theory of wave-particle interaction

Consider the evolution of the bump-on-tail instability (see Sec. 2.I.e and Fig. 2.I.1). There is a range of unstable wave numbers corresponding to an interval of phase velocities with resonant velocities falling on the velocity distribution function where the slope is positive, i.e.,

$$\left|\frac{\omega_p}{k^2}\Delta k\right| = \left|\Delta\left(\frac{\omega_p}{k}\right)\right| = |\Delta u| \quad \rightarrow \quad \left|\frac{\Delta k}{k}\right| \sim \left|\frac{\Delta u}{u}\right| \tag{6.E.1}$$

Concentrate on waves that grow to appreciable amplitude: waves with finite growth rates $\gamma_k > 0$. As a matter of convention, consider only positive frequencies for positive or negative wavenumbers so that the sign of the wavenumber determines the sign of the phase velocity $V_k$. The electric potential can be represented by

$$\phi(x,t) = \sum_{j=1}^{N} \phi_j \cos\left(k_j x - \omega_{k,j} t + \alpha_j\right) \tag{6.E.2}$$

---

[28] H.L. Berk, K.V. Roberts. Methods Comput. Phys., 9, 88 ( 1970).

What is the evolution of the particle velocity in the presence of a spectrum of waves with initial condition $\{x_0, v_0\}$? To answer this question we compute $\langle [\Delta v(x_0, v_0; t)]^2 \rangle$ at fixed $x_0$. Particles interact weakly with each wave component, but none can trap because of the competition with the other waves. We will show that a random walk in velocity space occurs with attributes:

$$\left\langle \Delta v \right\rangle_{x_0} = 0 \qquad \left\langle (\Delta v)^2 \right\rangle_{x_0} (v_0; t) \propto t \tag{6.E.3}$$

<u>Definition</u>: The linear growth in time of the variance in the velocity perturbation defines this as a "diffusion" process.

[*Editor's note: Allan Kaufman published two fundamental papers on quasilinear diffusion in plasmas in 1972.*[29]]

### 6.E.a Diffusion equations, e.g., Fick's law

The mathematical characterization of a diffusion process as defined in Eq.(6.E.3) shares many of the same results as random walk processes associated with Brownian motion. Brownian motion in a colloidal suspension with diffusion into a less dense region has the property that the diffusive flux density of particles is proportional to the gradient in the density.

<u>Theorem</u>: (Fick's law of diffusion) $\vec{\Gamma} = -D_x \nabla n$ and the conservation of particle number is embodied in the continuity equation

$$\frac{\partial}{\partial t} n(\mathbf{x}; t) = -\nabla \cdot \vec{\Gamma} = \nabla \cdot \left( D_x \nabla n \right) \to D_x \nabla^2 n \tag{6.E.4}$$

if $D_x$ is a constant, and $D_x$ is given by $D_x = \lim_{\Delta t \to \infty} \langle \frac{(\Delta x)^2}{2\Delta t} \rangle$ where $\langle (\Delta x)^2 \rangle$ is a function of $\Delta t$.

<u>Definition</u>: The velocity diffusion coefficient can be defined analogously

$$D_v = \lim_{\Delta t \to \infty} \langle \frac{(\Delta v)^2}{2\Delta t} \rangle \tag{6.E.5}$$

This is the diffusivity in velocity space.

---

[29] A. N. Kaufman, J. Plas. Phys. **8**, 1 (1972); A.N. Kaufman, Physics of Fluids **15**, 1063 (1972).

Theorem: (Velocity-space diffusion equation) The diffusive flux density in velocity space is

$$\langle \vec{\Gamma}(v) \rangle_x = -D_v \frac{\partial \langle f(v) \rangle_x}{\partial v} \tag{6.E.6}$$

and the velocity-space continuity equation yields the kinetic diffusion equation:

$$\frac{\partial \langle f \rangle}{\partial t} = -\frac{\partial \langle \Gamma_v \rangle}{\partial v} = \frac{\partial}{\partial v}\left( D_v \frac{\partial \langle f \rangle}{\partial v} \right) \tag{6.E.7}$$

The velocity space diffusion in Eq.(6.E.7) will deform a bump-on-tail velocity distribution function asymptotically to a uniform plateau over the interval in velocity space that is resonant with the unstable phase velocities.

The wave-particle interaction with a spectrum of waves, absent particle trapping, has the property that $\langle \Delta v \rangle_{x_0} \cong 0$ but only to lowest order in $\phi$ and not to second order. The phase average with respect to $x_0$ of $\Delta v^2$ is finite and determines the velocity diffusion. Let us calculate the diffusion coefficient from consideration of $\Delta v$:

$$\Delta v(x_0, v_0, t) = \frac{e}{m}\int_0^t dt' \sum_k k\phi_k \sin\left[k(x_0 + v_0 t') - \omega_k t' + \alpha_k\right]$$

$$= \frac{e}{m}\int_0^t dt' \sum_k k\phi_k \sin\left[(kv_0 - \omega_k)t' + kx_0 + \alpha_k\right] \tag{6.E.8}$$

$$= \frac{e}{m}\sum_k k\phi_k \frac{\cos\left[(kv_0 - \omega_k)t + kx_0 + \alpha_k\right] - \cos\left[kx_0 + \alpha_k\right]}{-(kx_0 - \omega_k)}$$

From the definition of the diffusion coefficient in Eq.(6.E.5) and using the identity $1 - \cos\theta = 2\sin^2 \theta/2$, then

$$\langle \Delta v^2 \rangle_{x_0} = \left(\frac{e}{m}\right)^2 \sum_k \frac{k^2\phi_k^2\left(\frac{1}{2} + \frac{1}{2} - \cos((kv_0 - \omega_k)t)\right)}{(kv_0 - \omega_k)^2}$$

$$= \left(\frac{e}{m}\right)^2 \sum_k \frac{k^2\phi_k^2 2\sin^2\frac{1}{2}((kv_0 - \omega_k)t)}{(kv_0 - \omega_k)^2} \tag{6.E.9}$$

Theorem: The velocity diffusion coefficient is then

$$D(v_0) = \lim_{t\to\infty} \frac{\langle \Delta v^2 \rangle_{x_0}}{2t} = \frac{1}{2t}\left(\frac{e}{m}\right)^2 \sum_k \frac{k^2\phi_k^2 2\sin^2\frac{1}{2}((kv_0 - \omega_k)t)}{(kv_0 - \omega_k)^2} \tag{6.E.10}$$

The evaluation of the limit with respect to large $t$ in Eq.(6.E.10) is clarified in the next several equations. The numerator of the diffusion coefficient of $\langle \Delta v^2 \rangle$ grows at $t^2$ for small argument of the sin function, i.e., for resonant particles $\omega_k - kv_0 \approx 0$, and otherwise is oscillatory.

We next pass to the limit of a continuum of modes. In a box with length $L$ and with periodic boundary conditions: $k = \frac{2\pi}{L}n, n = \pm 1, \pm 2, \ldots$ In an infinite plasma $L \to \infty$ and $n$ goes to a continuum. Then

$$\sum_k \equiv \sum_{n=-\infty}^{\infty} \to \int_{-\infty}^{\infty} dn = \frac{L}{2\pi}\int_{-\infty}^{\infty} dk = L\int_{-\infty}^{\infty} \frac{dk}{2\pi} \tag{6.E.11}$$

The average energy density of the electric field becomes

$$\left\langle \text{Field Energy Density} \right\rangle = \left\langle \frac{E^2}{8\pi} \right\rangle = \frac{1}{8\pi}\sum_k k^2\phi_k^2 \left\langle \cos^2\theta_k \right\rangle$$

$$= \frac{1}{8\pi}\sum_k \frac{k^2\phi_k^2}{2} \to \frac{L}{16\pi}\int_{-\infty}^{\infty} \frac{dk}{2\pi} k^2\phi_k^2 \tag{6.E.12}$$

<u>Definition</u>: $E(k) \equiv \frac{1}{2}Lk^2\phi_k^2$, the energy density per unit $k$ interval, i.e., the spectral density; and $\left\langle \frac{E^2}{8\pi} \right\rangle = \frac{1}{8\pi}\int_{-\infty}^{\infty} \frac{dk}{2\pi}E(k)$.

Hence,

$$D(v) = \lim_{t\to\infty} \frac{1}{t}\left(\frac{e}{m}\right)^2 \int_{-\infty}^{\infty} \frac{dk}{2\pi} \frac{E(k)2\sin^2\frac{1}{2}\left((kv-\omega_k)t\right)}{(kv-\omega_k)^2} \tag{6.E.13}$$

We define $\beta \equiv \frac{1}{2}(kv - \omega_k)$ and use the relation $\int_0^{\infty} dx \frac{\sin^2 x}{x^2} = \pi$ to demonstrate that

$$\lim_{t\to\infty} \frac{\sin^2\beta t}{\beta^2 t} = \pi\delta(\beta) \tag{6.E.14}$$

The diffusion coefficient becomes

$$D(v) = \left(\frac{e}{m}\right)^2 \int_{-\infty}^{\infty} \frac{dk}{2\pi} E(k)\pi\delta(kv-\omega_k) \tag{6.E.15}$$

The quasilinear diffusion coefficient in (6.E15) describes resonant diffusion in velocity space driven by a spectrum of waves and is used in the velocity-space diffusion equation (6.E.7). The result is valid to lowest order in the perturbed electric potential and only for weak growth and damping rates.

### 6.E.b Irreversibility and the H theorem

We introduce the concept of the entropy associated with the velocity distribution as a function of time $S(t)$:

Definition: Entropy (Boltzmann) $S(t) \equiv -\int dv \langle f \rangle (v;t) ln \langle f \rangle (v;t)$ and (Gibbs) $S(t) \equiv -\int dv \int dx \langle f \rangle (x,v;t) ln \langle f \rangle (x,v;t)$.

Irreversibility is a macroscopic property and requires coarse-grained averaging. We associate irreversibility with $dS/dt > 0$.

Theorem: (H Theorem) The time derivative of the coarse-grained entropy is

$$\frac{d}{dt}S = -\int dv \left[ \frac{\partial \langle f \rangle}{\partial t} \ln\langle f \rangle + \frac{\langle f \rangle}{\langle f \rangle} \frac{\partial \langle f \rangle}{\partial t} \right] = -\int dv \left[ \frac{\partial \langle f \rangle}{\partial t} \ln\langle f \rangle - \frac{\partial}{\partial v} \Gamma \right]$$

$$= -\int dv \left[ \left( -\frac{\partial}{\partial v} \Gamma \right) \ln\langle f \rangle \right] = -\int dv \left[ \frac{1}{\langle f \rangle} \frac{\partial \langle f \rangle}{\partial v} \Gamma \right] = \int dv \left[ \frac{D(v)}{\langle f \rangle} \left( \frac{\partial \langle f \rangle}{\partial v} \right)^2 \right] > 0$$

(6.E.16)

using Eqs.(6.E.6) and (6.E.7), and integrating by parts. That the right side of Eq.(6.E.16) is positive is clear given that $D>0$ from Eqs.(6.E.13-15), $\langle f \rangle > 0$, and $(\partial \langle f \rangle / \partial v)^2 > 0$.

The entropy $S$ is always increasing, but is bounded from above using energy conservation. Hence, $S$ has an asymptotic steady state at its maximum value. In consequence, the integrand on the right side of Eq.(6.E.16) must vanish everywhere in velocity space asymptotically in time. This implies that the velocity derivative $\partial \langle f \rangle / \partial v$ vanishes asymptotically over the range of resonant velocities, i.e., the velocity distribution flattens. With the use of the relations $|d(kv - \omega_k)|\delta(kv - \omega_k) = |dk|\delta\left(k - \frac{\omega_k}{v}\right)$ (or generally $\delta(u(x)) = \delta(x - x_r)\left|\frac{du}{dx}\right|^{-1}$ where $x_r$ is a root of $u$) and $\frac{d(kv-\omega_k)}{dk} = v - \frac{d\omega_k}{dk}$, an alternative form for the velocity diffusion coefficient Eq.(6.E.15) is

$$D(v) = \left(\frac{e}{m}\right)^2 \int_{-\infty}^{\infty} \frac{dk}{2\pi} E(k) \pi \frac{\delta(k - \frac{\omega_k}{v})}{\left|v - \frac{d\omega}{dk}\right|} = \frac{1}{2}\left(\frac{e}{m}\right)^2 \frac{E(\frac{\omega_k}{v})}{|v - V_g(k)|}$$

(6.E.17)

We note that the waves evolve as linear waves while the particles interact with the waves quadratically in the wave amplitudes. Thus, the spectral density E(k) grows or damps linearly.

Theorem: Recalling Eq.(2.I.7) the linear evolution of the spectral density is

$$\frac{\partial E(k;t)}{\partial t} = 2\gamma(k;t)E(k;t) \text{ where } \gamma(k;t) = -\frac{\varepsilon''}{\frac{\partial \varepsilon'}{\partial \omega}} = \frac{e^2}{mk} \frac{4\pi}{\frac{\partial \varepsilon'}{\partial \omega}} \int_{-\infty}^{\infty} dv \frac{\partial \langle f \rangle}{\partial v} \pi \delta(\omega_k - kv)$$

(6.E.18)

and $<f>$ evolves according to Eq(6.E.7) in the presence of the quasilinear diffusion from the wave-particle interactions.

### 6.E.c Validity of the quasi-linear treatment and conservation of energy and momentum

The conditions for the validity of quasilinear theory are that the waves are linear, i.e., small amplitude with no trapping of particles and no mode coupling. Energy and momentum conservation tells us nothing new but affords a check that things have been done right. The total energy density is a sum of the particle kinetic energy and the field energy densities.

<u>Theorem</u>: The total energy density is given by

$$U = \int dv \frac{1}{2}mv^2 \langle f \rangle (v;t) + \frac{1}{8\pi} \int \frac{dk}{2\pi} E(k) \omega_k \frac{\partial \epsilon'}{\partial \omega}\Big|_{\omega_k} \qquad (6.E.19)$$

The second term on the right side of Eq.(6.E.19) contains both the electric field energy density (if $\omega_k \frac{\partial \epsilon'}{\partial \omega}\big|_{\omega_k} = 1$) and the mechanical or sloshing energy of the particles in the wave (if $\omega_k \frac{\partial \epsilon'}{\partial \omega}\big|_{\omega_k} > 1$).

<u>Theorem</u>: Energy conservation is demonstrated by calculating the time derivative of Eq.(6.E.19) and using expressions derived in Sec. 6.E.b:

$$\frac{d}{dt}U = \int dv \frac{1}{2}mv^2 \frac{\partial}{\partial v}\left(D \frac{\partial \langle f \rangle}{\partial v}\right) + \frac{1}{8\pi}\int \frac{dk}{2\pi} 2\gamma E(k) \omega_k \frac{\partial \epsilon'}{\partial \omega}\Big|_{\omega_k}$$

$$= -\int dv mv \left(\frac{e^2}{m^2}\int \frac{dk}{2\pi} E(k) \pi \delta(\omega_k - kv) \frac{\partial \langle f \rangle}{\partial v}\right) \qquad (6.E.20)$$

$$+ \int \frac{dk}{2\pi} \frac{e^2}{mk} \frac{1}{\frac{\partial \epsilon'}{\partial \omega}} \int_{-\infty}^{\infty} dv \frac{\partial \langle f \rangle}{\partial v} \pi \delta(\omega_k - kv) E(k) \omega_k \frac{\partial \epsilon'}{\partial \omega}\Big|_{\omega_k} = 0$$

<u>Theorem</u>: The total momentum density is given by the sum of the particle and wave momentum densities

$$P = \int dv mv \langle f \rangle (v;t) + \frac{1}{8\pi}\int \frac{dk}{2\pi} E(k) k \frac{\partial \epsilon'}{\partial \omega}\Big|_{\omega_k} \qquad (6.E.21)$$

<u>Theorem</u>: Momentum conservation is demonstrated by calculating the time derivative of Eq.(6.E.21) and using expressions derived in Sec. 6.E.b:

$$\frac{d}{dt}P = \int dv\, mv\, \frac{\partial}{\partial v}\left(D\frac{\partial \langle f\rangle}{\partial v}\right) + \frac{1}{8\pi}\int \frac{dk}{2\pi} 2\gamma E(k) k \frac{\partial \varepsilon'}{\partial \omega}\bigg|_{\omega_k}$$

$$= -\int dv\, m\left(\frac{e^2}{m^2}\int \frac{dk}{2\pi} E(k)\pi\delta(\omega_k - kv)\frac{\partial \langle f\rangle}{\partial v}\right) \tag{6.E.22}$$

$$+ \int \frac{dk}{2\pi}\frac{e^2}{mk}\frac{1}{\frac{\partial \varepsilon'}{\partial \omega}}\int_{-\infty}^{\infty} dv\, \frac{\partial \langle f\rangle}{\partial v}\pi\delta(\omega_k - kv)E(k)k\frac{\partial \varepsilon'}{\partial \omega}\bigg|_{\omega_k} = 0$$

We return to a discussion of the validity of the quasilinear diffusion equations. We consider three issues:
  (a) How large should Δt be in the definition of the diffusion coefficient Eq.(6.E.5)?
  (b) How small must E be so that Δv can be calculate just to O($\phi$)?
  (c) How small must $\gamma$Δt be in order that the growth or damping of waves during Δt is negligible so that the diffusion coefficient can be calculated to lowest order?

(a) Δt is long on a microscopic time scale of the wave-particle resonant interaction, but short on the macroscopic time scale over which the spectral density changes by a finite amount. The limit taken in Eq.(6.E.5) must converge for $\gamma$Δt<<1. Examine the integral in the expression for D in Eq.(6.E.13). Assume that $E \sim E(k_0) exp - \frac{(k-k_0)^2}{2\delta k^2}$ and expand $\beta \equiv (\omega_k - kv) \approx (\omega_{k_0} - k_0 v) + \left(\frac{d\omega}{dk} - v\right)(k-k_0) \approx \left(\frac{d\omega}{dk} - v\right)(k-k_0) = \delta\beta$, then

$$D(v) \propto \lim_{\Delta t \to \infty}\left(\frac{e}{m}\right)^2 \int_{-\infty}^{\infty}\frac{dk}{2\pi} E(k) \frac{\sin^2 \beta \Delta t}{\beta^2 \Delta t} \approx \left(\frac{e}{m}\right)^2 \int_{-\infty}^{\infty}\frac{dk}{2\pi} E(k)\pi\delta(\beta) \tag{6.E.23}$$

where $\Delta t >> 1/|\delta\beta| = 1/\left(\delta k \left|\frac{d\omega}{dk} - v\right|\right)$ to obtain the result in (6.E.23), which can be restated as the time Δt must be long compared to the time required by the particles to experience phase decorrelation $\delta k \delta x > 1$. The particles have to see many waves. If the particles are phase-correlated to one wave, they can get trapped.

(b) The spectral density E should be so small that there is no trapping over the time step Δt. There will be no trapping if the trapping time $\tau_T$ is much longer than the correlation time of the particle with a single wave:

$$\tau_T \equiv \left(k\sqrt{\frac{e\phi}{m}}\right)^{-1} >> \frac{\delta x}{v} = \frac{1}{k\delta v} \Rightarrow \sqrt{\frac{m}{e\phi}} >> \frac{1}{\delta v} \tag{6.E.24}$$

In the example of the bump-on-tail instability, the condition (6.E.24) implies that the electric potential must be weak and the bump relatively broad in order to avoid particle trapping.

(c) There needs to be weak growth or damping in the time over which the wave-particle diffusion coefficient converges, i.e., the growth time is long compared to $\Delta t$ which is long compared to the diffusion and correlation time.

$$\frac{1}{\gamma} \gg \Delta t \gg \frac{\delta x}{v} \tag{6.E.25}$$

The validity conditions for quasilinear diffusion can be summarized as follows:

$$\begin{aligned} |e\phi| &\ll mv^2 & \text{small-amplitude waves} \\ \gamma &\ll v\delta k = k\delta v & \text{weak growth or damping} \\ \frac{1}{k}\sqrt{\frac{m}{e\phi}} &\gg \frac{1}{v\delta k} = \frac{1}{k\delta v} & \text{no trapping} \end{aligned} \tag{6.E.26}$$

In Eq.(6.E.26), we have assumed that Landau growth or damping is weak, $\frac{\partial \omega_k}{\partial k} \ll v$, and from consideration of the resonance conditions $\frac{|\delta k|}{k} = \frac{|\delta v|}{v}$.

Example: The two-stream instability is not amenable to a quasilinear diffusion treatment because $\delta v = 0$, and we cannot satisfy the validity conditions.

Example: Bump-on-tail instability
Consider the descriptions in Secs. 2.I.e and 6.B.d  The validity conditions for the application of quasilinear diffusion theory are as follows.
(1) Weak growth – Define the interval of positive slope $g'(v) > 0$ to be $\delta v$. Then the condition for weak growth is

$$\frac{\gamma}{\omega_k} \approx \frac{\omega_p^2}{k^2}\frac{\pi g'(v)}{2} \rightarrow \frac{\gamma}{\omega_p} \approx O(1)\frac{\Delta g}{g_0}\frac{v}{\delta v}\frac{v}{v_{th}} \ll \frac{k\delta v}{\omega_p} \Rightarrow \left(\frac{\delta v}{v}\right)^2 \frac{v_{th}}{v} \gg \frac{\Delta g}{g_0} \tag{6.E.27}$$

(2) No trapping – The trapping time must be long compared to the correlation time:

$$\frac{\omega_T}{\omega_p} \sim \frac{k\sqrt{\frac{e\phi}{m}}}{\omega_p} \sim \frac{O(1)\gamma}{\omega_p} \ll \frac{k\delta v}{\omega_p} \Rightarrow \left(\frac{\delta v}{v}\right)^2 \frac{v_{th}}{v} \gg \frac{\Delta g}{g_0} \tag{6.E.28}$$

where we have used as an estimate of the wave amplitude at saturation $\omega_T \sim O(1)\gamma$ based on the considerations in Sec. 6.B.d.  Thus, the conditions in Eqs.(6.E.27) and (6.E.28) are identical.
(3) Linear waves – That the wave amplitude is small $e\phi \ll mv^2$ is easily satisfied:

$$e\phi/m \ll v^2 \Rightarrow \frac{v_T^2}{v^2} \approx \frac{\omega_T^2}{\omega_p^2} = \frac{O(1)\gamma^2}{\omega_p^2} \ll 1 \tag{6.E.29}$$

Example: For the bump-on-tail instability and representative plasma conditions, one might find very approximately
$\Delta g / g_0 \sim 10^{-4}, \delta v / v \sim 10^{-1}, v / v_{th} \sim 10, \gamma / \omega_p \sim 10^{-2}, \omega_T / \omega_p \sim 10^{-1.2}$

Exercises: Check the validity of quasilinear theory for the (i) weak-cold-beam instability and (ii) the ion acoustic instability.

One can make some general comments and estimates regarding the growth and saturation of instabilities based on quasilinear theory. The electric potential grows at first exponentially at a growth rate that begins at the linear value and then evolves in time until it saturates. The instantaneous growth rate $\gamma(t)$ remains very close to the linear growth rate until that last e-folding or two before saturation:

$$\phi(t^{sat}) = \phi_0 \exp\left[\int_0^{t^{sat}} dt' \gamma(t')\right] \Rightarrow \frac{\phi^{sat}}{\phi_0} = \exp\left[\int_0^{t^{sat}} dt' \gamma(t')\right]$$

$$\Rightarrow \ln\frac{\phi^{sat}}{\phi_0} = \int_0^{t^{sat}} dt' \gamma(t') \Rightarrow t^{sat} \sim \frac{1}{\gamma_L}\frac{1}{2}\ln\frac{k^2|\phi^{sat}|^2}{k^2|\phi_0|^2} \sim \frac{1}{2}\frac{1}{\gamma_L}\ln\frac{\text{turbulent wave energy}}{\text{thermal wave energy}} \quad (6.E.30)$$

The ratio of the turbulent wave energy to the thermal wave energy is typically very large. An estimate for the thermal wave energy can be obtained by assuming that the thermal wave energy is set by discreteness effects and that the ratio of kinetic energy to field energy is determined by plasma parameter, i.e., the ratio of the electric field energy to kinetic energy is $\sim nT/\Lambda$ where $\Lambda \sim n\lambda_D^3$. Based on observational experience (experiments, simulations, and analytical theories) we can estimate the turbulent wave energy at saturation as $\sim nT/\mu$, where $\mu \gg 1$. Hence,

$$t^{sat} \sim \frac{1}{2\gamma_L}\left(\ln\Lambda - \ln\mu\right) \tag{6.E.31}$$

In typical plasma conditions we find that $10 < \ln\Lambda < 20$ and $5 < \ln\mu < 8$, so that $t^{sat} \sim (2 \text{ to } 15)/2\gamma_L$.

Quasilinear theory has been extended in many directions. Here are a few examples of extensions:

1. Three dimensions instead of one

2. The inclusion of a finite applied magnetic field

3. Inclusion of magnetic as well as electric field perturbations (an electromagnetic theory)

4. Inhomogeneous medium

5. Formal derivation of quasi-linear diffusion theory

*6.E.d Formal derivation of quasi-linear theory using canonical variables and the Liouville equation*

Here we present a formal derivation of quasi-linear theory using canonical variables that satisfy Hamilton's equations. The value of this formulation is that the resulting equations capture the physics in a more efficient and elegant manner that retains mathematical rigor. In this formulation for a three-dimensional Coulomb plasma with no applied magnetic field, the phase-space variables become $\{x, y, z, \dot{x}, \dot{y}, \dot{z}\} \rightarrow \{q_1, q_2, q_3, p_1, p_2, p_3\}$ With a finite $\mathbf{B}_0$ one might employ canonical momentum $\mathbf{p} = m\mathbf{v} + \frac{e}{c}\mathbf{A}$. However, we will not use the canonical momentum and instead express the independent phase-space variables in terms of invariants of the motion as much as possible.

A constant of the motion $p$ is determined by the absence of dependence of the Hamiltonian on the variable $q$ conjugate to $p$, i.e.,

$$H_0(q,p) = H_0(\cancel{q},p) \qquad \dot{p} = -\frac{\partial H_0}{\partial q} = 0$$

For example, in a uniform magnetic field in the z direction, with no zero order electric fields, $v_\perp^2, v_\parallel$ are invariants as well as the guiding-center positions X and Y. In this situation, the conjugate variables to $v_\perp^2$ and $v_\parallel$ are the gyrophase $\phi$ and the position z, respectively. The guiding-center positions X and Y are conjugates to one another. If $\{p_1, p_2, p_3\}$ are three invariants of the zero order motion corresponding to $\{v_\parallel, v_\perp^2, X\}$ then we can construct the zero-order phase-space distribution function from $f_0(v_\parallel, v_\perp^2, X)$. Because Y is also a constant of the zero-order motion, then $\frac{\partial H_0}{\partial p_3} = 0$ and $H_0(\mathbf{p}) \rightarrow H_0(p_1, p_2)$.

Quite generally, the phase-space distribution function will be a function of all of the canonical variables and time:

$$f(p_1, p_2, ..., q_1, q_2, ...; t) \quad \text{and} \quad \frac{\partial f}{\partial t} + \frac{\partial}{\partial p_1}(\dot{p}_1 f) + ... + \frac{\partial}{\partial q_1}(\dot{q}_1 f) + ... = 0 \qquad (6.E.32)$$

in the collisionless limit. A formal average of $f$ over $q_j$ can be performed by integrating the kinetic equation in (6.E.32).

<u>Theorem</u>: Using the property of the perfect differential and appropriate boundary conditions (either zero at $\pm\infty$ or periodic on 0 to $2\pi$), $\int dq_j \frac{\partial}{\partial q_j}(\dot{q}_j f) = (\dot{q}_j f)|_{-\infty,0}^{+\infty,2\pi} = 0$, the kinetic equation for $\langle f \rangle(\mathbf{p}; t)$ becomes

$$\frac{\partial \langle f \rangle(\mathbf{p};t)}{\partial t} = -\frac{\partial}{\partial \mathbf{p}}(\dot{\mathbf{p}}\langle f \rangle) \qquad \dot{\mathbf{p}} = -\frac{\partial H}{\partial \mathbf{q}} = -\frac{\partial \delta H}{\partial \mathbf{q}} \quad H = H_0(\mathbf{p}) + \delta H(\mathbf{p}, \mathbf{q}; t) \qquad (6.E.33)$$

where we have used $\langle \dot{\mathbf{p}}\langle f \rangle\rangle = \langle \dot{\mathbf{p}}\rangle\langle f \rangle = 0$. Thus,

$$f(\mathbf{q},\mathbf{p};t) = \langle f \rangle_\mathbf{q}(\mathbf{p};t) + \delta f(\mathbf{q},\mathbf{p};t) \quad \text{and} \quad \langle \dot{\mathbf{p}} f \rangle = \langle \dot{\mathbf{p}} \delta f \rangle \qquad (6.E.34)$$

We use the Liouville equation form of the Vlasov equation (6.E.32) to obtain

$$0 = \frac{df(\mathbf{p},\mathbf{q};t)}{dt} = \frac{d\langle f \rangle}{dt} + \frac{d\delta f}{dt} \to \frac{d\delta f}{dt} = -\left[\frac{\partial \langle f \rangle}{\partial t} + \frac{\partial \langle f \rangle}{\partial \mathbf{p}} \cdot \dot{\mathbf{p}}\right] \quad (6.\text{E}.35)$$

Given the Hamiltonian H=$p^2$/2m+$e\delta\phi$(**x**,t) with $e\delta\phi$ identified as a first-order perturbing field, the first term on the right side of the last expression in Eq.(6.E.35) is second order and the second term is first order. Then to first order in the perturbed electric potential

$$\frac{d\delta f}{dt} = -\frac{\partial \langle f \rangle}{\partial \mathbf{p}} \cdot \dot{\mathbf{p}} \quad (6.\text{E}.36)$$

We next integrate Eq.(6.E.36) from $t-\Delta t$ to $t$ along the particle phase-space characteristics to first order using

$$\left[\frac{\partial}{\partial t} + \dot{\mathbf{q}} \cdot \frac{\partial}{\partial \mathbf{q}} + \dot{\mathbf{p}} \cdot \frac{\partial}{\partial \mathbf{p}}\right]\delta f \approx \left[\frac{\partial}{\partial t} + \dot{\mathbf{q}} \cdot \frac{\partial}{\partial \mathbf{q}}\right]\delta f \quad (6.\text{E}.37)$$

because $\dot{\mathbf{p}}$ is first order and $\dot{\mathbf{p}} \cdot \frac{\partial}{\partial \mathbf{p}}\delta f$ is second order. Hence,

$$\int_{t-\Delta t}^{t} dt' \frac{d\delta f(\mathbf{q},\mathbf{p};t)}{dt} = \delta f(\mathbf{q}^t,\mathbf{p}^t;t) - \delta f(\mathbf{q}^{t-\Delta t},\mathbf{p}^{t-\Delta t};t-\Delta t) \approx \delta f(\mathbf{q}^t,\mathbf{p}^t;t) - \delta f(\mathbf{q}^{t-\Delta t},\mathbf{p}^t;t-\Delta t)$$

$$\Rightarrow \delta f(\mathbf{q}^t,\mathbf{p}^t;t) = \delta f(\mathbf{q}^{t-\Delta t},\mathbf{p}^t;t-\Delta t) - \int_{t-\Delta t}^{t} dt' \dot{\mathbf{p}} \cdot \frac{\partial \langle f \rangle}{\partial \mathbf{p}} \quad (6.\text{E}.38)$$

We average the Liouville equation, use Eqs.(6.E.35), use Eq.(6.E.38) to evaluate $\delta f(t)$, and subtract off the first order terms to identify second-order terms:

$$\frac{\partial \langle f \rangle}{\partial t} + \frac{\partial}{\partial \mathbf{p}} \cdot \langle \dot{\mathbf{p}}\delta f \rangle = 0 \to \frac{\partial \langle f \rangle}{\partial t} = -\frac{\partial}{\partial \mathbf{p}} \cdot \langle \dot{\mathbf{p}}(t)\delta f(t) \rangle =$$

$$-\frac{\partial}{\partial \mathbf{p}} \cdot \{\langle \dot{\mathbf{p}}(t)\delta f(t-\Delta t) \rangle\} + \frac{\partial}{\partial \mathbf{p}} \cdot \left\{\int_{t-\Delta t}^{t} dt' \langle \dot{\mathbf{p}}(t)\dot{\mathbf{p}}(t') \rangle \cdot \frac{\partial \langle f \rangle}{\partial \mathbf{p}}(t')\right\} \quad (6.\text{E}.39)$$

It can be shown that the first term on the right side of Eq.(6.E.39) falls off with $\Delta t$ and is negligible compared to the second term for $\Delta t \gg \delta x/v$ (the correlation time).

<u>Theorem</u>: Eq.(6.E.39) leads to a quasilinear diffusion equation for the slowly varying average distribution function where we have defined $t - \tau = t'$:

$$\frac{\partial \langle f \rangle}{\partial t} = \frac{\partial}{\partial \mathbf{p}} \cdot \left\{\int_{0}^{\Delta t} d\tau \langle \dot{\mathbf{p}}(t)\dot{\mathbf{p}}(t-\tau) \rangle \cdot \frac{\partial \langle f \rangle}{\partial \mathbf{p}}(t-\tau)\right\} \quad (6.\text{E}.40)$$

<u>Definition</u>: The quasilinear diffusion tensor is

$$\mathbf{D}(\mathbf{p},t) \equiv \int_{0}^{\Delta t} d\tau \langle \dot{\mathbf{p}}(t)\dot{\mathbf{p}}(t-\tau) \rangle \quad (6.\text{E}.41)$$

The integrand on the right side of Eq.(6.E.41) is the two-time momentum correlation function which is largest for small $\tau$ and falls off for large $\tau$ compared to the correlation time δx/v. If we expand

$$\frac{\partial \langle f \rangle}{\partial \mathbf{p}}(t-\tau) = \frac{\partial \langle f \rangle}{\partial \mathbf{p}}(t) - \tau \frac{\partial}{\partial \mathbf{p}} \frac{\partial \langle f \rangle}{\partial t}(t) \approx \frac{\partial \langle f \rangle}{\partial \mathbf{p}}(t) + O(\tau^2)$$

and just keep the lowest order first term for use in Eq.(6.E.40), then Eq.(6.E.41) becomes

$$\frac{\partial \langle f \rangle}{\partial t} = \frac{\partial}{\partial \mathbf{p}} \cdot \left\{ \mathbf{D}(\mathbf{p};t) \cdot \frac{\partial \langle f \rangle (\mathbf{p};t)}{\partial \mathbf{p}} \right\} \tag{6.E.42}$$

### 6.E.e Diffusion tensor

We have derived the diffusion tensor in Eq.(6.E.41). For a uniform magnetic field parallel to the z axis, the generalized momentum vector in canonical variables is $\mathbf{p} = (m v_z, \frac{1}{2} \frac{m v_\perp^2}{\Omega}, \frac{eB_0}{c} Y)$. One of the components of the diffusion tensor is then

$$D_\perp(\mathbf{p},t) \equiv \int_0^{\Delta t} d\tau \langle \dot{p}_\perp(t) \dot{p}_\perp(t-\tau) \rangle \tag{6.E.43}$$

Theorem: The standard form for a diffusion coefficient is

$$D_\perp(\mathbf{p},t) \equiv \lim_{\Delta t \to \infty} \frac{\langle (\Delta p_\perp)^2 \rangle}{2\Delta t} = \lim_{\Delta t \to \infty} \frac{\left\langle \left[ \int_0^{\Delta t} dt \dot{p}_\perp(t) \right]^2 \right\rangle}{2\Delta t} = \lim_{\Delta t \to \infty} \frac{\int_0^{\Delta t} dt \int_0^{\Delta t} dt' \langle \dot{p}_\perp(t) \dot{p}_\perp(t') \rangle}{2\Delta t} \tag{6.E.44}$$

Only near t=t' is the integrand in the double integral finite on the right side. Hence,

$$D_\perp(\mathbf{p},t) \approx \frac{\int_0^{\Delta t} dt \int_{-\infty}^{\infty} d\tau \langle \dot{p}_\perp(t) \dot{p}_\perp(t-\tau) \rangle}{2\Delta t} \approx \frac{2\Delta t \int_0^{\infty} d\tau \langle \dot{p}_\perp(t) \dot{p}_\perp(t-\tau) \rangle}{2\Delta t} = \int_0^{\infty} d\tau \langle \dot{p}_\perp(t) \dot{p}_\perp(t-\tau) \rangle$$

(6.E.45)

as long at Δt is large compared to the correlation time. We note that the integrand is an even function of $\tau$ which is justified by microscopic reversibility arguments.

Example: For a Coulomb model,

$$H = \frac{p_z^2}{2m} + p_\perp \Omega + e \delta \phi(\mathbf{r},t) \qquad \rho(\mathbf{x}|\mathbf{r}) \equiv e \delta(\mathbf{x}-\mathbf{r})$$

$$\delta H(p,q;t) = \int d^3 \mathbf{x} \, \delta \phi(\mathbf{x},t) \rho(\mathbf{x}|\mathbf{r})$$

$$\mathbf{r}(\mathbf{p},\mathbf{q}) \quad \mathbf{q} = (z, \phi, \frac{eB_0}{c} Y) \quad \mathbf{R} = (x,y,z) \tag{6.E.46}$$

With a continuum of waves present

$$\delta\phi(\mathbf{x},t) = \int \frac{d^3\mathbf{k}}{(2\pi)^3} \delta\phi(\mathbf{k},t) e^{i\mathbf{k}\cdot\mathbf{x}} \quad \delta H = \int \frac{d^3\mathbf{k}}{(2\pi)^3} \delta\phi(\mathbf{k},t) \int d^3\mathbf{x} e^{i\mathbf{k}\cdot\mathbf{x}} \rho(\mathbf{x}|\mathbf{p},\mathbf{q})$$

$$\int d^3\mathbf{x} e^{i\mathbf{k}\cdot\mathbf{x}} \rho(\mathbf{x}|\mathbf{p},\mathbf{q}) = \rho(\mathbf{k}|\mathbf{p},\mathbf{q}) = e \int d^3\mathbf{x} e^{i\mathbf{k}\cdot\mathbf{x}} \delta(\mathbf{x}-\mathbf{r}) = e\, e^{i\mathbf{k}\cdot\mathbf{r}(\mathbf{p},\mathbf{q})}$$

$$= e \exp\left[ik_z z + ik_x\left(X + r_\perp \cos\theta\right) + ik_y\left(Y + r_\perp \sin\theta\right)\right]$$

$$= e\, e^{i\mathbf{k}\cdot\mathbf{R}} e^{ik_\perp r_\perp \cos(\theta-\alpha)} = e\, e^{i\mathbf{k}\cdot\mathbf{R}} \sum_{\ell=-\infty}^{\ell=\infty} i^\ell J_\ell(k_\perp r_\perp) e^{i\ell(\theta-\alpha)}$$

(6.E.47)

Hence,

$$\delta H = e \int \frac{d^3\mathbf{k}}{(2\pi)^3} \delta\phi(\mathbf{k},t) e^{i\mathbf{k}\cdot\mathbf{R}} \sum_{\ell=-\infty}^{\ell=\infty} i^\ell J_\ell(k_\perp r_\perp) e^{i\ell(\theta-\alpha)} \quad \text{and}$$

$$\dot{p}_\perp = -e \int \frac{d^3\mathbf{k}}{(2\pi)^3} \delta\phi(\mathbf{k},t) e^{i\mathbf{k}\cdot\mathbf{R}} \sum_{\ell=-\infty}^{\ell=\infty} i^{\ell+1} \ell J_\ell(k_\perp r_\perp) e^{i\ell(\theta-\alpha)}$$

(6.E.48)

using $\dot{p}_\perp = -\partial \delta H / \partial q_\perp = -\partial \delta H / \partial \theta$.

<u>Theorem</u>: The integrand in the right side of Eq.(6.E.45) becomes

$$\langle \dot{p}_\perp(t) \dot{p}_\perp(t-\tau) \rangle = < e^2 \int \frac{d^3\mathbf{k}}{(2\pi)^3} \int \frac{d^3\mathbf{k}'}{(2\pi)^3} \delta\phi(\mathbf{k},t) \delta\phi^*(\mathbf{k},t-\tau) e^{i\mathbf{k}\cdot\mathbf{R}^t} e^{i\mathbf{k}'\cdot\mathbf{R}^{t-\tau}}$$

$$\times \sum_{\ell=-\infty}^{\ell=\infty} \sum_{\ell'=-\infty}^{\ell'=\infty} \ell\ell'(-i)^{\ell'} i^\ell J_\ell(k_\perp r_\perp^t) J_{\ell'}(k_\perp' r_\perp^{t-\tau}) e^{i\ell(\theta^t-\alpha)} e^{-i\ell'(\theta^{t-\tau}-\alpha')} >$$

(6.E.49)

We use the zero-order orbit for $z^{t-\tau} = z^t - v_z \tau$ and average over phases:

$$\lim_{L\to\infty} \frac{1}{L} \int_{-L/2}^{L/2} dz \exp\left[i(k_z - k'_z)z\right] = \frac{2\pi}{L} \delta(k_z - k'_z) \quad \text{and similarly in } x, y, \text{ and/or } \theta \text{ to obtain}$$

the following:

$$\langle \dot{p}_\perp(t) \dot{p}_\perp(t-\tau) \rangle = e^2 \int_0^\infty \frac{d\tau}{Vol} \int \frac{d^3\mathbf{k}}{(2\pi)^3} \delta\phi(\mathbf{k},t) \delta\phi^*(\mathbf{k},t-\tau) e^{i\mathbf{k}\cdot\mathbf{R}^t} e^{i\mathbf{k}'\cdot\mathbf{R}^{t-\tau}} \sum_{\ell=-\infty}^{\ell=\infty} \ell^2 J_\ell(k_\perp r_\perp)^2 < e^{i\mathbf{k}\cdot\Delta\mathbf{R}(\tau)+i\ell\Delta\phi(\tau)} >$$

(6.E.50)

We limit the ensemble of waves to only normal modes:

$$\delta\phi(\mathbf{k},t) = \delta\tilde{\phi}(\mathbf{k},t) e^{-i\omega_k t} \quad \delta\phi\delta\phi^* = \delta\tilde{\phi}(\mathbf{k},t) \delta\tilde{\phi}^*(\mathbf{k},t-\tau) e^{-i\omega_k \tau} \quad (6.E.51)$$

and ignore the slow temporal variation of the mode amplitudes $\delta\tilde{\phi}(\mathbf{k},t)$. Use of (6.E.50), (6.E.51), (6.E.45) and the definition of the spectral density

$$S_\phi(\mathbf{k},t) \equiv \lim_{Vol \to \infty} \frac{|\delta\tilde{\phi}(\mathbf{k},t)|^2}{Vol} \tag{6.E.52}$$

yields the expression for the diffusion coefficient

$$\begin{aligned}D_\perp &\approx e^2 \int \frac{d^3\mathbf{k}}{(2\pi)^3} S_\phi(\mathbf{k},t) \sum_\ell \ell^2 J_\ell^2 \int_0^\infty d\tau e^{-i\omega_k \tau + ik_z v_z \tau + i\ell\Omega\tau} \\ &= e^2 \int \frac{d^3\mathbf{k}}{(2\pi)^3} S_\phi(\mathbf{k},t) \sum_\ell \ell^2 J_\ell^2 \left[ \pi\delta(k_z v_z + \ell\Omega - \omega_k) + \frac{P}{i(\omega_k - k_z v_z - \ell\Omega)} \right] \end{aligned} \tag{6.E.53}$$

<u>Exercise</u>: Show that the imaginary part of the diffusion coefficient in Eq.(6.E.53) vanishes using the convention $\omega_{-\mathbf{k}} = -\omega_\mathbf{k}$ and summing over **k** in the integral.

<u>Exercise</u>: i. Show that the other important terms in the diffusion tensor satisfy

$$\begin{bmatrix} D_{zz} & D_{z\perp} \\ D_{\perp z} & D_{\perp\perp} \end{bmatrix} \propto \begin{bmatrix} k_z^2 & k_z \ell \\ k_z \ell & \ell^2 \end{bmatrix}$$

ii. Derive the diffusion tensor for electromagnetic waves, $\delta\phi \to \delta A$.

### 6.E.f Self-consistent quasi-linear diffusion equation and energy conservation

Here quasi-linear theory is further elaborated. In terms of momentum variables $(p_z, p_\perp)$ the diffusion tensor is

$$\vec{\mathbf{D}} = e^2 \int \frac{d^3\mathbf{k}}{(2\pi)^3} \sum_\alpha S_\phi^\alpha(\mathbf{k},t) \sum_\ell J_\ell^2(k_\perp r_\perp) \pi\delta(k_z v_z + \ell\Omega - \omega_k^\alpha) \mathbf{k}\mathbf{k}, \quad \mathbf{k} = k_z \hat{\mathbf{e}}_1 + \ell\hat{\mathbf{e}}_2 \tag{6.E.54}$$

where $\alpha$ indicates the branches of the dispersion relation, e.g., for electrostatic waves propagating parallel to the magnetic field this corresponds to electron plasma waves and ion acoustic waves. The diffusion equation for the slowly varying distribution function for a specific branch is then

$$\begin{aligned}\frac{\partial \langle f \rangle}{\partial t}(p_z, p_\perp; t) &= e^2 \int \frac{d^3 k}{(2\pi)^3} \sum_\alpha S_\phi^\alpha(\mathbf{k},t) \sum_\ell \left( k_z \frac{\partial}{\partial p_z} + \ell \frac{\partial}{\partial p_\perp} \right) J_\ell^2(k_\perp r_\perp) \\ &\quad \times \pi\delta(k_z v_z + \ell\Omega - \omega_k) \left( k_z \frac{\partial}{\partial p_z} + \ell \frac{\partial}{\partial p_\perp} \right) \langle f \rangle (p_z, p_\perp; t) \end{aligned} \tag{6.E.55}$$

<u>Definition</u>: We can define an invariant $I$ for a wave with characteristics $(k, \ell, \omega_k,)$ as follows:

$$I = \frac{\left( p_z - m\frac{\omega}{k_z} \right)^2}{2m} + \Omega p_\perp \tag{6.E.56a}$$

with momentum-space gradient

$$\nabla I = \hat{\mathbf{e}}_1\left(v_z - \frac{\omega}{k}\right) + \hat{\mathbf{e}}_2\Omega \tag{6.E.56b}$$

From Eqs.(6.54) and (6.E.56b) one concludes the following inner product vanishes:

$$\mathbf{k}\cdot\nabla I = k_z\left(v_z - \frac{\omega}{k_z}\right) + \ell\Omega = 0$$

as a consequence of the resonance condition. Hence, $\mathbf{k}$ and $\nabla I$ are perpendicular: $\nabla I$ is directed across the invariant contours, and $\mathbf{k}$ is in the direction of the invariant curve. $\mathbf{k}\cdot\nabla f$ should become uniform because particles diffuse along the invariant curves until $\nabla_\mathbf{k}\cdot\langle f\rangle = 0$.

At this point we return to consideration of the development of Eqs.(6.E.50-6.E.53). In this derivation there was an assumed limitation on the rapidity of the temporal variation of slowly varying coefficients: $\tilde{\phi}(\mathbf{k},t)\tilde{\phi}^*(\mathbf{k},t-\tau) \approx \tilde{\phi}(\mathbf{k},t)\tilde{\phi}^*(\mathbf{k},t)$. Inclusion of the next term in a Taylor series looks like

$\tilde{\phi}^*(\mathbf{k},t-\tau) \approx \tilde{\phi}^*(\mathbf{k},t) - \tau\frac{\partial}{\partial t}\tilde{\phi}^*(\mathbf{k},t) + ...$, where $\tau$ is the order of the correlation time. The linear correction term in the Taylor series is clearly small if $\gamma \ll 1/\tau_{corr}$, i.e., if the correlation time is assumed short compared to the linear growth time. This does *not* imply that $\tilde{\phi}(\mathbf{k},t)\tilde{\phi}^*(\mathbf{k},t-\tau) \approx \tilde{\phi}(\mathbf{k},t)\tilde{\phi}^*(\mathbf{k},t)$ is time independent: slow variations are explicitly included, and the growth or damping rate of $S_\phi^\alpha$ is $2\gamma$ to good approximation. Hence, the quasilinear diffusion equation includes slow temporal variations in both $\langle f\rangle = f_0 + f_2$ and $\mathbf{D}$:

$$\frac{\partial\langle f\rangle}{\partial t} = \frac{\partial}{\partial \mathbf{p}}\cdot\left(\mathbf{D}\cdot\frac{\partial\langle f\rangle}{\partial \mathbf{p}}\right) \tag{6.E.57}$$

The temporal variation in $\langle f\rangle$ is a second-order resonant perturbation. We can construct a closed set of equations based on the diffusion equation and energy conservation as follows. The partial time derivative of $\langle f\rangle$ is proportional to $S_\phi$, and the spectral density evolves to lowest order according to "linear theory" determining the growth or damping rate $\gamma$ but using the total perturbed distribution function that evolves in time.

<u>Theorem</u>: Self-consistent quasilinear theory

$$\begin{aligned}\frac{\partial\langle f\rangle}{\partial t} &= \frac{\partial}{\partial \mathbf{p}}\cdot\left(\mathbf{D}\cdot\frac{\partial}{\partial \mathbf{p}}\langle f\rangle\right) \quad \mathbf{D} = ...S_\phi(\mathbf{k},t) \\ \frac{\partial S_\phi(\mathbf{k},t)}{\partial t} &= 2\gamma(\mathbf{k},t)S_\phi(\mathbf{k},t) \quad \gamma(\mathbf{k},t) = ...\langle f\rangle\end{aligned} \tag{6.E.58}$$

Theorem: Energy conservation

$$\frac{d}{dt}\int d^3\mathbf{p}\int d^3\mathbf{q}\, H_0(\mathbf{p})\langle f\rangle(\mathbf{p},t) + \frac{d}{dt}(\text{field energy}) = 0$$

$$\Rightarrow \sum_s \int dp_z \int dp_\perp \int dz \int d\phi \int dX\, H_0(p_z,p_\perp)\frac{\partial}{\partial \mathbf{p}}\cdot\left(\mathbf{D}\cdot\frac{\partial}{\partial \mathbf{p}}\langle f\rangle\right) \quad (6.E.59)$$

$$+ Vol\cdot \int d^3\mathbf{k}\,\omega\frac{\partial \varepsilon'}{\partial \omega}\frac{k^2 S_\phi(\mathbf{k},t)}{8\pi}2\gamma = 0$$

We integrate the first term in Eq.(6.E.59) and use Eqs.(6.E.54) and (6.E.55) to obtain

$$-\sum_s m^3\int d^3\mathbf{v}\, e^2 \int \frac{d^3\mathbf{k}}{(2\pi)^3}S_\phi(\mathbf{k},t)\sum_\ell J_\ell^2(k_\perp r_\perp)\pi\delta(k_z v_z + \ell\Omega - \omega_k)\mathbf{K}\mathbf{K}\cdot\frac{\partial}{\partial \mathbf{p}}\langle f\rangle$$

$$+ \int \frac{d^3\mathbf{k}}{(2\pi)^3}\omega\left.\frac{\partial \varepsilon'}{\partial \omega}\right|_{\omega_k}\frac{k^2 S_\phi(\mathbf{k},t)}{8\pi}2\gamma = 0 \quad (6.E.60)$$

where we have divided out a common volume factor. We note that $m^3\langle f(\mathbf{p})\rangle \to \langle f(\mathbf{p})\rangle = n_0\langle g\rangle$ and $\int d\mathbf{v}^3\, g(\mathbf{v}) = 1$. Equation (6.E.60) can then be used to solve for $\gamma$ inside the integral over $\mathbf{k}$ space (and the spectral density $S_\phi$ divides out):

$$\gamma(\mathbf{k}) = \frac{1}{\left.\frac{\partial \varepsilon'}{\partial \omega}\right|_{\omega_k}}\sum_s \frac{4\pi e_s^2 n_{0s}}{m_s k^2}\int d^3\mathbf{v}\sum_\ell J_\ell^2(k_\perp r_\perp)\pi\delta(k_z v_z + \ell\Omega - \omega_k)m_s\mathbf{K}\cdot\frac{\partial}{\partial \mathbf{p}}\langle g\rangle$$

$$= \frac{1}{\left.\frac{\partial \varepsilon'}{\partial \omega}\right|_{\omega_k}}\sum_s \frac{\omega_s^2}{k^2}\int d^3\mathbf{v}\sum_\ell J_\ell^2(k_\perp r_\perp)\pi\delta(k_z v_z + \ell\Omega - \omega_k)\left(k_z\frac{\partial}{\partial v_z} + \frac{\ell\Omega}{v_\perp}\frac{\partial}{\partial v_\perp}\right)\langle g\rangle(v_z,v_\perp)$$

(6.E.61)

In Eqs.(6.E.60) and (6.E.61), $\mathbf{K} \equiv k_z\hat{\mathbf{e}}_1 + \ell\hat{\mathbf{e}}_2$. Now we can refer to the relation Eq.(2.I.7)

$$\gamma(\mathbf{k}) = -\frac{\varepsilon''}{\left.\frac{\partial \varepsilon'(\omega,\mathbf{k})}{\partial \omega}\right|_{\omega_k}}$$

to obtain an explicit formula for $\varepsilon''$, use the Kramers-Kronig relation to determine $\varepsilon'$, and can compare to the results derived in Sec. 2.I.e . Equations (6.E.58-6.E.61) provide a formalism based on quasi-linear theory for evolving both the distribution function and the amplitudes of the waves self-consistently.

**6.F Particle subject to gyroresonance -- instabilities and diffusion, invariants of the motion, loss-cone effects**

In the presence of an applied, static, uniform magnetic field, the wave-particle resonance is generalized from Landau resonance to the cyclotron resonance that appears mathematically in the denominator of the susceptibility in Eq.(4.F.12). The physics of the cyclotron resonance is that the Doppler-shifted frequency of the wave matches a cyclotron harmonic so that a cyclotron harmonic component of the wave can steadily accelerate or decelerate the particle: in the particle frame $\omega' = \omega - k_z v_z = \ell\Omega$. It is important to take into account the physical effects of a finite spatial variation of the wave field in the plane of the gyro-motion. Consider for example the special case of $k_\perp r \sim \pi/2$ where $r = v_\perp/\Omega$. In Fig. 6.F.1 is diagrammed the cyclotron orbit of a charged particle in its zero-order drift frame and the phase of the electric field it sees. The resonance condition is such that the particle's perpendicular velocity vector is co-aligned with the linearly polarized, perturbed electric field $\delta E_y$ at both 0 and 2, and so can continue to be accelerated, while the particle's velocity vector is perpendicular to $\delta\mathbf{E}$ at points 1 and 3.

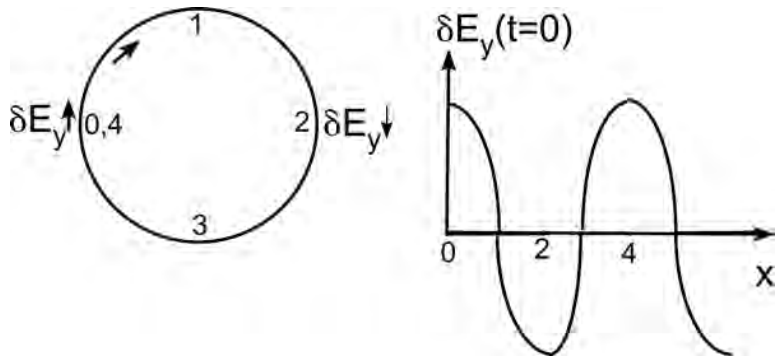

Fig. 6.F.1 A particle with $\omega' = \omega - k_z v_z = 2\Omega$ and with $k_\perp r = \pi/2$ will rotate around the gyro-radius from 0 to 2 in one-half a cyclotron period and see an electric field $\delta E_y$ whose phase has advanced in time by $2\pi$ and in $x$ by $\pi$.

We will describe the gyro-resonant wave-particle interaction in terms of increments to the particle energy, momentum, and angular momentum. We use canonical variables. In a quantum mechanical picture when a photon interacts with a particle, the change in energy satisfies $\Delta H = \hbar\omega$; the change in momentum satisfies $\Delta m v_z = \bar{h} k$; and the change in angular momentum satisfies $\Delta p_\perp = \ell\hbar$. We calculate the analogous classical changes in energy and in linear and angular momenta in the following analysis.

Theorem: For a gyrating particle,

$$H = \frac{1}{2}mv_z^2 + \frac{1}{2}mv_\perp^2 = \frac{1}{2}mv_z^2 + p_\perp\Omega, \quad p_\perp = \frac{\frac{1}{2}mv_\perp^2}{\Omega} \tag{6.F.1}$$

$$\Delta H = mv_z \Delta v_z + \Omega \Delta p_\perp$$

Definition: Canonical variables and guiding-center coordinates (see Sec. 6.E.d)

$$\{x,y,z,p_x,p_y,p_z\} \qquad p_x = mv_x + \frac{e}{c}A_x^0, \text{ etc.}$$

$$\Rightarrow \left\{z, p_z = mv_z, p_\perp = \frac{\tfrac{1}{2}mv_\perp^2}{\Omega}, \phi, X, aY\right\}$$

(6.F.2)

Theorem: We describe the particle gyration in terms of canonical variables:

$$x = X + r_\perp \cos\phi$$
$$y = Y + r_\perp \sin\phi$$
$$\dot{x} = -\Omega r_\perp \sin\phi \qquad (6.F.3)$$
$$\dot{y} = \Omega r_\perp \cos\phi$$
$$p_\perp \equiv \frac{\tfrac{1}{2}mv_\perp^2}{\Omega} \qquad r_\perp \equiv \frac{v_\perp}{\Omega}$$

In Eq.(6.F.3) $X$ and $Y$ are guiding center positions with the magnetic field oriented in the $z$ direction. $(p_\perp, \phi)$ is a canonically conjugate pair. $(X, aY)$ is also a canonically conjugate pair if $a = eB/c$, which choice is consistent with the conditions on the canonical variables: $\{p_i, q_j\} = \delta_{ij}$ and $\{q_i, q_j\} = 0$ where {f,g} is the Poisson bracket.

We next extend the equations in Eq.(6.F.1-6.F.3) to describe gyro-resonance in the presence of a longitudinal wave. The gyro-resonance condition is

$$\omega - k_z v_z = \ell\Omega \qquad (6.F.4)$$

where $\ell$ is an integer. $\ell = 0$ corresponds to Landau resonance analyzed earlier. Gyro-resonance can occur for $|\ell| > 0$ for $k_\perp \neq 0$. From a quantum physics point of view, when a particle interacts with a photon, the conservation of energy, momentum, and angular momentum leads to the following relations, and implies an invariant $I$:

$$\Delta H = \hbar\omega \qquad \Delta p_z = \hbar k_z \qquad \Delta p_\perp = \ell\hbar$$
$$\Delta(k_z H - \omega p_z) = \hbar\omega k_z - \hbar k_z \omega = 0 \quad \Rightarrow \quad I \equiv k_z H - \omega p_z = \text{invariant}$$

(6.F.5)

We now construct a quantity $I_w$ related to the particle kinetic energy in the frame of a wave:

$$V_z \equiv \frac{\omega}{k_z} \qquad I_w = \frac{1}{2}m(v_z - V_z)^2 + \frac{1}{2}mv_\perp^2 - \frac{1}{2}mV_z^2 \qquad (6.F.6)$$

However, $I_w$ needs to include the potential energy due to the wave to be useful in constructing the classical particle Hamiltonian and deducing an invariant of the motion. The classical Hamiltonian including a longitudinal wave is

$$H = \frac{p_z^2}{2m} + p_\perp \Omega + e\phi_0 \sin[k_z z + k_\perp x - \omega t] = \frac{p_z^2}{2m} + p_\perp \Omega + e\phi_0 \sin\left[k_z z + k_\perp(X + r_\perp \cos\phi) - \omega t\right]$$

(6.F.7)

Theorem: Using Hamilton's equations and $H$ in Eq.(6.F.7)

$$\dot{H} = \frac{\partial}{\partial t} H = -\omega e \phi_0 \cos\left[k_z z + k_\perp \left(X + r_\perp \cos\phi\right) - \omega t\right]$$

$$\dot{p}_z = -\frac{\partial}{\partial z} H = -k_z e \phi_0 \cos\left[k_z z + k_\perp \left(X + r_\perp \cos\phi\right) - \omega t\right]$$

(6.F.8)

It follows from Eq. (6.F.8) that there is an invariant $I$ that is analogous to that in Eq.(6.F.5):

$$I = H - \frac{\omega}{k_z} p_z, \quad \dot{I} = \dot{H} - \frac{\omega}{k_z} \dot{p}_z = 0 \tag{6.F.9}$$

In the wave frame $z' = z - \frac{\omega}{k_z} t$, there is no explicit time dependence in $H^{(W)}$, i.e., $\omega = 0$; and $H^{(W)}$ is then a constant independent of time.

Particles follow trajectories on families of curves in the phase space of $(v_z, v_\perp)$ determined by $I$=constant. The curves are half ellipses ($v_\perp > 0$) displaced in $v_z$ by $V_z$ from the origin, for infinitesimal wave amplitude. Finite wave amplitude distorts the invariant curves from exact ellipses. Particles with velocities

$$v_z^R = \frac{(\omega - \ell\Omega)}{k_z} = V_z - \frac{\ell\Omega}{k_z} \tag{6.F.10}$$

are resonant with the wave. Non-resonant particles have an adiabatic response to the wave. Resonant particles can have an irreversible response. A trapped particle will oscillate around the resonant velocity $v_z^R$ with characteristics that depend on the finite amplitude of the wave. Nearly resonant velocities $v_z$ whose magnitudes are less than $|v_z^R|$ will gain energy from the wave and accelerate, and nearly resonant velocities whose magnitudes are greater than $|v_z^R|$ will lose energy to the wave and decelerate. Depending on the sign of the slope of the distribution function, which determines if there are more or fewer particles with velocities greater or less than the resonant velocity to exchange energy with the wave, the resonant wave-particle interaction will lead to growth or damping of the wave similar to that discussed earlier in Secs. 2 and 3.

Theorem: i. If $V_z > 0$ and $v_z > 0$, then linear exponential growth $\gamma > 0$ is expected if $\partial f/\partial p_z > 0$; and linear exponential damping $\gamma < 0$ is expected if $\partial f/\partial p_z < 0$. ii. If $V_z > 0$ and $v_z < 0$, then linear exponential growth $\gamma > 0$ is expected if $\partial f/\partial p_z < 0$; and linear exponential damping $\gamma < 0$ is expected if $\partial f/\partial p_z > 0$.

Definition: In mirror machines there is a "loss cone" determined by the ratio of the axial magnetic field at its maximum to its value at its minimum (the mirror ratio), i.e., particles at the mid-plane with $|v_z/v_\perp| \geq \cotan(\theta_c)$ will not be confined by the magnetic mirror. The loss cone establishes velocity gradients of the distribution function, and the stability theorem then indicates that instabilities arise. The instabilities produce velocity-space diffusion into the loss cone that exacerbates particle loss out of the mirror, i.e., loss-cone instability.[30]

Exercise: Use conservation of energy and magnetic moment to derive the loss-cone condition on $|v_z/v_\perp|$ in a simple magnetic mirror configuration.

Consider a set of waves such that $V_z \equiv \omega(k_\perp, k_z)/k_z$ is the same for the range of $k_z$ and $k_\perp$ in the set. Consider particles with a broad distribution of $v_\perp$ values extending to $v_\perp = 0$ and a narrow band of resonant velocities in $v_z$ satisfying Eq.(6.F.10) with $\ell = -1$, i.e.,

$$\Delta v_z^R = \Omega \Delta\left(\frac{1}{k_z}\right), \quad v_z^R = V_z - \frac{\Omega}{k_z} \tag{6.F.11}$$

On any invariant curve there can be quasi-linear diffusion in the vicinity of the resonance (assuming the presence of a decorrelation mechanism) that manifests itself by a random walk. The random walks can occur throughout the band of resonant particles. The net result of the diffusion is uniformity of the velocity distribution in the resonant band. If there is a loss cone present, then the resonance band can cross the loss-cone boundary at $v_\perp = v_z^R/\cotan(\theta_c)$; and the diffusion can drive particles into the loss region. An increase in the amplitude of the turbulent waves will increase the diffusion rate into the loss region.

If we consider other cyclotron harmonics $\ell$, there are additional resonant bands for the same $V_z$; and there will be more resonant bands in phase space. Furthermore, if the wave spectrum contains waves with more than one value of $V_z$, the likelihood of diffusion increases. For every value of $V_z$ and infinitesimal wave amplitude, there is a new family of invariant curves that are displaced in $v_z$ by a unique value of $V_z$. A resonant particle can diffuse in its resonant band associated with one value of $V_z$, encounter an invariant curve, and then move along the invariant curve a short distance before encountering another unique resonant band in which it diffuses.

[*Editor's note: The preceding picture is significantly altered for a spectrum of finite wave amplitudes because the independence of the invariant curves for distinct values of the phase velocity $V_z$ becomes invalid. The solution of the invariant curves should come from Eqs.(6.F.7-6.F.9) summing over all waves with finite amplitude. However, the solution for "independent" invariant curves for each wave in a set of waves with distinct phase velocities and infinitesimal amplitudes should be valid and useful for understanding the dynamics in the near vicinity of where an invariant curve for a distinct phase velocity crosses a resonant velocity band for the same phase velocity.*]

---

[30] M. N. Rosenbluth and R. F. Post, Phys. Fluids **8**, 547 (1965).

*Resonant wave diffusion was an active area of research by Kaufman and his students and post-doctoral researchers in the next ten years after these lectures.[31]*]

With a wave spectrum of multiple distinct phase velocities there are bands of resonant velocity. The width of the resonant bands increases with increased wave amplitudes. The possibility of resonance overlap occurs. The consequence of resonance overlap is increased velocity-space transport. If the particle ends up farther from the origin in $(v_z, v_\perp)$, then heating has occurred (cooling if the particle ends up closer to the origin). For many waves, more of phase space becomes accessible to diffusion, and $f(\mathbf{v})$ tends to uniformity in the diffusive regions subject to the constraint that the overall energy is bounded. As uniformity of $f$ is approached, i.e., $f$ flattens, the growth or damping rate $\gamma$ tends to zero; and waves no longer exchange energy with the particles. The velocity distribution also tends to isotropy. The resulting modifications to $f$ render the system increasingly stable.

### 6.G Gardiner's theorem for stability of a Vlasov plasma – sources of instability and examples

In a plasma described by the Vlasov-Poisson system of equations it is possible to make a general statement on the stability of the plasma.

Theorem: (Gardiner's theorem) For the Vlasov-Poisson system of equations, if $f$ is isotropic and the derivative of $f$ with respect to energy $\partial f / \partial \varepsilon < 0$, then the Vlasov-Poisson system is linearly and nonlinearly stable. A spatially uniform, isotropic and monotonic velocity distribution function (no relative drifts, of course) is stable.

Sources that drive instability:
1. Nonuniformity – instability derives energy from redistributing plasma in space
2. Anisotropy – instability derives energy from redistributing plasma in velocity space, as in magnetic mirror devices
3. Non-monoticity – like bump on tail, energy becomes available as a plateau forms in velocity space
4. Relative drift – ordered drift motion can contribute kinetic energy to waves

A common feature in the preceding list is that instability is driven by some redistribution of the plasma in configuration or velocity space, which allows the conversion of kinetic energy in the plasma into wave energy while relaxing the source of free energy. We have already encountered a few examples of this in earlier lectures, e.g., Secs. 2.F and 6.B.

---

[31] G. R. Smith and A. N Kaufman, Phys. Rev. Lett. **34**, 1613 (1975); G. R. Smith and A. N. Kaufman, Phys. Fluids **21**, 2230 (1978); G. R. Smith and N. R. Pereira, Phys. Fluids **21**, 2253 (1978); G. R. Smith and B. I. Cohen, Phys. Fluids **26**, 238 (1983).

Example: The dispersion relation in the ion frame for a longitudinal ion-cyclotron wave is

$$\omega = \Omega_i \left[1 + \tau e^{-b_i} I_1(b_i)\right], \quad \tau = T_e/T_i, \quad b_i = \left(k_\perp r_\perp^i\right)^2 \tag{6.G.1}$$

The ions are assumed to be a Maxwellian, and the electron response is close to adiabatic in the low-frequency limit with $k\lambda_e \ll 1$. We further assume that the ion damping is weak, i.e., $\omega - \Omega_i \neq k_z v_z^i$ and $k_z \ll \frac{(\omega-\Omega_i)}{v_{th}^i} \sim O\left(\frac{\Omega_i}{v_{th}^i}\right) = \frac{1}{r_\perp^i} \sim k_\perp$. Thus, **k** is nearly perpendicular to **B**. The conditions for electron resonance are $\omega - \ell\Omega_e = k_z v_z^e$ which requires $\ell = 0$ and admits the possibility of a simple Landau resonance: $\omega/k_z = v_z^e \gg \Omega_i r_\perp^i \sim v_{th}^i$. It is difficult for the resonant electron velocity to fall in the heart of the electron velocity distribution. Does the electron Landau resonance contribute Landau damping or growth? Using the methods in Sec. 2 one obtains:

$$\gamma(k_z, k_\perp) = \frac{1}{\left.\frac{\partial \varepsilon'}{\partial \omega}\right|_{\omega_k}} \frac{\omega_e^2}{k^2} \int_{-\infty}^{\infty} dv_z \int_0^{\infty} 2\pi v_\perp dv_\perp J_0^2(k_\perp r_\perp) \pi \delta\left(v_z - \frac{\omega_k}{k_z}\right) \frac{\partial g(v_z, v_\perp)}{\partial v_z}$$

$$= \frac{1}{\left.\frac{\partial \varepsilon'}{\partial \omega}\right|_{\omega_k}} \frac{\omega_e^2}{k^2} \int_0^{\infty} 2\pi v_\perp dv_\perp J_0^2(k_\perp r_\perp) \pi \left.\frac{\partial g(v_z, v_\perp)}{\partial v_z}\right|_{v_z = \frac{\omega_k}{k_z}} \tag{6.G.2}$$

$$= (\omega - \Omega_i) \frac{k^2 \lambda_e^2 \omega_e^2}{k^2} \pi \left.\frac{\partial g(v_z, v_\perp)}{\partial v_z}\right|_{v_z = \frac{\omega_k}{k_z}} \cong \Omega_i v_{the}^2 \left.\frac{\partial g(v_z, v_\perp)}{\partial v_z}\right|_{v_z = \frac{\omega_k}{k_z}}$$

where $k_\perp r_\perp \ll 1$ and $J_0^2(k_\perp r_\perp) \approx 1$ because the electron Larmor radius is so small and we have assumed $g = g(v_z)g(v_\perp)$. If $\partial g(v_z)/\partial v_z < 0$ everywhere in $v_z$ there is electron damping. If there is an electron drift $v_{de}$ in the ion frame and the electron velocity distribution is symmetric about $v_{de}$ in the ion frame and peaked at $v_{de}$, then $\partial g(v_z)/\partial v_z > 0$ for $0 < \frac{\omega}{k_z} < v_{de}$; and the electrons are destabilizing. For net instability, the electron destabilizing effects must overcome ion damping. This is the Drummond-Rosenbluth instability,[32] with condition for instability: $v_{de} > 5\tau v_{thi}$, which is not difficult to achieve. For this instability one can readily construct a theory for quasi-linear diffusion following the procedure in Sec. 6.E, and one expects plateau formation in $g_e(v_z)$ driven by the ion cyclotron wave turbulence.

Exercise: Consider a wave with $\Omega_i, \gamma \ll \omega \ll \Omega_e$ driven unstable by an anisotropic ion velocity distribution. Treat the ions as unmagnetized and ignore electron Landau damping. Consider oblique propagation with $k_z \ll k_\perp$. Project the ion velocity distribution onto **k** to render the analysis one-dimensional. Instability results from the projected velocity distribution having an interval with positive slope. Qualitatively examine quasi-linear diffusion.

---

[32] W. E. Drummond, and M. N. Rosenbluth, Phys. Fluids **5**, 1507 (1962).

**6.H Nonlinear three-wave interactions**

Energy transfer between modes is dominated by resonant three-wave decay in a non-turbulent plasma, i.e., a plasma in which the energy in waves is still small compared to the particle kinetic energy. From a quantum mechanical point of view, three-wave decay can be described efficiently in terms of conservation laws:

$$\begin{aligned}\text{energy conservation:} \quad & \hbar\omega_1 = \hbar\omega_2 + \hbar\omega_3 \\ \text{momentum conservation:} \quad & \hbar k_1 = \hbar k_2 + \hbar k_3\end{aligned} \quad (6.\text{H}.1)$$

A few examples of three-wave decay processes are as follows. An electron plasma wave (Langmuir wave) can decay into another electron plasma wave and an ion acoustic wave. A transverse wave can decay into another transverse wave and either an electron plasma wave or an ion acoustic wave. In special circumstances a transverse wave can decay into two electron plasma waves. An Alfvén wave can decay into another Alfvén and an ion acoustic wave. There are many other examples that have been identified experimentally and studied theoretically; and a rich literature exists.

*6.H.a Resonance conditions derived from phase matching*

There is more than one approach to analyzing nonlinear three-wave interactions. There are direct methods using perturbation theory to solve the Maxwell equations and either the Vlasov equations (e.g., Davidson[33]) or fluid equations (e.g., Tsytovich[34], Drake et al.[35], Forslund et al.[36]) to model the plasma response. Some of these approaches involve a good deal of analytical brute force. There are more elegant theories based on a formal Hamiltonian theory for the plasma fluid response (e.g., Davidson) and very powerful Lagrangian theories for a fluid or Vlasov plasma response (e.g., Dewar, Dougherty, Suranlishvili, Galloway, and others).

[*Editor's note: In the more than forty years since these lectures the literature on three-wave interactions in plasmas has grown immensely. A literature survey is beyond the scope of these lecture notes.*]

To understand how quadratic nonlinearities in the plasma dynamical equations lead to three-wave coupling and to illustrate how three-wave resonance

---

[33] R. C. Davidson, *Methods in Nonlinear Plasma Theory* (Academic Press), 1972.
[34] V. N. Tsytovich, Sov. Phys. Uspekhi **9**, 805–836 (1967)
[35] JF Drake, PK Kaw, YC Lee, G. Schmidt, CS Liu, and MN Rosenbluth, Phys. Fluids **17**, 778 (1974).
[36] D. W. Forslund, J. M. Kindel, and E. L. Lindman, Phys. Fluids **18**, 1002 (1975).

conditions are derived from phase matching, consider the fluid continuity equation in one dimension:

$$\frac{\partial n}{\partial t} = -\frac{\partial}{\partial x}(nu) \tag{6.H.2}$$

where

$$n(x,t) = \sum_{j=1}^{3} |\tilde{n}_j|(t)\cos(k_j x - \omega_j t + \alpha_j(t)) = \text{Re}\sum_{j=1}^{3} \tilde{n}_j(t)\exp[i(k_j x - \omega_j t)] \text{ and } \tilde{n}_j = |\tilde{n}_j|\exp(i\alpha_j(t))$$

(6.H.3)

We assume that $u$ has the same form as $n$. We expand in the wave decomposition in Eq.(6.H.3) and look for the beating of distinct waves on the right side of Eq.(6.H.2) to couple to the remaining of the three waves on the left side:

$$e^{ik_2 x - i\omega_2 t}\left(-i\omega_2 + \frac{\partial}{\partial t}\right)\tilde{n}_2(t) = -\frac{\partial}{\partial x}\left(\tilde{n}_1 \tilde{u}_3 e^{ik_1 x - i\omega_1 t} e^{-ik_3 x + i\omega_3 t} + \tilde{n}_0 \tilde{u}_2 e^{ik_2 x - i\omega_2 t}\right) \tag{6.H.4}$$

We introduce a power-series expansion in the wave amplitudes for $n$ and $u$ to facilitate the bookkeeping in the coupling of terms:

$$n_j(x,t) = \sum_k \varepsilon^k n_{jk} = n_{j0} + \varepsilon n_{j1} + \varepsilon^2 n_{j2} + \ldots \tag{6.H.5}$$

There is an equilibrium density $n_0$ with no fast phase variation. The lowest-order wave terms are at first order in $\varepsilon$. Thus, there is a first-order (linear) contribution to $n_{21}$ from $n_0 u_{21}$ on the right side of Eq.(6.H.4) and a second-order (quadratic) contribution to $n_{22}$ from the beat of bilinear quantities $n_{11} u_{31}$. From the continuity equation we see that there is a linear relation for all three waves:

$$-i\omega_j n_j = -n_0 ik_j u_j \tag{6.H.6}$$

Using Eq.(6.H.6) in (6.H.4) and assuming perfect phase matching, we deduce that there is a second order contribution to $(\partial/\partial t)n_2$

$$\frac{\partial}{\partial t}n_{22} = -\frac{ik_2 \omega_3}{k_3 n_0} n_1 n_3^* \tag{6.H.7}$$

We note that the quadratic nonlinearity $q(\mathbf{E} + \mathbf{v}\times\mathbf{B}/c)\cdot\partial f/\partial \mathbf{v}$ in which the fields beat with the perturbed velocity distribution function $f$ has the same structure as the quadratic term in the continuity equation (and, for that matter, the quadratic terms in the fluid momentum balance equation). If the fast phase variation of the quadratic beat on the right sides of Eq.(6.H.4) does not match the fast phase variation on the left side, then the nonlinear interaction will be nonresonant. If we divide out the phase factor on the left side, the average of the right side over a few wavelengths in space or a few wave periods in time will phase mix to zero unless there is good phase matching, i.e.,

$$\omega_1 = \omega_2 + \omega_3 \qquad \mathbf{k}_1 = \mathbf{k}_2 + \mathbf{k}_3 \tag{6.H.8}$$

<u>Exercise</u>: Recover the three-wave phase-matching conditions Eq.(6.H.8) by considering Bragg scattering by a moving grating.

We return to the one-dimensional analysis leading to Eq.(6.H.7) and use Poisson's equation to relate the linear quantities, $ik_j E_{j1} = 4\pi q n_{j1}$ so that the following equation coupling electric field amplitudes results:

$$\frac{\partial}{\partial t} E_{22} = g_2 E_1 E_3^* \tag{6.H.9}$$

If we systematically reduce the fluid equations or the Vlasov equation in conjunction with the Poisson equation in the same manner as is done for obtaining linear dispersion relations but retaining all bilinear nonlinearities, one obtains

$$\varepsilon(\omega)E(\omega) = \left[\varepsilon(\omega_2) + (\omega - \omega_2)\frac{\partial \varepsilon}{\partial \omega}\bigg|_{\omega_2}\right]E(\omega_2) = \left[\varepsilon(\omega_2) + \frac{\partial \varepsilon}{\partial \omega}\bigg|_{\omega_2} i\frac{\partial}{\partial t}\right]E(\omega_2) = \text{nonlinear driving}$$

(6.H.10)

Given the inherent symmetries in the dynamical equations, the composite coupled-mode equations for three-wave interactions can be shown to take the form

$$\frac{\partial}{\partial t} E_1 = -\frac{g}{\frac{\partial \varepsilon}{\partial \omega}\big|_{\omega_1}} E_2 E_3 \quad \frac{\partial}{\partial t} E_2 = \frac{g}{\frac{\partial \varepsilon}{\partial \omega}\big|_{\omega_2}} E_1 E_3^* \quad \frac{\partial}{\partial t} E_3 = \frac{g}{\frac{\partial \varepsilon}{\partial \omega}\big|_{\omega_3}} E_1 E_2^* \tag{6.H.11}$$

where $g$ is real.

### 6.H.b Conservation laws – wave action

When photons scatter there is a quantum mechanical conservation law that the number of photons is conserved, although the outgoing photon may be shifted in frequency. The number of photons $N_j$ of a given frequency is given by the energy $W_j$ divided by $\hbar \omega_j$, $N_j = W_j / \hbar \omega_j$; and the action $J_j$ is defined by the product $J_j = \hbar N_j$ i.e., $\hbar$ is the quantum mechanical unit of action. The classical wave action has the same definition: $J_j = W_j / \omega_j$.

<u>Definition</u>: The classical wave energy, momentum, and wave action densities are

$$W_j = \frac{\omega_j \bar{\varepsilon}_j}{8\pi} E_j^2 \quad \mathbf{K}_j = \frac{\mathbf{k}_j \bar{\varepsilon}_j}{8\pi} E_j^2 \quad J_j = \frac{W_j}{\omega_j} \tag{6.H.12}$$

where $\bar{\varepsilon} \equiv \partial \varepsilon / \partial \omega$. We will omit the factor of $8\pi$ in the denominators in Eq.(6.H.12) in the following to reduce clutter in the equations.

<u>Theorem</u>: (Action Conservation) Using the definitions in Eq.(6.H.12) and the coupled-mode equations in Eq.(6.H.11) one can show

$$I_2 \equiv J_1 + J_2 = \text{const.} \quad I_3 \equiv J_1 + J_3 = \text{const.} \implies \dot{J}_1 = -\dot{J}_2 = -\dot{J}_3 \tag{6.H.13}$$

Theorem: (Energy and Momentum Conservation) Using Eqs.(6.H.12) and (6.H.13), and frequency and wavenumber matching Eq.(6.H.8)

$$\frac{d}{dt}(W_1 + W_2 + W_3) = \omega_1 \dot{J}_1 + \omega_2 \dot{J}_2 + \omega_3 \dot{J}_3 = (\omega_1 - \omega_2 - \omega_3)\dot{J}_1 = 0$$
$$\frac{d}{dt}(\mathbf{K}_1 + \mathbf{K}_2 + \mathbf{K}_3) = \mathbf{k}_1 \dot{J}_1 + \mathbf{k}_2 \dot{J}_2 + \mathbf{k}_3 \dot{J}_3 = (\mathbf{k}_1 - \mathbf{k}_2 - \mathbf{k}_3)\dot{J}_1 = 0$$

(6.H.14)

Definition: Introduce the slowly varying wave phase $\theta_j$ with definition $E_j = |E_j|e^{i\theta_j}$ and the interaction energy $H'$, and use the definitions in Eq.(6.H.12):

$$H' \equiv \frac{2g}{(\bar{\varepsilon}_1 \bar{\varepsilon}_2 \bar{\varepsilon}_3)^{1/2}} \sqrt{J_1 J_2 J_3} \sin(\theta_1 - \theta_2 - \theta_3)$$
$$= 2g|E_1 E_2 E_3|\sin(\theta_1 - \theta_2 - \theta_3) = 2g\operatorname{Im}(E_1 E_2^* E_3^*)$$

(6.H.15)

Theorem: Using Eqs.(6.H.11) and (6.H.15) the interaction energy is a constant in time

$$\dot{H}' = 2g\operatorname{Im}(\dot{E}_1 E_2^* E_3^* + E_1 \dot{E}_2^* E_3^* + E_1 E_2^* \dot{E}_3^*)$$
$$= 2g\operatorname{Im}\left(-\frac{g}{\bar{\varepsilon}_1}|E_2|^2|E_3|^2 + \frac{g}{\bar{\varepsilon}_2}|E_1|^2|E_3|^2 + \frac{g}{\bar{\varepsilon}_3}|E_1|^2|E_2|^2\right) \equiv 0$$

(6.H.16)

because Im(Real)≡0.

Definition: The total electric field is given as $E(x,t) = \sum_j |E_j|\cos(k_j x - \omega_j t + \theta_j(t))$

From this definition and Eq.(6.H.11) we calculate the following relations:

$$\dot{J}_1 = \bar{\varepsilon}_1 E_1^* \dot{E}_1 + c.c. = -gE_1^* E_2 E_3 + c.c. = -g|E_1 E_2 E_3|\left(e^{-i\theta_1 + i\theta_2 + i\theta_3} + e^{i\theta_1 - i\theta_2 - i\theta_3}\right)$$
$$= -g\sqrt{\frac{J_1 J_2 J_3}{\bar{\varepsilon}_1 \bar{\varepsilon}_2 \bar{\varepsilon}_3}} 2\cos(\theta_1 - \theta_2 - \theta_3) \equiv -\bar{g}\sqrt{J_1 J_2 J_3} \cos(\theta_1 - \theta_2 - \theta_3)$$

(6.H.17)

and we have introduced the definition of $\bar{g} = 2g/\sqrt{\bar{\varepsilon}_1 \bar{\varepsilon}_2 \bar{\varepsilon}_3}$. We note that the action $J_j$ is defined so that it always has the same sign as $\bar{\varepsilon}_j$, and the square roots in Eq.(6.H.17) can be considered to have strictly positive arguments. Given Eqs.(6.H.15) and (6.H.17) we can construct the square of the time derivative of $J_1$ and identify a pseudo-potential:

$$\tfrac{1}{2}\dot{J}_1^2 = \tfrac{1}{2}\bar{g}^2 |J_1 J_2 J_3|\left[1 - \frac{H'^2}{\bar{g}^2 |J_1 J_2 J_3|}\right] = \tfrac{1}{2}\bar{g}^2 |J_1 J_2 J_3| - \tfrac{1}{2}H'^2$$

(6.H.18)

Definition: Define a pseudo-potential $\Phi$,

$$\Phi(J_1) = -\tfrac{1}{2}\bar{g}^2 J_1 J_2 J_3 = -\tfrac{1}{2}\bar{g}^2 J_1 (I_2 - J_1)(I_3 - J_1)$$

(6.H.19)

Theorem: Given that $H'$ is a constant in time, Eq.(6.H.18) leads directly to an equation of motion for the pseudo-particle action $J_1$, subject to an action conservation law:

$$\tfrac{1}{2}\dot{J}_1^2 + \Phi(J_1) = -\tfrac{1}{2}H'^2 \tag{6.H.20}$$

Figure 6.H.1 illustrates Eq.(6.H.20). Initial conditions determine $I_2$, $I_3$ and $H'$. The solution of $J_1$ as a function of time can be expressed in terms of Jacobi elliptic functions. $J_1$ is a periodic function of time. The larger the value of $g$, the faster is the oscillation. The exchange of action among the three waves continues unabated in the absence of damping.

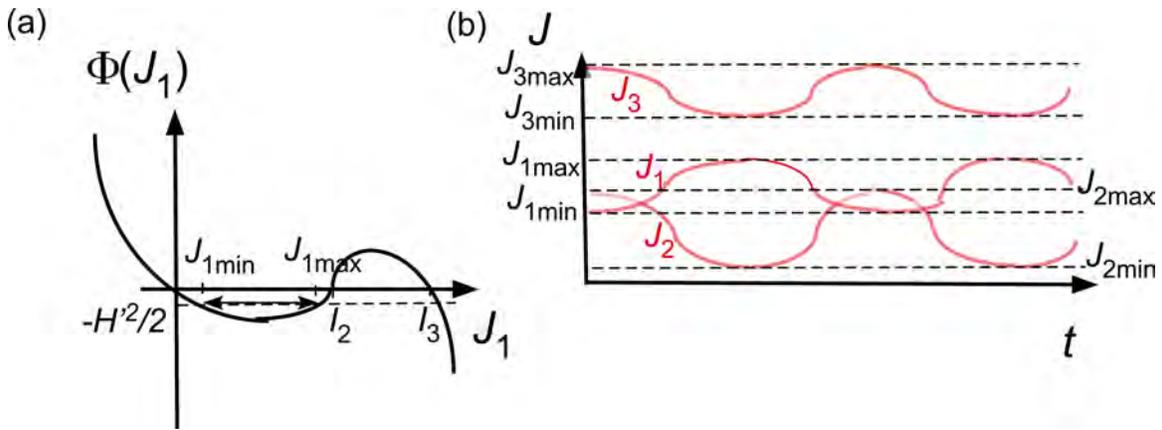

Fig. 6.H.1 Schematic of action conservation in three-wave interactions. (a) Pseudo potential $\Phi(J_1)$ vs. $J_1$. (b) Three-wave-interaction actions vs. time.

### *6.H.c Parametric and explosive instabilities*

In this section we discuss two special three-wave interactions cases of interest. In both examples the initial conditions are comprised of a finite-amplitude high-frequency wave and two lower-frequency waves with infinitesimal initial amplitudes.

Example: (Parametric Instability) Assume that wave #1 is a transverse wave with frequency very close to the plasma frequency, but somewhat larger. Wave #1 (the pump wave) can decay into a Langmuir wave (wave #2) and an ion-acoustic wave (wave #3): $\omega_1^t \approx \omega_{pe} + \omega_{ion\,acoustic}$ This is called parametric decay and provides a mechanism for anomalous absorption of the transverse wave in a plasma. Given the initial conditions that the decay product waves have infinitesimal initial amplitudes compared to the pump wave, we can use the coupled-mode equations Eq.(6.H.11) with $E_1$ approximately constant early in time to derive the exponential growth of the decay product waves:

$$\frac{\partial}{\partial t}E_2 = \frac{g}{\bar{\varepsilon}_2}E_1 E_3^* \quad \frac{\partial}{\partial t}E_3 = \frac{g}{\bar{\varepsilon}_3}E_1 E_2^* \tag{6.H.21}$$

If we assume exponentially growing solutions $E_{2,3} = E_{2,3}(0)e^{\gamma t}$ then Eq.(6.H.21) admits the solution for the growth rate of the parametric instability:

$$\gamma = \frac{g|E_1|}{\sqrt{\bar{\varepsilon}_2 \bar{\varepsilon}_3}} \tag{6.H.22}$$

Of course this derivation is only formal and schematic as we have not given explicit expressions for $g$ and $\bar{\varepsilon}_j \equiv \partial \varepsilon / \partial \omega |_{\omega_j}$

[*Editor's note: In a plasma with no applied magnetic field, three of the classic references on parametric instabilities are Drake, et al.,[37] Liu, et al.[38] and Kruer.[39] In the presence of an applied magnetic field, some representative references on parametric instability theory are by Porkolab,[40] Kaw,[41] Cohen,[42] and Stefan, et al.[43]*]

Example: (Explosive Instability) Here we consider the decay of wave #1 into waves #2 and #3 where wave #1 is a negative energy wave and all three waves have infinitesimal initial amplitude. In this case all three waves can grow in amplitude while the total energy remains infinitesimal. The frequency and wave number relations Eq.(6.H.8) and action conservation Eq.(6.H.13) are unchanged, but $\bar{\varepsilon}_3 < 0$ and $J_1 < 0$. Thus, as all three waves grow in amplitude, $\Delta J_1 = -|\Delta J_1| = -\Delta J_2 = -\Delta J_3$ and $\Delta W_1 + \Delta W_2 + \Delta W_3 = -|\Delta W_1| + \Delta W_2 + \Delta W_3 \approx 0$. We return to the relation for the pseudo-potential for the three-wave interaction in Eq.(6.H.19):

$$\Phi(J_1) = -\tfrac{1}{2}\bar{g}^2 |J_1 J_2 J_3| = -\tfrac{1}{2}\bar{g}^2 |J_1|(I_2 + |J_1|)(I_3 + |J_1|) \approx -\tfrac{1}{2}\bar{g}^2 |J_1|^3 \tag{6.H.23}$$

for $I_2, I_3 << |J_1|$ and the conserved total pseudo-energy on the right side of Eq.(6.H.20) is likewise small as the instability develops. In this limit Eq.(6.H.20) simplifies to

$$\left|\dot{J}_1\right|^2 = \bar{g}^2\left|J_1\right|^3 \quad \Rightarrow \quad \frac{d\left|J_1\right|}{\left|J_1\right|^{3/2}} = \bar{g}dt \quad \Rightarrow \quad \left|J_1\right|^{-1/2} = -\frac{1}{2}\bar{g}(t+const)$$

$$\Rightarrow \quad \left|J_1\right|(t) = \frac{1}{\left(\frac{1}{2}\bar{g}\right)^2}\frac{1}{(t-t_0)^2} \quad t_0^2 = \frac{1}{\left|J_1(0)\right|^2\left(\frac{1}{2}\bar{g}\right)^2}$$

---

[37] JF Drake, PK Kaw, YC Lee, G. Schmidt, CS Liu, and MN Rosenbluth, Phys. Fluids **17**, 778 (1974).

[38] C. S. Liu and P. K. Kaw, *Advances in Plasma Physics*, Vol. 6, Part-I, p.83 (1976).

[39] W. L. Kruer, *The physics of laser plasma interactions,* (AddisonWesley Publishing Co., Reading, MA), 1988.

[40] M. Porkolab, Phys. Fluids **17**, 1432 (1974).

[41] P. K. Kaw, *Advances in Plasma Physics*, Vol. 6, Part-I, p.179 (1976); P.K. Kaw, *Advances in Plasma Physics*, Vol. 6, Part-I, p.207 (1976).

[42] B.I. Cohen, Physics of Fluids **30**, 2676 (1987).

[43] V. Stefan, N. A. Krall, and J. B. McBride, Physics of Fluids **30**, 3703 (1987).

(6.H.24)

The solution for the growth of the action amplitude in Eq.(6.H.24) approaches infinity at a finite time $t_0$, i.e., the amplitude "explodes". This is a mathematical singularity and not physically realizable: other physics will come into play at large amplitudes to saturate the instability at finite amplitude.

### 6.H.d Including damping and frequency mismatch

The inclusion of damping in the three-wave interactions can significantly change the dynamics from that in the undamped limit. Equations(6.H.11) acquire additional terms to model dissipation:

$$\frac{\partial}{\partial t}E_1 = -\frac{g}{\bar{\varepsilon}_1}E_2 E_3 - \gamma_1 E_1 \quad \frac{\partial}{\partial t}E_2 = \frac{g}{\bar{\varepsilon}_2}E_1 E_3^* - \gamma_2 E_2 \quad \frac{\partial}{\partial t}E_3 = \frac{g}{\bar{\varepsilon}_3}E_1 E_2^* - \gamma_3 E_3 \quad (6.H.25)$$

The inclusion of damping breaks all of the conservation laws in Eq.(6.H.13) and (6.H.14).

Consider the simple example in which $\gamma_3$ is large and $\gamma_1, \gamma_2 \to 0$. Such a circumstance can apply to the stimulated scattering of a transverse wave by a longitudinal plasma wave into another transverse wave, $t \to t + l$, where the longitudinal wave is heavily damped. In this limit Eqs.(6.H.25) become

$$E_3 = \frac{g}{\gamma_3 \bar{\varepsilon}_3}E_1 E_2^* \quad \frac{\partial}{\partial t}E_1 = -\frac{g}{\bar{\varepsilon}_1}E_2 E_3 = -\frac{g^2}{\gamma_3 \bar{\varepsilon}_3 \bar{\varepsilon}_1}\left|E_2\right|^2 E_1 \quad \frac{\partial}{\partial t}E_2 = \frac{g}{\bar{\varepsilon}_2}E_1 E_3^* = \frac{g^2}{\gamma_3 \bar{\varepsilon}_3 \bar{\varepsilon}_2}\left|E_1\right|^2 E_2$$

(6.H.26)

From Eqs.(6.H.26) one concludes that

$$\dot{J}_1 = -\dot{J}_2 \quad J_1 + J_2 = I_2 = const \quad (6.H.27a)$$

$$\dot{J}_1 = -GJ_1 J_2 = -GJ_1(I_2 - J_1) \quad G = \frac{2g^2}{\gamma_3 \bar{\varepsilon}_1 \bar{\varepsilon}_2 \bar{\varepsilon}_3} \quad (6.H.27b)$$

$$J_1 = \frac{I_2}{1+\exp(I_2 Gt')} \quad J_2 = I_2\frac{\exp(I_2 Gt')}{1+\exp(I_2 Gt')} \quad t' = t - t_0 \quad (6.H.27c)$$

where at $t = t_0$ $J_1 = J_2 = \frac{1}{2}I_2$. Wave action and energy feed steadily from the high-frequency pump wave with action $J_1$ to the lower frequency wave with action $J_2$, conserving $I_2$. One can displace the time axis to allow for any desired ratio of $J_2/J_1$ at a given "initial" time. Thus, beat-wave heating or stimulated Raman scattering, or parametric instability can be modeled.

Consider a second example in which $E_1$ is large and $E_2$, $E_3$ are initially small and have finite damping rates $\gamma_2$, $\gamma_3$. We assume that $E_2, E_3 \sim e^{\Gamma t}$ and $E_1 \sim$ const in Eqs.(6.H.25) to obtain

$$\Gamma = \frac{1}{2}\left[-(\gamma_1+\gamma_2)+\sqrt{(\gamma_1-\gamma_2)^2+4G|E_1|^2}\right] \quad G \equiv \frac{g^2}{\bar{\varepsilon}_2\bar{\varepsilon}_3} \qquad (6.H.28)$$

From the expression for $\Gamma$ we determine that the amplitude $E_1$ must exceed a threshold value to overcome the damping in the decay product waves in order that $\Gamma > 0$, i.e.,

$$\Gamma > 0 \quad \to \quad |E_1|^2 > |E_1|_c^2 = \frac{\gamma_2\gamma_3}{G} = \frac{\bar{\varepsilon}_2\bar{\varepsilon}_2\gamma_2\gamma_3}{g^2} \qquad (6.H.29)$$

We next consider an example in which there is negligible damping but the three-wave resonance condition is subject to a finite frequency mismatch. Action remains conserved, i.e., $J_1+J_2 = I_2 = const \quad J_1+J_3 = I_3 = const$. The total Hamiltonian is

$$H = \sum_{j=1}^{3}\omega_j J_j + \bar{g}\sqrt{|J_1 J_2 J_3|}\sin(\theta_1-\theta_2-\theta_3) \qquad (6.H.30)$$

From $H$ in Eq.(6.H.30) and Hamilton's equations we deduce

$$\dot{H} = \frac{\partial}{\partial t}H = 0$$

$$\dot{J}_1 = -\frac{\partial H}{\partial \theta_1} = -\bar{g}\sqrt{|J_1 J_2 J_3|}\cos(\theta_1-\theta_2-\theta_3), \quad \dot{J}_2 = -\frac{\partial H}{\partial \theta_2} = -\dot{J}_1, \quad \dot{J}_3 = -\frac{\partial H}{\partial \theta_3} = -\dot{J}_1$$

$$\dot{\theta}_1 = \frac{\partial H}{\partial J_1} = \omega_1 + \frac{1}{2}\bar{g}\sqrt{\left|\frac{J_2 J_3}{J_1}\right|}\sin(\theta_1-\theta_2-\theta_3) \qquad (6.H.31)$$

$$\dot{\theta}_2 = \frac{\partial H}{\partial J_2} = \omega_2 + \frac{1}{2}\bar{g}\sqrt{\left|\frac{J_1 J_3}{J_2}\right|}\sin(\theta_1-\theta_2-\theta_3)$$

$$\dot{\theta}_3 = \frac{\partial H}{\partial J_3} = \omega_3 + \frac{1}{2}\bar{g}\sqrt{\left|\frac{J_1 J_2}{J_3}\right|}\sin(\theta_1-\theta_2-\theta_3)$$

The equations of motion in Eq.(6.H.31) admit oscillatory solutions in terms of Jacobi elliptic functions, and we can derive the following equation for a pseudo-particle:

$$\frac{1}{2}\dot{J}_1^2 + \Phi(J_1) = -\frac{1}{2}(H-I_2\omega_2-I_3\omega_3)^2$$

$$\Phi = -\frac{1}{2}\left[\bar{g}^2|J_1 J_2 J_3| - J_1^2(\Delta\omega)^2 + 2J_1\Delta\omega(H-I_2\omega_2-I_3\omega_3)\right] \qquad (6.H.32)$$

$$\Delta\omega = \omega_1-\omega_2-\omega_3$$

Exercise: Compare the structure of Eqs.(6.H.30-6.H.32) to (6.H.18-6.H.20).

Example: Explosive instability with finite mismatch can be deduced from Eq.(6.H.32). Consider $J_1<0$ and $J_2, J_3>0$ all with infinitesimal amplitudes at $t=0$. In this case $I_2\omega_2 + I_3\omega_3$ and $H$ are negligible in Eq.(6.H.32). For no frequency mismatch $\Delta\omega=0$, $\Phi = -\frac{1}{2}\bar{g}^2|J_1 J_2 J_3| \approx -\frac{1}{2}\bar{g}^2|J_1|^3$, i.e., $\Phi$ is negative and can monotonically increase in magnitude while the magnitudes of all three actions and their time derivatives increase: $\frac{1}{2}\dot{J}_1^2 = -\Phi \approx \frac{1}{2}\bar{g}^2|J_1|^3$. With a finite frequency mismatch, $\Phi = -\frac{1}{2}\bar{g}^2|J_1 J_2 J_3| + \Delta\omega^2 J_1^2 \approx -\frac{1}{2}\bar{g}^2|J_1|^3 + \Delta\omega^2 J_1^2$, and $\Phi$ is positive for small values of $|J_1|$ before turning negative at large values of $|J_1|$. Thus, the value of the action $|J_1|$ can remain trapped for values of $|J_1|$ less than some critical value, i.e., a threshold value $|J_1|_c$, above which $\Phi$ is monotonically decreasing and the quasi-particle can accelerate downhill in $\Phi$ and grow explosively. The threshold value for explosive instability corresponds to the value of $|J_1|$ at the peak of $\Phi$, which is given by $|J_1|_c = \frac{2}{3}\frac{\Delta\omega^2}{\bar{g}^2}$. The difference between the two cases, with and without the frequency mismatch, is that the frequency mismatch degrades the strength of the coupling leading to a threshold condition.

Example: Here we consider parametric instability including a finite frequency mismatch by analyzing Eq.(6.H.25) for infinitesimal $E_2, E_3$ and $E_1$ approximately constant:

$$\frac{\partial}{\partial t}E_2 = \frac{g}{\varepsilon_2}E_1 E_3^* e^{-i\Delta\omega t} \qquad \frac{\partial}{\partial t}E_3 = \frac{g}{\varepsilon_3}E_1 E_2^* e^{-i\Delta\omega t} \qquad (6.H.33)$$

We assume a time dependence $E_2, E_3 \sim e^{-i\Omega t}$ where $\Omega$ is complex. With this ansatz Eq.(6.H.33) yields a quadratic dispersion relation for $\Omega$ with solution:

$$\Omega = \frac{1}{2}\Delta\omega \pm i\sqrt{\frac{g^2|E_1|^2}{\varepsilon_2 \varepsilon_3} - \left(\frac{1}{2}\Delta\omega\right)^2} \qquad (6.H.34)$$

The frequency mismatch reduces the growth rate by reducing the effective strength of the coupling and leads to an amplitude threshold for instability $|E_1| > E_{1c} = |\Delta\omega/2g|\sqrt{\varepsilon_2 \varepsilon_3}$.

If there is damping in the parametric decay product waves and a frequency mismatch, the solution procedure for solving Eq.(6.H.25) is not so simple as that used in obtaining Eq.(6.H.34). The coupled nonlinear differential equations have time-dependent coefficients, and Mathieu equations result. Nishikawa addressed such a case.[44] Both damping and frequency mismatch conspire to reduce instability growth rates and introduce a threshold.

### *6.H.e Extension to many-wave system with three-wave interactions – wave kinetic equation*

Here we extend the analysis of nonlinear wave-wave interactions to a sum over many waves supporting three-wave and four-wave interactions (at least formally). We postulate random phases and employ a statistical approach and an action-angle representation. We specify initial conditions with $\{J_i\}$ given and $\{\theta_i\}$ unknown. We extend the Hamiltonian in Eq.(6.H.30) to

$$H = \sum_j \omega_j J_j + H'(\vec{J},\vec{\theta}) = \sum_j \omega_j J_j + \sum_{\pm} \sum_{ijk} \bar{g}_{ijk} \sqrt{|J_i J_j J_k|} \cos(\pm\theta_i \pm \theta_j \pm \theta_k)$$
$$+ \sum_{\pm} \sum_{ijkl} \bar{g}_{ijkl} \sqrt{|J_i J_j J_k J_l|} \cos(\pm\theta_i \pm \theta_j \pm \theta_k \pm \theta_l)$$
(6.H.35)

which includes resonant and off-resonant three-wave and four-wave interactions. Four-wave interactions can satisfy the frequency matching condition $\omega_1 + \omega_2 = \omega_3 + \omega_4$. The equations of motion for the action-angle pairs are

$$\dot{J}_i = -\frac{\partial H}{\partial \theta_i} = O(g) \qquad \dot{\theta}_i = \frac{\partial H}{\partial J_i} = \omega_i + \frac{\partial H'}{\partial J_i} = \omega_i + O(g)$$
(6.H.36)

---

[44] K. Nishikawa, J. Phys. Soc. Jpn. **24**, 1152 (1968).

<u>Definition</u>: In keeping with taking a statistical approach, we introduce a Liouville probability distribution in $\{\vec{J},\vec{\theta}\}$ space:

$$\rho(\vec{J},\vec{\theta}) = \rho_0(\vec{J};t) + \delta\rho(\vec{J},\vec{\theta};t), \quad \delta\rho << \rho_0$$
(6.H.37)

We recall the derivation of the theory of quasilinear diffusion in canonical variables in Sec. 6.E.d, Eqs.(6.E.40-6.E.42). in the weak turbulence limit the probability distribution satisfies the equations:

$$\frac{\partial}{\partial t}\rho_0(\vec{J};t) = \frac{\partial}{\partial \vec{J}} \cdot \left(\mathbf{D} \cdot \frac{\partial \rho_0}{\partial \vec{J}}\right) \qquad \mathbf{D}(\vec{J}) = \int_0^\infty d\tau \left\langle \dot{\vec{J}}(t)\dot{\vec{J}}(t-\tau)\right\rangle_{\vec{\theta}}$$
(6.H.38)

This equation describes the diffusion of the probability distribution function in action space. The averaging bracket $< >_\theta$ denotes averaging with respect to the phase angles at the specified time t. Double bracket averaging $<< >>$ denotes averaging with respect to ensembles of all possible initial phases. Hence,

$$\left\langle\left\langle \vec{J} \right\rangle\right\rangle(t) \equiv \int d^N J \, \rho_0(J;t)\vec{J}$$
(6.H.39)

We can calculate the $\frac{\partial}{\partial t}\left\langle\left\langle \vec{J} \right\rangle\right\rangle(t)$ using Eqs.(6.H.38) and (6.H.39):

$$\frac{\partial}{\partial t}\left\langle\left\langle \vec{J} \right\rangle\right\rangle(t) = \int d^N J \, \vec{J} \, \frac{\partial}{\partial \vec{J}} \cdot \left(\mathbf{D} \cdot \frac{\partial \rho_0}{\partial \vec{J}}\right) = -\int d^N J \, \mathbf{I} \cdot \mathbf{D} \cdot \frac{\partial \rho_0}{\partial \vec{J}}$$
$$= -\int d^N J \, \frac{\partial \rho_0}{\partial \vec{J}} \cdot \mathbf{D} = \int d^N J \, \rho_0 \frac{\partial}{\partial \vec{J}} \cdot \mathbf{D}(\vec{J}) = \left\langle\left\langle \frac{\partial}{\partial \vec{J}} \cdot \mathbf{D}(\vec{J}) \right\rangle\right\rangle$$
(6.H.40)

Note that **D** is symmetric.

Because the $\theta_i$ are each cyclic, we can expand $H'(\vec{J},\vec{\theta})$ in a Fourier series in $\theta_i$:

$$H'(\vec{J},\vec{\theta}) = \sum_{l=-\infty}^{\infty} H_{\vec{l}}(\vec{J})\exp(i\vec{l}\cdot\vec{\theta}), \quad \vec{l} = \{l_i\}, \quad H_{\vec{l}}(\vec{J}) \equiv \int \frac{d^N\theta}{(2\pi)^N} H'(\vec{J},\vec{\theta})\exp(-i\vec{l}\cdot\vec{\theta}) \quad (6.H.41)$$

We note that for $\cos(\theta_1 - \theta_2 - \theta_3)$ one has $l_1 = 1$, $l_2 = -1$, $l_3 = -1$, $l_4 = 0$. The values of $\{l_i\}$ control the couplings that are included. We use Eqs.(6.H.36) and (6.H.38), and the Fourier series representation to evaluate **D**:

$$\mathbf{D}(\vec{J}) = \int_0^\infty d\tau \sum_{\vec{l}\vec{l}'} (-i\vec{l})(i\vec{l}')\langle H_{\vec{l}}(\vec{J}^t) H_{\vec{l}'}^*(\vec{J}^{t-\tau})\exp(i\vec{l}\cdot\vec{\theta}^t - i\vec{l}'\cdot\vec{\theta}^{t-\tau})\rangle_{\vec{\theta}} \quad (6.H.42)$$

The time dependences of $\vec{\theta}$ and $\vec{J}$ are determined by Hamilton's equations Eqs.(6.H.36). To the lowest order needed in Eq.(6.H.42), $\vec{J}^{t-\tau} \approx \vec{J}^t$, $\vec{\theta}^{t-\tau} = \vec{\theta}^t - \tau\dot{\vec{\theta}} \approx \vec{\theta}^t - \tau\vec{\omega}$, and $\exp(i\vec{l}\cdot\vec{\theta}^t - i\vec{l}'\cdot\vec{\theta}^{t-\tau}) \approx \exp(i\vec{l}\cdot\vec{\theta}^t - i\vec{l}'\cdot\vec{\theta}^t + i\vec{l}'\cdot\vec{\omega}\tau)$. Hence, the following results are obtained.

Theorem:

$$\frac{\partial}{\partial t}\rho_0(\vec{J};t) = \frac{\partial}{\partial \vec{J}} \cdot \left(\mathbf{D} \cdot \frac{\partial \rho_0}{\partial \vec{J}}\right)$$

$$\mathbf{D}(\vec{J}) = \int_0^\infty d\tau \langle \dot{\vec{J}}(t)\dot{\vec{J}}(t-\tau)\rangle_{\vec{\theta}} = \int_0^\infty d\tau \sum_{\vec{l}} \vec{l}\vec{l} |H_{\vec{l}}(\vec{J})|^2 e^{i\vec{l}\cdot\vec{\omega}\tau}$$

$$= \sum_{\vec{l}} \vec{l}\vec{l} |H_{\vec{l}}(\vec{J})|^2 \left[\pi\delta(\vec{l}\cdot\vec{\omega}) - \frac{1}{i\vec{l}}\right] = \sum_{\vec{l}} \vec{l}\vec{l} |H_{\vec{l}}(\vec{J})|^2 \pi\delta(\vec{l}\cdot\vec{\omega}) \quad (6.H.43)$$

[↑ even in $l$    odd in $l$ ↑    $|H_{\vec{l}}(\vec{J})|^2$ is Hermitian]

$$\frac{\partial}{\partial t}\langle\langle\vec{J}\rangle\rangle(t) = \sum_{\vec{l}} \left\langle\left\langle \vec{l}\cdot\frac{\partial}{\partial\vec{J}}\cdot|H_{\vec{l}}(\vec{J})|^2\right\rangle\right\rangle \vec{l}\pi\delta(\vec{l}\cdot\vec{\omega})$$

Example: For $\vec{l} = [1,-1,-1,0,...]$ then $\vec{l}\cdot\vec{\omega} = \omega_1 - \omega_2 - \omega_3 = 0$ is the resonance condition.

We note that $|H_{\vec{l}}(\vec{J})|^2 \propto g^2$ is the coupling term representing the strength of the interaction. Consider the case in which there are resonant three-wave interactions. From Eq.(6.H.43) one obtains

$$\frac{\partial}{\partial t}\langle\langle\vec{J}\rangle\rangle(t) = \sum_{\vec{l}} \left\langle\left\langle \vec{l}\cdot\frac{\partial}{\partial\vec{J}}\cdot|H_{\vec{l}}(\vec{J})|^2\right\rangle\right\rangle \vec{l}\pi\delta(\vec{l}\cdot\vec{\omega})$$

$$\rightarrow \frac{\partial}{\partial t}\langle J_1\rangle = -\frac{\partial}{\partial t}\langle J_2\rangle = -\frac{\partial}{\partial t}\langle J_3\rangle = \pi\delta(\omega_1 - \omega_2 - \omega_3)\left\langle\left(\frac{\partial}{\partial J_1} - \frac{\partial}{\partial J_2} - \frac{\partial}{\partial J_3}\right)g_{123}^2 J_1 J_2 J_3\right\rangle \quad (6.H.44)$$

$$= \pi\delta(\omega_1 - \omega_2 - \omega_3)g_{123}^2\left(\langle J_2 J_3\rangle - \langle J_1 J_3\rangle - \langle J_1 J_2\rangle\right)$$

We next introduce a decomposition of each of the wave actions in terms of an integral over wavenumber space. We assign **k** to <$J_1$>(**k**), **k'** to <$J_2$>(**k'**), and **k"** to <$J_3$>(**k"**). Then Eq.(6.H.44) becomes a wave kinetic equation:

$$\frac{\partial}{\partial t}\langle J_1\rangle = \int \frac{d^3k'}{(2\pi)^3}\int \frac{d^3k''}{(2\pi)^3}\pi\delta(\omega_1^{\mathbf{k}}-\omega_2^{\mathbf{k'}}-\omega_3^{\mathbf{k''}})$$
$$\times g_{123}^2(\mathbf{k},\mathbf{k'},\mathbf{k''})\delta(\mathbf{k}-\mathbf{k'}-\mathbf{k''})\left(\langle J_2(\mathbf{k'})J_3(\mathbf{k''})\rangle-\langle J_1(\mathbf{k})J_3(\mathbf{k''})\rangle-\langle J_1(\mathbf{k})J_2(\mathbf{k'})\rangle\right)$$
$$\Rightarrow \int \frac{d^3k'}{(2\pi)^3}\int \frac{d^3k''}{(2\pi)^3}\pi\delta(\omega_1^{\mathbf{k}}-\omega_2^{\mathbf{k'}}-\omega_3^{\mathbf{k''}})$$
$$\times g_{123}^2(\mathbf{k},\mathbf{k'},\mathbf{k''})\delta(\mathbf{k}-\mathbf{k'}-\mathbf{k''})\left(\langle J_2(\mathbf{k'})\rangle\langle J_3(\mathbf{k''})\rangle-\langle J_1(\mathbf{k})\rangle\langle J_3(\mathbf{k''})\rangle-\langle J_1(\mathbf{k})\rangle\langle J_2(\mathbf{k'})\rangle\right)$$

(6.H.45)

In obtaining Eq.(6.H.45) we have made assumptions that many waves are present and statistical independence: $\langle J_2(\mathbf{k'})J_3(\mathbf{k''})\rangle = \langle J_2(\mathbf{k'})\rangle\langle J_3(\mathbf{k''})\rangle$.

Consider a quantum mechanics perspective on three-wave interactions. The equation for the time derivative of the number of quanta $N_1$ generated by stimulated scattering into $N_1$ from $N_2$ and $N_3$, and stimulated scattering into $N_2$ and $N_3$ from $N_1$ takes the form:

$$\dot{N}_1 = g^2\left[N_2N_3(N_1+1)-N_1(N_2+1)(N_3+1)\right] = g^2\left[N_2N_3-N_1N_3-N_1N_2-N_1\right] \quad (6.H.46)$$

after cancellations. The number of quanta is related to the wave action through the relation $N_l = J_l/\hbar$. Hence, the last term in Eq.(6.H.46), which is purely spontaneous, is $O(\hbar)$ smaller than the others and can be thrown away with result

$$\dot{J}_1 = \frac{g^2}{\hbar}g^2\left[J_2J_3-J_1J_3-J_1J_3\right] \quad (6.H.47)$$

We see that Eq.(6.H.47) has the same form as Eq.(6.H.45).

<u>Theorem</u>: Eqs.(6.H.44-6.H.45) and the frequency and wavenumber resonance conditions give rise to the following conservation laws.

(i) Action conservation $\quad \frac{\partial}{\partial t}\left(<J_1>+<J_2>\right) = \frac{\partial}{\partial t}\left(<J_1>+<J_3>\right) = 0 \quad$ (6.H.48a)

(ii) Energy conservation $\quad U \equiv \sum_l \int \frac{d^3\mathbf{k}}{(2\pi)^3}J_l(\mathbf{k})\omega_l(\mathbf{k}), \quad \dot{U}=0 \quad$ (6.H.48b)

(iii) Momentum conservation $\quad \mathbf{K} \equiv \sum_l \int \frac{d^3\mathbf{k}}{(2\pi)^3}J_l(\mathbf{k})\mathbf{k}, \quad \dot{\mathbf{K}}=0 \quad$ (6.H.48c)

We next introduce a reasonable definition of the wave entropy and show that in the absence of phase correlations there is entropy production and irreversibility.

<u>Definition</u>: Wave entropy $S = \ln\left(\Pi_l J_l\right) = \sum_l \ln J_l$ (6.H.49)

We then use the definition of the wave entropy Eq.(6.H.49) and Eq.(6.H.45), sum over the three modes, and integrate over **k**-space to evaluate the time derivative of the entropy:

$$\frac{\partial}{\partial t} S = \sum_l \int \frac{d^3\mathbf{k}}{(2\pi)^3} \frac{\dot{J}_s(\mathbf{k})}{J_s(\mathbf{k})} = \sum_l \int \frac{d^3\mathbf{k}}{(2\pi)^3} \int \frac{d^3\mathbf{k'}}{(2\pi)^3} \int \frac{d^3\mathbf{k''}}{(2\pi)^3} \pi\delta(\omega_1^\mathbf{k} - \omega_2^\mathbf{k'} - \omega_3^\mathbf{k''})$$

$$\times g_{123}^2(\mathbf{k},\mathbf{k'},\mathbf{k''}) \delta(\mathbf{k} - \mathbf{k'} - \mathbf{k''}) \left[ (J_2 J_3 - J_1 J_3 - J_1 J_2)\left(\frac{1}{J_1} - \frac{1}{J_2} - \frac{1}{J_3}\right) = \left|J_1 J_2 J_3\right| \left(\frac{1}{J_1} - \frac{1}{J_2} - \frac{1}{J_3}\right)^2 \right]$$

(6.H.50)

The right side of Eq.(6.H.50) is non-negative, $\partial S / \partial t \geq 0$, and tends to zero asymptotically in time with the asymptotic state satisfying $\partial S / \partial t = 0$:

$$\frac{1}{J_1} = \frac{1}{J_2} + \frac{1}{J_3} \quad (6.H.51)$$

We next introduce a quantity analogous to the Helmholtz free energy, i.e., $S - \beta U$. Consider the variation of this quantity about the asymptotic steady state determined in Eq.(6.H.51):

$$\sum_l \delta\left(\ln J_l - \beta\omega_l J_l\right) = \sum_l \delta J_l \left(\frac{1}{J_l} - \beta\omega_l\right) = 0 \quad \Rightarrow \quad \frac{1}{J_l} = \beta\omega_l \quad (6.H.52)$$

If we substitute the result in Eq.(6.H.52) into the asymptotic condition in Eq.(6.H.51), we recover the three-wave frequency resonance condition $\omega_1 = \omega_2 + \omega_3$. Furthermore, if we use the result in Eq.(6.H.52) in the wave energy, we deduce that the asymptotic wave energy satisfies a Rayleigh-Jeans law, i.e.,

$$U_j = \omega_j J_j = \frac{1}{\beta} \equiv T_{wave-kinetic} \quad (6.H.53)$$

Thus, all modes have the same asymptotic energy, which implies that this classical theory can lead to an ultra-violet catastrophe.

<u>Example</u>: Explosive instability – If the three-wave interaction involves $J_1$<0 and $J_2, J_3$>0 as in Sec. 6.H.c, then Eq.(6.H.51) is never satisfied,

$$\frac{1}{J_1} \neq \frac{1}{J_2} + \frac{1}{J_3} \quad (6.H.54)$$

and $|J_1|, J_2, J_3$, and $S$ can all continue to grow without bound.

# LECTURES ON THEORETICAL PLASMA PHYSICS – PART 3A

*Allan N. Kaufman*

## 7. Plasma collisional and discreteness phenomena

This section examines plasma collision processes and discreteness phenomena. This includes test-particle theory of single particles and many particles, radiation transport, and Dupree's theory of phase-space granulation and clump formation.

### 7.A Test-particle formulation in the electrostatic limit

We begin with a theoretical treatment based on a Coulomb model. In the Coulomb model we recall that

$$\phi(\mathbf{k},\omega) = \frac{\phi_0(\mathbf{k},\omega)}{\varepsilon(\mathbf{k},\omega)} \tag{7.A.1}$$

To support self-consistent electromagnetic fields within the context of Maxwell's equations the longitudinal dielectric $\varepsilon \to \mathbf{K}$, the conductivity tensor. Consider Poisson's equation for a single charged particle:

$$\rho_0(\mathbf{r},t) = e_0 \delta(\mathbf{r} - \mathbf{r}_0(t))$$
$$\rho_0(\mathbf{k},\omega) \equiv \int d^3\mathbf{r}\, e^{-i\mathbf{k}\cdot\mathbf{r}} \int dt\, e^{i\omega t} \rho_0(\mathbf{r},t) \tag{7.A.2}$$
$$k^2 \phi_0(\mathbf{k},\omega) = 4\pi \rho_0(\mathbf{k},\omega)$$

Using the lowest-order particle trajectory $\mathbf{r}_0(t) = \mathbf{r}_0(0) + \mathbf{v}_0 t + \ldots$ we can evaluate the Fourier-transformed charge density in Eq.(7.A.2):

$$\rho_0(\mathbf{k},\omega) \equiv \int d^3\mathbf{r}\, e^{-i\mathbf{k}\cdot\mathbf{r}} \int dt\, e^{i\omega t} \rho_0(\mathbf{r},t) = e_0 \int dt\, e^{i\omega t} e^{-i\mathbf{k}\cdot\mathbf{r}_0(t)}$$
$$= e_0 e^{-i\mathbf{k}\cdot\mathbf{r}_0(0)} \int_{-\infty}^{\infty} dt\, e^{i\omega t - i\mathbf{k}\cdot\mathbf{v}_0} = e_0 e^{-i\mathbf{k}\cdot\mathbf{r}_0(0)} 2\pi \delta(\omega - \mathbf{k}\cdot\mathbf{v}_0) \tag{7.A.3}$$

We substitute the expression in Eqs.(7.A.3) into Eqs.(7.A.1) and (7.A.2) to obtain

$$\phi(\mathbf{k},\omega) = \frac{4\pi}{k^2} \frac{1}{\varepsilon(\mathbf{k},\omega)} 2\pi e_0 \delta(\omega - \mathbf{k}\cdot\mathbf{v}_0) e^{-i\mathbf{k}\cdot\mathbf{r}_0(0)} \tag{7.A.4}$$

whose inverse Fourier transform is

$$\phi(\mathbf{r},t) = \int \frac{d\omega}{2\pi} \int \frac{d^3\mathbf{k}}{(2\pi)^3} e^{i\mathbf{k}\cdot\mathbf{r}-i\omega t} \phi(\mathbf{k},\omega)$$

$$= \int \frac{d\omega}{2\pi} \int \frac{d^3\mathbf{k}}{(2\pi)^3} e^{i\mathbf{k}\cdot\mathbf{r}-i\omega t} \frac{4\pi}{k^2} \frac{1}{\varepsilon(\mathbf{k},\omega)} 2\pi e_0 \delta(\omega - \mathbf{k}\cdot\mathbf{v}_0) e^{-i\mathbf{k}\cdot\mathbf{r}_0(0)} \quad (7.A.5)$$

$$= \int \frac{d^3\mathbf{k}}{(2\pi)^3} e^{i\mathbf{k}\cdot[\mathbf{r}-(\mathbf{r}_0(0)+\mathbf{v}_0 t)]} \frac{1}{\varepsilon(\mathbf{k},\mathbf{k}\cdot\mathbf{v}_0)} \frac{4\pi e_0}{k^2} \equiv \int \frac{d^3\mathbf{k}}{(2\pi)^3} e^{i\mathbf{k}\cdot\mathbf{s}} \frac{1}{\varepsilon(\mathbf{k},\mathbf{k}\cdot\mathbf{v}_0)} \frac{4\pi e_0}{k^2}$$

where $\mathbf{s} = \mathbf{r} - (\mathbf{r}_0(0) + \mathbf{v}_0 t)$, the spatial coordinate in the particle frame to lowest order. To proceed further, one choose a form for the velocity distribution function $g(\mathbf{u})$ as in Sec. 2.E.d with which to calculate the dielectric response $\varepsilon$. One can obtain analytic results for square distributions and Cauchy distributions, but not for a Maxwellian. We note that $\mathrm{Re}\,\varepsilon(-\mathbf{k}) = \mathrm{Re}\,\varepsilon(\mathbf{k})$ and $\mathrm{Im}\,\varepsilon(-\mathbf{k}) = -\mathrm{Im}\,\varepsilon(\mathbf{k})$.

Exercise: From Eq.(7.A.3) and the machinery in Eq.(7.A.5) calculate $\rho(\mathbf{s})$ and show that the total charge and dipole moment are both zero, i.e.,

$$e_{total} = \int d^3\mathbf{s}\, \rho(\mathbf{s}) = 0 \qquad \mathbf{p} = \int d^3\mathbf{s}\, \mathbf{s}\rho(\mathbf{s}) = 0 \quad (7.A.6)$$

However, the quadrupole moment is finite: $|\mathbf{Q}| \sim e_0 \lambda_D^2 (v_0/v_e)^2$ Pick a specific velocity distribution function $g$ and calculate $\mathbf{Q}$.

### 7.A.a Dynamic friction and wave emission at long wavelengths

We return to Eq.(7.A.5) and examine its properties. We expect that if $v_0 > \omega/k$ there will be a Cerenkov effect. What is the shielded force on a single particle in the frame of the lowest-order particle trajectory (an inertial frame)?

$$\mathbf{F} = e_0 \mathbf{E}(s=0) = e_0^2 \int \frac{d^3\mathbf{k}}{(2\pi)^3} 4\pi \frac{(-i\mathbf{k})}{k^2} \frac{1}{\varepsilon(\mathbf{k},\mathbf{k}\cdot\mathbf{v}_0)} \qquad \frac{1}{\varepsilon} = \frac{\varepsilon' - i\varepsilon''}{|\varepsilon|^2} \quad (7.A.7)$$

We note that $\varepsilon'$ is even with respect to changing the sign of $\mathbf{k}$, while $\varepsilon''$ and $\mathbf{k}/k^2$ are odd. Thus, Eq.(7.A.7) yields the following result.

Theorem: (Dynamic friction) $\quad \mathbf{F} = e_0 \mathbf{E}(s=0) = -e_0^2 \int \frac{d^3\mathbf{k}}{(2\pi)^3} 4\pi \frac{\mathbf{k}}{k^2} \frac{\varepsilon''}{|\varepsilon|^2} \quad (7.A.8)$

The dynamic friction has different characteristics for long and short wavelengths. (i) For $k < k_D$ ($\lambda > \lambda_D$), waves are weakly damped and weakly emitted by the Cerenkov effect $\mathbf{k}\cdot\mathbf{v}_0 = \omega_\mathbf{k}$. There is a radiation reaction, i.e., particles lose energy in emitting waves. (ii) For $k > k_D$ ($\lambda < \lambda_D$), shielding is insufficient and ineffective. Binary interactions (Coulomb collisions) produce friction. We consider the long and short wavelength limits separately, make appropriate approximations, and obtain analytic results to evaluate the dynamic friction and wave emission.

We examine Eq.(7.A.8) in the long wavelength limit:

$$\mathbf{F}_W = -\mathbf{F} = e_0^2 \int_{k<k_D} \frac{d^3\mathbf{k}}{(2\pi)^3} 4\pi \frac{\mathbf{k}}{k^2} \int d\omega\, \delta(\omega - \mathbf{k}\cdot\mathbf{v}_0) \text{Im} \frac{1}{\varepsilon(\mathbf{k},\omega)} \qquad (7.A.9)$$

where the force on the waves due to the particle is opposite in sign to the force of the waves on the particle. We make use of

$$\gamma_k = -\frac{\varepsilon''\big|_{\omega_k}}{\dfrac{\partial \varepsilon'}{\partial \omega}\bigg|_{\omega_k}}$$

for small $k$ and

$$\lim_{y \to 0} \frac{1}{x+iy} = P\left(\frac{1}{x}\right) - i\pi \delta(x) \text{sgn}(y)$$

Hence, if $|\varepsilon''| \ll |\omega \partial \varepsilon / \partial \omega|$, then

$$\text{Im} \frac{1}{\varepsilon' + i\varepsilon''} = -\pi \delta(\varepsilon') \text{sgn}(\varepsilon'') \qquad \delta(\varepsilon') = \delta(\varepsilon'(\mathbf{k},\omega)) = \sum_l \frac{\delta(\omega - \omega_\mathbf{k}^l)}{\left|\dfrac{\partial \varepsilon'}{\partial \omega}\right|_{\omega_\mathbf{k}^l}} \qquad \varepsilon'(\mathbf{k},\omega_\mathbf{k}^l) = 0$$

Thus Eq.(7.A.9) leads to the following result for long wavelengths.

<u>Theorem</u>: 
$$\mathbf{F}_W = -e_0^2 \pi \int_{k<k_D} \frac{d^3\mathbf{k}}{(2\pi)^3} 4\pi \frac{\mathbf{k}}{k^2} \sum_l \frac{1}{\left|\dfrac{\partial \varepsilon'}{\partial \omega}\right|_{\omega_\mathbf{k}^l}} \delta(\omega_\mathbf{k}^l - \mathbf{k}\cdot\mathbf{v}_0) \text{sgn}(\omega_\mathbf{k}^l)$$
(7.A.10)

We next derive the rate of energy loss by the particle, which is equal to the rate of energy in wave emission:

$$\dot{W} = -\mathbf{F}_W \cdot \mathbf{v}_0 = e_0^2 \pi \int_{k<k_D} \frac{d^3\mathbf{k}}{(2\pi)^3} \frac{4\pi}{k^2} \sum_l \frac{|\omega_\mathbf{k}^l|}{\left|\dfrac{\partial \varepsilon'}{\partial \omega}\right|_{\omega_\mathbf{k}^l}} \delta(\omega_\mathbf{k}^l - \mathbf{k}\cdot\mathbf{v}_0) \qquad (7.A.11)$$

The rate of energy in wave emission is positive for either stable or unstable waves. With no loss of generality we can limit the sum over $l$ in Eq.(7.A.11) to branches for which $\omega > 0$ and introduce a factor of 2 to account for the negative frequencies.

<u>Definition</u>: The rate of energy in wave emission per unit volume as a function of $\mathbf{k}$ for a particular branch $l$ (with positive frequency -- summing over both positive and negative frequencies for the branch will introduce another factor of 2)

$$\dot{W}^l(\mathbf{k}) = \frac{4\pi^2 e_0^2}{k^2} \frac{|\omega_\mathbf{k}^l|}{\left|\frac{\partial \varepsilon'}{\partial \omega}\right|_{\omega_\mathbf{k}^l}} \delta(\omega_\mathbf{k}^l - \mathbf{k} \cdot \mathbf{v}_0) \qquad (7.A.12)$$

We integrate Eq.(7.A.12) over the velocity distribution functions and sum over species to obtain the wave emission power density per unit volume in **k** space for all particles ignoring particle-radiation correlations (restriction to incoherent radiation):

$$\dot{W}^l(\mathbf{k}) = \sum_s \int d^3\mathbf{v}\, g_s(\mathbf{v}) \frac{\omega_s^2}{k^2} \frac{2\pi m_s |\omega_\mathbf{k}^l|}{\left|\frac{\partial \varepsilon'}{\partial \omega}\right|_{\omega_\mathbf{k}^l}} \delta(\omega_\mathbf{k}^l - \mathbf{k} \cdot \mathbf{v}_0)$$

$$= \frac{2\pi}{k^3} \frac{|\omega_\mathbf{k}^l|}{\left|\frac{\partial \varepsilon'}{\partial \omega}\right|_{\omega_\mathbf{k}^l}} \sum_s g_s(u = V_\mathbf{k}^l)_s m_s \omega_s^2 \qquad (7.A.13)$$

where $V_\mathbf{k}^l = \frac{\omega_\mathbf{k}^l}{k}$. Equation (7.A.13) is the spontaneous emission power density per unit volume in **k** space. [*Editor's note: At this point Kaufman commented that this derivation seems to be in error for negative energy waves for which $\partial \varepsilon'/\partial \omega < 0$. He did not explain the nature of the error and left the remediation of the derivation for negative energy waves unresolved.*]

The total time derivative of the energy density per unit **k** volume is a sum including the spontaneous emission, Landau damping or growth, and nonlinear effects:

<u>Theorem:</u> $\qquad \frac{d}{dt} W^l(\mathbf{k}) = \dot{W}_\mathbf{k}^l + 2\gamma_\mathbf{k}^l W_\mathbf{k}^l + \text{nonlinear} \qquad (7.A.14)$

If the wave amplitudes are small, then nonlinear effects can be ignored. If we further assume that the plasma is stable, i.e., $\gamma_\mathbf{k}^l < 0$, then Eq.(7.A.14) admits a simple steady-state relation for the energy density per unit **k** volume in branch *l*:

$$W_\mathbf{k}^l = \frac{\dot{W}_\mathbf{k}^l}{-2\gamma_\mathbf{k}^l} \qquad (7.A.15)$$

If we use the relation for $\gamma_\mathbf{k}^l = \frac{\pi}{k^2} \frac{1}{\left|\frac{\partial \varepsilon'}{\partial \omega}\right|_{\omega_\mathbf{k}^l}} \sum_s \omega_s^2 g_\mathbf{k}^{s\,\prime}(V_\mathbf{k}^l)$ in Eq.(7.A.15), we obtain

$$W^l_{\mathbf{k},s} = \frac{|V^l_{\mathbf{k}}| m_s g^s_{\mathbf{k}}(V^l_{\mathbf{k}})}{-g^s_{\mathbf{k}}{}'(V^l_{\mathbf{k}})} \qquad (7.A.16)$$

for a single resonant species.

Example: For a Maxwellian plasma $g(u) = const\ \exp(-mu^2/2T)$, then
$$g'(u) = -(mu/T)g \qquad (7.A.17a)$$
and
$$W^l(\mathbf{k}) = T_s \qquad (7.A.17b)$$

for a single resonant species. This is the Rayleigh-Jeans law for a plasma (equipartition law in statistical mechanics). This is only valid for weak damping, i.e., $k < k_D = \lambda_D^{-1}$. We integrate Eq.(7.A.17) over wavenumbers to compute the energy density $W$ in a quiescent plasma:

$$W = \int_{k<k_D} \frac{d^3k}{(2\pi)^3} \sum_l W^l(\mathbf{k}) \approx 2T k_D^3 \sim T/\lambda_D^3 \qquad (7.A.18)$$

An estimate of the kinetic energy density in the plasma is the pressure, i.e., $K \sim nT$. Hence,

$$\frac{W}{K} \sim \frac{T}{\lambda_D^3} \frac{1}{nT} \sim \frac{1}{n\lambda_D^3} \equiv \frac{1}{\Lambda} \ll 1 \qquad (7.A.19)$$

We note that the plasma parameter $\Lambda$ is typically large, e.g., $\Lambda \sim 10^4$-$10^8$.

Example: In a non-Maxwellian plasma we define an effective temperature $T_{\text{eff}}$ so that in analogy to Eq.(7.A.17a) one has
$$g'(u) = -\left(mu/T_{\text{eff}}(u)\right)g \qquad (7.A.20a)$$
and
$$W^l(\mathbf{k}) = T_{s,\text{eff}}(V^l_{\mathbf{k}}) \qquad (7.A.20b)$$

Equation (7.A.20b) retains the form of the Rayleigh-Jeans formula. The results in this sub-section can be believed as long as the wave energies are not so large as to invalidate our ignoring the nonlinear terms in Eq.(7.A.14). The extension of Eq.(7.A.19) to the non-Maxwellian limit gives

$$\frac{W}{K} \sim \frac{T_{\text{eff}}}{\lambda_D^3} \frac{1}{nT} \sim \frac{1}{\Lambda} \frac{T_{\text{res}}}{T_{\text{avg}}} \sim 10^{-5} \times 10^3 \sim 10^{-2} \ll 1 \qquad (7.A.21)$$

At this relative amplitude the wave energy density should be quite measurable.

Exercises: (i) Examine the generalization to the electromagnetic wave energy densities based on Maxwell's equations. (ii) Calculate the wave energy density in ion acoustic waves for $T_e \gg T_i$. Show that for ion waves one can deduce that the plasma parameter has to be quite large, i.e., $\Lambda > (m_i/m_e)^2$, in order that the wave energy density be small enough to avoid trapping electrons. Electron trapping in ion waves invalidates the neglect of nonlinearities in the theory.

### 7.A.b Superthermal radiation – Cerenkov radiation and radiation reaction

In Eq.(7.A.10) we calculated the radiation reaction $\mathbf{F}_W$ to a test particle based on the linear plasma response for long wavelengths modes. If we assume that the plasma is isotropic and unmagnetized ($\mathbf{B}=0$), then we can project the radiation reaction force onto the preferred direction of the test-particle velocity:

$$\hat{\mathbf{v}}_0 \cdot \mathbf{F}_W = -\frac{2e_0^2 \omega_p^2}{2\mathrm{v}_0} \int_{k<k_D} \frac{k^2 dk}{(2\pi)^3} \frac{4\pi^2}{k^2} 2\pi \int_{-1}^{1} d\mu\, \delta(\omega_p - k\mathrm{v}_0 \mu) = -\frac{e_0^2 \omega_p^2}{\mathrm{v}_0^2} \int_{0+}^{k_D} \frac{dk}{k} \qquad (7.A.22)$$

where we only include electron plasma waves for which $\left|\partial \varepsilon'/\partial \omega\right|_{\omega_k} = 2/\omega_p$, $\mu \equiv \cos\theta$ due to the projection of $\mathbf{k}$ on $\mathbf{v}_0$, there are two branches of waves (positive and negative frequencies), and $\delta(\omega_p - k\mathrm{v}_0 \mu) = (1/k\mathrm{v}_0)\delta(\mu - \omega_p/k\mathrm{v}_0)$. The Cerenkov resonance between the wave and the particle is embodied in the relation $\mathrm{v}_0 \mu = \omega_p/k$, which can only be satisfied if $\mu = \omega_p/k\mathrm{v}_0 < 1 \rightarrow \omega_p/\mathrm{v}_0 < k$. This implies the following result.

Theorem: Wave-particle resonance and emission can exist if

$$\frac{\omega_p}{\mathrm{v}_{0,res}} < k < k_D \quad \Rightarrow \quad \frac{\omega_p}{\mathrm{v}_{0,res}} < k_D \quad \text{or} \quad \mathrm{v}_{th}^e = \frac{\omega_p}{k_D} < \mathrm{v}_{0,res} \qquad (7.A.23)$$

Hence, only superthermal particle particles can emit waves; and the lower limit of the integral with respect to $k$ in Eq.(7.A.22) is $\omega_p/\mathrm{v}_0$.

Theorem: The magnitude of the resonant radiation reaction force is

$$\hat{\mathbf{v}}_0 \cdot \mathbf{F}_W = -\frac{e_0^2 \omega_p^2}{\mathrm{v}_0^2} \int_{\frac{\omega_p}{\mathrm{v}_0}}^{k_D} \frac{dk}{k} = -\frac{e_0^2 \omega_p^2}{\mathrm{v}_0^2} \ln\frac{\mathrm{v}_0}{\mathrm{v}_{th}^e} \qquad \left|\hat{\mathbf{v}}_0 \cdot \mathbf{F}_W\right| \approx \frac{e_0^2}{\lambda_D^2}\left(\frac{\mathrm{v}_{th}^e}{\mathrm{v}_0}\right)^2 \ln\frac{\mathrm{v}_0}{\mathrm{v}_{th}^e} \qquad (7.A.24)$$

### 7.A.c Dynamic friction and wave emission at short wavelengths

We return to Eq.(7.A.9) in the short-wavelength limit:

$$\mathbf{F}_D = -e_0^2 \int_{k>k_D} \frac{d^3\mathbf{k}}{(2\pi)^3} 4\pi \frac{\mathbf{k}}{k^2} \frac{\varepsilon''(\mathbf{k}, \mathbf{k}\cdot\mathbf{v}_0)}{|\varepsilon|^2} \qquad (7.A.25)$$

where

$$\varepsilon = 1 - \sum_s \frac{\omega_s^2}{k^2} \int du\, g_{\hat{\mathbf{k}}}^{s\,\prime}(u)\left[\frac{P}{u - \hat{\mathbf{k}}\cdot\mathbf{v}_0} + i\pi\delta(u - \hat{\mathbf{k}}\cdot\mathbf{v}_0)\right] \qquad (7.A.26a)$$

$$\varepsilon'' = -\pi \sum_s \frac{\omega_s^2}{k^2} \int du\, g_{\mathbf{k}}^{s\prime}(\hat{\mathbf{k}}\cdot\mathbf{v}_0) \quad \mathrm{Im}\frac{1}{\varepsilon} = -\frac{\varepsilon''}{|\varepsilon|^2} \quad g_{\mathbf{k}}^{s\prime} \equiv \frac{\partial}{\partial u}g_{\mathbf{k}}^s(u) \qquad (7.A.26b)$$

For short-wavelength modes, e.g., electron plasma waves, the phase velocities fall in the bulk of the electron velocity distribution, which gives rise to strong Landau damping. Thus, the normal mode frequencies $\omega_{\mathbf{k}}^l$ have large imaginary parts so that $\varepsilon(\mathrm{Re}\,\omega_{\mathbf{k}}^l)\neq 0$ and $\varepsilon = 1 + O\left(\frac{1}{k^2\lambda_D^2}\right)$.

<u>Theorem</u>: For $k > k_D$ we assume $|\varepsilon|^2 \to O(1)$, in which limit the dynamic friction is

$$\mathbf{F}_D = 4\pi^2 e_0^2 \sum_s \omega_s^2 \int_{k>k_D} \frac{d^3\mathbf{k}}{(2\pi)^3} \frac{\mathbf{k}}{k^4} g_{\mathbf{k}}^{s\prime}(\hat{\mathbf{k}}\cdot\mathbf{v}_0) \qquad (7.A.27)$$

We next assume that $g$ is isotropic. Then the only preferred direction is the velocity $\mathbf{v}_0$. Hence, Eq.(7.A.27) becomes

$$\hat{\mathbf{v}}_0\cdot\mathbf{F}_D^s = -4\pi^2 e_0^2 \omega_s^2 \int_{k>k_D} \frac{k^2 dk}{(2\pi)^3} \frac{2\pi}{k^3} \int_{-1}^{1} d\mu\, \mu g_{\mathbf{k}}^{s\prime}(\mathrm{v}_0\mu) \quad \mu\equiv\cos(\hat{\mathbf{k}},\hat{\mathbf{v}}_0) \qquad (7.A.28)$$

The lower limit of the $k$-integral in Eq.(7.A.28) is $k_D$. The upper limit cutoff is set by physical considerations. Classical considerations become invalid and quantum mechanics must be applied for wavelengths shorter than the deBroglie wavelength, viz., $k > m\mathrm{v}/\hbar$. Another constraint is set by the assumption of small-amplitude electric fields so that linear theory is valid: $e^2/r \ll m\mathrm{v}^2$ or $r > e^2/T \to k < T/e^2$. Hence, we obtain

$$k_{max} = \left(\frac{m\mathrm{v}}{\hbar}, \frac{T}{e^2}\right)_< \qquad (7.A.29)$$

where the former condition is set by quantum mechanics and prevails for $T > 10$ eV, and the latter condition is set by nonlinearity and prevails for $T < 10$ eV. For $k > k_{max}$ one must employ the quantum mechanical Boltzmann equation. In terms of interaction distances, Eq.(7.A.28) pertains to interaction distances $r$ satisfying $r_{min} \sim 1/k_{max} < r < \lambda_D$ where $k_{max}$ is given in Eq.(7.A.29).

The results here capture simultaneous, superposable, small-angle collisions. There are no waves here and no zeroes of $\varepsilon$. The lower limit of the integral over $k$ in Eq.(7.A.28) is $k_D$, so that the $k$ integral $\int dk/k = \ln(k_{max}/k_D)$. In the low-temperature classical limit for $k_{max}$, Eq.(7.A.28) yields

$$\hat{\mathbf{v}}_0\cdot\mathbf{F}_D^s = -\frac{e_0^2\omega_s^2}{2\pi}\int_{k_D}^{k_{max}}\frac{dk}{k} 2\pi\int_{-1}^{1} d\mu\,\mu g_{\mathbf{k}}^{s\prime}(\mathrm{v}_0\mu) = -e_0^2\omega_s^2 \ln\frac{k_{max}}{k_D}\int_{-1}^{1} d\mu\,\mu g_{\mathbf{k}}^{s\prime}(\mathrm{v}_0\mu)$$

$$\ln\frac{k_{max}}{k_D} = \ln\frac{T\lambda_D}{e^2} = \ln\frac{4\pi nT\lambda_D}{4\pi ne^2} = \ln 4\pi n\lambda_D^3 = \ln\Lambda + \ln 4\pi$$

(7.A.30)

10% corrections to the upper and lower limits on the *k*-integration are possible but are not important because collisions are already a minor effect compared to collective effects for $n\lambda_D^3 \gg 1$. Hence, we approximately evaluate Eq.(7.A.30)

$$\hat{\mathbf{v}}_0 \cdot \mathbf{F}_D^s \approx -e_0^2 \omega_s^2 \ln\Lambda \int_{-1}^{1} d\mu\, \mu g_{\hat{\mathbf{k}}}^{s\prime}(v_0\mu) = F_D^s(v_0) \sim \frac{e_0^2 \omega_s^2}{v_{th,s}^2}\ln\Lambda \sim \frac{e_0^2}{\lambda_{Ds}^2}\ln\Lambda \quad (7.A.31)$$

We compare the result in Eq.(7.A.31) with the result in Eq.(7.A.24) for the radiation reaction force: $F_W \sim e^2/\lambda_D^2 \ll F_D \sim \ln\Lambda\, e^2/\lambda_D^2$. $F_D$ is dominant.

Example: Consider a Maxwellian velocity distribution function

$$g^s(u) = \frac{1}{\sqrt{2\pi}\, v_s}\exp\left(-\frac{u^2}{2v_s^2}\right) \quad \lambda_s \equiv \frac{v_s}{\omega_s} \quad v_s \equiv \sqrt{\frac{T_s}{m_s}} \quad (7.A.32)$$

Then the dynamic friction is

$$F_D^s(v_0) \approx -\frac{e_0^2}{\lambda_s^2}\ln\Lambda\, \frac{v_0}{v_s}\frac{1}{\sqrt{2\pi}}\int_{-1}^{1}d\mu\,\mu^2 \exp\left(-\frac{\mu^2}{2\mu_0^2}\right) \quad \mu_0 \equiv \frac{v_s}{v_0} \quad (7.A.33)$$

The integral in Eq.(7.A.33) can be evaluated analytically in two opposite limits:

$$\int_{-1}^{1} d\mu\,\mu^2 \exp\left(-\frac{\mu^2}{2\mu_0^2}\right) \to \int_{-1}^{1} d\mu\, \mu^2 = \frac{2}{3} \quad \text{for } v_0 \ll v_s,\ \mu_0 \gg 1$$

$$\to \mu_0^3 \int_{-\frac{1}{\mu_0}}^{\frac{1}{\mu_0}} dx\, x^2 \exp(-\tfrac{1}{2}x^2) \approx \mu_0^3 \int_{-\infty}^{\infty} dx\, x^2 \exp(-\tfrac{1}{2}x^2) = \sqrt{2\pi}\mu_0^3 \quad \text{for } v_0 \gg v_s,\ \mu_0 \ll 1$$

Thus, Eq.(7.A.33) is evaluated asymptotically as

$$F_D^s(v_0) \approx \frac{e_0^2}{\lambda_s^2}\ln\Lambda\left\{\frac{2}{3\sqrt{2\pi}}\frac{v_0}{v_s} \approx 0.25\frac{v_0}{v_s} \text{ for } v_0 \ll v_s,\ \left(\frac{v_s}{v_0}\right)^2 \text{ for } v_0 \gg v_s\right\} \quad (7.A.34)$$

Exercise: Sketch $F_D^s$ using Eq.(7.A.34)) and $F_W$ based on Eq.(7.A.24) as functions of $v_0$.

### *7.A.d Calculation of classical collisional resistivity*

In Eq.(7.A.34) we calculated the dynamic friction on a test particle as a function of the test particle speed $v_0$ due to its collisional interaction with the plasma for a Maxwellian electron distribution. Consider ions and electrons subject to an electric field, and the ions and electrons collisionally drag on one another. Newton's law for each particle is

$$\frac{d}{dt}m_s\mathbf{v} = e_s\mathbf{E} + \mathbf{F}_D^{s'} \tag{7.A.35}$$

For singly charged ions and a quasi-neutral plasma $n_e \approx n_i = n$, the fluid current is

$$\mathbf{j} = ne(\mathbf{u}_i - \mathbf{u}_e) \equiv ne\mathbf{u}_D \equiv \sigma\mathbf{E} \tag{7.A.36}$$

where we have introduced the conductivity $\sigma$ which will be evaluated in the following analysis, $\sigma = neu_D/E$ and $\mathbf{u}_D$ is a drift of the ions relative to the electrons. The ions are cold, and $u_D > 0$ is assumed small compared to the electron thermal speed so that the ions are essentially at rest compared to the electrons. Due to Newton's third law there is no drag force of the ion fluid on itself. The equation of motion for the ion fluid is

$$\frac{d}{dt}(n_i m_i u_i) = en_i E - |F_i^e| n_i \tag{7.A.37}$$

There are more electrons slower than the ions than faster, so that the ion drag on the electrons decelerates the ions. The equation of motion for the electron fluid is

$$\frac{d}{dt}(n_e m_e u_e) = -en_e E + |\langle F_e^i \rangle| n_e \tag{7.A.38}$$

and in consequence of Newton's third law, $|F_i^e| = |<F_e^i>|$. At steady state, $D/Dt = 0$ in both Eqs.(7.A.37) and (7.A.38). The force of the electrons on the ions at steady state if $u_D \ll v_e$ is

$$|F_D^s(v_0)| = \frac{2}{3}\frac{1}{\sqrt{2\pi}}\frac{v_0}{v_s}\frac{e_0^2}{\lambda_s^2}\ln\Lambda = \frac{2}{3}\frac{1}{\sqrt{2\pi}}\frac{u_D}{v_e}\frac{e^2}{\lambda_s^2}\ln\Lambda = eE \tag{7.A.39}$$

where $\Lambda \equiv n\lambda_e^3$, $v_e = \sqrt{T_e/m_e}$, and $\lambda_e \equiv \sqrt{T_e/4\pi ne^2}$. Hence, from Eq.(7.A.38-7.A.39) the conductivity is

$$\sigma = \frac{neu_D}{E} = \frac{3n\sqrt{2\pi}v_e\lambda_e^3}{2\lambda_e\ln\Lambda} = \frac{3\sqrt{2\pi}}{2}\frac{\Lambda}{\ln\Lambda}\omega_p \tag{7.A.40}$$

Definition: The resistivity is defined as

$$\eta \equiv \frac{1}{\sigma} = \frac{1}{\omega_p}\frac{1}{3}\sqrt{\frac{2}{\pi}}\frac{\ln\Lambda}{\Lambda} \tag{7.A.41}$$

We note that the conductivity is nearly independent of density except for $\ln\Lambda$,

$$\sigma = \frac{3\sqrt{2\pi}}{2}\sqrt{\frac{4\pi ne^2}{m_e}}\frac{nT_e^{3/2}}{(4\pi)^{3/2}n^{3/2}e^3}\frac{1}{\ln\Lambda} = \frac{3\sqrt{2}}{8\sqrt{\pi}}v_e\frac{T_e}{e^2}\frac{1}{\ln\Lambda} \tag{7.A.42}$$

### *7.A.e Definition of collision-dominated parameter regime*

We can introduce the concept of the collision frequency $\nu$ using the analogy

$$\dot{v}=-\frac{eE}{m}-\nu v=0 \Rightarrow 0=-\frac{eE}{m}-\nu\bar{v} \Rightarrow 0=-\frac{eE}{m}-\nu\bar{u}_D \Rightarrow \sigma=\frac{neu_D}{E}=\frac{ne^2}{m\nu}=\frac{\omega_p^2}{4\pi\nu_e} \quad (7.A.43)$$

From Eqs.(7.A.42) and (7.A.43) we can evaluate the collision frequency:

$$\sigma=\frac{\omega_p^2}{4\pi\nu_e} \Rightarrow \nu_e=\frac{\omega_p}{\Lambda}\ln\Lambda\frac{1}{6\pi\sqrt{2\pi}}=\frac{8}{3}\sqrt{\frac{\pi}{2}}\frac{n_e e^4 \ln\Lambda}{m_e^{1/2}T_e^{3/2}} \quad (7.A.44)$$

An alternate heuristic derivation of Eq.(7.A.44) is as follows:

$$\nu = n(\text{cross section}) v \ln\Lambda = n\left(\frac{e^2}{T}\right)^2 v\ln\Lambda = \frac{n_e e^4 \ln\Lambda}{m_e^{1/2}T_e^{3/2}} \quad (7.A.45)$$

where we have used the Rutherford cross-section in Eq.(7.A.45) and have recovered Eq.(7.A.44) except for the numerical factors out in front. The characteristic slowing-down time due to electron collisions is $\nu_e^{-1} \sim \Lambda\omega_p^{-1}$.

<u>Definition</u>: The characteristic mean-free-path is

$$\ell_{coll} \sim \frac{v_e}{\nu_e} \sim \frac{\Lambda}{\ln\Lambda}\frac{v_e}{\omega_p}=\frac{\Lambda}{\ln\Lambda}\lambda_D \quad (7.A.46)$$

The strong interaction distance (in the classical sense) is

$$r_0 = \frac{\lambda_D}{\Lambda} \approx \frac{e^2}{T}$$

The collisionless Vlasov approximation is so good because $\nu_e << \omega_p$ by $\ln\Lambda/\Lambda$.

### 7.A.f Anomalous dynamic friction due to, for example, instabilities

We recall the steady-state relation used to reduce Eq.(7.A.38):

$$eE = F_i^e < E_0 \equiv 0.22\frac{e^2}{\lambda_D^2}\ln\Lambda \quad (7.A.47)$$

for $u_D \leq v_e$. If $E > E_0$ there can be no balance, and the electrons can runaway because the electrons will continue to accelerate. $E_0$ is the runaway electric field limit.[45]

In Sec. 7.A.e we calculated collisional friction due to Coulomb field interactions between test particles and shielding clouds given a relative drift $u_D$ between ions and electrons. A relative drift can be the source of free energy that excites collective modes of instability, which in turn provide an anomalous friction that can relax the drift. Here we describe some of the instabilities that can arise as a function of $u_D$. For $u_D > v_e$ electron plasma waves are destabilized by a modified two-stream instability or Buneman instability.[46] At much smaller drift velocities, ion-acoustic waves can be destabilized. For $u_D > 5v_i$ ion-acoustic waves are unstable only if $T_e >> T_i$, else Landau damping can stabilize the modes. For velocities $u_D > c_s = (T_e/m_i)^{1/2}$ there can be a strong ion-acoustic instability if $T_e >> T_i$. The wave turbulence due to the instabilities can produce quasilinear diffusion and dynamic friction that affect the particle velocity distributions (see Sec. 6.E). As the velocity distributions evolve due to quasilinear diffusion (for example, heating), the Landau resonance effects and Coulomb collisions are changed. The self-consistent evolution of the turbulent fields and the velocity distributions are closely coupled.

---

[45] H. Dreicer, Phys. Rev., 115, 238 (1959).
[46] O. Buneman, Phys. Rev. Lett., 10, 285 (1963).

Consider the following model problem. Assume there is a given relative drift $u_D$ and a driving electric field at $t=0$. At $t=0$ let $E=0$ and turn on the electric field until it exceeds the runaway threshold in magnitude. An approximate power balance relation for the electrons can be expressed as

$$eEu_D = \frac{\partial}{\partial t}\left[\frac{1}{2}m\left(u_D^2 + v_e^2\right)\right] \approx \frac{\partial}{\partial t}\left[\frac{1}{2}m2u_D^2\right] \approx \frac{\partial}{\partial t}\left[mu_D^2\right]$$

$$\Rightarrow eE = 2m\dot{u}_D \quad \Rightarrow u_D \approx u_D(0) + \frac{eEt}{2m}$$

(7.A.47)

We have assumed that collisional friction between the electrons and ions deposits energy into electron thermal energy. We further assume that the thermal and directed energy grow in equal amounts in Eq.(7.A.47). There is no steady current, only acceleration. In this model the ions do not do much. The electron heating is dominantly parallel to the drift, and $T_\parallel$ grows. The heated distribution function develops an anisotropy ($T_\parallel > T_\perp$) that can excite an electromagnetic instability like the Weibel instability.[47]

**7.B Extension of test-particle theory to many-particle phenomena**

*7.B.a Incorporating dynamic friction into Vlasov quasilinear theory to extend the Vlasov equation to the Boltzmann equation*

Here we extend our earlier treatment of quasi-linear diffusion (Sec. 6.E) to include the dynamic friction also due to collisions. Recall the Vlasov equation extended to include velocity diffusion:

$$\frac{\partial}{\partial t}f(\mathbf{r},\mathbf{v};t) + \frac{\partial}{\partial \mathbf{r}}(\dot{\mathbf{r}}f) + \frac{\partial}{\partial \mathbf{v}}(\dot{\mathbf{v}}f) = 0 \Rightarrow \frac{\partial}{\partial \mathbf{v}}\cdot\left(\mathbf{D}(\mathbf{v})\cdot\frac{\partial f}{\partial \mathbf{v}}\right)$$

(7.B.1)

This analysis will ignore large-angle (Boltzmann) collisions and quantum effects. The diffusion tensor $\mathbf{D}(\mathbf{v})$ was derived in Sec. 6 based either on a random-walk argument or by iterating the Vlasov equation. In a Coulomb model,

$$\dot{\mathbf{v}} = \frac{e}{m}\mathbf{E} \quad \rightarrow \quad \mathbf{D}(\mathbf{v}) = \left(\frac{e}{m}\right)^2 \int_0^\infty d\tau \left\langle \mathbf{E}(\mathbf{r}^t,t)\mathbf{E}(\mathbf{r}^{t-\tau},t-\tau)\right\rangle$$

(7.B.2)

The electric field in the theory of quasilinear diffusion is the Vlasov electric field associated with macroscopic waves and collective effects. In collisional diffusion the electric field is the "noise" electric field associated with a microscopic picture. Of course, the particles just respond to the total electric field. The relative field energy in microscopic noise fields is $W/nT \approx O(\Lambda^{-1}) \sim 10^{-6}$, for example. The relative field energy in saturated turbulent fields might be $W/nT \sim 10^{-2}$, for example. We note that one does *not* put the same electric field in both the diffusion tensor and in the acceleration term on the left side of Eq.(7.B.1), which would double count its effects.

---

[47] E.S. Weibel, Phys. Rev. Lett. **2**, 83 (1959).

### *7.B.b Correlation functions, Bogoliubov approximation, and irreversibility*

We want to calculate the two-point auto-correlation function for the electric field used in **D(v)** in Eq.(7.B.2). It is useful from a pedagogic perspective to reduce the two-point auto-correlation tensor in Eq.(7.B.2) to a scalar with respect to the electric potential:

$$\langle \phi(\mathbf{r},t)\phi(\mathbf{r}',t') \rangle = \langle \phi(\mathbf{r},t)\phi(\mathbf{r}-\mathbf{s},t-\tau) \rangle \tag{7.B.3}$$

What is meant by the average in Eq.(7.B.3)? For $s \gg \lambda_D$ we are looking at effects due to waves (collective effects). If $s \ll \lambda_D$, then discrete particle effects are dominant. Similarly, $\tau \gg 1/\omega_p$ corresponds to waves, while $\tau \gg 1/\omega_p$ is dominated by discrete particle effects.

Definition: The two-point auto-correlation function for the electric potential is

$$C_\phi \equiv C_\phi(\mathbf{s},\tau) \equiv \langle \phi(\mathbf{r},t)\phi(\mathbf{r}-\mathbf{s},t-\tau) \rangle_{\mathbf{r},t} \tag{7.B.4}$$

and $C_\phi(\mathbf{s}=0,\tau=0) \equiv \langle |\phi(\mathbf{r},t)|^2 \rangle$ is related to the field energy density.

Averaging in Eqs.(7.B.3) and (7.B.4) over position only makes sense in a uniform plasma, $L \gg \lambda_D$. Time averaging makes sense only if the system is stationary in time, $T \gg \omega_p^{-1}$. However, if the system is uniform and stationary on the microscopic scales, time and space averaging over the fast scales is allowed while accommodating variations in space and time that are much slower than the microscopic scales. We assume that the fast and slow time scales are well separated (Bogoliubov approximation).

Definition: (Truncated Fourier transforms) We define the following truncated Fourier transforms

$$\phi(\mathbf{k},\omega;\mathbf{r},t;V,T) \equiv \int_V d^3\mathbf{s} \int_{-T/2}^{T/2} d\tau\, \phi(\mathbf{r}+\mathbf{s},t+\tau)\exp(-i\mathbf{k}\cdot\mathbf{s}+i\omega\tau) \tag{7.B.5}$$

Later we will let $V,T \to \infty$ and will average with respect to **r** and *t*. We construct

$$\lim_{V,T\to\infty}\left\langle\phi(\mathbf{k},\omega;\mathbf{r},t;V,T)\phi^*(\mathbf{k}',\omega';\mathbf{r},t;V,T)\right\rangle_{\mathbf{r},t}$$

$$=\lim_{V,T\to\infty}\left\langle\int d^3\mathbf{s}\int d^3\mathbf{s}'\int d\tau\int d\tau'\,\phi(\mathbf{r}+\mathbf{s},t+\tau)\phi^*(\mathbf{r}+\mathbf{s}',t+\tau')\exp(-i\mathbf{k}\cdot\mathbf{s}+i\omega\tau)\exp(+i\mathbf{k}'\cdot\mathbf{s}'-i\omega'\tau')\right\rangle_{\mathbf{r},t}$$

$$=\lim_{V,T\to\infty}\int d^3\mathbf{s}\int d^3\mathbf{s}'\int d\tau\int d\tau'\,\left\langle\phi(\mathbf{r}+\mathbf{s},t+\tau)\phi^*(\mathbf{r}+\mathbf{s}',t+\tau')\right\rangle_{\mathbf{r},t}\exp(-i\mathbf{k}\cdot\mathbf{s}+i\omega\tau)\exp(+i\mathbf{k}'\cdot\mathbf{s}'-i\omega'\tau')$$

$$=\lim_{V,T\to\infty}\int d^3\mathbf{s}\int d^3\mathbf{s}'\int d\tau\int d\tau'\,C_\phi(\mathbf{s}-\mathbf{s}',\tau-\tau')\exp(-i\mathbf{k}\cdot\mathbf{s}+i\omega\tau)\exp(+i\mathbf{k}'\cdot\mathbf{s}'-i\omega'\tau')$$

$$=\lim_{V,T\to\infty}\int d^3\mathbf{s}\int d^3\mathbf{s}''\int d\tau\int d\tau''\,C_\phi(\mathbf{s}'',\tau'')\exp(-i\mathbf{k}\cdot\mathbf{s}+i\omega\tau)\exp(+i\mathbf{k}'\cdot(\mathbf{s}-\mathbf{s}'')-i\omega'(\tau-\tau''))$$

(7.B.6)

Note that because we let $V,T\to\infty$ there is no problem with the limits of integration when changing the variables. We next make use of the following identities

$$\int d^3s\,\exp(-i\mathbf{k}\cdot\mathbf{s}+i\mathbf{k}'\cdot\mathbf{s})=(2\pi)^3\delta(\mathbf{k}-\mathbf{k}')\qquad\int d\tau\,\exp(i\omega\tau-i\omega'\tau)=2\pi\delta(\omega-\omega')$$

to reduce Eq.(7.B.6) to

$$\lim_{V,T\to\infty}\left\langle\phi(\mathbf{k},\omega;\mathbf{r},t;V,T)\phi^*(\mathbf{k}',\omega';\mathbf{r},t;V,T)\right\rangle_{\mathbf{r},t}$$
$$=(2\pi)^4\delta(\mathbf{k}-\mathbf{k}')\delta(\omega-\omega')\int d^3s\int d\tau\,C_\phi(\mathbf{s},\tau)\exp(-i\mathbf{k}\cdot\mathbf{s}+i\omega\tau) \qquad (7.B.7)$$
$$=(2\pi)^4\delta(\mathbf{k}-\mathbf{k}')\delta(\omega-\omega')S_\phi(\mathbf{k},\omega)$$

<u>Definition</u>: In Eq.(7.B.7) we have identified the spectral density $S_\phi$ which is the Fourier transform of the auto-correlation function for the electric potential

$$S_\phi(\mathbf{k},\omega)=F_{\mathbf{s},\tau}\!\left(C_\phi(\mathbf{s},\tau)\right) \qquad (7.B.8)$$

Equation (7.B.8) is a statement of the Wiener-Khinchin-Einstein theorem. The implication of (7.B.8) is that space-time averaging is equivalent to phase-space (ergodic) averaging.

The analysis leading to the results in this section employs the linear plasma response and stationarity assumptions. This requires a stable plasma (an unstable plasma violates the stationarity assumption). Large-amplitude phenomena and trapping invalidate the analysis. Moreover, we have also assumed that the plasma is uniform. Consider the shielded field (Coulomb model) of a collection of bare particles:

$$\phi(\mathbf{k},\omega)=\sum_i\phi^i(\mathbf{k},\omega)=\frac{1}{\varepsilon(\mathbf{k},\omega)}\frac{4\pi}{k^2}e_i 2\pi\delta(\omega-\mathbf{k}\cdot\mathbf{v}_i)\exp(-i\mathbf{k}\cdot\mathbf{r}_i) \qquad (7.B.9)$$

The rigorous methods to calculate (7.B.9) are difficult. We have taken an intuitive test-particle point of view. From Eq.(7.B.9) we calculate

$$\langle\phi(\mathbf{k},\omega)\phi^*(\mathbf{k}',\omega')\rangle$$
$$=\sum_i\sum_j\frac{1}{\varepsilon^*(\mathbf{k}',\omega')}\frac{1}{\varepsilon(\mathbf{k},\omega)}\frac{4\pi}{k'^2}\frac{4\pi}{k^2}e_je_i\langle(2\pi)^2\delta(\omega'-\mathbf{k}'\cdot\mathbf{v}_j)\exp(i\mathbf{k}'\cdot\mathbf{r}_j)\delta(\omega-\mathbf{k}\cdot\mathbf{v}_i)\exp(-i\mathbf{k}\cdot\mathbf{r}_i)\rangle$$

(7.B.10)

We make a critical assumption in (7.B.10) that the particle positions are statistically independent of one another. However, the particle position and its corresponding shielding cloud are correlated with one another.

Theorem (Ergodic Theorem)
$$\langle\exp(-i\mathbf{k}\cdot\mathbf{r}_i)\exp(i\mathbf{k}'\cdot\mathbf{r}_j)\rangle=\langle\exp(-i\mathbf{k}\cdot\mathbf{r}_i)\rangle\langle\exp(i\mathbf{k}'\cdot\mathbf{r}_j)\rangle=0 \quad (7.B.11)$$
unless $\mathbf{r}_i=\mathbf{r}_j$.

So far the irreversibility associated with the Bogoliubov hypothesis has been buried. The goal of the analysis here is the derivation of the diffusion tensor and the kinetic equation including dynamic friction and diffusion associated with the shielded plasma fluctuations linked to the test particles. The resulting diffusion equation implies irreversibility (Sec. 6.E). To continue the derivation we consider disturbances that are turned on at $t=0$ and restrict the Fourier transforms to the positive $\omega$ half-plane. From Eqs.(7.B.7), (7.B.9), (7.B.10), and (7.B.11) we obtain

$$\langle\phi(\mathbf{k},\omega)\phi^*(\mathbf{k}',\omega')\rangle$$
$$=\sum_i\frac{1}{\varepsilon^*(\mathbf{k}',\omega')}\frac{1}{\varepsilon(\mathbf{k},\omega)}\frac{4\pi}{k^2}\frac{4\pi}{k'^2}e_i^2\langle(2\pi)^2\delta(\omega'-\mathbf{k}'\cdot\mathbf{v}_i)\delta(\omega-\mathbf{k}\cdot\mathbf{v}_i)\rangle\langle\exp(-i(\mathbf{k}-\mathbf{k}')\cdot\mathbf{r}_i)\rangle$$
$$=\sum_i\frac{1}{\varepsilon^*(\mathbf{k}',\omega')}\frac{1}{\varepsilon(\mathbf{k},\omega)}\frac{4\pi}{k^2}\frac{4\pi}{k'^2}e_i^2\langle(2\pi)^2\delta(\omega'-\mathbf{k}'\cdot\mathbf{v}_i)\delta(\omega-\mathbf{k}\cdot\mathbf{v}_i)\rangle\frac{1}{V}\int d^3\mathbf{r}\exp(-i(\mathbf{k}-\mathbf{k}')\cdot\mathbf{r}_i)$$
$$=\sum_i\frac{1}{\varepsilon^*(\mathbf{k}',\omega')}\frac{1}{\varepsilon(\mathbf{k},\omega)}\frac{4\pi}{k^2}\frac{4\pi}{k'^2}e_i^2\langle(2\pi)^2\delta(\omega'-\mathbf{k}'\cdot\mathbf{v}_i)\delta(\omega-\mathbf{k}\cdot\mathbf{v}_i)\rangle\frac{(2\pi)^3}{V}\delta(\mathbf{k}-\mathbf{k}')$$
$$=\sum_i\frac{1}{\varepsilon^*(\mathbf{k},\omega')}\frac{1}{\varepsilon(\mathbf{k},\omega)}\left(\frac{4\pi}{k^2}\right)^2 e_i^2(2\pi)^2\delta(\omega'-\omega)\langle\delta(\omega-\mathbf{k}\cdot\mathbf{v}_i)\rangle\frac{(2\pi)^3}{V}\delta(\mathbf{k}-\mathbf{k}')$$
$$=\sum_i\frac{1}{|\varepsilon(\mathbf{k},\omega)|^2}\left(\frac{4\pi}{k^2}\right)^2 e_i^2(2\pi)^2\langle\delta(\omega-\mathbf{k}\cdot\mathbf{v}_i)\rangle\frac{(2\pi)^3}{V}\delta(\omega'-\omega)\delta(\mathbf{k}-\mathbf{k}')$$
$$=(2\pi)^4\delta(\mathbf{k}-\mathbf{k}')\delta(\omega-\omega')S_\phi(\mathbf{k},\omega)$$

(7.B.12)

Hence,

$$S_\phi(\mathbf{k},\omega) = \frac{1}{|\varepsilon(\mathbf{k},\omega)|^2}\left(\frac{4\pi}{k^2}\right)^2 \frac{2\pi}{V}\sum_i e_i^2 \langle\delta(\omega-\mathbf{k}\cdot\mathbf{v}_i)\rangle$$

$$= \frac{1}{|\varepsilon(\mathbf{k},\omega)|^2}\left(\frac{4\pi}{k^2}\right)^2 2\pi\sum_s n_s e_s^2 \int d^3\mathbf{v}\, g_s(\mathbf{v})\delta(\omega-\mathbf{k}\cdot\mathbf{v}) \quad (7.B.13)$$

$$= \sum_s n_s \left|\frac{4\pi e_s}{k^2\varepsilon(\mathbf{k},\omega)}\right|^2 \int d^3\mathbf{v}\, g_s(\mathbf{v}) 2\pi\delta(\omega-\mathbf{k}\cdot\mathbf{v})$$

If there is no shielding, i.e., no polarization correlation, then $\varepsilon=1$ in Eq.(7.B.13).

We recall the expression for the velocity-space diffusion Eq.(7.B.2) which we can relate to the analysis in Eqs.(7.B.3-7.B.13):

$$\mathbf{D}(\mathbf{v}) = \left(\frac{e}{m}\right)^2 \int_0^\infty d\tau \langle \mathbf{E}(\mathbf{r}^t,t)\mathbf{E}(\mathbf{r}^{t-\tau},t-\tau)\rangle_{\mathbf{r}^t,t} = \left(\frac{e}{m}\right)^2 \int_0^\infty d\tau \langle C_\mathbf{E}(\mathbf{r}^t-\mathbf{r}^{t-\tau},\tau)\rangle_{\mathbf{r}^t,t}$$

$$= \left(\frac{e}{m}\right)^2 \int_0^\infty d\tau \langle C_\mathbf{E}(\mathbf{s}(\tau)=\mathbf{v}\tau,\tau)\rangle_{\mathbf{r}^t,t} = \left(\frac{e}{m}\right)^2 \int_0^\infty d\tau \int \frac{d^3\mathbf{k}}{(2\pi)^3}\int \frac{d\omega}{2\pi}\exp\left(i(\mathbf{k}\cdot\mathbf{s}-\omega\tau)\right)\mathbf{k}\mathbf{k}S_\phi(k,\omega)$$

$$= \left(\frac{e}{m}\right)^2 \int_0^\infty d\tau \int \frac{d^3\mathbf{k}}{(2\pi)^3}\int \frac{d\omega}{2\pi}\exp\left(i(\mathbf{k}\cdot\mathbf{v}-\omega)\tau\right)\mathbf{k}\mathbf{k}S_\phi(k,\omega)$$

$$= \left(\frac{e}{m}\right)^2 \int \frac{d^3\mathbf{k}}{(2\pi)^3}\int \frac{d\omega}{2\pi}\mathbf{k}\mathbf{k}S_\phi(k,\omega)\left(\pi\delta(\mathbf{k}\cdot\mathbf{v}-\omega)+\frac{i}{\mathbf{k}\cdot\mathbf{v}-\omega}\right)$$

$$\to \left(\frac{e}{m}\right)^2 \int \frac{d^3\mathbf{k}}{(2\pi)^3}\int \frac{d\omega}{2\pi}\mathbf{k}\mathbf{k}S_\phi(k,\omega)\pi\delta(\mathbf{k}\cdot\mathbf{v}-\omega)$$

(7.B.14)

owing to the symmetries of the integrals and integrand in (7.B.14). We note that the field fluctuations arise from the test particles satisfying the resonance $\omega=\mathbf{k}\cdot\mathbf{v}'$, and the fluctuations then interact with the other particles satisfying the resonance $\omega=\mathbf{k}\cdot\mathbf{v}$. Hence, $\mathbf{k}\cdot(\mathbf{v}-\mathbf{v}')=0$. Given that the spectral density in Eq.(7.B.13) is additive over species, so also is the diffusion tensor, which we rewrite for the diffusion of the momenta in terms of the un-normalized distribution function $f_s$:

$$\mathbf{D}_s^{s'}(\mathbf{p}) = e_s^2 \int \frac{d^3\mathbf{k}}{(2\pi)^3}\int \frac{d\omega}{2\pi}S_\phi^{s'}(k,\omega)\mathbf{k}\mathbf{k}\pi\delta(\mathbf{k}\cdot\mathbf{v}-\omega)$$

$$= \int \frac{d^3\mathbf{k}}{(2\pi)^3}\left|\frac{4\pi e_s e_{s'}}{k^2\varepsilon(\mathbf{k},\omega)}\right|^2 \int d^3\mathbf{p}'\,\pi\delta(\mathbf{k}\cdot\mathbf{v}-\mathbf{k}\cdot\mathbf{v}')\mathbf{k}\mathbf{k}f_s(p')$$

(7.B.15)

Only the $1/|\varepsilon|^2$ evidences the plasma shielding. We also note that $S_\phi$ and **D** diverge at large $k$ where linear theory breaks down.

### 7.B.c Derivation of the Lenard-Balescu equation

In this section we derive the extension of the Vlasov equation to a collisional kinetic equation that includes both dynamic friction and velocity-space diffusion. We recall the expression for the dynamic friction

$$\mathbf{F}_s(\mathbf{v}) = e_s^2 \int \frac{d^3\mathbf{k}}{(2\pi)^2} \frac{4\pi}{k^2} \mathbf{k}\, \text{Im}\,\varepsilon^{-1}(\mathbf{k},\mathbf{k}\cdot\mathbf{v}) = -e_s^2 \int \frac{d^3\mathbf{k}}{(2\pi)^2} \frac{4\pi}{k^2} \mathbf{k}\, \frac{\text{Im}\,\varepsilon}{|\varepsilon|^2} \qquad (7.B.16)$$

and we can sum over species assuming that the dynamic friction is simply additive in the other particle species. We can express the dynamic friction in terms of the momentum and evaluate the imaginary part of the dielectric response in terms of the species distribution functions as follows:

$$\mathbf{F}_s^{s'}(\mathbf{p}) = \int \frac{d^3\mathbf{k}}{(2\pi)^3} \left| \frac{4\pi e_s e_{s'}}{k^2 \varepsilon(\mathbf{k},\mathbf{k}\cdot\mathbf{v})} \right|^2 \int d^3\mathbf{p}'\, \pi \delta(\mathbf{k}\cdot\mathbf{v}-\mathbf{k}\cdot\mathbf{v}')\mathbf{k}\mathbf{k}\cdot\frac{\partial}{\partial\mathbf{p}'} f_{s'}(\mathbf{p}') \qquad (7.B.17)$$

We now return to the Vlasov equation including friction Eq.(7.B.17) and diffusion Eq.(7.B.15) terms on the right side.

Theorem: The Lenard-Balescu equation is

$$\frac{d}{dt} f_s(\mathbf{p};t) = -\frac{\partial}{\partial \mathbf{p}}\cdot\left(\mathbf{F}_s(\mathbf{p})f_s\right) + \frac{\partial}{\partial \mathbf{p}}\cdot\left(\mathbf{D}_s(\mathbf{p})\cdot\frac{\partial}{\partial \mathbf{p}} f_s\right) = -\frac{\partial}{\partial \mathbf{p}}\cdot\mathbf{J}_s(\mathbf{p}), \quad \mathbf{F}_s \equiv \sum_{s'}\mathbf{F}_s^{s'}, \quad \mathbf{D}_s \equiv \sum_{s'}\mathbf{D}_s^{s'}$$

(7.B.18)

Definition: $\mathbf{J}_s$ is defined as a flux density in momentum space:

$$\mathbf{J}_s(\mathbf{p}) \equiv \mathbf{F}_s(\mathbf{p}) f_s - \mathbf{D}_s(\mathbf{p})\cdot\frac{\partial}{\partial \mathbf{p}} f_s \qquad (7.B.19)$$

The second term on the right side of Eq.(7.B.19) is Fick's law in momentum space.

Theorem: From Eqs.(7.B.15-7.B.19) one derives

$$\mathbf{J}_s^{s'}(\mathbf{p}) = \int \frac{d^3\mathbf{k}}{(2\pi)^3} \left| \frac{4\pi e_s e_{s'}}{k^2 \varepsilon(\mathbf{k},\mathbf{k}\cdot\mathbf{v})} \right|^2 \int d^3\mathbf{p}'\, \pi \delta(\mathbf{k}\cdot\mathbf{v}-\mathbf{k}\cdot\mathbf{v}')\mathbf{k}\mathbf{k}\cdot\left(\frac{\partial}{\partial\mathbf{p}'}-\frac{\partial}{\partial\mathbf{p}}\right) f_s(\mathbf{p}) f_{s'}(\mathbf{p}')$$

$$\equiv \int d^3\mathbf{p}'\, \mathbf{Q}_{ss'}(\mathbf{v},\mathbf{v}')\cdot\left(\frac{\partial}{\partial\mathbf{p}'}-\frac{\partial}{\partial\mathbf{p}'}\right) f_s(\mathbf{p}) f_{s'}(\mathbf{p}')$$

(7.B.20)

where

$$\mathbf{Q}_{ss'}(\mathbf{v},\mathbf{v}') \equiv \int \frac{d^3\mathbf{k}}{(2\pi)^3} \mathbf{k}\mathbf{k}\left(\frac{4\pi e_s e_{s'}}{k^2}\right)^2 \int \frac{d\omega}{2\pi}(2\pi)^2 \frac{\delta(\omega-\mathbf{k}\cdot\mathbf{v})\delta(\omega-\mathbf{k}\cdot\mathbf{v}')}{|\varepsilon(\mathbf{k},\omega)|^2} \qquad (7.B.21)$$

and the collisional flux on the right side of Eq.(7.B.20) is in the Landau form.[48] We note that $\mathbf{Q}_{ss'}$ is symmetric in $\mathbf{v}$ and $\mathbf{v}'$ owing to the symmetry of the effective force between two particles, and is also symmetric in species $s$ and $s'$. However, the difference in momentum derivatives in Eq.(7.B.20) does not identically vanish.

We return to the analysis leading to Eqs.(7.B.12) and (7.B.13) to provide some additional details. Consider the auto-correlation function for the electric field used in the diffusion tensor in Eq.(7.B.14):

$$\left\langle \mathbf{E}(\mathbf{r}^t,t)\mathbf{E}(\mathbf{r}^{t-\tau},t-\tau) \right\rangle = \int \frac{d^3\mathbf{k}}{(2\pi)^3} \int \frac{d\omega}{2\pi} \exp\left(i(\mathbf{k}\cdot\mathbf{s}-\omega\tau)\right) \mathbf{k}\mathbf{k} S_\phi(k,\omega) \qquad (7.B.22)$$

From Eq.(7.B.22) it follows that

$$\left\langle \mathbf{E}(\mathbf{r}^t,t)^2 \right\rangle_{\mathbf{r},t} = \int \frac{d^3\mathbf{k}}{(2\pi)^3} \int \frac{d\omega}{2\pi} \mathbf{k}\mathbf{k} S_\phi(\mathbf{k},\omega) \equiv \int \frac{d^3\mathbf{k}}{(2\pi)^3} \mathbf{k}\mathbf{k} S_\phi(\mathbf{k})$$

$$S_\phi(\mathbf{k},\omega) = \frac{1}{|\varepsilon(\mathbf{k},\omega)|^2} \left(\frac{4\pi}{k^2}\right)^2 2\pi \sum_s e_s^2 \int d^3\mathbf{v}\, f_s(\mathbf{v})\delta(\omega-\mathbf{k}\cdot\mathbf{v})$$

(7.B.23)

Let us discuss the behavior of $S_\phi$. We note that $|\varepsilon|^2 = \varepsilon'(\mathbf{k},\omega)^2 + \varepsilon''(\mathbf{k},\omega)^2$. For $k > k_D$, the linear normal modes are not weakly damped in general; and $\varepsilon = 1 + O(1/k^2\lambda_D^2)$. For $k < k_D$, there are weakly damped, linear normal modes such that $\varepsilon'(\mathbf{k},\omega_k) \approx 0$, $\varepsilon''(\mathbf{k},\omega_k)$ is small; and the growth rate or damping rate is deduced from $\gamma_\mathbf{k} = -\varepsilon''/\frac{\partial \varepsilon'}{\partial \omega}$. However, these formulae are only good for a stable plasma: so $\gamma_\mathbf{k} \leq 0$. In general,

$$|\varepsilon|^2 = \left[\varepsilon'(\mathbf{k},\omega) + (\omega-\omega_k)\frac{\partial \varepsilon'}{\partial \omega}\bigg|_{\omega_k} + ...\right]^2 + \gamma_\mathbf{k}^2 \left|\frac{\partial \varepsilon'}{\partial \omega}\bigg|_{\omega_k}\right|^2 \approx \left|\frac{\partial \varepsilon'}{\partial \omega}\bigg|_{\omega_k}\right|^2 \left((\omega-\omega_k)^2 + \gamma_\mathbf{k}^2\right)$$

$$\rightarrow \frac{1}{|\varepsilon|^2} \approx \frac{1}{\left|\frac{\partial \varepsilon'}{\partial \omega}\bigg|_{\omega_k}\right|^2 \left((\omega-\omega_k)^2 + \gamma_\mathbf{k}^2\right)}$$

(7.B.24)

For velocity distributions that are Gaussian we can sketch $S_\phi$ based on Eqs.(7.B.23) and (7.B.24), and the foregoing discussion. For $k > k_D$, $S_\phi(\omega)$ vs. $\omega$ for fixed $k$ is approximately a Gaussian centered at $\omega=0$. For $k < k_D$, $S_\phi(\omega)$ vs. $\omega$ for fixed $k$ is approximately a Gaussian centered at $\omega=0$ for small $\omega$ with Lorentzian peaks at $\pm\omega_\mathbf{k}$, and the width of the peaks scales as $\gamma_\mathbf{k}$.

From Eqs.(7.B.23) and (7.B.24) we can derive

---

[48] L.D. Landau, Zh. Eksper. i Teoret. Fiz. **7**, 203 (1937)

$$S_\phi(\mathbf{k}) = \int \frac{d\omega}{2\pi} S_\phi(\mathbf{k},\omega) \approx \int \frac{d\omega}{2\pi} \frac{1}{|\varepsilon(\mathbf{k},\omega)|^2} \left(\frac{4\pi}{k^2}\right)^2 2\pi \sum_s e_s^2 \int d^3\mathbf{v}\, f_s(\mathbf{v}) \delta(\omega_\mathbf{k} - \mathbf{k}\cdot\mathbf{v})$$

$$= \frac{\pi}{|\gamma_k|} \frac{1}{\left(\left.\frac{\partial \varepsilon}{\partial \omega}\right|_{\omega_\mathbf{k}}\right)^2} \left(\frac{4\pi}{k^2}\right)^2 \sum_s e_s^2 \int d^3\mathbf{v}\, f_s(\mathbf{v}) \delta(\omega_\mathbf{k} - \mathbf{k}\cdot\mathbf{v}) \quad (7.B.25)$$

With the use of Eqs.(7.B.23) and (7.B.25) we can evaluate the wave energy and its time derivative. Recall that

$$W(\mathbf{k}) = \frac{\omega_\mathbf{k}^\ell}{4\pi} \frac{\partial \varepsilon'}{\partial \omega} k^2 S_\phi(\mathbf{k})$$

$$\dot{W}(\mathbf{k}) = -|2\gamma_\mathbf{k}| W(\mathbf{k}) = -\frac{4\pi}{k^2} \frac{|\omega_\mathbf{k}^\ell|}{\left|\frac{\partial \varepsilon'}{\partial \omega}\right|_{\omega_\mathbf{k}^\ell}} \sum_s e_s^2 \int d^3\mathbf{v}\, 2\pi \delta(\omega_\mathbf{k}^\ell - \mathbf{k}\cdot\mathbf{v}) n_s g_s(\mathbf{v}) \quad (7.B.26)$$

The correlation energy density is the electric field energy density minus the electric field energy of the vacuum ($\varepsilon=1$), which scales as $nT/\Lambda$ times a number. Thus, the Vlasov approximation is very good if $\Lambda \gg 1$ because particle-particle correlations can be ignored. However, we note that as $\gamma_k \to 0$ from below ($\gamma_k < 0$), correlations become very important; and the approximations in our derivations break down.

### 7.B.d Consideration of large-angle collisions and the Landau equation

We now focus on large-angle collisions, which corresponds to $k > k_D$. We ignore $k < k_D$ in

$$\mathbf{D}_s^{s'}(\mathbf{p}) = \int d^3\mathbf{p}'\, f_{s'}(\mathbf{p}') \mathbf{Q}_{ss'}(\mathbf{v},\mathbf{v}') \quad (7.B.27)$$

where

$$\mathbf{Q}_{ss'}(\mathbf{v},\mathbf{v}') \equiv \int \frac{d^3\mathbf{k}}{(2\pi)^3} \mathbf{k}\mathbf{k} \left(\frac{4\pi e_s e_{s'}}{k^2 \varepsilon(\mathbf{k},\mathbf{k}\cdot\mathbf{v})}\right)^2 \pi \delta(\mathbf{k}\cdot\mathbf{w}), \quad \mathbf{w} \equiv \mathbf{v} - \mathbf{v}' \quad (7.B.28)$$

For $k > k_D$ $\epsilon \to 1$, and we integrate the right side of Eq.(7.B.28) from $k_D$ to $k_{max}$:

$$\mathbf{Q}_{ss'}(\mathbf{v},\mathbf{v}') = 2(e_s e_{s'})^2 \int d^3\mathbf{k}\, \frac{\mathbf{k}\mathbf{k}}{k^4} \delta(\mathbf{k}\cdot\mathbf{w}) \quad (7.B.29)$$

We note that $\mathbf{w}\cdot\mathbf{Q} = 0$ and $\delta(k_\| w) = \delta(k_\|)/w$ where $k_\|$ is parallel to $\mathbf{w}$. Thus, we can express $\mathbf{Q}(\mathbf{w}) = Q(\vec{\mathbf{I}} - \hat{\mathbf{w}}\hat{\mathbf{w}})$ and $\mathrm{Tr}\,\mathbf{Q}(\mathbf{w}) = 2\,Q(w)$. Given this representation of $\mathbf{Q}(\mathbf{w})$ we can evaluate Eq.(7.B.29)

$$\mathbf{Q}(\mathbf{w}) = Q(w)\left(\ddot{\mathbf{I}} - \hat{\mathbf{w}}\hat{\mathbf{w}}\right)$$

$$Q \approx \tfrac{1}{2} 2 e_s^2 e_{s'}^2 \int dk_{\parallel} 2\pi k_{\perp} dk_{\perp} \frac{1}{k_{\perp}^2 + k_{\parallel}^2} \frac{\delta(k_{\parallel})}{w} \to \frac{2\pi e_s^2 e_{s'}^2}{w} \int_{k_D}^{k_{max}} \frac{dk_{\perp}}{k_{\perp}} = \frac{2\pi e_s^2 e_{s'}^2}{w} \ln \frac{k_{max}}{k_D} = \frac{2\pi e_s^2 e_{s'}^2}{w} \ln \Lambda$$
(7.B.30)

Note: The following identity is useful in dealing with the tensor $\mathbf{Q}(\mathbf{w}) = Q(\ddot{\mathbf{I}} - \hat{\mathbf{w}}\hat{\mathbf{w}})$,

$$\ddot{\mathbf{I}} - \hat{\mathbf{w}}\hat{\mathbf{w}} = w \frac{\partial}{\partial \mathbf{w}} \frac{\partial}{\partial \mathbf{w}} |\mathbf{w}|$$
(7.B.31)

Given $\mathbf{Q}$ we can evaluate $\mathbf{D}$ in Eq.(7.B.27) and the dynamic friction

$$\mathbf{F}_s^{s'}(\mathbf{p}) = \int d^3\mathbf{p}' \mathbf{Q}_{ss'}(\mathbf{v},\mathbf{v}') \cdot \frac{\partial}{\partial \mathbf{p}'} f_s(\mathbf{p}')$$
(7.B.32)

If the scatterers are isotropic, i.e., if $f_{s'}$ is isotropic, then $\mathbf{D}$ only depends on the vector information in $\mathbf{p}$: $\mathbf{D} = D_1(v)\mathbf{I} + D_2(v)\hat{\mathbf{v}}\hat{\mathbf{v}}$. We take traces and work with scalar equations.

<u>Theorem</u>: For $f_{s'}$ a Maxwellian then Eq.(7.B.27) yields

$$\mathbf{D}^{s'} = \frac{\sqrt{2\pi} n_{s'} e_s^2 e_{s'}^2 \ln\Lambda}{v_{s'}} \left[ \frac{\text{erf}(x)}{x}\left(\mathbf{I} - \hat{\mathbf{v}}\hat{\mathbf{v}}\right) + \frac{1}{2x}\frac{d}{dx}\left(\frac{\text{erf}(x)}{x}\right)\left(\mathbf{I} - 3\hat{\mathbf{v}}\hat{\mathbf{v}}\right) \right]$$
(7.B.33)

where $x \equiv v/\sqrt{2}v_{s'}$, $v_{s'} \equiv \sqrt{T_{s'}/m_{s'}}$, and

$$\text{erf}(x) \equiv \frac{2}{\sqrt{\pi}} \int_0^x dt\, e^{-t^2} \to \frac{2}{\sqrt{\pi}}\left(x - \frac{x^3}{3} + ...\right) \quad \text{for } x \ll 1$$

$$\to 1 - \frac{e^{-x}}{\sqrt{\pi x}}(1 + ...) \quad \text{for } x \gg 1$$
(7.B.34)

[use tables for the error function at intermediate values of $x$]. The dynamic friction Eq.(7.B.32) becomes

$$\mathbf{F}^{s'}(\mathbf{p}) = \hat{\mathbf{v}} \frac{e_s^2 \ln\Lambda}{2\lambda_{s'}^2} \frac{d}{dx}\left(\frac{\text{erf}(x)}{x}\right), \quad \lambda_{s'}^2 = \frac{T_{s'}}{4\pi n_{s'} e_{s'}^2}$$
(7.B.35)

We summarize the asymptotic forms for the diffusion tensor:

i. $v \ll v_s$ $\quad \mathbf{D}^{s'} = \frac{8}{3}\sqrt{\frac{\pi}{2}} \frac{n_{s'} e_s^2 e_{s'}^2 \ln\Lambda}{v_{s'}} \mathbf{I}$ velocity independent and isotropic
(7.B.36)

For $s'$=electrons $\mathbf{D}^{s'} = (m_e v_e)^2 v_e \mathbf{I}$

ii. $v \gg v_s$, $x \to \infty$, $\mathbf{w} = \mathbf{v} - \mathbf{v}' \approx \mathbf{v}$ in $\mathbf{Q}$ $\quad \mathbf{D}^{s'} \to n_{s'} \mathbf{Q}(\mathbf{v}) = n_{s'} \frac{2\pi e_s^2 e_{s'}^2 \ln\Lambda}{v}\left(\mathbf{I} - \hat{\mathbf{v}}\hat{\mathbf{v}}\right)$
(7.B.37)

The faster the particle, the weaker is the diffusion.

Exercise: Fill in the intermediate steps in the derivations of Eqs.(7.B.33), (7.B.36), and (7.B.37).

These simplified asymptotic forms Eqs.(7.B.32-7.B.37) for the dynamic friction and diffusion derived in the large-$k$ approximation (large-angle scattering) transform the Lenard-Balescu to the Landau equation. The Lenard-Balescu equation in Sec. 7.B.c is good up to some $k_{max}$ with a ~10% error by ignoring large-angle collisions. In the Landau equation derivation there is no $1/|\epsilon|^2$ and $\mathbf{Q}$ is given in Eq.(7.B.30). Equations (7.B.18-7.B.20) still describe the collisional kinetic equation. Using Eq.(7.B.31), Eq.(7.B.27), and Eq.(7.B30) we obtain the following alternative form for $\mathbf{D}$ in the Landau limit:

$$\mathbf{D}_s^{s'}(\mathbf{p}) = \int d^3\mathbf{p}' f_{s'}(\mathbf{p}')\mathbf{Q}_{ss'}(\mathbf{v},\mathbf{v}') = 2\pi e_s^2 e_{s'}^2 \ln\Lambda \int d^3\mathbf{p}' f_{s'}(\mathbf{p}') \frac{\partial^2}{\partial \mathbf{v}\partial \mathbf{v}}|\mathbf{v}-\mathbf{v}'|$$

$$= 2\pi e_s^2 e_{s'}^2 \ln\Lambda \frac{\partial^2}{\partial \mathbf{v}\partial \mathbf{v}}\int d^3\mathbf{p}' f_{s'}(\mathbf{p}')|\mathbf{v}-\mathbf{v}'|$$

(7.B.38)

### 7.B.e Derivation of the Fokker-Planck Equation from the Landau equation and model for Brownian motion

For electron speeds v<<$v_e$ the electron diffusion tensor was given in Eq.(7.B.36), $\mathbf{D}^e = (m_e v_e)^2 \nu_e \mathbf{I}$; and the dynamic friction of electrons on electrons for v<<$v_e$ is given by $\mathbf{F}^e = -m_e \nu_e \mathbf{v}_e$. The corresponding Fokker-Planck equation is then

$$\frac{d}{dt}f_e = -\frac{\partial}{\partial \mathbf{p}}\cdot\left(\mathbf{F}(\mathbf{p})f_e\right) + \frac{\partial}{\partial \mathbf{p}}\cdot\left(\mathbf{D}(\mathbf{p})\cdot\frac{\partial}{\partial \mathbf{p}}f_e\right)$$

(7.B.39)

This model for Brownian motion of electrons assumes that the asymptotic forms are good for all cases. This permits analytic solution of the Fokker-Planck equation in closed form.

Theorem: (Brownian motion model)

$$\frac{d}{dt}f_e = m\nu_e \frac{\partial}{\partial \mathbf{p}}\cdot\left(\mathbf{v}f_e + T_e\frac{\partial}{\partial \mathbf{p}}f_e\right)$$

(7.B.40)

If $f_e \sim \exp(-p^2/2m_e T_e)$ then $df_e/dt = 0$, i.e., the collisions preserve a Maxwellian.

### 7.B.f BGK (Bhatnagar-Gross-Krook) model – simplest collision model that conserves particles, momentum, and energy

The simplest collision operator that conserves particles, momentum, and energy was introduced by Bhatnagar, Gross, and Krook[49]

$$\left(\frac{\partial}{\partial t}f\right)_c = -\nu(f-f_0) \tag{7.B.41}$$

The simplicity of the BGK collision model lends itself to quickly obtaining some general idea as to the influence of collisional relaxation on the velocity distribution. However, it omits all velocity dependence of the collision frequency and large-angle scattering. It also cannot treat delicate things like the effect of collisions on resonant phenomena.

### 7.B.g Generalizations and applications of plasma collision theory, e.g., fast processes involving waves, slow processes, resonance broadening due to collisions

In this section we touch on a collection of topics associated with collision theory. For some topics only a brief commentary is given, and for others there is significant analysis.

**i**. Collision theory in a magnetized plasma

Norman Rostoker published a treatment of collisions in a plasma with a constant applied magnetic field in 1960.[50] Consider a Landau collision model for electrons with impact parameter $b$ and Larmor radius $r_g$ less than the Debye length $\lambda_D$. If the impact parameter $b$ exceeds the Debye length, then the collision is shielded and the electron does not experience a collisional Coulomb force. If the magnetic field is strong and $r_g$ is less than the impact parameter, the electron stays on its field line and does not undergo the collision. Only if $b < r_g$ can a collision take place. Hence, we can take $\max(b) \to r_g$ and $\ln\Lambda = \ln\lambda_D/r_o \to \ln r_g/r_o$ where $r_o \equiv e^2/T$ instead of using Rostoker's formulae. If $r_g > \lambda_D$ the magnetic field is weak; so weak that with respect to collision theory, it is as if there is no magnetic field at all, in which case there are no changes in the collision theory from the unmagnetized case. However, if there are shielding effects included, e.g., as in Eq.(7.B.28), the dielectric function is modified in the presence of the magnetic field.

**ii**. Collisions including electromagnetic effects

---

[49] P. L. Bhatnagar, E. P. Gross, and M. Krook, Phys. Rev. **94**, 511 (1954)
[50] N. Rostoker, Phys. Fluids **3**, 922 (1960).

When discrete charged particles interact, the most general treatment is based on Maxwell's equations.[51] For short-range collisions ($b < \lambda_D$), an electromagnetic treatment is needed instead of electrostatic. However, if the plasma is non-relativistic, the electromagnetic corrections to the electrostatic collision theory are small in $v^2/c^2$; and the electromagnetic corrections are insignificant. Electromagnetic effects are significant for charged-particle collisions in a relativistic plasma, e.g., consider bremsstrahlung. Theory based on Lienard Wiechert potentials is commonly employed, and the collision of two charged particles is sometimes described in terms of photon exchange. In any case, the hotter the plasma the weaker the collisional effects; and collisions in a relativistically hot plasma may not be important in many contexts.

**iii**. Large-angle scattering

In the foregoing we have developed collision theories for small and large angle scattering. The Lenard-Balescu theory is valid for larger impact parameters $b > r_o$. The Lenard-Balescu theory is accurate to approximately 10%. For $r_o < b < \lambda_D$ the Landau collision operator is useful and is accurate to approximately 15%. A Boltzmann collision model ignoring plasma shielding effects is appropriate for $b < \lambda_D$ absent quantum mechanical effects. Figure 7.B.1 diagrams the regimes for the three classical collision theories.

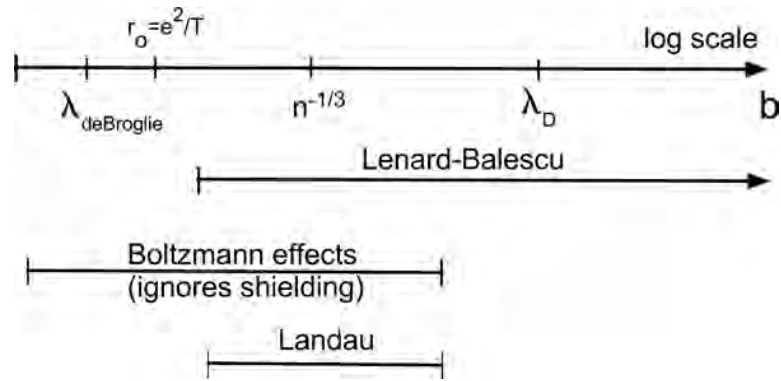

Fig. 7.B.1 Diagram of ranges in collisional impact parameter *b* for classical collision theories in plasmas

The region of overlap and the ranges of applicability invite the following practical recipe:

$$\left(\frac{\partial}{\partial t}f\right)_c = \left(\frac{\partial}{\partial t}f\right)_{Len-Bal} + \left(\frac{\partial}{\partial t}f\right)_{Boltz} - \left(\frac{\partial}{\partial t}f\right)_{Landau} \quad (7.B.42)$$

The Lenard-Balescu theory cancels with either the Boltzmann or the Landau theories in the intermediate zone where the theories overlap.

---

[51] J.D. Jackson, *Classical Electrodynamics* (3rd ed.). New York: John Wiley & Sons (1999).

**iv**. Quantum mechanical effects

For $b \leq \lambdabar_{dB} \equiv \hbar/\mu v$ quantum mechanical effects affect the collisions. When $T > 1$ Rydberg = 13.6 eV, then $\lambdabar_{dB} > e^2/T$. This parameter regime is very important in astrophysics. Consideration of quantum mechanical effects on collisions in plasmas is beyond the scope of these lectures. Some of the researchers who have worked in this area are DeWitt, Gould, Lampe, and Del Rio.

**v**. Applications – slow processes involving collisons

If the electron velocity distribution has finite temperature but is a non-Maxwellian, then electron-electron collisions will relax the velocity distribution toward a Maxwellian on a time scale $\tau^e_{Maxw} \sim 1/\nu_e$. Similarly, a non-Maxwellian ion velocity distribution will be relaxed by ion-ion collisions on a time scale $\tau^i_{Maxw} \sim 1/\nu_i$. For singly charged ions, $\nu_i$ is $\nu_e$ with $m_e \to m_i$ and $T_e \to T_i$. We note that the characteristic frequencies or rates fundamentally scale as $\nu \sim n\sigma v$, where n is the number density of field particles, $\sigma \sim (e^2/T)^2$ is the Coulomb cross section, and v is the relative speed $v \sim \sqrt{T/m}$. Electron collision frequencies are generally faster than ion collision frequencies due to the $1/\sqrt{m}$ scaling.

When there are two charged species present with different temperatures, collisions will relax the two velocity distributions to have a common temperature. Among collisional relaxation processes, temperature relaxation is very slow. Consider ions and electrons with Maxwellian velocity distributions such that initially $T_e > T_i$. Electrons will cool, and ions will heat. The kinetic equation for the ions (assumed singly charged, so $n_e = n_i = n$) is

$$\frac{\partial f^i(\mathbf{p};t)}{\partial t} = -\frac{\partial}{\partial \mathbf{p}} \cdot \mathbf{J}^e_i(\mathbf{p};t), \text{ where } f^i(\mathbf{p};t) = const \ \exp\left(-\frac{p^2}{2m_i T_i(t)}\right)$$

$$\to \int d^3\mathbf{p} \frac{p^2}{2m_i} \frac{\partial f^i(\mathbf{p};t)}{\partial t} = \frac{\partial}{\partial t}\left[n\frac{3}{2}T_i(t)\right] = -\int d^3\mathbf{p} \frac{p^2}{2m_i}\frac{\partial}{\partial \mathbf{p}} \cdot \mathbf{J}^e_i(\mathbf{p};t) = \int d^3\mathbf{p}\, \mathbf{v} \cdot \mathbf{J}^e_i(\mathbf{p};t)$$

$$= \int d^3\mathbf{p}\, \mathbf{v} \cdot \left[\mathbf{F}^e_i f_i - \mathbf{D}^e_i(\mathbf{p}) \cdot \frac{\partial f_i}{\partial \mathbf{p}}\right] = -m_e \nu_e \int d^3\mathbf{p}\, v^2 f_i + m_e \nu_e \frac{T_e}{T_i}\int d^3\mathbf{p}\, v^2 f_i = 3\frac{m_e}{m_i}\nu_e n(T_e - T_i)$$

$$\to \frac{dT_i}{dt} = 2\frac{m_e}{m_i}\nu_e(T_e - T_i)$$

(7.B.43)

Equation (7.B.43) describes the temperature relaxation process. The integration of Eq.(7.B.43) yields

$$\frac{dT_i}{dt} = 2\frac{m_e}{m_i}\nu_e(T_e - T_i) = -\frac{dT_e}{dt} \rightarrow \frac{d(T_e - T_i)}{T_e - T_i} = -4\frac{m_e}{m_i}\nu_e dt$$

$$\rightarrow T_e - T_i = (T_{e0} - T_{i0})\exp\left[-4\frac{m_e}{m_i}\nu_e(t - t_0)\right]$$

(7.B.44)

**vi**. Applications – fast collision processes involving waves

In Sec. 4.B we constructed a model for collisional damping of waves. In a simple relaxation model we made the replacement $\omega^2 \rightarrow \omega(\omega + i\nu)$ in the linear dispersion relations. For Langmuir waves the imaginary part of the mode frequency due to collisions is $\gamma_c = -\frac{1}{2}\nu_e$. For a transverse wave in an unmagnetized plasma $\gamma_c = -\frac{1}{2}\nu_e\left(\frac{\omega_p}{\omega}\right)^2$

Collisional diffusion alters the wave-particle resonance. In Sec. 7.B.b we derived velocity diffusion in a Coulomb model. In a one-dimensional model the velocity diffusion is given by

$$D(v) = \left(\frac{e}{m}\right)^2 \int_0^\infty d\tau\, C_E(s(\tau),\tau), \quad C_E(s(\tau),\tau) = \int\int \frac{dkd\omega}{(2\pi)^2} k^2 S_\phi(k,\omega)\exp(iks - i\omega\tau)$$

$$\rightarrow D(v) = \left(\frac{e}{m}\right)^2 \int\int \frac{dkd\omega}{(2\pi)^2} k^2 S_\phi(k,\omega)\int_0^\infty d\tau\, \exp(-i\omega\tau)\exp(iks(\tau))$$

(7.B.45)

What is the spatially averaged velocity diffusion? The average of Eq.(7.B.45) with respect to the particle position is

$$\langle D(v)\rangle = \left(\frac{e}{m}\right)^2 \int\int \frac{dkd\omega}{(2\pi)^2} k^2 S_\phi(k,\omega)\int_0^\infty d\tau\, \exp(-i\omega\tau)\langle\exp(iks(\tau))\rangle, \quad s(\tau) = v\tau + s'(\tau)$$

$$\rightarrow \langle D(v)\rangle = \left(\frac{e}{m}\right)^2 \int\int \frac{dkd\omega}{(2\pi)^2} k^2 S_\phi(k,\omega)\int_0^\infty d\tau\, \exp(-i\omega\tau + ikv\tau)\langle\exp(iks'(\tau))\rangle$$

(7.B.46)

where $s'(\tau)$ is the perturbed particle displacement. For fixed $\tau$ consider the perturbed displacement and its probability of occurring $P(s'|\tau)$. If $X \equiv \sum_{i=1}^N x_i$ and if the $x_i$ are independent, then in the limit $N \rightarrow \infty$, $X$ will be distributed normally (Central Limit Theorem).

<u>Theorem</u>: $\quad P(s'|\tau) = \frac{1}{\sqrt{2\pi\sigma^2(\tau)}}\exp\left(\frac{\left[s' - \langle s'\rangle(\tau)\right]^2}{2\sigma^2(\tau)}\right)$ (7.B.47)

Returning to Eq.(7.B.46), consider the time integral. If $s'=0$, then the Landau resonance integral results:

$$R = \text{Re} \int_0^\infty d\tau \exp(-i(\omega - k\text{v})\tau) \langle \exp(iks'(\tau) \to 0) \rangle \to \pi \delta(\omega - k\text{v}) \qquad (7.B.48)$$

Instead with *s'* distributed as a Gaussian, as in Eq.(7.B.47), one has

$$\langle \exp(iks'(\tau)) \rangle \equiv \int_{-\infty}^\infty ds' \exp(iks') P(s'|\tau) = \exp\left[ik\langle s' \rangle(\tau) - \frac{1}{2}k^2 \sigma^2(\tau)\right]$$

and  (7.B.49)

$$R(\omega, k, \text{v}) = \text{Re} \int_0^\infty d\tau \exp\left(-i(\omega - k\text{v})\tau + ik\langle s' \rangle(\tau) - \frac{1}{2}k^2 \sigma^2(\tau)\right)$$

The term $ik <s'> (\tau)$ inside the exponential in Eq.(7.B.49) contributes a frequency shift, which we will ignore, but it can be evaluated easily. Next consider the last term inside the exponential:

$$\sigma^2 \equiv \left\langle \left[s' - <s'>\right]^2 \right\rangle = <s'^2> - <s'>^2 \approx <s'^2> \gg <s'>^2 \qquad (7.B.50)$$

and

$$s'(\tau) \equiv \int_0^\tau dt\, \delta \text{v}(t) = \int_0^\tau dt \int_0^t dt'\, \dot{\text{v}}(t)$$

$$\left[s'(\tau)\right]^2 = \int_0^\tau dt_1 \int_0^\tau dt_2 \int_0^{t_1} dt_1' \int_0^{t_2} dt_2' \langle \dot{\text{v}}(t_1') \dot{\text{v}}(t_2') \rangle \equiv \int_0^\tau dt_1 \int_0^\tau dt_2 \int_0^{t_1} dt_1' \int_0^{t_2} dt_2'\, C_{\dot{\text{v}}}\left(|t_1' - t_2'|\right)$$

(7.B.51)

where $C_{\dot{\text{v}}}$ is the two-time auto-correlation function for the velocity derivative. Examining the innermost area integral with respect to $t_1'$ and $t_2'$ over the rectangle 0 to $t_1$ and 0 to $t_2$, $C_{\dot{\text{v}}}$ is significant only over the diagonal where $|t_1' - t_2'|$ is small because of the randomness of the perturbing fields. Hence,

$$\left[s'(\tau)\right]^2 = \int_0^\tau dt_1 \int_0^\tau dt_2 (t_1, t_2)_{<} \int_{-\infty}^\infty d\tau'\, C_{\dot{\text{v}}}(\tau') = \frac{1}{3}\tau^3 \int_{-\infty}^\infty d\tau'\, C_{\dot{\text{v}}}(\tau') \qquad (7.B.52)$$

However, from Eq.(7.B.2) $D(\text{v}) = \int_0^\infty d\tau \langle \dot{\text{v}}(t) \dot{\text{v}}(t - \tau) \rangle = \int_0^\infty d\tau\, C_{\dot{\text{v}}}(\tau)$. Hence,

$$\sigma^2(\tau) = \frac{2}{3} D(\text{v}) \tau^3 \quad \text{and} \quad R(\omega, k, \text{v}: D) = \int_0^\infty d\tau \exp\left(-\frac{1}{3}k^2 D(\text{v})\tau^3\right) \cos\left[(\omega - k\text{v})\tau\right] \quad (7.B.53)$$

In Eqs.(7.B.45-7.B.53) *D* is the diffusion in velocity space and not momentum space:

$$D_e^e(\text{v}) = \text{v}_e^2 \nu_e \qquad (7.B.54)$$

It then follows that

$$k^2 D\tau^3 = (k\mathrm{v}_e)^2 \nu_e \tau^3 = \left(\frac{\tau}{\tau_\nu}\right)^3, \quad \tau_\nu^{-3} \equiv (k\mathrm{v}_e)^2 \nu_e = k^2 D$$

$$\to \tau_\nu \equiv D^{-1/3} k^{-2/3} = \nu_e^{-1/3}(k\mathrm{v}_e)^{-2/3} \sim (k_D \mathrm{v}_e)^{-2/3} \nu_e^{-1/3} = \nu_e^{-1} \left(\frac{\nu_e}{\omega_p}\right)^{2/3} \quad (7.B.55)$$

for $k \sim k_D$, where $\tau_\nu$ is a characteristic time for collisions. For $\nu_e \sim \omega_p / \Lambda$ then $\tau_\nu^{-1} \sim \nu_e \Lambda^{2/3}$ or $\tau_\nu^{-1} \sim \omega_p / \Lambda^{1/3}$ Hence, the relative frequency broadening width for an electron plasma wave is $\tau_\nu^{-1}/\omega_p \sim 1/\Lambda^{1/3}$. We return to Eq.(7.B.53):

$$R(\omega, k, \mathrm{v}; \tau_\nu) = \int_0^\infty d\tau \exp\left(-\frac{1}{3} k^2 D(\mathrm{v})\tau^3\right) \cos\left[(\omega - k\mathrm{v})\tau\right] = \tau_\nu \int_0^\infty dx \exp\left(-\frac{1}{3} x^3\right) \cos(\lambda x)$$

(7.B.56)

where $\lambda \equiv (\omega - k\mathrm{v})\tau_\nu$ and $x \equiv \tau/\tau_\nu$. The integral in Eq.(7.B.56) is not an elementary integral. We recall from Eq.( 7.B.48) that as $D \to 0$ $R \to \pi\delta(\omega - k\mathrm{v})$. For small $\lambda \ll 1$ the integral on the right side of Eq.(7.B.56) is dominated by the exponential and converges quickly to a number of O(1). [Note that $\int_0^\infty dx \exp(-x^\mu) = \frac{1}{\mu}\Gamma(\frac{1}{\mu})$]. $R/\tau_\nu$ peaks as a function of $\lambda$ at $\lambda=0$ at a value O(1) and decreases for increasing $\lambda$. At large $\lambda$ the integral in Eq.(7.B.56) can be approximated using steepest descents yielding $R/\tau_\nu \sim \exp(-\lambda^{3/2})\sin(\lambda^{3/2})$.

Exercise: Evaluate the integral in Eq.(7.B.56) for large $\lambda$ using steepest descents.

We see that as $D$ increases the resonance integral $R$ spreads. However, the area under the curve $R$ vs. $\omega$, $\int_{-\infty}^\infty d\omega R = \pi$ is conserved, which can be verified by integrating $R$ with respect to $\omega$ in Eq.(7.B.56) to obtain $\pi\delta(\tau)$ and then doing the $\tau$ integral. We note that $R$ peaks at $\tau_\nu O(1)$ and has a full width at half-maximum $\sim O(1)\tau_\nu^{-1}$. We also observe that

$$\lambda = \left(\frac{\omega - k\mathrm{v}}{\omega_\nu}\right) = \left(\frac{\mathrm{v} - \omega/k}{\omega_\nu/k}\right), \quad \omega_\nu \equiv 1/\tau_\nu \quad (7.B.57)$$

Thus, the broadening of the resonance can be considered in the frequency domain for fixed velocity or in the velocity domain for fixed frequency.

Theorem: The width in velocity space is

$$w \equiv \frac{\omega_\nu}{k} = \left(\frac{D}{k}\right)^{1/3} = \left(\frac{\mathrm{v}_e^2 \nu_e}{k}\right)^{1/3} = \mathrm{v}_e \left(\frac{\nu_e}{k\mathrm{v}_e}\right)^{1/3} \sim \mathrm{v}_e \left(\frac{\nu_e}{\omega_p}\right)^{1/3} = \mathrm{v}_e / \Lambda^{1/3} \quad (7.B.58)$$

Similarly, for fixed frequency and velocity there is a resonance in $k$; and there is a corresponding width due to the collisional broadening:

$$\lambda = \left(\frac{\omega - k\text{v}}{\omega_\nu}\right) = \left(\frac{\omega/\text{v} - k}{\omega_\nu/\text{v}}\right) \rightarrow \text{for } \text{v} \sim \text{v}_e \quad k_\nu \sim \frac{k_D}{\Lambda^{1/3}} \qquad (7.B.59)$$

Thus, the scaling of the collisional spreading of the resonances in the frequency, velocity, and wavenumber domains is $1/\Lambda^{1/3}$.

Collisions are just one cause of resonance broadening. The relative width of the resonance in frequency or velocity is $(\nu_e/\omega_p)^{1/3}$ for collisional diffusion. Quasilinear diffusion in the presence of a spectrum of resonant waves is another cause of resonance broadening. The width in velocity space (particle velocity or phase velocity) scales as $\sqrt{e\phi_{rms}/m}$. Particle trapping in a monochromatic wave broadens the resonance in velocity space by $O(\text{v}_{\text{trap}}) \sim \sqrt{e\phi/m}$. The adiabatic interaction of many non-resonant waves can broaden the resonance of a particle with a specific wave. In general, the time interval for a resonant interaction is limited by the amount of time it takes for diffusion to drive the interaction off resonant; define it as $\tau_D$.

For collisional diffusion, the resonance is detuned when the rms diffusion displacement is comparable to the wavelength of the wave:

$$\sqrt{<s'(\tau_D)^2>} \sim \lambda \rightarrow <s'(\tau_D)^2> \sim \lambda^2 \sim \frac{1}{k^2}$$

$$<s'(\tau_D)^2> \sim (D\tau_D)\tau_D^2 \rightarrow D\tau_D^3 \sim \frac{1}{k^2} \rightarrow k^2 D\tau_D^3 \sim O(1) \qquad (7.B.60)$$

which is when the exponential begins to limit the integrand in Eq.(7.B.56) and $\tau_\nu$ in Eq.(7.B.55).

In quasilinear diffusion driven resonance broadening, the estimates for the diffusion time and the scaling of the resonance width proceed as follows:

$$\dot{\text{v}} = \frac{e}{m} E \rightarrow <(\text{v}')^2> = \left(\frac{e}{m}\right)^2 <E^2> \tau_D^2$$

$$s' \sim \text{v}' \tau_D \rightarrow <s'^2> \sim \left(\frac{e}{m}\right)^2 <E^2> \tau_D^4 \sim \lambda^2 \rightarrow \frac{e}{m} E_{rms} \tau_D^2 \sim \lambda \qquad (7.B.61)$$

$$\rightarrow \frac{e}{m} k^2 \phi_{rms} \sim \frac{1}{\tau_D^2} \sim \delta\omega^2 \rightarrow \frac{e}{m}\phi_{rms} \sim w^2 = \frac{\delta\omega^2}{k^2} \rightarrow \sqrt{\frac{e}{m}\phi_{rms}} \sim w$$

<u>Example</u>: Resonance broadening due to quasilinear diffusion for a discrete spectrum of waves. For quasilinear diffusion we earlier derived the velocity diffusion coefficient in Sec. 6.E

$$D(\text{v}) = \left(\frac{e}{m}\right)^2 \int \frac{dk}{2\pi} \int \frac{d\omega}{2\pi} S_E(k,\omega) R(\omega - k\text{v}, \delta\omega) \qquad (7.B.62)$$

In Eq.(7.B.62) the resonance function is $R = \pi\delta(\omega - k\mathrm{v})$ in the absence of broadening. With broadening, the resonance function acquires a finite width and height with respect to frequency. Consider a discrete spectrum of normal modes such that

$$S_E(k,\omega) = 2\pi S(k)\delta(\omega - \omega_k) \quad S_E(k) = \sum_{i=1}^{N} S_i 2\pi\delta(k - k_i) \qquad (7.B.63)$$

Despite the discrete spectrum, there are no infinities in the integrand in $D(\mathrm{v})$ in Eq.(7.B.62), and $\sum_i S_i = <E^2> = \sum_i k^2 \phi_{rms}^2$ if there is broadening. In velocity space the structure of $D$ acquires finite-width resonances for each value of $\omega_k(k_i)/k_i$. Thus, $D$ is a smooth function of velocity with no singularities, which is useful for evaluating $\dfrac{\partial f}{\partial t} = \dfrac{\partial}{\partial \mathrm{v}}\left(D\dfrac{\partial f}{\partial \mathrm{v}}\right)$.

# LECTURES ON THEORETICAL PLASMA PHYSICS – PART 3B

*Allan N. Kaufman*

### 7.C Radiation transport

In this section we consider the emission of electromagnetic waves in a plasma. We include the effects of gyration in a magnetic field and particle collisions on wave emission. We also examine the scattering of waves in a plasma and derive a general theory for the scattering of particles, fluctuations, and waves, including nonlinear Landau damping. We summarize the important processes in radiation transport in a plasma and conclude with a discussion of WKB theory in the context of radiation transport. The consideration of radiation transport here does not include any processes associated with atomic physics, and the theory is purely classical and excludes quantum mechanical effects.

### 7.C.a Calculation of emission from Maxwell's equations with arbitrary current sources

We recall from Sections 3 and 4 that the linearized Vlasov-Maxwell system yields the following relation for the electromagnetic response of a plasma to an arbitrary current density:

$$\mathbf{K}(\mathbf{k},\omega)\cdot\mathbf{E}(k,\omega) = \frac{4\pi}{i\omega}\mathbf{j}_0(\mathbf{k},\omega) \text{ where } \mathbf{K} = \vec{\vec{\varepsilon}} - n^2(\vec{\mathbf{I}} - \hat{\mathbf{k}}\hat{\mathbf{k}}) \text{ and } n = \frac{kc}{\omega} \quad (7.C.1)$$

When the electric fields do work on the plasma currents, energy comes out of the electromagnetic fields:

$$-\dot{W} = \int d^3\mathbf{x}\, \mathbf{E}(\mathbf{x},t)\cdot\mathbf{j}_0(\mathbf{x},t) = \int \frac{d^3\mathbf{k}}{(2\pi)^3}\mathbf{j}_0^*(\mathbf{k},t)\cdot\mathbf{E}(\mathbf{k},t) \quad (7.C.2)$$

If we invert the relation in Eq.(7.C.1) to solve for the electric field then

$$\mathbf{E}(\mathbf{k},\omega) = \frac{4\pi}{i\omega}\mathbf{K}^{-1}(\mathbf{k},\omega)\cdot\mathbf{j}_0(\mathbf{k},\omega) \quad (7.C.3)$$

from which one obtains

$$\mathbf{E}(\mathbf{k},\omega) = \int_{-\infty}^{\infty} d\tau\, \mathbf{G}(\mathbf{k},\tau)\cdot\mathbf{j}_0(\mathbf{k},t-\tau) \quad (7.C.4)$$

where

$$\mathbf{G}(\mathbf{k},\tau) = \int \frac{d\omega}{2\pi}\exp(-i\omega\tau)\frac{4\pi}{i\omega}\mathbf{K}^{-1}(\mathbf{k},\omega) \quad (7.C.5)$$

We assume that there are no poles of the integrand in Eq.(7.C.5) in the upper half of the complex $\omega$ plane (no exponentially growing linear modes), and we use analytic continuation to depress the integration contour while keeping the contour above poles on the real $\omega$ axis. Causality insures that **G**=0 for $\tau < 0$. We use Eqs.(7.C.3-7.C.5) in Eq.(7.C.2) to obtain

$$-\dot{W}(t) = \int \frac{d^3\mathbf{k}}{(2\pi)^3} \int_{-\infty}^{\infty} d\tau\, \mathbf{j}_0^*(\mathbf{k},t) \cdot \mathbf{G}(\mathbf{k},\tau) \cdot \mathbf{j}_0(\mathbf{k},t-\tau) \qquad (7.C.6)$$

Consider the product of the plasma currents in Eq.(7.C.6) and posit a sufficient number of sources to average over so that we can calculate the two-time correlation tensor for the plasma currents.

<u>Definition</u>: $\mathbf{C} = \mathbf{C}(\mathbf{k},\tau) \equiv \langle \mathbf{j}_0^*(\mathbf{k},t)\mathbf{j}_0(\mathbf{k},t-\tau) \rangle$. We assume that **C** is not a function of time, i.e., it is statistically stationary.

Hence, from Eq.(7.C.6)

$$-\dot{W} = \int \frac{d^3\mathbf{k}}{(2\pi)^3} \int_{-\infty}^{\infty} d\tau\, \mathbf{C}_{\mathbf{j}_0}^*(\mathbf{k},\tau) : \mathbf{G}(\mathbf{k},\tau) = \int \frac{d^3\mathbf{k}}{(2\pi)^3} \int_{-\infty}^{\infty} \frac{d\omega}{2\pi} \mathbf{S}^{\mathbf{j}_0 *}(\mathbf{k},\omega) : \mathbf{K}^{-1}(\mathbf{k},\omega) \frac{4\pi}{i\omega} \qquad (7.C.7)$$

We take the limit that there are no linearly unstable waves, Im $\omega \to 0$. We note that **K**=**K**'+i**K**" where **K**' is the reactive part and **K**" is the dissipative part. **K** is not Hermitian, and $\mathbf{K}^{-1} = \mathbf{K}^{-1}{}' + i\mathbf{K}^{-1}{}''$ where $\mathbf{K}^{-1}{}'$ is the Hermitian part and $i\mathbf{K}^{-1}{}''$ is the anti-Hermitian part.

<u>Theorem</u>: **S**\* = **S**' is Hermitian which follows from **C** being independent of time.

<u>Exercise</u>: Prove that **S**\* is Hermitian.

$$\begin{aligned}\mathbf{S}^{\mathbf{j}_0 *} &= \int_{-\infty}^{\infty} d\tau\, \mathbf{C}_{\mathbf{j}_0^*} = \int_{-\infty}^{\infty} d\tau \langle \mathbf{j}_0(\mathbf{k},t)\mathbf{j}_0^*(\mathbf{k},t-\tau) \rangle = -\int_{-\infty}^{\infty} d\tau \langle \mathbf{j}_0(\mathbf{k},t)\mathbf{j}_0^*(\mathbf{k},\tau-t) \rangle \\ &= -\int_{-\infty}^{\infty} d\tau \lim_{T\to\infty} \frac{1}{2T} \int_{-T}^{T} dt\, \mathbf{j}_0(\mathbf{k},t)\mathbf{j}_0^*(\mathbf{k},\tau-t) = -\lim_{T\to\infty}\frac{1}{2T}\int_{-T}^{T} d\tau \int_{-\infty}^{\infty} d\omega\, e^{-i\omega\tau} \mathbf{j}_0(\mathbf{k},\omega)\mathbf{j}_0^*(\mathbf{k},\omega) \\ &= -\lim_{T\to\infty}\frac{1}{2T}\int_{-T}^{T} d\tau \int_{-\infty}^{\infty} d\omega\, e^{-i\omega\tau} \mathbf{j}_0^*(\mathbf{k},\omega)\mathbf{j}_0(\mathbf{k},\omega) \Rightarrow \mathbf{S}^* = \mathbf{S}^T\end{aligned} \qquad (7.C.8)$$

where **S**$^T$ is the transpose of **S**.

In Eq.(7.C.7) $\mathbf{K}^{-1}$ appears, which we evaluate as follows:

$$\mathbf{K}^{-1} = \frac{\mathbf{K}^{adj}}{\det|\mathbf{K}|} = \frac{\mathbf{K}^{adj}}{D(\mathbf{k},\omega)} = \frac{\mathbf{K}^{adj}}{D'(\mathbf{k},\omega) + iD''(\mathbf{k},\omega)} = \mathbf{K}^{adj}\left(P\frac{1}{D'} - i\pi\delta(D')\mathrm{sgn}D''\right) \qquad (7.C.9)$$

for small $D''$. Substituting Eq.(7.C.9) in Eq.(7.C.7), we obtain

$$\dot{W} = \int \frac{d^3\mathbf{k}}{(2\pi)^3} \int_{-\infty}^{\infty} \frac{d\omega}{2\pi} \mathbf{S}^{j_0*}(\mathbf{k},\omega) : \mathbf{K}^{adj}(\mathbf{k},\omega) \frac{4\pi}{\omega} \pi\delta(D'(\mathbf{k},\omega)) \operatorname{sgn} D'' \qquad (7.C.10)$$

where

$$\pi\delta(D(\mathbf{k},\omega)) = \sum_{\ell,\pm} \frac{\delta(\omega - \omega_k^\ell)}{\left|\partial D'(\mathbf{k},\omega)/\partial \omega\right|_{\omega_k^\ell}} \qquad (7.C.11)$$

and the sum is over $\ell$ branches and $\pm|\omega|$. In the absence of an applied external magnetic field there are three electron branches (immobile ions): two branches for transverse modes and one branch for longitudinal modes. With mobile ions there are additional branches, and with an applied magnetic field there are even more branches. Equation (7.C.10) becomes

$$\dot{W} = \int \frac{d^3\mathbf{k}}{(2\pi)^3} \sum_{\ell,\pm} \frac{2\pi}{\omega_k^\ell} \mathbf{S}^{j_0*}(\mathbf{k},\omega_k^\ell) : \mathbf{K}^{adj}(\mathbf{k},\omega_k^\ell) \frac{\operatorname{sgn} D''}{\left|\frac{\partial D}{\partial \omega}\right|_{\omega_k^\ell}} \equiv \int \frac{d^3\mathbf{k}}{(2\pi)^3} \sum_{\ell,\pm} \dot{W}^\ell(\mathbf{k})$$

$$\dot{W}^\ell(\mathbf{k}) = \frac{2\pi}{\omega_k^\ell} \mathbf{S}^{j_0*}(\mathbf{k},\omega_k^\ell) : \mathbf{K}^{adj}(\mathbf{k},\omega_k^\ell) \frac{\operatorname{sgn} D''}{\left|\frac{\partial D}{\partial \omega}\right|_{\omega_k^\ell}} \qquad (7.C.12)$$

We next eliminate $\mathbf{K}^{adj}$ using the following relations:

$$\mathbf{K}^{-1} = \frac{\mathbf{K}^{adj}}{D} \quad \mathbf{K}^{-1} \cdot \mathbf{K} = \mathbf{I} = \mathbf{K} \cdot \mathbf{K}^{-1} \quad \mathbf{K}^{adj}(\mathbf{k},\omega_k^\ell) \cdot \mathbf{K}(\mathbf{k},\omega_k^\ell) = D\mathbf{I} = \mathbf{K} \cdot \mathbf{K}^{adj}$$

$$\frac{\partial}{\partial \omega}\left[\mathbf{K}^{adj}(\mathbf{k},\omega_k^\ell) \cdot \mathbf{K}(\mathbf{k},\omega_k^\ell)\right] = \frac{\partial \mathbf{K}^{adj}}{\partial \omega} \cdot \mathbf{K} + \mathbf{K}^{adj} \cdot \frac{\partial \mathbf{K}}{\partial \omega} = \frac{\partial D}{\partial \omega} \mathbf{I} \qquad (7.C.13)$$

and the matrix identity det($\mathbf{K}^{adj}$)=det($\mathbf{K}$)$^{n-1}$=$D^2$ where n=3 is the rank of the matrix $\mathbf{K}$. Thus, as $D \to 0$, det($\mathbf{K}^{adj}$) goes to zero even faster. We recall that the linear normal modes satisfy the following relations:

$$\mathbf{K} \cdot \mathbf{E} = 0 \Rightarrow D(\mathbf{k},\omega_k^\ell) = 0$$
$$\mathbf{K}(\mathbf{k},\omega_k^\ell) \cdot \hat{\mathbf{e}}_k^\ell = 0 \Rightarrow \text{polarization} \qquad (7.C.14)$$
$$\therefore \mathbf{K}(\mathbf{k},\omega_k^\ell) = K\,\hat{\mathbf{e}}_k^{\ell*}\hat{\mathbf{e}}_k^\ell \quad \text{and} \quad \mathbf{K}^{adj}(\mathbf{k},\omega_k^\ell) = K\,\hat{\mathbf{e}}_k^\ell \hat{\mathbf{e}}_k^{\ell*}$$

using that to lowest order $\mathbf{K}$ is Hermitian. Given Eq.(7.C.14) then from Eq.(7.C.13)

$$\hat{\mathbf{e}}^* \cdot \frac{\partial \mathbf{K}^{adj}}{\partial \omega} \cdot \mathbf{K} \cdot \hat{\mathbf{e}} + \hat{\mathbf{e}}^* \cdot \mathbf{K}^{adj} \cdot \frac{\partial \mathbf{K}}{\partial \omega} \cdot \hat{\mathbf{e}} = \hat{\mathbf{e}}^* \cdot \mathbf{K}^{adj} \cdot \frac{\partial \mathbf{K}}{\partial \omega} \cdot \hat{\mathbf{e}} = \frac{\partial D}{\partial \omega} \hat{\mathbf{e}}^* \hat{\mathbf{e}} = \frac{\partial D}{\partial \omega}$$

$$\Rightarrow K = \frac{\partial D/\partial \omega}{\hat{\mathbf{e}}^* \cdot \frac{\partial \mathbf{K}}{\partial \omega} \cdot \hat{\mathbf{e}}} \qquad (7.C.15)$$

We can now return to our evaluation of Eq.(7.C.12) using Eqs.(7.C.14) and (7.C.15) for $\mathbf{K}^{adj}$, and Eq.(7.C.8) for $\mathbf{S}$:

$$\dot{W}^\ell(\mathbf{k}) = \frac{2\pi}{\omega_k^\ell} \text{sgn} D'' \int_{-\infty}^{\infty} d\tau \exp(-i\omega_k^\ell \tau) \frac{\left\langle \mathbf{j}_0^*(\mathbf{k},t) \cdot \hat{\mathbf{e}}_k^\ell \, \hat{\mathbf{e}}_k^{\ell*} \cdot \mathbf{j}_0(\mathbf{k},t-\tau) \right\rangle}{\hat{\mathbf{e}}_k^{\ell*} \cdot \left. \frac{\partial \mathbf{K}}{\partial \omega} \right|_{\omega_k^\ell} \cdot \hat{\mathbf{e}}_k^\ell} \qquad (7.C.16)$$

<u>Definition</u>: Define the scalar current density as the projection of the current density onto the unit vector for the electric field polarization determined in Eq.(7.C.14):

$$j_0^{(k,\ell)} \equiv \hat{\mathbf{e}}_k^{\ell*} \cdot \mathbf{j}_0(\mathbf{x},t) \qquad (7.C.17)$$

Using the scalar current density we can construct the spectral density for the current density from

$$\int_{-\infty}^{\infty} d\tau \exp(-i\omega_k^\ell \tau) \left\langle j_0^{(k,\ell)*}(\mathbf{k},t) j_0^{(k,\ell)}(\mathbf{k},t-\tau) \right\rangle \equiv \int_{-\infty}^{\infty} d\tau \exp(-i\omega_k^\ell \tau) C_{j_k^\ell}^* \equiv S_{j_k^\ell} \qquad (7.C.18)$$

We note that the spectral density for the scalar current is real so that the conjugate can be dropped. With Eqs.(7.C.17) and (7.C.18), Eq.(7.C.16) can be rewritten as

$$\left| \dot{W}^\ell(\mathbf{k}) \right| = \frac{4\pi}{\omega_k^\ell} \frac{S_{j_{k,\ell}}(\mathbf{k},\omega_k^\ell)}{\hat{\mathbf{e}}_k^{\ell*} \cdot \left. \frac{\partial \mathbf{K}}{\partial \omega} \right|_{\omega_k^\ell} \cdot \hat{\mathbf{e}}_k^\ell}, \quad \omega_k^\ell > 0 \qquad (7.C.19)$$

where we have dropped sgn$D''$ and introduced a factor of 2 in the numerator because $\omega_k^\ell > 0$. A useful relation for $S_j$ is given by

$$\left\langle j_0^{(k,\ell)}(\mathbf{k},\omega) j_0^{(k,\ell)*}(\mathbf{k}',\omega') \right\rangle = (2\pi)^4 \delta(\omega-\omega') \delta(\mathbf{k}-\mathbf{k}') S_{j_k^\ell} \qquad (7.C.20)$$

### 7.C.b Emission by a particle gyrating in a magnetic field

We illustrate the results in the previous sub-section by calculating the electromagnetic wave emission from a gyrating particle in an externally applied magnetic field. The current density from a single particle is

$$\mathbf{j}(\mathbf{r},t) = e\mathbf{v}_0(t)\delta(\mathbf{r}-\mathbf{r}_0(t)) \rightarrow \mathbf{j}(\mathbf{k},t) = e\mathbf{v}_0(t)\exp\left(-i\mathbf{k}\cdot\mathbf{r}_0(t)\right)$$
$$j_x(\mathbf{k},t) = e v_x(t)\exp\left(-ik_x x(t) - ik_y y(t) - ik_z z(t)\right) \qquad (7.C.21)$$

where the equations of motion for the charged particle in a field $\mathbf{B} = B_0 \hat{\mathbf{z}}$ in the absence of any other fields yield

$$x(t) = X + r_\perp \cos\phi(t) \quad y(t) = Y + r_\perp \sin\phi(t) \quad z(t) = v_\parallel t + z_0 \quad v_x = -v_\perp \sin\phi \qquad (7.C.22)$$

The wave number is represented generally by

$$k_x = k_\perp \cos\alpha \quad k_y = k_\perp \sin\alpha \quad k_z = k_\parallel \qquad (7.C.23)$$

We substitute Eqs.(7.C.22) and (7.C.23) into Eq.(7.C.21) and use the Bessel function identity for a cosine inside the exponential to obtain an expression for $S_j$ from Eq.(7.C.8):

$$S_j(\mathbf{k},\omega) \propto J_\ell^2(k_\perp r_\perp)\delta(\omega - k_\parallel v_\parallel - \ell\Omega) \tag{7.C.24}$$

ignoring collisional resonance broadening. The expression in Eq.(7.C.24) is averaged over the initial gyro phase, $X$, $Y$, and $z_0$, and summed over particles.

Exercise: Work out the details in Eq.(7.C.24).

Having calculated the spectral density for the current from Eq.(7.C.24) one can then evaluate the right side of Eq.(7.C.19). To make progress one must evaluate the denominator, which depends on the wave characteristics (response tensor, dispersion relation, and polarization). For $\mathbf{k}||\mathbf{B}_0$ then $\hat{\mathbf{e}}_k^\ell = \frac{1}{\sqrt{2}}(\hat{\mathbf{x}} \pm \hat{\mathbf{y}})$, and as examples one can consider whistlers, Alfvén waves, or other waves as described in Sec. 4.B. For $\mathbf{k} \perp \mathbf{B}_0$ and $k_\parallel=0$ one might consider ordinary modes (electric field polarized parallel to $\mathbf{B}_0$), extraordinary modes (electric field polarized perpendicular to $\mathbf{B}_0$), or other modes described in Sec. 4.C. An important parameter influencing the intensity of the electromagnetic wave emission is the ratio of the particle energy to its rest mass energy. The more relativistic the particle is the stronger the emission. Relativity affects the equations of motion for the particles doing the emission, and relativity also affects the response tensor. Synchrotron radiation is an interesting example that has received much attention. By considering the emission of a particular electromagnetic wave and the kinetic linear damping of the same mode, as in Sec. 7.A.a for electrostatic modes, one can arrive at a balance equation between emission and damping, and recover the Rayleigh-Jeans relation that each wave has energy $W=T$ where $W$ is the energy per unit $\Delta^3 k \Delta^3 x$.

Exercise: Outline and sketch the details of one or two of the example calculations in the preceding paragraph. Try some of the algebra. Recover the Rayleigh-Jeans relation for electromagnetic emission and damping.

At this point we digress to consider the question of the influence of including collisions on electromagnetic wave emission in the presence of an applied magnetic field. We will conclude that the magnetic field effectively weakens collisional diffusion of the particle trajectories. We sketch what the analysis would be with the inclusion of collision-induced perturbations to the zero-order particle trajectories. The analysis includes the following relations:

$$\begin{aligned}\langle j_x(\mathbf{k},t)j_x^*(\mathbf{k},t-\tau)\rangle &\to \exp(-i\mathbf{k}\cdot\mathbf{r}(t)+i\mathbf{k}\cdot\mathbf{r}(t-\tau)) \\ &\equiv \exp(-i\mathbf{k}\cdot\mathbf{s}(\tau)) = \exp(-ik_x\Delta x(\tau)+...) \\ \Delta x(\tau) &= \Delta X(\tau) + r_\perp\left[\cos\phi(t)-\cos\phi(t-\tau)\right] \quad \Delta y = ... \quad \Delta z = ... \\ \Delta\phi(\tau) &\equiv \phi(t)-\phi(t-\tau) = \Omega\tau + \Delta\phi'(\tau)\end{aligned} \tag{7.C.25}$$

After $\Delta x$ is substituted into the exponential and the Bessel function identity is used, the following expression emerges

$$\langle\exp(i\ell\Delta\phi'(\tau))\rangle \to \exp(-\ell^2\sigma_\phi^2(\tau)) \quad \sigma_\phi^2(\tau) \equiv \langle(\Delta\phi)^2(\tau)\rangle \tag{7.C.26}$$

Recall the analysis of collision-induced resonance broadening and the use of the central limit theorem in Sec. 7.B.g. In fact collisions produce diffusive changes to all of the terms in

$$x(t) = X + \frac{v_\perp}{\Omega}\cos\phi(t) \Rightarrow \delta x(t) = \delta X + \frac{\delta v_\perp}{\Omega}\cos\phi(t) - \frac{v_\perp}{\Omega}\delta\phi\sin\phi(t) \qquad (7.C.27)$$

We can estimate (perhaps naively) the electron velocity diffusion and the diffusion of the various terms in Eq.(7.C.27):

$$\left\langle(\delta v)^2\right\rangle = \left\langle(\delta v_x)^2\right\rangle + \left\langle(\delta v_y)^2\right\rangle + \left\langle(\delta v_z)^2\right\rangle = 3\left\langle(\delta v_x)^2\right\rangle = 2D_v \tau \qquad (7.C.28)$$

and similarly for $\left\langle(\delta x)^2\right\rangle$ and $\left\langle(\delta\phi)^2\right\rangle$:

$$\left\langle(\delta x)^2\right\rangle = r_\perp^2 v_e \tau \quad \left\langle(\delta\phi)^2\right\rangle = 2v_e \tau \quad v_e = D_e^e / v_e^2 \qquad (7.C.29)$$

The first expression in Eq.(7.C.29) is just the classical collisional diffusion across the magnetic field, which scales as the square of the electron Larmor radius. The collisionless magnetized resonance morphs into a resonance function with collisions:

$$\delta(\omega - k_\parallel v_\parallel - \ell\Omega) \Rightarrow R(\omega - k_\parallel v_\parallel - \ell\Omega), \quad \frac{\delta\omega}{\omega} \sim \frac{v_e}{\omega}(k_\perp r_\perp)^2 \qquad (7.C.30)$$

The magnetic field significantly reduces the collisional resonance broadening because of the linear scaling in $v_e$ (rather than $v_e^{1/3}$) and typically $(k_\perp r_\perp)^2 \ll 1$ for electrons. We conclude that the effect of collisions on electron cyclotron and synchrotron radiation is significantly weakened by the applied magnetic field.

### 7.C.c Emission due to collisions - Bremsstrahlung

In this section we calculate collision-induced electromagnetic wave emission, i.e., bremsstrahlung, in the absence of an applied magnetic field. This amounts to calculating the two-time correlation function for the plasma current density including the influence of collisions. The analysis builds on the earlier material in Secs. 7.C.a and 7.C.b.

<u>Definitions:</u> $\quad \bar{\varepsilon} \equiv \omega_k^\ell \hat{\mathbf{e}}^* \cdot \left.\frac{\partial K}{\partial \omega} \cdot \hat{\mathbf{e}}\right|_{\omega_k^\ell} = \frac{\text{Total wave energy}}{\text{Electric field energy}} = O(1) \qquad (7.C.31)$

From Eq.(7.C.20)

$$\left\langle \hat{\mathbf{e}}_k^{\ell*} \cdot j(\mathbf{k},\omega_k^\ell) \hat{\mathbf{e}}_k^\ell \cdot j^*(\mathbf{k}',\omega_k^\ell{}') \right\rangle = (2\pi)^4 \delta(\omega-\omega')\delta(\mathbf{k}-\mathbf{k}')S^j(\mathbf{k},\omega) \qquad (7.C.32)$$

Then Eq.(7.C.19) becomes

$$\left|\dot{W}^\ell(\mathbf{k})\right| = \frac{4\pi}{\bar{\varepsilon}}S^j(\mathbf{k},\omega_k^\ell), \quad \omega_k^\ell > 0 \qquad (7.C.33)$$

The current density in Eq.(7.C.32) is a sum over the particles:

$$\mathbf{j}(\mathbf{x},t) = \sum_i e_i \mathbf{v}_i(t) \delta(\mathbf{x}-\mathbf{r}_i(t)) \quad \rightarrow \quad \mathbf{j}(\mathbf{k},t) = \sum_i e_i \mathbf{v}_i(t) \exp\left[-i\mathbf{k}\cdot\mathbf{r}_i(t)\right] \qquad (7.C.34)$$

where $\mathbf{r}_i = \mathbf{r}_0 + \mathbf{s}_i$. The time derivative of the current density is

$$\frac{\partial}{\partial t}\mathbf{j}(\mathbf{k},t) = \sum_i e_i \left[\dot{\mathbf{v}}_i(t) - i(\mathbf{k}\cdot\mathbf{v}_i(t))\mathbf{v}_i(t)\right]\exp\left[-i\mathbf{k}\cdot\mathbf{r}_i(t)\right] \qquad (7.C.35)$$

We restrict our calculation of bremsstrahlung to long wavelength, non-resonant modes: $k \ll \omega/v$ (dipole approximation). Thus, the second term in the square bracket in Eq.(7.C.35) is negligible; and the particle displacement compared to the wavelength is small inside the exponential in Eq.(7.C.35), i.e.,

$$-i\omega \mathbf{j}(\mathbf{k},\omega) \approx \sum_i e_i \dot{\mathbf{v}}_i(\omega)\exp(-i\mathbf{k}\cdot\mathbf{r}_{0i}) \approx \exp(-i\mathbf{k}\cdot\mathbf{r}_0)\sum_i e_i \dot{\mathbf{v}}_i(\omega) \qquad (7.C.36)$$

and we further assume that the spread of scattering centers is small compared to a wavelength. We next introduce some additional notation.

<u>Definition:</u> $\quad \mathbf{S}_i(t) \equiv \mathbf{b} + \mathbf{v}(t-t_0) \quad \Pi(t) \equiv \sum_i e_i \mathbf{S}_i(t) \quad \ddot{\Pi}(t) = \sum_i e_i \dot{\mathbf{v}}_i(t) \equiv \sum_i \ddot{\Pi}_i \qquad (7.C.37)$

where $\mathbf{b}$ is the impact parameter and is perpendicular to $\mathbf{v}$.

Hence,

$$-i\omega \mathbf{j}(\mathbf{k},\omega) = \exp(-i\mathbf{k}\cdot\mathbf{r}_0)\ddot{\Pi}(\omega) \qquad (7.C.38)$$

We note that e-i collisions produce dipole radiation. However, e-e collisions do not result in a net dipole moment in the collision; but the e-e collisions can produce quadrupole radiation. We will only consider the bremsstrahlung from electron-ion collisions. We calculate the Coulomb force on an electron colliding with an ion:

$$\ddot{\Pi}_i = -e\dot{\mathbf{v}}_i = \frac{e^2}{m_e}\mathbf{E}_e^i = \frac{e^2 e^{ion}}{m_e}\frac{\mathbf{S}_i(t)}{S_i^3(t)} \qquad (7.C.39)$$

and Fourier transforming

$$\ddot{\Pi}_i(\omega) = \frac{e^2 e^{ion}}{m_e}\exp(i\omega t_0)\left[\int_{-\infty}^{\infty}dt'\exp(i\omega t')\frac{\mathbf{b}+\mathbf{v}t'}{(b^2+v^2 t'^2)^{3/2}}\right] \qquad (7.C.40)$$

This expression is valid for small-angle scattering. For smaller $b$ and large angles, the electron trajectory is hyperbolic (see Panofsky and Phillips), but we will ignore this correction. We scale the variables inside the integral in Eq.(7.C.40) as follows:

$$\tau_c \equiv \frac{b}{v} \quad \omega_c \equiv \frac{1}{\tau_c} = \frac{v}{b} \quad \tau' \equiv \frac{t'}{\tau_c} \quad \omega' \equiv \frac{\omega}{\omega_c} = \frac{\omega b}{v} \qquad (7.C.41)$$

Using Eq.(7.C.41), Eq.(7.C.40) becomes

$$\ddot{\Pi}_i(\omega) = \frac{e^2 e^{ion}}{m_e}\exp(i\omega t_0)\left[\frac{\tau_c}{b^2}\int_{-\infty}^{\infty}d\tau'\exp(i\omega'\tau')\frac{\hat{\mathbf{b}}+\hat{\mathbf{v}}\tau'}{(1+\tau'^2)^{3/2}}\right] \qquad (7.C.42)$$

<u>Lemmas:</u>

$$\int_{-\infty}^{\infty} d\tau' \frac{\exp(i\omega'\tau')}{\left(1+\tau'^2\right)^{3/2}} = 2\omega' K_1(\omega') \qquad \int_{-\infty}^{\infty} d\tau' \frac{\tau'\exp(i\omega'\tau')}{\left(1+\tau'^2\right)^{3/2}} = 2i\omega' K_0(\omega') \qquad (7.C.43)$$

where $K_0$ and $K_1$ are modified Bessel functions of the second kind. The asymptotic forms for $K_0$ and $K_1$ are

$$K_0(\omega') \approx \ln\frac{1}{\omega'} \quad \omega' \ll 1$$

$$K_1(\omega') \approx \frac{1}{\omega'}, \quad \omega' \ll 1 \qquad (7.C.44)$$

$$K_n(\omega') \approx \sqrt{\frac{\pi}{2\omega'}} \exp(-\omega'), \quad \omega' \gg 1$$

Equation (7.C.42) becomes

$$\ddot{\Pi}_i(\omega) = \frac{e^2 e^{ion}}{\frac{1}{2}m_e v^2} \exp(i\omega t_0) \left[ \hat{\mathbf{b}}\omega' K_1(\omega') + \hat{\mathbf{v}} i\omega' K_0(\omega') \right] \qquad (7.C.45)$$

We introduce the definition $b_0 \equiv e^2/(\frac{1}{2}m_e v^2)$ and use Eqs.(7.C.38-7.C.45) to obtain

$$\langle j(\mathbf{k},\omega) j^*(\mathbf{k}'',\omega'') \rangle = \left( \frac{e^2 e^{ion}}{\frac{1}{2}m_e v^2} \right)^2 \frac{(2\pi)^4}{VT} \delta(\mathbf{k}-\mathbf{k}'')\delta(\omega-\omega'')$$

$$\times \left\langle \left| \hat{\mathbf{e}}^* \cdot \hat{\mathbf{b}} K_1(\omega') + \hat{\mathbf{e}}^* \cdot \hat{\mathbf{v}} i K_0(\omega') \right|^2 \right\rangle_{\hat{\mathbf{b}},\hat{\mathbf{v}}} \qquad (7.C.46)$$

In deriving Eq.(7.C.46) we made use of the independence of the averaging with respect to $\mathbf{r}_0$ and $t_0$ which leads to the following reductions:

$$\langle \exp(-i(\mathbf{k}\cdot\mathbf{r}_0 - \omega t_0) + i(\mathbf{k}'\cdot\mathbf{r}_0 - \omega' t_0)) \rangle_{\mathbf{r}_0,t_0} = \langle \exp(-i(\mathbf{k}-\mathbf{k}')\cdot\mathbf{r}_0 + i(\omega-\omega')t_0) \rangle_{\mathbf{r}_0,t_0}$$

$$= \langle \exp(-i(\mathbf{k}-\mathbf{k}')\cdot\mathbf{r}_0) \rangle_{\mathbf{r}_0} \langle \exp(i(\omega-\omega')t_0) \rangle_{t_0}$$

$$\langle \exp(-i(\mathbf{k}-\mathbf{k}')\cdot\mathbf{r}_0) \rangle_{\mathbf{r}_0} = \frac{1}{V}\int d^3r\, e^{i(\mathbf{k}-\mathbf{k}')\cdot\mathbf{r}} = \frac{(2\pi)^3 \delta(\mathbf{k}-\mathbf{k}')}{V} \qquad (7.C.47)$$

$$\langle \exp(i(\omega-\omega')t_0) \rangle_{t_0} = \frac{1}{T}\int d^3t\, e^{i(\omega-\omega')t} = \frac{(2\pi)^3 \delta(\omega-\omega')}{T}$$

We use Eqs.(7.C.20) and (7.C.47), and sum over the collisions $N_c$ that take place in the volume $V$ and time period $T$ with identical values of $\mathbf{b}$ and $\mathbf{v}$ to deduce

$$S^j(\mathbf{k},\omega) = \left( \frac{e^2 e^{ion}}{\frac{1}{2}m_e v^2} \right)^2 \frac{N_c(V,T)}{VT} \left\langle \left| \hat{\mathbf{e}}^* \cdot \hat{\mathbf{b}} K_1(\omega') + \hat{\mathbf{e}}^* \cdot \hat{\mathbf{v}} i K_0(\omega') \right|^2 \right\rangle_{\hat{\mathbf{b}},\hat{\mathbf{v}}} \qquad (7.C.48)$$

We now replace $N_c/(VT)$ with $n_c dv db$ in Eq.(7.C.48) and integrate over v and b. In doing this integral, we make use of two identities:

$$\langle \hat{\mathbf{e}}^* \cdot \hat{\mathbf{b}}\, \hat{\mathbf{e}} \cdot \hat{\mathbf{b}}^* \rangle = \frac{1}{3} \qquad \langle \hat{\mathbf{e}}^* \cdot \hat{\mathbf{b}}\, \hat{\mathbf{e}} \cdot \hat{\mathbf{v}}^* \rangle = 0 \qquad (7.C.49)$$

Exercise: Verify the relations in Eq.(7.C.49). Note that **b** and **v** are always perpendicular to one another, and **b** has no preferred direction.

Hence,

$$S^j(\mathbf{k},\omega) = \int d\mathbf{v} \int db \left( \frac{e^2 e^{ion}}{\frac{1}{2} m_e \mathbf{v}^2} \right)^2 n_c(\mathbf{v},b) \left[ \frac{1}{3} K_1^2(\omega') + \frac{1}{3} K_0^2(\omega') \right] \qquad (7.C.50)$$

The number of collisions per unit volume per unit time with speed v and impact parameter b in intervals dv and *db* is determined by the collision rate and is given by:

$$n_c d\mathbf{v}\, db = n_i \sigma \mathbf{v} = n_i \mathbf{v}\, 2\pi b\, db$$
$$\Rightarrow \int\int db\, d^3\mathbf{v}\, f_e(\mathbf{v}) n_i \mathbf{v}\, 2\pi b = \int\int db\, d^3\mathbf{v}\, g_e(\mathbf{v}) n_e n_i \mathbf{v}\, 2\pi b = \int\int db\, d\mathbf{v}\, 4\pi \mathbf{v}^2 g_e(\mathbf{v}) n_e n_i \mathbf{v}\, 2\pi b$$
(7.C.51)

where we have then integrated over velocities and impact parameters. With the use of Eq.(7.C.41) and (7.C.51), Eq.(7.C.50) becomes

$$S^j(\mathbf{k},\omega) = \frac{32}{3}\pi^2 \frac{(e^2 e^{ion})^2}{m_e^2} n_e n_i \int \frac{d\mathbf{v}}{\mathbf{v}} \int b\, db \left[ K_1^2(\omega') + K_0^2(\omega') \right]$$

$$= \frac{32}{3}\pi^2 \frac{(e^2 e^{ion})^2}{m_e^2} n_e n_i \int \frac{d\mathbf{v}}{\mathbf{v}} \int_{\omega'_{max}}^{\omega'_{min}} \omega'\, d\omega'\, \mathbf{v}^2 \left[ K_1^2(\omega') + K_0^2(\omega') \right] \qquad (7.C.52)$$

We use the identity:

$$K_0^2 + K_1^2 = -\frac{1}{\omega'} \frac{d}{d\omega'} \left( \omega' K_0 K_1 \right) \qquad (7.C.53)$$

to do the $\omega'$ integral in Eq.(7.C.53) and include the effects of Debye shielding to restrict the upper bound of the integral to $b < \lambda_D$, i.e., $\omega' < \frac{\omega}{\mathrm{v}} \lambda_D$. For large $\omega'$ we have

$$K_n \sim \sqrt{\frac{\pi}{2\omega'}} e^{-\omega'} \ll 1$$

The upper limit $\omega'_{min}$ of the $\omega'$ integral in Eq.(7.C.52) (minimum impact parameter) is determined by either the bound electron limit $b \sim e^2/m\mathrm{v}^2$ or quantum mechanics $b \sim \lambdabar \sim \hbar/m v$ whichever is larger. For temperatures less than 1 Rydberg (~13.6 eV) the former limit pertains, and for higher temperatures the deBroglie wavelength sets the minimum impact parameter. For high-temperature plasmas, quantum mechanics sets the lower limit on *b*: hence,

$$\frac{1}{\omega'_{min}} \sim \frac{v}{\omega}\frac{mv}{\hbar} = \frac{mv^2}{\hbar\omega} \gg 1 \tag{7.C.54}$$

For small $\omega'$ the small argument limits in Eq.(7.C.44) determine that the upper limit of the $\omega'$ integral in Eq.(7.C.52) is dominant:

$$S^j(\mathbf{k},\omega) = \frac{32}{3}\pi^2 \frac{(e^2 e^{ion})^2}{m_e^2 \omega^2} n_e n_i \int v\, dv\, g_e(v) \ln\frac{mv^2}{\hbar\omega} \tag{7.C.55}$$

We see that the integral over impact parameter is insensitive to the maximum impact parameter cutoff; thus, the plasma shielding is irrelevant.

Example: For a Maxwellian plasma we evaluate the integral in Eq.(7.C.55) as follows

$$g_e(v) = \left(\frac{m}{2\pi T_e}\right)^{3/2} e^{-\frac{mv^2}{2T_e}} \quad\Rightarrow\quad \varepsilon \equiv \frac{\tfrac{1}{2}mv^2}{T_e}$$

$$\int_0^\infty d\varepsilon\, e^{-\varepsilon} \ln\left(\varepsilon \frac{2T_e}{\hbar\omega}\right) = \int_0^\infty d\varepsilon\, e^{-\varepsilon}\left[\ln\varepsilon + \ln\frac{2T_e}{\hbar\omega}\right] = -0.577 + \ln\frac{2T_e}{\hbar\omega} = \ln\frac{2T_e}{1.781\hbar\omega} \approx \ln\frac{T_e}{\hbar\omega}$$

(7.C.56)

We note that 0.577 is Euler's constant. Equation (7.C.55) then becomes

$$S^j(\mathbf{k},\omega) = \frac{1}{2\pi} v_e T_e \frac{\omega_p^2}{\omega^2} \frac{\ln\left(\frac{T_e}{\hbar\omega}\right)}{\ln\Lambda} \qquad v_e \equiv \frac{4}{3}\sqrt{2\pi}\frac{n_i(ee^{ion})^2 \ln\Lambda}{\sqrt{m_e}\, T_e^{3/2}} \tag{7.C.57}$$

In this treatment we see that Bremsstrahlung involves radiation of high-frequency, long-wavelength modes in order that the radiation is weakly damped by the plasma. Given Eq.(7.C.57) the wave emission is

$$\dot{W}_k^\ell = \frac{4\pi}{\bar{\varepsilon}} S^j(\mathbf{k},\omega_\mathbf{k}^\ell) \tag{7.C.58}$$

where $\bar{\varepsilon}$ is given by Eq.(7.C.31). For emission of longitudinal waves $\bar{\varepsilon} \to \omega\, \partial\varepsilon/\partial\omega$.

Example: The Langmuir wave (electron plasma wave = EPW) emission just due to collisions (ignoring Landau emission for the moment) is:

$$\dot{W}_k^{EPW} \approx \frac{\omega_p^4}{\omega^4} v_e T_e \frac{\ln\left(\frac{T_e}{\hbar\omega_p}\right)}{\ln\Lambda} \tag{7.C.59}$$

where $\bar{\varepsilon} \to \omega\, \partial\varepsilon/\partial\omega \approx 2\omega_p^2/\omega^2 \approx 2$, $\omega \approx \omega_p$, and $\Lambda \approx \frac{\lambda_D}{\lambdabar} = \sqrt{\frac{T_e}{4\pi n_e e^2}}\sqrt{\frac{2m_e T_e}{\hbar^2}} \approx \frac{T_e}{\hbar\omega_p}$.

Hence,

$$\dot{W}_k^{EPW} \approx \frac{\omega_p^4}{\omega^4} \nu_e T_e \approx \nu_e T_e \tag{7.C.60}$$

If resonant effects are negligible, i.e., $g\left(v = \frac{\omega}{k}\right) \ll 1$, then the wave damping is solely due to collisions: $|\gamma_k| \approx \frac{1}{2}\nu_e$; and collisional damping of the Langmuir waves can balance emission giving a Rayleigh-Jeans law

$$W_k = \frac{\dot{W}_k}{2|\gamma_k|} = \frac{\dot{W}_k}{\nu_e} \approx T_e \tag{7.C.61}$$

Exercise: Calculate the ratio of the collisional wave emission for Langmuir waves to the corresponding Cerenkov emission (see Sec. 7.A.a).

Example: For transverse waves in a cold plasma

$$K \to \varepsilon - \frac{k^2 c^2}{\omega^2} = 1 - \frac{\omega_p^2}{\omega^2} - \frac{k^2 c^2}{\omega^2} = 1 - \frac{\omega_p^2 + k^2 c^2}{\omega^2}$$

$$\to \omega \frac{\partial K}{\partial \omega} = 2\frac{\omega_p^2 + k^2 c^2}{\omega^2} = 2 \tag{7.C.62}$$

The ratio of the wave energy to electric field energy for transverse waves is two, the same as for Langmuir waves; and this ratio is also two in a vacuum where the electric field energy and the magnetic energy are the same. In the plasma there is finite mechanical energy in the particles, but the magnetic energy in the transverse wave decreases. The transverse wave emission is given by

$$\dot{W}_k^{trans} \approx \frac{\omega_p^2}{\omega_k^2} \nu_e T_e \frac{\ln\left(\frac{T_e}{\hbar\omega_k}\right)}{\ln\left(\frac{T_e}{\hbar\omega_p}\right)} \tag{7.C.63}$$

If we amend our definition of the collision frequency in Eq.(7.C.59) as follows:

$$\nu_e(\omega_\mathbf{k}) \equiv \nu_e \frac{\ln\left(\frac{T_e}{\hbar\omega_\mathbf{k}}\right)}{\ln\Lambda = \ln\frac{T_e}{\hbar\omega_p}} = \frac{4}{3}\sqrt{2\pi} \frac{n_i\left(ee^{ion}\right)^2 \ln\frac{T_e}{\hbar\omega_\mathbf{k}}}{\sqrt{m_e}T_e^{3/2}} \tag{7.C.64}$$

Then by adjusting the logarithmic factors the collisional wave damping can be written compactly as

$$\left|\gamma_\mathbf{k}^{coll}\right| = \frac{1}{2}\nu_e(\omega_\mathbf{k})\frac{\omega_p^2}{\omega_\mathbf{k}^2} \tag{7.C.65}$$

and one recovers the Rayleigh-Jeans law:

$$0 = \frac{dW_k}{dt} = \dot{W}_k - |2\gamma_k|W_k \quad \to \quad W_k = T_e \tag{7.C.66}$$

Exercise: Calculate the collisional wave emission in a magnetized plasma and check whether the Rayleigh-Jeans law is recovered once again.

Example: Generalization to non-Maxwellian velocity distribution functions – In going from Eq.(7.C.51) to (7.C.56), there is a velocity-space integration over an isotropic distribution function:

$$S^j(\mathbf{k},\omega) \propto \int d^3\mathrm{v}\, g_e(\mathbf{v}) \ln\frac{m\mathrm{v}^2}{\hbar\omega} \sim \int d\mathrm{v}\, \mathrm{v}\, g_e(\mathrm{v}) \ln\frac{m\mathrm{v}^2}{\hbar\omega}$$

$$\sim \int d^3\mathrm{v}\, \frac{1}{\mathrm{v}} g_e(\mathrm{v}) \ln\frac{2T_e}{\hbar\omega} \sim \ln\frac{2T_e}{\hbar\omega}\left\langle \frac{1}{\mathrm{v}}\right\rangle \tag{7.C.67}$$

Hence, the generalization of Bremsstrahlung to a non-Maxwellian velocity distribution is

$$\dot{W}_{Bremss} = \left(\dot{W}_{Bremss}\right)^{Maxw} \frac{\langle \mathrm{v}^{-1}\rangle}{\langle \mathrm{v}^{-1}\rangle^{Maxw}} \tag{7.C.68}$$

How is the collisional damping of the emitted waves modified by a non-Maxwellian velocity distribution? For Coulomb collisions

$$\gamma_{coll} \propto \nu_{ei} \approx n_i\langle \sigma \mathrm{v}\rangle \propto n_i\left\langle \frac{1}{\mathrm{v}^4}\mathrm{v}\right\rangle \propto n_i\langle \mathrm{v}^{-3}\rangle$$

$$\rightarrow \gamma_{coll} \propto \nu_{ei} = \left(\nu_{ei}\right)^{Maxw} \frac{\langle \mathrm{v}^{-3}\rangle}{\langle \mathrm{v}^{-3}\rangle^{Maxw}} \tag{7.C.69}$$

The <v$^{-3}$> moment would be logarithmically divergent at its lower limit were it not for being cutoff at v$_{th}$/Λ instead of 0:

$$\langle \mathrm{v}^{-3}\rangle = 4\pi\int_0^\infty \mathrm{v}^2 d\mathrm{v}\, \mathrm{v}^{-3} g_e(\mathrm{v}) = 4\pi\int_0^\infty d\mathrm{v}\, \mathrm{v}^{-1} g_e(\mathrm{v}) \rightarrow 4\pi\int_{\mathrm{v}_{th}/\Lambda}^\infty d\mathrm{v}\, \mathrm{v}^{-1} g_e(\mathrm{v}) \sim \ln\Lambda \tag{7.C.70}$$

We then use the results of Eqs.(7.C.68) and (7.C.69) to obtain

$$W_k = \frac{\dot{W}_{Bremss}}{2|\gamma_{coll}|} = \frac{\langle \mathrm{v}^{-1}\rangle}{\langle \mathrm{v}^{-3}\rangle}\left[\frac{\langle \mathrm{v}^{-1}\rangle^{Maxw}}{\langle \mathrm{v}^{-3}\rangle^{Maxw}}\right]^{-1} T_e \tag{7.C.71}$$

Thus, $W_k$ cannot differ significantly from $T_e$.

We return to Eqs.(7.C.63) and (7.C.64) to calculate the integrated Bremsstrahlung transverse radiation:

$$\dot{W}_{Bremss} = \int \frac{d^3\mathbf{k}}{(2\pi)^3} \dot{W}_{\mathbf{k}} = \int \frac{d^3\mathbf{k}}{(2\pi)^3} \nu_e(\omega) \frac{\omega_p^2}{\omega^2} T_e \times 2(\text{polarizations}) =$$

$$= \frac{\omega_p^2}{(2\pi)^3} 2T_e \int \frac{4\pi k^2 dk}{\omega^2} \nu_e(\omega) = \frac{1}{\pi^2} \nu_e T_e \left(\frac{\omega_p}{c}\right)^3 \int \frac{d\omega}{\omega} \sqrt{\frac{\omega^2}{\omega_p^2} - 1} \frac{\ln\left(\frac{\omega_\hbar}{\omega}\right)}{\ln\left(\frac{\omega_\hbar}{\omega_p}\right)} \quad (7.C.72)$$

where we have made use of $\omega^2 = k^2 c^2 + \omega_p^2$ and the quantum cutoff at high frequencies: $\hbar\omega_\hbar \equiv T_e$, $\frac{\omega_\hbar}{\omega_p} = \Lambda = \frac{\lambda_D}{\lambda}$. Ordinarily, $\omega_\hbar \gg \omega_p$. Introducing the notation $\dot{W}_{Bremss} \equiv \int d\omega \, \dot{W}_{Bremss}(\omega)$, we see from Eq.(7.C.72) that $\dot{W}_{Bremss}(\omega)$ is zero below $\omega = \omega_p$ because the transverse waves are evanescent, becomes positive for higher frequencies, and then is relatively flat until it cuts off at $\omega_\hbar$. For the frequencies near the cutoff a Bethe-Heitler calculation is needed.[52] To do the integral in Eq.(7.C.72) we note that

$$\int_1^\Lambda \frac{d\Omega}{\Omega} \sqrt{\Omega^2 - 1} \left[1 - \frac{\ln\Omega}{\ln\Lambda}\right] \approx \frac{\Lambda}{\ln\Lambda}, \quad \Lambda \gg 1, \quad \Omega \equiv \frac{\omega}{\omega_p} \quad (7.C.72)$$

Finally, the Bremsstrahlung transverse-wave emission rate can then be expressed as

$$\dot{W}_{Bremss} \approx \frac{n_e n_i T_e^{1/2} e^4 (e^{ion})^2}{\hbar m_e^{3/2} c^3} \quad (7.C.73)$$

times a numerical factor due to important quantum mechanical corrections that we have ignored. For an optically thin medium one can observe the Bremsstrahlung spectrum. For a thick medium, quantum mechanics takes over, and one gets black-body radiation and the Stefan-Boltzmann law.

### 7.C.d Dawson-Oberman theory of resistivity including scattering due to waves and particle discreteness

The calculation of the frequency dependence of the plasma resistivity in the presence of electric fields must include wave scattering by discrete fluctuations and particles. We follow the theory of Dawson and Oberman.[53] Consider a very long wavelength wave:

$$\mathbf{E}(\mathbf{x},t) = \mathbf{E}_0 \sin(\mathbf{k}_0 \cdot \mathbf{x} - \omega_0 t + \theta) \approx \mathbf{E}_0 \sin(\omega_0 t), \quad k_0 \ll k_D \quad (7.C.74)$$

---

[52] H. A. Bethe and L. C. Maximon, Phys. Rev. 93, 768 (1954).
[53] J. M. Dawson and C. Oberman, Phys. Fluids **5**, 517 (1962).

All of the electrons oscillate together in the field, and the ions oscillate the other way. Define the oscillatory frame of the electrons by the velocity coordinate $\mathbf{w}(t)$:

$$\mathbf{w}(t) = \frac{e}{m_e \omega_0} \mathbf{E}_0 \cos\omega_0 t \quad m_e \dot{\mathbf{w}} = -e\mathbf{E}_0 \sin\omega_0 t \tag{7.C.75}$$

The electron acceleration in the oscillatory frame will see the real forces and a pseudo force:

$$m_e \dot{\mathbf{v}} = -m_e \dot{\mathbf{w}} - e\mathbf{E}_0 \sin\omega_0 t - e\mathbf{E}^{other} = -e\mathbf{E}^{other} \tag{7.C.76}$$

For the ions we have

$$m_i \dot{\mathbf{v}} = -m_i \dot{\mathbf{w}} + e\mathbf{E}_0 \sin\omega_0 t + e\mathbf{E}^{other}$$

$$\rightarrow \quad \dot{\mathbf{v}} = -\dot{\mathbf{w}} + \frac{e}{m_i}\mathbf{E}_0 \sin\omega_0 t + \frac{e}{m_i}\mathbf{E}^{other} = \frac{e}{m_e}\mathbf{E}_0 \sin\omega_0 t + \frac{e}{m_i}\mathbf{E}_0 \sin\omega_0 t + \frac{e}{m_i}\mathbf{E}^{other} \tag{7.C.77}$$

$$\approx \frac{e}{m_e}\mathbf{E}_0 \sin\omega_0 t + \frac{e}{m_i}\mathbf{E}^{other} + O\left(\frac{m_e}{m_i}\right)$$

In the laboratory frame of reference the equations of motion for the electrons yield

$$m_e \ddot{\mathbf{r}}_e = -e\mathbf{E}_0 \sin\omega_0 t \quad \rightarrow \quad \mathbf{r}_e(t) = \frac{e}{m_e \omega_0^2}\mathbf{E}_0 \sin\omega_0 t \tag{7.C.78}$$

while the ion motion in the oscillatory frame of reference is

$$\mathbf{r}_i(t) = -\frac{e}{m_e \omega_0^2}\mathbf{E}_0 \sin\omega_0 t \tag{7.C.79}$$

absent "other fields" and microscopic fields. The "other fields" satisfy Maxwell's equations combined with the Vlasov or Lenard-Balsescu equations. In the electron oscillating frame the ions move up and down, with a dipole moment that is $2(e^2 E_0/m_e\omega_0^2)$. There is dipole radiation in all directions, but due to the randomness of positions (say) the radiation is incoherent. Thus, the original wave has been scattered or partially scattered into dipole radiation in the random-phase approximation.

Suppose one ion wave is present. Then over and above graininess there is a density correlation. Thus, there is some coherence and enhancement of the radiation. An electron wave can scatter off the ion wave into a scattered electron wave. If instead there is turbulence (due to an instability or other mechanism) so that there are many waves present, there will be multi-enhanced scattering in all directions. This is observed in laboratory and ionospheric experiments.

We return to the ion motion in the oscillating frame, Eq.(7.C.79), and construct the Fourier-transformed charge density of a singly charged ion:

$$\rho_0(\mathbf{k},t) = e\exp(-i\mathbf{k}\cdot\mathbf{r}_i(t)) \approx e(1 - i\mathbf{k}\cdot\mathbf{r}_i(t)) \tag{7.C.80}$$

where we have expanded the exponential for $kr_i \ll 1$ in the dipole approximation. For electrostatic modes, the shielded electric potential can be calculated from the test-particle charge density using Poisson's equation:

$$\phi(\mathbf{k},\omega) = \frac{4\pi}{k^2}\frac{\rho_0(\mathbf{k},\omega)}{\varepsilon(\mathbf{k},\omega)} = \frac{4\pi}{k^2}\frac{e^2}{2m_e\omega_0^2}\frac{\mathbf{k}\cdot\mathbf{E}_0}{\varepsilon(\mathbf{k},\omega)}2\pi\left[\delta(\omega+\omega_0)-\delta(\omega-\omega_0)\right] \qquad (7.C.81)$$

The electric field at the ion position is then

$$\begin{aligned}\mathbf{E}(\mathbf{r}_0(t),t) &= \int\frac{d^3\mathbf{k}}{(2\pi)^3}\int\frac{d\omega}{2\pi}e^{i\mathbf{k}\cdot\mathbf{r}_0(t)}e^{-i\omega t}(-i\mathbf{k})\phi(\mathbf{k},\omega)\\ &\approx \int\frac{d^3\mathbf{k}}{(2\pi)^3}\int\frac{d\omega}{2\pi}\left(1+i\mathbf{k}\cdot\mathbf{r}_0(t)\right)e^{-i\omega t}(-i\mathbf{k})\phi(\mathbf{k},\omega) \qquad (7.C.82)\\ &\approx \frac{4\pi e^2}{2m_e\omega_0^2}\int\frac{d^3\mathbf{k}}{(2\pi)^3}\left(-i\hat{\mathbf{k}}\hat{\mathbf{k}}\cdot\mathbf{E}_0\right)\left[\frac{e^{i\omega_0 t}}{\varepsilon(\mathbf{k},-\omega_0)} - \frac{e^{-i\omega_0 t}}{\varepsilon(\mathbf{k},\omega_0)}\right]\end{aligned}$$

We assume that the unperturbed plasma is isotropic in order to simplify and combine terms in Eq.(7.C.82). Then $\varepsilon(\mathbf{k},-\omega_0) = \varepsilon(-\mathbf{k},-\omega_0) = \varepsilon^*(\mathbf{k},\omega_0)$. Hence,

$$\begin{aligned}\mathbf{E}(\mathbf{r}_0(t),t) &\approx -\frac{4\pi e^2}{m_e\omega_0^2}\int\frac{d^3\mathbf{k}}{(2\pi)^3}\left(\hat{\mathbf{k}}\hat{\mathbf{k}}\cdot\mathbf{E}_0\right)\mathrm{Im}\left[\frac{e^{-i\omega_0 t}}{\varepsilon(\mathbf{k},\omega_0)}\right]\\ &= -\frac{4\pi e^2}{m_e\omega_0^2}\int\frac{d^3\mathbf{k}}{(2\pi)^3}\left(\tfrac{1}{3}\mathbf{I}\cdot\mathbf{E}_0\right)\mathrm{Im}\left[\frac{e^{-i\omega_0 t}}{\varepsilon(k,\omega_0)}\right] = \frac{4\pi e^2}{3m_e\omega_0^2}\int\frac{d^3\mathbf{k}}{(2\pi)^3}\mathbf{E}_0\,\mathrm{Im}\left[\frac{e^{-i\omega_0 t}}{\varepsilon(k,\omega_0)}\right]\end{aligned} \qquad (7.C.83)$$

The work done on the plasma by the fields is

$$-\dot{W} \equiv e\mathbf{E}(\mathbf{r}_0(t),t)\cdot\dot{\mathbf{r}}_0(t) \Rightarrow \dot{W} = -\frac{2\pi e^4}{3m_e^2\omega_0^3}\mathbf{E}_0^2\int\frac{d^3\mathbf{k}}{(2\pi)^3}\mathrm{Im}\left[\frac{1}{\varepsilon(k,\omega_0)}\right] \qquad (7.C.84)$$

The resonant part in Eq.(7.C.84) is interpreted as the wave-emission part (normal modes), while the non-resonant part is interpreted as the resistive component. We first examine the resonant contribution:

$$\mathrm{Im}\left[\frac{1}{\varepsilon(k,\omega_0)}\right] = \mathrm{Im}\left[\frac{1}{\varepsilon'+i\varepsilon''}\right] = -\pi\delta(\varepsilon')\,\mathrm{sign}\,\varepsilon'' = -\pi\delta(\varepsilon'(k,\omega_0)) \qquad (7.C.85)$$

taking the positive sign $\varepsilon''$ for stable modes. The integral over wavenumber space in (7.C.84) can be performed by using the relations:

$$\delta(\varepsilon'(k,\omega_0)) = \frac{\delta(k-k(\omega_0))}{\left|\frac{\partial\varepsilon'(k,\omega)}{\partial k}\right|_{k(\omega_0)}} \quad\text{and}\quad \omega_0^2 = \omega_p^2 + 3k^2 v_e^2 \Rightarrow k(\omega_0) = \frac{1}{\sqrt{3}\,v_e}\sqrt{\omega_0^2-\omega_p^2} \qquad (7.C.86)$$

Hence, the energy radiated into the weakly damped, resonant radiation fields derived from Eqs.(7.C.84-7.C.86) is

$$\dot{W}_{rad} = \frac{1}{18\sqrt{3}}\left(\frac{e^2}{T_e}\right)^2\left(\frac{\omega_p}{\omega_0}\right)^3\sqrt{\left(\frac{\omega_p}{\omega_0}\right)^2-1}\,v_e E^2,\quad \omega_0 \geq \omega_p \qquad (7.C.87)$$

In the oscillating frame we have radiated waves travelling out in all directions. In the limit of small amplitude fields, the response field and the radiated waves can be superposed.

## 7.C.e Scattering cross-sections for waves calculated from Dawson-Oberman theory

The scattering cross-section for the incident waves scattering into the emitted radiation can be defined and calculated as follows:

$$\sigma \equiv \frac{\text{energy radiated per unit time}}{\text{incident energy flux density}} = \frac{\dot{W}_{rad}}{\frac{1}{2}\frac{E_0^2}{8\pi}\bar{\varepsilon}V_g^{inc}} \quad (7.C.88)$$

where the ½ in the denominator takes into account a time average of $(\mathbf{E}_0\sin\omega_0 t)^2$ and $V_g = 3k^{inc}v_e^2/\omega_p$ for incident Langmuir waves and $V_g = c\sqrt{1 - \frac{\omega_p^2}{\omega_0^2}}$ for transverse waves.

The scattering cross-section for transverse waves scattered into longitudinal waves is

$$\sigma_{t\to\ell}(\omega_0) = \frac{4\pi}{9\sqrt{3}}\left(\frac{e^2}{T_e}\right)^2\left(\frac{\omega_p}{\omega_0}\right)^2 \frac{v_e}{c} \quad (7.C.89)$$

For longitudinal waves scattered into longitudinal wave the cross-section is

$$\sigma_{\ell\to\ell}(\omega_0) = \frac{4\pi}{27}\left(\frac{e^2}{T_e}\right)^2\left(\frac{\omega_p}{\omega_0}\right)^3 \quad (7.C.90)$$

In order that the scattered longitudinal wave be weakly damped $\frac{\omega_p}{\omega} \approx O(1)$. We note that for comparison the cross-section for particle scattering in a plasma is $O(1)(e^2/T_e)^2\ln\Lambda$.

The cross-section for transverse waves scattering into transverse waves is calculated with the help of the formalism presented in previous lectures yielding the Thomson cross-section:

$$\sigma_{t\to t}(\omega_0) = \frac{8\pi}{3}\left(\frac{e^2}{m_e c^2}\right)^2 \quad (7.C.91)$$

The cross-section for scattering of longitudinal waves into transverse waves is

$$\sigma_{\ell\to t}(\omega_0) = \frac{8\pi}{3}\left(\frac{e^2}{m_e c^2}\right)^2 \frac{c}{\sqrt{3}v_e}\frac{\omega_p}{\omega_0} \quad (7.C.92)$$

We return to Eq.(7.C.84) and consider the non-resonant contributions:

$$\dot{W} = -\frac{2\pi e^4}{3m_e^2\omega_0^3} E_0^2 \int \frac{d^3\mathbf{k}}{(2\pi)^3} \operatorname{Im}\left[\frac{1}{\varepsilon(\mathbf{k},\omega_0)}\right] = -\frac{2\pi e^4}{3m_e^2\omega_0^3} E_0^2 \int \frac{d^3\mathbf{k}}{(2\pi)^3}\left[\frac{-\varepsilon''}{|\varepsilon|^2} \to 1\right]$$

$$= \frac{1}{3\sqrt{2\pi}} \frac{e^4 \omega_p^2}{m_e^2 \omega_0^2} \frac{E_0^2}{v_e^3} \int_0^{k_{max} \sim \tilde{\lambda}_{dB}^{-1}} \frac{dk}{k} \exp\left(-\frac{\omega_0^2}{2k^2 v_e^2}\right) = \frac{1}{3\sqrt{2\pi}} \frac{e^4 \omega_p^2}{m_e^2 \omega_0^2} \frac{E_0^2}{v_e^3} \ln\left(\frac{T_e}{\hbar\omega_0}\right) \quad (7.C.93)$$

$$= m_e \bar{w}^2 v_e \frac{\ln\left(\frac{T_e}{\hbar\omega_0}\right)}{\ln\Lambda} \equiv m_e \bar{w}^2 v_e(\omega), \qquad \bar{w}^2 \equiv \frac{1}{2}\left(\frac{eE_0}{m_e\omega_0}\right)^2$$

From Eq.(7.C.93) we can immediately deduce the cross-section for absorption from Eq.(7.C.88):

$$\sigma_{abs} = \frac{\dot{W}}{\frac{1}{2}\frac{E_0^2}{8\pi}\bar{\varepsilon}V_g^{inc}} = \frac{v_e(\omega)}{nV_g^{inc}}\frac{\omega_p^2}{\omega^2} \quad (7.C.94)$$

where $V_g^{inc}$ is the group velocity of the incident longitudinal or transverse wave and we have specialized to the case of a singly charge ion in an approximately quasi-neutral plasma. We note that the cross-section for collisional absorption is dominant over the other cross-sections by a term of order $\ln\Lambda$. The damping rate for the incident wave energy flux for either longitudinal or transverse waves is

$$2|\gamma| = n\sigma_{abs} V_g^{inc} = v_e(\omega)\frac{\omega_p^2}{\omega^2} \quad (7.C.95)$$

This calculation has ignored correlations of the ions due to ion waves present, which Dawson and Oberman have examined.

### 7.C.f More general derivation of scattering theory for particles, fluctuations, or waves

Here we develop a more general theory of the scattering of an incident wave by particles, shielding clouds, fluctuations, or waves, including wave turbulence. If the scattering involves an outgoing wave we have a formalism for wave emission:[54]

$$\dot{W}^b(\mathbf{k}) = \frac{4\pi}{\bar{\varepsilon}} S^{\mathbf{j}\hat{\mathbf{e}}^*}(\mathbf{k},\omega_\mathbf{k}^b), \qquad \mathbf{j}(\mathbf{x},t) = \sum_s e_s n_s \mathbf{v}_s = \sum_s e_s \int d^3\mathbf{v}\, \mathbf{v} F_s(\mathbf{x},\mathbf{v};t) \quad (7.C.96)$$

where $b$ denotes a branch of the dispersion relation for the emitted waves. We use the Klimontovich representation for the velocity distribution function assembled from the particles:

$$F_s(\mathbf{x},\mathbf{v};t) = \sum_i \delta\left(\mathbf{x} - \mathbf{x}_i^s(t)\right)\delta\left(\mathbf{v} - \mathbf{v}_i^s(t)\right) \quad (7.C.97)$$

which satisfies the Vlasov equation. The time derivative of the current is then

---

[54] T D. A. Tidman and T. H. Dupree, Phys. Fluids **8**, 1860 (1965).

$$\frac{\partial}{\partial t}\mathbf{j}(\mathbf{x},t)=\sum_s e_s\int d^3\mathbf{v}\,\mathbf{v}\left(-\mathbf{v}\cdot\frac{\partial F_s}{\partial \mathbf{x}}-\frac{\partial}{\partial \mathbf{v}}\cdot\left(\frac{e_s}{m_s}\mathbf{E}F_s\right)\right)=-\frac{\partial}{\partial \mathbf{x}}\cdot\sum_s e_s\int d^3\mathbf{v}\,\mathbf{v}\mathbf{v}F_s+\sum_s\frac{e_s^2}{m_s}n_s\mathbf{E}(\mathbf{x},t)$$

(7.C.98)

We next Fourier transform Eq.(7.C.98) in time and space:

$$-i\omega\mathbf{j}(\mathbf{k},\omega)=-i\mathbf{k}\cdot\sum_s e_s\int d^3\mathbf{v}\,\mathbf{v}\mathbf{v}F_s(\mathbf{k},\mathbf{v};\omega)$$

$$+\sum_s\frac{e_s^2}{m_s}\int\frac{d^3\mathbf{k}_1 d\omega_1}{(2\pi)^4}\int\frac{d^3\mathbf{k}_2 d\omega_2}{(2\pi)^4}n_s(\mathbf{k}_2,\omega_2)\mathbf{E}(\mathbf{k}_1,\omega_1)(2\pi)^4\delta(\mathbf{k}_1+\mathbf{k}_2-\mathbf{k})\delta(\omega_1+\omega_2-\omega)$$

(7.C.99)

The first term on the right side of Eq.(7.C.99) is of order $i(\mathbf{k}\cdot\mathbf{v})\mathbf{j}$, and for $\omega\gg\mathbf{k}\cdot\mathbf{v}$ this term is negligible compared to the second term. If $\omega/k\gg v$ is not satisfied then one must consider nonlinear Landau damping.[55]

We recall the development in Eqs.(7.C.20 – 7.C.34) to obtain an expression for the wave emission. Given the relations $-i\omega\mathbf{j}(\mathbf{k},\omega)\cdot\hat{\mathbf{e}}^*=-i\omega\hat{\mathbf{e}}^*\cdot\mathbf{j}(\mathbf{k},\omega)$ and from Eq.(7.C.20)

$$\left\langle\hat{\mathbf{e}}^*\cdot\mathbf{j}(\mathbf{k},\omega)\hat{\mathbf{e}}\cdot\mathbf{j}^*(\mathbf{k}',\omega')\right\rangle=(2\pi)^4\delta(\omega-\omega')\delta(\mathbf{k}-\mathbf{k}')S^{\hat{\mathbf{e}}^*\cdot\mathbf{j}}(\mathbf{k},\omega) \qquad (7.C.100)$$

The analysis can be simplified in special cases. For example, there are no density perturbations for electric fields due to transverse waves. Hence, the electric field and the density perturbations are uncorrelated; and following ensemble average can be simplified:

$$\left\langle\hat{\mathbf{e}}\cdot\mathbf{E}(\mathbf{k}_1,\omega_1)\hat{\mathbf{e}}\cdot\mathbf{E}^*(\mathbf{k}_1^{*'},\omega_1')n_s(\mathbf{k}_2,\omega_2)n_{s'}^*(\mathbf{k}_2^{*'},\omega_2')\right\rangle=$$
$$\left\langle\hat{\mathbf{e}}\cdot\mathbf{E}(\mathbf{k}_1,\omega_1)\hat{\mathbf{e}}\cdot\mathbf{E}^*(\mathbf{k}_1^{*'},\omega_1')\right\rangle\left\langle n_s(\mathbf{k}_2,\omega_2)n_{s'}^*(\mathbf{k}_2^{*'},\omega_2')\right\rangle$$

(7.C.101)

Then using Eqs.(7.C.99-7.C.101) with the assumption $\omega\gg\mathbf{k}\cdot\mathbf{v}$ we obtain the following relation after doing the integrations and making use of the three four-fold delta-functions:

$$S^{\hat{\mathbf{e}}^*\cdot\mathbf{j}}(\mathbf{k},\omega)\propto\int\frac{d^3\mathbf{k}_1 d\omega_1}{(2\pi)^4}S^{\mathbf{E}\cdot\hat{\mathbf{e}}^*}(\mathbf{k}_1,\omega_1)S^{n_s n_{s'}}(\mathbf{k}_2=\mathbf{k}-\mathbf{k}_1,\omega_2=\omega-\omega_1) \qquad (7.C.102)$$

The electron clouds surrounding ions constitute scattering centers. When an electron wave comes by, the ions cannot keep up; and the oscillating electron clouds induce positive electron holes that oscillate with the electron clouds. No net oscillating dipole moment results. However, ion fluctuations are accompanied by electron cloud oscillations; and there is a net oscillating dipole moment that radiates.

---

[55] Last chapter in R. C. Davidson, *Methods in Nonlinear Plasma Theory* (Academic Press, New York, 1972).

Example: Consider the scattering of transverse waves by a longitudinal wave or a fluctuation. The transverse waves have no accompanying density perturbation (at small amplitude). In an unmagnetized plasma the wave emission given in Eq.(7.C.96) is

$$\dot{W}^{\hat{e}}(\mathbf{k}) = \frac{4\pi}{\varepsilon} S^{\mathbf{j}\hat{e}}(\mathbf{k},\omega), \qquad \mathbf{j}(\mathbf{x},t) = \sum_s e_s n_s(\mathbf{x},t) \mathbf{u}_s(\mathbf{x},t) \tag{7.C.103}$$

Here $\hat{\mathbf{e}} = \hat{\mathbf{e}}^*$ and

$$\hat{\mathbf{e}} \cdot \mathbf{j}(\mathbf{k},\omega) = \sum_s e_s \int \frac{d^4\mathbf{k}_1}{(2\pi)^4} \int \frac{d^4\mathbf{k}_2}{(2\pi)^4} \delta^{(4)}(\mathbf{k}_1 + \mathbf{k}_2 - \mathbf{k}) n_s(\mathbf{k}_2^{(4)}) \hat{\mathbf{e}} \cdot \mathbf{u}_s(\mathbf{k}_1^{(4)}) (2\pi)^4 \tag{7.C.104}$$

where

$$d^4\mathbf{k}_1 \equiv d^3\mathbf{k}_1 d\omega_1, \ \delta^{(4)}(\mathbf{k}_1 + \mathbf{k}_2 - \mathbf{k}) \equiv \delta(\mathbf{k}_1 + \mathbf{k}_2 - \mathbf{k})\delta(\omega_1 + \omega_2 - \omega), \text{ and } \mathbf{k}_1^{(4)} = (\mathbf{k}_1, \omega_1). \tag{7.C.105}$$

Using Eq.(7.C.104) we can now form

$$\begin{aligned}\left\langle \hat{\mathbf{e}} \cdot \mathbf{j}(\mathbf{k}^{(4)}) \hat{\mathbf{e}} \cdot \mathbf{j}^*(\mathbf{k}'^{(4)}) \right\rangle &= \sum_{s,s'} e_s e_{s'} \int \frac{d^4\mathbf{k}_1}{(2\pi)^4} \int \frac{d^4\mathbf{k}_2}{(2\pi)^4} \int \frac{d^4\mathbf{k}_1'}{(2\pi)^4} \int \frac{d^4\mathbf{k}_2'}{(2\pi)^4} (2\pi)^8 \delta^{(4)}(\mathbf{k}_1 + \mathbf{k}_2 - \mathbf{k}) \\ &\quad \times \delta^{(4)}(\mathbf{k}_1' + \mathbf{k}_2' - \mathbf{k}') \left\langle n_s(\mathbf{k}_2^{(4)}) n_{s'}(\mathbf{k}_2^{(4)}{}') \hat{\mathbf{e}} \cdot \mathbf{u}_s(\mathbf{k}_1^{(4)}) \hat{\mathbf{e}} \cdot \mathbf{u}_{s'}(\mathbf{k}_1^{(4)}{}') \right\rangle \\ &= \sum_{s,s'} e_s e_{s'} \int \frac{d^4\mathbf{k}_1}{(2\pi)^4} \int \frac{d^4\mathbf{k}_2}{(2\pi)^4} \int \frac{d^4\mathbf{k}_1'}{(2\pi)^4} \int \frac{d^4\mathbf{k}_2'}{(2\pi)^4} (2\pi)^8 \delta^{(4)}(\mathbf{k}_1 + \mathbf{k}_2 - \mathbf{k}) \\ &\quad \times \delta^{(4)}(\mathbf{k}_1' + \mathbf{k}_2' - \mathbf{k}') \left\langle n_s(\mathbf{k}_2^{(4)}) n_{s'}(\mathbf{k}_2^{(4)}{}') \right\rangle \left\langle \hat{\mathbf{e}} \cdot \mathbf{u}_s(\mathbf{k}_1^{(4)}) \hat{\mathbf{e}} \cdot \mathbf{u}_{s'}(\mathbf{k}_1^{(4)}{}') \right\rangle \end{aligned} \tag{7.C.106}$$

The ensemble averages $\left\langle n_s(\mathbf{k}_2^{(4)}) n_{s'}(\mathbf{k}_2^{(4)}{}') \right\rangle$ and $\left\langle \hat{\mathbf{e}} \cdot \mathbf{u}_s(\mathbf{k}_1^{(4)}) \hat{\mathbf{e}} \cdot \mathbf{u}_{s'}(\mathbf{k}_1^{(4)}{}') \right\rangle$ force $\mathbf{k}_1^{(4)} = \mathbf{k}_1^{(4)}{}'$ and $\mathbf{k}_2^{(4)} = \mathbf{k}_2^{(4)}{}'$, which when combined with the two delta functions in Eq.(7.C.106) constrain $\mathbf{k}^{(4)} = \mathbf{k}^{(4)}{}'$. Hence, all but one of the integrations in Eq.(7.C.106) can be performed yielding

$$\left\langle \hat{\mathbf{e}} \cdot \mathbf{j}(\mathbf{k}^{(4)}) \hat{\mathbf{e}} \cdot \mathbf{j}^*(\mathbf{k}'^{(4)}) \right\rangle = \sum_{s,s'} e_s e_{s'} (2\pi)^4 \delta^{(4)}(\mathbf{k} - \mathbf{k}') \int \frac{d^4\mathbf{k}_1}{(2\pi)^4} S^{n_s n_{s'}}(\mathbf{k} - \mathbf{k}_1) S^{\hat{\mathbf{e}} \cdot \mathbf{u}_s, \hat{\mathbf{e}} \cdot \mathbf{u}_{s'}}(\mathbf{k}_1) \tag{7.C.107}$$

Now we can evaluate the wave emission rate in Eq.(7.C.103):

$$\dot{W}^{\hat{\mathbf{e}}}(\mathbf{k}) = \frac{4\pi}{\varepsilon} \sum_{s,s'} e_s e_{s'} \int \frac{d^4\mathbf{k}_1}{(2\pi)^4} S^{n_s n_{s'}}(\mathbf{k} - \mathbf{k}_1) S^{\hat{\mathbf{e}} \cdot \mathbf{u}_s, \hat{\mathbf{e}} \cdot \mathbf{u}_{s'}}(\mathbf{k}_1) \tag{7.C.108}$$

Because the basic nonlinearity in the current has been incorporated in terms of the bilinearity in $n$ and $\mathbf{u}$, we can use linear theory to evaluate $n$ and $\mathbf{u}$. The velocity in the transverse wave is

$$-i\omega_1 \mathbf{u}^s = \frac{e_s}{m_s} \mathbf{E} \qquad \hat{\mathbf{e}} \cdot \mathbf{u}^s = \frac{1}{-i\omega_1} \frac{e_s}{m_s} \hat{\mathbf{e}} \cdot \hat{\mathbf{e}}_1 E \tag{7.C.109}$$

Because of the inverse mass dependence in the velocity, only the electron contribution to the scattering is important. Hence, Eq.(7.C.108) becomes

$$\dot{W}^{\hat{e}}(\mathbf{k}) = \frac{4\pi}{\bar{\varepsilon}} e^2 \int \frac{d^4 \mathbf{k}_1}{(2\pi)^4} S^{nn}(\mathbf{k}-\mathbf{k}_1) S^{\hat{e}_1 \cdot \mathbf{u}\hat{e}_1 \cdot \mathbf{u}}(\mathbf{k}_1)$$

$$= \frac{4\pi}{\bar{\varepsilon}} \frac{e^4}{m^2} \left|\hat{\mathbf{e}} \cdot \hat{\mathbf{e}}_1\right|^2 \int \frac{d^3 \mathbf{k}_1 d\omega_1}{(2\pi)^4} S^{nn}(\mathbf{k}-\mathbf{k}_1) \frac{1}{\omega_1^2} S^E(\mathbf{k}_1)$$

(7.C.110)

Suppose the incident wave "1" has a sharp frequency distribution:

$$S_E(\mathbf{k}_1, \omega_1) = \pi \left[ S_E(\mathbf{k}_1) \delta(\omega_1 - \omega_{\mathbf{k}1}) + S_E(-\mathbf{k}_1) \delta(\omega_1 + \omega_{\mathbf{k}1}) \right]$$

(7.C.111)

which delta functions are a good representation for weak damping $\gamma/\omega_{k1} \ll 1$. We can then perform the integral over frequency in (7.C.110) to obtain

$$\dot{W}^{\hat{e}}(\mathbf{k}) = \frac{4\pi}{\bar{\varepsilon}} \frac{e^4}{m^2} \left|\hat{\mathbf{e}} \cdot \hat{\mathbf{e}}_1\right|^2 \int \frac{d^3 \mathbf{k}_1}{(2\pi)^3} S^{nn}(\mathbf{k}-\mathbf{k}_1)$$

$$\times \frac{1}{2} \left[ \frac{S^E(\mathbf{k}_1)}{\omega_{\mathbf{k}1}^2} S^{nn}(\mathbf{k}-\mathbf{k}_1, \omega-\omega_{\mathbf{k}1}) + \frac{S^E(-\mathbf{k}_1)}{\omega_{-\mathbf{k}1}^2} S^{nn}(\mathbf{k}+\mathbf{k}_1, \omega+\omega_{\mathbf{k}1}) \right]$$

(7.C.112)

The first term in the square bracket in Eq.(7.C.112) describes the scattering of the transverse wave with wavenumber $\mathbf{k}_1$ being scattered by a density fluctuation with wavenumber $\mathbf{k}_2$ into a scattered transverse wave with wavenumber $\mathbf{k}$, satisfying the matching condition $\mathbf{k}=\mathbf{k}_1+\mathbf{k}_2$ Waves "1" and "2" can be considered as incoming waves in a three-wave interaction with the outgoing wave having wavenumber $\mathbf{k}$. The second term in the square bracket in Eq.(7.C.112) describes the relatively rare process of induced emission of the wave with wavenumber $\mathbf{k}$ in the presence of transverse wave "1" mediated by a density fluctuation with wavenumber $\mathbf{k}_2=\mathbf{k}_1+\mathbf{k}$. Thus, the first term in the square bracket is assumed dominant.

In terms of an action representation:

$$j^{\hat{e}}(\mathbf{k}) = \frac{\dot{W}^{\hat{e}}(\mathbf{k})}{\omega_{\mathbf{k}}} = \left(\frac{4\pi e^2}{m}\right)^2 \int \frac{d^3 \mathbf{k}_1}{(2\pi)^3} \frac{\left|\hat{\mathbf{e}} \cdot \hat{\mathbf{e}}_1\right|^2}{\omega_{\mathbf{k}} \omega_{\mathbf{k}1} \bar{\varepsilon}_{\mathbf{k}} \bar{\varepsilon}_{\mathbf{k}_1}} S^{n_e}(\mathbf{k}_2, \omega_2) J^{\hat{e}_1}(\mathbf{k}_1)$$

(7.C.113)

where

$$J^{\hat{e}_1}(\mathbf{k}_1) = (2\pi)^3 \delta(\mathbf{k}_1 - \mathbf{k}_{inc}) J_{inc}$$

(7.C.114)

Theorem: The macroscopic cross-section for scattering of a transverse wave into a scattered transverse wave is defined

$$\sigma_{t \to t} \equiv \frac{\int \frac{d^3 k}{(2\pi)^3} \sum_{\hat{\mathbf{e}}} \omega_{\mathbf{k}} j^{\hat{e}}(\mathbf{k})}{V_g^{inc} \omega_1 J_{inc}} = \int \frac{d^2 \hat{\mathbf{k}}}{4\pi} \sum_{\hat{\mathbf{e}}} \left|\hat{\mathbf{e}} \cdot \hat{\mathbf{e}}_1\right|^2 \left(\frac{e^2}{m_e c^2}\right)^2 \int_0^\infty d\omega \frac{k\omega}{k_1 \omega_1} S^{n_e}(\mathbf{k}_2, \omega_2)$$

(7.C.115)

The sum is over two polarizations. For transverse waves in an unmagnetized plasma $\omega_k^2 = \omega_p^2 + k^2 c^2$ and $\bar{\varepsilon} = 2$. The quantity $\frac{k\omega}{k_1 \omega_1} = O(1)$. We note that the macroscopic cross-section as defined here does not have the units of area, but instead has units of inverse length, i.e., like the product of a density and a microscopic cross-section having units of area. This result is discussed in Sec. 12.9 of Clemmow and Dougherty.[56]

Exercise: Modify the result for $\sigma_{t \to t}$ to obtain $\sigma_{t \to \ell}$

Example: Scattering off bare electrons, i.e., for situations in which the Debye length $\lambda_D$ is so large that shielding can be ignored.

$$n_e(\mathbf{k},\omega) = \sum_i 2\pi\delta(\omega - \mathbf{k}\cdot\mathbf{v})e^{-i\mathbf{k}\cdot\mathbf{r}_i} \Rightarrow S^0_{n_e}(\mathbf{k}_2,\omega_2) = 2\pi n \int d^3\mathbf{v}\, g_e(\mathbf{v})\delta(\omega_2 - \mathbf{k}_2\cdot\mathbf{v}) = \frac{2\pi n}{k_2} g^e_{\hat{\mathbf{k}}}\left(\frac{\omega_2}{k_2}\right)$$

(7.C.116)

For a Maxwellian $g^e_{\hat{\mathbf{k}}}(\omega_2/k_2) \propto e^{-\frac{1}{2}\left(\frac{\omega_2}{k_2 v_e}\right)^2} \to e^{-\frac{1}{2}\left(\frac{\omega-\omega_1}{k_2 v_e}\right)^2}$ for outgoing waves. Thermal effects produce a spread in frequencies of the scattered radiation of order: $\Delta\omega/\omega_1 \sim k_2 v_e/\omega_1$, which is small in a non-relativistic plasma for which $v_e/(\omega_1/k_1) \sim v_e/c \ll 1$. We note that for $k_1 \approx k$, $k_2 = 2k_1 \sin(\theta/2)$ where $\theta$ is the angle between $\mathbf{k}_1$ and $\mathbf{k}$. From Eqs.(7.C.115) and (7.C.116) one obtains

$$\sigma_{t \to t} \equiv \frac{\int \frac{d^3k}{(2\pi)^3} \sum_{\hat{\mathbf{e}}} \omega_{\mathbf{k}} j^{\hat{\mathbf{e}}}(\mathbf{k})}{V_g^{inc}\omega_1 J_{inc}} = n_e \frac{8\pi}{3} r_e^2 = n_e \sigma_{\text{Thomson}}$$

(7.C.117)

Example: Scattering off shielded particles
We use the linearized Vlasov equation to calculate the response of the electron shielding cloud and the ion perturbations:

$$n_e^{cloud}(\mathbf{k},\omega) = \frac{n_e e_e}{m_e} Z'_e(v)\phi(\mathbf{k},\omega) \quad Z_s(v) = \int_{-\infty}^{\infty} du \frac{g^s_{\hat{\mathbf{k}}}(u)}{u-v}$$

$$\phi(\mathbf{k},\omega) = \frac{4\pi}{k^2} \frac{\sum_s e_s n^0_s(\mathbf{k},\omega)}{\varepsilon(\mathbf{k},\omega)} \quad \varepsilon = 1 + \chi_e + \chi_i \quad \chi_s \equiv -\frac{\omega_s^2}{k^2} Z'_s(v)$$

(7.C.118)

The total electron charge density perturbation is the sum over bare electrons, their shielding clouds, and the electron shielding clouds surrounding the ion perturbed charge density:

---

[56] P. C. Clemmow and J. P. Dougherty, *Electrodynamics of Particles and Plasmas* (Perseus Books, New York, NY, 1989). ISBN: 9780201515008.

$$n_e(\mathbf{k},\omega) = n_e^0(\mathbf{k},\omega) + n_e^{cloud} = n_e^0(\mathbf{k},\omega)\left[1 + \frac{\omega_{pe}^2}{k^2}\frac{Z_e'(v)}{\varepsilon(\mathbf{k},\omega)}\right] + n_i^0(\mathbf{k},\omega)\left[-\frac{\omega_{pe}^2}{k^2}\frac{Z_e'(v)}{\varepsilon(\mathbf{k},\omega)}\right] \quad (7.C.119)$$

$$= \frac{1}{\varepsilon}\left[n_e^0(\varepsilon - \chi_e) + n_i^0 \chi_e\right] = \frac{1}{\varepsilon}\left[n_e^0(1 + \chi_i) + n_i^0 \chi_e\right]$$

We assume that the bare ion and bare electron charge density perturbations are uncorrelated and obtain from Eqs.(7.C.116) and (7.C.120)

$$S_{n_e}(\mathbf{k},\omega) = \frac{1}{|\varepsilon|^2}\left[S_{n_e}^0 |1 + \chi_i|^2 + S_{n_i}^0 |\chi_e|^2\right]$$

$$S_{n_i}(\mathbf{k},\omega) = \frac{1}{|\varepsilon|^2}\left[S_{n_i}^0 |1 + \chi_e|^2 + S_{n_e}^0 |\chi_i|^2\right] \quad (7.C.120)$$

$$\frac{S_{n_e}}{S_{n_i}} = \frac{g_e|1 + \chi_i|^2 + g_i|\chi_e|^2}{g_e|\chi_i|^2 + g_i|1 + \chi_e|^2}$$

where $S^0 \equiv \frac{2\pi n}{k}g$. We can compare $g_e$ and $g_i$ as a function of the phase velocity v. First we recall

$$g_s = \frac{1}{\sqrt{2\pi}\, v_s} e^{-\frac{v^2}{2v_e^2}}$$

For v~$v_i$<<$v_e$, then $g_e$<<$g_i$ in Eq.(7.C.120) and $\chi_e = \frac{1}{k^2\lambda_e^2}$. Hence, $S_{n_e}/S_{n_i} \cong (1 + k^2\lambda_e^2)^{-2}$. Thus, the electron and ion fluctuations are comparable in this limit. For v~$v_e$>>$v_i$ and $\omega \gg \omega_{pi}$ then $g_i$ is exponentially small, $g_i$<<$g_e$, and $|\chi_i| \ll 1$. Hence, $S_{n_e}/S_{n_i} \gg 1$ for high frequencies.

Our calculations of scattering so far have assumed that the underlying plasma is an isotropic, non-drifting Maxwellian and there is no externally applied magnetic field. With the inclusion of a drift, the approach of instability thresholds alters the dielectric response and the dispersion relations, which modify the density correlation functions. If there is an applied magnetic field, then particle gyration must be included, and the more general **K** response tensor must be used instead of the scalar longitudinal dielectric function.

Example: Wave-wave scattering. Consider an incident transverse wave scattering off a longitudinal wave into a scattered transverse wave. We recall Eq.(7.C.114) for the time derivative of the scattered transverse wave action rewritten as

$$\dot{J}^{\hat{\mathbf{e}}}(\mathbf{k}) = \frac{\dot{W}^{\hat{\mathbf{e}}}(\mathbf{k})}{\omega_\mathbf{k}} = \frac{1}{n^2}\int\frac{d^3\mathbf{k}_1}{(2\pi)^3}\frac{\omega_p^4}{\omega_\mathbf{k}\omega_{\mathbf{k}1}}\frac{|\hat{\mathbf{e}}\cdot\hat{\mathbf{e}}_1|^2}{\bar{\varepsilon}_\mathbf{k}\bar{\varepsilon}_{\mathbf{k}_1}}S^{n_e}(\mathbf{k}_2,\omega_2)J^{\hat{\mathbf{e}}_1}(\mathbf{k}_1) \quad (7.C.121a)$$

where

$$S^{n_e}(\mathbf{k}_2,\omega_2) = \pi\left[S^{n_e}(\mathbf{k}_2)\delta(\omega_2-\omega_{\mathbf{k}_2}) + S^{n_e}(-\mathbf{k}_2)\delta(\omega_2+\omega_{\mathbf{k}_2})\right] \quad (7.C.121b)$$

For an electron longitudinal wave present

$$4\pi e\delta n_e = i\mathbf{k}_2\cdot\mathbf{E} \to S^{n_e}(\mathbf{k}_2) = \left(\frac{k_2}{4\pi e}\right)^2 S^E(\mathbf{k}_2) \to \frac{8\pi}{\overline{\varepsilon}_2}W(\mathbf{k}_2) \to \omega_{\mathbf{k}_2}J^{e\ell}(\mathbf{k}_2) \quad (7.C.122)$$

Hence, from Eq.(7.121a)

$$J^{\hat{\mathbf{e}}}(\mathbf{k}) = \int\frac{d^3\mathbf{k}_1}{(2\pi)^3}\int\frac{d^3\mathbf{k}_2}{(2\pi)^3}(2\pi)^2\delta(\mathbf{k}_1+\mathbf{k}_2-\mathbf{k})\delta(\omega_1+\omega_2-\omega)8\pi^2\left(\frac{e}{m_e}\right)^2\frac{k_2^2\omega_2}{\omega\omega_1}\frac{|\hat{\mathbf{e}}\cdot\hat{\mathbf{e}}_1|^2}{\overline{\varepsilon}_{\mathbf{k}}\overline{\varepsilon}_{\mathbf{k}_1}\overline{\varepsilon}_{\mathbf{k}2}}J^{\hat{\mathbf{e}}_1}(\mathbf{k}_1)J^{e\ell}(\mathbf{k}_2)$$

(7.C.123)

The inverse process is the induced decay of the transverse wave with $(\omega,\mathbf{k})$ into decay product waves $(\omega_1,\mathbf{k}_1)$ and $(\omega_2,\mathbf{k}_2)$. We include the additional induced decay terms to obtain

$$J^{\hat{\mathbf{e}}}(\mathbf{k}) = \int\frac{d^3\mathbf{k}_1}{(2\pi)^3}\int\frac{d^3\mathbf{k}_2}{(2\pi)^3}(2\pi)^2\delta(\mathbf{k}_1+\mathbf{k}_2-\mathbf{k})\delta(\omega_1+\omega_2-\omega)8\pi^2\left(\frac{e}{m_e}\right)^2\frac{k_2^2\omega_2}{\omega\omega_1}\frac{|\hat{\mathbf{e}}\cdot\hat{\mathbf{e}}_1|^2}{\overline{\varepsilon}_{\mathbf{k}}\overline{\varepsilon}_{\mathbf{k}_1}\overline{\varepsilon}_{\mathbf{k}2}} \quad (7.C.124)$$
$$\times\left[J^{\hat{\mathbf{e}}_1}(\mathbf{k}_1)J^{e\ell}(\mathbf{k}_2) - J^{\hat{\mathbf{e}}}(\mathbf{k})J^{e\ell}(\mathbf{k}_2) - J^{\hat{\mathbf{e}}}(\mathbf{k})J^{\hat{\mathbf{e}}_1}(\mathbf{k}_1)\right]$$

The expression in the square bracket is the coupling coefficient we did not know in Sec. 6.H for three-wave coupling in the random-phase limit.

### 7.C.g Nonlinear Landau damping

Nonlinear Landau damping is the name given to the nonlinear scattering of an incident wave by particles. [Note that this is not the nonlinear modification of Landau resonance due to particle trapping.] In this example the incident and scattered waves are transverse waves. Recall the expression in Eq.(7.C.120)

$$S_{n_e}(\mathbf{k}_2,\omega_2) = \frac{1}{|\varepsilon(\mathbf{k}_2)|^2}\left[S^0_{n_e}|1+\chi_i|^2 + S^0_{n_i}|\chi_e|^2\right] \quad (7.C.125)$$

where $S^0_{n_s}$ is given in Eq.(7.C.116). Then from Eq.(7.C.121a) one obtains

$$J^{\hat{\mathbf{e}}}(\mathbf{k}) = \int\frac{d^3\mathbf{k}_1}{(2\pi)^3}\frac{2\pi}{|\varepsilon(\mathbf{k}-\mathbf{k}_1,\omega-\omega_1)|^2}\frac{1}{n^2}\frac{\omega_p^4}{\omega_{\mathbf{k}}\omega_{\mathbf{k}1}}\frac{|\hat{\mathbf{e}}\cdot\hat{\mathbf{e}}_1|^2}{\overline{\varepsilon}_{\mathbf{k}}\overline{\varepsilon}_{\mathbf{k}_1}} \quad (7.C.126)$$
$$\times\int d^3\mathbf{v}\,\delta\left(\omega-\omega_1-(\mathbf{k}-\mathbf{k}_1)\cdot\mathbf{v}\right)\left[|1+\chi_i|^2 f_e(\mathbf{v})J^{\hat{\mathbf{e}}_1}(\mathbf{k}_1) + |\chi_e|^2 f_i(\mathbf{v})J^{\hat{\mathbf{e}}_1}(\mathbf{k}_1)\right]$$

The quantum mechanics conservation laws for momentum and energy that describe wave scattering by a particle are

$$\mathbf{p}+\hbar\mathbf{k}_1 = \mathbf{p'}+\hbar\mathbf{k} \implies \mathbf{p'}-\mathbf{p} \equiv \Delta\mathbf{p} = \hbar(\mathbf{k}_1-\mathbf{k})$$

$$\frac{p^2}{2m}+\hbar\omega_1 = \frac{p'^2}{2m}+\hbar\omega \implies \Delta\left(\frac{p^2}{2m}\right) \approx \mathbf{v}\cdot\Delta\mathbf{p} = \hbar(\omega_1-\omega) \quad (7.\text{C}.127)$$

for differential changes in the particle energy due to the scattering. These conservation laws directly lead to the condition

$$\omega_1-\omega = (\mathbf{k}_1-\mathbf{k})\cdot\mathbf{v} \quad (7.\text{C}.128)$$

which we recognize as the Landau particle resonance condition for wave scattering by a particle. The inverse (not reverse) scattering process has the same coupling due to hermiticity and detailed balance.

In the quantum mechanics calculation the time derivative of the quanta is given by

$$\frac{d}{dt}N(\mathbf{k}) = C_{\mathbf{kk}_1 s}\left\{f(\mathbf{p})N(\mathbf{k}_1)[1+N(\mathbf{k})]-f(\mathbf{p'})N(\mathbf{k})[1+N(\mathbf{k}_1)]\right\}$$

$$\qquad\qquad \uparrow \qquad\quad \uparrow \quad \uparrow \qquad\qquad\quad \uparrow \quad (7.\text{C}.129)$$

$$\quad \text{\# scatters} \quad \text{spont.} \ \text{induced in} \qquad \text{scattering out}$$

$$= C_{\mathbf{kk}_1 s}\left\{N(\mathbf{k}_1)N(\mathbf{k})[f(\mathbf{p})-f(\mathbf{p'})]+N(\mathbf{k}_1)f(\mathbf{p})-N(\mathbf{k})f(\mathbf{p'})\right\}$$

where $N\equiv J/\hbar$ and the rules of Bose statistics apply. We use Eq.(7.C.127) and linearly expand

$$f(\mathbf{p})-f(\mathbf{p'}) \approx (\mathbf{p}-\mathbf{p'})\cdot\frac{\partial f}{\partial \mathbf{p}} = \hbar(\mathbf{k}-\mathbf{k}_1)\cdot\frac{\partial f}{\partial \mathbf{p}} \quad (7.\text{C}.130)$$

Hence,

$$\frac{1}{\hbar}\frac{d}{dt}J(\mathbf{k}) = C_{\mathbf{kk}_1 s}\left\{\frac{J(\mathbf{k}_1)}{\hbar}J(\mathbf{k})(\mathbf{k}-\mathbf{k}_1)\cdot\frac{\partial f}{\partial \mathbf{p}}+\frac{J(\mathbf{k}_1)}{\hbar}f(\mathbf{p})-\frac{J(\mathbf{k})}{\hbar}f(\mathbf{p'})\right\}$$

$$\propto \left\{J(\mathbf{k}_1)J(\mathbf{k})(\mathbf{k}-\mathbf{k}_1)\cdot\frac{\partial f}{\partial \mathbf{p}}+[J(\mathbf{k}_1)-J(\mathbf{k})]f(\mathbf{p})+O\left[(\mathbf{p'}-\mathbf{p})\cdot\frac{\partial f}{\partial \mathbf{p}}\right]\right\} \quad (7.\text{C}.131)$$

$$\times \delta\left(\hbar\omega+\frac{p'^2}{2m}-(\hbar\omega_1+\frac{p^2}{2m})\right)\delta(\hbar\mathbf{k}+\mathbf{p'}-(\hbar\mathbf{k}_1+\mathbf{p}))$$

We integrate Eq.(7.C.131) over the spectrum of waves $\mathbf{k}_1$ and over the distribution function of momentum $\mathbf{p}$ to get an expression for $dJ(\mathbf{k})/dt$. Given Eq.(7.C.126) the coefficient $C_{\mathbf{kk}_1 s}$ can be evaluated:

$$C_{\mathbf{kk}_1 s} = \frac{2\pi}{|\varepsilon(\mathbf{k}-\mathbf{k}_1,\omega-\omega_1)|^2}\frac{1}{n^2}\frac{\omega_p^4}{\omega_\mathbf{k}\omega_{\mathbf{k}1}}\frac{|\hat{\mathbf{e}}\cdot\hat{\mathbf{e}}_1|^2}{\bar{\varepsilon}_\mathbf{k}\bar{\varepsilon}_{\mathbf{k}_1}}\left[|1+\chi_i|^2 \text{ electrons},\ |\chi_e|^2 \text{ ions}\right] \quad (7.\text{C}.132)$$

where $\mathbf{k}_2=\mathbf{k}-\mathbf{k}_1$.

The scattering considered here involves small changes in the particle momentum consistent with Eq.(7.C.127) and the linear expansion in (7.C.130).

Combined with the assumption of random phases in the scattering processes, the momentum scattering satisfies the conditions for a Fokker-Planck description:

$$\frac{\partial f^s(\mathbf{p};t)}{\partial t} = -\frac{\partial}{\partial \mathbf{p}} \cdot \left(\mathbf{F}(\mathbf{p})f(\mathbf{p})\right) + \frac{\partial}{\partial \mathbf{p}} \cdot \left(\mathbf{D}(\mathbf{p}) \cdot \frac{\partial f(\mathbf{p})}{\partial \mathbf{p}}\right) \quad (7.C.133)$$

The first momentum moment of the Fokker-Planck equation summed over species yields the time derivative of the total particle momentum in the plasma:

$$\frac{d}{dt}\mathbf{P}_{particles} = \frac{d}{dt}\sum_s \int d^3\mathbf{p}\, \mathbf{p}\, f_s(\mathbf{p};t) = \sum_s \int d^3\mathbf{p}\left[\mathbf{F}f - \mathbf{D}\cdot\frac{\partial f}{\partial \mathbf{p}}\right] \quad (7.C.134)$$

where an integration by parts has been performed. The time derivative of the total wave momentum (fields and particle sloshing in the wave) is

$$\frac{d}{dt}\mathbf{P}_{waves} = \frac{d}{dt}\left[\int \frac{d^3\mathbf{k}}{(2\pi)^3}\mathbf{k}\, J^{\hat{\mathbf{e}}}(\mathbf{k};t) + \int \frac{d^3\mathbf{k}_1}{(2\pi)^3}\mathbf{k}_1\, J^{\hat{\mathbf{e}}_1}(\mathbf{k}_1;t)\right] \quad (7.C.135)$$

The right side of Eq.(7.C.135) is evaluated using Eqs.(7.C.131) and (7.C.132). One can show that the conservation of the total number of quanta implies that the total wave action is conserved. Conservation of total momentum is dictated by Eq.(7.C.127) and built into the equations. Thus, one can sum Eqs.(7.C.134) and (7.C.135) and assert

$$\frac{d}{dt}\left(\mathbf{P}_{particles} + \mathbf{P}_{waves}\right) = 0 \quad (7.C.136)$$

from which we infer $\mathbf{F}(\mathbf{p})$ and $\mathbf{D}(\mathbf{p})$ consistent with conservation of total momentum:

$$\mathbf{D}^s(\mathbf{p}) = \int \frac{d^3\mathbf{k}}{(2\pi)^3}\int \frac{d^3\mathbf{k}_1}{(2\pi)^3}(\mathbf{k}-\mathbf{k}_1)C_{\mathbf{k}\mathbf{k}_1 s} J^{\hat{\mathbf{e}}_1}(\mathbf{k}_1) J^{\hat{\mathbf{e}}}(\mathbf{k})\delta\left(\omega-\omega_1-(\mathbf{k}-\mathbf{k}_1)\cdot\mathbf{v}\right)(\mathbf{k}-\mathbf{k}_1)$$

$$\mathbf{F}^s(\mathbf{p}) = \int \frac{d^3\mathbf{k}}{(2\pi)^3}\int \frac{d^3\mathbf{k}_1}{(2\pi)^3}(\mathbf{k}-\mathbf{k}_1)C_{\mathbf{k}\mathbf{k}_1 s}\left(J^{\hat{\mathbf{e}}}(\mathbf{k})-J^{\hat{\mathbf{e}}_1}(\mathbf{k}_1)\right)\delta\left(\omega-\omega_1-(\mathbf{k}-\mathbf{k}_1)\cdot\mathbf{v}\right)(\mathbf{k}-\mathbf{k}_1)$$

(7.C.137)

Exercise: Derive Eq.(7.C.137) filling in the intermediate steps.

### 7.C.h Radiation transport: summary of important processes

This lecture serves as a summary of the important processes in radiation transport.

Emission processes
1. Bremsstrahlung (discreteness) – Particles scatter off one another or particles scatter off fluctuations, and in so doing produce a time-dependent dipole moment and current, which lead to emission of radiation.
2. Cerenkov (resonance) – A particle with velocity satisfying wave-particle resonance $\omega = \mathbf{k}\cdot\mathbf{v}$ can emit a wave.

3. Beat resonance (scattering) – The beat of an incident and scattered wave are resonant with particles $\omega_1 - \omega_2 = (\mathbf{k}_1 - \mathbf{k}_2) \cdot \mathbf{v}$, which can enhance the scattering.
4. Gyro-resonance (cyclotron emission) – The classical acceleration of a charged particle gyrating in a magnetic field emits radiation satisfying the gyro-resonance condition: $\omega - k_\parallel v_\parallel = \ell \Omega$.
5. Three-wave (or more) scattering – An incident wave is scattered by another wave into a third wave subject to the three-wave resonance conditions: $\omega_1 = \omega_2 + \omega_3 \quad \mathbf{k}_1 = \mathbf{k}_2 + \mathbf{k}_3$
6. Induced emission – Wave emission (single wave and two waves) by the plasma can reinforce itself leading to an instability.
7. Radiation emission due to atomic transitions (not covered in these lectures)

Conditions affecting propagation
1. Damping or instability
    a. Anisotropy in velocity space
    b. Non-monotonicity in energy
    c. Inhomogeneity in space
    d. Non-stationarity due to time dependence, e.g., parametric instability
2. Refraction in a plasma with weak inhomogeneity $L \gg \lambda$
3. Reflection in a plasma with sharp inhomogeneity $L \ll \lambda$ or sudden change in polarization of the wave
4. Diffraction when $L \sim \lambda$

Absorption processes
1. Wave scattering
2. Three or more wave interactions
3. Mode conversion from one branch to another
4. Inverse Cerenkov effect, i.e., Landau damping
5. Collisional absorption, i.e., inverse Bremsstrahlung
6. Atomic transitions
7. Gyro-resonant absorption
8. Two-wave absorption

A complication in most of the resonant processes is resonance broadening. Particle trapping is a particularly important nonlinear effect and influences the saturation of many instabilities.

### *7.C.i WKB theory in the context of radiation transport*

We return to the WKB theory to include WKB wave propagation in a general theory of wave emission and absorption. Our WKB theory uses the following representation:

$$\mathbf{E}(\mathbf{x},t) = \tilde{\mathbf{E}}(\mathbf{x},t)\, e^{i\Phi(\mathbf{x},t)} \quad \Phi = \int^{x}\!\left(\mathbf{k}\cdot d\mathbf{x} - \omega\, dt\right) \tag{7.C.138a}$$

and

$$\mathbf{k}(\mathbf{x},t) = \frac{\partial \Phi}{\partial \mathbf{x}} \quad \omega(\mathbf{x},t) = -\frac{\partial \Phi}{\partial t} \;\Rightarrow\; \frac{\partial}{\partial t}\mathbf{k}(\mathbf{x},t) = -\frac{\partial}{\partial \mathbf{x}}\omega(\mathbf{x},t) \tag{7.C.138b}$$

We use Maxwell's equations to obtain the local linear dispersion relation:

$$D(\omega, \mathbf{k}; \mathbf{x}, t) = D(-\frac{\partial \Phi}{\partial t}, \frac{\partial \Phi}{\partial \mathbf{x}}; \mathbf{x}, t) \to \omega\!\left(\mathbf{k}(\mathbf{x},t); \mathbf{x}, t\right) \tag{7.C.139}$$

Equation (7.C.140) is a first-order quasilinear partial differential equation which is solved by the method of characteristics. The equation of motion for the wave packet characteristic determined from Eq.(7.C.138b) is

$$\frac{d\mathbf{x}}{dt} = \frac{\partial}{\partial \mathbf{k}}\omega(\mathbf{k};\mathbf{x},t) \tag{7.C.140}$$

and there is the companion equation:

$$\begin{aligned}
\frac{d}{dt}\mathbf{k} &= \left(\frac{\partial}{\partial t} + \left.\frac{\partial \omega}{\partial \mathbf{k}}\right|_{\mathbf{x},t}\cdot\frac{\partial}{\partial \mathbf{x}}\right)\mathbf{k} = -\frac{\partial \omega(\mathbf{x},t)}{\partial \mathbf{x}} + \left.\frac{\partial \omega}{\partial \mathbf{k}}\right|_{\mathbf{x},t}\cdot\frac{\partial \mathbf{k}}{\partial \mathbf{x}} \\
&= -\left.\frac{\partial \omega(\mathbf{x},t)}{\partial \mathbf{x}}\right|_{\mathbf{k},t} - \left.\frac{\partial \omega(\mathbf{x},t)}{\partial \mathbf{x}}\right|_{\mathbf{x},t}\cdot\frac{\partial \mathbf{k}}{\partial \mathbf{x}} + \left.\frac{\partial \omega}{\partial \mathbf{k}}\right|_{\mathbf{x},t}\cdot\frac{\partial \mathbf{k}}{\partial \mathbf{x}} = -\left.\frac{\partial \omega(\mathbf{x},t)}{\partial \mathbf{x}}\right|_{\mathbf{k},t}
\end{aligned} \tag{7.C.141}$$

If there is no emission or absorption, then there should be an action conservation theorem. This is entirely analogous to the conservation of photons or plasmons in quantum mechanics if there is no absorption or emission. Recall that the number of photons or plasmons is related to the action by the relation $N = J/\hbar$. Dougherty used a covariant treatment to develop a classical theory of wave-action conservation for small-amplitude waves based on a Lagrangian.[57]

The total action is derived from the sum over branches of the integral over configuration and $\mathbf{k}$ space of the action density in $\mathbf{x}$ and $\mathbf{k}$ space:

$$J_{tot} = \sum_{\alpha} J^{\alpha}(\mathbf{x},\mathbf{k};t)\frac{d^{3}\mathbf{x}\, d^{3}\mathbf{k}}{(2\pi)^{3}} \tag{7.C.142}$$

The Liouville theorem for action density is

$$\begin{aligned}
\frac{\partial J^{\alpha}(\mathbf{x},\mathbf{k};t)}{\partial t} &+ \frac{\partial}{\partial \mathbf{x}}\cdot\left(\dot{\mathbf{x}} J^{\alpha}\right) + \frac{\partial}{\partial \mathbf{k}}\cdot\left(\dot{\mathbf{k}} J^{\alpha}\right) = \frac{\partial J^{\alpha}(\mathbf{x},\mathbf{k};t)}{\partial t} + \frac{\partial}{\partial \mathbf{x}}\cdot\left(\frac{\partial \omega}{\partial \mathbf{k}} J^{\alpha}\right) + \frac{\partial}{\partial \mathbf{k}}\cdot\left(-\frac{\partial \omega}{\partial \mathbf{x}} J^{\alpha}\right) \\
&= \frac{\partial J^{\alpha}(\mathbf{x},\mathbf{k};t)}{\partial t} + J^{\alpha}\!\left(\frac{\partial}{\partial \mathbf{x}}\cdot\frac{\partial \omega}{\partial \mathbf{k}} - \frac{\partial}{\partial \mathbf{k}}\cdot\frac{\partial \omega}{\partial \mathbf{x}}\right) + \left(\dot{\mathbf{x}}\cdot\frac{\partial}{\partial \mathbf{x}} + \dot{\mathbf{k}}\cdot\frac{\partial}{\partial \mathbf{k}}\right)J^{\alpha} \\
&= \left(\frac{\partial}{\partial t} + \dot{\mathbf{x}}\cdot\frac{\partial}{\partial \mathbf{x}} + \dot{\mathbf{k}}\cdot\frac{\partial}{\partial \mathbf{k}}\right)J^{\alpha}(\mathbf{x},\mathbf{k};t) = \frac{d}{dt}J^{\alpha}(\mathbf{x},\mathbf{k};t) = 0
\end{aligned} \tag{7.C.143}$$

---

[57] J. P. Dougherty, J. Plasma Phys. 4, 761 (1970).

Equation (7.C.143) describes the advection of the action density along the characteristics in (**x**,**k**) space determined by the WKB equations of motion.

The generalization of Eq.(7.C.143) to include emission, absorption, and independent damping effects (but do not double count) is then

$$\frac{d}{dt}J^\alpha = j^\alpha(\text{emission-absorption}) - 2\gamma_{dpg}(\mathbf{k};\mathbf{x},t) \qquad (7.C.144)$$

The damping rate $\gamma_{dpg}$ is a function of wave amplitude if particle trapping effects are included. Otherwise, $\gamma_{dpg}$ may be independent of wave amplitude in many instances. However, we note that diffraction, coherent effects, and departures from WKB are all lost in this formulation.

What about polarization effects? To include polarization effects we must go back to the response tensor:

$$\mathbf{K}(\omega,\mathbf{k};\mathbf{x},t)\cdot\mathbf{E}=0 \Rightarrow \det|\mathbf{K}|\equiv D=0 \Rightarrow \omega^\alpha(\mathbf{k};\mathbf{x},t) \quad \mathbf{K}(\omega,\mathbf{k};\mathbf{x},t)\cdot\hat{\mathbf{e}}^\alpha(\mathbf{x},t)=0 \quad (7.C.145)$$

As we track a wave packet, we solve continuously for the frequency, wavenumber, and polarization evolution.

In a non-uniform plasma, solutions of the dispersion relation for the wavenumber as a function of frequency can exhibit resonances in some instances, i.e., the solution for the index of refraction exhibits a singularity, $n^2(\omega,\boldsymbol{x}) = k^2c^2/\omega^2 \to \infty$ (see Chapter 10 in Stix' book).[58] The resonant frequency $\omega_{res}$ typically depends on the local values of the cyclotron frequency, the plasma frequency, and the angle of the wave number with respect to the magnetic field. Consider the special case of a stationary medium. For a given frequency $\omega$ and location **x** such that the solution for $n^2$ is finite, the wave simply propagates. Propagation continues until the solution for $\omega_{res}(\mathbf{x})=\omega$ at which location $n^2(\omega,\boldsymbol{x}) \to \infty$ and the wave suffers either strong absorption or is evanescent. Including finite temperature and collisional effects in the dispersion relations makes the complex parts of $n^2$ non-negligible and removes the infinities. Because finite temperature effects introduce significant analytical complexity, here we will illustrate how the inclusion of collisionality resolves resonances. Clearly WKB theory is no good near a resonance because $k$ is not slowly varying in space as it diverges.

Consider a simple model equation for the index of refraction in a plasma with spatially varying resonance frequency with resonance at $x=0$:

$$n^2(\omega,x) = \frac{k^2c^2}{\omega^2} = \frac{c_1}{\omega_{res}(x)-\omega} \to \frac{c_1}{\omega + \left.\frac{d\omega_{res}}{dx}\right|_{x=0} x - i\nu - \omega} = \frac{c_1}{\left.\frac{d\omega_{res}}{dx}\right|_{x=0} x - i\nu} \qquad (7.C.146)$$

for $c_1>0$. Because WKB theory is invalid near resonance we require a field equation, e.g., a Helmholtz equation with a singular point at $x=0$:

---

[58] T. H. Stix, *The Theory of Plasma Waves* (McGraw-Hill, New York, NY) 1962.

$$\frac{d^2E}{dx^2}+k^2(x)E(x) = \frac{d^2E}{dx^2}+\frac{c_1}{\left.\frac{d\omega_{res}}{dx}\right|_{x=0}x-i\nu}\frac{\omega^2}{c^2}E(x)=0 \qquad (7.C.147)$$

The collision rate is assumed small and is important only in the neighborhood of $x=0$. WKB is good for $x \neq 0$:

$$k^2(x)=\frac{c_1}{\omega_R' x-i\nu}\frac{c^2}{\omega^2}\equiv\frac{\mu^2}{x-i\nu'}, \quad \nu'\equiv\frac{\nu}{\omega_R'}, \quad \mu^2\equiv\frac{c_1}{\omega_R'}\frac{c^2}{\omega^2}$$

$$\frac{d^2E}{dx^2}+k^2(x)E(x)=0 \qquad E(x)=\sum_\pm c_\pm\frac{1}{\sqrt{k}}e^{\pm i\int k\,dx} \qquad (7.C.148)$$

Let $z \equiv x - i\nu'$ so that Eq.(7.C.146) becomes

$$\frac{d^2E}{dz^2}+\frac{\mu^2}{z}E(z)=0 \qquad (7.C.149)$$

which has solutions in terms of Bessel functions:

$$E \sim z^{1/2}J_1(2\mu z^{1/2}),\ z^{1/2}Y_1(2\mu z^{1/2}) \to c_1 z^{1/2}J_1(2\mu z^{1/2})+c_2 z^{1/2}Y_1(2\mu z^{1/2}) \quad (7.C.150a)$$

Although these forms are good for $z<0$, there is some ambiguity because of the $z^{1/2}$. Therefore we introduce Bessel functions of imaginary argument:

$$E(z<0)=-c_1(-z)^{1/2}I_1(2\mu(-z)^{1/2})+c_2(-z)^{1/2}\left[iI_1(2\mu(-z)^{1/2})-\frac{2}{\pi}K_1(2\mu(-z)^{1/2})\right] \quad (7.C.150b)$$

$I_1$ grows exponentially for large argument while $K_1$ decays exponentially. For simplicity assume that there are no other propagating regions for $x<0$. Then the two factors with $I_1$ must cancel one another at large negative $z$, which leaves only the factor involving $K_1$:

$$E(x<0)=\frac{2i}{\pi}(-z)^{1/2}K_1(2\mu(-z)^{1/2}) \qquad (7.C.151)$$

For large negative $x$, the solution in (7.C.151) matches nicely onto the decaying WKB solution:

$$E(x<0) \to \frac{i}{\sqrt{\pi}}\frac{1}{|k|^{1/2}}e^{-\int_0^{|z|}|k|^{1/2}dz} \qquad (7.C.152)$$

In the propagating region we have the solution in Eq.(7.C.149), which for $|\mu z^{1/2}|<<1$ has the following limiting forms

$$z^{1/2}J_1(2\mu z^{1/2}) \propto \mu z \quad z^{1/2}Y_1(2\mu z^{1/2}) \propto \frac{1}{\pi\mu}\left[-1+\mu^2 z\left(\ln(\mu^2 z)+2\gamma-1\right)\right],\ \gamma=0.5772$$

$$(7.C.153)$$

We use these limiting forms in Eq.(7.C.150) and then assert continuity of $E$ at $x=0$ to evaluate constants by equating Eqs.(7.C.150) and (7.C.151) (note that there is one free constant fixing the amplitude of the incident wave from positive $x$). For large positive $x$, there is only an incoming wave and no outgoing wave:

$$E(x \to \infty) \sim z^{1/2}[-J_1 - iY_1] \to \frac{1}{\sqrt{\pi k}} \exp -i\left(\int_0^z k\, dz + \frac{\pi}{4}\right) \qquad (7.\text{C}.154)$$

From the constructed solutions we see that there is an incoming wave from positive $x$ that propagates through $x=0$ (without reflection) to negative $x$ where it decays. The absorption length is set by $\Delta x \sim v' = \nu |d\omega_R/dx|^{-1}$. The weaker the collision rate the narrower the absorption length near resonance. This is in contrast to the effect of collisions on absorption far from resonance for which $\text{Im } k \sim \frac{\nu}{V_g} \sim \frac{1}{\Delta x}$ so that the absorption length increases with decreasing collision rate. Stix's book has more to say on the subject of resonances.

### 7.D Dupree's theory of phase-space granulation and clump formation

Trapped particles moving with velocity near the phase velocity of the wave in which they are trapped tend to move as a clump. If there are many waves present with sufficient amplitude to produce trapping, the many waves can dissipate this clump. The particles can radiate in going from trapped to untrapped. The clump does not travel together for a long enough time to radiate at a sharp frequency. However, there is sufficient coherence in the clump to produce some enhancement of the radiation. Some of the early important work on clump theory was due to Dupree[59,60] and Kadomtsev and Pogutse.[61] Radiation produced by one clump tends to form other clumps, i.e., clumps tend to be "self-sustaining." Waves can form and sustain a clump, while other waves tend to diffuse and destroy the clump. Dynamic friction and velocity diffusion due to the waves are clearly present. One must use a finite-width resonance function in the theory and not $\delta(\omega - \mathbf{k} \cdot \mathbf{v})$. The theory includes wave emission by clumps, Landau damping (stable plasma), clump formation by wave trapping, and clump destruction by other waves.

The clump phenomena depend on the auto-correlation time and the trapping time. The auto-correlation time is the time required for phase-mixing to become complete:

$$\tau_{ac} \sim \frac{\delta x}{V} \sim \frac{1}{\delta k\, V} \qquad (7.\text{D}.1)$$

where $\delta k$ is the spectral width and V is the phase velocity $\sim$ the resonant particle velocity. Because the spectral width $\delta k$ induces a spread of phase velocities: $\delta V/V = \delta k/k$, then $\tau_{ac} \sim 1/k\delta V$ alternatively. The broader the spectrum, the shorter is $\tau_{ac}$. We expect that the coherence of the clump radiation to scale as $\Delta \omega \sim \frac{1}{\tau_{ac}}$. Consider three regimes of behavior for clumps:

---

[59] T. H. Dupree, Phys. Rev. Lett. **25**, 789 (1970).
[60] T. H. Dupree, Phys. Fluids **15**, 334 (1972).
[61] B. B. Kadomtsev and O. P. Pogutse, Phys. Rev. Lett. **25**, 1155 (1970).

(1) $\tau_{ac} \ll \tau_B \equiv 2\pi/\omega_B$ where $\omega_B = k\sqrt{e\phi/m}$. The fields experienced by the particles are changing so fast that there is no trapping. There is only quasi-linear diffusion.

(2) $\tau_{ac} \sim \tau_B$. There is a diffusion in velocity space with a tendency to trap. Clumps are short lived.

(3) $\tau_{ac} \gg \tau_B$. There is well-defined trapping. Eventually trapping is destroyed by diffusion in waves. Clumps are long lived (long-lived eddies).

The velocity distribution function can be decomposed as the sum of the unperturbed lowest-order distribution plus a second term comprising the linear coherent perturbation and the quasi-linear perturbation and a third term that is the nonlinear perturbation due to trapping. We can make the following scaling arguments. The velocity diffusion due to scattering by waves is given by

$$D \sim \frac{\delta v^2}{\tau} \tag{7.D.2}$$

where $\delta v$ is the change in velocity during a time $\tau$ due to scattering by all the waves. Over the lifetime of the clump $\tau_{clump}$, there is a change in velocity $w \sim \delta v_{rms} \cong (D\tau_{clump})^{1/2}$. As derived earlier the spatial diffusion due to velocity diffusion grows as $s'^2 \sim D\tau^3$. When the displacement is comparable to a wavelength of the wave, $\lambda^2 \sim s'^2$, then the clump has moved out of the trap terminating the clump. Hence, $\tau_{clump}^3 = s'^2/D \sim \lambda^2/D$ and

$$\tau_{clump} \sim \frac{w^2}{D} \sim \left(\frac{\lambda^2}{D}\right)^{1/3} \rightarrow w \sim (D\lambda)^{1/3} \tag{7.D.3}$$

where $w$ is the clump width in velocity if the spatial and velocity diffusion times are equal. In general, if the spatial diffusion determines $\tau_{clump}$, then $w \cong (D\tau_{clump})^{1/2}$. Given an estimate for $w$, we can then estimate the corresponding velocity distribution function perturbation $\delta f \sim \left|\frac{\partial f_0}{\partial v}\right| w$. We now use the the Liouville theorem to calculate the perturbed velocity distribution in more detail:

$$\frac{df}{dt} = \frac{d}{dt}(f_0 + \tilde{f}) = 0 \rightarrow \frac{d\tilde{f}}{dt} = -\frac{df_0}{dt} = -\frac{\partial f_0}{\partial t} - \dot{v}\frac{\partial f_0}{\partial v} = -\dot{v}\frac{\partial f_0}{\partial v} \rightarrow \tilde{f} \sim -\int dt\, \dot{v}\frac{\partial f_0}{\partial v} \tag{7.D.4}$$

because $f_0$ is assumed time independent. We can now form an expression for the two-time correlation function for the perturbed velocity distribution function:

$$\langle \tilde{f}(1)\tilde{f}(2) \rangle \sim w^2 \left(\frac{\partial f_0}{\partial v}\right)^2 \Delta\left(\frac{x_1 - x_2}{\lambda}\right) \Delta\left(\frac{v_1 - v_2}{w}\right), \quad \Delta(y) \equiv \{1\ |y|<1,\ 0\ |y|>1\} \tag{7.D.5}$$

Although it does violence to the detailed physics, the following simplifications are adequate for subsequent integrations:

$$\Delta\left(\frac{x_1 - x_2}{\lambda}\right) \rightarrow \lambda\delta(x_1 - x_2) \quad \Delta\left(\frac{v_1 - v_2}{w}\right) \rightarrow w\delta(v_1 - v_2) \tag{7.D.6}$$

Using Eqs.(7.D.3) - (7.D.6) we can estimate that

$$\langle \tilde{f}(x_1,v_1)\tilde{f}(x_2,v_2)\rangle \sim \lambda^2 D\left(\frac{\partial f_0}{\partial v}\right)^2 \delta(x_1-x_2)\delta(v_1-v_2) \qquad (7.D.7)$$

and hence,

$$\langle \tilde{n}(x_1)\tilde{n}(x_2)\rangle \sim \lambda^2 \delta(x_1-x_2)\int dv D(v)\left(\frac{\partial f_0}{\partial v}\right)^2 \qquad (7.D.8)$$

Next we Fourier transform from $x_1 - x_2 \to k$ and make use of $\int dx e^{ikx}\delta(x) = 1$ to obtain the spectral density for density fluctuations:

$$S_{\tilde{n}}(k) \sim \langle \tilde{n}\tilde{n}\rangle(k) \sim \frac{1}{k^2}\int dv D(v)\left(\frac{\partial f_0}{\partial v}\right)^2 \qquad (7.D.9)$$

We note that the correlation functions were derived earlier to be evaluated at the same time. If instead they involve two times $t_1$ and $t_2$, then

$$\langle \tilde{f}(x_1,v_1)\tilde{f}(x_2,v_2)\rangle \sim \lambda^2 D\left(\frac{\partial f_0}{\partial v}\right)^2 \delta(x_1-x_2)\delta(v_1-v_2)\Delta\left(\frac{t_1-t_2}{\tau_{clump}}\right) \qquad (7.D.10)$$

If we allow for diffusion and resonance broadening then $\delta(\omega - kv) \to R(\omega - kv; D)$. The generalization of Eq.(7.D.9) is then

$$S_{\tilde{n}}(k,\omega) \sim \langle \tilde{n}\tilde{n}\rangle(k,\omega) \sim \frac{1}{k^2}\int dv D(v)\left(\frac{\partial f_0}{\partial v}\right)^2 R(\omega-kv;D) \qquad (7.D.11)$$

Now use $S_{\tilde{\rho}}(k,\omega) = e^2 S_{\tilde{n}}(k,\omega)$ and Poisson's equation with plasma shielding $\phi = \frac{4\pi\tilde{\rho}}{k^2\varepsilon}$ to obtain the field spectral density dropping various numerical constants along the way

$$S_E(k,\omega) = k^2 S_\phi(k,\omega) \sim \left(\frac{4\pi}{k^2}\right)^2 \frac{e^2}{|\varepsilon|^2}\int dv D(v)\left(\frac{\partial f_0}{\partial v}\right)^2 R(\omega-kv;D) \qquad (7.D.12)$$

From our earlier development of quasi-linear theory of diffusion

$$D_{waves}(v) = \left(\frac{e}{m}\right)^2 \iint \frac{dkd\omega}{(2\pi)^2} S_E(k,\omega) R(\omega-kv;D) \qquad (7.D.13)$$

The equations are not closed without equations for $f_0$ and $\frac{df_0}{dt}$:

$$\frac{\partial f_0}{\partial t} = -\frac{\partial}{\partial v}(F f_0) + \frac{\partial}{\partial v}\left(D(v)\frac{\partial f_0}{\partial v}\right) \qquad (7.D.14)$$

where $F$ is the dynamic friction (in this case the friction per unit mass, or deceleration) and Eq.(7.D.14) is the Fokker-Planck equation for the quasi-linear diffusion of the clumps.

We return to Eq.(7.D.12) and (7.D.13) to obtain a consistency relation for $f_0$. Approximate the resonance function with a $\delta$ function. Then

$$D(v) = \left(\frac{e}{m}\right)^2 \iint \frac{dk d\omega}{(2\pi)^2} S_E(k,\omega) \delta(\omega - kv)$$

$$= \iint \frac{dk d\omega}{(2\pi)^2} \left(\frac{4\pi e^2}{k^2 m}\right)^2 \int dv' \frac{D(v')}{|\varepsilon(k,\omega)|^2} \left(\frac{\partial f_0}{\partial v}\right)^2 \frac{1}{k} \delta(v' - \frac{\omega}{k}) \delta(\omega - kv) \quad (7.D.15)$$

$$\sim \left(\frac{4\pi e^2}{m}\right)^2 \int \frac{dk}{2\pi} \frac{1}{k^5} \frac{D(v' = \frac{\omega}{k} = v)}{|\varepsilon(k,\omega = kv)|^2} \left(\frac{\partial f_0}{\partial v}\right)^2$$

We define $f_0 = n_0 g$ and divide both sides of Eq.(7.D.15) by $D(v)$ to obtain a consistency relation for $g$:

$$1 \sim \omega_p^4 \int \frac{dk}{k^5} \frac{1}{|\varepsilon(k,\omega = kv)|^2} \left(\frac{\partial g}{\partial v}\right)^2 \approx \omega_p^4 \left(\frac{\partial g}{\partial v}\right)^2 \int \frac{dk}{k^5} \frac{1}{\left(\frac{\partial \varepsilon'}{\partial \omega}\right)^2 \left[(kv - \omega_k)^2 + \gamma_k^2\right]^2} \quad (7.D.16)$$

where we used $|\varepsilon|^2 = \varepsilon'^2 + \varepsilon''^2 \approx \left(\frac{\partial \varepsilon'}{\partial \omega}\right)^2 \left[(kv - \omega_k)^2 + \gamma_k^2\right]^2$ and expanded around resonance. For Langmuir waves Eq.(7.D.16) yields

$$1 \sim \omega_p^6 \frac{1}{v^2} \left(\frac{\partial g}{\partial v}\right)^2 \int \frac{dk}{k^5} \frac{1}{\left[(k - \frac{\omega_p}{v})^2 + \frac{\gamma_k^2}{v^2}\right]^2} \sim \omega_p^6 \frac{1}{v^2} \left(\frac{\partial g}{\partial v}\right)^2 \frac{1}{k^5} \pi \frac{v}{\gamma_k}$$

$$\sim \omega_p^6 \frac{1}{v^2} \left(\frac{\partial g}{\partial v}\right)^2 \left(\frac{v}{\omega_p}\right)^5 \pi \frac{v}{\gamma_k} \sim \frac{\omega_p v^4}{\gamma_k} \left(\frac{\partial g}{\partial v}\right)^2 \quad (7.D.17)$$

In Sec. 2.I we derived the relation $\gamma_k \sim \frac{\omega_p^3}{k^2} g'(v = \omega/k) = \frac{\omega_p^3}{(\omega_p/v)^2} g'(v = \omega/k)$ aside from numerical factors. Introduce the definition $A \equiv -v^2 dg/dv$ where $A$ is a number. Then $dg/dv = A/v$.

Finally we can calculate the radiation in Langmuir waves due to clumps.

$$\dot{W}(k) = S_j(k,\omega_k) = \left(\frac{\omega}{k}\right)^2 S_{\tilde{\rho}}(k,\omega_k) = \frac{\omega_p^2 e^2}{k^5} D(v = \frac{\omega_k}{k}) \left(\frac{\partial f_0}{\partial v}\right)^2_{\frac{\omega_k}{k}} \quad (7.D.18)$$

The expression in Eq.(7.D.18) can be integrated over $k$ and weighted by $k/\omega_k$ to calculate the growth of wave momentum to conserve momentum with the dynamic friction:

$$\int \frac{dk}{2\pi} \frac{k}{\omega_k} \dot{W}(k) = \int d\text{v}\, m |F| f_0 \qquad (7.D.19)$$

$$\rightarrow \int \frac{d\text{v}}{2\pi} \frac{k}{\text{v}^2} \dot{W}(k=\frac{\omega_k}{\text{v}}) = \int d\text{v}\, m|F|f_0 \rightarrow m|F|f_0 = \frac{1}{2\pi}\frac{k}{\text{v}^2}\dot{W}(k=\frac{\omega_k}{\text{v}})$$

Equating the integrands in the integrals over velocity and using Eq.(7.D.18) we deduce

$$\dot{\text{v}} = F(\text{v}) = -\text{v}^2 D(\text{v}) \frac{(g')^2}{g(\text{v})} \qquad (7.D.20)$$

Before returning to the Fokker-Planck equation we compare the dynamic friction calculated in Eq.(7.D.20) to the deceleration due to collisions using the resonance width due to clumps $w \sim (D\lambda)^{1/3}$:

$$\frac{\dot{\text{v}}}{\text{v}} \sim \left\{ \nu \sim \frac{\omega_p}{\Lambda} \text{ due to collisons}, \quad \omega_p \left(\frac{w}{\text{v}}\right)^3 \text{ due to dynamic friction from clumps} \right\}$$

(7.D.21)

The dynamic friction is much larger than collisional deceleration if the resonance width is sufficiently large.

Exercise: Fill in the numerical factors in this section and extend the formalism to three dimensions in velocity and configuration space.

The derivation in Sec. 7.D yields a steady state for $f_0$, i.e., $\partial f_0 / \partial t = 0$. For whatever value of $A$ introduced after Eq.(7.D.17) is used, the plasma is stable. Nowhere have we established the turbulence level. We have only determined the velocity range where resonant waves are strong enough to give clumps. In this range, $g \sim A/\text{v}$. The characteristic rate for the wave dynamics (for Langmuir waves) is $\sim \omega_p$. The characteristic rate for clump evolution is $\sim \omega_p w/\text{v}$. The characteristic rate of evolution for $g$ is $\sim \omega_p (w/\text{v})^3$.

Ion acoustic waves can clump electron and ions more so. Both ion and electron clumps have dynamic friction. Dupree calculated anomalous resistivity associated with clumps and obtained $\sigma \sim 10\omega_p / k\lambda_D$ where $k$ is the dominant wave number of the ion acoustic instability responsible for the waves.[59]

# LECTURES ON THEORETICAL PLASMA PHYSICS – PART 4A

*Allan N. Kaufman*

## 8. Non-uniform plasmas: adiabatic invariance, local instabilities driven by nonuniformity, and configurational instabilities

In Sec. 8 we address non-uniform plasmas. We present a theory of adiabatic invariance, a Lagrangian approach to guiding-center drifts, guiding-center theory and hydromagnetic equations, and an introduction to the theory of the stability of drift waves.

### 8.A Adiabatic invariance

Consider the adiabatic invariance of the magnetic moment using the model problem of a one-dimensional harmonic oscillator. Then we extend the model to a nonlinear, anharmonic oscillator. The approach taken here is due to R. L. Dewar.[62] The equation of a harmonic oscillator with a time-dependent restoring force in one dimension is

$$\ddot{x}(t) + \Omega^2(t) x(t) = 0, \quad |\dot{\Omega}|/\Omega \ll \Omega \tag{8.A.1}$$

<u>Definition</u>: Action $J \equiv \dfrac{H}{\Omega} = \dfrac{\tfrac{1}{2}\dot{x}^2 + \tfrac{1}{2}\Omega^2 x^2}{\Omega}$ with the mass $m \equiv 1$.

$J$ is an approximate invariant. In what sense is this true? This is a classic Rayleigh-Lorentz pendulum problem posed by Lorentz and solved by Einstein at the beginning of the 20th century at a Solvay Conference. The WKB solution to Eq.(8.A.1) is

$$x(t) \approx \dfrac{a}{\sqrt{\Omega(t)}} \sin\left[\int_0^t dt'\, \Omega(t') + \alpha\right] \tag{8.A.2}$$

Using the definition of $J$ in the preceding, one concludes that $J = \Omega <x^2> \approx \dfrac{a^2}{2}$ is a constant in time. Now suppose we introduce a WKB form for $x(t)$ in terms of a new unknown frequency $\omega(t)$ so that

$$x(t) \approx \dfrac{a}{\sqrt{\omega(t)}} \sin\left[\int_0^t dt'\, \omega(t') + \alpha\right] \equiv \dfrac{a}{\sqrt{\omega(t)}} \sin\theta(t) \tag{8.A.3}$$

---

[62] R. L. Dewar, http://people.physics.anu.edu.au/~rld105/Dewar_Docs/RLD-Publications.html

We use Eq.(8.A.3) to evaluate the first and second time derivatives of $x$ and then the equation of motion Eq.(8.A.1) to obtain

$$-a\left[\frac{3}{4}\omega^{-5/2}\dot{\omega}^2 - \frac{1}{2}\omega^{-3/2}\ddot{\omega} + \omega^{-1/2}\left(\Omega^2 - \omega^2\right)\right] = 0 \qquad (8.A.4a)$$

or

$$\frac{3}{4}\omega^{-2}\dot{\omega}^2 - \frac{1}{2}\omega^{-1}\ddot{\omega} + \left(\Omega^2 - \omega^2\right) = 0 \rightarrow \omega^2 - \Omega^2(t) = \omega^{1/2}\frac{d^2}{dt^2}(\omega^{-1/2}) \qquad (8.A.4b)$$

which is just as difficult to solve. However, what is gained is that Eq.(8.A.4b) gives a recipe for an iterative solution:

$$\omega^2 = \Omega^2(t) + \Omega^{1/2}\frac{d^2}{dt^2}(\Omega^{-1/2}) + \ldots \qquad (8.A.5)$$

We see the first correction to $\omega^2 \approx \Omega^2(t)$ on the right side of Eq.(8.A.5), which is an asymptotic series. Unfortunately the series diverges. In what sense is this result useful? We answer in an example. Consider the action

$$\left\langle \omega(t)x^2(t) \right\rangle_\theta = a^2\left\langle \sin^2\theta(t) \right\rangle_\theta = \frac{a^2}{2} \qquad (8.A.6)$$

If we use a linear expansion of Eq.(8.A.5) for $\omega(t)$, which is just a function of $t$,

$$\omega \approx \Omega(t)\left[1 + \frac{1}{2}\Omega^{-3/2}\frac{d^2}{dt^2}(\Omega^{-1/2}) + \ldots\right] \qquad (8.A.7)$$

and divide both sides of Eq.(8.A.6) by this expression, we obtain an auxiliary action $I_0(t)$:

$$I_0(t) = \left\langle \Omega x^2(t) \right\rangle_\theta \approx \frac{a^2}{2}\left[1 - \frac{1}{2}\Omega^{-3/2}\frac{d^2}{dt^2}(\Omega^{-1/2}) + \ldots\right] \qquad (8.A.8)$$

The second term on the right side of Eq.(8.A.8) is useful in providing a quantitative estimate of the relative constancy of the lowest-order expression for the action. We note that

$$\ln I_0(t) = \ln\frac{a^2}{2} - \frac{1}{2}\Omega^{-3/2}\frac{d^2}{dt^2}(\Omega^{-1/2}) + \ldots \rightarrow \frac{\dot{I}_0}{I_0} = -\frac{d}{dt}\left[\frac{1}{2}\Omega^{-3/2}\frac{d^2}{dt^2}(\Omega^{-1/2}) + \ldots\right] \ll \frac{\dot{\Omega}}{\Omega} \qquad (8.A.9)$$

Thus, the relative rate of change in the action is much smaller than the relative rate of change in $\Omega$.

Exercise: Assume $\Omega(t) = \Omega_0(1 + \varepsilon e^{\gamma t})$ and $\gamma \ll \Omega_0$. Evaluate Eq.(8.A.9) and confirm its conclusion.

Example: An example of an asymptotic series. At this point we pause to discuss an example of a potentially divergent asymptotic power series. Useful references for this and related matters are Erdelyi et al.,[63] Dwight,[64] and Abramowitz and Stegun.[65] Consider the function $f(x)$

$$f(x) = \int_0^\infty dt \frac{\exp(-t)}{1+xt}, \quad \text{Re}(x) \geq 0 \tag{8.A.10}$$

At $x=0$, $f(0)=1$; and $f$ is monotonically decreasing for increasing $x$. For $x>>1$, $f(x>>1) \approx \frac{\ln(x-\gamma)}{x}\exp(-x)$, $\gamma = 0.577$ (Euler's constant). Note that $\frac{1}{1+xt} = 1 - xt + x^2t^2 - x^3t^3 + \ldots$, which converges only for $xt<1$. Also $\exp(-x) = \sum_n g_n = \sum_n (-1)^n \frac{x^n}{n!}$. The power series representation for $f(x)$ is then

$$f(x) = \sum_n f_n(x) = 1 - x1! + x^2 2! - x^3 3! + \ldots (-1)^n x^n n! + \ldots \tag{8.A.11}$$

The ratio of the nth term to the (n-1)th term is $(-1)nx$, the magnitude of which is greater than unity for $n>1/x$. The series is clearly oscillating and diverging. The power series expansion of $f(x)$ only makes sense for $x<<1$. However, the truncated asymptotic power series has an error that is less than the next successive term in the truncated series. For example, if we retain four terms in Eq.(8.A.11), the next term is $x^4 4!$. By choosing $n$ and $x$ carefully, we can minimize the error in the truncated power series. Consider $x=0.1$ for which $f(0.1)=0.91563$ The partial sums are

$$S_0 = 1, S_1 = 0.9, S_2 = 0.92, S_3 = 0.914, S_4 = 0.9164, S_5 = 0.9152, S_6 = 0.91592,$$
$$S_7 = 9158700, S_8 = 0.9158736, S_9 = 0.9158700, S_{10} = 0.9158736$$

<u>Definition</u>: The *Error* $\equiv |f(x) - S_N| < |f_{n+1}| = |x^n n!|$

In this series the error due to truncation is minimal when the ratio of successive terms is O(1), i.e., $nx=1$. Hence, for $x=0.1$, we truncate the series at $n=10$, for which $x^n n! = 0.00036$ is the error bound. The actual error is 0.00024 Thus, there is always an error, but the error can be made tolerably small for a wide range of argument.

---

[63] A. Erdélyi, W. Magnus, F. Oberhettinger, and F. Tricomi, *Tables of Integral Transforms*, Vols. 1 and 2, McGraw-Hill, New York, (1954).
[64] H. B. Dwight, *Tables of Integrals and Other Mathematical Data*, Fourth Edition (Macmillan), (1964).
[65] Milton Abramowitz and Irene Ann Stegun, eds. *Handbook of Mathematical Functions with Formulas, Graphs, and Mathematical Tables*. Washington D.C.; New York: United States Department of Commerce, National Bureau of Standards (1964); reprinted by Dover Publications.

### 8.A.a Harmonic oscillator model – use of WKB eikonal theory, asymptotic theory, and action-angle variables to derive an approximate constant of the motion

Consider the following model problem with a time-dependent magnetic field and a Lagrangian description with a particular choice of gauge:

$$\mathbf{B} = B(t)\hat{\mathbf{z}} \quad B_z = \frac{\partial}{\partial x} A_y \quad \mathbf{A} = B(t)x\hat{\mathbf{y}} \quad \mathbf{E}^T = -\frac{1}{c}\frac{\partial}{\partial t}\mathbf{A} = -\frac{1}{c}\frac{\partial}{\partial t}\dot{B}x\hat{\mathbf{y}} \quad (8.A.12)$$

Motion in $z$ is ignored. The Lagrangian per unit mass for this system is

$$\begin{aligned}\frac{L}{m} &= \frac{1}{2}\dot{x}^2 + \frac{1}{2}\dot{y}^2 + \frac{e}{mc}\mathbf{v}\cdot\mathbf{A} - \frac{e\phi(x,t)}{m} \\ &= \frac{1}{2}\dot{x}^2 + \frac{1}{2}\dot{y}^2 + \frac{eB(t)x\dot{y}}{mc} - \frac{e\phi(x,t)}{m} = \frac{1}{2}\dot{x}^2 + \frac{1}{2}\dot{y}^2 + \Omega(t)x\dot{y} - \frac{e\phi(x,t)}{m}\end{aligned} \quad (8.A.13)$$

Definition: The canonical momentum is

$$p_y \equiv \frac{\partial L/m}{\partial \dot{y}} = \dot{y} + \Omega(t)x. \quad (8.A.14)$$

From the Lagrangian equations

$$\frac{d}{dt}p_y = \frac{\partial L}{\partial y} = 0 \rightarrow p_y = \text{constant} \equiv \Omega\langle x\rangle = \Omega x_{g.ctr.} \quad (8.A.15a)$$

$$p_x = \dot{x} \quad \ddot{x} = \frac{\partial L}{\partial x} = \Omega(t)\dot{y} - \frac{e}{m}\frac{\partial \phi}{\partial x} = \Omega(t)[p_y - \Omega(t)x] - \frac{e}{m}\frac{\partial \phi}{\partial x} \quad (8.A.15b)$$

The evaluation of $p_y$=constant in Eq.(8.A.15a) follows from time averaging the two oscillatory terms on the right side of Eq.(8.A.14) which obviously sum to the same constant and using $\langle\dot{y}\rangle = 0$ so that only the average of the second term remains. Hence, one arrives at the following equation of motion:

$$\ddot{x} + \Omega^2(t)x = \Omega(t)p_y - \frac{e}{m}\frac{\partial \phi}{\partial x} \Rightarrow \ddot{x} + \Omega^2(t)x = 0 \quad (8.A.16)$$

by choosing the guiding center on the $y$ axis ($p_y$=0) and with no perturbing electric potential ($\phi = 0$). With arbitrary initial conditions for the guiding center position

$$\ddot{x} + \Omega^2(t)x = \lambda(t) = \Omega(t)p_y \quad (8.A.17)$$

This is a linear inhomogeneous differential equation with non-constant coefficients.

One possible solution of Eq.(8.A.17) can be constructed iteratively:

$$\begin{aligned}\left(D_t^2 + \Omega^2(t)\right)x &= \Omega(t)p_y \rightarrow x = \left(D_t^2 + \Omega^2(t)\right)^{-1}\Omega(t)p_y = \Omega^{-1}(t)p_y(1+...) \\ \rightarrow \quad x &= \frac{\lambda(t)}{\Omega^2(t)} - \frac{\ddot{x}}{\Omega^2(t)} = \frac{\lambda(t)}{\Omega^2(t)} - \frac{1}{\Omega^2(t)}\frac{d^2}{dt^2}\left(\frac{\lambda(t)}{\Omega^2(t)}\right) + ...\end{aligned} \quad (8.A.18)$$

Example: Consider a perturbing electric potential $\phi(x,t) = \phi_0(t)\cos kx$ and $kx$<<1 so that Eq.(8.A.16) becomes

$$\ddot{x} + \Omega^2(t)x = \frac{ek}{m}\phi_0(t)\sin kx \approx \frac{ek^2}{m}\phi_0(t)x$$

$$\Rightarrow \ddot{x} + \left[\Omega^2(t) - \frac{ek^2}{m}\phi_0(t)\right]x = \ddot{x} + \left[\Omega^2(t) - \omega_B^2(t)\right]x = 0 \quad (8.A.19)$$

where $\omega_B$ is the bounce frequency associated with trapping. We recall Eq.(8.A.7)

$$\omega(t) \approx \Omega(t)\left[1 + \frac{1}{2}\Omega^{-3/2}\frac{d^2}{dt^2}(\Omega^{-1/2}) + ...\right] \quad (8.A.20)$$

which is an asymptotic series. For $\dot{\Omega}/\Omega$ small, the size of the $n^{th}$ term is small; and if the series is asymptotic we can prescribe an error limit and how many terms to retain in the series for $\omega(t)$.

Definition: A power series is formally asymptotic if it satisfies $|f(x) - S_n(x)| < |S_{n+1}(x) - S_n(x)|$. The series can be converging or diverging.

Example: $\Omega(t) = \Omega_0(1 + \varepsilon e^{\gamma t})$, $\varepsilon e^{\gamma t} << 1$, $\gamma << \Omega_0$. Then analogous to Eq.(8.A.8)

$$I_0(t) = I_0(0)\left[1 + \frac{\varepsilon}{4}\left(\frac{\gamma}{\Omega_0}\right)^2 e^{\gamma t}\right] \quad (8.A.21)$$

Thus, the action is not conserved: there is a small, exponentially growing perturbation.

Example: Consider a $\Omega(t)$ that is initially constant and then grows smoothly and slowly on the time scale $T >> \Omega_0^{-1}$ and saturates at some higher constant value. The expression for $I_0$ from Eq.(8.A.8) is

$$I_0(t) = \langle \Omega x^2(t) \rangle_\theta \approx \frac{a^2}{2}\left[1 - \frac{1}{2}\Omega^{-3/2}\frac{d^2}{dt^2}(\Omega^{-1/2}) + ...\right]$$

and shows that the action remains essentially constant, although at first order there is a small correction that is at first negative and then positive before relaxing to zero. There is an adiabatic transition of the action from one stable asymptotic state to another. In Clemow and Dougherty's book it is shown that

$$\left|\frac{I_0(\infty) - I_0(-\infty)}{I_0}\right| \sim e^{-\frac{\Omega T}{2\pi}} \quad (8.A.22)$$

Thus, there is an exponentially small change in the action. In fact, the action is conserved to all powers in $2\pi/\Omega T$ (we showed only through the first correction in $I_0$). In the $(x,\dot{x})$ phase space, the particle orbit is an ellipse with radius $a \sim 1/\sqrt{\Omega}$ in $x$ from Eq.(8.A.2) and radius $\Omega a \sim \sqrt{\Omega}$ in $\dot{x}$, but the area of the ellipse $\pi\Omega(t)a^2$ is a constant. Hence, as $\Omega(t)$ varies the particle rides on an evolving ellipse whose eccentricity changes, but the area remains invariant.

### 8.A.b Phase-space dynamics for trapped and passing particles in a single electrostatic wave

Consider a perturbing electric potential $\phi(x,t) = \phi_0(t)\cos kx$ in the wave frame with no applied magnetic field. For $kx \ll 1$ the analysis in Eqs.(8.A.20-8.A.22) shows that the phase-space orbits for the trapped particles lie on ellipses. For particle energies such that the particle excursion in the wave is larger, the electric potential cannot be expanded for small $kx$; and the orbits distort from ellipses. For sufficiently large energies, the particles are no longer confined in the potential trough; and the particle trajectories are unconfined (untrapped). The Hamiltonian for particles in the wave frame is

$$H = \frac{1}{2}\dot{x}^2 + \frac{e\phi_0(t)}{m}\cos kx \tag{8.A.23}$$

For slowly varying potential amplitudes relative to the bounce time, there is a conserved action. For trapped particles the action can be constructed from

$$J \equiv \frac{1}{2\pi}\oint \dot{x}(x;H,\phi_0)dx = J(H,\phi_0) \tag{8.A.24}$$

With $J$ an invariant, as $\phi_0$ varies, then $H$ must also vary. For an untrapped particle, we define the action based on one pass across the change in the electric potential:

$$J \equiv \frac{1}{2\pi}\int_0^{2\pi/k} \dot{x}(x;H,\phi_0)dx \tag{8.A.25}$$

We use Eq.(8.A.23) to solve for $\dot{x}$ and then evaluate the actions in Eqs.(8.A.24) and (8.A.25). The untrapped particles have action

$$J(H,\phi_0) = \frac{4}{\pi k}\sqrt{\frac{mH + e\phi_0}{2m}}E(\kappa), \quad \kappa = \sqrt{\frac{2e\phi_0}{mH + e\phi_0}} \equiv \frac{1}{\mu} \tag{8.A.26}$$

$\kappa = 0$ corresponds to infinite energy, while $\kappa = 1$ corresponds to a particle on the separtrix between passing and trapped, and $E$ is the complete elliptic integral of the second kind. For $\kappa \ll 1, E \approx \frac{\pi}{2}\left(1 - \frac{\kappa^2}{8} + \cdots\right)$. For $\kappa \to 1$, $E \approx 1 + \frac{1}{2}(1-\kappa^2)(\ln\frac{4}{\sqrt{1-\kappa^2}} - \frac{1}{2})$. The trapped particles have action

$$J(H,\phi_0) = \frac{8}{\pi k}\sqrt{\frac{e\phi_0}{m}}\left[E(\mu) - (1-\mu^2)K(\mu)\right] \tag{8.A.27}$$

The argument of the complete elliptic integrals of the first and second kind in Eq.(8.A.27) are the inverse of that in Eq.(8.A.26). For $\mu = \frac{1}{\kappa} \to 0$, corresponding to the bottom of the potential well, $E(\mu) \approx \frac{\pi}{2}(1 + \frac{\mu^2}{4} + \cdots)$ and $K(\mu) \approx \ln\frac{4}{\sqrt{1-\mu^2}}$. The action is a linearly increasing function of $H$ for the trapped particle region, $-\frac{e\phi_0}{m} < H < \frac{e\phi_0}{m}$, until approaching the separatrix at $H = e\phi_0$ where $J$ swings up sharply. For $H \gg e\phi_0$ the action increases with $H$ asymptotically as $\sqrt{H}$. We note that at the bottom of the potential well, the particle motion is that of a harmonic oscillator and $J = H/\Omega$. We also note the following relations:

$$\omega = \dot{\theta} = \frac{\partial H}{\partial J} = \omega(J) \quad \dot{J} = -\frac{\partial H}{\partial \theta} = 0 \quad \tau \equiv \frac{2\pi}{\omega(J)} = 2\pi \frac{dJ}{dH} \tag{8.A.28}$$

As the separtrix is approached, $\tau \to \infty$ and $dJ/dH \to \infty$.

We can prove conservation of action quite generally. Consider a particle Hamiltonian in terms of canonically conjugate variables with a slowly varying parameter $\lambda(t)$:

$$H(q,p;\lambda(t)), \quad p(q,H;\lambda(t)) \tag{8.A.29}$$

We define the action $J$ as

$$J \equiv \oint dq\, p(q,H;\lambda) \tag{8.A.30}$$

Then the time derivative of $J$ is calculated as follows:

$$\dot{J} = \frac{\partial J}{\partial H}\dot{H} + \frac{\partial J}{\partial \lambda}\dot{\lambda}, \quad \left.\frac{dH}{dt}\right|_{p,q} = \frac{\partial H}{\partial t} = \frac{\partial H}{\partial \lambda}\dot{\lambda} \quad \Rightarrow \quad \dot{J} = \left(\frac{1}{\omega}\frac{\partial H}{\partial \lambda} + \frac{\partial J}{\partial \lambda}\right)\dot{\lambda} \tag{8.A.31}$$

At this point the following lemma is helpful:

$$\left.\frac{\partial p}{\partial \lambda}\right|_{q,H} = \frac{-\left.\frac{\partial H}{\partial \lambda}\right|_{p,q}}{\left.\frac{\partial H}{\partial p}\right|_{q,\lambda}} \tag{8.A.32}$$

We can then evaluate $\partial J/\partial \lambda|_H$

$$\begin{aligned}\left.\frac{\partial J}{\partial \lambda}\right|_H &= \frac{1}{2\pi}\oint dq\, \left.\frac{\partial p(q,H,\lambda)}{\partial \lambda}\right|_{q,H} = -\frac{1}{2\pi}\oint \frac{dq}{\dot{q}(q,p,\lambda)}\left.\frac{\partial H}{\partial \lambda}\right|_{p,q} \\ &= -\frac{\tau}{2\pi}\oint \frac{dt}{\tau}\left.\frac{\partial H}{\partial \lambda}\right|_{p,q} = -\frac{1}{\omega}\left\langle \frac{\partial H}{\partial \lambda}\right\rangle(H,\lambda)\end{aligned} \tag{8.A.33}$$

where the average on the right side of Eq.(8.A.33) is over the orbit. We then substitute the result of Eq.(8.A.33) in Eq.(8.A.31) to obtain

$$\dot{J} = \frac{\dot{\lambda}}{\omega}\left(\frac{\partial H}{\partial \lambda} - \left\langle\frac{\partial H}{\partial \lambda}\right\rangle\right) \quad \to \quad \left\langle \dot{J}\right\rangle \approx \frac{\dot{\lambda}}{\omega}\left(\left\langle\frac{\partial H}{\partial \lambda}\right\rangle - \left\langle\frac{\partial H}{\partial \lambda}\right\rangle\right) = 0 \tag{8.A.34}$$

for $\left|\dot{\lambda}/\omega\lambda\right| \ll 1$. Thus, $\left\langle \dot{J}\right\rangle_{orbit} = 0$, although $\dot{J} \neq 0$ due to slow variations in $\lambda$ and $H$.

Example: For the Hamiltonian in Eq.(8.A.23), as $\phi_0(t)$ changes, $H$ changes; and the separatrix in $(x,v)$ phase space changes its width. A particle will tend to remain on a curve of constant action. Although the average of the time derivative of the action over the phase in Eq.(8.A.34) is zero, the time (not orbit) averaged $\left\langle dJ/dt\right\rangle = O(\dot{\lambda})$ from Eq.(8.A.31) is not zero. As a parameter is slowly changed causing an adiabatic transition from one stable orbit to another stable orbit, there is an exponentially small change in $J$. When a particle crosses the separatrix, the change in $J$ is still perturbative.[66]

Example: The examples considered here have implications for the saturation of instabilities. Consider a weakly growing bump-on-tail instability. The phase velocity of an unstable wave falls on a region of the velocity distribution function $f(v)$ with positive slope due to the bump on the tail. As the growing wave traps resonant particles, there are more particles with velocities faster than the wave than slower. Hence, the resonant particles lose momentum; and the non-resonant particles that account for the wave momentum must gain momentum to conserve total momentum. The net deceleration of resonant particles and acceleration of non-resonant particles does not continue indefinitely: the resonant momentum loss and non-resonant momentum gain reach limits. A more careful analysis shows that the energy of the resonant particles changes. Conservation of energy implies that the wave energy must evolve in consequence. However, in the wave frame the wave frequency and energy would be zero were it not for the fact that trapping leads to a frequency shift $\delta\omega$.[67] Dewar[68] shows that the wave continues to grow until the trapping frequency exceeds the linear growth rate by a certain amount:

$$\omega_B \equiv \sqrt{\frac{k^2 e\phi_0}{m}} = \frac{256}{9\pi^2}\gamma_L = 2.88\gamma_L \tag{8.A.35}$$

B. D. Fried, C. S. Liu, R. W. Means, and R. Z. Sagdeev found in numerical simulation a factor of 3.2 rather than 2.88 in the saturation.[69]

Exercise: Suppose that the parameter in the Hamiltonian $\lambda(t)$ has variation $\lambda(t) = \lambda_0 + \lambda_1 \sin\Omega_1 t$ with $\lambda_1 \ll \lambda_0$ and consider three interesting situations in which the particle Hamiltonian has a sinusoidal electrostatic wave with frequency $\omega$:
1. $\Omega_1 \approx \omega$ parametric resonance
2. $\Omega_1 \approx \omega \frac{\ell}{m}$ rational relation
3. $\Omega_1 \ll \omega$ adiabatic

Example: Anharmonic oscillator – Consider a Hamiltonian

$$H = \frac{1}{2}\dot{x}^2 + V(x) = \frac{1}{2}\dot{x}^2 + a_0 + \frac{1}{2}a_1(t)x^2 + \frac{1}{4}a_2 x^4 \tag{8.A.36}$$

with $a_1(t)<0$ and $|a_1|$ increases in magnitude. With $a_1(t)=0$ there is a single potential well that is symmetric about $x=0$. For finite $|a_1|$ there are two wells symmetric

---

about x=0 that become deeper as $|a_1|$ increases in magnitude. With the inclusion of a term $\frac{1}{3}a_3 x^3$ in $V$, there is no longer symmetry with respect to $x=0$. There can still be two potential wells that can trap particles, but the wells are asymmetric. The probability of trapping in one or the other well is proportional to the phase-space area encompassed by each well. Under adiabatic changes, the total probability is conserved.

### 8.A.c Magnetic mirror geometry and derivation of adiabatic invariant using canonical transformations

Consider particle motion in a simple magnetic mirror configuration with two degrees of freedom, i.e., the magnetic field has $x$ and $z$ components and varies in $x$ and $z$ (no $y$ variation). The Lagrangian for a charged particle in this case is

$$L = \frac{1}{2}\dot{x}^2 + \frac{1}{2}\dot{y}^2 + \frac{1}{2}\dot{z}^2 + \frac{e}{mc}\mathbf{v}\cdot\mathbf{A} \tag{8.A.37}$$

The magnetic field is represented by

$$\mathbf{B} = \nabla\times\mathbf{A} \quad \mathbf{A} = \hat{\mathbf{y}}A(x,z) = \hat{\mathbf{y}}xB_0(z) \quad B_z = \frac{\partial A}{\partial x} \quad B_x = -\frac{\partial A}{\partial z} = -x\frac{\partial B_0}{\partial z} \tag{8.A.38}$$

Assume that $B_0(z) = B_0(1+z^2/L^2)$. Because the Lagrangian in Eq.(8.A.37) has no dependence on $y$, $p_y$ is a constant of the motion:

$$p_y \equiv \frac{\partial L}{\partial \dot{y}} = \dot{y} + \frac{e}{mc}A(x,z) = \dot{y} + \Omega(z)x, \quad \Omega(z) \equiv \frac{eB_0(z)}{mc} \tag{8.A.39}$$

$p_y$ specifies which field line the particle is gyrating around and transiting on. Energy conservation is determined by

$$E = \frac{1}{2}\dot{x}^2 + \frac{1}{2}\dot{y}^2 + \frac{1}{2}\dot{z}^2 = L = \frac{1}{2}\dot{x}^2 + \frac{1}{2}(p_y - \Omega(z)x)^2 + \frac{1}{2}\dot{z}^2 \tag{8.A.40}$$

We choose $p_y=0$ and obtain

$$H(x,z;p_x,p_z) = E = \frac{1}{2}\dot{x}^2 + \frac{1}{2}\Omega^2(z)x^2 + \frac{1}{2}v_\parallel^2 \equiv \frac{1}{2}v_\perp^2 + \frac{1}{2}v_\parallel^2 \tag{8.A.41}$$

$E$ is clearly a constant of the motion because $H$ has no explicit time dependence. In consequence, only three of the four variables $(x,z;p_x,p_z)$ are independent, which defines a volume in the 4D phase space. From Eqs.(8.A.40) and (8.A.41)

$$p_x^2 + \Omega^2(z)x^2 = 2E - p_z^2 \le 2E \tag{8.A.42}$$

The surface defined by $p_x^2 + \Omega^2(z)x^2 = 2E$ acts as a bounding surface in $(x,z,p_x)$ space, which contains the particle trajectory. The particle trajectory may be a closed curve (it is exactly periodic) or it may fill the bounded volume eventually, i.e., the orbit may be ergodic. We define the orbit to be ergodic if for any point in the volume one can construct a small neighborhood around the point and the particle trajectory/orbit will eventually intersect the volume.

Let us construct a conserved action from the particle motion in the simple mirror. Assume that the gyroradius is much smaller than the axial scale length of the magnetic well: $\rho \ll L$. Now freeze $(z, p_z)$. From Eq.(8.A.30) we construct the action

$$J_x \equiv \oint dx\, p_x(x; H, z, p_z) \to \mu \tag{8.A.43}$$

which is the magnetic moment except for factors of *e*, *m*, and *c*. Consider $dJ_x/dt$:

$$\dot{J}_x = \frac{\partial J_x}{\partial z}\dot{z} + \frac{\partial J_x}{\partial p_z}\dot{p}_z \tag{8.A.44}$$

and

$$\frac{\partial J_x}{\partial z} = \oint dx\, \frac{\partial p_x}{\partial z}\bigg|_{x,H,p_z} = \oint dx\, \frac{-\frac{\partial H}{\partial z}\big|_{x,p_x,p_z}}{\frac{\partial H}{\partial p_x}\big|_{x,z,p_z}} = \oint dx\, \frac{\dot{p}_z}{\dot{x}} \equiv \tau_x \left\langle \dot{p}_z \right\rangle_{x-orbit}$$

$$\frac{\partial J_x}{\partial p_z} = \oint dx\, \frac{\partial p_x}{\partial p_z}\bigg|_{x,z,H} = \oint dx\, \frac{-\frac{\partial H}{\partial p_z}\big|_{x,z,p_x}}{\frac{\partial H}{\partial p_x}\big|_{x,z,p_z}} = -\oint dx\, \frac{\dot{z}}{\dot{x}} \equiv -\tau_x \left\langle \dot{z} \right\rangle_{x-orbit} \tag{8.A.45}$$

Hence, $dJ_x/dt$ from Eq.(8.A.44) becomes

$$\dot{J}_x = \frac{\partial J_x}{\partial z}\dot{z} + \frac{\partial J_x}{\partial p_z}\dot{p}_z = \tau_x \left[ \dot{z}\left\langle \dot{p}_z \right\rangle_{x-orbit} - \dot{p}_z \left\langle \dot{z} \right\rangle_{x-orbit} \right]$$

$$\Rightarrow \left\langle \dot{J}_x \right\rangle_{x-orbit} = 0 \tag{8.A.46}$$

The average over *x* orbits is equivalent to an average over many particles with different phases in their orbits within the accessible volume of orbits. As in Eqs.(8.A.31-8.A.34) the orbit average over phases is zero but the time average $< dJ_x/dt > = 0 + O(\epsilon^2)$, where $\epsilon = \rho_g/L_z$ is the ratio of the Larmor radius to the axial scale length of the magnetic field.

We return to Eq.(8.A.43) to calculate the action. We use $p_x = \left(2E - p_z^2 - \Omega^2(z)x^2\right)^{1/2}$ and integrate $J_x = \oint dx\, p_x$ for fixed *z* and $p_z$. One obtains

$$J_x(H, z, p_z) = \oint dx \left(2E - p_z^2 - \Omega^2(z)x^2\right)^{1/2} = \oint dx \left(2(\tfrac{1}{2}mv_\perp^2) - \Omega^2(z)x^2\right)^{1/2}$$

$$= \oint dx \left(2W_\perp - \Omega^2(z)x^2\right)^{1/2} \approx \sqrt{2W_\perp} \oint dx \left(1 - \frac{\Omega^2(z)x^2}{2W_\perp}\right)^{1/2} \tag{8.A.47}$$

$$= \frac{2W_\perp}{\Omega} \oint d\theta \sin^2\theta = 2\pi \frac{W_\perp}{\Omega} \propto \frac{W_\perp}{\Omega} \equiv \mu$$

where $W_\perp$ is approximately constant over the gyro orbit, $dx = -\frac{\sqrt{2W_\perp}}{\Omega}\sin\theta d\theta$, and the integration is over the gyro orbit. Action and energy conservation constrain the particle orbit in the $(x,p_x)$ plane to be an ellipse with radii $(2E - p_z^2)^{1/2}/\Omega(z) = (2J_x/\Omega(z))^{1/2}$ in $x$ and $(2E - p_z^2)^{1/2} = (2J_x\Omega(z))^{1/2}$ in $p_x$.

How well is the magnetic moment conserved over the bounce motion of the particle in the magnetic well? Suppose $\epsilon = \rho_g/L_z \sim 10^{-2}$. An estimate of the relative change in the action $\mu = J_x \equiv J_g$ over a bounce is

$$\frac{\Delta\langle J_g\rangle}{J} \sim \tau_{bounce} \frac{\langle \dot J_g\rangle}{J} = O\left(\frac{\tau_g}{\varepsilon}\right)\frac{\langle \dot J_g\rangle}{J} \sim O\frac{(\varepsilon^2)}{\varepsilon} \sim O(\varepsilon) \sim 10^{-2} \tag{8.A.48}$$

If the drift time is one order longer in $\epsilon^{-1}$, i.e., 100 bounces in this example, then the relative change in the action is O(1). Thus, a detailed calculation is needed to quantify the change in action over many bounces.

The Hamiltonian in Eq.(8.A.41) is

$$H(x,\dot x,z,\dot z) \to H(x,p_x;z,p_z) = \frac{1}{2}p_x^2 + \frac{1}{2}\Omega^2(z)x^2 + \frac{1}{2}p_z^2 \tag{8.A.49}$$

We introduce a canonical transformation to represent the system using the old coordinates but with new momentum variables, so that the new Hamiltonian will be written in terms of action-angle variables: $(x,p_x;z,p_z) \to (\theta_g, J_g; Z, P_z)$. The generating function for the canonical transformation is

$$S(x,z;J_g,P_z) = \int_0^x dx'\sqrt{2J_g\Omega(z) - \Omega^2(z)x'^2} + zP_z$$

$$\to \theta_g = \frac{\partial S}{\partial J_g}$$

$$Z = \frac{\partial S}{\partial P_z} = z \tag{8.A.50}$$

$$p_x = \frac{\partial S}{\partial x} = \sqrt{2J_g\Omega(z)}\cos\theta_g$$

$$p_z = \frac{\partial S}{\partial z} = P_z + O(\varepsilon) = P_z + \int_0^x dx' \frac{J_g\Omega' - \Omega\Omega' x'^2}{\sqrt{2J_g\Omega(z) - \Omega^2(z)x'^2}} = P_z + \frac{\Omega'}{2\Omega}J\sin 2\theta_g$$

where $\Omega' = d\Omega/dz = O(\varepsilon)\Omega$ and $x \equiv \sqrt{\frac{2J_g}{\Omega(z)}}\sin\theta_g$ from the action orbit in the $(x,p_x)$ plane. The Hamiltonian in the new coordinates is

$$H(\theta_g, J_g, Z, P_z) = \Omega(z)J_g + \tfrac{1}{2}[P_z + \frac{\Omega'}{2\Omega}J_g\sin 2\theta_g]^2 \tag{8.A.51}$$

From the Hamiltonian in Eq.(8.A.51) we can evaluate the time derivative of the action:

$$\dot{J}_g = -\frac{\partial H}{\partial \theta_g} = -[P_z + \frac{\Omega'}{2\Omega} J_g \sin 2\theta_g] \frac{\Omega'}{\Omega} J_g \cos 2\theta_g + O(\varepsilon^2) = O(\varepsilon) \qquad (8.A.52)$$

We see that the time derivative of the action has a small $O(\Omega'/\Omega) = O(\epsilon)$ rapid variation. To lowest order the time integral of Eq.(8.A.52) ignoring other derivatives yields

$$\Delta J_g = -P_z \frac{\Omega'}{2\Omega^2} J_g \sin 2\theta_g + O(\varepsilon^2) = O(\varepsilon) \qquad \dot{\theta}_g = \frac{\partial H}{\partial J_g} = \Omega + O(\varepsilon) \qquad (8.A.53)$$

We define a new approximately conserved quantity

$$I_g \equiv J_g - \Delta J_g \quad \Rightarrow \quad \dot{I}_g \equiv \dot{J}_g - \Delta \dot{J}_g = O(\varepsilon^2) \qquad (8.A.54)$$

We next introduce a new canonical transformation to remove the high-frequency jitter in Eq.(8.A.52): $(\theta_g, J_g; Z, P_z) \to (\Theta_g, I_g; Z, P_z)$ using

$$S(\theta_g, Z; I_g, P_z) = I_g \theta_g + P_z \frac{\Omega'}{4\Omega^2} I_g \cos 2\theta_g + ZP_z$$

$$\to \quad J_g = \frac{\partial S}{\partial \theta_g} = I_g - P_z \frac{\Omega'}{2\Omega^2} I_g \sin 2\theta_g$$

$$Z = \frac{\partial S}{\partial P_z} = Z + \frac{\Omega'}{4\Omega^2} I_g \cos 2\theta_g \qquad (8.A.55)$$

$$P_z = \frac{\partial S}{\partial z} = P_z \left[ 1 + \frac{d}{dz}\left(\frac{\Omega'}{\Omega^2}\right) \frac{I_g}{4} \cos 2\theta_g \right]$$

$$\Theta_g = \frac{\partial S}{\partial I_g} = \theta_g + P_z \frac{\Omega'}{4\Omega^2} \cos 2\theta_g$$

With this new canonical transformation the Hamiltonian can be rewritten in terms of the new variables:

$$H = I_g \left( 1 - P_z \frac{\Omega'(Z)}{2\Omega^2(Z)} \sin 2\theta_g \right) \Omega(Z) + \frac{1}{2}\left[ P_z(1+...) + ... \right]^2$$

$$\Omega(Z) = \Omega(Z - \frac{\Omega'(Z)}{4\Omega^2} I_g \cos 2\theta_g) = \Omega(Z) + O(\varepsilon) \qquad (8.A.56)$$

We expand the terms in Eq.(8.A.56) for $H$ and find that the $O(\varepsilon)$ terms cancel leaving

$$H(\Theta_g, I_g, Z, P_z) = I_g \Omega(Z) + \frac{1}{2} P_z^2 + O(\varepsilon^2) \qquad (8.A.57)$$

In terms of these action-angle variables the following equations of motion are obtained:

$$\dot{I}_g = -\frac{\partial H}{\partial \Theta_g} = 0 + O(\varepsilon^2) \qquad \dot{P}_z = -\frac{\partial H}{\partial Z} = -\Omega'(Z) I_g + O(\varepsilon^2)$$

$$\dot{\Theta}_g = \frac{\partial H}{\partial I_g} = \Omega(Z) + O(\varepsilon^2) \qquad \dot{Z} = \frac{\partial H}{\partial P_z} = P_z + O(\varepsilon^2) \qquad (8.A.58)$$

The finite time derivative of $P_z$ represents the mirroring forces on particles as they transit or bounce back and forth along the field line.

Through second order in $\epsilon$ the Hamiltonian in terms of action-angle variables is

$$H(\Theta_g, I_g, Z, P_z) = I_g \Omega(Z) + \frac{1}{2} P_z^2 - \frac{\Omega'^2}{4\Omega^2} I_g^2 \cos 2\theta_g + \frac{1}{8} \frac{\Omega'^2}{\Omega^2} I_g^2 \sin^2 2\theta_g$$
$$+ \frac{1}{4} P_z^2 I_g \left[ \frac{d}{dz}\left(\frac{\Omega'}{\Omega^2}\right) \cos 2\theta_g - \frac{\Omega'^2}{\Omega^3} \sin^2 2\theta_g \right] \tag{8.A.59}$$

where

$$\theta_g = \Theta_g - P_z \frac{\Omega'}{4\Omega^2} \cos 2\theta_g \quad z = Z - I_g \frac{\Omega'}{4\Omega^2} \cos 2\theta_g \tag{8.A.60}$$

The correction terms in $\theta_g$ and $z$ only contribute $O(\epsilon^3)$ and $O(\epsilon^4)$ terms in $H$, so they are superfluous through second order in $\epsilon$ in affecting $H$. We note that

$$I_g \Omega(Z) = \frac{mc}{e} \mu \frac{eB(z)}{mc} = \mu B(z) = \frac{1}{2} m v_\perp^2 \quad \rightarrow \quad H^{(0)} = \frac{1}{2}\dot{z}^2 + \mu B(z) \tag{8.A.61}$$

The second term in $H^{(0)}$ gives the standard bouncing motion along the field lines in a magnetic mirror configuration.

<u>Example</u>: Quadratic magnetic well – We make an explicit assumption about the magnetic field:

$$B(z) = B_0 \left(1 + \frac{1}{2} \frac{z^2}{L^2}\right) \quad \rightarrow \quad \Omega(Z) = \Omega_0 \left(1 + \frac{1}{2} \frac{Z^2}{L^2}\right) \tag{8.A.62}$$

In this magnetic well the equations of motion yield

$$\dot{I}_g = -\frac{\partial H}{\partial \Theta_g} = O(\varepsilon^2) \qquad \dot{P}_z = -\frac{\partial H}{\partial Z} = O(\varepsilon) \tag{8.A.63}$$

and

$$H^{(0)} = I_g \Omega_0 + \frac{1}{2} I_g \Omega_0 \frac{Z^2}{L^2} + \frac{1}{2} P_z^2 \tag{8.A.64}$$

where the axial motion in the quadratic well satisfies

$$Z = \sqrt{\frac{2J_b}{\omega_b}} \sin\phi_b \quad P_z = \sqrt{2J_b \omega_b} \cos\phi_b \quad \omega_b \equiv \frac{\sqrt{I_g \Omega_0}}{L} \tag{8.A.65}$$

We introduce yet another canonical transformation

$$\left(I_g, \Theta_g; P_z, Z\right) \rightarrow \left(\mu, \phi_g; J_b, \phi_b\right) \tag{8.A.66}$$

with the generating function

$$S(\Theta_g, Z; \mu, J_b) = \int_0^Z dz' \sqrt{2J_b \omega_b(\mu) - \omega_b^2(\mu) z'^2} + \mu \Theta_g \tag{8.A.67}$$

The transformation from old to new variables is then

$$I_g = \frac{\partial S}{\partial \Theta_g} = \mu \quad P_z = \frac{\partial S}{\partial Z} = \sqrt{2J_b\omega_b}\cos\phi_b \quad \phi_b = \frac{\partial S}{\partial J_b}$$

$$\phi_g = \frac{\partial S}{\partial \mu} = \Theta_g + \frac{J_b}{2\mu}\sin 2\phi_b \quad Z = \sqrt{\frac{2J_b}{\omega_b}}\sin\phi_b \tag{8.A.68}$$

We then rewrite the Hamiltonian:

$$H(I_g,\Theta_g;P_z,Z) = \mu\Omega_0 + J_b\omega_b(\mu) - \frac{\Omega'^2}{4\Omega^2}\mu^2\cos 2\theta_g + \ldots \tag{8.A.69}$$

We substitute for $\Omega, \Omega',$ and $\theta_g$ in Eq.(8.A.69) and expand terms. To make some estimates, we simplify, and use a model Hamiltonian for certain quantities. We have

$$\omega_b = \frac{\sqrt{\mu\Omega_0}}{L} \quad \dot\mu = -\frac{\partial H}{\partial \phi_g} = O(\varepsilon^2) \quad \dot\phi_g = \frac{\partial H}{\partial \mu} = \Omega_0\left(1 + \frac{1}{2}\frac{J_b}{\sqrt{\mu L}}\right) + O(\varepsilon^2) \tag{8.A.70a}$$

$$\dot J_b = -\frac{\partial H}{\partial \phi_b} = O(\varepsilon^2) \quad \dot\phi_b = \frac{\partial H}{\partial J_b} = \omega_b(\mu) + O(\varepsilon^2) \tag{8.A.70b}$$

We note that $\dot\phi_g = \langle\Omega(z)\rangle_b$ and $\dot\Theta_g = \Omega(z)$ with $O(\epsilon^2)$ corrections. We can select units so that $m = 1, \Omega_0 = 1,$ and $E = 1$. Hence, the Larmor radius $\rho = \frac{v_\perp}{\Omega} \sim 1$, $\mu \sim 1, \omega_g \sim 1, \omega_b \sim \epsilon, J_b \sim \frac{1}{\epsilon},$ and $\epsilon \equiv \frac{1}{L}$. Generally, the second order terms in the Hamiltonian can be represented as

$$H^{(2)}(\mu,J_b;\phi_b,\phi_g) = \sum_{\ell_b,\ell_g} H_{\ell_b,\ell_g}(\mu,J_b)e^{i(\ell_b\phi_b+\ell_g\phi_g)} \tag{8.A.71}$$

The Hamilton has the following orderings through $O(\epsilon^2)$:

$$H = \mu + J_b\frac{\sqrt{\mu}}{L} - \frac{\Omega'^2}{4\Omega^2}\mu^2\cos 2\theta_g + \frac{1}{4}P_z^2\mu[\ldots] + \frac{1}{8}\left(\frac{\Omega'}{\Omega}\right)^2\sin^2 2\theta_g \tag{8.A.72}$$

and after cancellations $H_{\ell_b,\ell_g} \sim \frac{1}{32}\varepsilon^2$. Consider a model Hamiltonian with a typical $O(\epsilon^2)$ term:

$$H = \mu + J_b\omega_b + \frac{1}{32}\varepsilon^2\sin 2\theta_g = \mu + J_b\omega_b + \frac{1}{32}\varepsilon^2\sin 2(\phi_g - \frac{J_b}{2\mu}\sin 2\phi_b)$$

$$= \mu + J_b\omega_b + \frac{1}{32}\varepsilon^2\text{Im}\left(e^{i2\phi_g}e^{-i\frac{J_b}{\mu}\sin 2\phi_b}\right) = \mu + J_b\omega_b + \frac{1}{32}\varepsilon^2\text{Im}\left(e^{i2\phi_g}\sum_{\ell=-\infty}^{\ell=\infty}J_\ell\left(\frac{J_b}{\mu}\right)e^{-\ell 2\phi_b}\right)$$

$$\tag{8.A.73}$$

where $J_\ell$ is the Bessel function with argument $J_b/\mu$. Thus, at $O(\epsilon^2)$ the Hamiltonian in Eq.(8.A.73) has a phase factor

$$\sin 2\theta_g = \sum_{\ell=-\infty}^{\ell=\infty} J_\ell\left(\frac{J_b}{\mu}\right)\sin 2(\phi_g - \ell\phi_b) \tag{8.A.74}$$

We use the model Hamiltonian in Eq.(8.A.73) with Eq.(8.A.74) to construct the equations of motion:

$$\dot{\mu} = -\frac{\partial H}{\partial \phi_g} = -\frac{2\varepsilon^2}{32}\sum_\ell J_\ell \cos 2(\phi_g - \ell\phi_b)$$

$$\dot{\phi}_g = \frac{\partial H}{\partial \mu} = \omega_g(\mu, J_b) + O(\varepsilon^2) \tag{8.A.75}$$

$$\dot{\phi}_b = \frac{\partial H}{\partial J_b} = \omega_b(\mu) + O(\varepsilon^2)$$

To lowest order Eq.(8.A.75) implies:

$$\dot{\mu} = -\frac{\varepsilon^2}{16}\sum_\ell J_\ell \cos 2(\omega_g - \ell\omega_b)t \tag{8.A.76}$$

Given Eq.(8.A.76) we ask under what circumstances there are oscillatory or systematic variations in $\mu$.

<u>Theorem</u>: If for some integer $\ell_0$, $\omega_g - \ell_0\omega_b = 0$ then $\Delta\mu \sim -\frac{\varepsilon^2}{16}J_{\ell_0}t$, i.e., there is a systematic secular variation in the magnetic moment.[70] If $\ell_0 = \omega_g/\omega_b \sim 1/\epsilon$, e.g., $\ell_0 = 100$, then $J_{\ell_0}\left(\frac{J_b}{\mu}\right) \sim J_{1/\epsilon}(1/\epsilon) \sim \epsilon^{1/3}$ From Eq.(8.A.73) $H^{(2)} \to \frac{1}{32}\epsilon^{7/3}$

<u>Corollary</u>: For situations close to resonance but not at precise resonance,

$$\Delta\mu = -\frac{\varepsilon^2}{16}\sum_\ell J_\ell \frac{\sin[2(\omega_g - \ell\omega_b)t]}{2(\omega_g - \ell\omega_b)} \tag{8.A.77}$$

<u>Example</u>: For the model Hamiltonian with many possible resonances

$$H(\mu,\phi_g;J_b,\phi_b) = H^{(0)}(\mu,J_b) + H^{(2)}(\phi_g,\phi_b)$$

$$H^{(0)}(\mu,J_b) = \mu + J_b\omega_b(\mu), \quad \omega_b(\mu) = \varepsilon\sqrt{\mu}, \quad \varepsilon \equiv \frac{1}{L}$$

$$H^{(2)}(\phi_g,\phi_b) = \frac{\varepsilon^{7/3}}{32}\sum_\ell \sin[2(\phi_g - \ell\phi_b)] \approx \frac{\varepsilon^{7/3}}{32}\sum_\ell \sin[2(\omega_g t - \ell\omega_b t)] \tag{8.A.78}$$

$$\Omega_0 = 1 \quad \omega_g \equiv \dot{\phi} = \frac{\partial H}{\partial \mu} = 1 + \frac{\varepsilon J_b}{2\sqrt{\mu}}$$

---

[70] R. H. Cohen, G. Rowlands, and J. H. Foote, Phys. Fluids **21**, 627 (1978).

We select a fixed energy $E=1$, so that $H^{(0)} \sim 1$. Then the choice of $\mu$ approximately fixes the action $J_b$. Hence, for $E=1$

$$\varepsilon J_b \approx \frac{1-\mu}{\sqrt{\mu}} = O(1) \tag{8.A.79}$$

As a function of $\mu$, $\mu \in [0,1]$, $\epsilon J_b = 1$ for $\mu = 0$; and $\epsilon J_b$ drops monotonically to 0 as $\mu \to 1$, while $\omega_b$ increases as $\sqrt{\mu}$ but is $O(\epsilon)$ compared to $\omega_g$ which is $O(1)$ and decreases as $\mu \to 1$. Examine the resonance condition in $H^{(2)}$:

$$\omega_g = \ell \omega_b \implies 2\ell\varepsilon = \frac{1 + \frac{1}{\mu}}{\sqrt{\mu}} \tag{8.A.80}$$

For example, if $\epsilon = 1/100$ and we choose $<\mu> \approx 1/2$ then the resonance occurs for

$$\ell\varepsilon = \frac{3\sqrt{2}}{2} \approx 2.12 \to \ell \approx 212 \tag{8.A.81}$$

For a small change in the integer $\ell$ the resonance occurs at a slightly different value of $\mu$ obtained by calculating the differential of both sides of Eq.(8.A.80):

$$2\varepsilon\delta\ell = -\frac{7}{\sqrt{\mu}}\delta\mu \to \delta\mu = -\frac{2\sqrt{\mu}}{7}\varepsilon\delta\ell = \mp\frac{\sqrt{2}}{7}\varepsilon \to \mp 0.00202, \quad \delta\ell = \pm 1 \tag{8.A.82}$$

in this example. Thus, the resonances are fairly dense with respect to $\mu$. If the change in magnetic moment $\mu$ from a single near-resonant term in Eq.(8.A.77) is large enough to overcome the spacing to the next resonance in $\mu$ and $\ell$ based on Eq.(8.A.82), then one can expect a significant diffusive change in $\mu$. This is an example of Chirikov's resonance overlap criterion; if not satisfied, then the magnetic moment is conserved:

$$\frac{1}{2}\left|\mu_\ell - \mu_{\ell\pm 1}\right| \sim \frac{\sqrt{\mu_\ell}}{7}\varepsilon > \left|\Delta\mu(t)\right| = \left|\frac{\varepsilon^{7/3}}{16} \sum_\ell \frac{\sin[2(\omega_g t - \ell\omega_b t)]}{2(\omega_g - \ell\omega_b)}\right| \sim \frac{\varepsilon^{7/3}}{32}\frac{1}{\frac{1}{2}\omega_b} \sim \frac{\varepsilon^{7/3}}{32}\frac{1}{\frac{1}{2}\varepsilon\sqrt{\mu}} \tag{8.A.83}$$

Hence, for this example, the magnetic moment is conserved for

$$1 > \frac{7}{16}\frac{\varepsilon^{1/3}}{\mu} \to \frac{7}{8}\varepsilon^{1/3} \tag{8.A.84}$$

We note that the pre-factor of ½ out front of the left side of Eq.(8.A.83) is not precise. If the factor were 1/3 instead, then the factor 7/8 on the right side of (8.A.85) would become 21/16. If $\epsilon$ is small then the condition in Eq.(8.A.84) is readily satisfied.

Example: For the model Hamiltonian in Eq.(8.A.78) with a single resonance the change in magnetic moment and action can be calculated from

$$\dot{\mu} = -\frac{\varepsilon^{7/3}}{16}\cos 2\psi \quad \dot{J}_b = \frac{\ell\varepsilon^{7/3}}{16}\cos 2\psi \quad \psi \equiv \phi_g - \ell\phi_b \tag{8.A.85}$$

We note that for a single resonance there is an invariant $I = \ell\mu + J_b$ with zero time derivative based on Eq.(8.A.85). Hence, the $\phi_b$ and $\phi_g$ degrees of freedom are reduced to the single degree of freedom $\psi$ by using the invariant $I$. We introduce another canonical transformation for the variables and the Hamiltonian:

$$S(\phi_g, \phi_b; \mu', I) = (\mu_\ell + \mu')\phi_g - \ell\mu'\phi_b \tag{8.A.86}$$

from which

$$\omega_g(\mu_\ell) = \ell\omega_b(\mu_\ell) \quad \mu' \equiv \mu - \mu_\ell \quad \mu = \frac{\partial S}{\partial \phi_g} = \mu_\ell + \mu' \quad \psi = \frac{\partial S}{\partial \mu'} = \phi_g - \ell\phi_b \quad \mu_\ell = \text{const} \tag{8.A.87}$$

Now add $I\phi_b$ to $S$ in Eq.(8.A.86) so that

$$J_b = \frac{\partial S}{\partial \phi_b} = I - \ell\mu' \quad \text{and} \quad \chi \equiv \frac{\partial S}{\partial I} = \phi_b \tag{8.A.88}$$

Then

$$H^{(0)} = \mu_\ell + \mu' + (I - \ell\mu')\varepsilon\sqrt{\mu_\ell + \mu'} \approx \text{const} - \frac{7}{4}\mu'^2 \tag{8.A.89}$$

and

$$H^{(2)} = \frac{\varepsilon^{7/3}}{32}\sin 2\psi \tag{8.A.90}$$

for $\mu_\ell = 1/2$ and $\ell\epsilon = 3\sqrt{2}/2$ from Eqs.(8.A.78), (8.A.79), (8.A.80), and (8.A.81). With this Hamiltonian the variables $\psi$ and $\mu'$ are conjugate and satisfy

$$\dot{\mu}' = -\frac{\varepsilon^{7/3}}{16}\cos 2\psi \quad \dot{\psi} = -\frac{7}{2}\mu' \rightarrow \ddot{\psi} = \frac{7}{32}\varepsilon^{7/3}\cos 2\psi \tag{8.A.91}$$

Recall the physics in the definition of the phase variable $\psi$. We are following the motion in phase space over many gyrations. The equation for the second time derivative of $\psi$ is identical to the equation of motion for a quasi-particle in a well defined by

$$U(\psi) = -\frac{7}{64}\varepsilon^{7/3}\sin 2\psi \tag{8.A.92}$$

Quasi-particles can be trapped in the periodic wells of $U(\psi)$, or with more "energy" the quasi-particles are untrapped and pass over the wells. If passing, then the excursions in $\psi$ grow in time without bound. The trapped quasi-particles have limited, periodic excursions in $\psi$. Given the definition of $\psi$ in Eq.(8.A.87) then

$$\Delta\psi = \Delta\phi_g - \ell\Delta\phi_b \tag{8.A.93}$$

For changes in the bounce phase corresponding to an integer number of bounces and for $\ell$ an integer, then $\Delta\psi = \Delta\phi_g$ modulo $2\pi$. Hence, in the median plane of the mirror $\Delta\psi = \Delta\phi_g = \Delta\theta_g$.

We note that in the phase space $(\mu', \psi)$ the periodic excursion of the quasi-particle in $\mu'$ in Eq.(8.A.91) is $\Delta\mu = \epsilon^{7/6}/4$. The separation between resonances in $\mu_\ell$ and $\mu_{\ell\pm1}$ for $\mu_\ell \approx 1/2$ from Eq.(8.A.83) is $|\mu_\ell - \mu_{\ell\pm1}| = 2\sqrt{\mu_\ell}/7 \sim \epsilon/5$. Then $\mu$ will be conserved if the excursion in $\mu$ is less than the separation between resonances:

$$\Delta\mu < |\mu_\ell - \mu_{\ell\pm1}| \rightarrow \epsilon^{\frac{7}{6}}/4 < \epsilon/5 \rightarrow \epsilon^{\frac{1}{6}} < 4/5 \tag{8.A.94}$$

A more careful calculation yields 0.77 instead of 4/5, from which one concludes that if $\epsilon < 0.77^6 = 0.21 \sim 1/5$ the magnetic moment is conserved. If all possible rational resonances are included, the resonances are more densely packed; and $\epsilon$ must be smaller to conserve the magnetic moment. For example, if the spacing of resonances in Eq.(8.A.94) were a factor of two smaller, then $\epsilon^{\frac{1}{6}} < 0.4$ and $\epsilon < 0.004$ to conserve magnetic moment.

Consider a slightly different model to explain the adiabatic invariant $\mu$. We postulate a Hamiltonian evaluated in median plane of the mirror

$$H(\mu,\phi_g;J,\phi_b) = J\omega_b + \frac{1}{2}\mu^2 + \lambda \sum_{\ell=-\infty}^{\ell=\infty} \cos(\phi_g - \ell\phi_b)$$

$$= J\omega_b + \frac{1}{2}\mu^2 + \lambda 2\pi \cos\phi_g \sum_{n=-\infty}^{n=\infty} \delta(\phi_b - 2\pi n) \tag{8.A.95}$$

$$\dot{\phi}_b = \frac{\partial H}{\partial J} = \omega_b \quad \dot{\mu} = -\frac{\partial H}{\partial \phi_g} = \lambda 2\pi \sin\phi_g \sum_{n=-\infty}^{n=\infty} \delta(\phi_b - 2\pi n) \quad \dot{\phi}_g = \frac{\partial H}{\partial \mu} = \mu$$

where $\lambda = \epsilon^{7/4}/32$ for example. We integrate the equations of motion from bounce to bounce through the median plane, $\Delta\phi_g = \mu 2\pi/\omega_b$:

$$\begin{aligned}
\mu_{n+1} &= \mu_n + \frac{2\pi\lambda}{\omega_b}\sin\phi_g(n) \\
&\rightarrow \mu'_{n+1} = \mu'_n + \Lambda\sin\phi_g(n) \\
\phi_{g,n+1} &= \phi_{g,n} + \mu'_{n+1} \\
\mu' &= \mu\frac{2\pi}{\omega_b} \quad \Lambda \equiv \left(\frac{2\pi}{\omega_b}\right)^2 \lambda
\end{aligned} \tag{8.A.96}$$

The equations in Eq.(8.A.96) are mapping equations for any point in the $(\phi_g,\mu')$ plane into another point on the next crossing of the median plane of the mirror. This is an area-preserving, standard mapping known as the Chirikov-Taylor or Chirikov standard map. The map is periodic in both $\phi_g$ and $\mu'$ with periodicity equal to $2\pi$. For small values of $\Lambda$ the mapping populates the phase space with a periodic, regular pattern that is not chaotic. When $\Lambda$ approaches unity the phase space acquires regions in which the mapping is chaotic, which ergodic regions become relatively larger with increasing $\Lambda$. From the dependency of the ergodicity on the value of $\Lambda$, we can deduce the corresponding value of $\epsilon$.

Exercise: A fixed point for the mapping in Eq.(8.A.96) is a point that maps onto itself. (i) Find two fixed points for Eq.(8.A.96). (ii) Linearize the mapping about a fixed point and compute the ratio of increments to $\phi_g$ and $\mu'$ on successive steps to show stability for $\Lambda<4$.

### 8.A.d Perturbation of motion in a magnetic mirror due to an electrostatic wave – Rosenbluth's concept of superadiabaticity and resonance overlap leading to stochasticity

In a seminal paper, Marshall Rosenbluth analyzed adiabatic invariance of the magnetic moment for particle orbits in a magnetic mirror configuration in the presence of an electrostatic wave.[71] Rosenbluth introduced the concept of superadiabtaticity to describe the circumstance in which the amplitude of the perturbing electrostatic wave modifies the magnetic moment but does not lead to random changes and particle loss. For larger wave amplitudes chaotic motion can ensue, and the magnetic moment is no longer conserved. Here only an outline of Rosenbluth's calculation is given, and details are left as an exercise.

Consider the Hamiltonian of a charged particle in a magnetic mirror field as in Sec.8.A.c but ignore $O(\epsilon^2)$ effects. Include an oscillating electric field perturbation with frequency near the cyclotron frequency. Is there a new invariant? The outline of the calculation is as follows:

1) Add to the Hamiltonian $e\phi(z,t) = e\phi_0 \sin(k_\perp x - \omega_0 t)$, $\omega_0 = \Omega(\pm z_0)$
2) Transform to gyro-variables $\mu, \theta_g$
3) Expand $e\phi(z,t)$ in Bessel functions using the Bessel function identity
4) Extract the resonant term from the series, $\sin(\theta_g - \omega_0 t) \sim \sin(\Omega(z)t - \omega_0 t)$
5) Define a canonical transformation $S$ similar to that in Eqs.(8.A.86-8.A.90) to introduce $\psi = \theta_g - \omega_0 t$ and find the new Hamiltonian such that

$$K = H + \frac{\partial S}{\partial t}, \quad \frac{\partial H}{\partial t} \neq 0, \quad \frac{\partial K}{\partial t} = 0 \tag{8.A.97}$$

6) In analogy to Eqs.(8.A.95-8.A.96), find the mapping of $\mu, \psi$ from one bounce through the median plane of the mirror to the next and obtain the new $\Lambda$. Investigate the stability of the mapping. Stability corresponds to superadiabaticity, while instability is associated with ergodic orbits and diffusion of the magnetic moment.

[*Editor's note: Diffusion of the magnetic moment in the presence of cyclotron resonant electrostatic turbulence has received considerable attention.*[72,73,74] *The study by Smith and Cohen*[74] *employed a Hamiltonian approach much like that in Kaufman's lectures and research.*]

### 8.B Lagrangian theory of guiding-center drifts

---

[71] Marshall N. Rosenbluth, Phys. Rev. Lett. **29**, 408 (1972).
[72] D. E. Baldwin, H. L. Berk, and L. D. Pearlstein, Phys. Rev. Lett. **36**, 1051 (1976).
[73] H. L. Berk and J. J. Stewart, Phys. Fluids **20**, 1080 (1977).
[74] G. R. Smith and B. I. Cohen, Phys. Fluids **26**, 238 (1983).

In this section is presented a Lagrangian theory of guiding-center drifts. Bob Dewar's methodology making use of Whitham averaging is used.[75] Equations of motion and approximate constants of the motion are derived.

### 8.B.a Dewar's Lagrangian theory relying on Whitham averaging

Dewar systematically derived the guiding-center Lagrangian[76] from the particle Lagrangian. The particle Lagrangian is

$$L(\mathbf{x},\dot{\mathbf{x}};t) = \frac{1}{2}|\dot{\mathbf{x}}|^2 - \frac{e\phi(\mathbf{x},t)}{m} + \frac{e}{mc}\dot{\mathbf{x}}\times\mathbf{A}(\mathbf{x},t) \quad (8.B.1)$$

$\mathbf{x}(t)$ is the particle location; $\mathbf{R}(t)$ is the guiding-center location; and $\mathbf{r}=\mathbf{R}-\mathbf{x}$. The displacement $\mathbf{r}$ is related to the Larmor radius according to

$$\mathbf{r}(t) = \rho(t)\left[\hat{\mathbf{e}}_2\cos\theta(t) + \hat{\mathbf{e}}_1\sin\theta(t)\right] + O(\varepsilon^2) \quad (8.B.2)$$

The time derivative of Eq.(8.B.2) is

$$\dot{\mathbf{r}} = \rho\dot{\theta}\left[-\hat{\mathbf{e}}_2\sin\theta + \hat{\mathbf{e}}_1\cos\theta\right] + \varepsilon\frac{\partial\rho}{\partial(\varepsilon t)}\left[\hat{\mathbf{e}}_2\cos\theta + \hat{\mathbf{e}}_1\sin\theta\right] + O(\frac{d}{dt}\hat{\mathbf{e}}_1,\frac{d}{dt}\hat{\mathbf{e}}_2) \quad (8.B.3)$$

The unit vectors $\hat{\mathbf{e}}_1$ and $\hat{\mathbf{e}}_2$ change if the direction of the magnetic field changes, but this leads to unnecessary complications in the analysis and adds no new physics. The analysis makes use of

$$\phi(\mathbf{x}) = \phi(\mathbf{R}+\mathbf{r}) = e^{\mathbf{r}\cdot\nabla}\phi(\mathbf{R}) \quad \mathbf{r}\cdot\nabla = \rho\left[\cos\theta\hat{\mathbf{e}}_2\cdot\nabla + \sin\theta\hat{\mathbf{e}}_1\cdot\nabla\right] \quad (8.B.4)$$

If $\phi$ varies spatially as $\exp(i\mathbf{k}\cdot\mathbf{x})$ then

$$\mathbf{r}\cdot\nabla = \rho\cos(\theta-\psi)\sqrt{\partial_1^2+\partial_2^2} = ik_\perp\rho\cos(\theta-\psi), \quad \tan\psi = \frac{\partial_1}{\partial_2} \to \frac{k_1}{k_2} \quad (8.B.5)$$

and

$$e^{\mathbf{r}\cdot\nabla} = e^{ik_\perp\rho\cos(\theta-\psi)} = \sum_{\ell=-\infty}^{\ell=\infty} J_\ell(k_\perp\rho)e^{i\ell(\theta-\psi+\pi/2)} \quad (8.B.6)$$

Then the gyro-average of $\phi(x)$ is

$$\langle\phi(\mathbf{x})\rangle_\theta = \langle e^{\mathbf{r}\cdot\nabla}\rangle_\theta \phi(\mathbf{R}) \quad (8.B.7)$$

and only the $\ell = 0$ term in Eq.(8.B.6) survives when used in Eq.(8.B.7).

We can return to the Lagrangian in Eq.(8.B.1) and calculate the gyro-average:

$$\langle L\rangle = \frac{1}{2}\langle|\dot{\mathbf{x}}|^2\rangle - \frac{e\langle\phi(\mathbf{x},t)\rangle}{m} + \frac{e}{mc}\langle\dot{\mathbf{x}}\times\mathbf{A}(\mathbf{x},t)\rangle \quad (8.B.8)$$

Given $\dot{\mathbf{x}} = \dot{\mathbf{R}} + \dot{\mathbf{r}}$ then

$$\langle|\dot{\mathbf{x}}|^2\rangle = |\dot{\mathbf{R}}|^2 + \langle|\dot{\mathbf{r}}|^2\rangle + 2\langle\dot{\mathbf{r}}\cdot\dot{\mathbf{R}}\rangle = |\dot{\mathbf{R}}|^2 + \langle\rho^2\dot{\theta}^2\rangle + 2\langle\dot{\mathbf{r}}\rangle\cdot\dot{\mathbf{R}} = |\dot{\mathbf{R}}|^2 + \langle\rho^2\dot{\theta}^2\rangle + O(\varepsilon^2) \quad (8.B.9)$$

---

[75] G. B. Whitham, *Linear and Nonlinear Waves* (John Wiley & Sons, 1974).
[76] J. B. Taylor, Phys. Fluids **7**, 767 (1964).

The surviving Bessel function can be expanded as $J_0(k_\perp \rho) \approx 1 - \frac{1}{4}(k_\perp \rho)^2$. The last term on the right side of Eq.(8.B.8) is evaluated with the help of

$$\dot{\mathbf{x}} \cdot \mathbf{A}(\mathbf{x}) = (\dot{\mathbf{R}} + \dot{\mathbf{r}}) \cdot \mathbf{A}(\mathbf{R} + \mathbf{r}) = (\dot{\mathbf{R}} + \dot{\mathbf{r}}) \cdot \left[ \mathbf{A}(\mathbf{R}) + \mathbf{r} \cdot \nabla \mathbf{A}(\mathbf{R}) + \frac{1}{2}(\mathbf{r} \cdot \nabla)^2 \mathbf{A}(\mathbf{R}) + \ldots \right]$$

$$\langle \dot{\mathbf{x}} \cdot \mathbf{A}(\mathbf{x}) \rangle = \dot{\mathbf{R}} \cdot \mathbf{A}(\mathbf{R}) + \langle \dot{\mathbf{r}} \mathbf{r} \rangle : \nabla \mathbf{A}(\mathbf{R}) + \frac{1}{2} \langle (\mathbf{r} \cdot \nabla)^2 \rangle \mathbf{A}(\mathbf{R}) \cdot \dot{\mathbf{R}}$$

(8.B.10)

Consider $\mathbf{r}\mathbf{r} : \nabla \mathbf{A} = \mathbf{r}\mathbf{r} : (\nabla \mathbf{A})^s$ where $(\nabla \mathbf{A})^s$ is the symmetric part of $\nabla \mathbf{A}$, because the asymmetric part of $\nabla \mathbf{A}$ does not contribute when taking the double dot product. The time derivative of $\mathbf{r}\mathbf{r} : \nabla \mathbf{A}$ is

$$\frac{d}{dt}\left(\mathbf{r}\mathbf{r} : \nabla \mathbf{A}\right) = 2\dot{\mathbf{r}}\mathbf{r} : (\nabla \mathbf{A})^s + \mathbf{r}\mathbf{r} : \left(\nabla\left[\frac{\partial \mathbf{A}}{\partial t} + \dot{\mathbf{R}} \cdot \nabla \mathbf{A}\right]\right)^s, \quad \frac{\partial \mathbf{A}}{\partial t} = -c\mathbf{E}^T \quad (8.B.11)$$

and hence,

$$2\dot{\mathbf{r}}\mathbf{r} : (\nabla \mathbf{A})^s = \frac{d}{dt}\left(\mathbf{r}\mathbf{r} : \nabla \mathbf{A}\right) - \mathbf{r}\mathbf{r} : \left(\nabla\left[-c\mathbf{E}^T + \dot{\mathbf{R}} \cdot \nabla \mathbf{A}\right]\right)^s \quad (8.B.12)$$

Upon calculating the time average over the cyclotron period and using periodic boundary conditions on the gyro-phase, the first term on the right side of Eq.(8.B.12) will not contribute to the averaged Lagrangian. We also need the contribution from $\dot{\mathbf{r}}\mathbf{r} : (\nabla \mathbf{A})^a$ in Eq.(8.B.10):

$$\dot{\mathbf{r}}\mathbf{r} : (\nabla \mathbf{A})^a = \frac{1}{2}(\mathbf{r} \times \dot{\mathbf{r}}) \cdot \mathbf{B} \quad \rightarrow \quad \langle \dot{\mathbf{r}}\mathbf{r} \rangle : (\nabla \mathbf{A})^a = \frac{1}{2}\rho^2 \dot{\theta} B \quad (8.B.13)$$

The remaining terms in Eq.(8.B.10) and (8.B.11) are $\mathbf{r}\mathbf{r} : \nabla\left[\nabla(\mathbf{A} \cdot \dot{\mathbf{R}}) + c\mathbf{E}^T - \dot{\mathbf{R}} \cdot \nabla \mathbf{A}\right]$, which after a little more algebra leads to

$$\langle L \rangle(\rho; \dot{\theta}; R, \dot{R}; t) = \frac{1}{2}|\dot{\mathbf{R}}|^2 + \frac{1}{2}\rho^2 \dot{\theta}^2 - \frac{e\phi(\mathbf{R}, \varepsilon t)}{m} + \frac{e}{mc}\dot{\mathbf{R}} \cdot \mathbf{A}(\mathbf{R}, \varepsilon t) - \frac{e}{2mc}\rho^2 \dot{\theta} B$$

$$+ \frac{e}{4m}\rho^2 \ddot{\mathbf{I}}^\perp : \nabla\left(\mathbf{E}^T + \mathbf{E}^L + \frac{1}{c}\dot{\mathbf{R}} \times \mathbf{B}\right) \quad (8.B.14)$$

The "T" and "L" in Eq.(8.B.14) are transverse (divergence-free) and longitudinal (curl-free) vector field components. $\ddot{\mathbf{I}}^\perp \equiv \ddot{\mathbf{I}} - \hat{\mathbf{b}}\hat{\mathbf{b}}$; hence, $\ddot{\mathbf{I}}^\perp \cdot \nabla = \nabla_\perp$. The last term on the right side of Eq.(8.B.14) is $O(\epsilon^2)$. Large-amplitude fields are allowed, but they must be slowly varying.

Definition: We introduce two momentum-like variables:

$$p_\rho \equiv \frac{\partial <L>}{\partial \dot{\rho}} = 0 + O(\varepsilon^2) \quad (8.B.15a)$$

$$p_\theta \equiv \frac{\partial <L>}{\partial \dot{\theta}} = \rho^2 \dot{\theta} - \frac{e}{2mc}\rho^2 B = \rho^2\left(\dot{\theta} - \frac{1}{2}\Omega\right) \quad (8.B.15b)$$

From the definition in Eq.(8.B.15a) we have $p_\rho = 0$ and, hence, from the Lagrangian equation:

$$0 = \dot{p}_\rho = \frac{\partial \langle L \rangle}{\partial \rho} = \rho \dot{\theta}^2 - \frac{e}{mc}\rho\dot{\theta}B(\mathbf{R};t) \tag{8.B.16}$$

which we divide by $\rho\dot{\theta}$ to obtain

$$\dot{\theta} = \frac{eB(\mathbf{R};t)}{mc} \equiv \Omega + O(\varepsilon^2) \tag{8.B.17}$$

With the use of Eq.(8.B.17) in Eq.(8.B.15b) one obtains

$$p_\theta \equiv \frac{\partial <L>}{\partial \dot{\theta}} = \rho^2\left(\dot{\theta} - \frac{1}{2}\Omega\right) = \frac{1}{2}\Omega\rho^2 \propto \mu \sim \frac{1}{2}\frac{mv_\perp^2}{B} \quad \dot{p}_\theta = \frac{\partial}{\partial \theta}\langle L \rangle = 0 \tag{8.B.18}$$

which is a constant of the motion through $O(\epsilon)$.

### 8.B.b Orderings, forces, mirroring, drifts, definition of flux tubes

We now introduce a set of orderings as chosen by Dewar.

Postulate: Orderings

$$\begin{aligned} O(\varepsilon^{-1}): &\quad e, c \\ O(1): &\quad m, v, \phi, E_\perp, L \\ O(\varepsilon): &\quad \rho, E_\parallel \end{aligned} \tag{8.B.19}$$

With these orderings <L> is

$$\langle L \rangle(\rho;\dot{\theta};R,\dot{R};t) = \frac{1}{2}|\dot{\mathbf{R}}|^2 + \frac{1}{2}\rho^2\dot{\theta}^2 - \frac{e\phi(\mathbf{R},t)}{m} + \frac{e}{mc}\dot{\mathbf{R}}\cdot\mathbf{A}(\mathbf{R},t) - \frac{e}{2mc}\rho^2\dot{\theta}B + O(\varepsilon^2)$$
$$\quad O(1) \quad O(1) \quad O(\varepsilon^{-1}) \quad O(\varepsilon^{-1}) \quad O(1) \tag{8.B.20}$$

We note that <L> has no $O(\epsilon)$ term. We also note that the term $\rho^2\dot{\theta}^2/2$ involves the product of $1/B^2$ and $B^2$, which cancel and leaves only constants, i.e., this term in the averaged Lagrangian is independent of $R$ and $t$; and we can set it aside. From the Lagrangian equations, Eqs.(8.B.16-8.B.18), we deduced the constancy of the gyrofrequency $\Omega$ and the magnetic moment $\mu$ through $O(\epsilon)$.

Definition: Based on the orderings and Eq.(8.B.20), we define the guiding-center Lagrangian

$$\begin{aligned} L_{gc} &= L(\varepsilon^{-1}) + L(1) = -\frac{e\phi(\mathbf{R},t)}{m} + \frac{e}{mc}\dot{\mathbf{R}}\cdot\mathbf{A}(\mathbf{R},t) + \frac{1}{2}|\dot{\mathbf{R}}|^2 - \frac{e}{2mc}\rho^2\dot{\theta}B \\ &= -\frac{e\phi(\mathbf{R},t)}{m} + \frac{e}{mc}\dot{\mathbf{R}}\cdot\mathbf{A}(\mathbf{R},t) + \frac{1}{2}|\dot{\mathbf{R}}|^2 + \mu B \end{aligned} \tag{8.B.21}$$

We recognize the $\mu B$ term as a potential energy $-\boldsymbol{\mu}\cdot\mathbf{B}$, where $\boldsymbol{\mu}=-\mu\hat{\mathbf{b}}$ the magnetic dipole moment of a gyrating particle in a magnetic field.

From the Lagrangian equations applied to the guiding center Lagrangian in Eq.(8.B.21) one can derive the properties of the guiding center motion at $O(\epsilon^{-1})$ and $O(1)$.

<u>Definition</u>: $\mathbf{P}^{(\varepsilon^{-1})} = \dfrac{\partial L^{(\varepsilon^{-1})}}{\partial \dot{\mathbf{R}}} = \dfrac{e}{mc}\mathbf{A}(\mathbf{R},t) \quad \dot{\mathbf{P}}^{(\varepsilon^{-1})} = \dfrac{e}{mc}\left(\dfrac{\partial}{\partial t}\mathbf{A}+\dot{\mathbf{R}}\cdot\nabla\mathbf{A}\right)$ (8.B.22)

<u>Theorem</u>:  From the Lagrangian equation at $O(\epsilon^{-1})$

$$\dot{\mathbf{P}}^{(-1)} = \dfrac{\partial L^{(-1)}}{\partial \mathbf{R}} = \dfrac{e}{m}\mathbf{E}^L + \dfrac{e}{mc}\nabla\mathbf{A}\cdot\dot{\mathbf{R}} \tag{8.B.23}$$

Equating expression for $\dot{\mathbf{P}}^{(-1)}$ in Eqs.(8.B.22) and (8.B.23) leads to

$$0 = \dfrac{e}{m}\mathbf{E}^L + \dfrac{e}{mc}\nabla\mathbf{A}\cdot\dot{\mathbf{R}} - \dfrac{e}{mc}\left(\dfrac{\partial}{\partial t}\mathbf{A}+\dot{\mathbf{R}}\cdot\nabla\mathbf{A}\right) = \dfrac{e}{m}\mathbf{E} + \dfrac{e}{mc}\left(\nabla\mathbf{A}\cdot\dot{\mathbf{R}}-\dot{\mathbf{R}}\cdot\nabla\mathbf{A}\right) = \dfrac{e}{m}\mathbf{E} + \dfrac{e}{mc}\dot{\mathbf{R}}\times\left(\nabla\times\mathbf{A}\right)$$

$$\Rightarrow \quad 0 = \mathbf{E} + \dfrac{1}{c}\dot{\mathbf{R}}\times\mathbf{B} + O(\varepsilon) \quad \Rightarrow \quad \mathbf{E}_\perp + \dfrac{1}{c}\mathbf{V}_\perp\times\mathbf{B} = O(\varepsilon) \quad \mathbf{E}_\| = O(\varepsilon) \tag{8.B.24}$$

The results for the perpendicular and weaker parallel vector components of the electric field in Eq.(8.B.24) are consistent with the orderings in Eq.(8.B.19). Particles move more freely along the magnetic field lines and can neutralize strong parallel electric fields. Solving Eq.(8.B.24) for the perpendicular velocity

$$\mathbf{V}_\perp = c\dfrac{\mathbf{E}(\mathbf{R},t)\times\hat{\mathbf{b}}}{B} + O(\varepsilon) \tag{8.B.25}$$

The leading term in Eq.(8.B.25) is $O(1)$, which is the **ExB** drift velocity $\mathbf{u}_E$.

To next order,

$$\mathbf{P} = \dfrac{\partial L}{\partial \dot{\mathbf{R}}} = \dfrac{e}{mc}\mathbf{A}(\mathbf{R},t) + \dot{\mathbf{R}} \tag{8.B.26}$$

We take the time derivative of Eq.(8.B.26) and equate it to $\dot{\mathbf{P}} = \partial L_{gc}/\partial \mathbf{R}$ through $O(1)$ to obtain

$$\dfrac{1}{c}\left(\dfrac{\partial}{\partial t}\mathbf{A}+\dot{\mathbf{R}}\cdot\nabla\mathbf{A}\right) + \dfrac{m}{e}\ddot{\mathbf{R}} = \mathbf{E}^L + \dfrac{1}{c}(\nabla\mathbf{A})\cdot\dot{\mathbf{R}} - \dfrac{m}{e}\mu\nabla B$$

$$\Rightarrow \quad 0 = \dfrac{e}{m}\left(\mathbf{E}+\dfrac{1}{c}\dot{\mathbf{R}}\times\mathbf{B}\right) - \ddot{\mathbf{R}} - \mu\nabla B \tag{8.B.27}$$

The last two terms on the right side of Eq.(8.B.27) are perturbations. As a consequence of the orderings,

$$\dot{\mathbf{R}} = \dot{\mathbf{R}}^{(1)}+\dot{\mathbf{R}}^{(\varepsilon)}, \quad \dot{\mathbf{R}}^{(1)} = v_\|\hat{\mathbf{b}}+\mathbf{u}_E, \quad \dot{\mathbf{R}}^{(\varepsilon)} = \mathbf{v}_d \tag{8.B.28}$$

Hence, to the lowest required order $\dot{\mathbf{R}}^{(1)} = v_\|\hat{\mathbf{b}}+\mathbf{u}_E$ can be used in the $\ddot{\mathbf{R}}$ term in Eq.(8.B.27).

Theorem: The guiding-center drifts result from Eqs.(8.B.27) and (8.B.28), and (8.B.27) naturally decouples into components parallel and perpendicular to the applied magnetic field:

$$\hat{\mathbf{b}}: \quad m\dot{v}_{\parallel} = eE_{\parallel} - \mu\hat{\mathbf{b}}\cdot\nabla B - m\hat{\mathbf{b}}\cdot\dot{\mathbf{u}}_E \tag{8.B.29a}$$

$$\hat{\mathbf{b}}_\perp: \quad \mathbf{v}_\perp = \mathbf{u}_E + \mathbf{v}_d, \quad \mathbf{v}_d = \frac{1}{\Omega}\left(-\frac{\mu\nabla B}{m} - v_\parallel \dot{\hat{\mathbf{b}}} - \dot{\mathbf{u}}_E + \ldots\right)\times\hat{\mathbf{b}} \tag{8.B.29b}$$

The second term on the right side of Eq.(8.B.29a) is the mirroring force. The first term on the right side of $\mathbf{v}_d$ is the $\nabla B$ drift; the second term is the centrifugal drift; the third term is the polarization drift; and the remaining terms could accommodate other accelerations, e.g., gravity and collisional friction. We note $\mathbf{u}_E \cdot \hat{\mathbf{b}} = 0$. Hence

$$\dot{\mathbf{u}}_E \cdot \hat{\mathbf{b}} = -\dot{\hat{\mathbf{b}}}\cdot\mathbf{u}_E \quad \text{and} \quad m\dot{v}_\parallel = eE_\parallel - \mu\hat{\mathbf{b}}\cdot\nabla B + \dot{\hat{\mathbf{b}}}\cdot\mathbf{u}_E \tag{8.B.30}$$

We ascribe the notion of a moving field line to the time derivative (in a Lagrangian sense) of the magnetic unit vector as seen by the particle in its guiding center motion.

### 8.B.c Equations for field lines, Euler potentials and Clebsch representation

To lowest order, the particles are tied to the magnetic field lines; and their trajectories lie on surfaces that enclose the magnetic flux. We define a field line velocity $\mathbf{u}_L(\mathbf{r},t)$ such that flux moving with this velocity will be conserved. We will elaborate this notion in the following discussion. First there are some preliminaries. Consider an applied magnetic field in a region of plasma such that $\nabla\times\mathbf{B} = 0$. Introduce an artificial gravity g such that

$$\mathbf{v}_g = \mathbf{v}_{\nabla_\perp B} + \mathbf{v}_{curv} \quad \Rightarrow \quad \mathbf{g} = \frac{v_\parallel^2 + v_\perp^2}{R}\hat{\kappa} \tag{8.B.31}$$

where $\hat{\boldsymbol{\kappa}}$ is the magnetic curvature unit vector and $R$ is the radius of curvature of the magnetic field line.

Example: Consider a plasma in pressure balance with an applied magnetic field. When the magnetic field is a minimum (maximum) in the plasma, the plasma is stable (unstable) with respect to perturbations, as shown in the important paper by Rosenbluth and Longmire on interchange instability.[77]

Definition: Field lines are be mapped by $\dfrac{dx}{B_x} = \dfrac{dy}{B_y} = \dfrac{dz}{B_z}$, $\mathbf{B}(\mathbf{x},t) = (B_x, B_y, B_z)$ (8.B.32)

Definition: The magnetic flux $\Phi(C)$ is defined by the integral over an area bounded by the curve $C$:

---

[77] M. N. Rosenbluth and C. Longmire, Ann. Phys. **1**, 120 (1957).

$$\Phi(C) \equiv \oint_C \mathbf{B} \cdot d\vec{\sigma} \tag{8.B.33}$$

Introduce the differential line element $d\vec{\ell}$ on the bounding curve $C$. The time derivative of the surface integral in Eq.(8.B.33) has contributions from the time derivative of the magnetic field and from the time derivative of the contour $C$ ($d\vec{\ell} \times \mathbf{u}_L$ gives the time derivative of the differential parallelogram which can intercept magnetic fields lines and can contribute to the time derivative of the magnetic flux). We calculate the time derivative of Eq.(8.B.33) and constrain it to be zero:

$$\begin{aligned}\frac{d}{dt}\Phi(C) &= \oint_C \left[ (\mathbf{u}_L \times d\vec{\ell}) \cdot \mathbf{B} + \frac{\partial \mathbf{B}}{\partial t} \cdot d\vec{\sigma} \right] = \oint_C \left[ -d\vec{\ell} \cdot (\mathbf{u}_L \times \mathbf{B}) + \frac{\partial \mathbf{B}}{\partial t} \cdot d\vec{\sigma} \right] \\ &= \oint_C d\vec{\sigma} \cdot \left[ -\nabla \times (\mathbf{u}_L \times \mathbf{B}) + \frac{\partial \mathbf{B}}{\partial t} \right] \equiv 0 \end{aligned} \tag{8.B.34}$$

using the Stokes' theorem. The constraint that the time derivative of the magnetic flux vanishes imposes a condition that determines $\mathbf{u}_L$.

Theorem: (Field-line velocity) Setting the terms inside the area integral equal to zero in the final expression on the right side of Eq.(8.B.34) and use of Faraday's law yield

$$\begin{aligned}-\nabla \times (\mathbf{u}_L \times \mathbf{B}) + \frac{\partial \mathbf{B}}{\partial t} &= 0 = c\nabla \times \mathbf{E} + \frac{\partial \mathbf{B}}{\partial t} \\ \to \nabla \times \left( \mathbf{E} + \frac{1}{c}\mathbf{u}_L \times \mathbf{B} \right) &= 0 \quad \to \quad \mathbf{E} + \frac{1}{c}\mathbf{u}_L \times \mathbf{B} = -\nabla \chi(\mathbf{x},t) \end{aligned} \tag{8.B.35}$$

Equation (8.B.35) is only a condition on $\mathbf{u}_L \perp \mathbf{B}$. The component of $\mathbf{u}_L \| \mathbf{B}$ does not matter at all. So we only consider $\mathbf{u}_L \perp \mathbf{B}$ as a postulate. If we take the component of the last expression in Eq.(8.B.35) parallel to the magnetic field

$$E_{\|} = -\hat{\mathbf{b}} \cdot \nabla \chi(\mathbf{x},t) \quad \to \quad \chi(\ell) = -\int_0^\ell d\ell' E_{\|}(\ell') \tag{8.B.36}$$

integrated along the field line. This determines the order of $\chi$, viz., $\chi$ is the same order as $E_{\|} \sim O(\epsilon)$. The component of $\mathbf{u}_L$ perpendicular to the magnetic field is then

$$\begin{aligned}\mathbf{u}_L &= c\frac{\mathbf{E} \times \hat{\mathbf{b}}}{B} + c\frac{\nabla_\perp \chi \times \hat{\mathbf{b}}}{B} \\ \mathbf{u}_E &\sim O(1) \quad O(\varepsilon) \end{aligned} \tag{8.B.37}$$

We recall that $\mathbf{v}_\perp = \mathbf{u}_E + \mathbf{v}_d$. Hence, $\mathbf{v}_\perp - \mathbf{u}_L = O(\varepsilon)$; and the field lines are then tied to the fluid velocity at leading order. In this sense the plasma is tied to the field lines.

The field-line velocity is a construct. There is significant ambiguity. We know how to relate the values of $\chi$ along the same field line from Eq.(8.B.37), but not across the field lines. We can illustrate the ambiguity of the field-line velocity in the following examples.

Example: Suppose B is uniform and static, and $\mathbf{E} = -\nabla\phi$. Further assume that $\phi$ has no spatial derivative along the magnetic field lines so that only $\mathbf{E}_\perp \neq 0$. Two choices for $\chi$ can be made. (a) $\chi = 0$ everywhere, in which case $\mathbf{u}_L = \mathbf{u}_E$. (b) $\chi = \phi$, so that both are constant along the field lines. Then from Eq.(8.B.37) $\mathbf{u}_L = 0$. Thus, the possible ambiguity in constructing the field line velocity is obvious.

It is useful to introduce auxiliary functions to characterize the magnetic field lines. Consider labeling functions $\alpha$ and $\beta$ such that $d\Phi_{flux} = d\alpha d\beta$. $(\alpha, \beta)$ labels a field line emerging normally from a curvilinear $(\alpha, \beta)$ surface. A third coordinate $\sigma$ along the field line is needed. These labels (Euler potentials) can be used to calculate the magnetic field.

Definition: The differential length along a field line is $d\ell = h(\alpha, \beta, \sigma) d\sigma$. For example, if $\sigma = \theta$ then $h = r$, where $r$ is the local radius of curvature of the field line.

A magnetic field can be represented as
$$\mathbf{B} = (\nabla\alpha \times \nabla\beta) \Lambda(\alpha, \beta, \sigma) \tag{8.B.38}$$
where $\Lambda(\alpha, \beta, \sigma)$ is a scalar function. Given Eq.(8.B.38) then
$$\mathbf{B} \cdot \nabla\alpha = 0 \quad \mathbf{B} \cdot \nabla\beta = 0 \quad \frac{d\alpha}{B_\alpha} = \frac{d\beta}{B_\beta} \tag{8.B.39}$$
so that $\alpha$ and $\beta$ can be used as field line labels. The differential magnetic flux is
$$d\Phi \equiv \mathbf{B} \cdot d\mathbf{a} = dxdy \left( \hat{\mathbf{z}} \cdot \nabla\alpha \times \nabla\beta \right) \Lambda = dxdy\, J\left(\frac{\alpha,\beta}{x,y}\right) \Lambda = d\alpha d\beta \Lambda \tag{8.B.40}$$
where J is the Jacobian of $\alpha$ and $\beta$ with respect to $x$ and $y$. In this example we can take $\Lambda = 1$; and then Eq.(8.B.38) is the Clebsch representation for the magnetic field. The Clebsch representation guarantees that the magnetic field is divergence free, but does *not* guarantee that the magnetic field has no curl.

Examples: Magnetic field configurations (Figure 8.B.1)
1. Straight wire carrying current $I$
$$\alpha(r;I) = 2I\ln r \quad \beta = z \quad \sigma = \theta \quad \rightarrow \quad \mathbf{B} = \frac{2I}{r}\hat{\theta} \quad \Phi = 2Iz\ln\frac{r_2}{r_1} \tag{8.B.41}$$

2. Axisymmetric poloidal field – magnetic mirror
  Select the $z$ dependence of $B_z$, e.g., for a simple mirror $B_z = B_0(1 + \frac{z^2}{L^2})$ in the paraxial limit. Then
$$\alpha = \frac{1}{2}r^2 B_0 (1 + \frac{z^2}{L^2}) \quad \beta = \theta \quad \sigma = z \quad B_r = -rB_0 z/L^2 \quad B_\theta = 0 \tag{8.B.42}$$

Note that $\nabla \times \mathbf{B} = -\frac{rB_0}{L^2}\hat{\theta}$, which is non-zero, but is higher order in the paraxial limit.

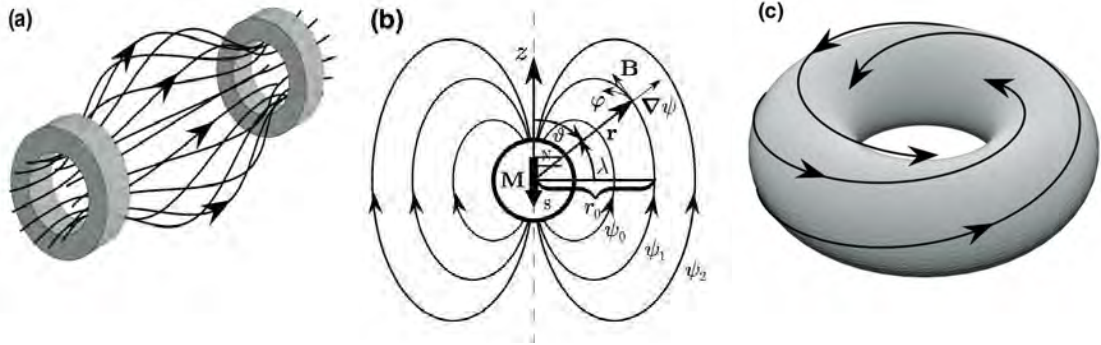

Figure 8.B.1 Magnetic field configurations: (a) mirror, (b) dipole, and (c) tokamak.

3. Dipole field
   $\alpha(r, z)$ enumerates the field lines. $\sigma = \theta$ is like a latitude along the field line. $\beta = \phi$ is the longitude. The Clebsch representation for the dipole magnetic field with magnetic moment $\mu$ in spherical polar coordinates is

$$\alpha = \frac{\sin\theta}{r^2}\mu \quad \beta = \phi \quad B_r = \frac{\cos\theta}{r^3}\mu \quad B_\theta = \frac{2\sin\theta}{r^3}\mu \quad B_\phi = 0 \tag{8.B.43}$$

4. Toroidal magnetic field
   In cylindrical coordinates a model toroidal field is described by

$$\alpha = \frac{\mu r \sin\theta}{(a^2 + 2ar\sin\theta + z^2)^{3/2}} \quad \beta = z \quad \sigma = \theta$$

$$\mathbf{B}(z \approx 0) \approx \mu\cos\theta\frac{(2a + r\sin\theta)}{(a + 2r\sin\theta)^{5/2}}\hat{\mathbf{r}} - \mu\sin\theta\frac{(2a + r\sin\theta)}{(a + 2r\sin\theta)^{5/2}} + B_z\hat{\mathbf{z}} \tag{8.B.44}$$

$$B_z(r \approx 0) \approx \frac{2\mu}{(a^2 + z^2)^{3/2}}$$

5. Tokamak magnetic field
   In cylindrical coordinates $(R, \phi, z)$ a model axisymmetric tokamak field[78] can be derived from

$$\mathbf{B} = B_\phi(R, z)\hat{\mathbf{e}}_\phi + \mathbf{B}^p(R, z) \quad \mathbf{B}^p = \nabla \times A_\phi \hat{\mathbf{e}}_\phi \quad A_\phi = -\frac{1}{R}\psi$$

$$\to B_R = \frac{1}{R}\frac{\partial \psi}{\partial z} \quad B_z = -\frac{1}{R}\frac{\partial \psi}{\partial R} \tag{8.B.45}$$

The toroidal magnetic field $B_\phi$ is determined by external current-carrying coils. The poloidal field $B^p$ is determined by toroidal plasma currents and an externally imposed vertical magnetic field.

<u>Exercise</u>: Show that $A_\phi \hat{\mathbf{e}}_\phi = \alpha\nabla\beta$, determine $\alpha$ and $\beta$, and check against Ref. 78.

---

[78] A. N. Kaufman, Phys. Fluids **15**, 1063 (1972).

We return to the analysis of field line motion leading to Eqs.(8.B.35-8.B.37) to demonstrate that with the Clebsch representation the field line labels advect with the field line velocity $u_L$ such that

$$\left.\frac{d\alpha}{dt}\right|_L = \frac{\partial \alpha(\mathbf{x},t)}{\partial t} + \mathbf{u}_L \cdot \nabla \alpha(\mathbf{x},t) = 0 \tag{8.B.46}$$

and similarly for $\beta$. Using the identity $\mathbf{B} = \nabla\alpha \times \nabla\beta = \nabla \times (\alpha\nabla\beta)$, then

$$E = -\nabla\phi - \frac{1}{c}\frac{\partial}{\partial t}(\alpha\nabla\beta) = -\nabla\phi - \frac{1}{c}\left(\frac{\partial\alpha}{\partial t}\nabla\beta + \alpha\nabla\frac{\partial\beta}{\partial t}\right) \tag{8.B.47}$$

We then use Eq.(8.B.46) and the equivalent for $\beta$ to evaluate

$$\frac{1}{c}\mathbf{u}_L \times \mathbf{B} = \frac{1}{c}\mathbf{u}_L \times (\nabla\alpha \times \nabla\beta) = \frac{1}{c}\left[\nabla\alpha\, \mathbf{u}_L \cdot \nabla\beta - \nabla\beta\, \mathbf{u}_L \cdot \nabla\alpha\right]$$

$$= -\frac{1}{c}\left[\nabla\alpha\frac{\partial\beta}{\partial t} - \nabla\beta\frac{\partial\alpha}{\partial t}\right] \tag{8.B.48}$$

Recalling Eq.(8.B.35)

$$\mathbf{E} = -\frac{1}{c}\mathbf{u}_L \times \mathbf{B} - \nabla\chi(\mathbf{x},t) = \frac{1}{c}\left[\nabla\alpha\frac{\partial\beta}{\partial t} - \nabla\beta\frac{\partial\alpha}{\partial t}\right] - \nabla\chi(\mathbf{x},t) = -\nabla\phi - \frac{1}{c}\left(\frac{\partial\alpha}{\partial t}\nabla\beta + \alpha\nabla\frac{\partial\beta}{\partial t}\right)$$

$$\Rightarrow -\nabla\chi(\mathbf{x},t) = -\nabla\phi - \frac{1}{c}\nabla\left(\alpha\frac{\partial\beta}{\partial t}\right) \Rightarrow \chi(\mathbf{x},t) = \phi + \frac{1}{c}\alpha\frac{\partial\beta}{\partial t}$$

(8.B.49)

up to an arbitrary constant. Equation (8.B.49) is quite general, and no ordering is involved. However, $\chi \sim O(\epsilon)$; so $\phi$ and $\left(\frac{\alpha}{c}\right)\partial\beta/\partial t$ nearly cancel.

Example: For an electrostatic model $\nabla \times \mathbf{E} = 0$ and $\mathbf{E} = -\nabla\phi$, there are two possible choices for the Euler potentials:

(1) $\chi = 0 \quad \frac{\partial\beta}{\partial t} = -c\frac{\phi}{\alpha}$

(2) $\chi = \phi \quad \frac{\partial\beta}{\partial t} = 0$

However, for a proper ordering only (1) is consistent; and (2) is inconsistent unless $\phi$ is sufficiently weak.

### 8.B.d Guiding-center Lagrangian using Euler potentials and Clebsch representation recovering guiding-center drifts

We return to the general consideration of field-line motion as in Eq.(8.B.46) and the guiding-center Lagrangian:

$$\frac{d^{g.c.}\beta}{dt} = \frac{\partial \beta(\mathbf{x},t)}{\partial t} + \mathbf{V}\cdot\nabla\beta(\mathbf{x},t)$$

$$L_{g.c.} = -e\phi + \frac{e}{c}\mathbf{v}\cdot\mathbf{A} + \frac{1}{2}mV^2 - \mu B(\mathbf{R},t)$$

$$O(\varepsilon^{-1})? \quad O(\varepsilon^{-1}) \quad O(\varepsilon^{-1}) \quad O(1) \quad O(1)$$

$$\rightarrow L_{g.c.} = -e\left(\chi - \frac{\alpha}{c}\frac{\partial \beta}{\partial t}\right) + \frac{e\alpha}{c}\left(\dot{\beta} - \frac{\partial \beta}{\partial t}\right) + \frac{1}{2}mV^2 - \mu B(\mathbf{R},t) \quad (8.B.50)$$

$$= -e\chi + \frac{e\alpha}{c}\dot{\beta} + \frac{1}{2}mV^2 - \mu B(\mathbf{R},t)$$

$$O(1) \quad O(1) \quad O(1) \quad O(1)$$

We recall from Sec. 8.B.C that

$$E_{\parallel} = -\hat{\mathbf{b}}\cdot\nabla\chi \quad \mathbf{V} = \mathbf{V}_{g.c.} = \mathbf{V}_\perp + V_\parallel\hat{\mathbf{b}} = \mathbf{u}_L + O(\varepsilon) + \hat{\mathbf{b}}h\dot{\sigma} \quad V^2 = u_L^2 + h^2\dot{\sigma}^2 + O(\varepsilon) \quad (8.B.51)$$

From Eqs.(8.B.50) and (8.B.51) we have

$$L_{g.c.} = -e\chi(\alpha,\beta,\sigma,t) + \frac{e\alpha}{c}\dot{\beta} + \frac{1}{2}m u_L^2(\alpha,\beta,\sigma,t) - \mu B(\alpha,\beta,\sigma,t) \quad (8.B.52)$$

where every term is O(1) and there is no mixing of velocity and coordinates. The first, third, and fourth terms on the right side of Eq.(8.B.52) are potential energy terms (with a minus sign); and the other terms are kinetic energy terms.

There is a problem with the Lagrangian in Eq.(8.B.52), which is revealed when we calculate the $\alpha$ momentum from the Lagrangian equation:

$$p_\alpha = \frac{\partial L}{\partial \dot{\alpha}} = 0 \quad (8.B.53)$$

This is erroneous. Consider the Hamiltonian derived from the same Lagrangian:

$$H = p_\alpha\dot{\alpha} + p_\beta\dot{\beta} + p_\sigma\dot{\sigma} - L_{g.c.} = \frac{e\alpha}{c}\dot{\beta} + mhV_\parallel\dot{\sigma} - \frac{e\alpha}{c}\dot{\beta} + e\chi$$

$$= mV_\parallel^2 - \frac{1}{2}mh^2\dot{\sigma}^2 - \frac{1}{2}mu_L^2 + \mu B + e\chi = \frac{p_\sigma^2}{2mh^2} + \mu B + e\chi - \frac{1}{2}mu_L^2 \quad (8.B.54)$$

The last term on the right side of Eq.(8.B.54) is a constraint associated with the moving field line. This $H$ is *not* equal to the energy associated with the guiding center motion.

If instead of the guiding center Lagrangian in Eq.(8.B.52), we use

$$L_{g.c.} = -e\chi - \mu B + \frac{e\alpha}{c}\dot{\beta} + \frac{1}{2}mV^2 \quad (8.B.55)$$

where

$$\mathbf{V} = \frac{d\mathbf{R}}{dt}(\alpha,\beta,\sigma,t) = \left.\frac{\partial \mathbf{R}}{\partial t}\right|_{\alpha,\beta,\sigma} + \dot{\alpha}\frac{\partial \mathbf{R}}{\partial \alpha} + \dot{\beta}\frac{\partial \mathbf{R}}{\partial \beta} + \dot{\sigma}\frac{\partial \mathbf{R}}{\partial \sigma} \quad (8.B.56)$$

The first term on the right side of Eq.(8.B.57) is $\mathbf{u}_L$ the field line velocity and is O(1); the second and third terms are O($\varepsilon$) drifts off the field line; and the last term is the velocity along the field line which is also O(1). Now proceed to obtain the equations of motion from the Lagrangian equations:

$$p_\beta \equiv \frac{\partial L}{\partial \dot{\beta}} = \frac{e}{c}\alpha + m\mathbf{V}\cdot\frac{\partial \mathbf{R}}{\partial \beta}$$

$$p_\alpha = m\mathbf{V}\cdot\frac{\partial \mathbf{R}}{\partial \alpha} \tag{8.B.57}$$

$$p_\sigma = mh^2\dot{\sigma}$$

We see that $p_\alpha$ is non-zero, and this degree of freedom cannot be removed. Furthermore,

$$\dot{p}_\beta = \frac{d}{dt}\left(\frac{e}{c}\alpha + m\mathbf{V}\cdot\frac{\partial \mathbf{R}}{\partial \beta}\right) = \frac{\partial L}{\partial \beta} = -e\frac{\partial \chi}{\partial \beta} - \mu\frac{\partial B}{\partial \beta} + \frac{1}{2}m\frac{\partial V^2}{\partial \beta}$$

$$\dot{\alpha} = \frac{c}{e}\frac{\partial}{\partial \beta}\left(\frac{1}{2}mV^2 - e\chi - \mu B\right) - \frac{mc}{e}\frac{d}{dt}\left(\mathbf{V}\cdot\frac{\partial \mathbf{R}}{\partial \beta}\right) \tag{8.B.58}$$

$$\dot{\beta} = -\frac{c}{e}\frac{\partial}{\partial \beta}\left(\frac{1}{2}mV^2 - e\chi - \mu B\right) + \frac{mc}{e}\frac{d}{dt}\left(\mathbf{V}\cdot\frac{\partial \mathbf{R}}{\partial \alpha}\right)$$

The second and third equations are $O(\epsilon)$ equations and include the polarization drifts. One can recognize terms that derive from the Hamiltonian equations in canonical variables using $H^{(0)} \equiv \frac{1}{2}mV^2 - e\chi - \mu B$.

Exercise: Interpret all the relevant drifts that appear in Eqs.(8.B.57) and (8.B.58).

At O(1) the equation of motion along the field line is

$$m\frac{d}{dt}(h^2\dot{\sigma}) = \frac{\partial}{\partial \sigma}(\tfrac{1}{2}mV^2 - e\chi - \mu B) \tag{8.B.59}$$

In Cartesian form it is straightforward to relate Eq.(8.B.59) to earlier formulae. Once again we note that the motion across the magnetic field lines is $O(\epsilon)$ and $O(1)$ along the field line. So we freeze the motion across the field lines and only study the motion along the field line for the moment. Through $O(1)$

$$L_\sigma(\sigma,\dot{\sigma};t;\alpha,\beta) = -e\chi - \mu B + \frac{1}{2}mu_L^2 + \frac{1}{2}mh^2\dot{\sigma}^2, \quad p_\sigma \equiv \frac{\partial L}{\partial \dot{\sigma}} = mh^2\dot{\sigma} \tag{8.B.60}$$

Theorem: The Hamiltonian through O(1) is then

$$H_\sigma = \dot{\sigma}p_\sigma - L_\sigma = \frac{p_\sigma^2}{2mh^2} + e\chi(\sigma,t) + \mu B(\sigma,t) - \frac{1}{2}mu_L^2(\sigma,t) \tag{8.B.61}$$

We require that $H_\sigma$ is essentially time independent through O(1), i.e., we require

$$\frac{\partial}{\partial t}\left(B,\chi,u_L,h\right) \sim O(\varepsilon), \text{ i.e., } \frac{2\pi}{\omega_B} << \frac{B}{\left(\frac{\partial B}{\partial t}\right)} \tag{8.B.62}$$

Thus, we postulate that $\frac{d}{dt}H = \frac{\partial}{\partial t}H = O(\epsilon)$; so $H$ is a constant of the motion through $O(1)$. This energy is the sum of the parallel kinetic energy, the electrical potential energy, and the perpendicular kinetic energy ($\mu B$), less the constraint energy $-\frac{1}{2}mu_L^2(\sigma,t)$ associated with the moving field line.

### 8.B.e Canonical transformations, orderings, and derivation of approximate constants of the motion

To further simplify the analysis, we assume that $u_L \sim O(\epsilon)$, which loses the polarization drift and is a great restriction. Under this assumption the $O(1)$ guiding center Lagrangian is

$$L_{g.c.} = -e\chi - \mu B + \frac{e}{c}\alpha\dot{\beta} + \frac{1}{2}mh^2\dot{\sigma}^2 \qquad (8.B.63)$$

which leads to

$$p_\beta = \frac{\partial L}{\partial \dot{\beta}} = \frac{e}{c}\alpha \quad p_\alpha = \frac{\partial L}{\partial \dot{\alpha}} = 0 \quad H_{g.c.}(\alpha,\beta;\sigma,p_\sigma;t) = \frac{p_\sigma^2}{2mh^2} + e\chi + \mu B \qquad (8.B.64)$$

There are two degrees of freedom for the motion perpendicular to the field lines, $\alpha$ and $\beta$ (which are conjugate), and two degrees of freedom for the motion parallel to the field line, $\sigma$ and $p_\sigma$, plus a degree of freedom for the bounce motion, $t$. We have thrown away the motion associated with the polarization drift. From the Lagrangian equations of motion

$$O(\varepsilon): \quad \dot{\alpha} = \frac{c}{e}\dot{p}_\beta = -\frac{c}{e}\frac{\partial H}{\partial \beta} \quad \dot{\beta} = \frac{\partial H}{\partial p_\beta} = \frac{c}{e}\frac{\partial H}{\partial \alpha}$$

$$O(1): \quad \dot{\sigma} = \frac{\partial H}{\partial p_\sigma} = \frac{p_\sigma}{mh^2} \quad \dot{p}_\sigma = -\frac{\partial H}{\partial \sigma} \qquad (8.B.65)$$

We note that if we freeze $\alpha, \beta, t$ then $\partial H/\partial t = O(\epsilon)$. With the two degrees of freedom $\sigma$ and $p_\sigma$, and other quantities slowly varying, there is an action that is an adiabatic invariant, which we have previously demonstrated.

<u>Definition:</u> $\quad J_\sigma(H,\alpha,\beta) \equiv \frac{1}{2\pi}\oint p_\sigma d\sigma = \frac{1}{2\pi}\oint \sqrt{2mh^2(H-e\chi-\mu B)}d\sigma \qquad (8.B.66)$

With $\alpha$ and $\beta$ slowly varying, i.e., $2\pi/\omega_B \ll \left|B/\left(\partial B/\partial t\right)\right|$, the bounce action $J_\sigma$ is conserved through $O(1)$: $\dot{J}_\sigma = O(\varepsilon) \quad \langle \dot{J}_\sigma \rangle = O(\varepsilon^2)$. The hierarchy of time scales is as follows: $t/\epsilon^{-1}$ is the gyro time scale; $t$ is the bounce time scale; and $\epsilon^{-1}t$ is the drift time scale, which is the slowest. Over the drift time scale the bounce action is an invariant to $O(\epsilon)$: $\langle \dot{J}_\sigma \rangle \tau_{drift} \equiv \langle \dot{J}_\sigma \rangle \frac{t}{\varepsilon} = \Delta\langle J_\sigma \rangle = O(\varepsilon)$. Now we can introduce action-angle variables for $H(J_\sigma, \theta_\sigma; \alpha, \beta)$ by constructing a generating function $S$ to perform the canonical transformation as we have done in earlier lectures (see Sec. 8.A.c).

# LECTURES ON THEORETICAL PLASMA PHYSICS – PART 4B

*Allan N. Kaufman*

**8.C Guiding center theory and hydromagnetic equations**

This section of the lecture notes presents the systematic derivation of a fluid theory of a plasma beginning with the guiding-center equations of motion and distribution functions. The formalism will include considerations of magnetohydrodynamic (MHD), ballooning, and resistive MHD stability.

*8.C.a Derivation of distribution functions and lowest-order velocity moments from guiding-center theory*

We construct particle distribution functions in a six-dimensional phase space representing guiding-center motion: $(\mu, \theta_g; \sigma, p_\sigma; \alpha, \beta)$. The actual particle positions corresponding to $(x, y, z, v_x, v_y, v_z)$ in phase space are not in the same place! However, we can calculate the number of particles in a small volume of 6D phase space centered around a specific location in either representation:

$$f(x,y,z;v_x,v_y,v_z)d^6V = f(x,y,z;v_x,v_y,v_z)dxdydzdv_xdv_ydv_z$$
$$= F(\mu,\theta_g,\sigma,p_\sigma;\alpha,\beta)d^6\mathcal{V} = F(\mu,\theta_g,\sigma,p_\sigma;\alpha,\beta)d\alpha d\beta dp_\sigma d\mu d\theta_g \quad (8.C.1)$$

where

$$\frac{d^6\mathcal{V}}{d^6V} = \mathcal{J}\left(\frac{\alpha\beta\sigma p_\sigma \mu \theta_g}{xyzv_xv_yv_z}\right) = const \quad (8.C.2)$$

for canonical transformations, and the constant in Eq.(8.C.2) has factors like $c/e$. Hence, $f=F$ apart from some numerical constants. We can calculate moments of $f$ and relate them to moments of $F$. For example, the number density is given by

$$n(x,y,z) = \int dv_x dv_y dv_z f \quad (8.C.3)$$

and similarly for the current density $j(x,y,z)$ with the inclusion of $e\mathbf{v}$ inside the integrand on the right side of Eq.(8.C.3). It is significant that $f$ includes the rapid variation due to the gyro motion, but $F$ should have a much slower variation because we expect that the gyro phase will be randomly distributed. In order that the moments and fields have a slow temporal variation, we expect $F$ to have a slow variation.

We can compute the number density using $F$ as follows:

$$n(\mathbf{x}) = \int d^6\mathcal{V}\delta\bigl(\mathbf{x} - \mathbf{x}(\alpha\beta\sigma p_\sigma \mu \theta_g)\bigr)F \quad (8.C.4)$$

using the $\delta$ function to obtain a density in the configuration space of the particles and $\mathbf{x}(\alpha,\beta,\sigma,p_\sigma,\mu,\theta_g) = \mathbf{R}(\alpha,\beta,\sigma) + \mathbf{r}(\mu,\theta_g)$. We expand inside the integrand in Eq.(8.C.4) in a Taylor series for small $\mathbf{r}$:

$$\delta(\mathbf{x}-\mathbf{R}-\mathbf{r}) = e^{-\mathbf{r}\cdot\frac{\partial}{\partial \mathbf{x}}}\delta(\mathbf{x}-\mathbf{R}) = \delta(\mathbf{x}-\mathbf{R}) - \mathbf{r}\cdot\frac{\partial}{\partial \mathbf{x}}\delta(\mathbf{x}-\mathbf{R}) + \frac{1}{2}\left(\mathbf{r}\cdot\frac{\partial}{\partial \mathbf{x}}\right)^2 \delta(\mathbf{x}-\mathbf{R}) + \ldots \quad (8.C.5)$$

to obtain

$$n(\mathbf{x}) = \int d^6 \mathcal{V}\, F\delta(\mathbf{x}-\mathbf{R}) - \frac{\partial}{\partial \mathbf{x}}\cdot\int d^6 \mathcal{V}\, \mathbf{r} F\delta(\mathbf{x}-\mathbf{R}) + \frac{1}{2}\nabla\nabla : \int d^6 \mathcal{V}\, \mathbf{r}\mathbf{r} F\delta(\mathbf{x}-\mathbf{R}) + \ldots \quad (8.C.6)$$

or, alternatively,

$$n(\mathbf{x}) = \int d^6 \mathcal{V}\, e^{-\mathbf{r}\cdot\frac{\partial}{\partial \mathbf{x}}} F\delta(\mathbf{x}-\mathbf{R}) \to N(\mathbf{x}) + \ldots = N(\mathbf{x}) + \frac{1}{4}\nabla_\perp^2 \left(N\langle\rho^2\rangle\right) \quad (8.C.7)$$

where $N(\mathbf{x})$ is the density of guiding centers at x, $\nabla_\perp^2 \equiv (\mathbf{I}-\hat{\mathbf{b}}\hat{\mathbf{b}}):\nabla\nabla$, and $\int d\mu\, \mathbf{r}\cdot\mathbf{r} F \to \langle\rho^2\rangle$. We note the alternative expression for $n(\mathbf{x})$ correct to leading order terms obtained by expanding the Bessel function after averaging over $\theta_g$:

$$n(\mathbf{x}) = \int d^6 \mathcal{V}\, e^{-\mathbf{r}\cdot\frac{\partial}{\partial \mathbf{x}}} F\delta(\mathbf{x}-\mathbf{R}) \to \int d^6 \mathcal{V}\, J_0(k_\perp \rho) F\delta(\mathbf{x}-\mathbf{R}) = \int d^6 \mathcal{V}\, J_0(-i\nabla_\perp \rho) F\delta(\mathbf{x}-\mathbf{R})$$
$$(8.C.8)$$

Now consider the current density. The flux density is
$$\vec{\Gamma}(\mathbf{x}) = \int d^6 \mathcal{V} F\delta(\mathbf{x}-\mathbf{x}(\ldots))\mathbf{v} = \int d^6 \mathcal{V} F\left(\delta(\mathbf{x}-\mathbf{R}) - \mathbf{r}\cdot\nabla\delta(\mathbf{x}-\mathbf{R})\right)(\mathbf{V}+\dot{\mathbf{r}})$$
$$= N\langle\mathbf{V}\rangle(\mathbf{x}) - \nabla\cdot N\langle\mathbf{r}\dot{\mathbf{r}}\rangle(\mathbf{x}) = N\langle\mathbf{V}\rangle(\mathbf{x}) - \tfrac{1}{2}\nabla\cdot\left[N<\rho^2>\Omega(\hat{\mathbf{e}}_2\hat{\mathbf{e}}_1 - \hat{\mathbf{e}}_1\hat{\mathbf{e}}_2)\right] + \ldots \quad (8.C.9)$$
$$= N\langle\mathbf{V}\rangle(\mathbf{x}) - \tfrac{1}{2}\nabla\times\left(N\Omega<\rho^2>\hat{\mathbf{b}}\right)$$

using $\dot{\mathbf{r}} = \rho\Omega(-\hat{\mathbf{e}}_2 \sin\theta_g + \hat{\mathbf{e}}_1 \cos\theta_g)$ and the inner product with the anti-symmetric dyadic $\hat{\mathbf{e}}_2\hat{\mathbf{e}}_1 - \hat{\mathbf{e}}_1\hat{\mathbf{e}}_2$ leads to the cross product. The first term in the final expression for the flux density in Eq.(8.C.9) is the product of the guiding-center density and the average guiding-center velocity. The second term leads to the diamagnetic current or magnetization current.

<u>Definition</u>: The magnetization is $\mathbf{M}(\mathbf{x}) \equiv N(\mathbf{x})\langle -\mu\hat{\mathbf{b}}\rangle(\mathbf{x})$ where $\mu = \dfrac{e}{2c}\Omega\rho^2$ \quad (8.C.10)

<u>Theorem</u>: The current density is given by
$$\mathbf{j}(\mathbf{x}) = e\vec{\Gamma}(\mathbf{x}) = eN<\mathbf{V}> + c\nabla\times\mathbf{M}(\mathbf{x}) \quad (8.C.11)$$
The first term in Eq.(8.C.11) is the guiding-center current density and the second term is identified as the diamagnetic current density.

<u>Examples</u>:
1. Consider a plasma with finite gradients $\nabla n$ and $\nabla T$ in the $x$ direction with uniform magnetic field in $z$. In consequence of the particle gyration and the gradient in $n$ there is a net current in the $y$ direction.

$$\vec{\Gamma}_{diamagnetic} = -\tfrac{1}{2}\nabla\times\left(N\Omega<\rho^2>\hat{\mathbf{b}}\right) = \tfrac{1}{2}\Omega\hat{\mathbf{b}}\times\nabla\left(N<\rho^2>\right) \neq 0 \qquad (8.C.12)$$

However, there is nothing implied with respect to the guiding-center flux $\vec{\Gamma}_{gc}$ which can be zero.

2. Assume $\nabla n = 0 = \nabla T$, but $\nabla\Omega \neq 0$. The gradient in the magnetic field does not produce a total net flux. However, there is a finite particle drift due to $\nabla B$, which yields a finite guiding-center flux and a finite diamagnetic flux Eqs.(8.C.9) and (8.C.12).

Exercise: Show the details.

3. A stress tensor can be calculated from

$$\ddot{\mathbf{p}}(\mathbf{x}) = nm\langle(\mathbf{v}-<\mathbf{v}>)(\mathbf{v}-<\mathbf{v}>)\rangle(\mathbf{x}) = p_\parallel \hat{\mathbf{b}}\hat{\mathbf{b}} + p_\perp(\ddot{\mathbf{I}}-\hat{\mathbf{b}}\hat{\mathbf{b}}), \quad p_\parallel = ..., \quad p_\perp = ... \qquad (8.C.13)$$

Exercise: Show that $\nabla\cdot\ddot{\mathbf{p}} = \dfrac{1}{c}\mathbf{j}\times\mathbf{B} + en\mathbf{E} - nm\dfrac{d}{dt}<\mathbf{u}>$, i.e., momentum balance, where $\mathbf{j} = e\Gamma = e(\Gamma_{gc} + \Gamma_{diamagnetic})$.

### 8.C.b Hamiltonian theory and Liouville's theorem

In Eq.(8.C.8) an expression was derived for $n(\mathbf{x}) = N(\mathbf{x}) + O(\epsilon^2)$ where $N(\mathbf{x})$ is the guiding center density at $\mathbf{x}$. Introducing a coordinate transformation we can express the number of guiding centers in a volume $dxdydz$ as

$$N(x,y,z)dxdydz = \mathscr{N}(\alpha,\beta,\sigma)d\alpha d\beta d\sigma \qquad (8.C.14)$$

The Jacobian relating $d\alpha d\beta d\sigma$ to $dxdydz$ is

$$\mathscr{J} = \frac{d\alpha d\beta d\sigma}{dxdydz} = \mathscr{J}\left(\frac{\alpha\beta\sigma}{xyz}\right) = \nabla\alpha\times\nabla\beta\cdot\nabla\sigma = B\hat{\mathbf{b}}\cdot\nabla\sigma = B\frac{d\sigma}{d\ell} = \frac{B}{h} \qquad (8.C.15)$$

where $h$ is the field-line metric. The guiding-center density is

$$\mathscr{N}(\alpha,\beta,\sigma) \equiv \int F(\mu,\sigma,p_\sigma,\alpha,\beta)d\mu dp_\sigma \qquad (8.C.16)$$

and

$$N(x,y,z) = \frac{B(\alpha,\beta,\sigma)}{h(\alpha,\beta,\sigma)}\mathscr{N}(\alpha,\beta,\sigma) \qquad (8.C.17)$$

Recall from Eqs(8.C.1) and (8.C.2) that although $F$ has no fast gyro time scale variation, while $f$ does; $F$ differs from $f$ only by some constants, which is a consequence of the canonical transformations. Under certain conditions, $F$ is just a function of canonical variables. Assuming that $\mathbf{u}_L$ is $O(\epsilon)$ as in Eq.(8.B.64) the lowest-order guiding center Hamiltonian is

$$H_{g.c.}(\alpha,\beta;\sigma,p_\sigma;t) = \frac{p_\sigma^2}{2mh^2} + e\chi + \mu B$$

through $O(1)$. With $\alpha$ and $\beta$ frozen, there is only one degree of freedom corresponding to the bounce motion; and $F(\sigma,p_\sigma,t)$ satisfies a Vlasov equation for one-dimensional motion.

Theorem: Liouville equation for a Hamiltonian system with one-dimensional motion:

$$\frac{d}{dt}F(\sigma,p_\sigma,t) \equiv \left[\frac{\partial}{\partial t} + \dot{\sigma}\frac{\partial}{\partial \sigma} + \dot{p}_\sigma\frac{\partial}{\partial p_\sigma}\right]F(\sigma,p_\sigma,t) = 0 \tag{8.C.18}$$

Equation(8.C.18) and Maxwell's equations with appropriate boundary and initial conditions provide a closed system of equations. With this system of equations one can look for time-independent solutions for equilibrium states. One can linearize about these solutions to assess waves and instabilities. One can also analyze nonlinear, time-dependent solutions, e.g., BGK modes, mode coupling equations, and quasi-linear equations.

Consider the limit in which there is no explicit time dependence, i.e., $\partial/\partial t = 0$. Then Eq.(8.C.18) yields Jeans' theorem for $F(H(\sigma,p_\sigma))$

$$\frac{dF}{dt} = \frac{\partial H}{\partial p_\sigma}\frac{dF}{dH}\frac{\partial H}{\partial \sigma} - \frac{\partial H}{\partial \sigma}\frac{dF}{dH}\frac{\partial H}{\partial p_\sigma} = 0 \tag{8.C.19}$$

In general, the equilibrium guiding-center distribution function is given by

$$F(\mu,\sigma;p_\sigma) = F(\mu,H(\mu,\sigma,p_\sigma)) \tag{8.C.20}$$

where $\alpha$ and $\beta$ are implicit. To lowest order the particle density $n(\mathbf{x})$ is the same as the guiding center density:

$$n(\mathbf{x}) = \frac{B}{h}\int_0^\infty d\mu \int_{-\infty}^\infty dp_\sigma F(\mu,H) = \frac{B}{h}\int_0^\infty d\mu \int_{e\chi+\mu B}^\infty dH\, p_\sigma \left(-\frac{\partial F(H,\mu)}{\partial H}\bigg|_\mu\right) \tag{8.C.21}$$

where we have integrated by parts: $F dp_\sigma = d(Fp_\sigma) - p_\sigma dF$, and used $\int_{-\infty}^\infty d(Fp_\sigma) = Fp_\sigma\big|_{-\infty}^\infty = 0$ and $dF\big|_\mu = \left(\frac{\partial F}{\partial H}dH\right)\big|_\mu$. With $p_\sigma = (2mh^2(H-e\chi-\mu B))^{1/2}$ Eq.(8.C.21) becomes

$$n(\mathbf{x}) = B(\sigma)\int_0^\infty d\mu \int_{e\chi+\mu B}^\infty dH (2m(H-e\chi-\mu B))^{1/2}\left(-\frac{\partial F(H,\mu)}{\partial H}\bigg|_\mu\right) \tag{8.C.22}$$

for a given species.

In the Coulomb plasma (electrostatics) Poisson's equation for a singly charged ion species is

$$\nabla^2 \phi = -4\pi e(n_i - n_e) \tag{8.C.23}$$

In the limit that $L \equiv \left|\phi/\sqrt{|\nabla^2\phi|}\right| \gg \lambda_D$, then $n_i \approx n_e$ to $O(\lambda_D/L)^2$, i.e., the plasma is quasi-neutral; and one can solve for $\chi(B)$ from $n_i(B,\chi) \approx n_e(B,\chi)$. From the solution for $\chi(B)$ one can calculate the parallel electric field:

$$E_\parallel = -\frac{d\chi(B)}{d\sigma} = -\frac{d\chi}{dB}\frac{\partial B}{\partial \sigma} \tag{8.C.24}$$

Example: Magnetic mirror configuration (Fig. 8.B.1a). Assume that in the midplane of the mirror ($\sigma = 0$)

$$f(\sigma=0;v_\perp,v_\parallel) \sim e^{-\frac{mv_\parallel^2}{2T_\parallel}} e^{-\frac{mv_\perp^2}{2T_\perp}} = e^{-\frac{H-\mu B_0 - e\chi}{T_\parallel}} e^{-\frac{\mu B_0}{T_\perp}} \tag{8.C.25}$$

$B_0$ is the magnetic field in the midplane. $\chi(\sigma=0)$ can be set to zero by choice. Recall that $F$ is a function of $H$ and $\mu$, but must be evaluated at a particular value of $\sigma$.

Theorem: $$F(H,\mu) \sim e^{-\frac{H-\mu B_0 - e\chi}{T_\parallel}} e^{-\frac{\mu B_0}{T_\perp}} \tag{8.C.26}$$

Definition: Define $\beta_{\parallel s} \equiv \dfrac{1}{T_{\parallel s}}$ and an anisotropy parameter $a_s \equiv \left.\dfrac{T_\parallel - T_\perp}{T_\perp}\right|_s$.

From Eqs.(8.C.22) and (8.C.26) we calculate

$$n(\sigma) = const\, e^{-\beta_\parallel e\chi(\sigma)} \frac{b(\sigma)}{a+b(\sigma)}, \quad b \equiv \frac{B(\sigma)}{B_0} \tag{8.C.27}$$

for each species. Then quasi-neutrality imposes

$$C_i\, e^{-\beta_{\parallel i} e\chi(\sigma)} \frac{b}{a_i + b} = C_e\, e^{\beta_{\parallel e} e\chi(\sigma)} \frac{b}{a_e + b} \tag{8.C.28}$$

Taking $\chi = 0$ at the midplane where $b=1$, then $c_i/(a_i+1) = c_e/(a_e+1)$: we choose $c_i = a_i + 1$ and $c_e = a_e + 1$. Elsewhere than the midplane $n_e/n_i = 1$, which leads to

$$\frac{a_e + b}{a_i + b} \frac{a_i + 1}{a_e + 1} = e^{(\beta_{\parallel e} + \beta_{\parallel i})e\chi} \rightarrow (\beta_{\parallel e} + \beta_{\parallel i})e\phi = \ln\left[\frac{a_e+b}{a_i+b}\frac{a_i+1}{a_e+1}\right] \tag{8.C.29}$$

which determines $\phi(b)$. We note that if $(T_\parallel/T_\perp)^i = (T_\parallel/T_\perp)^e \rightarrow a_e = a_i$ then Eq.(8.C.29) dictates $\phi(b) = 0 \rightarrow E_\parallel(b) = 0$. Thus, not only must there be anisotropy, but the anisotropies for the species must be different if there is a finite $E_\parallel$.

Theorem: (i) For $E_\parallel \neq 0$ then anisotropy is necessary and $\partial F(H,\mu)/\partial\mu \neq 0$ (local anisotropy). (2) For $E_\parallel \neq 0$ then different anisotropies for different species.

We further elaborate the solution for $F$ and $n(\sigma)$ in a magnetic mirror configuration. Consider

$$F(\mu,H) \sim e^{-\beta_\parallel (H + a\mu B_0)} \qquad a \equiv \frac{T_\parallel - T_\perp}{T_\perp}$$

$$n(\sigma) = n_B(\sigma) n_\phi(\sigma) \sim \frac{1 + b'(\sigma)}{1 + b'(\sigma)\frac{T_\parallel}{T_\perp}} e^{-\frac{e\chi(\sigma)}{T_\parallel}} \qquad b' = \frac{B(\sigma) - B_0}{B_0} \qquad (8.C.30)$$

$$n_B(\sigma) = \frac{1 + b'(\sigma)}{1 + b'(\sigma)\frac{T_\parallel}{T_\perp}} \qquad n_\phi(\sigma) = e^{-\frac{e\chi(\sigma)}{T_\parallel}}$$

A plot of $n_b(\sigma)$ is shown in Fig. 8.C.1 which illustrates that the density variation along the field line depends significantly on the relative anisotropy. We note that the duration of a charged particle in a segment $d\ell$ is $d\ell/v_\parallel$, which implies that the duration increases near the turning point where $v_\parallel \to 0$ and the density should increase. However, the smaller $T_\parallel$ is relative to $T_\perp$, the closer the turning points will be to the midplane of the mirror, which implies that the density will decrease away from the midplane. When the temperatures are isotropic, $n_b(\sigma)$ is uniform.

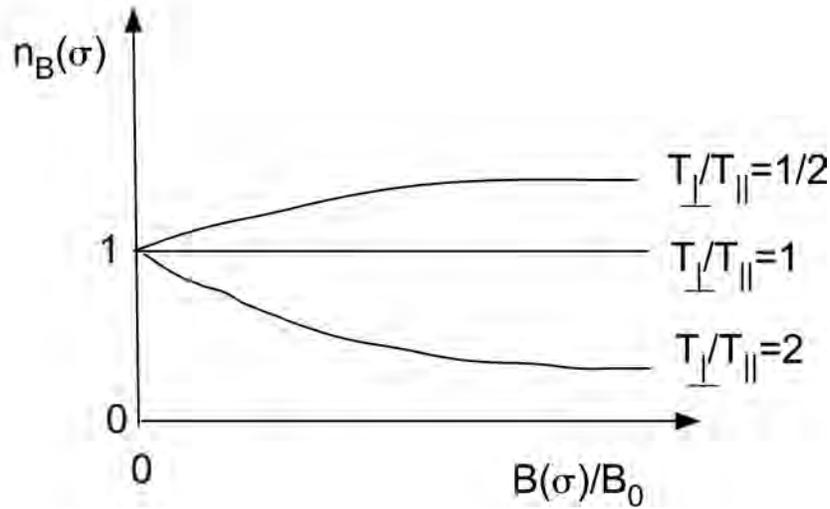

Figure 8.C.1 $n_b(\sigma)$ vs. $B(\sigma)/B_0$ from Eq.(8.C.30)

Exercise: Assume $f(z=0,\mathbf{v}) \sim \left(\frac{v_\perp^2}{v^2}\right)^\ell g(v^2)$ and construct $f(H,\mu) \to n(\sigma)$. Show that $n(\sigma) \sim B^{-\ell}$. Construct $p_\parallel(\sigma) = nm\langle v_\parallel^2 \rangle$ and evaluate $\partial p_\parallel(\sigma)/\partial \sigma$.

*8.C.c Derivation of lowest-order hydromagnetic equations (MHD)*

In this section we derive a reduced set of Maxwell and fluid equations, i.e., the lowest-order hydromagnetic equations, magnetohydrodynamics (MHD). We begin by postulating an ordering system. As a first pass, assume

$\epsilon^{-2}: j, p, \omega_p$
$\epsilon^{-1}: e, c, B, A, \Omega = eB/mc$
$\epsilon^0: \phi, E_\perp, m, v, u_L, L, T, \beta, m_e/m_i$ (temporarily), $\rho_c$ (8.C.31)
$\epsilon: E_\parallel, \chi, v_d, |\boldsymbol{u}^i - \boldsymbol{u}^e|, \Delta n$

At this point, $\omega_p$ and $\Omega$ are not necessarily the same order. However, if $\omega_p^2 \sim \Omega^2 \sim O(\epsilon^{-2})$, consistent with $ne^2/m \sim \omega_p^2 \sim \epsilon^{-2}$, then this implies $n \sim O(\epsilon^0)$. Now examine the relative ordering of terms in Ampere's law:

$$\nabla \times \mathbf{B} = \frac{4\pi}{c}\mathbf{J} + \frac{1}{c}\frac{\partial \mathbf{E}}{\partial t} \implies \mathbf{J} \sim O(\varepsilon^{-2}) \quad (8.C.32)$$
$O(\varepsilon^0)O(\varepsilon^{-1})\quad O(\varepsilon)O(?)\quad O(\varepsilon)O(1)O(1)$

The time dependence of all motion is assumed to be slow (adiabatic), and we will throw away the displacement current again as higher order. Thus, $\mathbf{J} \sim O(\epsilon^{-2})$, which is consistent with $n \sim O(\epsilon^{-2})$:

$$j \sim nev_d + c\nabla(n\mu) \sim O(\varepsilon^{-2})O(\varepsilon^{-1})O(\varepsilon) + O(\varepsilon^{-1})O(1)O(\varepsilon^{-2})O(\varepsilon) \sim O(\varepsilon^{-2}) \quad (8.C.33)$$

Consider the ordering of $\beta \equiv \frac{p}{B^2} \sim \frac{nmv^2}{B^2} \sim \frac{\epsilon^{-2}\epsilon^0}{\epsilon^{-2}} = O(1)$. Returning to the reduced Ampere's law Eq.(8.C.32):

$$\nabla \times \mathbf{B} = \frac{4\pi}{c}\mathbf{J} \implies \nabla \cdot \mathbf{J} = 0 \quad (8.C.34)$$

Given $\mathbf{J} \sim O(\epsilon^{-2})$, we can evaluate $u_\parallel^i - u_\parallel^e = \frac{j_\parallel}{ne} \sim \frac{\varepsilon^{-2}}{\varepsilon^{-2}\varepsilon^{-1}} \sim O(\varepsilon)$. Furthermore, the perpendicular fluid velocities are equal to the field-line velocity to leading order, i.e., $\mathbf{u}_\perp = \mathbf{u}_E + O(\varepsilon)$. Hence, $|\boldsymbol{u}^i - \boldsymbol{u}^e| \sim O(\epsilon)$.

<u>Theorem</u>: In MHD ordering, $ne \sim O(\epsilon^{-3})$ is appreciable, while the velocity difference between ions and electrons is small, $O(\epsilon)$, to give large currents, $O(\epsilon^{-2})$. Electrons, ions, and field lines have the same velocity perpendicular to the field lines to leading order; and the electrons and ions have the same parallel velocity to leading order. Hence, a single-fluid model is justified.

The fluid momentum balance equation is

$$n_s m_s \left(\frac{\partial}{\partial t} + \mathbf{u}^s \cdot \nabla\right)\mathbf{u}^s = -\nabla \cdot \mathbf{p}^s + n_s e_s \mathbf{E} + \frac{n_s e_s}{c}\mathbf{u}^s \times \mathbf{B} + \ldots \quad (8.C.35)$$

We sum Eq.(8.C.35) over species using $|\boldsymbol{u}^i - \boldsymbol{u}^e| \sim O(\epsilon)$, and the net charge density $\rho_c$ and the electric field $\mathbf{E}$ are both $O(1)$ to obtain the follow equation retaining only terms at $O(\epsilon^{-2})$:

$$\rho_m\left(\frac{\partial}{\partial t}+\mathbf{u}\cdot\nabla\right)\mathbf{u}=-\nabla\cdot\mathbf{p}+\frac{1}{c}\mathbf{j}\times\mathbf{B} \tag{8.C.36}$$

where $\rho_m$ is the mass density and $\mathbf{p}$ is the total pressure summed over species.

We next examine the ordering of the Poisson equation:

$$-\nabla^2\phi=4\pi\rho_c \sim \frac{1}{L^2}O(1)\sim O(1) \;\;\Rightarrow\;\; \rho_c\sim e(n_i-n_e)\sim O(1) \;\;\Rightarrow\;\; \Delta n\sim O(\frac{1}{e})\sim O(\varepsilon) \tag{8.C.37}$$

Hence, the plasma is quasi-neutral.

<u>Theorem</u>: Summary of lowest-order hydromagnetic equations

$$\rho_m\left(\frac{\partial}{\partial t}+\mathbf{u}\cdot\nabla\right)\mathbf{u}=-\nabla\cdot\mathbf{p}+\frac{1}{c}\mathbf{j}\times\mathbf{B}=-\nabla\cdot\mathbf{p}+\frac{1}{4\pi}(\nabla\times\mathbf{B})\times\mathbf{B}$$

$$\frac{\partial\mathbf{B}}{\partial t}=\nabla\times(\mathbf{u}_L\times\mathbf{B})\sim\nabla\times(\mathbf{u}\times\mathbf{B}) \tag{8.C.38}$$

$$\left(\frac{\partial}{\partial t}+\mathbf{u}\cdot\nabla\right)n=-n\nabla\cdot\mathbf{u} \;\;\rightarrow\;\; \left(\frac{\partial}{\partial t}+\mathbf{u}\cdot\nabla\right)\rho_m=-\rho_m\nabla\cdot\mathbf{u}$$

along with the constraint $\nabla\cdot\mathbf{B}=0$. Equations(8.C.38) constitute three equations in four unknowns ($\rho_m,\mathbf{u},\mathbf{p},\mathbf{B}$). A fourth equation must be added, which might be either quite ad hoc or derived from a collisional theory. For example, $\mathbf{p}$ might be assumed to be isotropic and satisfy an equation of state: $\frac{D}{Dt}(p\rho_m^{-\gamma})=0$. More generally, the pressure is a tensor: $\mathbf{p}=p_\|\hat{\mathbf{b}}\hat{\mathbf{b}}+p_\bot(\vec{\mathbf{I}}-\hat{\mathbf{b}}\hat{\mathbf{b}})$.

If we are willing to solve the Vlasov equation in guiding-center variables for the distribution function $F^s(\sigma,p_\sigma;\mu;\alpha,\beta;t)$, where $\alpha$ and $\beta$ are constant for motion fixed to a particular field line, then the density and pressure moments can be calculated from $F^s$:

$$(n,p_\|,p_\bot)^s\equiv\frac{B}{h}\int d\mu\int dp_\sigma\left(1,mv_\|^2=p_\sigma^2/mh^2,\frac{1}{2}mv_\bot^2=\mu B\right)F^s(\sigma,p_\sigma;\mu;\alpha,\beta;t) \tag{8.C.39}$$

$F^s$ satisfies the Vlasov equation:

$$\frac{\partial F}{\partial t}+\dot\sigma\frac{\partial F}{\partial\sigma}+\dot p_\sigma\frac{\partial F}{\partial p_\sigma}=0, \;\;\dot\sigma=\frac{\partial H^s}{\partial p_\sigma}, \;\;\dot p_\sigma=-\frac{\partial H^s}{\partial\sigma}$$

$$H^s=\frac{p_\sigma^2}{2m_s h^2}+\mu B+e\chi-\frac{1}{2}mu_L^2 \tag{8.C.40}$$

where

$$\chi=\phi+\frac{1}{c}\alpha\frac{\partial\beta}{\partial t} \;\;\;\; \frac{\partial(\alpha,\beta)}{\partial t}=-\mathbf{u}\cdot\nabla(\alpha,\beta) \tag{8.C.41}$$

and the quasi-neutrality condition determines $\phi$. The analytic solution of Eqs.(8.C.38-8.C.41) in the most general circumstances is not possible. The numerical solution can even prove difficult. Nevertheless, we wish to apply this formalism to non-uniform plasmas; and we want to do a stability analysis.

### 8.C.d Perturbation theory and variational principle applied to MHD

MHD theory has received much attention and is nicely reviewed in books by Friedberg,[79] Callen[80], and Bateman[81] authored since these lecture notes originated. The basic approach pursued in MHD stability theory is to linearize the equations in all of the variables with respect to a zero-order time-independent solution of the MHD equations. The zero-order equations are nonlinear and are by no means trivial to solve in general circumstances. The MHD equation set admits the application of a variational principle to investigate stability. Theories for ideal (non-resistive) MHD with no flows have been worked out in papers by Bernstein, Frieman, Kulsrud, Kruskal, Oberman, Newcomb, Taylor, Hastie, Chew, Goldberger and Low, and others. Frieman and Rotenberg extended previous work to include rotation. Coppi, Glasser, Greene, and Johnson extended MHD stability theory to include resistive effects and toroidal geometry.

### 8.C.e Derivation of an energy principle for MHD stability using self-adjointness

Consider a model problem in which the pressure is an isotropic pressure satisfying

$$\frac{1}{p}\frac{D}{dt}p = -\gamma \nabla \cdot \mathbf{u} \qquad \gamma \equiv \frac{c_p}{c_v} = \frac{5}{3} \qquad (8.C.42)$$

or equivalently $p \sim \rho^\gamma$. Such a relation can be derived from a Chapman-Enskog expansion of the kinetic equations. One can drop electron inertia effects in the electron momentum equation with no lowest-order flow or rotation and in the low-frequency limit to obtain an Ohm's law:

$$\mathbf{E} + \frac{1}{c}\mathbf{u}\times\mathbf{B} = \eta\mathbf{J} \qquad (8.C.43)$$

and order the resistivity $\eta$ to zero. There are four equations of evolution for $\mathbf{u}, p, \rho,$ and $\mathbf{B}$. The evolution equations lead to an energy conservation law:

$$\frac{dU}{dt} = 0 \quad U(t) = \int d^3\mathbf{x}\left[\frac{|\mathbf{B}(\mathbf{x},t)|^2}{8\pi} + \frac{p(\mathbf{x},t)}{\gamma-1} + \tfrac{1}{2}\rho(\mathbf{x},t)|\mathbf{u}(\mathbf{x},t)|^2\right] \qquad (8.C.44)$$

---

[79] J. P. Freidberg, *Ideal Magnetohydrodynamics* (Plenum Press, New York, 1987).
[80] J. D. Callen, *Fundamentals of Plasma Physics*, http://homepages.cae.wisc.edu/~callen/book.html
[81] G. Bateman, *MHD Instabilities* (MIT Press, Cambridge, Mass), 1978.

The first two terms on the right side of Eq.(8.C.44) contribute to a potential energy $W(t)$ and the third term is a kinetic energy $K(t)$. There is no electrical energy term related to $E^2/8\pi$ because the electric energy density is higher order.

The equilibrium is stationary by definition ($\partial/\partial t = 0$) and static ($\mathbf{u}_0 \equiv 0$) by assumption. Hence, $D/Dt = 0$ in Eqs.(8.C.38) and (8.C.42). The only remaining equation is

$$\nabla p = \frac{1}{c}\mathbf{J}_0 \times \mathbf{B}_0 = \frac{1}{4\pi}\left(\nabla \times \mathbf{B}_0\right) \times \mathbf{B}_0 \tag{8.C.45}$$

Equation (8.C.45) is a statement that the Lorentz force balances the pressure gradient in establishing an equilibrium. Furthermore, the pressure gradient must be perpendicular to both the equilibrium current and magnetic field:

$$\mathbf{B}_0 \cdot \nabla p_0 = 0 \quad \mathbf{J}_0 \cdot \nabla p_0 = 0 \tag{8.C.46}$$

and the equilibrium pressure is constant along the directions of the equilibrium magnetic field lines and the currents.

The linearization of the equations in the stability theory begins with

$$\mathbf{u} = \delta\mathbf{u} \quad \mathbf{B} = \mathbf{B}_0 + \delta\mathbf{B} \quad \mathbf{J} = \mathbf{J}_0 + \delta\mathbf{J} \quad p = p_0 + \delta p \quad \rho = \rho_0 + \delta\rho \tag{8.C.47}$$

The linearized versions of Eqs.(8.C.38) and (8.C.42) become

$$\rho_0 \frac{\partial}{\partial t}\delta\mathbf{u} = -\nabla \cdot \delta\mathbf{p} + \frac{1}{c}\delta\mathbf{j} \times \mathbf{B} + \frac{1}{c}\mathbf{j} \times \delta\mathbf{B}$$

$$\frac{\partial \delta\mathbf{B}}{\partial t} = \nabla \times \left(\delta\mathbf{u} \times \mathbf{B}_0\right) \tag{8.C.48}$$

$$\frac{\partial}{\partial t}\delta p + \delta\mathbf{u} \cdot \nabla p_0 = -\gamma p_0 \nabla \cdot \delta\mathbf{u}$$

The linearized Ampere's law $\nabla \times \delta\mathbf{B} = 4\pi\delta\mathbf{J}/c$ allows us to substitute for $\delta\mathbf{J}$ and reduce the dependent variables to the set ($\delta\mathbf{u}, \delta\mathbf{B}, \delta p$). Equation (8.C.48) leads to

$$\rho_0 \frac{\partial^2}{\partial t^2}\delta\mathbf{u} = -\nabla\left[-\gamma p_0 \nabla \cdot \delta\mathbf{u} - \delta\mathbf{u} \cdot \nabla p_0\right] - \frac{1}{4\pi}\left\{\nabla \times \left[\nabla \times \left(\delta\mathbf{u} \times \mathbf{B}_0\right)\right]\right\} \times \mathbf{B}_0$$

$$+ \frac{1}{c}\mathbf{j}_0 \times \left[\nabla \times \left(\delta\mathbf{u} \times \mathbf{B}_0\right)\right] \equiv \mathbf{F}(\delta\mathbf{u}) \tag{8.C.49}$$

$\mathbf{F}$ in Eq.(8.C.49) is a linear, second-order differential operator in space and is self-adjoint. Hence, its eigenvalues are real; and its eigenvectors can be made orthonormal.

<u>Definition</u>: A self-adjoint operator satisfies $\int d^3 vol\, \vec{\eta} \cdot \mathbf{F}(\vec{\xi}) = \int d^3 vol\, \vec{\xi} \cdot \mathbf{F}(\vec{\eta})$

The self-adjointness of the operator $\mathbf{F}$ was proven by Kulsrud and shown generally from the Lagrangian nature of the fluid equations.[82]

---

[82] R. M. Kulsrud, Astrophys. J. **152**, 1121 (1968).

We next introduce the linear displacement $\vec{\xi}(\mathbf{x},t)$ from the equilibrium position of the fluid element in $(\mathbf{x},t)$. In the absence of zero-order flows: $\dfrac{d\vec{\xi}}{dt} = \dfrac{\partial \vec{\xi}}{\partial t} = \delta \mathbf{u}$. We replace $\delta \mathbf{u}$ with $\delta \mathbf{u} = \dfrac{\partial \vec{\xi}}{\partial t}$ in Eq.(8.C.49) and integrate in time to obtain

$$\rho_0 \frac{\partial^2}{\partial t^2}\vec{\xi} = -\nabla\left[-\gamma p_0 \nabla\cdot\vec{\xi} - \vec{\xi}\cdot\nabla p_0\right] - \frac{1}{4\pi}\left\{\nabla\times\left[\nabla\times\left(\vec{\xi}\times\mathbf{B}_0\right)\right]\right\}\times\mathbf{B}_0 \\ + \frac{1}{c}\mathbf{j}_0\times\left[\nabla\times\left(\vec{\xi}\times\mathbf{B}_0\right)\right]$$

(8.C.50)

The right side of Eq.(8.C.50) is the force density. The inner product of the force density with the displacement $\vec{\xi}(\mathbf{x},t)$ integrated over volume is a quadratic in the displacement and is "virial." We will come back to this subsequently. We can construct an energy that is quadratic in $\vec{\xi}(\mathbf{x},t)$:

$$U(t) \equiv W(t) + K(t) \qquad K(t) = \int d^3x \tfrac{1}{2}\rho_0(\mathbf{x})\left[\dot{\vec{\xi}}(\mathbf{x},t)\right]^2$$

$$W(t) = W\left[\vec{\xi}(\mathbf{x},t)\right] = \sum_{n=0}^{\infty} a_n\left[\xi^n\right] \quad \text{a functional} \equiv W_0 + W_1(\vec{\xi}) + W_2(\vec{\xi},\vec{\xi}) + \ldots$$

(8.C.51)

From the conservation law Eq.(8.C.44) and the linear expansion with $\mathbf{u}_0 = 0$

$$0 = \dot{U} = \dot{W} + \dot{K} \quad 0 = \dot{W}_0 + \dot{W}_1 + \dot{W}_2 + \dot{K} + O(\varepsilon^2) = \dot{W}_1 + \dot{W}_2 + \dot{K} + O(\varepsilon^2)$$

$$K(t) \equiv \int d^3\mathbf{x}\tfrac{1}{2}\rho u^2 = \int d^3\mathbf{x}\tfrac{1}{2}\rho\dot{\vec{\xi}}^2 + O(\xi^3)$$

(8.C.52)

$$W(t) = \int d^3\mathbf{x}\left[\frac{B^2}{8\pi} + \frac{p}{\gamma-1}\right] = W_0 + W_1(\vec{\xi}) + W_2(\vec{\xi},\vec{\xi}) + O(\xi^3)$$

We can use Eq.(8.C.48) to express

$$\delta \mathbf{B}(\vec{\xi}) \equiv \nabla\times(\vec{\xi}\times\mathbf{B}_0)$$

(8.C.53)

and to rewrite Eq.(8.C.50). We note that

$$\dot{K} \equiv \int d^3\mathbf{x}\,\rho_0\dot{\vec{\xi}}\cdot\ddot{\vec{\xi}} = \int d^3\mathbf{x}\,\dot{\vec{\xi}}\cdot\mathbf{F}(\vec{\xi}) = -\dot{W}_1(\vec{\xi}) - W_2(\vec{\xi},\dot{\vec{\xi}}) - W_2(\dot{\vec{\xi}},\vec{\xi})$$

(8.C.54)

We choose as an initial condition $\vec{\xi}(\mathbf{x},t=0) = 0$, but with finite initial velocity $\dot{\vec{\xi}}$. Hence, $W_2(\vec{\xi},\dot{\vec{\xi}}) = 0$ and $W_2(\dot{\vec{\xi}},\vec{\xi}) = 0$, which leaves $\dot{W}_1(\vec{\xi}) = 0$ for arbitrary $\dot{\vec{\xi}}$. $\dot{W}_1(\vec{\xi}) = 0$ does not depend separately on $\vec{\xi}$. Thus, $W_1(\vec{\xi}) = 0$ for all time.

We introduce the definition $\vec{\eta} = \dot{\vec{\xi}}$ and note from Eq.(8.C.54)

$$-\int d^3\mathbf{x}\,\vec{\eta}\cdot\mathbf{F}(\vec{\xi}) = W_2(\vec{\xi},\vec{\eta}) + W_2(\vec{\eta},\vec{\xi}) = -\int d^3\mathbf{x}\,\vec{\xi}\cdot\mathbf{F}(\vec{\eta})$$

(8.C.55)

This explicitly demonstrates that the operator $\mathbf{F}(\vec{\xi})$ is self-adjoint to the order calculated. For self-adjoint operators any arbitrary displacement can be represented as a sum over normal modes:

$$\vec{\xi}(\mathbf{x},t) = \sum_n \vec{\xi}_n(\mathbf{x})e^{-i\omega_n t} \qquad -\omega_n^2\vec{\xi}_n = \mathbf{F}(\vec{\xi}_n)$$

(8.C.56)

where $-\omega_n^2$ is real and $\vec{\xi}_n$ are orthogonal, possibly complex-valued eigenfunctions. The eigenvalues $\omega_n$ are either real and the solutions for the displacement are oscillatory, or $\omega_n$ are imaginary such that the solutions for the displacement are purely growing or decaying. The set of eigenfunctions $\vec{\xi}_n$ are complete for square-integrable modes. Singular solutions of Eq.(8.C.50) are exceptions.[83]

We consider the terms in $W$ at second order in the displacement using Eqs.(8.C.50) and (8.C.52):

$$\delta W \equiv W_2(\vec{\xi},\vec{\xi}) = -\tfrac{1}{2}\int d^3\mathbf{x}\,\vec{\xi}\cdot\mathbf{F}(\vec{\xi}) = \int d^3\mathbf{x}\left[\frac{\delta\mathbf{B}^2}{8\pi} + \frac{1}{2}\delta\mathbf{B}\cdot\mathbf{J}_0\times\vec{\xi} + \frac{\gamma}{2}p_0(\nabla\cdot\vec{\xi})^2 + \frac{1}{2}(\nabla\cdot\vec{\xi})(\vec{\xi}\cdot\nabla p_0)\right]$$

(8.C.57)

If there is instability, then as the amplitude for $\vec{\xi}(\mathbf{x},t)$ grows exponentially, $K(t)$ also grows. Recall that $K(t)$ is positive definite. Hence, $W$ must decrease exponentially if there is instability, which implies that $\delta W$ is negative for some $\vec{\xi}(\mathbf{x},t)$ for instability when $K(t)$ becomes sufficiently large. This conclusion can be turned around to demonstrate stability. Conclusions about stability based on Eq.(8.C.57) and the self-adjointness of the operator $\mathbf{F}(\vec{\xi})$ are the essence of the MHD energy principle.

Theorem: Suppose $W_2$ is positive definite for all $\vec{\xi}(\mathbf{x},t)$, then the system is stable, i.e., $W_2(\vec{\xi},\vec{\xi}) > 0$ is sufficient for stability. We will prove subsequently that this is also a necessary condition.

It is useful to introduce the following definition:

Definition: $K_2(\vec{\xi}) \equiv \int d^3\mathbf{x}\,\tfrac{1}{2}\rho_0|\vec{\xi}|^2$ (moment of inertia).

Theorem: Virial theorem
$$\dot{K}_2(\vec{\xi}) \equiv \int d^3\mathbf{x}\,\rho_0\dot{\vec{\xi}}\cdot\vec{\xi}$$
$$\ddot{K}_2(\vec{\xi}) \equiv \int d^3\mathbf{x}\,\rho_0|\dot{\vec{\xi}}|^2 + \int d^3\mathbf{x}\,\vec{\xi}\cdot\mathbf{F}(\vec{\xi}) = 2K_2(\dot{\vec{\xi}}) - 2W_2(\vec{\xi},\vec{\xi}) \qquad (8.C.58)$$
$$\rightarrow \quad \tfrac{1}{2}\ddot{K}_2(\vec{\xi}) = K_2(\dot{\vec{\xi}}) - W_2(\vec{\xi},\vec{\xi})$$

Exercise: Use the virial theorem to show that without gravity **g**, a plasma cannot be self-contained.

---

[83] G. Laval, C. Mercier, and R. Pellat, Nucl. Fusion 5, 156 (1965).

Consider a displacement $\vec{\eta}(\mathbf{x})$ such that $W_2(\vec{\eta}) < 0$ and define $\gamma^2 \equiv \dfrac{-W_2(\vec{\eta})}{K_2(\vec{\eta})}$. Choose initial conditions:

$$\vec{\xi}(\mathbf{x},t=0) = \vec{\eta}(\mathbf{x}) \quad \dot{\vec{\xi}}(\mathbf{x},t=0) = \gamma\vec{\eta}(\mathbf{x}) \tag{8.C.59}$$

The energy then becomes

$$U = W_2 + K_2(\dot{\vec{\xi}}) = -\gamma^2 K_2(\vec{\eta}) + \gamma^2 K_2(\vec{\eta}) = 0 \tag{8.C.60}$$

at all times. Thus, if $W_2 < 0$ then $K_2 > 0$ for all time. From the virial theorem Eq.(8.C.58)

$$\ddot{K}_2(\vec{\xi}) = 4K_2(\dot{\vec{\xi}}) \tag{8.C.61}$$

We use the Schwarz inequality $(\mathbf{a} \cdot \mathbf{b})^2 \leq a^2 b^2$ and generalize to Hilbert space:

$$\left|\dot{K}_2\right|^2 = \left|\int d^3\mathbf{x}\,\rho_0 \vec{\xi}(\mathbf{x}) \cdot \dot{\vec{\xi}}(\mathbf{x})\right|^2 \leq \left[\int d^3\mathbf{x}\,\rho_0 \left|\vec{\xi}(\mathbf{x})\right|^2\right]\left[\int d^3\mathbf{x}\,\rho_0 \left|\dot{\vec{\xi}}(\mathbf{x})\right|^2\right] = 2K_2(\vec{\xi})\,2K_2(\dot{\vec{\xi}}) \tag{8.C.62}$$

Then using Eqs.(8.C.61) and (8.C.62)

$$\ddot{K}_2(\vec{\xi}) = 4K_2(\dot{\vec{\xi}}) \geq \frac{\left|\dot{K}_2(\vec{\xi})\right|^2}{K_2(\vec{\xi})} \geq 0 \quad \text{and} \quad K_2(\vec{\xi}) \geq \frac{\left|\dot{K}_2(\vec{\xi})\right|^2}{4K_2(\dot{\vec{\xi}})} \geq 0 \tag{8.C.63}$$

We next define $y(t) \equiv \ln\left[K_2(\vec{\xi})/K_2(\vec{\eta})\right] = 2\gamma(t)$ which removes the spatial dependence of the initial conditions. Given the definition of $y$, $y(0) = 0$ and $\dot{y}(0) = 2\gamma$; and in consequence of $W_2 < 0$ and Eq.(8.C.63), one can show $\ddot{y} > 0$ and $K_2(\vec{\xi}) \geq |W_2(\vec{\eta})|e^{2\gamma t}$ which implies $\gamma^2 > 0$ and the possibility of instability.

<u>Theorem</u>: If $W_2 > 0$ for all $\vec{\xi}$, the system is stable; and if $W_2 < 0$, it is possible to find instability for some choice of $\vec{\xi}$. Because of the energy principle for this system, $W_2 < 0$ is a necessary and sufficient condition for instability.

### *8.C.f Interchange instability*

Perhaps the most basic example of MHD instability is the interchange instability. We begin the analysis of the interchange by introducing a new expansion parameter $\beta \sim p_0/(B_0^2/4\pi) \ll 1$, i.e., we assume a low-$\beta$ plasma. We expand the relevant quantities in powers of $\beta$:

$$\mathbf{B}_0(\mathbf{x}) = \mathbf{B}_0^{(0)}(\mathbf{x}) + \mathbf{B}_0^{(1)}(\mathbf{x}) + \ldots$$
$$\mathbf{j}_0(\mathbf{x}) = \mathbf{j}_0^{(0)}(\mathbf{x}) + \mathbf{j}_0^{(1)}(\mathbf{x}) + \ldots$$
$$p_0(\mathbf{x}) = p_0^{(0)}(\mathbf{x}) + p_0^{(1)}(\mathbf{x}) + \ldots \quad (8.\text{C}.64)$$
$$\vec{\xi}(\mathbf{x}) = \vec{\xi}^{(0)}(\mathbf{x}) + \vec{\xi}^{(1)}(\mathbf{x}) + \ldots$$
$$W_2(\mathbf{x}) = W_2^{(0)}(\mathbf{x}) + W_2^{(1)}(\mathbf{x}) + \ldots$$

In this example we also assume

$$\nabla \times \mathbf{B}_0^{(0)} = \frac{4\pi}{c} \mathbf{j}_0^{(0)} = 0 \quad (8.\text{C}.65)$$

The plasma is contained by a vacuum magnetic field. Consistent with Eq.(8.C.65), $p_0^{(0)} = 0$ and

$$W_2^{(0)} = \int d^3\mathbf{x} \frac{\left|\nabla \times \delta \mathbf{B}^{(0)}\right|^2}{8\pi} = \int d^3\mathbf{x} \frac{\left|\nabla \times \left(\vec{\xi}^{(0)} \times \mathbf{B}^{(0)}\right)\right|^2}{8\pi} \quad (8.\text{C}.66)$$

We minimize $W_2^{(0)}$ with respect to $\vec{\xi}^{(0)}$ to zero order by requiring $\delta \mathbf{B}^{(0)} = 0$. The lowest-order energy is

$$W^{(0)} = \int d^3\mathbf{x} \frac{\left|\mathbf{B}^{(0)}\right|^2}{8\pi} \quad (8.\text{C}.67)$$

The zero-order displacement must satisfy $\nabla \times \left(\vec{\xi}^{(0)} \times \mathbf{B}^{(0)}\right) = 0$. Hence,

$$\vec{\xi}^{(0)} \times \mathbf{B}^{(0)} = \nabla \phi_B \quad \rightarrow \quad \mathbf{B}^{(0)} \cdot \nabla \phi_B = 0 \quad (8.\text{C}.68)$$

The first-order $W_2^{(1)}$ is then

$$W_2^{(1)} = \frac{1}{2} \int d^3\mathbf{x} \left\{ \gamma p_0^{(1)} \left|\nabla \cdot \vec{\xi}^{(0)}\right|^2 + \left(\vec{\xi}^{(0)} \cdot \nabla p_0^{(1)}\right) \nabla \cdot \vec{\xi}^{(0)} \right\} \quad (8.\text{C}.69)$$

Through this order in the $\beta$ expansion we do not need to know $\vec{\xi}^{(1)}$ or $\mathbf{B}_0^{(1)}$. Using Eq.(8.C.68) one can express the displacement as

$$\vec{\xi} = \frac{\mathbf{B}_0 \times \nabla \phi_B(\alpha, \beta)}{B_0^2} + \xi_\parallel \hat{\mathbf{b}}_0 \quad (8.\text{C}.70)$$

where $\xi_\parallel$ is arbitrary. For $\gamma = 5/3$ then $p/(\gamma - 1) = (3/2)p \sim (3/2)nk_B T$. $W_2^{(1)}$ is primarily associated with changes in the thermal energy content in a volume being converted to a macroscopic displacement. However, flux tubes can interchange with no thermal energy exchange in what we call an interchange instability. If there are no topological constraints, then minimization of $W_2^{(1)}$ with respect to $\xi_\parallel$ yields $\nabla \cdot \vec{\xi}$ that is not a function of $\sigma$ and is constant along a field line. We examine Eq.(8.C42) rewritten as

$$Dp = -\gamma p_0 \nabla \cdot \vec{\xi} \quad (8.\text{C}.71)$$

and conclude that $Dp$ is constant along the field line. This removes the possibility of sound waves along the field line, which waves would contradict our hypothesis of virtual slow (adiabatic) displacements of the plasma. The Eulerian version of Eq.(8.C.71) is

$$\delta p(\alpha,\beta) = -\gamma p_0 \nabla \cdot \vec{\xi} - \xi \cdot \nabla p_0 \qquad (8.C.72)$$

Because the magnetic field is a vacuum magnetic field, **B** can be represented by a gradient of a potential, i.e., $\mathbf{B} = \nabla\sigma$ by choice and is convenient. The differential volume element is given by

$$d^3\mathbf{x} = \frac{d\alpha\, d\beta\, d\sigma}{B^2(\alpha,\beta,\sigma)} = d\alpha\, d\beta\, \frac{d\ell}{B}$$

$$h \equiv \frac{d\ell}{d\sigma} = \text{metric} \quad \hat{\mathbf{b}}\cdot\mathbf{B} = \hat{\mathbf{b}}\cdot\nabla\sigma \quad B = \frac{d\sigma}{d\ell} = \frac{1}{h} \qquad (8.C.73)$$

<u>Theorem</u>: Using Eqs.( 8.C.72) and (8.C.73) in Eq.(8.C.69) one obtains

$$W_2^{(1)} = -\frac{1}{2}\int d\alpha\, d\beta (\nabla\cdot\vec{\xi})\delta p \int \frac{d\sigma}{B^2(\alpha,\beta,\sigma)} = -\frac{1}{2}\int d\alpha\, d\beta (\nabla\cdot\vec{\xi})\delta p \int \frac{d\ell}{B} \qquad (8.C.74)$$

The last integral factor on the right side of Eq.(8.C.74) has special significance as the volume per unit flux.

$$U_{vol} \equiv \int \frac{d\ell}{B} = \int \frac{d^3\mathbf{x}}{d\alpha\, d\beta} = \int \frac{d^3\mathbf{x}}{d\Phi_B} = \int \frac{d\text{vol}}{d\text{flux}} \qquad (8.C.75)$$

<u>Definition</u>: The specific volume is the volume per unit flux.

Consider pressure surfaces consistent with momentum balance to leading order. We note that $\hat{\mathbf{b}}\cdot\nabla p = 0$, $\mathbf{j}\cdot\nabla p = 0$, and $|j_\parallel| << |j_\perp|$ to leading order. A three-dimensional plot of the pressure surfaces, the magnetic fields lines, and the currents will show the following. **B** and **j** vector fields are perpendicular to one another and lie in surfaces of constant $p$ because $\nabla p$ is perpendicular to both **B** and **j**. Hence, **B** and **j** form a cage for the pressure (Fig. 8.C.2). The constant pressure surfaces are also magnetic flux surfaces, i.e., the pressure can be written as a function of the magnetic flux.

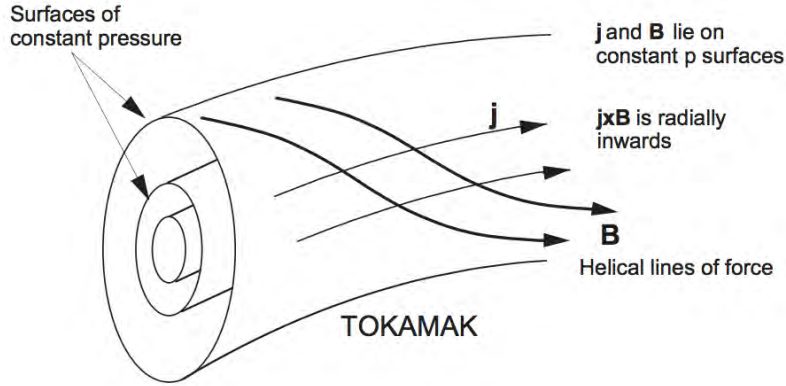

Figure 8.C.2 Plasma equilibrium in which the current density and magnetic field lines lie in nested surfaces of constant pressure which are also constant flux surfaces.[84]

Consider two pressure surfaces an infinitesimal apart, the volume contained within the two pressure surfaces, and the magnetic fluxes associated with the two surfaces: $vol(p)$ and $\Phi_B(p)$. In consequence, $dvol/d\Phi_B$ is a function of the pressure. One can also show that

$$\int_{\alpha,\beta} \frac{d\ell}{B} = \left[\int_{\alpha,\beta} \frac{d\ell}{B}\right]_{\Phi_B} \tag{8.C.76}$$

is a function of the magnetic flux. In guiding-center theory, the flux surface is also the drift surface. We use Eqs.(8.C.74) and (8.C.75), and cancel factors of $d\Phi_B$ in the numerator and denominator to express the energy $W_2^{(1)}$:

$$W_2^{(1)}(\vec{\xi}) = -\frac{1}{2}\int dvol\, \delta p \nabla\cdot\vec{\xi} = -\frac{1}{2}\int d\alpha d\beta\, U_{vol}\delta p \nabla\cdot\vec{\xi}$$

$$= -\frac{1}{2}\int d\alpha d\beta\, \delta p DU_{vol} = \int d\alpha d\beta\, w_2^{(1)}(\alpha,\beta) \tag{8.C.77}$$

where $U_{vol} \equiv dvol/d\Phi_B$ is the specific volume, $\nabla\cdot\vec{\xi} = DU_{vol}/U_{vol}$, and

$$w_2^{(1)}(\alpha,\beta) \equiv -\tfrac{1}{2}\delta p(\alpha,\beta) DU_{vol}(\alpha,\beta) \tag{8.C.78}$$

is the flux surface density of energy for a virtual displacement $\vec{\xi}$, i.e., the energy per unit flux. We use $Dp \equiv \delta p + (dp/d\Phi_B)D\Phi_B$ to express Eq.(8.C.78) as

$$w_2^{(1)}(\alpha,\beta) \equiv -\tfrac{1}{2}\left(Dp - \frac{dp}{d\Phi_B}d\Phi_B\right)DU_{vol}(\alpha,\beta) \tag{8.C.79}$$

---

[84] John Howard, *Introduction to Plasma Physics,* Fig. 6.2, Chapter 6, Australian National University, https://physics.anu.edu.au/prl/intranet/_files/c17/chap06.pdf

Next we use $DU_{vol} = \dfrac{dU_{vol}}{d\Phi_B}$ and the adiabatic law $\dfrac{Dp}{p} = -\gamma \dfrac{DU_{vol}}{U_{vol}}$ in Eq.(8.C.79) to obtain

$$w_2^{(1)}(\alpha,\beta) \equiv \tfrac{1}{2}(d\Phi_B)^2 \left( \frac{dU_{vol}}{d\Phi_B} \frac{dp}{d\Phi_B} + \gamma \frac{p}{U_{vol}} \left( \frac{dU_{vol}}{d\Phi_B} \right)^2 \right) \qquad (8.C.80)$$

or alternatively,

$$w_2^{(1)}(\alpha,\beta) \equiv \tfrac{1}{2}(d\Phi_B)^2 \gamma \frac{p}{U_{vol}} \left( \frac{dU_{vol}}{d\Phi_B} \right)^2 \left( 1 + \frac{dp}{dU_{vol}} \frac{U_{vol}}{\gamma p} \right) \qquad (8.C.81)$$

Stability (instability) is determined by $w_2^{(1)} > 0$ ($w_2^{(1)} < 0$). Hence, the stability condition is

$$\frac{dp}{dU_{vol}} \frac{U_{vol}}{\gamma p} > -1 \quad \text{or} \quad \frac{dp}{dU_{vol}} > -\frac{\gamma p}{U_{vol}} \qquad (8.C.82)$$

which is a statement about the configuration and its pressure gradient. For example, consider a typical situation in which

$$\frac{dp}{dU_{vol}} = \frac{dp/d\Phi_B < 0}{dU_{vol}/d\Phi_B > 0} \rightarrow \frac{dp}{dU_{vol}} < 0 \qquad (8.C.83)$$

but how large is $dp/dU_{vol}$ compared to $-\gamma p/U_{vol}$?

Theorem: A sufficient condition for stability is $dp/dU_{vol} > 0$.

Thus, if $p$ decreases with magnetic flux while $U_{vol}$ also decreases, then moving pressure from higher to lower down the gradient is accompanied by decreasing $U_{vol}$ with increasing flux. In this case the plasma is stable. If $U_{vol}$ can increase with increasing flux while the pressure decreases such that perturbations can grow as the expense of cooling under an adiabatic expansion, then the plasma is unstable. Note that $U_{vol}$ is always positive by definition of the right-hand coordinate system $\alpha, \beta, \sigma$.

The preceding arguments are predicated on considerations of confined plasmas, i.e., plasma pressure in the interior of a configuration with a vacuum magnetic field external to the plasma: $\frac{dp}{d\Phi_B} < 0$; and a sufficient condition for stability is $\frac{dU_{vol}}{d\Phi_B} < 0$.

Definition: $V'' \equiv \dfrac{dU_{vol}}{d\Phi_B} < 0$ for stability. $V''$ is a term seen in the MHD stability literature.

Definition: From Eq.(8.C.76)

$$U_{vol} \equiv \int_{\alpha,\beta} \frac{d\ell}{B} \equiv \frac{L}{\langle B \rangle} \tag{8.C.84}$$

Thus, in a stable confined plasma where the pressure and the volume per unit flux are maximal then <B> is minimal. Minimum average B is equivalent to V"<0 as a sufficient condition for stability and is well-established as a constraint in designing low-$\beta$ magnetic confinement experiments.

Theorem: In sum, $dpdU_{vol} > 0$, minimum average B, and V"<0 are equivalent conditions for stability with respect to interchange modes.

### 8.C.g Interpretation of interchange stability theory in magnetically confined plasmas

A physical interpretation of the interchange instability can be given as follows.
1) There can be an unstable expansion of plasma down a pressure gradient into a region of weaker magnetic field accompanied by a reduction in temperature.
2) There is a thermodynamic drive for the instability because the plasma is diamagnetic, i.e., the plasma magnetization is oppositely directed to the applied magnetic field, so that the plasma is attracted to weaker fields.
3) Field-line curvature can be stabilizing if the field lines are convex relative to the plasma pressure maximum or destabilizing if the field lines are concave relative to the plasma pressure maximum. The field-line curvature vector is $\vec{\kappa} \equiv \hat{\mathbf{b}} \cdot \nabla \hat{\mathbf{b}}$ and is perpendicular to **B**. Greene and Johnson[85] showed that the drive term in $w_2^{(1)}$ can be rewritten as

$$-\frac{1}{4\pi} \vec{\xi} \cdot \nabla p_0 \, \vec{\xi} \cdot \vec{\kappa} \tag{8.C.85}$$

When the curvature and the pressure gradient are anti-parallel, this term is positive and is stabilizing. When the pressure gradient has a component that is parallel to the curvature, there can be displacements such that the drive term is negative and destabilizing if the drive term is sufficiently strong enough to render $w_2^{(1)}$ negative.

We return to consideration of Eq.(8.C.84). Recall Eq.(8.C.73) and note that $B = d\sigma/d\ell$ so that it follows

$$\delta U_{vol} = \int \frac{d\sigma}{B^2} \rightarrow \frac{\partial \delta U_{vol}}{\partial \alpha} = \frac{\partial}{\partial \alpha} \int \frac{d\sigma}{B^2} = -2 \int \frac{d\sigma}{B^3} \frac{\partial B}{\partial \alpha} \tag{8.C.86}$$

However, $d\Phi_B = d\alpha d\beta$ and

---

[85] J. M. Greene and J. L. Johnson, *Hydromagnetic Equilibria and Stability*, in Advances in Theoretical Physics, 1965 (Academic Press, N.Y.); J. M Greene and J. L. Johnson, Plasma Phys. **10**, 729 (1968).

$$\alpha = \frac{\Phi_B}{2\pi} = \frac{B\pi r^2}{2\pi} = \frac{1}{2}Br^2 \quad d\alpha = Brdr \tag{8.C.87}$$

along a field line if $\beta$ is the azimuthal angle; and $\alpha$ is constant along the field line. In this cylindrical configuration $\partial B/\partial r = B/R$ where $R$ is the radius of curvature of the magnetic field line. Using Eq.(8.C.87) in Eq.(8.C.86) one obtains

$$\frac{\partial \delta U_{vol}}{\partial \alpha} = -2\int \frac{d\sigma}{B^3}\frac{\partial B}{\partial \alpha} = -2\int \frac{d\sigma}{B^4}\frac{\partial B}{r\partial r} = -2\int \frac{d\sigma}{B^3}\frac{1}{rR} = -2\int \frac{d\ell}{B^2}\frac{1}{rR} = -\frac{1}{2\alpha^2}\int \frac{d\ell r^3}{R} \tag{8.C.88}$$

If the region along the field line where $R<0$ dominates the integral along the field line in Eq.(8.C.88), then the right side of (8.C.88) is positive and the plasma is unstable. A positive radius of curvature promotes $V''<0$ and is stabilizing.

4) Consider the plasma drift motion and magnetic shear. There is a strong analogy between Rayleigh-Taylor instability in which a fluid accelerates due to gravity into a region of lower mass density and the expansion of a plasma into a region of decreasing magnetic field strength. If the plasma starts to flute along the field line, the concomitant electric field across the flutes can drive an **E**×**B** drift that enhances the amplitude of the flute perturbations. The positive contribution to the potential energy in magnetic tension decreases as the plasma expands into a weaker field region with bad curvature. Magnetic shear can provide a topological constraint that can balance this decrease. An interchange mode has flute structure along the field line. Shearing of the field causes the interchange mode to bend, which costs it energy to do so. Hence, magnetic shear has a stabilizing effect. An intuitive argument yields an order-of-magnitude estimate for stabilization when the magnetic shearing rate satisfies

$$\left|\frac{d\theta}{dz}\right| > \frac{2\gamma_0}{c_A} \tag{8.C.89}$$

where $\theta$ is the angle of inclination of the magnetic field, $\gamma_0 \sim c_s/L$ is the ideal MHD growth rate in the absence of shear, and $c_A$ is the Alfvén speed.

When the plasma pressure is finite and $\beta$ is no longer low, the pressure perturbation contributes a negative term to the energy:

$$W_2 = \int d^3\mathbf{x}\left[\frac{\delta B^2}{8\pi} - \delta p + ...\right] \tag{8.C.90}$$

Symbolically the contribution of finite pressure to the energy is negative and enhances the instability drive. Magnetic shear does not seem to help against ballooning instability when $\beta \sim 1$.

5) The inclusion of resistivity adds a qualitatively new character to hydrodynamic instability. Ohm's law becomes

$$\mathbf{E} + \frac{1}{c}\mathbf{u}\times\mathbf{B} = \eta\mathbf{j} \tag{8.C.91}$$

Faraday's law becomes

$$\frac{\partial \mathbf{B}}{\partial t} = -c\nabla \times \mathbf{E} = \nabla \times (\mathbf{u} \times \mathbf{B} - c\eta \mathbf{j}) = \nabla \times (\mathbf{u} \times \mathbf{B}) - 4\pi \nabla \times (\eta \nabla \times \mathbf{B}) \quad (8.C.92)$$

Now plasma can move across the field lines, and the magnetic field can diffuse. There is a large body of research on resistive MHD stability theory addressing instabilities such as the resistive interchange, resistive ballooning, and resistive tearing instabilities.

### 8.C.h Chew-Goldberger-Low double-adiabatic theory - accommodating a tensor pressure in MHD theory

A tensor pressure can be represented by
$$\mathbf{p} = p_\| \hat{\mathbf{b}}\hat{\mathbf{b}} + p_\perp (\vec{\mathbf{I}} - \hat{\mathbf{b}}\hat{\mathbf{b}}) \quad (8.C.93)$$

Chew, Goldberger, and Low (CGL)[86] derived a fluid model from consideration of the Vlasov equation in the collisionless limit and with certain assumptions on the heat flow. They calculated the second velocity moment of the Vlasov equation to obtain

$$\frac{\partial}{\partial t}\mathbf{p} = -\nabla \cdot [\mathbf{Q} + \mathbf{u}\mathbf{p}] - (\nabla \mathbf{u} + \mathbf{u}\overleftarrow{\nabla}) \cdot \mathbf{p} + \frac{e}{mc}(\mathbf{p} \times \mathbf{B} - \mathbf{B} \times \mathbf{p}) \quad (8.C.94)$$

where $\mathbf{Q}$ is the heat flow:
$$\mathbf{Q} = nm\langle (\mathbf{v}-\mathbf{u})(\mathbf{v}-\mathbf{u})(\mathbf{v}-\mathbf{u})\rangle_f \quad (8.C.95)$$

The parallel and perpendicular components of the time derivative of the pressure tensor $\mathbf{p}$ are obtained by operating on Eq.(8.C.94) with $\hat{\mathbf{b}}\hat{\mathbf{b}}$ and $\vec{\mathbf{I}}_\perp = (\vec{\mathbf{I}} - \hat{\mathbf{b}}\hat{\mathbf{b}})$. In so doing one must represent the heat flow:

$$\mathbf{Q} = Q_\| \hat{\mathbf{b}}\hat{\mathbf{b}}\hat{\mathbf{b}} + Q_\perp (\hat{\mathbf{b}}\mathbf{I}_\perp + \mathbf{I}_\perp \hat{\mathbf{b}} + \hat{\mathbf{e}}_1 \mathbf{b}\hat{\mathbf{e}}_1 + \hat{\mathbf{e}}_2 \mathbf{b}\hat{\mathbf{e}}_2) \quad (8.C.96)$$

The analysis leads to the following pair of evolution equations:

$$\frac{D}{Dt}\ln\left(\frac{p_\perp}{\rho B}\right) = \left(\frac{\partial}{\partial t} + \mathbf{u} \cdot \nabla\right)\ln\left(\frac{p_\perp}{\rho B}\right) = -\frac{B^2}{p_\perp}\frac{\partial}{\partial \ell}\left(\frac{Q_\perp}{B^2}\right)$$

$$\frac{D}{Dt}\ln\left(\frac{p_\| B^2}{\rho^3}\right) = \ldots(Q_\|) \quad (8.C.97)$$

where $Q_\perp \equiv \tfrac{1}{2}nm\langle v_\perp^2 v_\|\rangle$  $Q_\| \equiv \tfrac{1}{2}nm\langle v_\|^2 v_\|\rangle$. CGL assumed that the right sides of Eqs.(8.C.97) are negligible, i.e., negligible heat flow. With this assumption there are two adiabatic invariants that advect with the fluid: $\dfrac{p_\perp}{\rho B}$ and $\dfrac{p_\| B^2}{\rho^3}$.

---

[86] G. F. Chew, M. L. Goldberger, F. E. Low, Proc. R. Soc. Lond. A **236**, 112 (1956).

Example: ⊥ compression – Consider a compression of the plasma and the magnetic field lines perpendicular to the field that increase $B$ and $\rho$ together. We then note that $p_\perp = \dfrac{nm\mathrm{v}_\perp^2}{2} = n\mu B$  $\mu \approx \text{const} \to p_\perp \propto nB = \dfrac{\rho}{m}B \propto \rho^2$  $\therefore \dfrac{D}{Dt}\left(\dfrac{p_\perp}{\rho B} \propto \dfrac{\rho^2}{\rho^2}\right) = 0$ and

$p_\| \sim nm\mathrm{v}_\|^2 \propto \rho$  $\therefore \dfrac{p_\| B^2}{\rho^3} \propto \dfrac{\rho \rho^2}{\rho^3} \sim 1$.

Example: || compression – Consider a compression of the plasma along the field line but no compression of the magnetic field strength. Then

$p_\perp \propto \rho$  $\int p\,dq = \mathrm{v}_\| L = \text{const} \to \mathrm{v}_\| \propto \dfrac{1}{L} \propto \rho \to p_\| \sim \rho \mathrm{v}_\|^2 \propto \rho^3$  $\therefore \dfrac{p_\| B^2}{\rho^3} \propto \dfrac{\rho^3}{\rho^3} = 1$

An energy theorem like that in Secs. 8.C.e and 8.C.f can be derived with the CGL anisotropic pressure. One begins with the fluid equation of motion

$$\rho \dfrac{D\mathbf{u}}{Dt} = -\nabla \cdot \left[ p_\| \hat{\mathbf{b}}\hat{\mathbf{b}} + p_\perp \left( \vec{\mathbf{I}} - \hat{\mathbf{b}}\hat{\mathbf{b}} \right) \right] + \dfrac{1}{c} \mathbf{j} \times \mathbf{B} \qquad (8.C.98)$$

and linearizes about an equilibrium. The analysis leads to an equation of the form $\rho \ddot{\vec{\xi}} = \mathbf{F}(\vec{\xi})$. We can again show that the operator $\mathbf{F}(\vec{\xi})$ is self-adjoint. The analysis involved is three times the amount of formalism as for ideal MHD, where the pressure is assumed to be a scalar. The same theorems result, but the formulae are different. We can make comparisons between the CGL and the ideal MHD energy principles.[79]

Theorem: Comparison theorem – Assume $p_\|^0 = p_\perp^0$ but allow the perturbations to be anisotropic and render the perturbed pressure to be anisotropic in the CGL theory. The dynamics of $\vec{\xi}(t)$ are very different in the ideal MHD theory from that in the CGL theory. One finds

$$W_2^{CGL} \geq W_2^{MHD} \qquad (8.C.99)$$

Thus, if the energy theorem for ideal MHD stability indicates stability, then CGL theory is certain to predict stability. Similarly, if CGL theory indicates instability, then MHD theory is certain to do so as well. However, $W_2^{CGL}$ is relatively useless because it is so complicated.

Theorem: Comparison theorem for low-$\beta$ interchange in an axisymmetric system. In ideal MHD with $p_\|^0 = p_\perp^0 = p$ the analysis in Sec. 8.C.f can be used to show that the stability condition for interchange is

$$\int \dfrac{d\ell}{B^2 rR} \dfrac{dp}{d\alpha} < 0 \qquad \text{ideal MHD} \qquad (8.C.100)$$

where $R$ is the radius of curvature and $r$ is the radial distance from the axis of symmetry. Bad curvature (concave field lines facing the plasma) corresponds the $R<0$, while good curvature (convex field lines facing the plasma) corresponds to $R>0$. For anisotropy $p_\parallel^0 \neq p_\perp^0$ one can use CGL theory to derive the stability condition for interchange

$$\int \frac{d\ell}{B^2 rR} \frac{d(p_\parallel^0 + p_\perp^0)}{d\alpha} < 0 \quad \text{CGL} \quad (8.C.101)$$

Mirror plasmas are typically anisotropic, and stability of an axisymmetric mirror is difficult to achieve. Recall that $(p_\parallel^0 + p_\perp^0)/R \propto (v_\parallel^2 + \tfrac{1}{2}v_\perp^2)/R \sim v_d$. Thus, consideration of the guiding-center drifts can give the stability condition in Eq.(8.C.101) a physical interpretation.[87]

### 8.C.i Kulsrud and Kruskal-Oberman extensions to the energy principle

In addition to the CGL theory there have been other important extensions of ideal MHD theory leading to energy principles with added physics content. Some prominent examples of these are the works of Kulsrud,[82,88] Kruskal and Oberman,[89] and Newcomb.[90] Kulsrud derives an energy principle starting from a guiding-center Vlasov equation for the linearly perturbed guiding-center distribution function $\delta f^{G.C.}$. With considerable labor Kulsrud shows that the force $\mathbf{F}(\vec{\xi})$ is self-adjoint, and an energy principle is obtained.

Theorem: Kulsrud demonstrates that the minimum values of the energy satisfy $w_2^{CGL} \geq w_2^{G.C.} \geq w_2^{MHD}$. Thus, if the system is MHD stable, then guiding-center and CGL theories also indicate stability. Conversely, if the system is unstable according to CGL theory, then the other two theories also indicate instability. We retire this approach and go to J.B. Taylor's general ordering scheme in the next section.[91]

### 8.C.j Taylor's guiding-center theory and Taylor and Hastie's analysis of linear stability

The guiding-center Vlasov kinetic equation capturing the motion parallel to the magnetic field is

$$\frac{\partial F}{\partial t} + \dot{\sigma} \frac{\partial F}{\partial \sigma} + \dot{p}_\sigma \frac{\partial F}{\partial p_\sigma} = 0 \quad (8.C.102)$$

We introduce the expansion $F = F_0(H) + \delta F$. The linearized form of Eq.(8.C.103) becomes

---

[87] M. N. Rosenbluth and C. L. Longmire, Ann. Phys. **1**, 120 (1957).
[88] R. M. Kulsrud, Phys. Fluids **5**, 192 (1962).
[89] M. D. Kruskal and C. Oberman, Phys. Fluids **1**, 275 (1958).
[90] W. A. Newcomb, Annals of Physics **10**, 232 (1960).
[91] J. B. Taylor, Proc. Royal Society A **304**, 335 (1968).

$$\frac{\partial \delta F}{\partial t} + \delta\dot\sigma\frac{\partial F_0}{\partial \sigma} + \delta\dot p_\sigma \frac{\partial F_0}{\partial p_\sigma} + \dot\sigma\frac{\partial \delta F}{\partial \sigma} + \dot p_\sigma \frac{\partial \delta F}{\partial p_\sigma} = 0 \qquad (8.\text{C}.103)$$

where

$$H_{G.C.} = \frac{p_\sigma^2}{2mh^2} + \mu B + e\chi, \quad \delta H = \mu\delta B + e\delta\chi + ..., \quad \delta\dot\sigma = \frac{\partial \delta H}{\partial p_\sigma}, \quad \delta p_\sigma = -\frac{\partial \delta H}{\partial \sigma} \sim \delta E_\parallel \qquad (8.\text{C}.104)$$

and the guiding center drifts excluding the polarization drift are

$$\dot\alpha = -\frac{c}{e}\frac{\partial H}{\partial \beta} \qquad \dot\beta = \frac{c}{e}\frac{\partial H}{\partial \alpha} \qquad (8.\text{C}.105)$$

The theory assumes that $u_L \sim O(\varepsilon)$ and time derivatives are similarly ordered so that to lowest order the particles are confined to field lines, which field lines do not change much. The resulting system has two degrees of freedom with slow time dependence. Despite these simplifications there is lengthy mathematics in the theory.

We recall the longitudinal adiabatic invariant derived earlier:

$$J_\sigma = \oint d\sigma\, p_\sigma(H, \alpha, \beta, \sigma) \qquad (8.\text{C}.106)$$

One can show that $\dot J_\sigma = O(\varepsilon)$ because $\dot H, \dot\alpha, \dot\beta \sim O(\varepsilon)$. Moreover, it was demonstrated earlier that $\langle \dot J_\sigma \rangle = O(\varepsilon^2)$ where the average is over a bounce.

Theorem: $J_\sigma$ is effectively invariant on the drift time scale.

Hence, one can construct a bounce-averaged Hamiltonian such that

$$H(J_\sigma; \alpha, \beta) \quad \dot\alpha = -\frac{c}{e}\frac{\partial H}{\partial \beta}\bigg|_{\alpha, J_\sigma, t} \quad \dot\beta = \frac{c}{e}\frac{\partial H}{\partial \alpha}\bigg|_{\beta, J_\sigma, t} \quad \dot J_L = \frac{\partial H}{\partial \theta_L}\bigg|_{\alpha, \beta, t} = 0 \qquad (8.\text{C}.107)$$

One must use canonical transformations to obtain $H(J_\sigma; \alpha, \beta)$.

Before continuing with guiding center theory, we digress to identify some of the relevant literature and theoretical results. Taylor and Hastie introduced a theory of equilibrium and stability of a small Larmor-radius plasma in general geometry and derived multi-pole dispersion relations for electrostatic modes.[92,93] The modes examined satisfy $\omega_d, \omega_b \sim \omega \ll \Omega$ and allow for $\lambda \sim \rho_i$, but with weak scale lengths for the plasma and the magnetic field so that the magnetic moment is conserved. C. S. Liu examined the dynamics of the drift equations and analyzed two instabilities excited by temperature gradients.[94] In all of these guiding center theories there is a requirement on the distribution function $F$ in order to do the mathematics: one must assume

---

[92] J.B. Taylor and R.J. Hastie, Plasma Physics **10**, 479 (1968).
[93] R.J. Hastie and J.B. Taylor, Plasma Physics **13**, 265 (1971).
[94] C.S. Liu, Phys. Fluids **12**, 1489 (1969).

$$F_0\left(\mu;\alpha,\beta;H(\sigma,p_\sigma)\right), \quad \left.\frac{\partial F_0}{\partial H}\right|_{\mu,\alpha,\beta} < 0 \tag{8.C.108}$$

This restriction is rather unsatisfactory and limiting. However, there are two relevant and important examples of kinetic microstability in a collisionless Vlasov plasma.

Example: A uniform, one-dimensional, unmagnetized plasma, is stable if $f_0(H)$ satisfies $df_0/dH<0$. If the inequality is reversed then one can get an unstable bump-on-tail mode associated with Landau resonance when the wave phase velocity falls on an interval of $f_0$ with positive slope.

Example: In a uniform plasma with a uniform applied magnetic field, there is gyro-resonance when $\omega - k_\parallel v_\parallel = \ell\Omega$. Taylor and Hastie show that modes satisfying bounce resonance $\omega = \ell\omega_b(\mu,H)$ are stable if $\partial F_0/\partial H < 0$ because the waves transfer energy to the particles rather than vice versa with this sign of the derivative of the distribution function.

Given the values of the magnetic moment $\mu$ and the Hamiltonian $H$, the surfaces of constant $J$ in Eq.(8.C.106) contain the drift surfaces. The gradient of $J$ is normal to the drift surfaces:

$$\left.\nabla J\right|_{\mu,H} = \frac{\partial J}{\partial \alpha}\nabla\alpha + \frac{\partial J}{\partial \beta}\nabla\beta \tag{8.C.109}$$

where $\alpha$ and $\beta$ are the Euler potentials. Using the Hamiltonian, one can show after some algebra

$$\mathbf{v}_d = const\, \nabla J \times \hat{\mathbf{b}} \quad \rightarrow \quad m\mathbf{v}_d \equiv \langle m\mathbf{v}_d\rangle_b = \frac{\omega_b}{\langle\Omega\rangle_b}\nabla J \times \hat{\mathbf{b}} \tag{8.C.110}$$

where the average is over the bounce motion. Recall the derivation of the guiding-center drifts in Sec. 8.B.b and Eq.(8.B.29). Consider the limit $\omega \ll \omega_b$; in fact, let $\omega \to 0$. Freeze the bounce action and magnetic moment, and look just at the guiding-center drifts. In the $\alpha\beta$ plane there are nested contours of $H(\alpha,\beta) = \mathcal{E}$. These surfaces are characterized by constant energy, bounce action, and magnetic moment. The parameters of a guiding-center particle consist of $\mu, J, H$ for a given species (which specification sets the mass and charge). In this zero frequency limit, we will not worry about the drift phase, bounce phase, or gyro phase, because $\dot\mu, \dot J, \dot H$ depend on $\frac{\partial}{\partial\theta_d}, \frac{\partial}{\partial\theta_b}, \frac{\partial}{\partial\theta_g}$. Thus, the density in phase space better not be dependent on the phases else there will be time dependence: thus $F_0(\mu, J, H)$. Given the Hamiltonian $H$ we can calculate the magnetic flux:

$$\oint \alpha\, d\beta = \int\int d\alpha\, d\beta = \Phi(H) \tag{8.C.111}$$

We can invert Eq.(8.C.111) to obtain $H(\mu, J, \Phi)$. Because of the ignorable phases $\dot\mu, \dot J, \dot H$ all vanish.

Now we relax the constraint of no time dependence and allow for a weak time dependence: $\omega \ll \omega_b$ with $\dot{H} \neq 0$ while $\dot{\mu}, \dot{J}, \dot{\Phi}$ are all zero, i.e., flux is conserved; $J$ is conserved over a bounce; and $\mu$ is conserved over a gyration period. There are some options for how to parameterize the particle motion. We consider the following options: $(\mu, J, H), (\mu, J, \Phi), (\mu, H, \Phi)$, and $(B_{T.P.} = H/\mu, H, \Phi)$ where $B_{T.P.} = v^2/(v_\perp^2/B)$ at the bounce turning point. $\mu$ and $H$ can be measured, but $J$ is not immediately measurable. If we know the symmetry of the configuration and thus can describe the flux surface, then $\Phi$ is measurable. Consider $H(\mu, J, \Phi)$ and

$$\langle \Omega \rangle_{b,d} = \langle \dot{\theta}_g \rangle_{b,d} = \left.\frac{\partial H}{\partial \mu}\right|_{J,\Phi}, \quad \langle \omega_b \rangle_d = \langle \dot{\theta}_b \rangle_{g,d} = \left.\frac{\partial H}{\partial J}\right|_{\mu,\Phi}, \quad \omega_d = \langle \dot{\theta}_d \rangle_{g,b} = \left.\frac{\partial H}{\partial \Phi}\right|_{\mu,J} \tag{8.C.112}$$

with fixed $e, m$ and $c$, where the subscripts for the averaging are defined as $g$=gyro period, $b$=bounce period, and $d$=drift period. In more detail we have

$$H(\mu; J; \alpha, \beta; t), \quad \dot{\beta} = \frac{c}{e}\frac{\partial H}{\partial \alpha}, \quad \dot{\alpha} = -\frac{c}{e}\frac{\partial H}{\partial \beta}, \quad \dot{H} = \frac{\partial H}{\partial t} \tag{8.C.113}$$

where $\alpha$ and $\beta$ are Euler potentials for the magnetic field and serve as labels for the magnetic field lines: $\mathbf{A} = \alpha \nabla \beta$ and $\mathbf{B} = \nabla \times \mathbf{A}$; and

$$H(\mu; \alpha, \beta; \sigma, p_\sigma; t) = \mu B + e\chi + \frac{p_\sigma^2}{2mh^2} \tag{8.C.114}$$

Example: $B = B_0(\alpha, \beta)\left(1 + \frac{\sigma^2}{2L^2(\alpha, \beta)}\right)$, and we choose $\chi = 0$ and $h=1$ for simplicity. Hence,

$$H(\mu; \alpha, \beta; \sigma, p_\sigma; t) = \frac{p_\sigma^2}{2mh^2} + \mu B_0(\alpha, \beta) + \frac{\frac{1}{2}\mu B_0(\alpha, \beta)}{L^2(\alpha, \beta)}\sigma^2 \tag{8.C.115}$$

which has the structure of a harmonic oscillator. From Eq.(8.C.115) we deduce the bounce frequency

$$\omega_b^2 = \frac{\mu B_0(\alpha, \beta)}{mL^2(\alpha, \beta)} \tag{8.C.116}$$

which is a function of the field-line label. The bounce-averaged action is

$$J_b = \frac{1}{2\pi}\oint d\sigma p_\sigma = \frac{H - \mu B_0(\alpha, \beta)}{\omega_b(\alpha, \beta)} = J(H, \alpha, \beta) \tag{8.C.117}$$

With the use of Eq.(8.C.117) and a canonical transformation to action-angle variables we have

$$H(\mu, J, \alpha, \beta) = \mu B_0(\alpha, \beta) + \omega_b(\alpha, \beta, \mu)J \tag{8.C.118}$$

From Eq.(8.C.113) one derives

$$\dot{\beta} = \frac{c}{e}\left[\mu\frac{\partial B_0}{\partial \alpha} + J\frac{\partial \omega_b}{\partial \alpha}\right] \tag{8.C.119}$$

We now postulate a time-independent equilibrium with no electric potential:

$$H_0(\mu; J; \Phi) = H_0(\sigma, p_\sigma; \alpha, \beta; \mu), \quad \dot{\alpha} = 0, \quad \dot{\beta} = \dot{\beta}(\mu, J, \alpha) \tag{8.C.120}$$

The total Hamiltonian including a linear perturbation due to a small-amplitude electrostatic field is

$$H = H_0(\sigma, p_\sigma; \alpha, \beta; \mu) + \delta H = H_0(\sigma, p_\sigma; \alpha, \beta; \mu) + e\delta\phi(\alpha, \beta, \sigma, t) \quad (8.C.121)$$

The bounce action is

$$J_b = \frac{1}{2\pi}\oint d\sigma\, p_\sigma = \frac{1}{2\pi}\oint d\sigma \sqrt{2m(H - \mu B - e\phi)}$$

$$\rightarrow\; 0 = \delta J = \frac{1}{2\pi}\oint \frac{d\sigma}{\dot\sigma}(\delta H - e\delta\phi) \quad \therefore\; \delta H = e\langle\delta\phi\rangle_b (\alpha, \beta; \mu, J) \quad (8.C.122)$$

for $\omega \ll \omega_b$. The perturbed equations of motion lead to

$$\dot\alpha = -c\frac{\partial\langle\delta\phi\rangle_b}{\partial\beta} \quad \dot\beta = \frac{c}{e}\frac{\partial H_0}{\partial\alpha} + \ldots = \dot\beta_0 + \delta\dot\beta \quad (8.C.123)$$

The linearized Vlasov equation for the guiding center distribution function

$$F(\mu; J; \alpha, \beta; t) = F_0(\mu, J, \alpha) + \delta F(\mu, J, \alpha) e^{i\ell\beta - i\omega t} \quad (8.C.124)$$

is then

$$\frac{dF}{dt} = \frac{\partial F}{\partial t} + \dot\alpha\frac{\partial F}{\partial\alpha} + \dot\beta\frac{\partial F}{\partial\beta} = \frac{\partial \delta F}{\partial t} + \delta\dot\alpha\frac{\partial F_0}{\partial\alpha} + \dot\beta_0\frac{\partial \delta F}{\partial\beta}$$

$$= -i\omega\delta F + \delta\dot\alpha\frac{\partial F_0}{\partial\alpha} + \dot\beta_0 i\ell\delta F \quad (8.C.125)$$

Hence,

$$\delta F = \frac{-\delta\dot\alpha\frac{\partial F_0}{\partial\alpha}}{-i\omega + i\ell\dot\beta_0(\mu, J, \alpha)} = \frac{i\ell c\langle\delta\phi(\alpha,\sigma)\rangle_\sigma^s \frac{\partial F_0}{\partial\alpha}(\mu, J, \alpha)}{-i\omega + i\ell\dot\beta_0(\mu, J, \alpha)} \quad (8.C.126)$$

using

$$\delta\phi = \delta\phi(\alpha,\sigma)e^{i\ell\beta - i\omega t} \quad \dot\alpha = -c\frac{\partial\langle\delta\phi\rangle_b}{\partial\beta} \quad (8.C.127)$$

We next construct the density of guiding centers by integrating the distribution function:

$$N(\alpha, \beta, \sigma; t) = \int d\mu \int dp_\sigma\, F(\alpha, \beta; \sigma, p_\sigma; \mu; t)$$

$$= -\int d\mu \int p_\sigma\, dF = -\int d\mu \int dJ\, p_\sigma\frac{\partial F}{\partial J} \quad (8.C.128)$$

with an integration by parts. The linearization of Eq.(8.C.128) yields

$$\delta N(\alpha, \beta, \sigma; t) = -\int d\mu \int dJ \left(\delta p_\sigma\frac{\partial F_0}{\partial J} + p_\sigma^{(0)}\frac{\partial \delta F}{\partial J}\right)_s \quad (8.C.129)$$

where $p_\sigma^{(0)} = \sqrt{2m[H_0(\mu, J, \alpha) - \mu B(\alpha, \beta, \sigma) - e\phi^{(0)}(\alpha, \beta, \sigma)]}$ and remember the species labels: $\delta N = \sum_s \delta N_s$. Given that the particle number density is related to the guiding-center number density by $n = N + O(\frac{\rho}{\lambda})^2$, so $n \approx N$; and the quasi-neutrality condition becomes

$$\delta N_e = \delta N_i \qquad (8.C.130)$$

for singly charged ions. The quasi-neutrality condition is equivalent to Poisson's equation for $\rho \ll \lambda$. Using Eqs.(8.C.126-8.C.129) to evaluate (8.C.130), we obtain a messy integral equation for $\delta\phi(\sigma)$. $\alpha$ is only a parameter, and $\beta$ appears only through $\ell\dot{\beta}_0$. Equation (8.C.130) is an eigenvalue problem subject to the restriction $\omega \ll \omega_b$. We note that there is a pole in the $\delta F$ term on the right side of Eq.(8.C.126) that corresponds to the drift resonance: $\omega = \ell\dot{\beta}_0$. In the velocity-space integration we use the Landau prescription for the imaginary part of the contribution from the resonance and the principal value for the real part. We cannot directly solve the eigenvalue equation; so we construct a variational principle for it. We remark on the stability characteristics of two types of perturbations.

(i) Perturbations that conserve $\mu, J$: long-wavelength modes $\lambda \gg r_{gyro}$ to conserve $\mu$ and low frequency $\omega \ll \omega_b$ to conserve $J$. Taylor and Hastie[91,92] show

$$\left.\frac{\partial F_0}{\partial H}\right|_{\mu,J} < 0 \qquad (8.C.131)$$

is sufficient for stability and

$$\left.\frac{\partial F_0}{\partial H}\right|_{\Phi,\mu} < 0 \qquad (8.C.132)$$

ensures stability against bounce resonant modes.

Example: In the earth's radiation belt the electron density increases and then decreases. Which edge of the band is unstable? Do the gradients drive instabilities that can cause the aurora? Consider

$$\left.\frac{\partial F_0}{\partial \Phi}\right|_{\mu,J} \equiv \left.\frac{\partial F_0}{\partial H}\right|_{\mu,J} \left.\frac{\partial H}{\partial \Phi}\right|_{\mu,J} \qquad (8.C.133)$$

where

$$\left.\frac{\partial H}{\partial \Phi}\right|_{\mu,J} = -\left.\frac{\partial H}{\partial J}\right|_{\mu,\Phi} \left.\frac{\partial J}{\partial \Phi}\right|_{\mu,H} \qquad \left.\frac{\partial H}{\partial J}\right|_{\mu,\Phi} = \langle\omega_b\rangle_d > 0 \qquad (8.C.134)$$

using the properties of a cyclic chain rule. Hence, from Eqs.(8.C.131), (8.C.133), and (8.C.134) one obtains

$$\left.\frac{\partial F_0}{\partial H}\right|_{\mu,J} < 0 \;\Rightarrow\; \left.\frac{\partial F_0}{\partial \Phi}\right|_{\mu,J} \left.\frac{\partial J}{\partial \Phi}\right|_{\mu,H} > 0 \qquad (8.C.135)$$

for stability. A corollary to Eq.(8.C.135) is

$$dF_0 dJ > 0 \qquad (8.C.136)$$

for stability. With $\delta\Phi > 0$, i.e., moving outward in the magnetosphere, then one wants maximum density where there is a maximum in $J$. This is called "maximum $J$ stability" in the literature. From the relation

$$J = \oint d\sigma \sqrt{H - \mu B} \tag{8.C.137}$$

$J$ being a maximum in a region of space implies that $B$ is a minimum. Hence,

$$\left.\frac{\partial J}{\partial \Phi}\right| \propto -\left\langle\frac{\partial B}{\partial \Phi}\right\rangle_{b,d} \tag{8.C.138}$$

Theorem: Eq.(8.C.135-8.C.138) imply that

$$\max J \Leftrightarrow \min <B> \Leftrightarrow \left.\frac{\partial F_0}{\partial H}\right|_{\substack{\Phi,\mu \\ \mu,J}} < 0 \tag{8.C.139}$$

are equivalent stability conditions.

(ii) Perturbations that conserve $\mu$ but violate $J$ conservation: long-wavelength modes $\lambda \gg r_{gyro}$ to conserve $\mu$ and but with $\omega \gtrsim \omega_b$. No general conclusions are reached for arbitrary distribution functions, but for a so-called Taylor distribution function with two degrees of freedom:

$$F_0(\mu, J, \Phi) = F_0(\mu, H(J, \Phi)) \tag{8.C.140}$$

there is a stability condition

$$\left.\frac{\partial F_0}{\partial H}\right|_\mu < 0 \tag{8.C.141}$$

The Taylor distribution function is useful for applications to mirror confinement because its dependence on $\mu$ facilitates anisotropy. The species number density for a Taylor distribution is

$$N_0^s(\alpha, \beta, \sigma) = \int d\mu \int dp_\sigma F_0^s(\mu, H(\alpha, \beta, \sigma)) = \int d\mu \int \frac{dH}{\dot\sigma} F_0^s(\mu, H(\alpha, \beta, \sigma))$$
$$\to N_0^s(B_0(\alpha, \beta, \sigma)) = \int d\mu \int \frac{dH}{\sqrt{H - \mu B_0}} F_0^s(\mu, H) \tag{8.C.142}$$

In a minimum-$B$ configuration there are nested surfaces of constant density threaded by the magnetic field lines. Iso-field and iso-density surfaces are coincident.

Exercise: (i) Show that $p_\perp^s$ and $p_\parallel^s$ are functions of position only through $B$ for the Taylor distribution. (ii) Show that $N_0$ increases as $B$ decreases.

Nested magnetic surfaces facilitate good confinement. If there is also minimum average $B$, then there is stability as well. The analysis for perturbations that violate both $\mu$ and $J$ conservation is more complicated.[91,92] Returning to consideration of the magnetosphere, with conditions on the inner side of the layer such that $N_0$ is increasing as $B$ is decreasing and with minimum average $B$ there is stability. However, on the outer side of the layer with $N_0$ decreasing as $B$ is increasing there can be instability. This may be a mechanism for the aurora.

**8.D Introduction to the theory of the stability of drift waves**

The theory of drift waves has a very large literature. [*Editor's note: A good review paper that was published a few years after these lectures is Tang's paper,[95] but this review is now forty years old.*] Here we adopt a slab model and a uniform magnetic field $\mathbf{B} = B_0 \hat{\mathbf{z}}$. We assume that the equilibrium quantities vary only in $x$. We also assume that the equilibrium electric field vanishes (or we transform to a frame in which the electric field vanishes). The perturbed fields are assumed to be electrostatic here, i.e.,

$$\nabla \times \delta \mathbf{E} = 0 \quad \delta \mathbf{E} = -\nabla \delta \phi \quad \delta \mathbf{B} = 0 \tag{8.D.1}$$

We further assume $\lambda \gg \rho$ and $\omega \ll \Omega$ so that the magnetic moment is conserved. The equilibrium distribution function is $F_0(\mu, v_z, X)$. The linearly perturbed distribution function is $\delta F(\mu; v_z, z; X, Y; t)$. We choose an electric potential perturbation of the form

$$\delta \phi(X, Y, z; t) = \delta \phi(X) e^{-i\omega t + i k_z z + i k_y Y} \tag{8.D.2}$$

and a perturbed distribution function

$$\delta F = \delta F(\mu; v_z; X) e^{-i\omega t + i k_z z + i k_y Y} \tag{8.D.3}$$

We ignore finite-Larmor-radius (FLR) effects. The distribution function satisfies the collisionless Vlasov equation:

$$\frac{D}{Dt} F(\mu; v_z, z; X, Y; t) = \frac{\partial F}{\partial t} + \dot{z} \frac{\partial F}{\partial z} + \dot{v}_z \frac{\partial F}{\partial v_z} + \dot{X} \frac{\partial F}{\partial X} + \dot{Y} \frac{\partial F}{\partial Y} = 0 \tag{8.D.4}$$

The linearized Vlasov equation is

$$\frac{\partial \delta F}{\partial t} + v_z \frac{\partial \delta F}{\partial z} + \dot{v}_z \frac{\partial F_0}{\partial v_z} + \dot{X} \frac{\partial F_0}{\partial X} + \dot{Y} \frac{\partial F_0}{\partial Y} = \frac{\partial \delta F}{\partial t} + v_z \frac{\partial \delta F}{\partial z} + \dot{v}_z \frac{\partial F_0}{\partial v_z} + \dot{X} \frac{\partial F_0}{\partial X} = 0 \tag{8.D.5}$$

with solution

$$\delta F(v_z, X, \mu) = \frac{e}{m} \frac{\delta \phi}{v_z - \frac{\omega}{k_z}} \left( \frac{\partial}{\partial v_z} + \frac{1}{\Omega} \frac{k_y}{k_z} \frac{\partial}{\partial x} \right) F_0(X, v_z, \mu) \tag{8.D.6}$$

where $\Omega_s \equiv e_s B / m_s c$ with signs ($\Omega_e < 0$). We note the Landau-type denominator in Eq.(8.D.6). We use the quasi-neutrality condition or Poisson's equation to determine the electric potential self-consistently and to obtain dispersion relations:

$$n_s(x) = \int d\mu \int dv_z F_s(v_z, x, \mu) \tag{8.D.7}$$

In Eq.(8.D.7) we have dispensed with the distinction between $x$ and $X$ because we are ignoring FLR effects. The integral in $v_z$ in Eq.(8.D.7) will lead to a Hilbert transform and a Landau pole.

<u>Definition:</u>
$$Z^s \left( V_z = \frac{\omega}{k_z}, x \right) n_0(x) \equiv \int d\mu \int dv_z \frac{F_0^s(v_z, x, \mu)}{v_z - V_z} \tag{8.D.8}$$

---

[95] W. M. Tang, Nucl. Fusion **18**, 1089 (1978).

With the definition in Eq.(8.D.8) the linearized perturbed number density deduced from the velocity-space integral of Eq.(8.D.6) is

$$\delta n^s(x) = \frac{e_s}{m_s}\delta\phi(x)\left\{n_0(x)\frac{\partial}{\partial V_z}Z^s(V_z,x) + \frac{k_y}{k_z}\frac{1}{\Omega}\frac{\partial}{\partial x}\left[Z^s(V_z,x)n_0(x)\right]\right\}$$
$$= \frac{e_s}{m_s}\delta\phi(x)\left\{\frac{\partial}{\partial V_z} + \frac{k_y}{k_z}\frac{1}{\Omega}\frac{\partial}{\partial x}\right\}Z^s(V_z,x)n_0(x) \quad (8.D.9)$$

If we use Eq.(8.D.9) and invoke quasi-neutrality for $\omega \ll \omega_p$ and $\lambda \gg \lambda_D, r_{gyro}$, $\delta n^e(x) = \delta n^i(x)$, we obtain a dispersion relation for $\omega(k_z, k_y, x)$ for the frequency of a quasi-mode. This is a WKB-like expression for the local frequency of the quasi-mode. From Eq.(8.D.9) and the quasi-neutrality relation, we can form

$$\frac{\delta n^e(x)}{\delta\phi(x)} = \chi^e(x) = \frac{\delta n^i(x)}{\delta\phi(x)} = \chi^i(x) \quad (8.D.10)$$

from which we determine a local dispersion relation for $\omega(k_z, k_y, x)$. From this local dispersion relation one can deduce the group velocity for wave packets:

$$\mathbf{V}^g \equiv \left.\frac{\partial \omega}{\partial \mathbf{k}}\right|_x = \hat{\mathbf{y}}\frac{\partial \omega}{\partial k_y} + \hat{\mathbf{z}}\frac{\partial \omega}{\partial k_z} \quad (8.D.11)$$

Thus, the wave packets move in the surface containing the drifts. For shorter wavelength modes there is also a component of the group velocity across the slab in $x$. Corrections to the local dispersion relation and the group velocity are order $(k\lambda_D)^2 \ll 1$.

Next consider the role of Landau damping. For $v_{th} \sim V_z$ there is strong Landau damping. There are two limiting cases of interest in evaluating the Hilbert transform of a Gaussian unperturbed velocity distribution function.

Example: $v_{th} \ll V_z$ then $Z_{Gaussian} \to -\frac{1}{V_z} - \frac{v_{th}^2}{V_z^3} + ...,\quad v_{th}^2 \equiv \langle v_z^2 \rangle$

Example: $v_{th} \gg V_z$ then

$$\frac{\partial}{\partial V_z}Z_{Gaussian} \to -\frac{1}{v_{th}^2} + i\pi g'(V_z) \;\to\; Z_{Gaussian} \to O\left(\frac{V_z}{v_{th}^2}\right) + i\pi g(V_z),\quad g = \frac{1}{n_0}\int d\mu F_0, k_z > 0$$

Example: $v_{th,e} \gg V_z \gg v_{th,i}$ then selecting the dominant terms results in an electron response

$$\frac{\delta n_e(x)}{n_0(x)} = \frac{e\delta\phi(x)}{T_e},\quad T_e \equiv m_e v_{th,e}^2$$
$$\to n_e(x) = n_0 e^{\frac{e\phi}{T_e}} \quad (8.D.12)$$

We note that Eq.(8.D.12) implies that the results of the Vlasov equation agree with the Boltzmann equation including collisions, but this is somewhat of a coincidence and nothing profound. The ion response includes ion inertia, and the quasi-neutrality relation leads to

$$\omega^2 - \omega k_y u_*^e(x) = k_z^2 c_s^2(x), \quad c_s^2 = \frac{T_e(x)}{m_i}, \quad u_*^e(x) \equiv -\frac{cT_e(x)}{|e|B}\frac{d}{dx}\ln n_0(x) \quad (8.D.13)$$

where $u_*^e(x)$ is called the electron diamagnetic flow velocity. One can derive the diamagnetic flow velocity from consideration of force balance on the electrons:

$$\nabla p_e = \frac{1}{c}\mathbf{j}^e \times \mathbf{B} \quad \to \quad \hat{\mathbf{x}}\frac{d}{dx}nT_e = -\frac{1}{c}en\mathbf{u}^e \times \mathbf{B} \quad \to \quad \mathbf{u}^e = ... \quad (8.D.14)$$

The local dispersion relation in Eq.(8.D.13) leads to the following results in specific limits:

$$\omega = \pm k_z c_s + \tfrac{1}{2} k_y u_*^e \quad |k_z| \to \infty, \quad \omega \to 0, k_y u_*^e \quad |k_z| \to 0 \quad (8.D.15)$$

For propagation parallel to $\hat{\mathbf{y}}$, in the electron frame moving with $\mathbf{u}_*^e$ the wave has zero frequency, i.e., an electron drift wave. There is no comparable wave carried by the ions. The interpretation is that the pressure gradient accelerates or retards the ion acoustic wave and morphs it into a drift wave.

Including the Landau resonance effects contributes an imaginary part to $Z$ and $Z'$, i.e., $Z = ReZ + i\pi g$ and $Z' = ReZ' + i\pi g'$. We calculate the growth rates in the limit $v_{th,e} \gg V_z \gg v_{th,i}$ and using

$$\varepsilon = \varepsilon' + i\varepsilon'' = 0 \quad \to \quad \gamma \approx -\frac{\varepsilon''}{\frac{\partial \varepsilon'}{\partial \omega}}, \quad \omega\frac{\partial \varepsilon'}{\partial \omega} > 0 \quad (8.D.16)$$

The waves are positive energy. Consider two limits.

(i) Assume $\dfrac{dT_e}{dx} \neq 0$ and $u_z^e = 0$ (no current along magnetic field lines). After re-evaluating the local dispersion relation including imaginary parts of the electron response function in the small $k_z$ limit with $V_z^2 \ll 2v_e^2$, one obtains

$$\frac{\gamma}{\omega} = -\sqrt{\frac{\pi}{2}}\frac{V_z}{v_{th,e}}\left\{\tfrac{1}{2}\frac{d\ln T_e}{d\ln n} + \frac{V_z^2}{2v_{th,e}^2}\left(1 - \tfrac{1}{2}\frac{d\ln T_e}{d\ln n}\right)\right\} \ll 1, \quad \omega \sim k_y u_*^e \quad (8.D.17)$$

Equation (8.D.17) implies instability for

$$\frac{d\ln T_e}{d\ln n} < 0 \quad \to \quad \gamma > 0 \quad (8.D.18)$$

(ii) Assume $\dfrac{dT_e}{dx} = 0$ and $u_z^e \neq 0$. The local dispersion relation yields

$$\frac{\gamma}{\omega} = \sqrt{\frac{\pi}{2}}\left[\frac{u_z^e}{v_{th,e}} - \tfrac{1}{2}\left(\frac{V_z}{v_{th,e}}\right)^3\right] \quad (8.D.19)$$

and there is instability for

$$\frac{u_z^e}{v_{th,e}} > \frac{1}{2}\left(\frac{V_z}{v_{th,e}}\right)^3 \rightarrow \gamma > 0 \tag{8.D.20}$$

When ion Landau damping is included there is a threshold condition:

$$u_z^e > \frac{m_e}{m_i} v_{th,i} \tag{8.D.21}$$

This is a very small threshold for the electron flow velocity along the field line to lead to an unstable drift wave propagating across the field lines. There is a significant cancellation of the Landau damping in this limit.

Exercise: Derive the results in Eqs.(8.D.17-21).

There are many possible refinements to the theory of drift-wave instability, and there is a large literature.[94,96] Some examples of refinements are as follows.

1. Allow short wavelengths $\lambda \sim \rho_i$ and $\dot{\mu}_i \neq 0$. $\nabla n \neq 0$ leads to instability.
2. Shear in the magnetic field as a function of $x$ moving across the slab is strongly stabilizing due to ion Landau damping.
3. Collisions lead to new instabilities.
4. Electromagnetic extension $\delta \mathbf{B} \neq 0$ leads to new electromagnetic instabilities, e.g., drift-Alfvén waves.
5. Curvature of the magnetic field lines introduces magnetically trapped particles and new possible instabilities.
6. $\nabla_\perp B \neq 0$ may lead to new instabilities?
7. Higher frequency modes, e.g., $\omega \sim \Omega_i$, allows for gyro-resonance effects and drift cyclotron instability.
8. Higher frequency modes, e.g., $\omega \sim \Omega_i$, and a loss-cone distribution can lead to drift-cone instability.

Taylor and Hastie[91,92] have examined some of the physics issues in nos. 1, 2, 5, and 6.

We next present an example of how local theory can be extended to a nonlocal theory from which normal mode frequencies can be calculated. Consider the inclusion of space-charge effects by using Poisson's equation instead of quasi-neutrality:

$$-\nabla^2 \delta\phi = -\left(\frac{d^2}{dx^2} - k_y^2 - k_z^2\right)\delta\phi = 4\pi e\left(\delta n_i - \delta n_e\right) \tag{8.D.22}$$

We construct a WKBJ solution to Eq.(8.D.22) for a pure drift wave ($k_z=0$):

---

[96] AB Mikhaĭlovskiĭ, *Theory of Plasma Instabilities: Instabilities of a homogeneous plasma,* Consultants Bureau, 1974.

$$\delta\phi(x) = \frac{1}{\sqrt{k_x(x)}} e^{\pm i \int_{x_i}^{x} dx' k_x(x')}, \quad k_x(\omega, k_y, k_z = 0, x) = k_D^2(x)\left(\frac{k_y u_*^e}{\omega} - 1\right) - k_y^2 \qquad (8.D.23)$$

Next we use the Bohr-Sommerfeld prescription for calculating normal modes:

$$\int_{x_1}^{x_2} dx\, k_x(\omega, k_y, x) = (\ell + \tfrac{1}{2})\pi, \quad \ell = 0,1,2,\dots \quad \to \quad \omega_\ell(k_y, k_z = 0) \qquad (8.D.24)$$

where $x_1$ and $x_2$ are two turning points: $k_x(\omega, k_y, x_{1,2}) = 0$, and $\omega_\ell(k_y, k_z)$ is the normal mode frequency. When the size of the slab is long compared to the typical wavelengths, the set of discretely space $\omega_\ell$ is so finely spaced as to approach a continuum.

[*Editor's note: There is clearly much more that could be said about both drift waves and non-uniform plasmas in general. However, the series of lectures ended at this point.*]

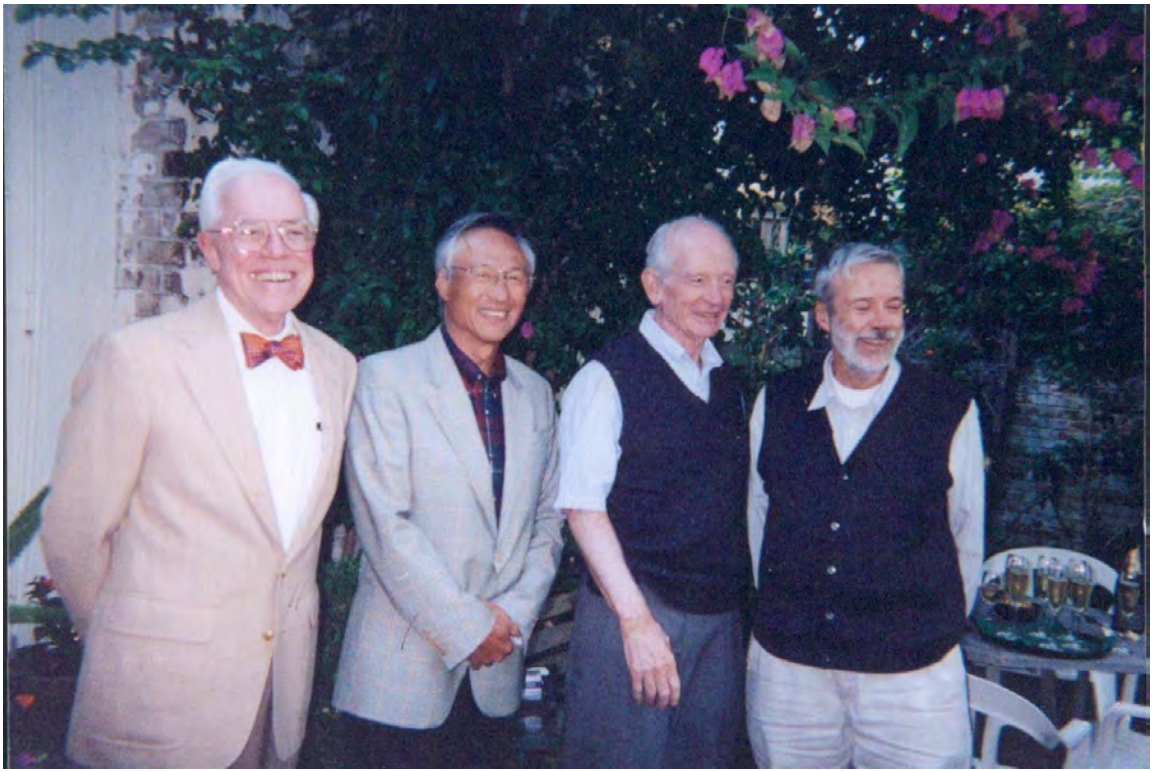

Allan Kaufman, Akira Hasegawa, Ned Birdsall, and Allan Lichtenberg